\documentclass[11pt, twoside, openright, usenames,dvipsnames]{scrbook}
\pdfoutput=1
\usepackage[english,german,ngerman]{babel}
\usepackage[a4paper]{geometry}
\usepackage{latexsym}
\usepackage{amssymb} 
\usepackage{amsmath}
\usepackage{array}
\usepackage{eucal}
\usepackage{mathrsfs}
\usepackage{makeidx}
\usepackage{graphicx}
\usepackage{subfigure}
\usepackage{float}
\usepackage{framed}
\usepackage{mdframed}
\usepackage{booktabs}
\usepackage{multirow}
\usepackage{tcolorbox}
\usepackage{lineno}
\usepackage{atlasphysics}
\usepackage{xcolor}
\usepackage{url}
\usepackage{comment}
\usepackage[figuresright]{rotating}
\usepackage{booktabs}
\usepackage{multirow}
\usepackage{float}
\usepackage{cite}
\usepackage{dsfont}
\usepackage{braket}
\usepackage{pdfpages}
\usepackage[titletoc]{appendix}
\usepackage[Bjornstrup]{fncychap}
\usepackage{siunitx}
\usepackage{dcolumn}
\usepackage[nottoc]{tocbibind}
\usepackage{imakeidx}
\usepackage{epigraph}

\colorlet{partbgcolor}{gray!30}
\colorlet{partnumcolor}{gray}
\colorlet{chapbgcolor}{gray!30}
\colorlet{chapnumcolor}{gray}

\renewcommand*\partformat{
  \fontsize{76}{80}\usefont{T1}{pzc}{m}{n}\selectfont
  \hfill\textcolor{partnumcolor}{\thepart}}

\makeatletter
\renewcommand*{\@part}{}
\def\@part[#1]#2{
  \ifnum \c@secnumdepth >-2\relax
    \refstepcounter{part}
    \@maybeautodot\thepart
    \addparttocentry{\thepart}{#1}
  \else
    \addparttocentry{}{#1}
  \fi
  \begingroup
    \setparsizes{\z@}{\z@}{\z@\@plus 1fil}\par@updaterelative
    \raggedpart
    \interlinepenalty \@M
    \normalfont\sectfont\nobreak
    \setlength\fboxsep{0pt}
    \colorbox{partbgcolor}{\rule{0pt}{40pt}
    \makebox[\linewidth]{
    \begin{minipage}{\dimexpr\linewidth+20pt\relax}
      \ifnum \c@secnumdepth >-2\relax
        \vskip-25pt
        \size@partnumber{\partformat}
      \fi      
      \vskip\baselineskip
      \hspace*{\dimexpr\myhi+10pt\relax}
      \parbox{\dimexpr\linewidth-2\myhi-20pt\relax}{\raggedleft\LARGE#2\strut}
      \hspace*{\myhi}\par\medskip
    \end{minipage}
      }
    }
    \partmark{#1}\par
  \endgroup
  \@endpart
}

\renewcommand\DOCH{
  \settowidth{\py}{\CNoV\thechapter}
  \addtolength{\py}{-10pt}
  \fboxsep=0pt
  \colorbox{chapbgcolor}{\rule{0pt}{40pt}\parbox[b]{\textwidth}{\hfill}}
  \kern-\py\raise20pt
  \hbox{\color{chapnumcolor}\CNoV\thechapter}\\
}

\makeatother

\colorlet{chapbgcolor}{Cerulean}
\colorlet{chapnumcolor}{black}

\definecolor{shadecolor}{rgb}{1, 0.8, 0.3}
\usepackage[colorlinks=true, pdfstartview=FitV, linkcolor=blue, 
           citecolor=blue, urlcolor=blue]{hyperref}

\def\ttbar{\ensuremath {t\bar{t}}}
\def\com{centre-of-mass}
\def\mtop{\ensuremath {m_{t}~}}

\newcommand{\dphi}{\ensuremath{\Delta \phi}}
\newcommand{\dphidQ}{\ensuremath{\Delta \phi(l, d)}}
\newcommand{\dphibQ}{\ensuremath{\Delta \phi(l, b)}}

\newcommand{\etmiss}{\ensuremath{\et^{\textrm{miss}}}}
\newcommand{\sumet}{\ensuremath{\Sigma{E_T}}}

\newcommand{\ejets}{\ensuremath{e+\textrm{jets}}}
\newcommand{\mujets}{\ensuremath{\mu+\textrm{jets}}}
\newcommand{\ljets}{\ensuremath{\ell+\textrm{jets}}}
\newcommand{\fsm}{\ensuremath{f^{\textrm{SM}}}}

\newcommand{\Wmt}{\ensuremath{W_{m^{T}}}}

\newcommand{\ptmax}{\ensuremath{\textrm{P}_{T}~\text{Max}}}
\newcommand{\dQ}{down-type quark}
\newcommand{\bQ}{$b$-quark}
\newcommand{\bjet}{$b$-jet}
\newcommand{\btag}{$b$-tag}

\newcommand{\bhad}{b_{\mathrm{had}}}
\newcommand{\blep}{b_{\mathrm{lep}}}
\newcommand{\lqone}{q_1}
\newcommand{\lqtwo}{q_2}
\newcommand{\lep}{l}
\newcommand{\nul}{\nu}
\newcommand{\mw}{m_W}
\newcommand{\gammaw}{\Gamma_W}
\newcommand{\gammatop}{\Gamma_{\mathrm{top}}}
\newcommand{\KLFitter}{{\tt KLFitter}}
\newcommand{\bat}{{\tt BAT}}
\newcommand{\mvone}{{\tt MV1}}
\newcommand{\etcone}{{\tt EtCone20}}
\newcommand{\ptcone}{{\tt PtCone30}}

\newcommand{\mcatnlo}{{\tt MC@NLO}}
\newcommand{\powheg}{{\tt POWHEG}}
\newcommand{\pythia}{{\tt PYTHIA}}
\newcommand{\herwig}{{\tt HERWIG}}
\newcommand{\jimmy}{{\tt JIMMY}}
\newcommand{\acermc}{{\tt AcerMC}}
\newcommand{\alpgen}{{\tt ALPGEN}}

\newcommand{\intlumi}{$4.6$~fb$^{-1}$}
\newcommand{\intlum}{\ensuremath{\int \mathcal{L}\,dt}}
\newcommand{\ttbarxsec}{\ensuremath{\sigma_{t\bar{t}}}}
\newcommand{\radlength}{\ensuremath{\textrm{X}_0}}

\newcommand{\newword}[1] {\textit{#1}\index{#1}}
\newcommand{\HRule}{\rule{\linewidth}{0.5mm}}

\makeatletter
\g@addto@macro\bfseries{\boldmath}
\makeatother

\title{Measurement of Spin Correlations in \ttbar\ Events From $pp$ Collisions at \rts = 7 TeV in the Lepton+Jets Final State with the ATLAS Detector Using a Kinematic Likelihood Fit}
\author{Boris Lemmer}

\makeindex[intoc]

\begin{document}

\frontmatter
\begin{titlepage}
\begin{center}
\begin{otherlanguage}{german}

\HRule \\[0.4cm]
{\Large \bfseries Measurement of Spin Correlations in $t\bar{t}$ Events from $pp$ Collisions at $\sqrt{s}$ = 7 TeV in the Lepton + Jets Final State with the ATLAS Detector \\[0.4cm] }

\HRule \\[1.5cm]
{\large
Dissertation\\[1.0cm]
zur Erlangung des mathematisch-naturwissenschaftlichen Doktorgrades\\
\glqq Doctor rerum naturalium\grqq\\
der Georg-August-Universit\"at G\"ottingen\\[1.0cm]
im Promotionsprogramm ProPhys\\
der Georg-August University School of Science (GAUSS)\\[2.0cm]
vorgelegt von\\[1.0cm]
Boris Lemmer\\[0.5cm]
aus Gie{\ss}en\\[2.0cm]

\vfill
G\"ottingen, 2014
}

\end{otherlanguage}
\end{center}

\end{titlepage}
\thispagestyle{empty}
{\small
\vfill
\noindent
\underline{Betreuungsausschuss}\\[0.2cm]
Prof. Dr. Ariane Frey\\
Prof. Dr. Kevin Kr\"oninger\\
Prof. Dr. Arnulf Quadt\\[1.0cm]

\noindent
\underline{Mitglieder der Pr\"ufungskommission:}\\[0.2cm]
\begin{tabular}{ll}
Referent: & Prof. Dr. Arnulf Quadt\\
{}&{\footnotesize II. Physikalisches Institut, Georg-August-Universit\"at G\"ottingen}\\[0.2cm]
Koreferentin: &Jun.-Prof. Dr. Lucia Masetti\\
{}&{\footnotesize Institut f\"ur Physik/ETAP, Johannes Gutenberg-Universit\"at Mainz}\\[0.2cm]
2. Koreferentin: &Prof. Dr. Ariane Frey\\
{}&{\footnotesize II. Physikalisches Institut, Georg-August-Universit\"at G\"ottingen}\\
\end{tabular}\\[1.0cm]

\noindent
\underline{Weitere Mitglieder der Pr\"ufungskommission:}\\[0.2cm]
PD Dr. J\"orn Grosse-Knetter\\
{\footnotesize II. Physikalisches Institut, Georg-August-Universit\"at G\"ottingen}\\[0.2cm]
Prof. Dr. Hans Hofs\"ass\\
{\footnotesize II. Physikalisches Institut, Georg-August-Universit\"at G\"ottingen}\\[0.2cm]
Prof. Dr. Wolfram Kollatschny\\
{\footnotesize Institut f\"ur Astrophysik, Georg-August-Universit\"at G\"ottingen}\\[0.2cm]
Jun.-Prof. Dr. Steffen Schumann\\
{\footnotesize II. Physikalisches Institut, Georg-August-Universit\"at G\"ottingen}\\[1.0cm]
}
Tag der m\"undlichen Pr\"ufung: 10.07.2014\\[1.0cm]

\noindent
Referenz: II.Physik-UniG\"o-Diss-2014/02

\clearpage
\thispagestyle{empty}
\vspace{20em}
\selectlanguage{ngerman}
\setlength{\epigraphwidth}{.6\textwidth}
\epigraph{\glqq So eine Arbeit wird eigentlich nie fertig,\\
man mu{\ss} sie f\"ur fertig erkl\"aren,\\
wenn man nach Zeit und Umst\"anden\\
das m\"oglichste getan hat.\grqq}%
{\textit{Goethe} }

\selectlanguage{english}

\cleardoublepage
\thispagestyle{empty}
\begin{center}

\HRule \\[0.2cm]
{\large \bfseries Measurement of Spin Correlations in $t\bar{t}$ Events from $pp$ Collisions at $\sqrt{s}$ = 7 TeV in the Lepton + Jets Final State with the ATLAS Detector \\[0.2cm] }

\HRule \\[1.0cm]

\textbf {Abstract}\\[1.0cm]
{\small
The top quark decays before it hadronises. Before its spin state can be changed in a process of strong interaction, it is directly transferred to the top quark decay products.
The top quark spin can be deduced by studying angular distributions of the decay products. The Standard Model predicts the top/anti-top quark (\ttbar) pairs to have correlated spins. The degree is sensitive to the spin and the production mechanisms of the top quark. Measuring the spin correlation allows to test the predictions. New physics effects can be reflected in deviations from the prediction. In this thesis the spin correlation of \ttbar\ pairs, produced at a \com\ energy of $\sqrt{s} = 7\,\TeV$ and reconstructed with the ATLAS detector, is measured. The dataset corresponds to an integrated luminosity of \intlumi. 
\ttbar\ pairs are reconstructed in the \ljets\ channel using a kinematic likelihood fit offering the identification of light up- and down-type quarks from the $t \rightarrow bW \rightarrow bq\bar{q}'$ decay. The spin correlation is measured via the distribution of the azimuthal angle \dphi\ between two top quark spin analyzers in the laboratory frame. It is expressed as the degree of \ttbar\ spin correlation predicted by the Standard Model, \fsm. The results of
\begin{align*}
&\fsm ( \dphi (\text{charged lepton, \dQ} )) &= 1.53 \pm 0.14\,\text{(stat.)} \pm 0.32\,\text{(syst.)},  \\ 
&\fsm ( \dphi(\text{charged lepton, \bQ} )) &= 0.53 \pm 0.18\,\text{(stat.)} \pm 0.49\,\text{(syst.)},  \\ 
&\fsm ( \dphi(\text{combined})) &= 1.12 \pm 0.11\,\text{(stat.)} \pm 0.22\,\text{(syst.)},  
\end{align*}
are consistent with the Standard Model prediction of $\fsm = 1.0$. }
\end{center}

\cleardoublepage

\cleardoublepage
\thispagestyle{empty}
\begin{center}

\HRule \\[0.2cm]
{\large \bfseries Messung von Spin-Korrelationen in $t\bar{t}$-Ereignissen aus $pp$-Kollisionen bei $\sqrt{s}$ = 7 TeV im Lepton + Jets Endzustand mit dem ATLAS Detektor\\[0.2cm] }

\HRule \\[1.0cm]

\textbf {Zusammenfassung}\\[1.0cm]
{\small
Das Top-Quark zerf\"allt, bevor es hadronisiert. Bevor die Spin-Konfiguration des Top-Quarks durch Prozesse der Starken Wechselwirkung ge\"andert werden kann, wird sie direkt an die Zerfallsprodukte des Top-Quarks weitergegeben. R\"uckschl\"usse auf den Spin des Top-Quarks k\"onnen \"uber Winkelverteilungen der Zerfallsprodukte gezogen werden. Die Spins von Top-/Anti-Top-Quark (\ttbar) Paaren sind, gem\"a{\ss} der Vorhersage durch das Standardmodell, korreliert. Der Grad der Korrelation ist sensitiv auf den Spin und die Produktionsmechanismen des Top-Quarks. Die Messung der Spin-Korrelation bietet einen Test der Vorhersagen. Effekte von Physik jenseits des Standardmodells k\"onnen sich in Abweichungen der vorhergesagten Spin-Korrelation manifestieren. In dieser Arbeit wird die Spin-Korrelation von Top-Quark Paaren, die bei einer Schwerpunktsenergie von  $\sqrt{s} = 7~\TeV$ produziert und mit dem ATLAS Detektor rekonstruiert wurden, gemessen. Der Datensatz entspricht einer integrierten Luminosit\"at von \intlumi. Die Top-Quarks wurden im Lepton+Jets Zerfallskanal mittels eines kinematischen Likelihood-Fits, der eine Trennung der leichten up- und down-Typ Quarks aus dem $t \rightarrow bW \rightarrow bq\bar{q}'$ Zerfall erlaubt, rekonstruiert. 
Die Spin-Korrelation wird \"uber die Verteilung des Azimutalwinkels \dphi\ zwischen zwei Top-Quark Spin-Analysatoren im Laborsystem gemessen. Sie wird als Grad \fsm\ der Spin-Korrelation, wie sie im Rahmen des Standardmodells berechnet wird, angegeben.
Die Messungen ergeben
\begin{align*}
&\fsm ( \dphi( \text{geladenes Lepton, down-Typ Quark} )) &= 1.53 \pm 0.14\,\text{(stat.)} \pm 0.32\,\text{(syst.)},  \\ 
&\fsm ( \dphi(\text{geladenes Lepton, }b\text{-Quark} )) &= 0.53 \pm 0.18\,\text{(stat.)} \pm 0.49\,\text{(syst.)},  \\ 
&\fsm ( \dphi(\text{kombiniert})) &= 1.12 \pm 0.11\,\text{(stat.)} \pm 0.22\,\text{(syst.)}. 
\end{align*}
Die Ergebnisse stimmen mit der Berechnung im Rahmen des Standardmodells, $\fsm = 1.0$, \"uberein.}
\end{center}
\cleardoublepage

\pagenumbering{Roman}
\setcounter{tocdepth}{1}
\tableofcontents

\mainmatter
\pagenumbering{arabic}
\setcounter{page}{1}
\pagestyle{headings}

\chapter[Preface]{Preface}

Curiosity is one of the fundamental driving forces of human kind. Without it, we would not have reached the high level of development in technology and health that we have nowadays and that we do not want to miss. Every little kid is equipped with curiosity and can decide how much it wants to know. Playing the game of asking ``Why is that?'' again and again �will finally end up in asking:  ``What are we made of?'', ``Where do we come from?'' and ``Why is everything working the way it does?''. 

The field of particle physics is addressing these questions. During the last two centuries, the knowledge of the fundamental building blocks of nature has developed rapidly, leading to changing ideas of what is really fundamental.  The current understanding of the elementary particles and their interactions is reflected in the \textit{Standard Model of Particle Physics}\index{Standard Model of Particle Physics} (\textit{SM}\index{SM | see {Standard Model of Particle Physics }}). This theory framework classifies the particles of matter -- the fermions -- in groups of quarks and leptons, and it describes the interactions among them via the exchange of gauge bosons. The power of the Standard Model has been more than just the description of particles and forces that are known so far. It also allows precision tests to check its self-consistency and to search for unknown physics effects. 

Only very few particles are stable and can be observed and analysed in the laboratory. The more massive the particles are, the earlier they decay into lighter ones. During the very first moments after the creation of our universe, the environment of very high energy density allowed a balanced production and decay of such heavy particles. The balance between creation and decay got lost during the expansion and cooling of the universe. Sufficient energy for the creation was no longer available.

Recreation of such very high energy densities in laboratories on earth is possible by accelerating particles, colliding them and using their kinetic energy to recreate massive particles. The more massive they are, the more energy is required. As technology kept evolving, more and more particles of the Standard Model were discovered. 

\begin{figure}[ht]
	\centering
	\subfigure[]{
		\includegraphics[width=0.45\textwidth]{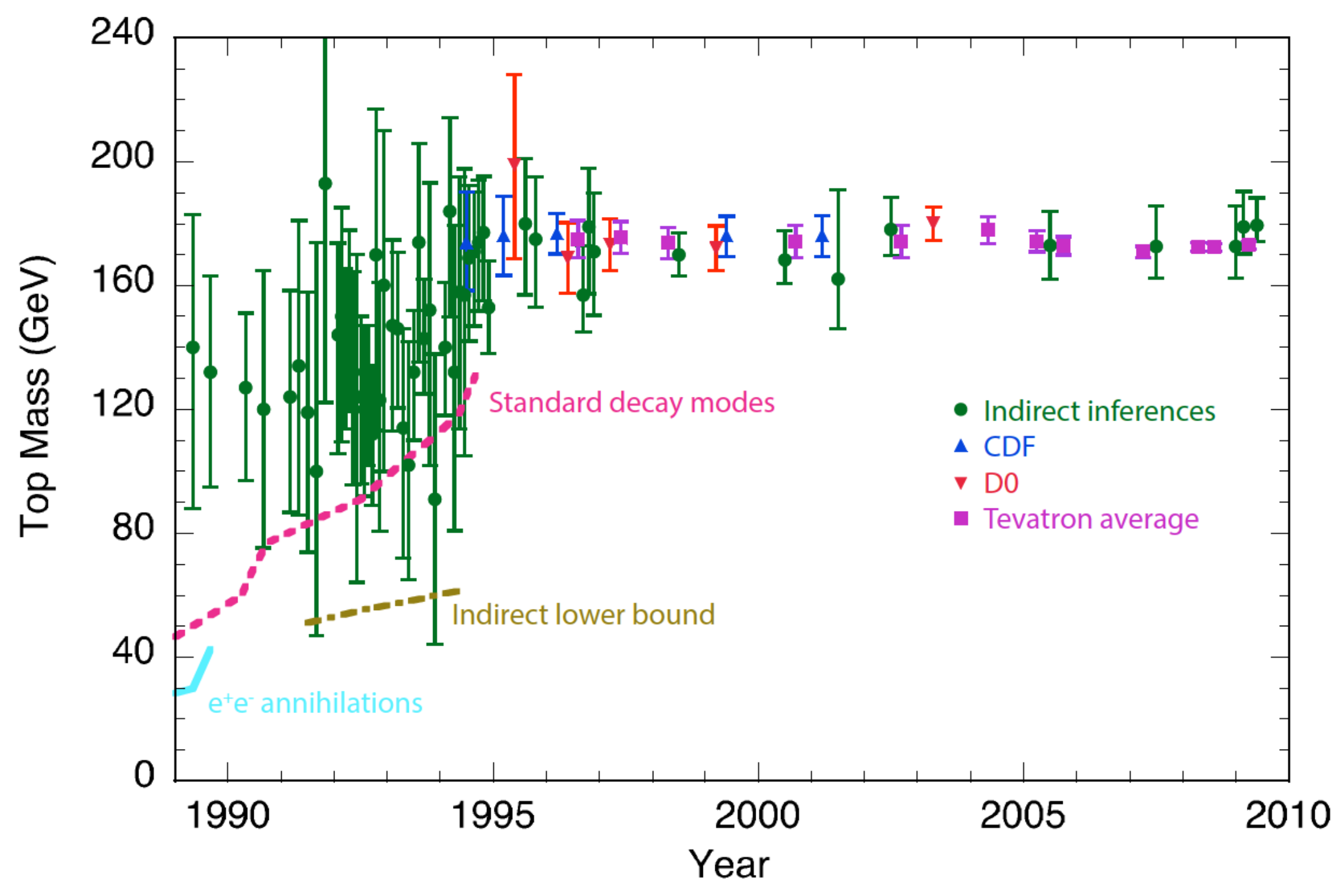}
			\label{fig:topmass_evolution}
		}
		\subfigure[]{
		\includegraphics[width=0.45\textwidth]{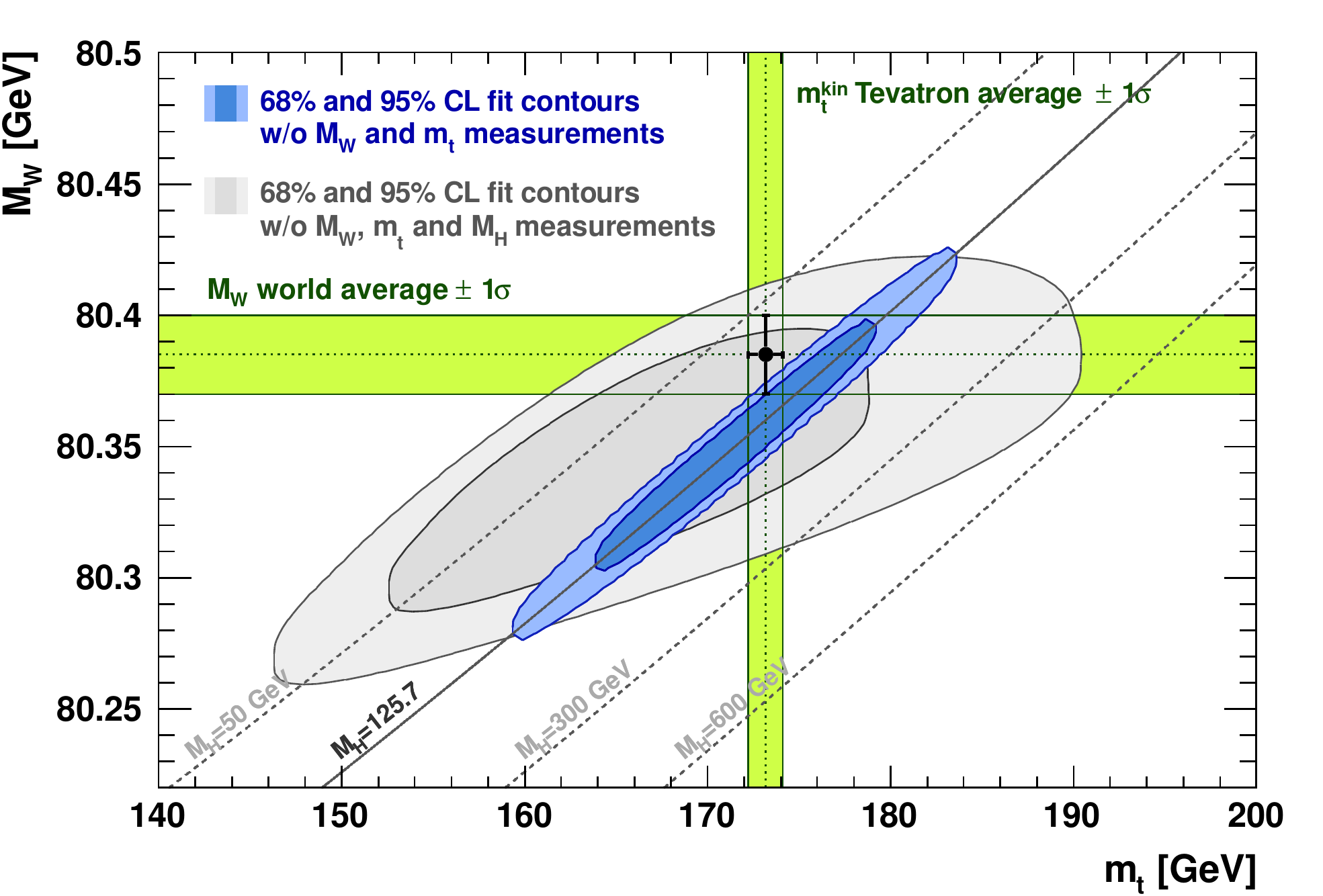}
			\label{fig:mw_vs_mtop}
		}
		\caption{\subref{fig:topmass_evolution} Indirect determinations of the top quark mass via fits to electroweak observables, results of direct measurements as well as lower bounds from direct searches and $W$ boson width analyses \cite{topmass_evolution}. \subref{fig:mw_vs_mtop} Measured masses of the $W$ boson ($M_W$) and the top quark ($m_t$), shown in green bands. These are compared to electroweak fit results excluding the direct measurements of $m_t$ and $M_W$ (blue area) and excluding $m_t$, $M_W$ and the measured Higgs boson mass $M_H$ (grey area) \cite{gfitter}.}
	\label{fig:introduction_indirect}
\end{figure}

Being the most massive of all quarks, the top quark has been discovered as the last missing quark in 1995 by the two experiments D0 and CDF, located at the Tevatron proton/anti-proton accelerator at Fermilab \cite{top_disc_D0, top_disc_CDF}. Before its discovery, the existence was already suggested to complete the third generation of quarks as a partner for the \bQ.
Precision measurements of the parameters of the Standard Model allowed to constrain the top quarks mass. Figure \ref{fig:topmass_evolution} shows the prediction and, after its discovery, the measured mass of the top quark as a function of time. 

Indirect searches and limit settings were not only performed for the top quark. Another important example is the search for the Higgs boson. Before its discovery, the Higgs boson's role in corrections to the masses of the $W$ boson and the top quark ($m_W$, $m_t$) was powerful enough to constrain the Higgs boson mass $m_H$ via electroweak fits. Figure \ref{fig:mw_vs_mtop} shows the directly measured masses of the $W$ boson and the top quark compared to electroweak fit results excluding the direct measurements. Measuring $m_t$ and $m_W$ shows the preferences for certain Higgs boson masses (diagonal lines). 
The particle under study in this thesis is the top quark. As the heaviest of all quarks it offers unique opportunities of physics studies. With a lifetime of about $5 \cdot 10^{-25}~\text{s}$, which is shorter than the time scale of forming bound hadronic systems, \textit{hadronisation}\index{Hadronisation}, the top quark transfers its spin to its decay products before the spin information is diluted. This makes the top quark the only quark whose spin is directly accessible. 

According to the Standard Model, top quarks produced via the strong interaction are almost unpolarized, but have correlated spins. The degree of correlation depends on the initial state of the production and the involved production processes. The degree of correlation which will be measured depends on the decay mechanisms as well. 

In this thesis, the degree of correlation is measured. This addresses the following questions: Does the top quark carry a spin of $\frac{1}{2}$? Does the production of top/anti-top quark (\ttbar) pairs  follow the rules given by the Standard Model? And in particular: Are the spins of top/anti-top quark pairs correlated as they are expected to be? Modifications of the Standard Model due to new physics effects can be reflected in deviations from the predicted spin correlation of \ttbar\ pairs. This allows the analysis presented in this thesis to constrain physics effects beyond the Standard Model in the same way that the masses of the top quark and the Higgs boson were constrained before their discovery. 

It is not only the result and the following conclusions that leave a message. The detailed studies of top quark reconstruction and the impact of systematic uncertainties guide the way to future measurements of the \ttbar\ spin correlation. 

\cleardoublepage

\chapter[Standard Model, Top Quarks and Spin Correlation]{Standard Model, Top Quarks and Spin Correlation}
\label{sec:spincorrelations}
What are we made of? What does the Universe consist of? And why does nature behave as it actually does? Physicists observe nature and analyse the underlying laws and principles. Particle physicists in particular study nature on the elementary level. The actual meaning of ``elementary'' has developed in time. It started with the \textit{elements}\index{Element}, the smallest units of a certain type of matter with unique properties. Dmitri Mendeleev and others started grouping these into the \textit{periodic system of elements}\index{Periodic System of Elements} \cite{periodictable}.
According to the approval of the International Union of Pure and Applied Chemistry (IUPAC) 114 elements are presently known \cite{elements}.\footnote{The discoveries of further elements have been reported, but not yet confirmed.} Along with the search for the truly fundamental building blocks of nature comes the search for underlying symmetries. Not only matter is, in terms of size,  supposed to be fundamental. Laws of nature can also have more fundamental principles. For the latter, the unification of electricity and magnetism to the electromagnetic force serves as an example \cite{maxwell}. Symmetries refer to such unified or more fundamental laws. 

A first important step in the simplification of the set of elements was made by J. J. Thomson who discovered the electron as being a constituent of all atoms \cite{electron}. H. Geiger and E. Marsden made important steps in their scattering experiments \cite{rutherford}, depicting the atom structure as heavy nuclei surrounded by light electrons. These measurements strengthened the idea of W. Prout who found the atomic masses being multiples of the hydrogen atom mass \cite{prout1, prout2}. The picture of atomic nuclei as a composition of hydrogen nuclei objects was established. The only flaw, the neutrality of some of these components, was finally resolved when J. Chadwick discovered the neutron in 1932 \cite{neutron}.  

The set of elementary particles seemed to be reduced from 114 elements to the proton, neutron and electron.\footnote{At the time of the discovery of the neutron, the number of known elements was smaller.} This small set of building blocks of nature did not last very long. 
Not only that the discovery of the positron \cite{positron} introduced anti-particles -- particles with equal masses but quantum numbers such as the electric charge multiplied by $-1$ -- and confused the simple picture of three basic particles. Also, new particles with masses and properties unknown at that time were discovered in studies of cosmic rays \cite{lattes1, lattes2, lattes3, muon}. Today they are known as pions and muons. 

These and further newly discovered particles were ordered by Murray Gell-Mann's \textit{eightfold way} \index{Eightfold way} \cite{eightfold1, eightfold2}. The idea came up that in fact \textit{quarks}\index{Quark}, a new type of particle, are the real fundamental building blocks of which protons, neutrons and several other newly discovered particles, are made of \cite{quarkmodel}. The experimental proof for the theory came along with the results of \textit{deep-inelastic scattering} \index{Deep-inelastic scattering} (\textit{DIS}\index{DIS | see {Deep-inelastic scattering }}) experiments. Results from these scattering experiments with electrons off protons were compatible with a model of point-like constituents, namely the quarks \cite{dis1, dis2, dis3}.

Today we have a consistent set of elementary particles including the building blocks of matter, the \textit{fermions}\index{Fermion}, as well as three of the four fundamental forces\footnote{Gravity is missing in the SM without breaking the self-consistency of the SM as it can be neglected at the mass scale of elementary particles.} and their mediating \textit{gauge bosons}\index{Gauge boson}: the Standard Model. It will be explained in Section \ref{sec:SM}. For a long time the mechanism of mass generation of the bosons and fermions has been a mystery. It was resolved in 2012 by the discovery of the Higgs boson by the ATLAS and CMS experiments \cite{ATLAS_higgs_discovery, CMS_higgs_discovery} confirming the Higgs mechanism\footnote{Even though the same idea was brought up by Brout, Englert, Guralnik, Hagen, Higgs and Kibble at about the same time, the name \textit{Higgs mechanism} has manifested.} \cite{Higgs1, Higgs2, Higgs3, Higgs4, Higgs5, Higgs6} as the process responsible for electroweak symmetry breaking and mass generation. Section \ref{sec:symmetrybreaking} explains this process and highlights the important role of the SM's most massive fermion, the \textit{top quark}\index{Top quark} in its study. 

The production and decay mechanisms of the top quark as well as its discovery and the study of  most of its properties are explained in Section \ref{sec:top}. A property of each elementary particle is its spin. During the production and decay of particles the spin information is propagated according to the rules of the conservation of angular momentum. The knowledge of the spin configuration of a final state demands the knowledge of the initial state, its spin configuration and the whole dynamics of the scattering process. Hence, measuring the spin configurations and comparing them to the predictions made by the SM leads to a validation of the latter one or to necessary extensions. The fact that the top quark is the only quark whose spin configurations can be probed directly and the way how a corresponding measurement can be realized is explained in Section \ref{sec:SC}. 

The measurement of the spin correlation of  top and anti-top quark pairs might indicate physics beyond the SM (BSM) in case of observing an incompatibility between prediction and measurement. Possible BSM scenarios and their effects on the \ttbar\ spin correlation are discussed in Section \ref{sec:BSM}. 
As the \ttbar\ spin correlation depends on the kinematics of the production process, variations of the initial state composition and its kinematics change the predicted correlation. Thus, measurements presented at the \newword{Tevatron} \cite{tevatron} collider and its two experiments \newword{D0} \cite{D0} and \newword{CDF} \cite{CDF} are complementary to the measurements at the LHC. The results of \ttbar\ spin correlation measurements at both the Tevatron and the LHC will be presented in Section \ref{sec:recentresults}. 

At the end of this chapter the reader is equipped with all necessary information about the motivation and the idea of a measurement of the \ttbar\ spin correlation.

\section{The Standard Model of Particle Physics}
\label{sec:SM}
The Standard Model of particle physics contains the present knowledge about elementary particles and their interactions. Fermions as matter particles with a spin $\frac{1}{2}$ interact via the mediation of gauge bosons (with spin 1). 
The underlying mathematical formulation of the SM is a renormalizable quantum field theory based on a local $SU(3) \times SU(2) \times U(1)$ gauge symmetry \cite{GWS1, GWS2, GWS3, georgi72, politzer73, asymp_freedom2, asymp_freedom, weinberg2004, hooft71, hooft72a, hooft72b}. While the $SU(3)$ subgroup describes the interaction with the gluon fields (\newword{Quantum Chromodynamics}, \textit{QCD}\index{QCD | see {Quantum Chromodynamics }}, also called \textit{strong interaction}\index{Strong Interaction}), $SU(2) \times U(1)$ is the representation of the electroweak interaction, unifying the electromagnetic and the weak interaction. 

\begin{figure}[ht]
	\centering
		\includegraphics[width=0.95\textwidth]{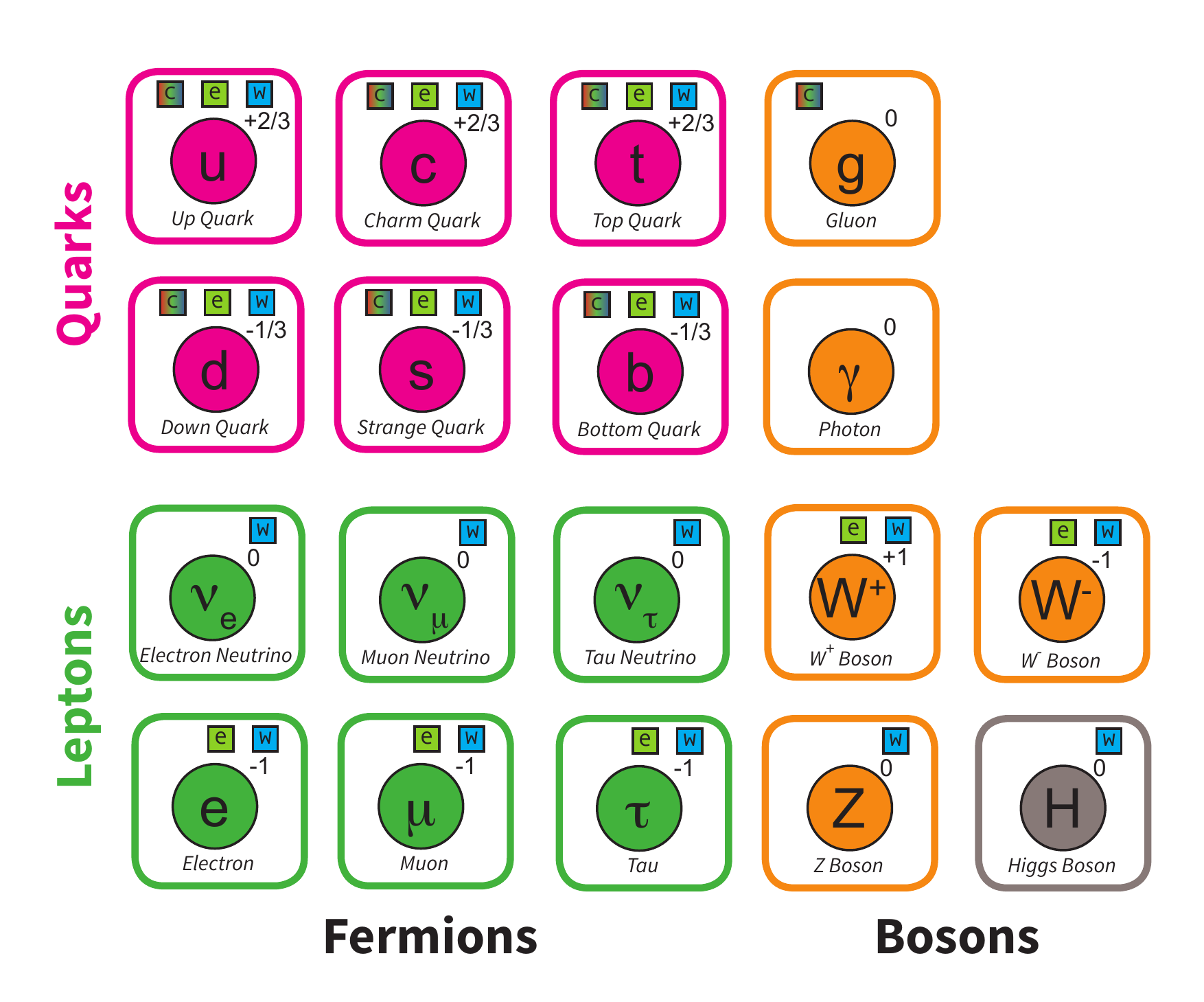}
		\caption{Fermions (quarks and leptons) and gauge bosons of the Standard Model and some of their basic properties. The small boxes indicate the fields to which the particles couple: colour (c), electromagnetic (e) and weak (w). The number in the upper right corner represents the electric charge.}
					\label{fig:SM}
\end{figure}

The main properties of the strong and the weak interaction are described in Sections \ref{sec:QCD} and \ref{sec:EWint}. 
All fermions and gauge bosons are introduced in Figure \ref{fig:SM}. 

Depending on how the fermions interact, they can be grouped into quarks (interacting via the strong interaction) and \textit{leptons}\index{Lepton} (not interacting via the strong interaction). An additional \textit{colour charge}\index{Colour charge} is assigned to particles interacting via the strong interaction.\footnote{Colour charge is the equivalent preserved quantity in QCD as is the electric charge in electrodynamics. Colour is just an additional degree of freedom needed to describe quarks. There is no relation to colour in the literal sense.}
Left-handed fermions have $T = \frac{1}{2}$ and are arranged in doublets of the weak isospin $T$, right-handed ones are singlets with $T=0$. Only left-handed fermions interact via the weak interaction. Thereby, the third component of the weak isospin, $T_3$, is conserved. 

Quarks with $T_3 = + \frac{1}{2}$ carry an electric charge of $+\frac{2}{3}$ in terms of the positron charge $\left| e\right|$, quarks with $T_3 = - \frac{1}{2}$ carry a charge of $-\frac{1}{3}$.\footnote{From now on all electric charges are quoted in terms of $\left| e\right|$.} In contrast to quarks, leptons with $T_3 = + \frac{1}{2}$ carry a charge of $0$. The ones with $T_3 = - \frac{1}{2}$ do carry a charge of $-1$. All weak isospin doublets appear in three generations. Their properties are the same with increasing masses as the only difference. Since heavy generations will decay into light ones, stable matter on earth is composed of $u$- and $d$-quarks as well as electrons. For all fermions corresponding anti-particles exist. The quantum numbers of the latter have opposite sign. Table \ref{tab:fermions} lists the fermion properties.\footnote{In this thesis natural units ($\hbar = c = 1$) are used if not stated otherwise. In particular, this concerns the units of masses which are quoted as \MeV\ instead of $\MeV / c^2$, for example.}

\begin{table}[htbp]
\begin{center}
\begin{tabular}{|c|r|r|c|c|}

\hline
Fermion &  \multicolumn{1}{c|}{$Q$} & \multicolumn{1}{c|}{$T_3$} & Colour Charge & Mass [MeV]\\
\hline
\hline
Up Quark ($u$)& $+ 2/3$ & $+ 1/2$  &yes &2.3 \\
Down Quark ($d$)& $- 1/3$ & $- 1/2$ &yes  &4.8 \\
Charm Quark ($c$)& $+ 2/3$ & $+ 1/2$  &yes &1275 \\
Strange Quark ($s$)& $- 1/3$ & $-1/2$ &yes  &95 \\
Top Quark ($t$)& $+2/3$ & $+ 1/2$  &yes &173340 \\
Bottom Quark ($b$)& $- 1/3$ & $- 1/2$ &yes  &4180 \\
\hline
Electron Neutrino ($\nu_e$)& $0${ ~} & $+ 1/2$ &no  &$< 2 \cdot 10^{-6}$\\
Electron ($e$)& $-1${ ~} & $-1/2$ &no &0.511\\
Muon Neutrino ($\nu_{\mu}$)& $0${~ } & $+ 1/2$  &no&  $<0.19$\\
Muon ($\mu$)& $-1${ ~} & $- 1/2$ &no  &105.7\\
Tau Neutrino ($\nu_{\tau}$)& $0${ ~} & $+ 1/2$  &no &$<18.2$\\
Tau ($\tau$)& $-1${ ~} & $-1/2$&no  &1776.8\\
\hline
\end{tabular}
\end{center}
\caption{Fermions of the Standard Model with their electric charges $Q$ and the third component of the weak isospin $T_3$ . The values refer to left-handed fermions. Right-handed ones have $T_3 = 0$ and $Q$ as the left-handed. The mass values are taken from \cite{PDG, m_top_world}.}
\label{tab:fermions}
\end{table}

The mediating gauge bosons for the interactions are the gluons for the strong interaction, the $W$ and $Z$ bosons for the weak interaction and the photons for the electromagnetic interaction. The gauge bosons with their most important properties are listed in Table \ref{tab:bosons}.

\begin{table}[htbp]
\begin{center}
\begin{tabular}{|c|c|c|c|c|c|c|}
\hline
\rule{0pt}{0.4cm}Boson &  Interaction &$Q$ & $T_3$ & Gauge Coupling $g$& Charges& Mass [MeV]\\
\hline
\hline
\rule{0pt}{0.4cm} $W^{+}$			&Weak			& $+ 1$ 	& $+ 1/2$ 	 		&$\sqrt{4 \pi \alpha} / \sin{\theta_W}$ $\left[\sqrt{4 \pi \alpha}\right]$		&w [e] &80385 \\
\rule{0pt}{0.4cm} $W^{-}$			&Weak			& $- 1$ 	& $- 1/2$ 	 		&$\sqrt{4 \pi \alpha} / \sin{\theta_W}$	 $\left[\sqrt{4 \pi \alpha}\right]	$	&w [e] &80385 \\
\rule{0pt}{0.4cm}$Z$				&Weak			& $0$ 	& $0$ 	 		&$\sqrt{4 \pi \alpha} / \left( \sin{\theta_W} \cos{\theta_W} \right)$		&w&91188 \\
\rule{0pt}{0.4cm}$\gamma$	&Electromag.	& $0$ 	& $0$			&$\sqrt{4 \pi \alpha}$		&---&$< 10^{-24}$\\
\rule{0pt}{0.4cm}$g$		&Strong			& $0$	&$0$ 			&$\sqrt{4 \pi \alpha_s}$		&c&0\\
\hline
\end{tabular}
\end{center}
\caption{The gauge bosons of the Standard Model with their corresponding interaction, electric charge $Q$, third component of the weak isospin $T_3$ and coupling constant $g$. The mass values are taken from \cite{PDG}. The couplings refer to the interaction strengths and the charges to colour ($c$), weak ($w$) and electric ($e$) charge. The values $\alpha$, $\alpha_s$ and $\theta_W$ are explained in Sections \ref{sec:QCD} and \ref{sec:EWint}.}
\label{tab:bosons}
\end{table}

Despite the fact that the SM is a powerful framework to calculate strong and electroweak interactions at high precision, it does not describe gravity. The former interactions are described in the following sections.

\subsection{Strong Interaction}
\label{sec:QCD}
The strong interaction and its field theory, QCD, are based on an $SU(3)$ gauge group. The eight generators of the group are represented by eight gluons. As the gauge group of QCD is non-Abelian, each gluon carries a colour and an anti-colour, allowing it to couple to other gluons. 

The strong interaction plays an important role in the regime of high energy physics. In particular, it is the main interaction responsible for \ttbar\ pair production process at hadron colliders (see Section \ref{sec:top}) and thus responsible for the spin configuration of the \ttbar\ pair. One should be careful not to take the word \textit{strong} too seriously. The actual strength of the strong coupling $\sqrt{4 \pi \alpha_s}$ depends on the energy scale $Q^2$ of the process of interest, making the strong coupling constant $\alpha_s$ everything but a constant.\footnote{$Q^2$ refers to the absolute value of the squared four-momentum transferred at a vertex ($Q^2 = \left| q^2\right|$).} 
For values of $\alpha_s$ which are significantly smaller than unity, QCD can be treated perturbatively. Corrections of higher orders lead to the modified effective coupling, calculated at a specific \textit{renormalization scale}\index{Renormalization scale} $\mu_R$. 

The dependence of $\alpha_s$ on the energy scale $Q^2$ and squared renormalization scale $\mu_R^2$ is given by \cite{halzenmartin}
\begin{align}
\alpha_s\left(Q^2, \mu_R^2 \right) =\frac{\alpha_s \left( \mu_R^2\right)}{1+ \frac{\alpha_s \left( \mu_R^2\right)}{12 \pi}\left(11 n_c - 2 n_f\right) \ln \left( Q^2 / \mu_R^2\right)}.
\label{eq:running_as}
\end{align}
Both corrections of fermionic and bosonic loops are included, giving a different sign to the change of $\alpha_s$: $n_c$ refers to the number of colours, $n_f$ to the number of light quark flavours ($m_q \muchless \mu_R$).
The equation can also be reformulated as 
\begin{align}
\alpha_s\left(Q^2, \Lambda^2 \right) &=\frac{12 \pi}{\left(11 n_c - 2 n_f\right) \ln \left( Q^2 / \Lambda^2\right)}
\label{eq:running_as2}
\end{align}
by the introduction of the cut-off parameter $\Lambda$, which is chosen in a way that it defines the scale where QCD cannot be calculated using perturbation theory. Depending on the number of fermions $n_f$ included in the renormalization the values of $\Lambda_{n_f}$ are \cite{PDG}
\begin{align}
\Lambda_5 &= 213 \pm 8 \MeV\ \\
\Lambda_4 &= 296 \pm 10 \MeV\ \\
\Lambda_3 &= 339 \pm 10 \MeV.
\end{align}
For a renormalization scale $\mu_R^{2}$ set to the energy scale $Q^2$ of the process of interest, Equation \ref{eq:running_as} describes the energy scale dependence of $\alpha_s$. As $n_c = 3$ and $n_f <6$ QCD becomes non-perturbative for $\alpha_s\left( Q^2 \rightarrow 0 \right)$ with \newword{quark confinement} as a consequence of the coupling increasing with distance. On the other hand, for short ranges and high energy scales, \textit{asymptotic freedom}\index{Asymptotic freedom} of QCD holds as $\alpha_s\left(Q^2 \rightarrow \infty \right) = 0$ \cite{asymp_freedom}. 

\begin{figure}[ht]
	\centering
		\includegraphics[width=0.75\textwidth]{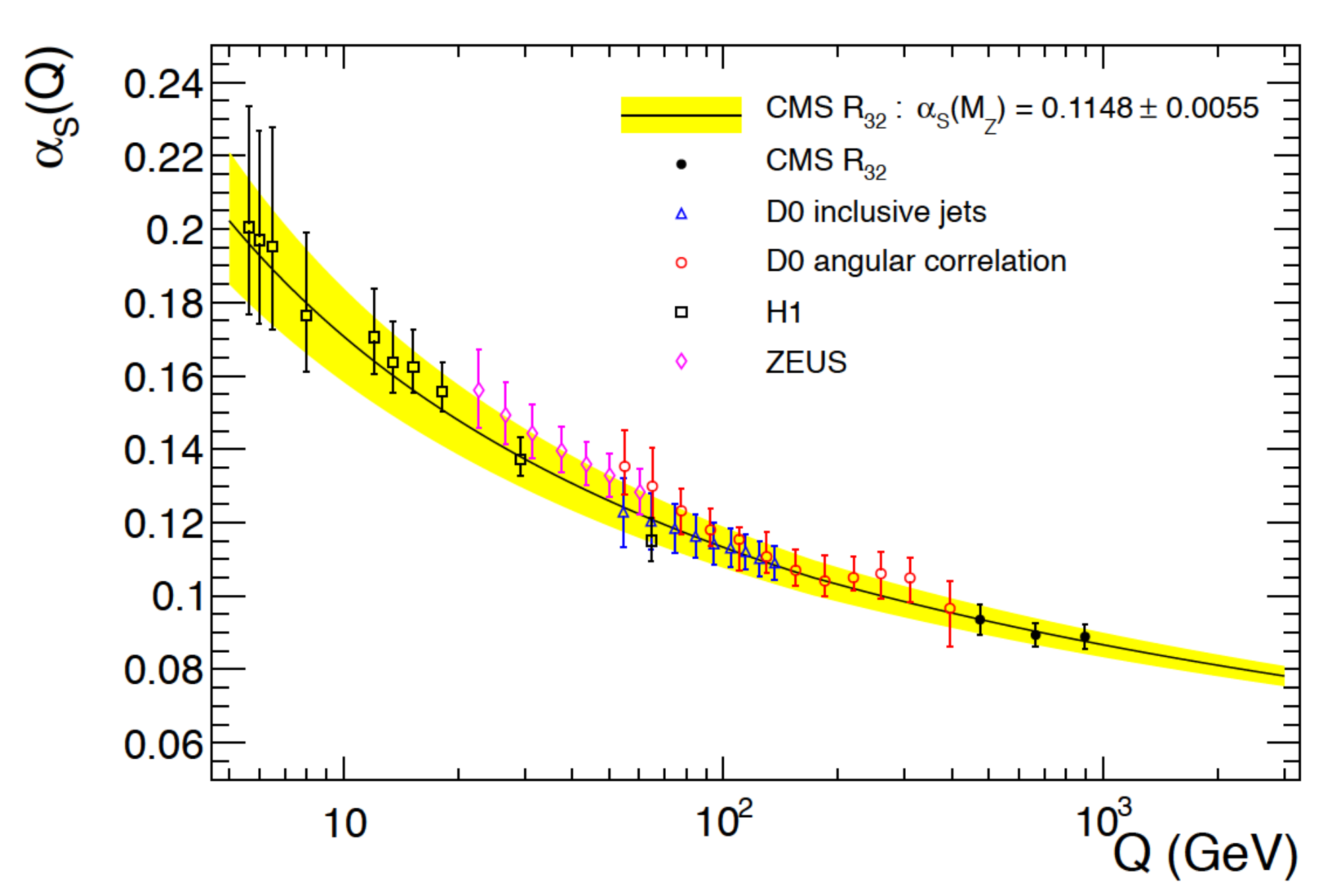}
		\caption{The measured dependence of $\alpha_s$ on the energy scale $\sqrt{Q^2}$ \cite{alpha_s}.}
	\label{fig:alpha_s}
\end{figure}

Experimental determinations of $\alpha_s$ show good agreement with the predicted behaviour. Figure \ref{fig:alpha_s} summarizes the measurements by the H1 \cite{alpha_s_h1a, alpha_s_h1b}, ZEUS \cite{alpha_s_zeus}, D0 \cite{alpha_s_d0a, alpha_s_d0b} and CMS \cite{alpha_s} collaborations.

A common reference for quoting the value of $\alpha_s$ is the mass of the $Z$ boson. The world average value was determined in \cite{PDG} as
\begin{align}
\alpha_s \left( m_Z^2 \right) = 0.1184 \pm 0.0007.
\label{eq:alpha_s_world}
\end{align}

\subsection{Electroweak Interaction}
\label{sec:EWint}
While in nature\footnote{Or more precisely: on energy scales we do observe in nature.}, the electromagnetic and the weak interaction appear as separate interactions with quite different properties, their underlying theoretical framework is the same. Due to the major contributions of S. Glashow, S. Weinberg and A. Salam \cite{GWS1, GWS2, GWS3} it is often referred to as the \textit{Glashow-Weinberg-Salam}\index{Glashow-Weinberg-Salam model} (\textit{GWS}\index{GWS model | see {Glashow-Weinberg-Salam model }}) model.

The electroweak (EW) symmetry, manifested in the $SU(2) \times U(1)$ gauge group, is spontaneously broken via the Higgs mechanism which will be described in detail in the next section. As a consequence of the EW symmetry breaking, the four massless bosons\footnote{A direct mass term is forbidden to preserve the local gauge invariance.} $W^{1}$, $W^{2}$, $W^{3}$ and $B$, generators of the $SU(2)$ and $U(1)$ gauge groups, mix to the observable gauge bosons $W^{+}$, $W^{-}$, $Z^{0}$ and $\gamma$:
\begin{align}
\left(\begin{array}{c}\gamma \\Z^{0}\end{array}\right) &= \left(\begin{array}{cc}\cos{\theta_W} & \sin{\theta_W} \\-\sin{\theta_W} & \cos{\theta_W}\end{array}\right)\left(\begin{array}{c}B \\W^{3}\end{array}\right)\\
\left(\begin{array}{c}W^{+} \\W^{-}\end{array}\right) &= \left(\begin{array}{cc}\frac{1}{\sqrt{2}} & \frac{-i}{\sqrt{2}} \\\frac{1}{\sqrt{2}} & \frac{i}{\sqrt{2}}\end{array}\right)\left(\begin{array}{c}W^{1} \\W^{2}\end{array}\right)
\end{align}
The photon as the mediator of the electromagnetic force remains massless, unlike the massive $W^{+}$, $W^{-}$ and $Z^{0}$ of the weak interaction. The mixing angle $\theta_W$, or rather its squared sine, is determined experimentally. The quoted value depends on the renormalization scheme and ranges from $\sin{\theta_W}^2 = 0.22295(28)$ to $\sin{\theta_W}^2 = 0.23116(12)$ \cite{PDG}.

\subsubsection{Electromagnetic Interaction}
The quantum field theory describing the electromagnetic part of the GWS model is called \newword{Quantum Electrodynamics}. It is based on the $U(1)$ part of the $SU(2) \times U(1)$ gauge symmetry of the electroweak interaction. Unlike QCD, QED is an Abelian gauge group. As a consequence, no photon-photon couplings exist. Thus, the QED equivalent to Equation \ref{eq:running_as} has no bosonic loop contribution with opposite effect as the fermionic ones \cite{halzenmartin} :
\begin{align}
\alpha\left(Q^2, \mu_R^2 \right) =\frac{\alpha \left( \mu_R^2\right)}{1- \frac{\alpha \left( \mu_R^2\right)}{3 \pi}\ln \left( Q^2 / \mu_R^2\right)}.
\label{eq:running_a}
\end{align}
Depending on the corrections considered\footnote{Here, only electron/positron loops are considered. This corresponds to $n_f=1$.} the factor in front of the logarithm may change, but the $Q^2$ dependence is the same: $\alpha$ increases with lower $Q^2$ and vice versa. Equation \ref{eq:running_a} holds for $Q^2 \gg \mu_R$ only. In the limit of $Q^2 \rightarrow 0$, $\alpha$ takes the numerical value of $\frac{1}{137}$, also known as the \textit{fine-structure constant}\index{Fine-structure constant}. 
The variations of $\alpha$ by $Q^2$ are rather low ($\alpha(M_Z)^{-1} = 127.944 \pm 0.014$ \cite{PDG}). 

\subsubsection{Weak Interaction}
\label{sec:weakinteraction}
Particles taking part in a weak interaction process are members of the same weak isospin doublet (see Figure \ref{fig:SM}). This means that the weak interaction does not cross different generations. 
However, it is observed in nature that weak interactions across quark generations do occur, for example in the decay of Kaons \cite{PDG}. 

This is possible as the weak doublet partners of the $T_3 = + \frac{1}{2}$ quarks are in fact superpositions of mass eigenstates ($d,c,b$). The linear combinations are described by the unitary \newword{CKM matrix}\footnote{Named after the editors of \cite{CKM}, M. Kobayashi and T. Maskawa, as well as N. Cabbibo on whose ideas \cite{CKM} is footing \cite{cabbibo}.} \cite{CKM}:

\begin{align}
\begin{pmatrix}  d^\prime  \\  s^\prime  \\  b^\prime  \end{pmatrix} = \begin{pmatrix} V_{ud} & V_{us} & V_{ub} \\ V_{cd} & V_{cs} & V_{cb} \\ V_{td} & V_{ts} & V_{tb} \end{pmatrix} \begin{pmatrix}  d  \\  s  \\  b  \end{pmatrix}.
\end{align}
The unitarity requirement reduces the nine parameters $V_{ij}$ to three mixing angles and a complex phase responsible for \newword{CP violation}.\footnote{The combined charge and parity symmetry is broken.}

As a consequence of the CKM matrix mixing, the weak interaction allows interactions across quark generations. It should be stressed that only left-handed fermions are part of the isospin doublets while the right-handed ones are singlets. Thus, right-handed particles do not interact via the charged weak interactions involving a $W^{\pm}$ boson.\footnote{As $Z$ bosons are linear combinations of the $W^{3}$ and $B$ and only the former one requires particles from the isospin doublet, neutral weak interactions do not have this restriction.} To account for the maximal parity violation of the weak interaction -- as observed in nature \cite{wu} -- the weak interaction vertex has a \textit{vector $-$ axialvector} (\textit{$V-A$}) structure\index{v@$V-A$ structure}:

\begin{align}
W^{\pm}&: \frac{-i g_W}{2\sqrt{2}} \gamma^\mu \left( 1- \gamma^5\right) V_{ij}\\
Z&: \frac{-i g_Z}{2} \gamma^\mu \left( c_V- c_A \gamma^5\right)
\end{align}
Here, $\gamma^{\mu} (\mu =1..4)$ represent the Dirac matrices, $\gamma^{5}=i\prod_{j=0..3}{\gamma^{j}}$, $g_{W/Z}$ the weak coupling constants as in Table \ref{tab:bosons}, $c_V = T_{3} - 2 Q \sin{\theta_W}$ the vector and $c_A =  T_3$ the axial vector part of the coupling. The $V-A$ structure is manifested in the term $\gamma^{\mu}(1-\gamma^{5})$ with the vector component $\gamma^{\mu}$ and the axialvector component $\gamma^{\mu}\gamma^5$. The $(1-\gamma^{5})$ can also be interpreted as a projection operator for the left-handed components of a fermion wave function.\footnote{The equivalent right-handed projection operator is $(1+\gamma^{5})$.} The $V-A$ structure of the weak interaction is of great importance for the propagation of the top quark's spin to its decay products (see Section \ref{sec:SC}).  

The CKM matrix is clearly diagonally dominant, stressing the favoured inter-isospin doublet interactions. The values for $V_{ij}$ are determined experimentally and can be found in \cite{PDG}. \textit{Flavour changing neutral currents}\index{Flavour changing neutral currents} (\textit{FCNC}\index{FCNC | see {Flavour changing neutral current }}) would lead to a change of quark flavour without changing the charge, such as a $c\rightarrow u$ transition. In the SM, flavour changing neutral currents are only possible at higher orders (double $W$ exchange) and are strongly suppressed by the \newword{GIM mechanism}\footnote{Named after S.L. Glashow, J. Iliopoulos and L. Maiani.} \cite{GWS3}.

\subsection{Electroweak Symmetry Breaking}
\label{sec:symmetrybreaking}
As predicted and also observed experimentally, the $W^{\pm}$ \cite{W_disc_UA1, W_disc_UA2} and $Z$ \cite{Z_disc_UA1, Z_disc_UA2} bosons are massive. But in order to preserve the local gauge invariance of the SM, masses may not be attributed to the gauge bosons explicitly. A dynamic mass generation mechanism is needed, such as the Higgs mechanism \cite{Higgs1, Higgs2, Higgs3, Higgs4, Higgs5, Higgs6}. 

Before electroweak symmetry breaking the situation is the following:
The gauge fields  $W^{1,2,3}$ belong to the $SU(2)$ group and couple with a strength $g_W$. The quantity which is invariant under $SU(2)$ transformations is the weak isospin.
$B$ is the corresponding gauge field of the $U(1)$ group with a coupling $g'$ and the \textit{weak hypercharge}\index{Weak hypercharge} $Y$ as the conserved quantity. All four fields are massless.

The GWS theory makes use of the Higgs mechanism by adding four scalar fields $\phi_i$ with special properties \cite{GWS2}. Arranged in a complex isospin doublet with hypercharge $Y=1$ it preserves the $SU(2) \times U(1)$ gauge invariance. By assigning a vacuum expectation value (VEV) $v$ to the real neutral component, the symmetry operations of the electroweak interaction are broken and their corresponding bosons get massive. $U_{EM}(1)$ as subgroup of $SU(2) \times U(1)$ remains invariant (as the Higgs field with a VEV is neutral). The conserved quantity is the electric charge $Q$, related to the third component of the weak isospin $T_3$ and the weak hypercharge by the \newword{Gell-Mann-Nishijima formula}\footnote{In its original version it was relating the electric charge to the hadronic isospin $I_3$, the baryon number $B$ and the strangeness $S$ via $Q=I_3 + \frac{1}{2}\left(B+S\right)$.} \cite{gmn1, gmn2}
\begin{align}
Q = T_3 + \frac{Y}{2}.
\end{align}
The gauge boson of the $U_{EM}(1)$ group (the photon) remains massless while the others ($W^{\pm}$ and $Z$) obtain masses. The masses of the gauge bosons depend on the weak couplings $g_W$ and $g'$ and the VEV $v$ of the Higgs field \cite{halzenmartin}:
\begin{align}
m_W &= \frac{1}{2} v\,g_W \\
m_Z &= \frac{1}{2} v\,\sqrt{g_W^2 + {g'}^2}
\end{align}
The weak couplings $g_W$ and $g'$ are related to the couplings of the gauge bosons via 
\begin{align}
g_{\gamma} &= g_W \cdot \sin{\theta_W} = g' \cdot \cos{\theta_W} = g_Z \cdot \sin{\theta_W} \cdot \cos{\theta_W}
\end{align}
Knowing $m_W$, $m_Z$ and their couplings allows to predicting the VEV of the Higgs field. The field itself acquires mass, depending on its VEV but also depending on the shape of its potential. There is no fixed choice for the Higgs potential, but a potential such as
\begin{align}
V(\phi) = - \mu^2 \phi^{\dagger} \phi  + \lambda \left( \phi^{\dagger} \phi \right)^2
\end{align}
serves all needs.\footnote{Higher orders in $\phi$ break renormalizability \cite{GWS2}.} The parameters $\mu$ and $\lambda$ determine the VEV $v$ via
\begin{align}
v = \sqrt{\frac{\mu^2}{\lambda}}
\end{align}
but they are in principle free. The field quantum of the Higgs field is the scalar {\newword{Higgs boson}}. No prediction on its mass, given by
\begin{align}
m_H = \sqrt{\frac{\lambda}{2}} v,
\end{align}
can be made unless the shape of the Higgs potential is known \cite{peskin}. 

Next to the gauge boson mass terms, the mass terms of the fermions would also break the local gauge invariance of the theory. Thus, also for fermion masses the Higgs mechanism can be used to take a workaround via symmetry breaking, but in a different way than for the gauge bosons. For each massive fermion $f$ -- excluding neutrinos\footnote{Within the SM, neutrinos are massless.} \cite{peskin} -- an additional \newword{Yukawa coupling} $y_f$ to the Higgs field is introduced. This relates the fermion masses $m_f$ to the Higgs field VEV:
\begin{align}
m_f = \frac{y_f}{\sqrt{2}} v
\label{eq:fermion_masses}
\end{align}
By using the relation 
\begin{align}
v^2 = \frac{1}{\sqrt{2} G_F}
\end{align}
and the value for the Fermi constant \mbox{$G_F = 1.1663787(6) \cdot 10^{-5} \GeV^{-2}$} (determined experimentally via measurement of the muon lifetime \cite{PDG}), the Higgs VEV $v$ turns out to be 246.22 GeV. The more massive a fermion, the higher its coupling to the Higgs field is.

The Higgs mechanism serves well in the GWS model. About 50 years after the proposal of the mechanism it could be experimentally confirmed in 2012 by the observation of the missing Higgs boson. The ATLAS \cite{ATLAS_higgs_discovery} and CMS \cite{CMS_higgs_discovery} experiment reported the observation of a new boson having the expected properties of the Higgs particle. First details could already be studied, leading to evidence of the spin-0 property and a strong preference to its positive parity \cite{higgs_ATLAS_spin_evidence}, as expected. The exact couplings to fermions will have to be studied in detail in the future. 

Equation \ref{eq:fermion_masses} shows that the higher a fermion's mass, the higher its Higgs coupling is. Using the world average top quark mass of $m_{t} = 173.34~\GeV$ \cite{m_top_world} yields a Yukawa coupling of $y_{t} = 0.996$. This makes the top quark as the most massive elementary particle a very important probe for studies of the Higgs mechanism.

\subsection{Limitations of the Standard Model}
The Standard Model is a powerful theory, providing the description of a broad variety of natural phenomena at high precision. However, observed limitations of the Standard Model indicate that it needs to be extended or embedded in a larger theory.
Such a theory could unify the strong and the electroweak interaction and also include gravity, which is not described by the SM. Astrophysical observations show distributions of non-baryonic matter interacting via the gravitational force, which cannot be explained with the matter particles contained in the SM (\newword{Dark Matter}, see e.g. \cite{darkmatter}). As observed in the context of \textit{neutrino oscillations}\index{Neutrino oscillation} \cite{neutrino_oscillations}, neutrinos have a non-zero mass. This is also contradicting the SM assumption of massless neutrinos. 

One example for a SM extension is \textit{supersymmetry}\index{Supersymmetry} (\textit{SUSY}\index{SUSY | see {Supersymmetry }}), introducing a symmetry between fermions and bosons \cite{susy1,susy2,susy3,susy4,susy5,susy6,susy7, susy8,susy9}. Such BSM scenarios include modifications of SM predictions. The \ttbar\ spin correlation, analysed in this thesis, is a possible way to probe BSM physics. The relation between BSM scenarios and \ttbar\ spin correlation is explained in Section \ref{sec:BSM}.

\section{Proton Structure}
\label{sec:proton}
For the prediction of final state configurations it is important to know production and decay mechanisms of the process of interest in detail. In the case of \ttbar\ production and decay, the process under study in this thesis, the details are explained in Section \ref{sec:top}. But furthermore, each process needs a well-defined initial state. 

Using a proton-proton collider such as the LHC introduces an undetermined initial state. The machine parameters provide a value for the momenta of the incoming protons. But the initial state of the hard scattering process requires two of the protons' constituents, namely either quarks or gluons (in general: \textit{partons}\index{Parton}). The density of quarks and gluons within the proton depends on two parameters: the fraction $x$ of the longitudinal proton momentum that the parton carries as well as the energy scale $Q^2$ of the scattering process. As the partons inside the proton interact via the strong interaction, gluon radiations are allowed as well as gluon to quark/antiquark and gluon to gluon splittings. Hence, the total quark density is a sum of the three valence quark densities and the virtual quarks from gluon splittings. 
In general, the density of a parton $a$ inside a proton is given by the \newword{Parton Distribution Function} (\textit{PDF}\index{PDF | see {Parton Distribution Function }}). QCD does not provide an a-priori prediction of quark ($q_i$) and gluon ($g$) PDFs. The evolution of a PDF with $Q^2$ for a fixed value of $x$  is described by the \textit{Dokshitzer-Gribov-Lipatov-Altarelli-Parisi (DGLAP) equations}\index{Dokshitzer-Gribov-Lipatov-Altarelli-Parisi equations} \index{DGLAP equations | see {Dokshitzer-Gribov-Lipatov-Altarelli-Parisi equations }} \cite{dglap1, dglap2, dglap3}:

\begin{align}
\frac{d q_i(x,Q^2)}{d \ln{Q^2}}~&=
~\frac{\alpha_s(Q^2)}{2\pi} \int_x^1 \left[ \frac{q_i(y,Q^2)}{y}\cdot P_{qq}\left(\frac{x}{y}\right)~+\frac{g(y,Q^2)}{y}\cdot P_{qg}\left(\frac{x}{y}\right)  \right] dy\\
\frac{d g(x,Q^2)}{d \ln{Q^2}}~&=~\frac{\alpha_s(Q^2)}{2\pi}  \int_x^1
 \left[ \sum_j \frac{q_j(y,Q^2)}{y}\cdot P_{gq}\left(\frac{x}{y}\right)~+\frac{g(y,Q^2)}{y}\cdot P_{gg}\left(\frac{x}{y}\right)  \right] dy
\end{align} 

The splitting functions $P_{ab}\left(\frac{x}{y}\right)$ describe the probability for a parton $b$ to emit a parton $a$ with a momentum fraction $\frac{x}{y}$. PDFs are determined experimentally via hadron-hadron and lepton-hadron collider measurements. For a parameterization in $x$ and an example of PDF determination see for instance \cite{herapdf}. 

\begin{figure}[ht]
	\centering
		\includegraphics[width=0.6\textwidth]{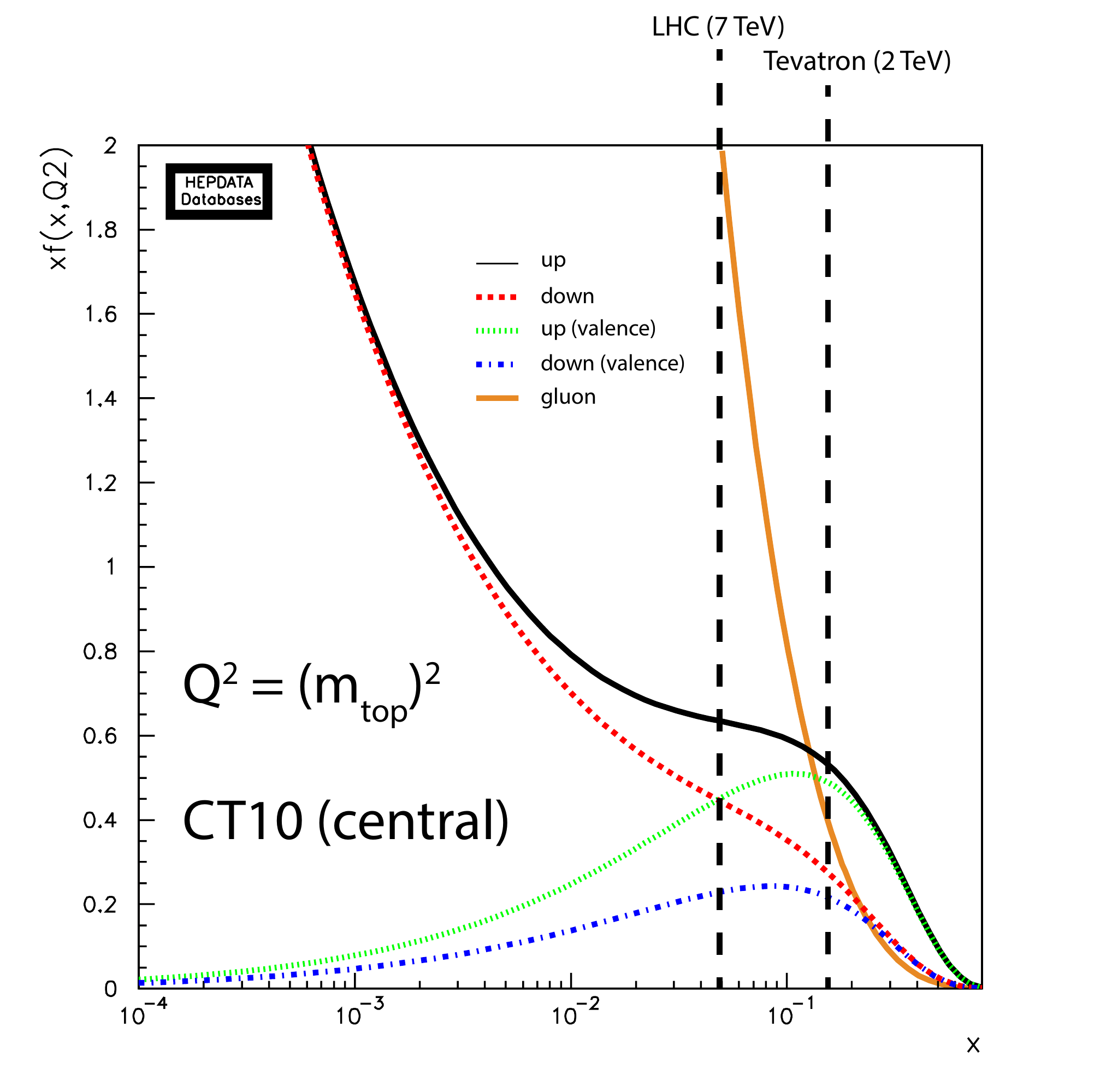}
		\caption{Parton distribution function of the nominal CT10 set at $Q^2 = m_t^2$. The minimal average proton momentum fractions $x$ for \ttbar\ production are shown for Tevatron ($\sqrt{s} = 2\,\TeV$) and the LHC ($\sqrt{s} = 7\,\TeV$).}
	\label{fig:PDF}
\end{figure}

Several collaborations are performing fits of $f_a$ to data and provide the respective PDF sets. These are for example \newword{HERAPDF} \cite{herapdf}, \newword{CTEQ} \cite{CT10}, \newword{NNPDF} \cite{NNPDF} and \newword{MSTW} \cite{MSTW}. An example plot for PDFs$(x, Q^2=m_t^2)$ is shown in Figure \ref{fig:PDF}.

Quark PDFs $q_i$ contain the sea quark distribution increasing for lower $x$ and -- in case of up and down quark PDFs -- a valence quark distribution which peaks at about $\frac{1}{6}$. For high values of $x$, quark densities are dominating while gluon densities dominate for lower $x$. This has an important consequence which should be kept in mind in the context of \ttbar\ spin correlation analyses. If \ttbar\ pairs are produced, the minimum amount of energy needed is $E=2\,m_t$. With the assumption that each of the incoming \mbox{(anti-)}protons provides a parton with the same energy, the minimum $x$ for \ttbar\ production at the Tevatron ($E_{\text{beam}}=0.98\,\TeV$) and the LHC ($E_{\text{beam}}=3.5\,\TeV$ for the analysed 2011 run) yields
\begin{align}
x_{\text{Tevatron}} &= \frac{173.5\,\GeV}{0.98\,\TeV} = 0.18\\
x_{\text{LHC}} &=  \frac{173.5\,\GeV}{3.5\,\TeV} = 0.05
\end{align}
These two values of $x$ are also indicated in Figure \ref{fig:PDF}. For the production of \ttbar\ pairs two different mechanisms exist: quark/antiquark annihilation and gluon fusion (see Section \ref{sec:top}). The parton with the higher density defines the dominating production mechanism. This has significant implications on the spin configuration of the \ttbar\ pair. In particular, measurements at the Tevatron and the LHC are complementary as different production mechanisms dominate. How this configuration is determined is discussed in Section \ref{sec:SC}.

\section{The Top Quark}
\label{sec:top}
Several hints suggested the existence of a top quark well in advance, before it was observed as the last quark of the SM. As V. Fitch and J. Cronin observed CP violation\index{CP violation} in 1964 \cite{croninfitch}, the need for a theoretical explanation came up. One way to establish this was suggested by Kobayashi and Maskawa in 1973 \cite{CKM} by the proposal of a third quark generation. This idea was strengthened by the discovery of the $\tau$ lepton \cite{tau_disc}, increasing the number of lepton generations to three. As there were only two quark generations, the GIM mechanism broke. With the discovery of the $\Upsilon$ -- a meson consisting of a $b$- and a $\bar{b}$-quark -- by the E288 experiment \cite{b_disc}, the door to a third generation opened. The need of a weak isospin partner of the \bQ\ was finally satisfied in 1995 with the discovery of the top quark at the Tevatron accelerator by the D0 \cite{top_disc_D0} and the CDF \cite{top_disc_CDF} collaborations. Electroweak precision measurements had already constrained the top quark's mass before it was finally measured. By fitting electroweak precision data without using direct top quark mass measurements, the top quark mass can today be determined as $175.8^{+2.7}_{-2.4}~\text{GeV}$ \cite{gfitter}. Former predictions were summarized in Figure \ref{fig:topmass_evolution}.
Indirect measurements have a great prediction power as this example shows. 

A whole set of unique measurement possibilities comes along with the properties of the top quark. It is the by far heaviest fermion with a Yukawa coupling close to unity (see Equation \ref{eq:fermion_masses}), making it a good probe for Higgs physics studies. In \cite{gamma_top2} the top quark decay width $\Gamma_t$ was calculated at NLO. Approximations for  $\beta \equiv \frac{m_W}{m_t}$ were provided as
\begin{align}
\Gamma_t^{\beta \rightarrow 0} &= \Gamma^0_t
\left[ 1- \frac{2 \alpha_s}{3 \pi} \left( \frac{2\pi^2}{3}-\frac{5}{2}\right)\right] \label{eq:gammatop1}\\
\Gamma_t^{\beta \rightarrow 1} &= \Gamma^0_t
\left[ 1- \frac{2 \alpha_s}{3 \pi} \left(3 \ln\left(1-\frac{m_W^2}{m_t^2}\right) + \frac{4\pi^2}{3} - \frac{9}{2} \right)\right] \label{eq:gammatop2}
\end{align}
using
\begin{align}
\Gamma^0_t &= \frac{G_F m_t^3 }{8 \pi \sqrt{2}} \left| V_{tb}\right|^2\left( 1- \frac{m_W^2}{m_t^2}\right)^2\left( 1+ 2\frac{m_W^2}{m_t^2}\right)
\end{align}
with the Fermi constant $G_F$ and the CKM matrix element $V_{tb}$.
As the ratio $\Gamma_t / \Gamma_t^0$ of NLO to LO top decay width versus the ratio $m_W / m_t$ stays almost constant for $0 < m_W/m_t < 0.6$ (see Figure 2 in \cite{gamma_top2}) Equation \ref{eq:gammatop1} is a valid approximation. 
Using $m_t = 173.34 \GeV$ \cite{m_top_world}, $m_W = 80.385 \GeV$ \cite{PDG} and $\alpha_s \left( m_Z^2 \right) = 0.1184$ \cite{PDG} leads to a top quark width of $\Gamma_t^{\beta \rightarrow 0} = 1.36 \GeV$. 

Using $\hbar = 6.58211928 \cdot 10^{-6} \text{eV s}$ \cite{CODATA}  leads to a predicted top quark lifetime of
\begin{align}
\tau_t &= \frac{\hbar }{\Gamma_t}= 4.85 \cdot 10^{-25}~\text{s}
\end{align}
Comparing the top quark lifetime to the time scale needed for hadronisation \cite{hadr_time_scale} given by
\begin{align}
t_{\text{had}} = \frac{\hbar}{\Lambda_{QCD}} = \frac{\hbar}{213 \MeV} \approx 3 \cdot 10^{-24} ~\text{s}
\end{align}
shows one order of magnitude difference. Thus, the top quark decays before forming bound states. This statement was also made in \cite{hadr_time_scale} with a quoted $t_{\text{had}} \approx 10^{-23}\,\text{s}$. The quoted value of $t_{\text{had}}$ depends on the used cutoff parameter $\Lambda_{QCD}$. 

It is important to realize the implications of this relatively short lifetime. In case the top quark decays before it hadronises, its spin properties would directly be transferred to the decay products. Measurements of the top quark decay width indicate that this is indeed the case (see Section \ref{sec:width}). In the literature, the time scale for hadronisation is used in many cases as the relevant quantity to compare the top quark lifetime to when arguing about the spin transfer to the decay products. However, the spin decorrelation time is in fact even longer than the hadronisation time as explained e.g. in \cite{top_depol_timescale}. In \cite{Mahlon2010} the depolarization time 
\begin{align}
t_{\text{depol}} = \frac{\hbar\,m_t}{\Lambda_{\text{QCD}}^2} \approx 3 \cdot 10^{-21}\,\text{s}
\end{align}
is quoted, which is longer than $t_{\text{had}} = \frac{\hbar}{\Lambda_{QCD}} \approx 3 \cdot 10^{-24}\,\text{s}$.

In the following sections a description of the production and the decay mechanisms of the top quark is given, followed by an overview of its properties. This provides the basis for discussing the spin correlation of \ttbar\ pairs and the access to it via measurements in Section \ref{sec:SC}.

\subsection{Top Quark Production and Decay}
At hadron colliders -- to which this discussion will be limited\footnote{So far, no lepton collider has sufficient energy to produce top quarks.} -- top quarks can be produced in two ways: as single top quarks via the electroweak interaction or in pairs via the strong interaction. 

In both cases, the production process can be factorized into two components: The initial state prescription via the PDFs of a parton $i$ in a proton $p$,  $f_{i/p}$, and the cross section $\hat{\sigma}$ of the partonic hard interaction process. This separation is called \textit{factorization theorem}\index{Factorization theorem} and is described in \cite{fact_theorem, fact_theorem2}. In order to factorize, two energy scales need to be defined. The first one is called \textit{factorization scale}\index{Factorization scale} $\mu_F$, separating the perturbative from the non-perturbative part. The second one, the renormalization scale $\mu_R$, has already been introduced in Section \ref{sec:QCD}. 

For inclusive top quark pair production in proton-proton collisions the factorized cross section at a \com\ enery $\sqrt{s}$ reads \cite{ttbar_xsec_theo}
\begin{align}
  \sigma_{pp \rightarrow {t\bar{t}X}}(s ,m_t^2) =
  \sum\limits_{i,j = q,{\bar{q}},g}
  \int\limits_{4m_t^2}^{s}  d \hat{s} ~L_{ij}(\hat{s}, s, \mu_f^2) ~
   \hat{\sigma}_{ij \to \ttbar} (\hat{s},m_t^2,\mu_f^2,\mu_r^2)\, .
\end{align}
with the partonic density
\begin{align}
  L_{ij}(\hat{s},s,\mu_f^2)= 
  \frac{1}{s} \int\limits_{\hat{s}}^s 
    \frac{d\tilde{s}}{\tilde{s}} f_{i/p}\left(\mu_f^2,{\frac{\tilde{s}}{s}}\right) 
    f_{j/p}\left(\mu_f^2,{\frac{\hat{s}}{\tilde{s}}}\right).
\end{align}
For \ttbar\ production one usually sets $\mu_R = \mu_F = m_t$, so to the mass scale of the process of interest. 

\subsubsection{\ttbar\ Production via Strong Interaction}
\begin{figure}[ht]
	\centering
			\subfigure[]{
		\includegraphics[width=0.4\textwidth]{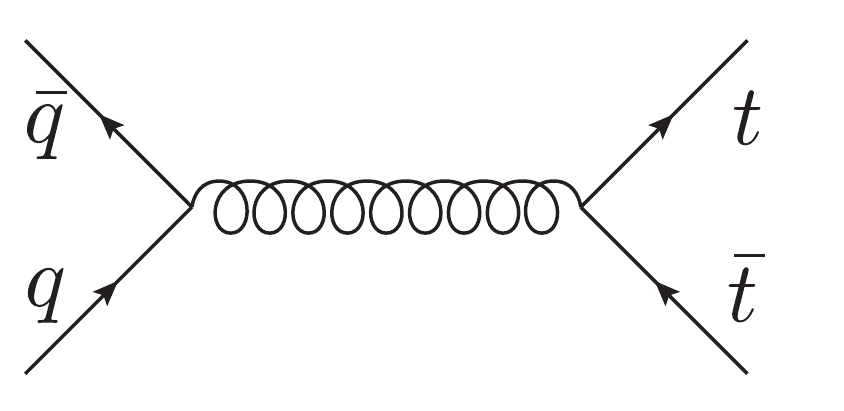}
			\label{fig:ttbar_prod_feyn_qq}
		}
					\subfigure[]{
		\includegraphics[width=0.4\textwidth]{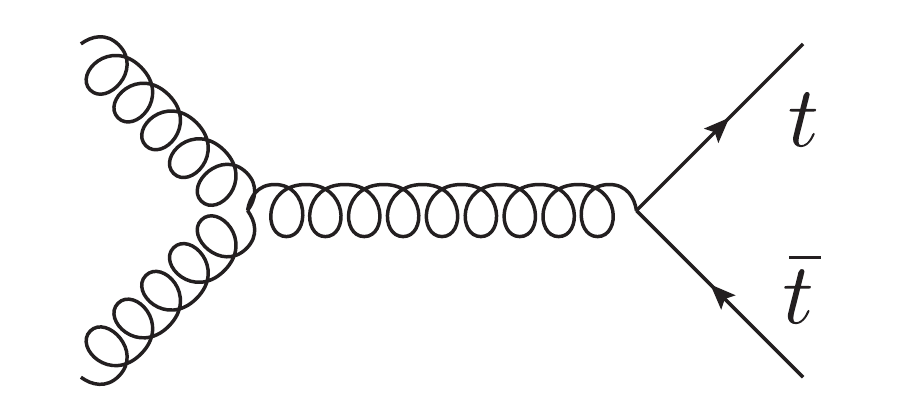}
			\label{fig:ttbar_prod_feyn_gg1}
		}\\
	\subfigure[]{
		\includegraphics[width=0.4\textwidth]{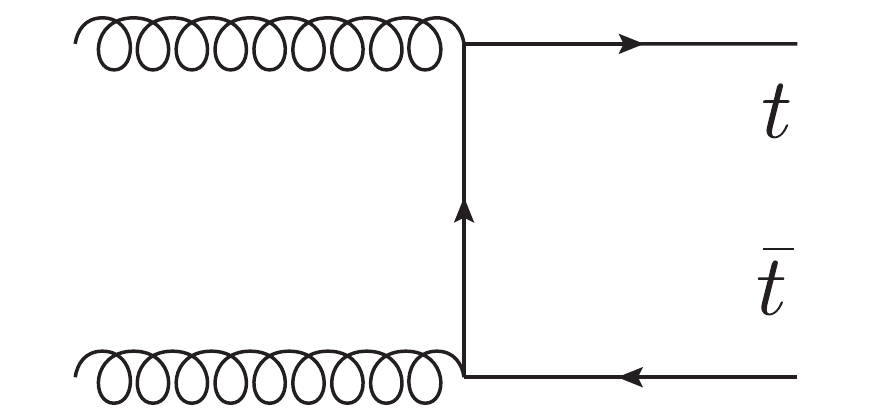}
			\label{fig:ttbar_prod_feyn_gg2}
		}
			\subfigure[]{
		\includegraphics[width=0.4\textwidth]{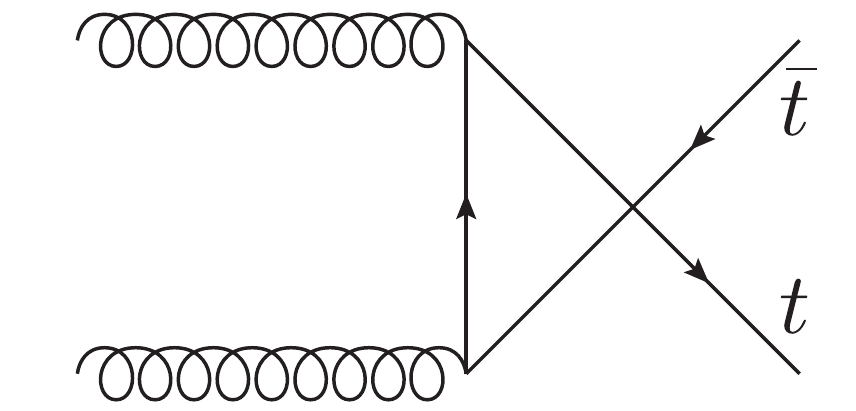}
			\label{fig:ttbar_prod_feyn_gg3}
		}

		\caption{\ttbar\ production via strong interactions. \subref{fig:ttbar_prod_feyn_qq} Quark/antiquark annihilation, \subref{fig:ttbar_prod_feyn_gg1}-\subref{fig:ttbar_prod_feyn_gg3} gluon fusion. }
\label{fig:ttbar_prod_feyn}
\end{figure}

At hadron colliders top quarks are dominantly produced in pairs via the strong interaction.
Figure \ref{fig:ttbar_prod_feyn} shows the different ways of \ttbar\ production at leading order. 

The PDFs determine the initial state and also the contributions of the different diagrams. By grouping into quark-antiquark annihilation (Figure \ref{fig:ttbar_prod_feyn_qq}) and gluon fusion (figures \ref{fig:ttbar_prod_feyn_gg1} - \ref{fig:ttbar_prod_feyn_gg3}), two statements can be made, which are of importance for the analysis of \ttbar\ spin correlation:
\begin{itemize}
\item The higher $\sqrt{s}$, the lower the needed $x$ for \ttbar\ production. The dominating partons for low $x$ are gluons. Hence, for high $\sqrt{s}$, in particular for LHC energies, the process $gg \rightarrow \ttbar$ is dominating. In contrast to the LHC, $q\bar{q}\rightarrow \ttbar$ is the dominating process at the Tevatron. Figure \ref{fig:PDF} illustrates this. 
\item As antiquarks are only available as sea quarks in the case of the LHC,  $q\bar{q}\rightarrow \ttbar$ is suppressed. In the case of the Tevatron antiquarks are present as valence quarks in the anti-proton. 
\end{itemize}

In \cite{ttbar_xsec_theo2}, the \ttbar\ cross sections have been computed at next-to-next-to-leading order (NNLO) using the MSTW2008nnlo68cl PDF set \cite{MSTW} and assuming a top quark mass of $m_t = 173.3 \GeV$ via the TOP++ \cite{Czakon:2011xx} program. Table \ref{tab:ttbar_xsec_theo} shows the results for the Tevatron and the LHC accelerators for different \com\ energies. 

{\renewcommand{\arraystretch}{1.2}
\begin{table}[htbp]
\begin{center}
\begin{tabular}{|c|c|c|}
\hline
Accelerator  & $\sqrt{s}~\left[\TeV\ \right]$ & $\sigma_{\ttbar}  \pm \text{scale unc.} \pm \text{PDF unc.}~\left[ \text{ pb }\right] $\\

\hline
\hline
Tevatron & 2 & $7.164~{}^{+0.110}_{-0.200}~{}^{+0.169}_{-0.122}$\\

\hline
\multirow{3}{*}{LHC} & 7&$172.0~{}^{+4.4}_{-5.8}~{}^{+4.7}_{-4.8}$\\

{}& 8&$245.8~{}^{+6.2}_{-8.4}~{}^{+6.2}_{-6.4}$\\

{}& 14&$953.6~{}^{+22.7}_{-33.9}~{}^{+16.2}_{-17.8}$\\

\hline

\end{tabular}
\end{center}
\caption{\ttbar\ production cross sections at NNLO+NNLL for different accelerators and \com\ energies calculated for a top quark mass of $m_t = 173.3 \GeV$ \cite{ttbar_xsec_theo2}.}
\label{tab:ttbar_xsec_theo}
\end{table}
}

\begin{figure}[ht]
	\centering
		\includegraphics[width=0.75\textwidth]{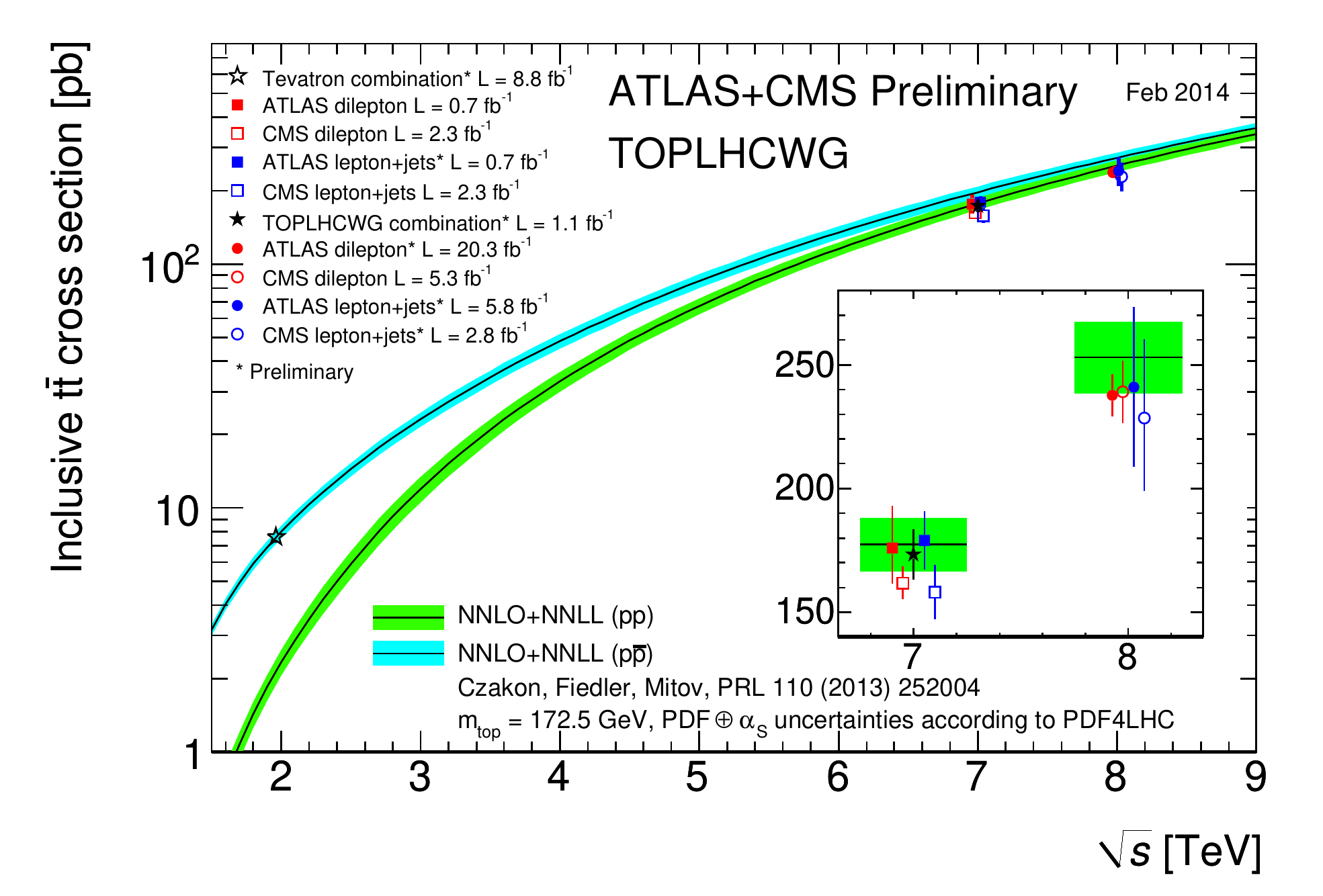}
		\caption{Comparison of the measured \ttbar\ cross sections at the Tevatron and the LHC using input values from \cite{top_xsec1, top_xsec2,top_xsec3,top_xsec4,top_xsec5,top_xsec6,top_xsec7,top_xsec8,top_xsec9,top_xsec10} and the predictions from \cite{ttbar_xsec_theo2}. The figure is taken from \cite{top_summary_plots}.}
	\label{fig:ttbar_xsec}
\end{figure}

Meanwhile all of the predicted \ttbar\ cross sections for up to $\sqrt{s} = 8\,\TeV$, listed in Table \ref{tab:ttbar_xsec_theo}, can be compared to measured values. An overview of all \ttbar\ cross section measurements and a comparison to the theory predictions is shown in Figure \ref{fig:ttbar_xsec}.
The measurements are in good agreement with the predictions.

\subsubsection{Single Top Production Via Weak Interaction}
\begin{figure}[ht]
	\centering
			\subfigure[]{
		\includegraphics[width=0.31\textwidth]{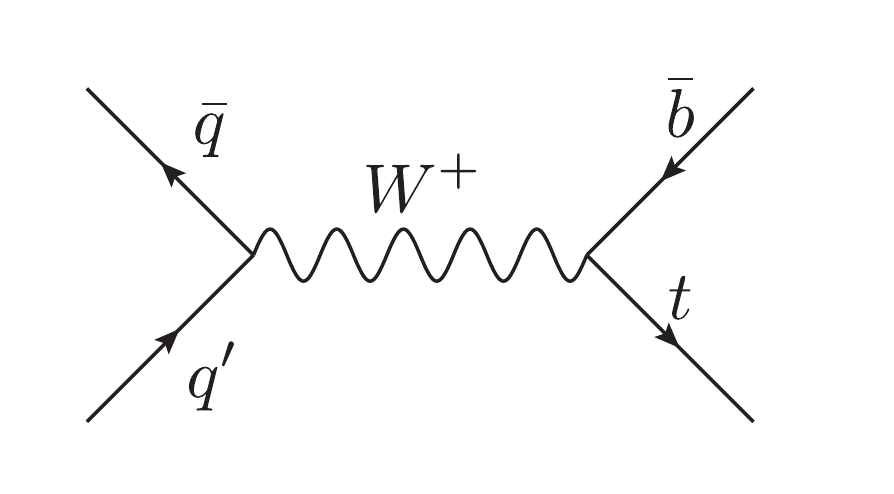}
			\label{fig:singletop_s}
		}
					\subfigure[]{
		\includegraphics[width=0.31\textwidth]{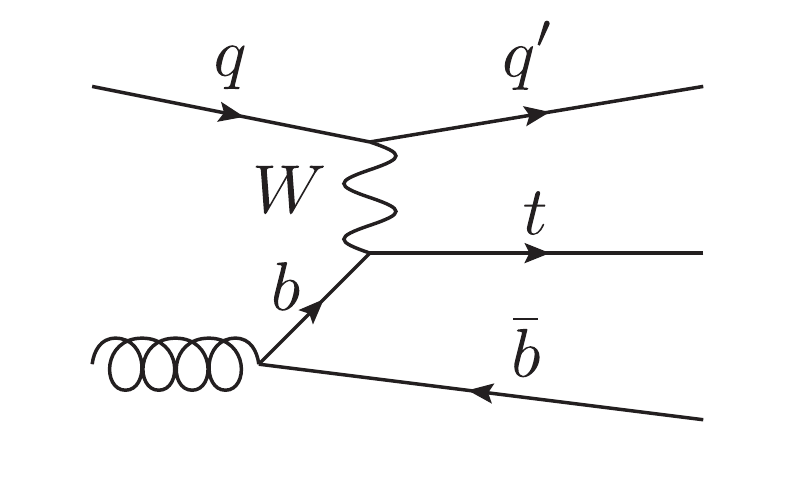}
			\label{fig:singletop_t}
		}
	\subfigure[]{
		\includegraphics[width=0.31\textwidth]{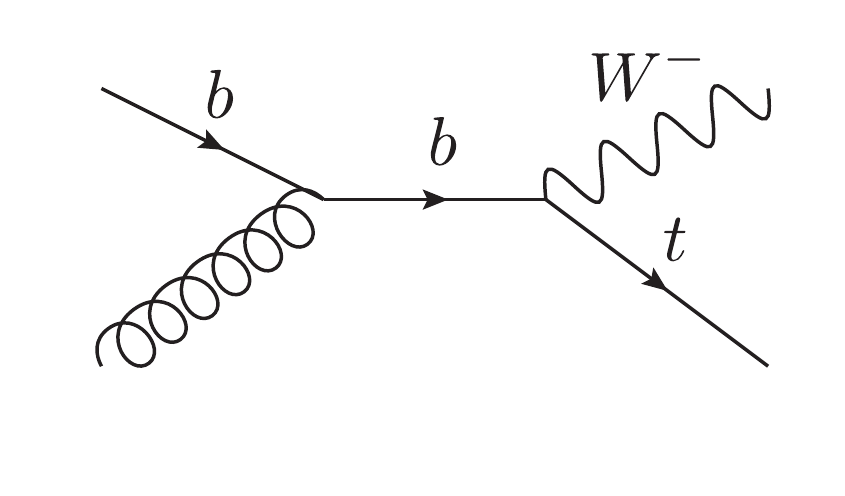}
			\label{fig:singletop_Wt}
		}
		\caption{Single top quark production via electroweak interactions in \subref{fig:singletop_s} the $s$-channel, \subref{fig:singletop_t} the $t$-channel and \subref{fig:singletop_Wt} produced in association with a $W$ boson ($Wt$ channel). }
		\label{fig:singletop}
\end{figure}

Single top quarks can be produced in several ways as illustrated in Figure \ref{fig:singletop}: via the $s$- or the $t$-channel or in association with a $W$ boson ($Wt$-channel). 
In contrast to the production of \ttbar\ pairs, the single top production channels can be measured individually. Predictions of the cross sections at NNLO were made in \cite{st_xsec_theo_s, st_xsec_theo_t, st_xsec_theo_Wt} and are listed in Table \ref{tab:st_xsec_theo}.

 {\renewcommand{\arraystretch}{1.2}
\begin{table}[htbp]
\begin{center}
\begin{tabular}{|c|c|c|c|c|c|c|c|}
\hline
{}&{} &\multicolumn{2}{c|}{$s$-channel}&\multicolumn{2}{c|}{$t$-channel}&\multicolumn{2}{c|}{$Wt$-channel}\\
Accelerator  & $\sqrt{s}~\left[\TeV\ \right]$& $\sigma_{t}~\left[ \text{pb}\right] $ & $\sigma_{\bar{t}} ~\left[ \text{pb}\right] $ &$\sigma_{t} \left[ \text{pb}\right] $ & $\sigma_{\bar{t}}~\left[ \text{pb}\right] $&$\sigma_{t}~\left[ \text{pb}\right] $ & $\sigma_{\bar{t}}~\left[ \text{pb}\right] $\\
\hline
\hline
Tevatron & 2 			& \multicolumn{2}{c|}{0.52}& \multicolumn{2}{c|}{1.04}& \multicolumn{2}{c|}{---}\\

\hline
\multirow{2}{*}{LHC} & 7	&3.17 &1.42& 41.7 & 22.5& \multicolumn{2}{c|}{7.8}\\

{}& 14				&7.93 &3.99& 151 & 91.6& \multicolumn{2}{c|}{41.8}\\

\hline

\end{tabular}
\end{center}
\caption{Calculated single top production cross sections at NNLO+NNLL for different accelerators and \com\ energies for $m_t = 173.3~\GeV$ \cite{st_xsec_theo_s, st_xsec_theo_t, st_xsec_theo_Wt}. The $s$- and $t$-channel cross sections are symmetric for $t$ and $\bar{t}$ at the Tevatron. The same is true for the $Wt$ cross section at the LHC. The quoted symmetric cross sections refer to both the $t$ and the $\bar{t}$ cross sections, not the sum.}
\label{tab:st_xsec_theo}
\end{table}
}
A variety of cross section measurements at both the LHC and the Tevatron exist, briefly summarized in Table \ref{tab:st_xsec_meas}.
 {\renewcommand{\arraystretch}{1.2}
\begin{table}[htbp]
\begin{center}
\begin{tabular}{|c|c|c|c|c|}
\hline
 {}&{}& \multicolumn{3}{c|}{$\sigma_{t} + \sigma_{\bar{t}} \text{ [pb]}$}\\
 $\sqrt{s} \left[\TeV\ \right]$&Experiment&$s$-channel&$t$-channel&$Wt$-channel\\
\hline
\hline
 \multirow{2}{*}{2}& CDF 			& \multirow{2}{*}{$1.29{\,}^{+0.26}_{-0.24}$ \cite{st_xsec1}} & $1.49{\, }^{+0.47}_{-0.42}$ \cite{st_xsec2} & ---\\
 {}& D0 										& {} & $3.07{\, }^{+0.54}_{-0.49}$ \cite{st_xsec3} & ---\\
\hline
  \multirow{2}{*}{7}	&ATLAS			& $<20.5$ \cite{st_xsec6}& $ 83{\, }^{+20}_{-19}$ \cite{st_xsec5} & $ 16.8 \pm 5.7$ \cite{st_xsec7}\\
 {}& CMS 										& --- & $ 67.2 \pm {\;\,}6.1$ \cite{st_xsec8} & $ 16{\, }^{+5}_{-4}$ \cite{st_xsec9}\\  
\hline
  \multirow{2}{*}{8}				&ATLAS			&  --- & $ 82.6 \pm 12.1$ \cite{st_tchan_ATLAS_new} & $ 27.2 \pm 5.8$ \cite{st_xsec11}\\
 {}& CMS 										& $<11.5$ \cite{st_xsec10} & $ 83.6 \pm {\;\, }7.7$ \cite{st_tchan_CMS_new} & $ 23.4 \pm 5.4$ \cite{st_xsec12}\\
\hline

\end{tabular}
\end{center}
\caption{Measured cross sections and limits on the single top and anti-top cross sections \cite{st_xsec1, st_xsec2,st_xsec3,st_xsec4,st_xsec5,st_xsec6,st_xsec7,st_tchan_CMS_new,st_xsec9,st_xsec10,st_xsec11,st_xsec12}. Limits are quoted at 95\,\% CL.}
\label{tab:st_xsec_meas}
\end{table}}
All measurements are in good agreement with the SM prediction at NNLO precision. Events where a single top quark is produced are one of the main backgrounds for the analysis of \ttbar\ spin correlation. 

\subsubsection{Top Quark Decay}
\label{sec:topdecay}
\begin{figure}[ht]
	\centering
		\includegraphics[width=0.7\textwidth]{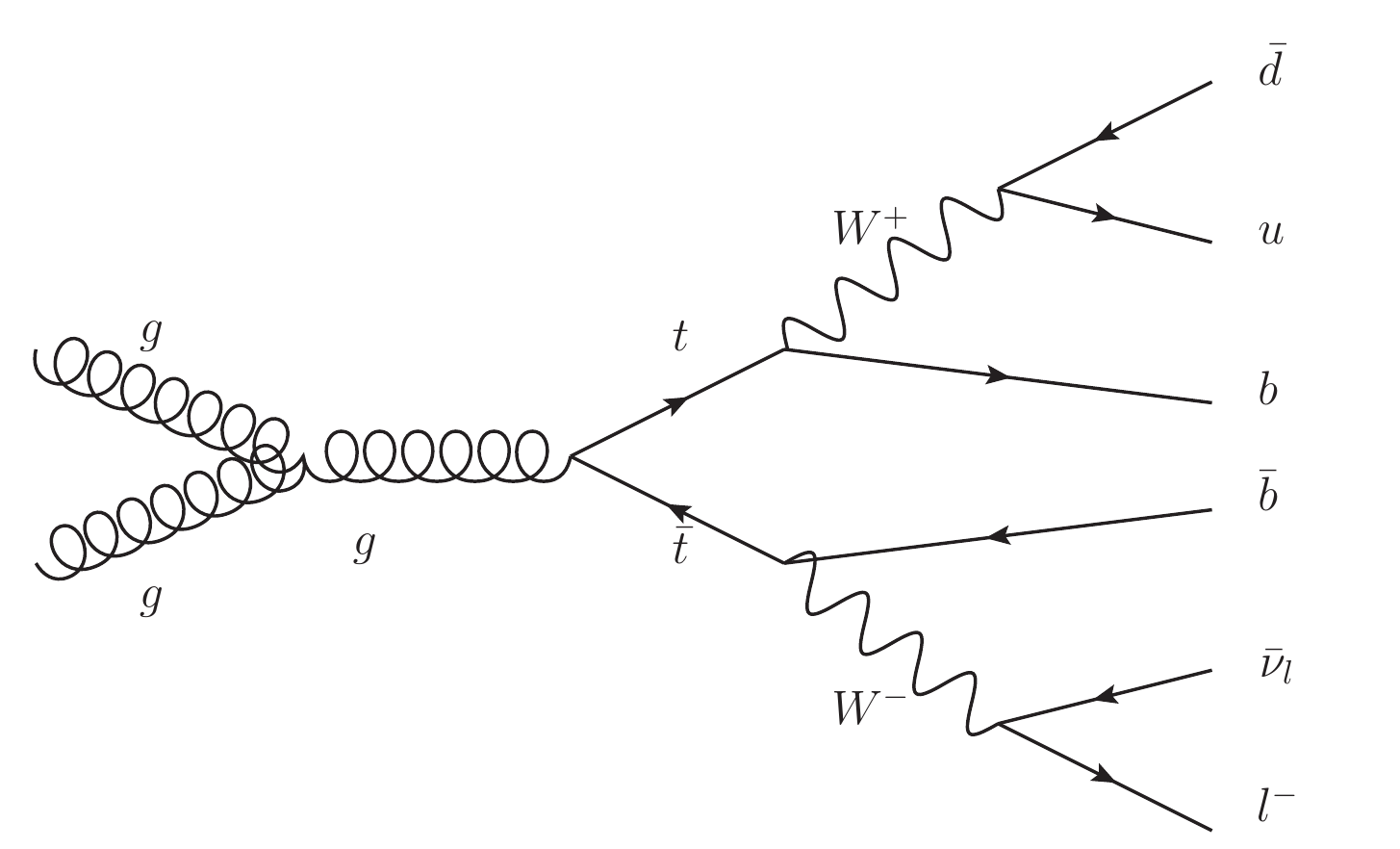}
		\caption{The decay of a \ttbar\ pair in the \ljets\ channel. }
		\label{fig:ttbar_decay_ljets}
\end{figure}

\begin{figure}[ht]
	\centering
			\subfigure[]{
		\includegraphics[width=0.7\textwidth]{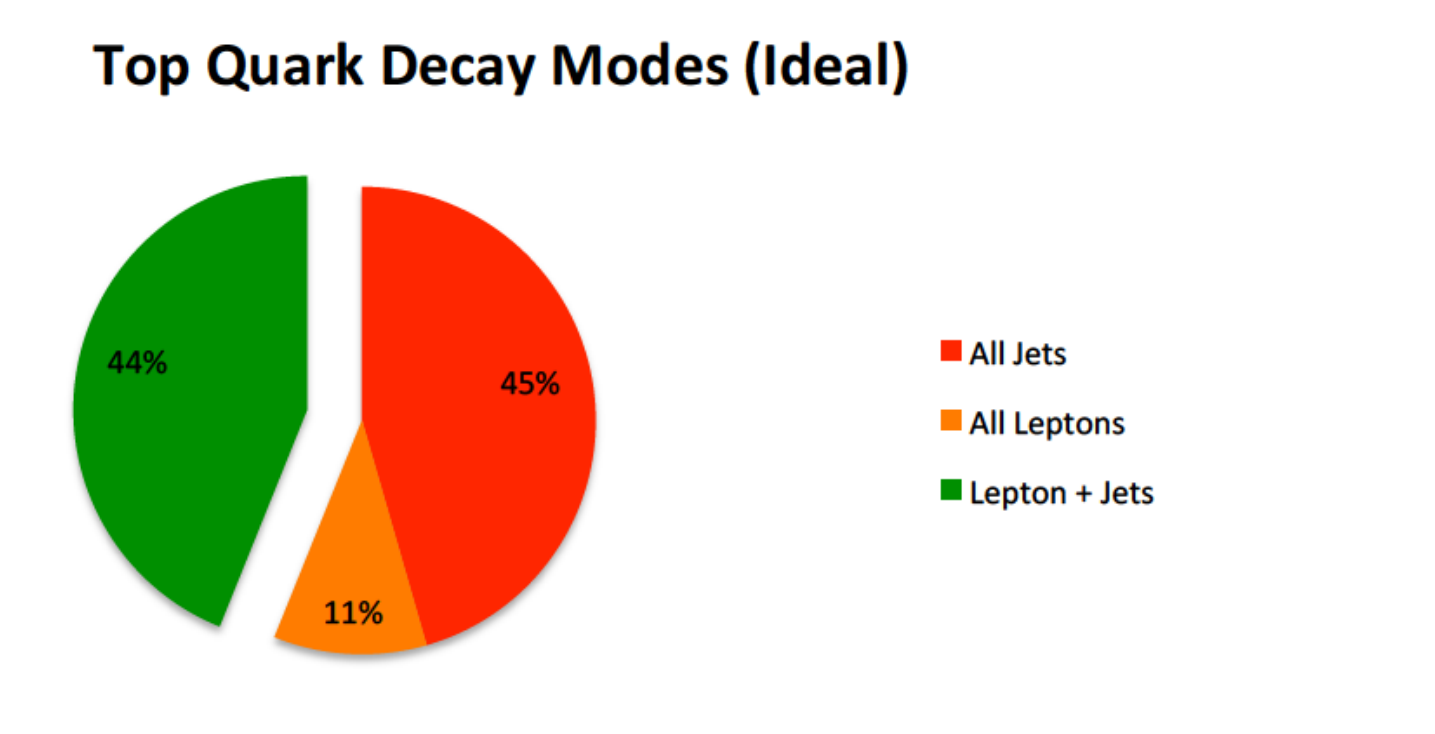}
			\label{fig:BR1}
		}
					\subfigure[]{
		\includegraphics[width=0.7\textwidth]{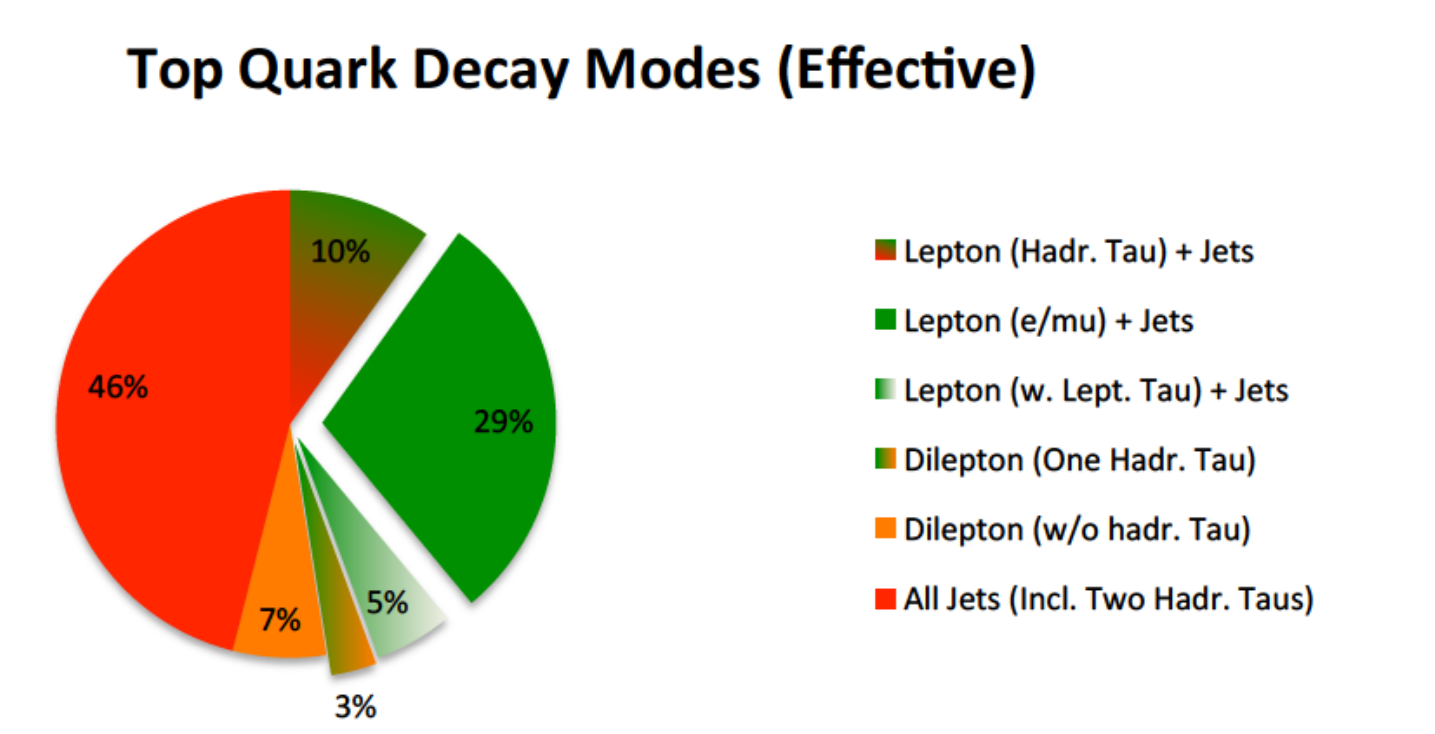}
			\label{fig:BR2}
		}
		\caption{\subref{fig:BR1} Decay channels of \ttbar\ pairs and their relative rates. \subref{fig:BR2} A more detailed view of \subref{fig:BR1} including migration effects caused by $\tau$ leptons. }
		\label{fig:BR}
\end{figure}

The top quark decays via the weak interaction. While decays via the weak interaction usually take place at larger time scales than via the strong interaction, top quark decays are still at a short time scale due to the top quark's high mass. As the decay rates of a top quark into a $W$ boson and a quark $q$ are proportional to the squared absolute value of the CMK matrix element $\left| V_{tq}\right|^2$ and $\left| V_{tb}\right| = 0.999146^{+0.000021}_{-0.000046}$ \cite{PDG}, the top quark can be considered as decaying uniquely via $t \rightarrow W^{+} b$. Hence, the final state of a top quark decay is determined by the decay of the $W^{+}$ boson. In 67.7\,\% of the cases, the $W^{+}$ boson decays into hadrons \cite{PDG}, leading to a \textit{hadronically decaying}\index{Hadronic decay} top quark. If not decaying into hadrons, the $W^{+}$ boson decays into a charged anti-lepton $\bar{\ell}$ and the according neutrino ($\bar{e}{\nu}_e$, $\bar{\mu} {\nu}_{\mu}$, $\bar{\tau} {\nu}_{\tau}$). For each of the charged leptons the probabilities are almost equal \cite{PDG}. 
At first order the \ttbar\ decay channels can be grouped into the following modes:
\begin{align}
\text{All jets} &: \begin{cases}\ttbar\ \rightarrow b\bar{b}W^{+}W^{-} \rightarrow b\bar{b}  q\bar{q}'q''\bar{q}'''\end{cases}\\
\text{Dilepton}&: \begin{cases}\ttbar\ \rightarrow b\bar{b}W^{+}W^{-} \rightarrow b\bar{b} ~\bar{\ell} \ell^{'} \nu_{\ell} \bar{\nu}_{\ell^{'}} \end{cases}\\
\text{Lepton + jets}&: \begin{cases}\ttbar\ \rightarrow b\bar{b}W^{+}W^{-} \rightarrow b\bar{b} ~\bar{\ell} \nu_{\ell}q\bar{q'}\\
\ttbar\ \rightarrow b\bar{b}W^{+}W^{-} \rightarrow b\bar{b} ~\ell \bar{\nu}_{\ell}q\bar{q}'
\end{cases}
\label{eq:top_decays}
\end{align}
Each decay channel of the \ttbar\ pair\footnote{Even though a channel usually refers to a single particle, this expression is commonly used.}  has advantages and disadvantages. The choice of the decay channel for an analysis is thus always a trade-off and there is no ad-hoc recommendation. Practical issues that should be considered when analysing \ttbar\ pairs are:
\begin{itemize}
\item \textbf{All jets channel} Due to the high branching fraction of $W \rightarrow \text{hadrons}$ this channel leads to a large sample of selected events. However, the signal-to-background ratio will be quite low as the sample will be contaminated by multijet background to a large extent. Furthermore, the resolution of the jets is worse than the one of leptons. Complex event reconstructions will suffer from this fact. A correct assignment of the six jets (two $b$-jets and four light jets from the hadronic $W$ decay) will be very difficult and the combinatorial background quite large and hard to suppress. 
\item \textbf{Dilepton channel} Two leptons with a good reconstruction efficiency and energy/momentum resolution can easily be reconstructed as \ttbar\ decay products. Even the charge of these objects can be determined, allowing for a correct assignment to the top and anti-top quark. Furthermore, the requirement of two charged leptons suppresses multijet background to a large extent. The dilepton channel has a smaller event yield than the all jets channel. Reconstructing the full events is a non-trivial issue as the two neutrinos cannot be reconstructed. Using momentum conservation in the transverse plane and calculating the missing transverse momentum (see Section \ref{sec:etmiss}) allows measuring the vectorial sum of the two neutrino's transverse momenta indirectly. Still, the longitudinal neutrino momenta stay undetermined. Complex reconstruction algorithms such as \newword{Neutrino Weighting} \cite{neutrinoweighting}, demanding kinematic assumptions as input, are required. 
\item \textbf{Lepton+jets channel} The lepton+jets (\ljets) channel is a compromise between the other two. Multijet background is suppressed due to the charged lepton in the event. The suppression is not as powerful as in the dilepton channel. As only one neutrino is present, the event kinematics are no longer underconstrained as in the case of dilepton events. This allows for a full event reconstruction. A light up-type quark jet replaces the second neutrino (compared to the dilepton channel) and can be reconstructed. But the hadronic equivalent to the second charged lepton, a down-type quark jet, suffers from worse reconstruction efficiency and a less precise measurement. 
\end{itemize}
The \ljets\ channel was chosen for the analysis presented in this thesis. It offers a full event reconstruction, a high event yield and a sufficiently low multijet background. The technique used for full event reconstruction is described in Section \ref{sec:KLFitter}. Figure \ref{fig:ttbar_decay_ljets} shows the decay of a \ttbar\ pair in the \ljets\ channel. 
The relative yields for each of the three channels is shown in Figure \ref{fig:BR1}. 
It should be mentioned at this point that the distinction into three different channels is idealized. When analysing events on the basis of reconstructed quantities, migration effects between the channels have to be taken into account. A crucial role is played by the $\tau$ lepton. It will decay into either another charged lepton and a neutrino or into hadrons within $\tau_{\tau} = 2.9 \cdot 10^{-13}~\text{s}$ \cite{PDG}.\footnote{This lifetime is too short for a direct detection in the detector but allows for the reconstruction of a secondary vertex.} In the latter case this leads to a wrong assignment of a leptonically decaying top quark to a hadronically decaying top quark. Thus, Figure \ref{fig:BR2} gives a more realistic picture of the channels that are actually reconstructed, including migration effects. These are:
\begin{itemize}
\item Events from the dilepton channel can migrate to the \ljets\ channel in case they contain one $\tau$ decaying into hadrons, misidentified as jet. 
\item Events from the \ljets\ channel can be identified as such, but the reconstructed lepton does not stem directly from a $W$ boson, but from a leptonically decaying $\tau$. The lepton properties do not match the expectations. This causes a part of the \ttbar\ signal being -- from the physics point of view -- in fact a background.  
\item  In case the lepton from the \ljets\ channel is a hadronically decaying $\tau$, reconstructed as a jet, the event will not pass the \ljets\ event selection. 
\end{itemize}

\subsection{Measured Top Quark Properties}
\label{sec:topprop}
Its properties make the top quark unique: As it is the heaviest fermion -- and even elementary particle -- its lifetime is too short to create any bound states. Hence, the spin configuration is directly transferred to its decay products. Further, the high mass implies a Yukawa coupling to the Higgs field of about one, making it an important probe for Higgs physics. This section gives an overview of several important top quark property measurements.
\subsubsection{Mass}
The mass of the top quark has been of great interest since the very beginning. On the one hand, loop corrections of the $W$ and $Z$ boson mass allowed for its prediction without a direct measurement. On the other hand, a direct measurement could provide an important input for electroweak fits and loop correction calculations.
 
\begin{figure}[ht]
	\centering
		\includegraphics[width=0.75\textwidth]{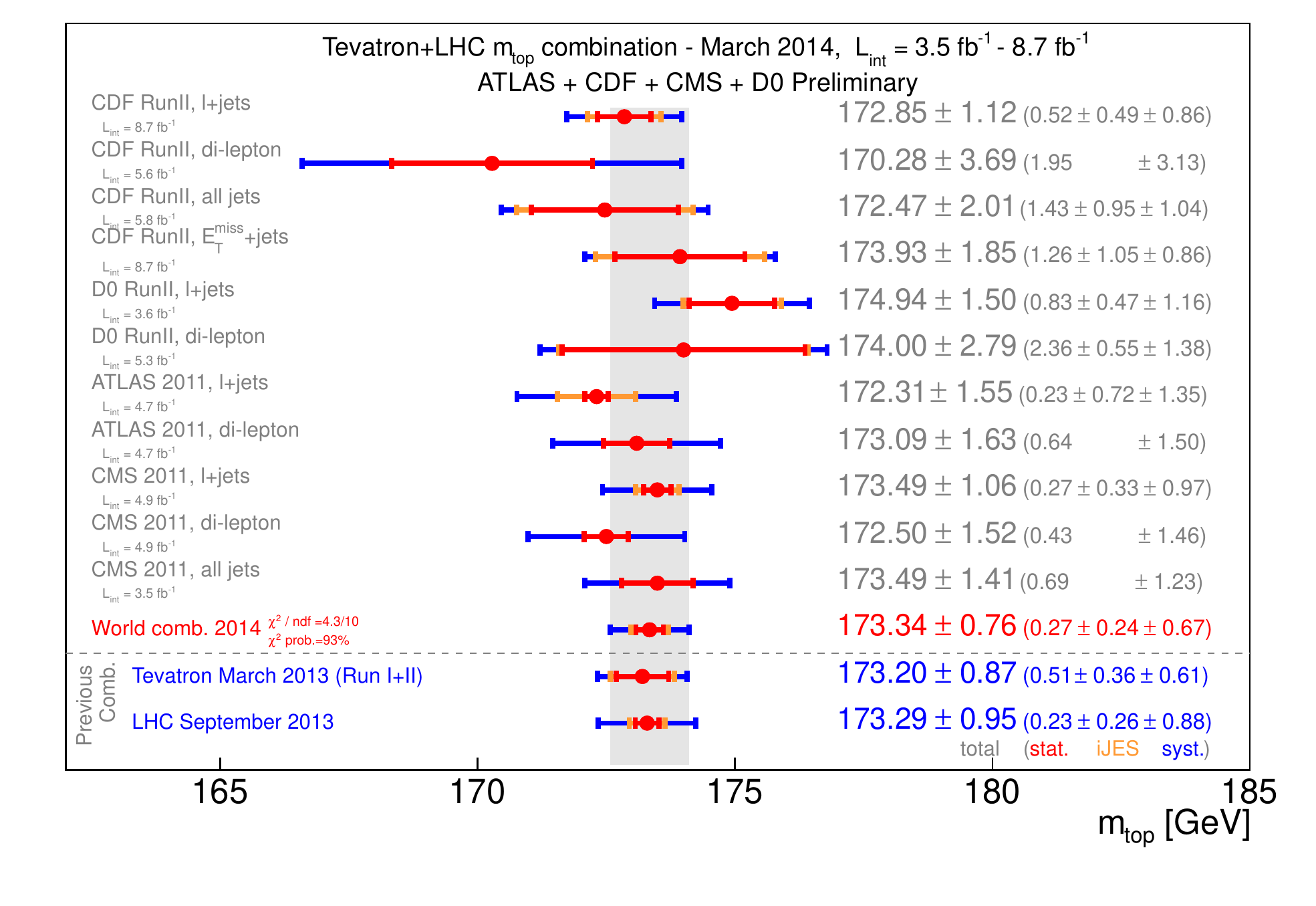}
		\caption{World combination of the top quark mass \cite{m_top_world} together with the input results \cite{m_top_world_input1,m_top_world_input2,m_top_world_input3,m_top_world_input4,m_top_world_input5,m_top_world_input6,m_top_world_input7,m_top_world_input8,m_top_world_input9,m_top_world_input10,m_top_world_input11} and the dedicated Tevatron \cite{m_top_tevatron} and LHC combinations \cite{m_top_LHC}.}
	\label{fig:m_top_world}
\end{figure}

Today, combinations of individual top mass measurements of the two Tevatron experiments \cite{m_top_tevatron}, the two LHC experiments \cite{m_top_LHC} as well as a world combination exist \cite{m_top_world}. The latter one yields $m_t = 173.34 \pm 0.76 \GeV$. Figure \ref{fig:m_top_world} gives an overview of the world combination and its input values taken from the individual measurements \cite{m_top_world_input1,m_top_world_input2,m_top_world_input3,m_top_world_input4,m_top_world_input5,m_top_world_input6,m_top_world_input7,m_top_world_input8,m_top_world_input9,m_top_world_input10,m_top_world_input11}. 

\subsubsection{Charge}
By using methods of measuring the charge of jets via the charges of their associated tracks, the top quark's charge has been studied extensively. Scenarios of exotic top quarks with a charge of $- \frac{4}{3}$ were excluded and a good agreement with the SM prediction of $q_t = \frac{2}{3}$ was found by the CDF \cite{top_charge_CDF}, D0 \cite{top_charge_D0} and CMS experiments  \cite{top_charge_CMS}. ATLAS has quoted a direct measurement of $q_t = 0.64 \pm 0.08$ \cite{top_charge_ATLAS}. All analyses were carried out in the \ljets\ channel.

\subsubsection{Top Charge Asymmetry}
In \ttbar\ production the kinematic distributions $\pt$, $\eta$ and $\phi$ of the top and the anti-top quark are equivalent \cite{top_asymm_theo} at LO. Interference terms at NLO, however, cause a difference of the top and anti-top rapidities $y$ in case of production via $q\bar{q}$ annihilation \cite{top_asymm_theo}. The Tevatron as $p\bar{p}$ collider has well-defined directions of the annihilating $q$ and $\bar{q}$.\footnote{This is true for the dominating case in which these quarks are valence quarks.}  Hence, an asymmetry 
\begin{align}
A_{\text{FB}} = \frac{N(y_t > y_{\bar{t}}) - N(y_t < y_{\bar{t}})}{N(y_t > y_{\bar{t}}) + N(y_t < y_{\bar{t}})}
\label{eq:AFB}
\end{align}
can be calculated \cite{top_asymm_theo2} and measured. A tension between the measurement and the SM prediction with a significance of more than $2\,\sigma$ has been observed by the CDF collaboration \cite{top_asymm_CDF}. In a similar fashion an asymmetry of leptons from top and anti-top quark decays $A_{\text{FB}}^{\text{lep}}$ was measured to be also more than $2\,\sigma$ above the SM prediction \cite{top_asymm_lep_CDF}. The D0 measurements of $A_{\text{FB}}$ \cite{top_asymm_D0} and $A_{\text{FB}}^{\text{lep}}$ \cite{top_asymm_lep_D0} were, in contrast to the CDF measurements, compatible with the SM expectation. 

\begin{figure}[ht]
	\centering
		\includegraphics[width=0.55\textwidth]{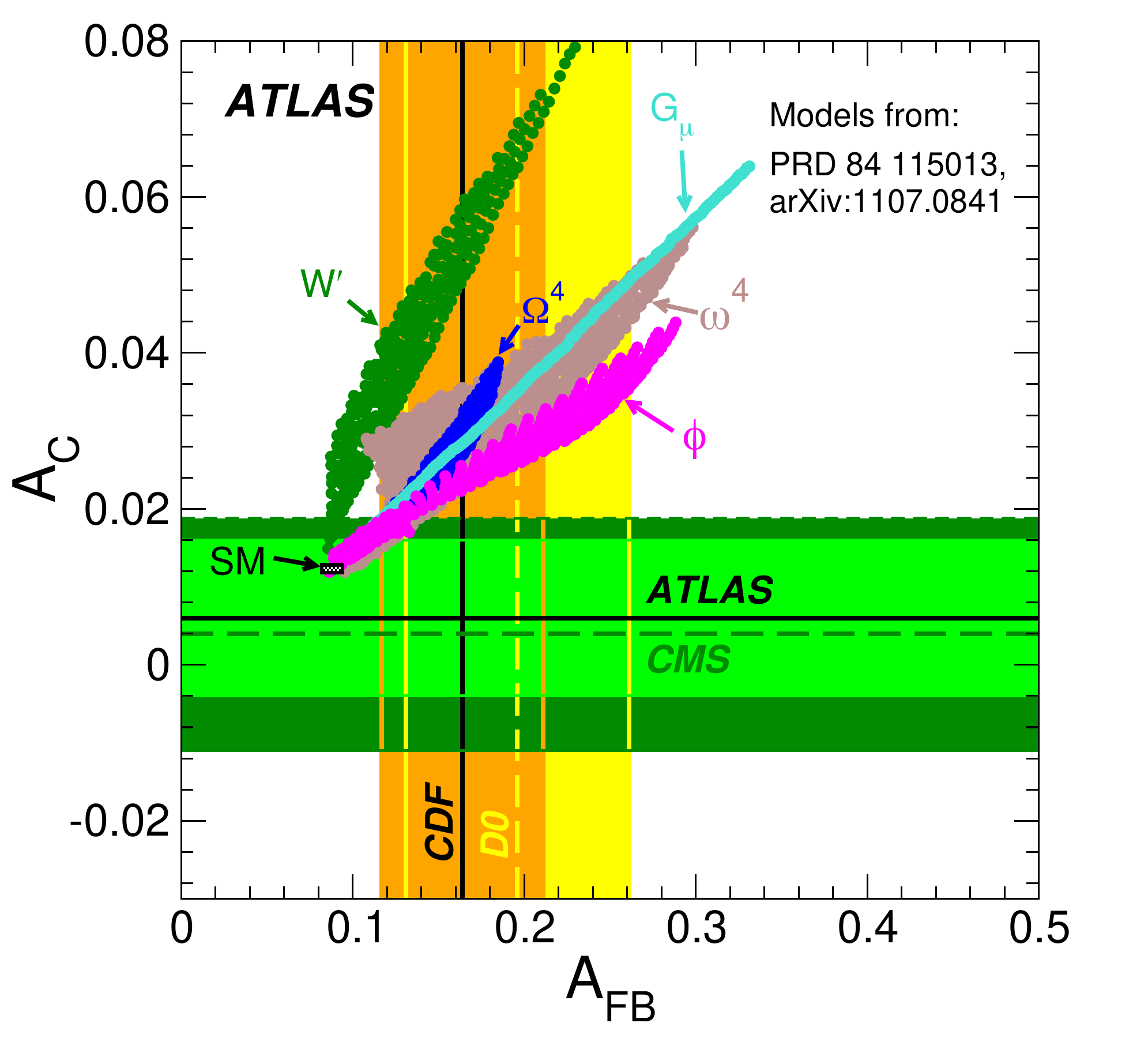}
		\caption{Comparison of the measurements of $A_{\text{FB}}$ and $A_{\text{C}}$ \cite{top_asymm_ATLAS, top_asymm_D0_old, top_asymm_CMS, top_asymm_CDF} to BSM scenarios from \cite{top_asymm_BSM1, top_asymm_BSM2} as shown in \cite{top_asymm_ATLAS}.}
			\label{fig:charge_asymm_BSM}
\end{figure}

At the LHC, a measurement of $A_{\text{FB}}$ is not possible due to the symmetric $pp$ production. However, as the valence quarks have on average a higher momentum than the sea antiquarks (see Figure \ref{fig:PDF}) a non-vanishing charge asymmetry 
\begin{align}
A_{\text{C}} = \frac{N(\left|y_t\right| - \left| y_{\bar{t}} \right| > 0) - N(\left|y_t\right| - \left| y_{\bar{t}} \right| < 0)}{N(\left|y_t\right| - \left| y_{\bar{t}} \right| > 0) + N(\left|y_t\right| - \left| y_{\bar{t}} \right| < 0)}
\end{align}
is predicted \cite{top_asymm_theo2}. The overall effect is expected to be small due to the charge symmetric $gg$ fusion process dominating at the LHC. The measurements of  $A_{\text{C}}$ by ATLAS and CMS at $\sqrt{s} = 7\,\TeV$ were combined in \cite{top_asymm_LHC} and reported to be consistent with the SM prediction. 
In \cite{top_asymm_BSM1, top_asymm_BSM2} BSM scenarios modifying both $A_{\text{FB}}$ and $A_{\text{C}}$ were calculated. Figure \ref{fig:charge_asymm_BSM} shows a comparison of these BSM predictions to the measurements of ATLAS, CMS, CDF and D0. For the latter measurement an older result \cite{top_asymm_D0_old} was used for which also a tension to the SM prediction was observed. 
Considering the updated D0 result \cite{top_asymm_D0}, a good agreement with the SM and an exclusion of a large parameter space for certain BSM scenarios was observed. 

\subsubsection{Branching Fractions / Couplings}
The assumption that $\left| V_{tb}\right| \approx 1$ is certainly sensible (see Section \ref{sec:topdecay}) and validated in global fits \cite{PDG}. Still, a direct measurement is well motivated. Such measurements make use of the orthogonality property of the CKM matrix. This way $\left| V_{tb}\right|$ can be measured as a ratio of the branching fractions $B$:
\begin{align}
B \left( t \rightarrow Wb \right) / \sum_{q=d,s,b}B \left( t \rightarrow Wq\right) = \left| V_{tb}\right|^2 / \sum_{q=d,s,b} \left| V_{tq}\right|^2 =  \left| V_{tb}\right|^2 .
\end{align}
By assuming unitarity of the CKM matrix and the existence of three quark generations CDF measures $\left| V_{tb}\right| = 0.97 \pm 0.05$ in the \ljets\ decay channel and $\left| V_{tb}\right| = 0.93 \pm 0.04$ in the dilepton decay channel.
The D0 experiment measures a limit of $\left| V_{tb}\right| > 0.92$ @95\,\% CL \cite{st_xsec3} in the combined $s$+$t$-channel single top cross section measurement. 
At the LHC, CMS combined the 7 and 8 TeV results of the $t$-channel single top results to $\left| V_{tb}\right| = 0.998 \pm 0.038\,\text{(exp.)}\,\pm 0.016\,\text{(theo.)}$ \cite{st_tchan_CMS_new} while ATLAS measured $\left| V_{tb}\right| = 0.97{}^{+0.09}_{-0.10}$ in the 8 TeV single top $t$-channel measurement \cite{st_tchan_ATLAS_new}.

To probe the weak and electromagnetic couplings of the top quark, $\ttbar V$ processes -- where $V$ represents an additional vector boson such as a photon, a $W$ or a $Z$ boson -- need to be investigated. It has to be stated that such studies need a very good understanding of the signal and background modelling. The inclusive measurements of $\ttbar V$ include initial and final state radiation of $V$ from other partons than the top quark as well as interference terms. 

At the Tevatron, CDF observed evidence of the $\ttbar\gamma$ process by measuring $\sigma_{\ttbar\gamma} = 0.18 \pm 0.08 \text{ pb}$ \cite{ttgamma_CDF}, also consistent with the SM prediction. Consistency with the SM was also measured by ATLAS ($\sigma_{\ttbar\gamma} = 2.0 \pm 0.9 \text{ pb}$) \cite{ttgamma_ATLAS} and CMS. CMS derived $\sigma_{\ttbar\gamma} = 2.4 \pm 0.6 \text{ pb}$ by measuring the ratio $\sigma_{\ttbar\gamma} / \sigma_{\ttbar}$ \cite{ttgamma_CMS}. 

Heavy gauge bosons in association with \ttbar\ events were investigated at the LHC by both CMS and ATLAS. ATLAS set an upper limit of $\sigma_{\ttbar Z} < 0.71 \text{ pb}$ @95\,\% CL by measuring final states with three leptons \cite{ttZ_ATLAS}. In the same final state CMS measured $\sigma_{\ttbar Z} = 0.28^{+0.15}_{-0.11} \text{ pb}$ \cite{ttV_CMS}. In the same publication the inclusive cross section $\sigma_{\ttbar Z} + \sigma_{\ttbar W}$ was also measured as $\sigma_{\ttbar Z} + \sigma_{\ttbar W}= 0.43^{+0.19}_{-0.17} \text{ pb}$.

\subsubsection{$W$ Boson Helicity}
One of the main properties of the top quark decay, that is used for spin correlation analyses, is the weak $V-A$ structure of its decay vertex. It requires \bQ s to be left-handed since $m_b \muchless m_t$. Hence, at LO the $W$ boson must either be left-handed as well or longitudinally polarized. At LO and by neglecting the \bQ\ mass, the fractions $\mathcal{F}$ of longitudinal, left- and right-handed $W$ polarizations read \cite{WHel_theo2, WHel_theo}
\begin{align}
\mathcal{F}_0 : \mathcal{F}_L : \mathcal{F}_R = \frac{1}{1+2x^2} : \frac{2x^2}{1+2x^2} : 0
\label{eq:helfractions}
\end{align} 
with $x = m_W / m_t$. Anomalous couplings caused by BSM physics will be reflected in deviations of the $W$ helicity fractions. Table \ref{tab:WHel} summarizes the measurements at the Tevatron and the LHC and compared to NLO SM predictions. All measurements agree with the SM predictions.

\begin{table}[htbp]
\begin{center}
\begin{tabular}{|c|c|c|c|}
\hline
{} &  Comb. CDF and D0 & Comb. ATLAS and CMS & NNLO SM\\
\hline
\hline
$\mathcal{F}_0$ 	& {\,\,\,\,\,}$0.722 \pm 0.081$	&$0.626 \pm 0.059$		& $0.687 \pm 0.005$\\
$\mathcal{F}_L$	& ---$^{*}$ 			&$0.359 \pm 0.035$		& $0.311 \pm 0.005$\\
$\mathcal{F}_R$	& $-0.033 \pm 0.046$ 	&$0.015 \pm 0.034$				& $0.0017 \pm 0.0001$\\
\hline
\end{tabular}
\end{center}
\caption{Measurements of the $W$ polarization fractions at the Tevatron \cite{WHel_tevatron} and the LHC \cite{WHel_LHC} together with the NLO SM predictions \cite{WHel_theo} (${}^{*}$No results of $\mathcal{F}_L$ were quoted for the Tevatron combination). }
\label{tab:WHel}
\end{table}

\subsubsection{Width / Lifetime}
\label{sec:width}
As stated in Section \ref{sec:top}, the large mass of the top quark, implying a large decay width, leads to a short top quark lifetime. As it is predicted to be shorter than the timescale of hadronisation, the direct measurement of the top quark spin, reflected in its polarization and \ttbar\ spin correlation, is possible. A short top quark lifetime is required to perform top quark spin measurements. Vice versa, dilutions in the spin measurements can be a sign for a top lifetime longer than the prediction. 

Concerning measurements at the Tevatron, neither CDF nor D0 have observed deviations from the NLO SM prediction of $\Gamma_t = 1.36 \GeV$. D0 measured $\Gamma_t = 2.00^{+0.47}_{-0.43} \GeV$ ($\tau_t =3.29^{+0.90}_{-0.63} \cdot 10^{-25}\text{ s}$) via the partial decay width $\Gamma_t(t\rightarrow bW)$ taken from the $t$-channel single top cross section measurement and the branching fraction $B(t\rightarrow bW)$ from $\ttbar$ events \cite{gammatop_D0}. While this measurement assumed SM couplings, CDF performed a direct measurement and obtained the 68\,\% CL interval of $1.1 < \Gamma_t < 4.04 \GeV$ ($1.6 \cdot 10^{-25} < \tau_t<  6.0 \cdot 10^{-25} \text{ s}$).
At the LHC, CMS has recently published a result with impressive precision. By combining a measurement of the ratio $B(t \rightarrow Wb)/B(t \rightarrow Wq)$ with the results from the single top $t$-channel cross section measurement \cite{st_xsec8} they measured $\Gamma_t = 1.36 \pm 0.02\,(\text{stat.}) {\,}^{+0.14}_{-0.11}\,(\text{syst.}) \GeV$ \cite{gammatop_CMS}.

These results justify measurements involving top quark spin and the transfer to its decay products. The following sections report expectations of the top quark polarization and the top quark spin correlation. Furthermore, they provide a prescription for accessing these quantities as well as an overview of their measurements. 

\section{Top Quark Polarization and Spin Correlation in \texorpdfstring{$t \bar{t}$}{Top/Anti-Top Quark} Events}
\label{sec:SC}
The spin of the top quark is determined by its production process and transferred to the decay products via the decay process. As $\Gamma_t \muchless m_t$, the \textit{leading pole approximation}\index{Leading pole approximation} \cite{Bernreuther2001, leadingpole} can be used to factorize the production and the decay process. By averaging over spin and colour configurations of the initial states, the squared matrix element can be expressed \cite{Baumgart2013} as
\begin{align}
 {\textstyle \Big(\, \frac{1}{3^2 \; {\rm or} \; 8^2}    \underset{\rm colours}{\sum}  \,\Big) \, \Big(\, \frac{1}{2^2} \underset{\rm spins}{\sum} \,\Big) } \,
\big|\mathcal{M}(\, q\bar q / gg \,\to\, t \bar t \,\to\, (f_1 \bar f_1' b) \, (\bar f_2 f_2' \bar b) \, )\big|^2  \; = \; \lambda_{ab} \, \rho_{ab,\bar{a}\bar{b}} \, \bar\lambda_{\bar{a}\bar{b}} , 
\label{eq:spinmatrix}
\end{align}
where $f_i$ represent the fermions of the $W$ boson decay, $a,b$ the top quark spins and $\lambda$ and $\rho$ the spin density matrix for the production and the decay, respectively. The bar on top of the variables indicates the corresponding values for the anti-top and the number used for colour averaging varies for $q\bar{q}$ annihilation (3) and $gg$ fusion (8). Using the Pauli matrices $\sigma$ the production density matrix can be expressed \cite{Baumgart2013} as
\begin{align}
\rho_{ab,\bar{a}\bar{b}} &\,\equiv\,  {\textstyle \Big(\, \frac{1}{3^2 \; {\rm or} \; 8^2}    \underset{\rm colours}{\sum}  \,\Big) \, \Big(\, \frac{1}{2^2} \underset{\rm initial \; spins}{\sum} \,\Big) } \, \mathcal{M}( q\bar q / gg \,\to\,  t_a \bar t_{\bar{a}}) \, \mathcal{M}( q\bar q / gg \,\to\, t_b \bar t_{\bar{b}})^*\\
& \,=\, \frac14 M^{\mu\bar{\mu}}\,\sigma^{\mu}_{ab}\,\sigma^{\bar{\mu}}_{\bar{a}\bar{b}} \nonumber \\
                     & \,=\, \frac14 \left( M^{00}\,\delta_{ab}\,\delta_{\bar{a}\bar{b}} + M^{i0}\,\sigma^{i}_{ab}\,\delta_{\bar{a}\bar{b}} + 
                               M^{0\bar{i}}\,\delta_{ab}\,\sigma^{\bar{i}}_{\bar{a}\bar{b}} + M^{i\bar{i}}\,\sigma^{i}_{ab}\,\sigma^{\bar{i}}_{\bar{a}\bar{b}} \right)  \nonumber \\
                     & \,\equiv\, \frac14 M^{00}\,\left( \delta_{ab}\,\delta_{\bar{a}\bar{b}} + P^{i}\,\sigma^{i}_{ab}\,\delta_{\bar{a}\bar{b}} + 
                                    \bar P^{\bar{i}}\,\delta_{ab}\,\sigma^{\bar{i}}_{\bar{a}\bar{b}} + \widehat{C}^{i\bar{i}}\,\sigma^{i}_{ab}\,\sigma^{\bar{i}}_{\bar{a}\bar{b}} \right) 
\label{eq:proddecomp}           
\end{align}
Here, $M^{00}$ represents the total, spin independent production rate, $P_i = \langle 2 S_i\rangle$ the polarization of the top quark and $\widehat{C}_{i\bar{i}} = \langle 4 S_i \bar{S}_{\bar{i}} \rangle$ the correlation between the top and the anti-top quark spin, using the top quark spin operators $S$. Examples for the spin correlation matrix $\widehat{C}_{i\bar{i}}$ were calculated in \cite{Baumgart2013} and are shown in Appendix \ref{sec:app_spincorrmat}.
The spin density matrix $\lambda$ of a top quark can be simplified by integrating the decay phase space except one decay product $i$, which serves as spin analyser of the top. With $\vec{e}_i$ as its direction of flight in the top quark rest frame one obtains \cite{Baumgart2013}
\begin{align}
\tilde{\lambda}\left( \vec{e}\right)_{ab} \sim \delta_{ab} + \alpha_i \vec{e}_i \cdot \vec{\sigma}_{ab}.
\end{align}
The degree to which the top quark spin is transferred to the decay product $i$ is quantified by the \textit{spin analysing power}\index{Spin analysing power} $\alpha_i$. This quantity, and in particular its numerical value for several spin analyser candidates, is further discussed in Section \ref{sec:analyser}. The analysing powers for the decay products of the anti-top have the same magnitude, but opposite sign \cite{Baumgart2013}.

By choosing one spin analyser for each top quark of a \ttbar\ event, $i$ from $t$ and $j$ from $\bar{t}$, Equation \ref{eq:spinmatrix} leads to
\begin{align}
\frac{d \sigma}{d^2 \vec{e}_i d \vec{{e}}_j} \sim 1 + \alpha_i \vec{P} \vec{e}_i  + \alpha_j \vec{\bar{P}} \vec{{e}}_j + \alpha_i {\alpha_j} \vec{e}_i \widehat{C} \vec{{e}}_j
\end{align}

Moving from these generalized quantities to measurable ones requires the definition of a spin quantization axis. One can define this spin axis as z-direction and use polar coordinates.\footnote{In polar coordinates, $\vec{e} = \left( \cos{\phi}\sin{\theta}, \sin{\phi}\sin{\theta}, \cos{\theta}\right)$.}
Differential distributions of $\cos{\theta}$ allow to access the top quark polarization $P^3$:
\begin{align}
\frac{1}{\sigma} \frac{d \sigma}{d \cos{\theta_i}} = \frac{1}{2}\left(1 + \alpha_i \cdot P^3 \cdot \cos{\theta} \right).
\label{eq:analyser_angle}
\end{align}
Here, $\theta$ denotes the angle of the spin analyser with respect to the spin basis in the top quark rest frame. 
In publications motivating spin correlation measurements (such as \cite{Bernreuther2010}) the following equation is often quoted for the double differential \ttbar\ cross section:
\begin{align}
\frac{1}{\sigma} \frac{d^2 \sigma}{d\cos{\theta_i} d\cos{\theta_j}} = \frac{1}{4} \left( 1 + \alpha_i B_1 \cos{\theta_i} + \alpha_j B_2 \cos{\theta_j} + \alpha_i \alpha_j C \cos{\theta_i} \cos{\theta_j}  
\right)
\label{eq:doublediff}
\end{align}
where $B_i$ are said to describe the polarization of the top and the anti-top quark and $C$ 
\begin{align}
C      &=  \frac{N(\uparrow \uparrow) + N(\downarrow \downarrow) - N(\uparrow \downarrow) - N(\downarrow \uparrow)}
     {N(\uparrow \uparrow) + N(\downarrow \downarrow) + N(\uparrow \downarrow) + N(\downarrow \uparrow)}
    \label{eq:spincorr_C}
\end{align}
the term used for \ttbar\ spin correlation.\footnote{In \cite{Bernreuther2001}, Equation \ref{eq:doublediff} has a minus sign in front of $C$. The reason is that the authors assign the same sign to the spin analysing power values for top and anti-top quarks and introduce the different sign in the spin decay density matrix \cite{Bernreuther2001}. In some definitions of $C$ the spin analysing powers are also already included, so careful reading is required.} 

It is sensible to call $C$, the relative difference between like and unlike spin configurations, correlation term. As $-1 \leq C \leq 1$ values between full anti-correlation $(C=-1)$, no correlation $(C=0)$ and full correlation $(C=1)$ are possible. But one should keep in mind that in fact $B \sim P^3$ and $C \sim \widehat{C}^{33}$, using the polarization and spin matrices from Equation \ref{eq:proddecomp}. Hence, only parts of these matrices are described by the angular distributions of Equation \ref{eq:doublediff}. For reasons of simplicity it is from now on referred to $B_i$ and $C$ if not stated otherwise. 

What degree of polarization and spin correlation can be expected at the LHC? In case of single top quark production the weak interaction with its $V-A$ structure leads to a strong polarization \cite{single_top_pol}. In contrast to single top quark production, the strong interaction is the dominating production process for \ttbar\ pairs. Parity invariance of QCD leads to almost unpolarized \ttbar\ pairs\footnote{In fact, a polarization transverse to the production plane is still allowed\cite{Baumgart2013}, but expected to be very small \cite{Bernreuther1996}.} -- and thus the coefficients $B_i$ in Equation \ref{eq:doublediff} vanish at leading order \cite{Bernreuther2001}. 

The remaining question is: What degree of \ttbar\ spin correlation can we expect for a QCD production and a weak decay according to the SM? Of course, the answer depends on the choice of spin quantization axis.

\subsection{Choice of the Spin Quantization Axis}
\label{sec:basis}
The measured spin correlation depends on the choice of the spin basis. Choosing a basis such that the correlation $C$ is maximal increases the separation between correlated and uncorrelated \ttbar\ pairs, which is certainly desirable. A first hint to a proper choice is given by kinematic limits of the two \ttbar\ production mechanisms: $gg$ fusion and $q\bar{q}$ annihilation. 

\subsubsection{Beam Line Basis}
\begin{figure}[ht]
	\centering
			\subfigure[]{
		\includegraphics[width=0.45\textwidth]{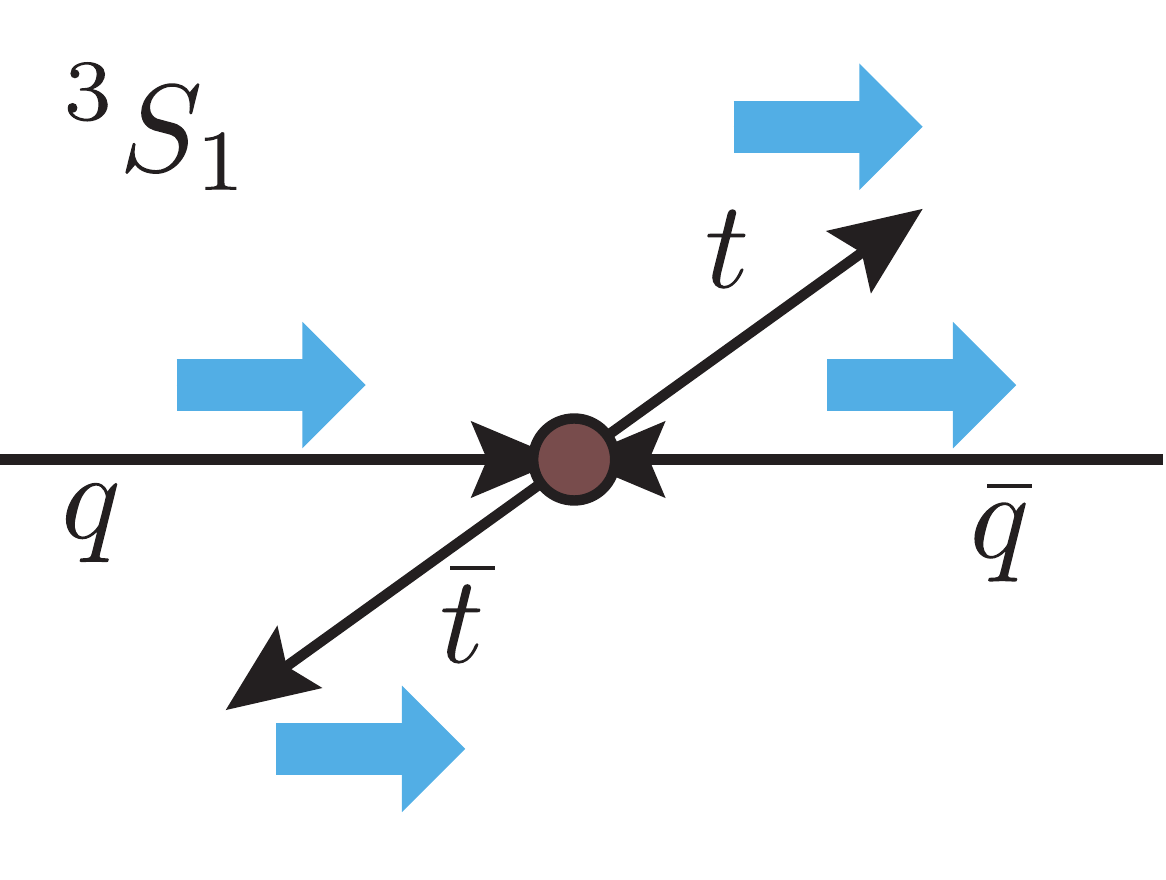}
			\label{fig:spinconfig_lowlim_qq}
		}
					\subfigure[]{
		\includegraphics[width=0.45\textwidth]{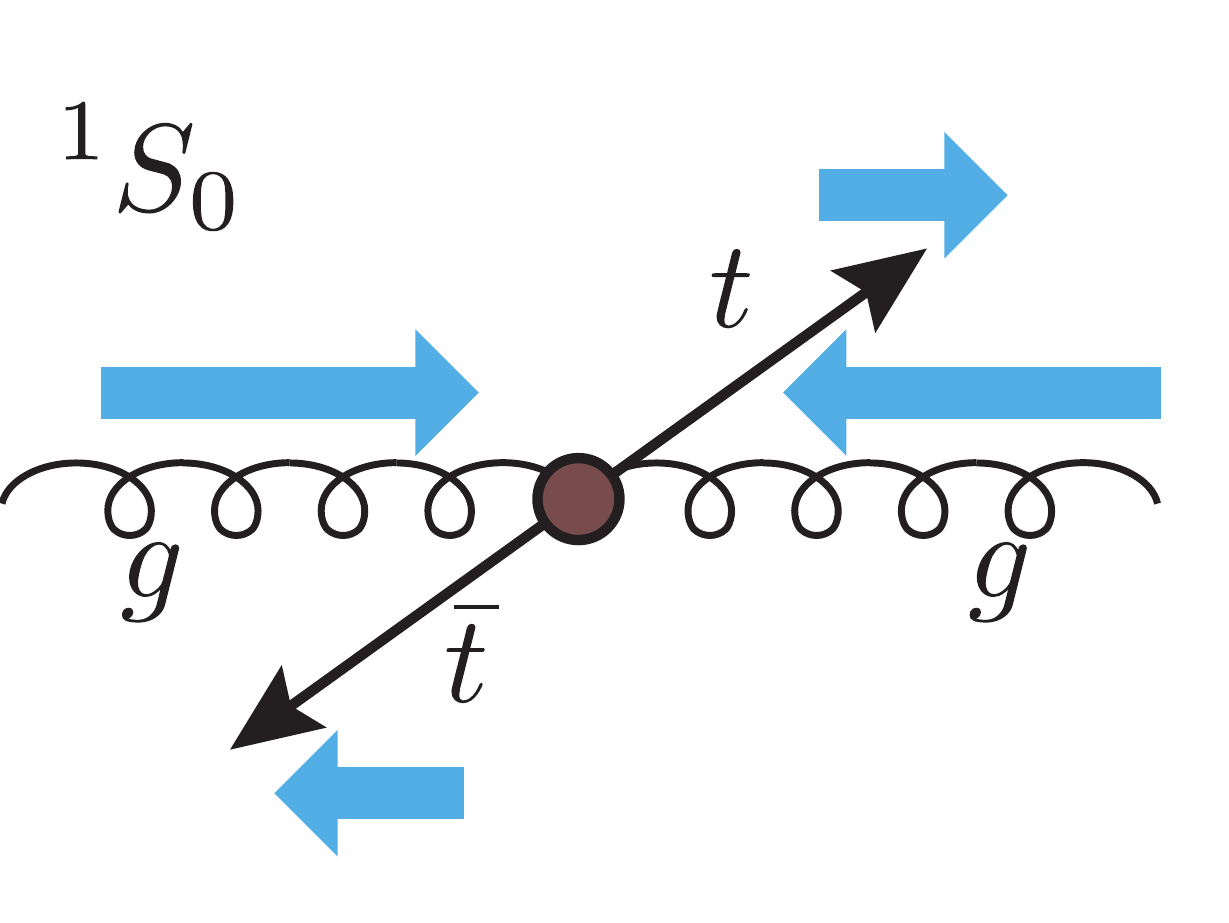}
			\label{fig:spinconfig_lowlim_gg}
		}
		\caption{\ttbar\ spin configuration at the production threshold limit for  \subref{fig:spinconfig_lowlim_qq} $q\bar{q}$ annihilation and \subref{fig:spinconfig_lowlim_gg} $gg$ fusion. }
		\label{fig:spinconfig_lowlim}
\end{figure}

\ttbar\ production at the kinematic threshold ($m_{\ttbar} = 2 m_t$) leads to a $^{3}S_{1}$ \ttbar\ spin state in the $q\bar{q} \rightarrow \ttbar$ channel (due to chirality conservation) and to a $^{1}S_{0}$ configuration in the $g\bar{g} \rightarrow \ttbar$ channel \cite{Mahlon1996, Hara1991, Arens1993}. This spin configuration is determined by the initial state. It is therefore useful to define the spin quantization axis for the top and the anti-top quark as the directions of the incoming partons. This is illustrated in Figure \ref{fig:spinconfig_lowlim}. 
This particular spin basis is referred to as \textit{beam line basis}\index{Beam line basis} \cite{Mahlon1996, Bernreuther2001, Bernreuther2004}. In this basis the spins of top and anti-top are aligned in opposite directions ($\uparrow \downarrow$\footnote{This notation includes $\downarrow \uparrow$ and represents opposite alignment of the spins. The same holds true for $\uparrow \uparrow$ and $\downarrow \downarrow$ for parallel alignments.}, due to different bases) for $q\bar{q} \rightarrow \ttbar$  and in the same direction ($\uparrow \uparrow$) for $gg \rightarrow \ttbar$ for top quark velocities\footnote{Velocities are quoted as fraction $\beta = v/c$ of the speed of light.} $\beta \rightarrow 0$. 
The beam line basis is useful in particular at the \ttbar\ production threshold where no additional angular momentum is added.

\subsubsection{Helicity Basis}
\begin{figure}[ht]
	\centering
			\subfigure[]{
		\includegraphics[width=0.34\textwidth]{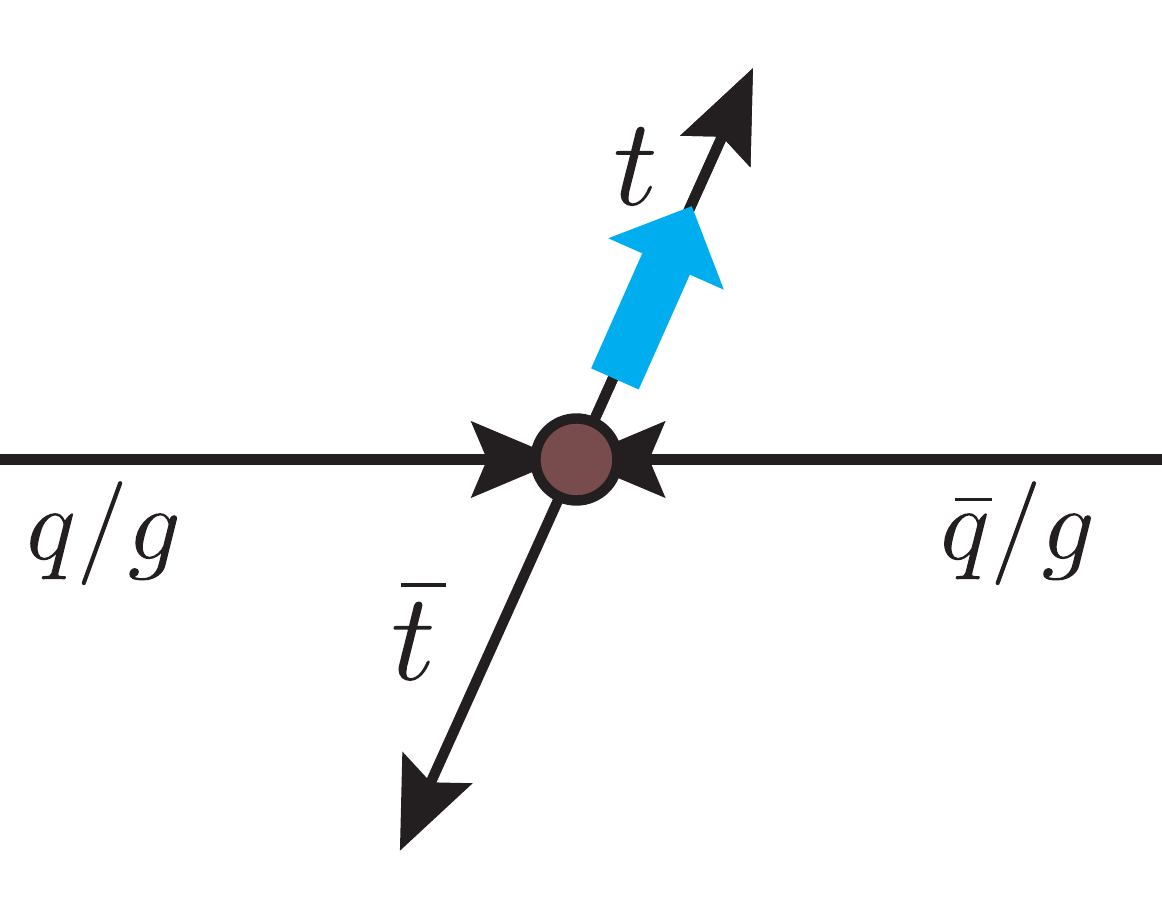}
			\label{fig:helicity_basis}
		}
					\subfigure[]{
		\includegraphics[width=0.34\textwidth]{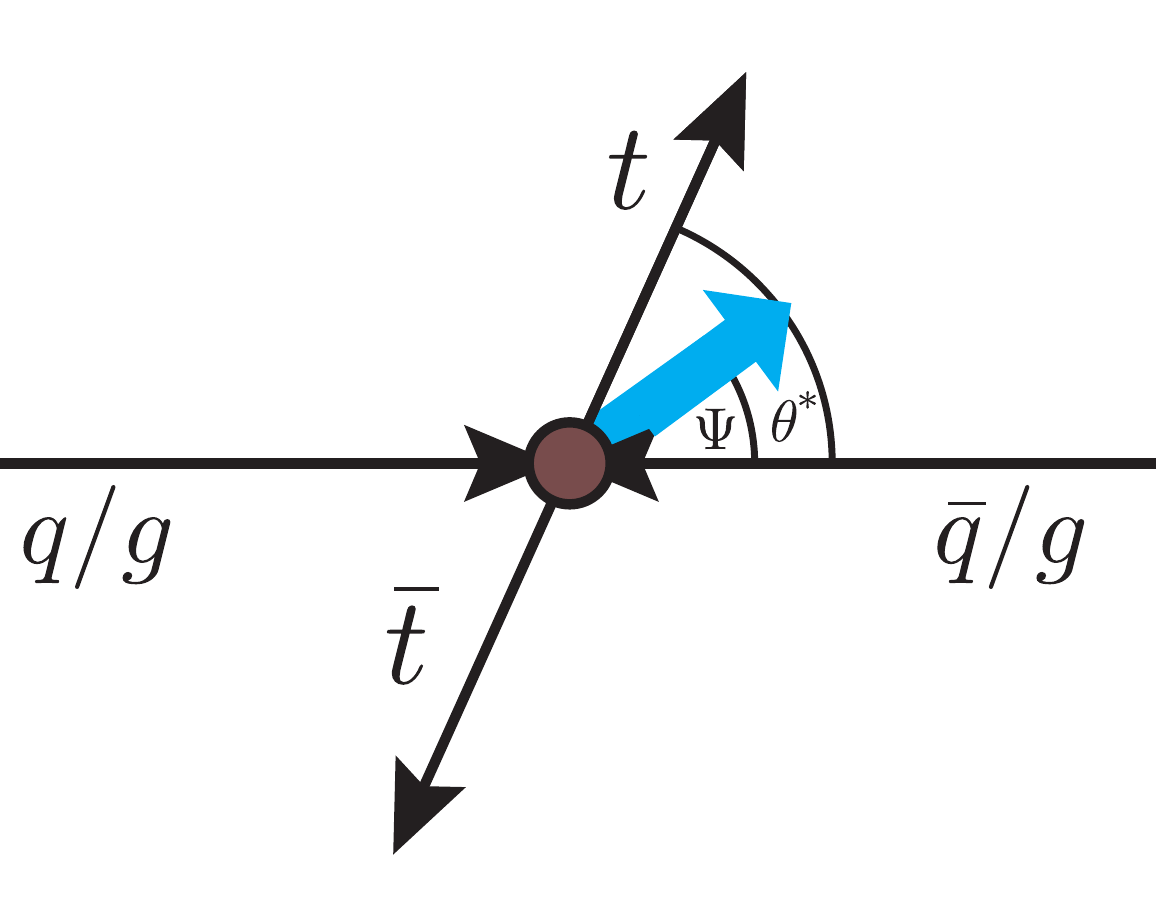}
			\label{fig:offdiag_basis}
		}
							\subfigure[]{
		\includegraphics[width=0.28\textwidth]{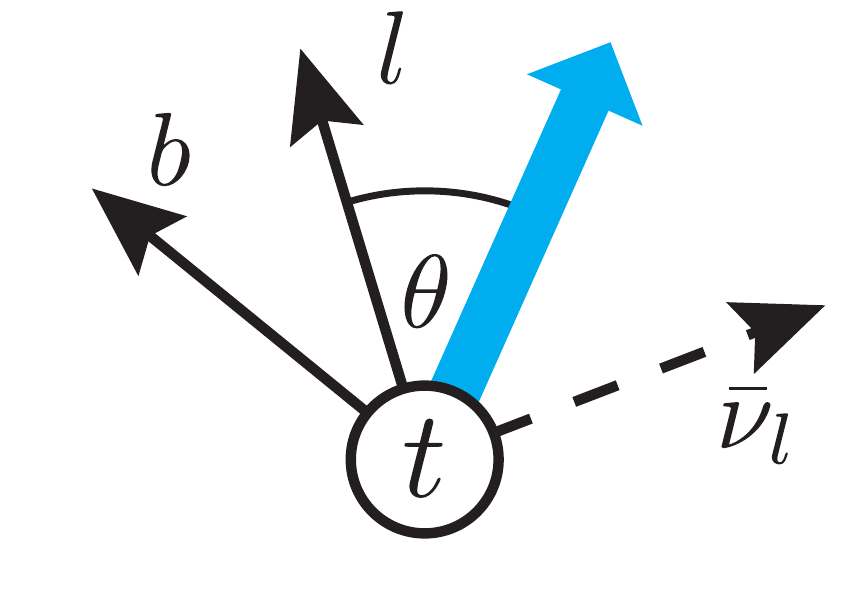}
			\label{fig:spinangle}
		}
		\caption{Illustration of \subref{fig:helicity_basis} the helicity basis and \subref{fig:offdiag_basis} the off-diagonal basis. \mbox{\subref{fig:spinangle} The angle} $\theta$ between a spin analyser (here: the charged lepton) and the spin basis in the top quark rest frame. The blue arrows indicate the spin quantization axes.}
		\label{fig:further_bases}
\end{figure}

Another basis of interest is the \textit{helicity basis}\index{Helicity basis}. Here, the top quark direction of flight in the \com\ frame\footnote{Also referred to as centre-of-momentum frame or zero-momentum frame (ZMF). It is the frame where the \ttbar\ pair is at rest.} is taken as top spin axis (see Figure \ref{fig:helicity_basis}). The anti-top spin axis is defined accordingly. 
In the ultrarelativistic limit of $\beta \rightarrow 1$ only $\uparrow \downarrow$ configurations are allowed for both $q\bar{q} \rightarrow \ttbar$ and $gg \rightarrow \ttbar$ events in the helicity basis due to chirality conservation of QCD. 

Kinematic limit considerations allow for a straightforward expression of the spin correlation $C$ in the helicity basis. Using Clebsch-Gordan coefficients (see for example \cite{PDG} for a list) leads to three representations of the $^{3}S_1$ spin state for $q\bar{q}$ annihilation at the production threshold \cite{Stelzer1996}:
\begin{align}
&\ket{++}\\
&\frac{1}{\sqrt{2}} \left( \ket{+-} + \ket{-+} \right)\\
&\ket{--}
\end{align}
The $\pm$ signs represent the spin eigenstates in a common basis. Same sign spin states of the top and anti-top quark imply opposite sign helicity states and vice versa.\footnote{$\pm$ signs were used not to confuse e.g. $\ket{++}$ with $\ket{\uparrow \uparrow}$. In the former case, a common basis is used whereas in the latter case the top and anti-top quark have individual bases.} Thus, for $\beta \rightarrow 0$ the spin correlation is $C = \frac{1}{3} - \frac{2}{3}= -\frac{1}{3}$. In the ultrarelativistic limit of $\beta \rightarrow 1$ the helicity conservation of QCD ensures both the initial $q\bar{q}$ pair and the \ttbar\ pair to have opposite helicities, so $C = -1$ \cite{Stelzer1996}. Between these two limits the fraction of opposite sign to same sign helicity fractions is determined by the invariant mass $m_{\ttbar}$ of the \ttbar\ system \cite{Stelzer1996}:
\begin{align}
\frac{N(\uparrow \downarrow) + N(\downarrow \uparrow)}{N(\uparrow \uparrow) + N(\downarrow \downarrow)} = 2\frac{m_{\ttbar}^2}{4m_t^2}
\end{align}

\begin{figure}[ht]
	\centering
			\subfigure[]{
		\includegraphics[width=0.45\textwidth]{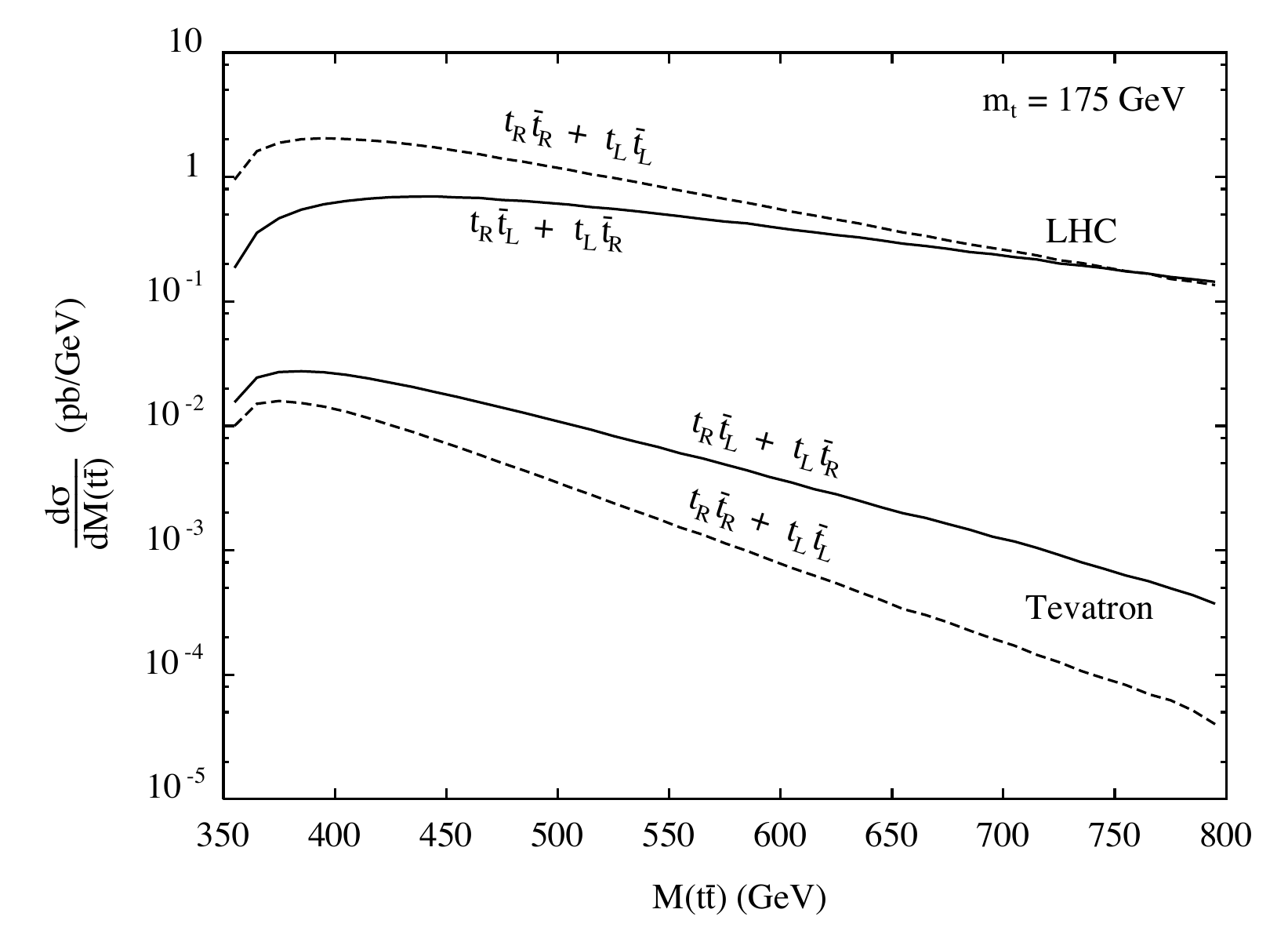}
			\label{fig:helicity_vs_mttbar_both}
		}
					\subfigure[]{
		\includegraphics[width=0.45\textwidth]{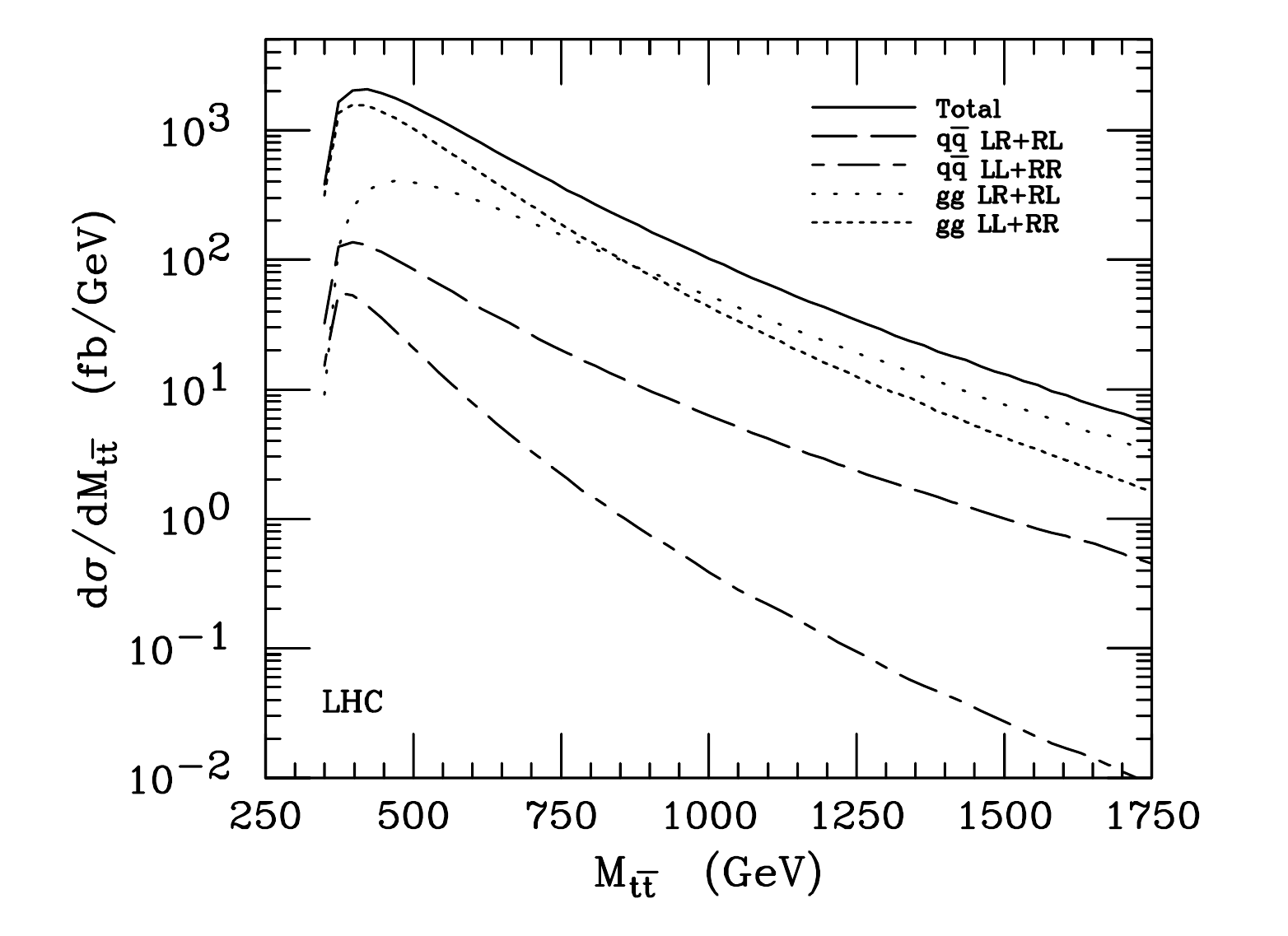}
			\label{fig:helicity_vs_mttbar_LHC}
		}
		\caption{\subref{fig:helicity_vs_mttbar_both} \ttbar\ production cross section as a function of the invariant mass $M_{\ttbar}$ of the \ttbar\ system for production at the LHC ($pp$ collisions at $\sqrt{s} = 14\,\TeV$) and the Tevatron ($p\bar{p}$ collisions at $\sqrt{s} = 2\,\TeV$) \cite{Stelzer1996}. \subref{fig:helicity_vs_mttbar_LHC} Decomposition of the LHC cross section ($\sqrt{s} = 14\,\TeV$) into $gg$ fusion and $q\bar{q}$ annihilation \cite{Mahlon1996}.}
		\label{fig:helicity_vs_mttbar}
\end{figure}

Concerning $gg$ fusion, the $\beta \rightarrow 0$ limit with the $^{1}S_0$ state and its $\frac{1}{\sqrt{2}}\left(\ket{+-} - \ket{-+}\right)$ configuration implies same sign helicity. The limit of $\beta \rightarrow 1$ again implies helicity conservation and thus opposite sign helicities. As the $gg$ fusion process is the dominant \ttbar\ production mode at the LHC, it leads to a particularly high spin correlation $C$ using the helicity basis. The \ttbar\ cross section as a function of the invariant mass of the \ttbar\ system is shown in Figure \ref{fig:helicity_vs_mttbar} for both the LHC ($pp$ collisions at $\sqrt{s} = 14\,\TeV$) and the Tevatron ($p\bar{p}$ collisions at $\sqrt{s} = 2\,\TeV$) \cite{Stelzer1996}. 

\subsubsection{Off-Diagonal Basis}
The spin configurations are purely of opposite sign in the case of $q\bar{q}$ annihilation for both the beam line basis for $\beta \rightarrow 0$ as well as in the helicity basis for $\beta \rightarrow 1$. Hence, a basis interpolating between these two limits can lead to pure oppositely signed spin states of the \ttbar\ pairs. Such a basis exists \cite{Mahlon1997, Parke1996} and is named \textit{off-diagonal basis}\index{Off-diagonal basis}. With $\theta^{*}$ as the top quark production angle and $\Psi$ as the angle between the beam axis and the off-diagonal basis (see Figure \ref{fig:offdiag_basis}) \cite{Mahlon1997}, the interpolation is done via
\begin{align}
\tan{\Psi} = \frac{\beta^2 \sin{\theta^{*}} \cos{\theta^{*}}}{1-\beta^2 \sin^2{\theta^{*}}}.
\end{align}
The limits $\beta \rightarrow 0$ and $\beta \rightarrow 1$ lead to the beam line and the helicity basis, respectively. 
As the off-diagonal basis maximizes the $q\bar{q}$ annihilation but not the $gg$ fusion, it is a preferred choice for measurements at the Tevatron, but not at the LHC. The exact expression of the spin correlation matrix $\widehat{C}_{i\bar{i}}$ in the off-diagonal basis is shown in Appendix \ref{sec:app_spincorrmat}.

\subsubsection{Maximal Basis}
Optimizations with respect to the helicity basis can be found at the LHC, as well. The off-diagonal basis does not only lead to purely oppositely signed \ttbar\ spins for $q\bar{q}$ annihilation, but also for $gg$ fusion with unlike-helicity gluons \cite{Mahlon2010}. However, this is not the case for like-helicity $gg$ fusion, which is dominating at the LHC \cite{Mahlon1996, Mahlon2010} for low invariant masses of the \ttbar\ system. A way of defining a \textit{maximal basis}\index{Maximal basis} for the general case of $gg$ fusion is described in \cite{Mahlon2010, Uwer2005}. 

\subsubsection{Expected Spin Correlations for Different Bases}
\label{sec:exp_correlations}
As mentioned, the explicit value of $C$ as defined in Equation \ref{eq:spincorr_C} depends on the production and the spin analysing basis. For several experimental setups and bases, these values have been calculated using two charged leptons as spin analysers. The results from \cite{Bernreuther2001, Bernreuther2004, Bernreuther2010, ueberpaper} are summarized in Table \ref{tab:exp_corr}. The extent of higher order corrections differs but does not change the overall picture. The numbers indicate that a proper choice of basis is crucial.
\begin{table}[htbp]
\begin{center}
\begin{tabular}{|c|c|c|c|c|c|}
\hline
Mode&$\sqrt{s}$& $C_{\text{beamline}}$&$C_{\text{off-diagonal}}$& $C_{\text{helicity}}$ & $C_{\text{maximal}}$ \\
\hline
\hline
$p\bar{p}$ & 2 TeV 	& {\,\,0.78 \cite{Bernreuther2004}} 	& {\,\,0.78 \cite{Bernreuther2004}} 	& {-0.35 \cite{Bernreuther2004}} & {---}\\
$pp$ & 7 TeV		& {---} 						& {---} 						& {\,\,0.31 \cite{Bernreuther2010}} & {0.44 \cite{ueberpaper}}\\
$pp$ & 8 TeV		& {---} 						& {---} 						& {\,\,0.32 \cite{Bernreuther2010}} & {---}\\
$pp$ & 14 TeV		& {-0.07 \cite{Bernreuther2001}}	& {-0.09 \cite{Bernreuther2001}} 	& {\,\,0.33 \cite{Bernreuther2004}} & {---}\\
\hline
\end{tabular}
\end{center}
\caption{Expected spin correlation $C$ as in Equation \ref{eq:spincorr_C}. Two charged leptons served as spin analysers.}
\label{tab:exp_corr}
\end{table}
\subsection{Spin Analysing Powers}
\label{sec:analyser}
Each decay product of the top quark carries information of its parent's top quark spin. The degree, the spin analysing power $\alpha$, is determined by the weak interaction and its $V-A$ structure. Large values of $\alpha$ lead to larger differences in the angular distributions of the top quark spin analysers between the scenarios of SM \ttbar\ spin correlation/polarization and vanishing spin correlation/polarization (see equations \ref{eq:analyser_angle} and \ref{eq:doublediff}).
Non-SM couplings and possible V+A structures will directly be reflected in changes of the predictions for $\alpha$ \cite{Jezabek1994}. Examples for such modifications are shown in Section \ref{sec:BSM_decay}.

In this section the numerical values of the top quark spin analysing powers are introduced and explained. Following the convention made in the previous sections, the spin analysing power of the corresponding anti-top decay products have a reversed sign.\footnote{One might argue that the same derivation could be repeated for the anti-top, leading to the same sign. But in this case the definition of the angular distribution from Equation \ref{eq:analyser_angle} would need the reversed sign for the anti-top quark.}

To understand the spin analysing power of the \bQ, $\alpha_b$, at leading order, one can boost into the rest frame of the top quark and see easily that the \bQ\ spin state depends on the helicity of the $W$ boson. For longitudinally polarized $W$ bosons the $b$ and $t$ spin are parallel. For left-handed $W$ bosons they are anti-parallel. The direction of flight of the \bQ\ is anti-parallel to its spin. Hence, by using Equation \ref{eq:helfractions}, $\alpha_b$ is determined as
\begin{align}
\alpha_b &= \frac{\mathcal{F}_L - \mathcal{F}_0}{\mathcal{F}_L + \mathcal{F}_0}=\frac{2 \left(\frac{m_W}{m_t} \right)^2 -1}{2 \left(\frac{m_W}{m_t} \right)^2 +1}.
\label{eq:alpha_b}
\end{align}
As the \bQ\ and the $W$ boson are emitted back-to-back in the top quark rest frame, the angle between the $W$ and the top spin axis is $\pi$ subtracted by the angle between the spin axis and the $b$. This leads to $\alpha_W = - \alpha_b$ when comparing Equation \ref{eq:analyser_angle} for both spin analysers. 

It is a remarkable feature of the $V-A$ structure of the top decay that leads to a maximal spin analysing power of $\alpha_l = 1$ for charged leptons as shown in \cite{Jezabek1989, Brandenburg2002}. The \textit{down-type quark}\index{Down-type quark} as the $T_3= -\frac{1}{2}$ component of the weak isospin doublets is the analogue to the charged lepton in terms of weak interactions. Hence, the same value of $\alpha$ is derived at leading order: $\alpha_d = \alpha_s = +1$ for down and strange quarks. This makes the down-type quark the most powerful hadronic analyser. Since it is much more challenging to identify the \dQ\ jets, advanced reconstruction techniques are necessary. These reconstruction techniques are described in Sections \ref{sec:KLFitter} and \ref{sec:udsep}.

The last analysers to be studied are the  $T_3= +\frac{1}{2}$ decay products from the $W$ boson, namely neutrinos, $u$- and $c$-quarks. The analytic form of the analysing power depends on $\frac{m_W}{m_t}$ and is listed in \cite{Mahlon1996}. Given the small values of $\left| \alpha \right| \approx 0.3$ and the low reconstruction efficiencies of neutrinos, $u$- and $c$-quark jets, these are no alternatives to charged leptons and down-type quarks in this analysis.
All spin analysing powers at LO and NLO are listed in Table \ref{tab:anapower}.
\begin{table}[htbp]
\begin{center}
\begin{tabular}{|c|c|c|c|c|c|}
\hline
\rule{0pt}{0.5cm} {} & \bQ & $W^+$ & $l^+$ & $\bar{d}$-quark or $\bar{s}$-quark& $u$-quark or $c$-quark\\
\hline
\hline
$\alpha_i$ (LO) & -0.41 & 0.41 & 1 & 1 & -0.31 \\
$\alpha_i$ (NLO) & -0.39 & 0.39 & 0.998 & 0.97 & -0.32 \\
\hline
\end{tabular}
\end{center}
\caption{Standard Model spin analysing power at LO and NLO 
for the decay products of the top quark from the decay $t \rightarrow bW^+$ and for the decay products of the $W$ boson \cite{Czarnecki1990, Brandenburg2002, Hubaut2005, Mahlon1996, Jezabek1994}.}
\label{tab:anapower}
\end{table}

While for the spin correlation analyses in the dilepton channel the choice of analyser is quite obvious (the two charged leptons), the \ljets\ channel offers two attractive possibilities: the down-type quarks due to their high spin analysing power and the \bQ s as being relatively easy to reconstruct. Both will be studied and used for individual measurements. A combined fit with both analysers will also be performed. In \cite{Stelzer1996} such a combination is suggested. It is further justified by specific checks for this analysis (see Section \ref{sec:correlation}). 

\subsection{Observables with Sensitivity to \ttbar\ Spin Correlation}
\label{sec:observables}
The most natural way to measure the \ttbar\ spin correlation $C$ is via the angular distributions of the decay products with respect to the corresponding beam axis, such as in Equation \ref{eq:doublediff}. However, there are several other kinematic distributions which are sensitive to the spin correlation. These will be briefly described in the following. 

\subsubsection{Distributions of $\cos{\theta_i} \cos{\theta_j}$}
\begin{figure}[ht]
	\centering
			\subfigure[]{
		\includegraphics[width=0.45\textwidth]{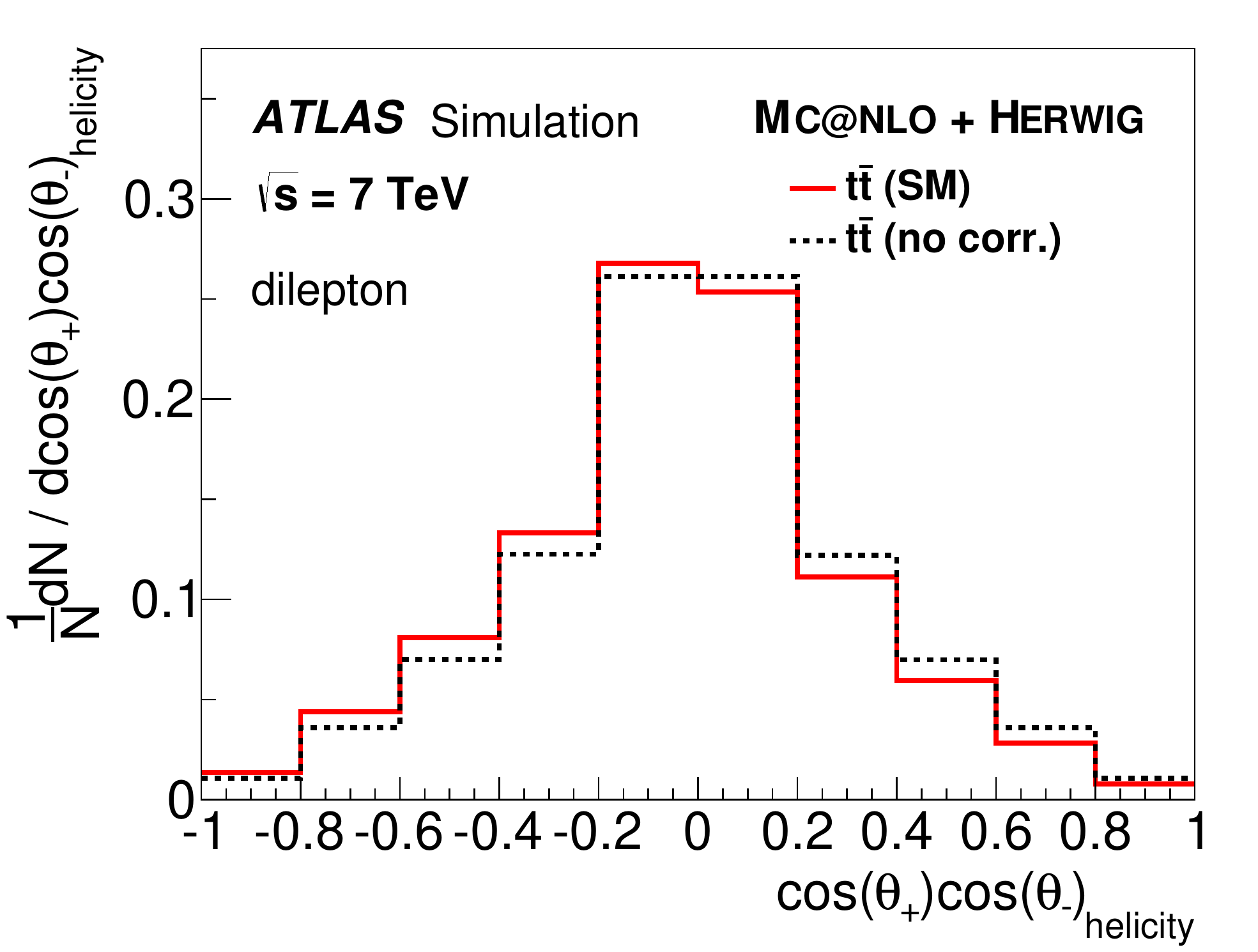}
			\label{fig:parton_dilep_coscos_helicity}
		}
					\subfigure[]{
		\includegraphics[width=0.45\textwidth]{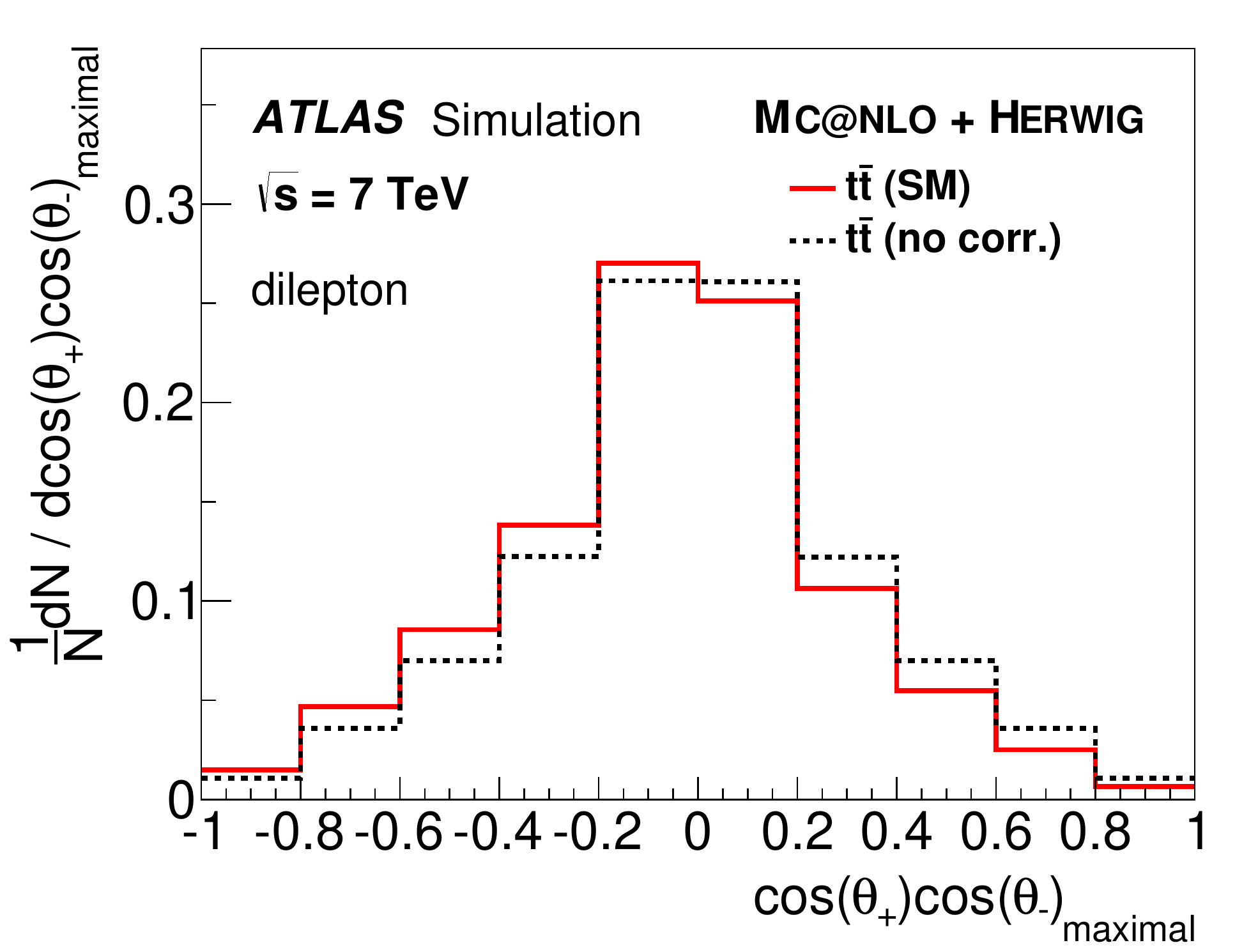}
			\label{fig:parton_dilep_coscos_maximal}
		}
		\caption{Distributions of $\cos{\theta_i}\cos{\theta_j}$ at parton level using \subref{fig:parton_dilep_coscos_helicity} the helicity  and the \subref{fig:parton_dilep_coscos_maximal} maximal  basis \cite{ueberpaper}. The charged leptons from the dilepton channel served as spin analysers. \mcatnlo\ was used generating events with SM spin correlation ($A=\text{SM}$) and uncorrelated \ttbar\ events ($A=0$). The notation $A$ used in the figure corresponds to the spin correlation $C$ used in this thesis.}
		\label{fig:basis1}
\end{figure}

To start with the distributions discussed in Section \ref{sec:basis}, Figure \ref{fig:basis1} shows the parton level distributions of $\cos{\theta_i}\cos{\theta_j}$ using two charged leptons in the helicity basis and the LHC maximal basis \cite{ueberpaper}. For these distributions the \mcatnlo\ generator was used, simulating both the SM spin correlation of \ttbar\ events as well as uncorrelated \ttbar\ events. Details about the signal sample are given in Section \ref{sec:signal}.

\subsubsection{S-Ratio}
\begin{figure}[ht]
	\centering
			\subfigure[]{
		\includegraphics[width=0.45\textwidth]{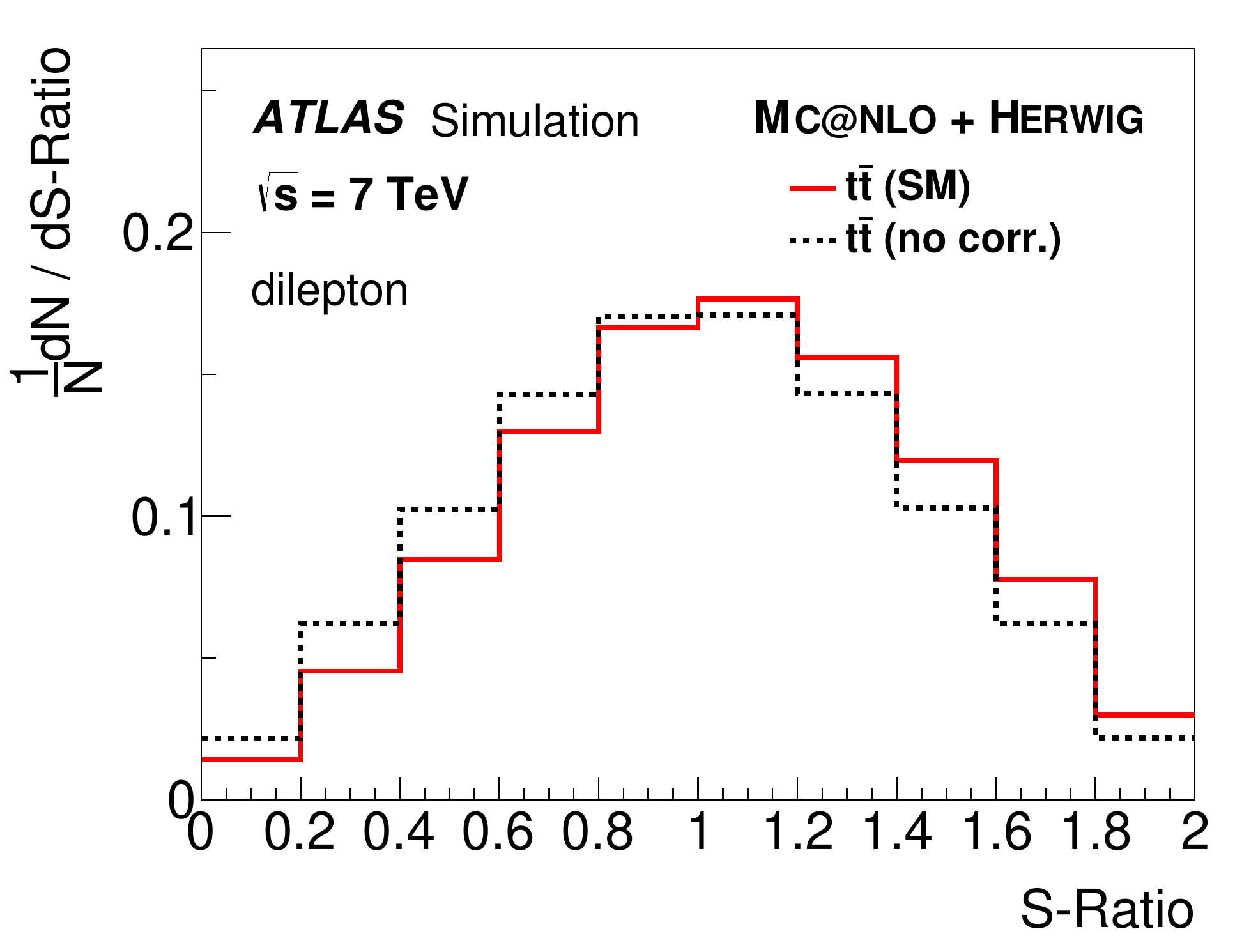}
			\label{fig:parton_sratio}
		}
					\subfigure[]{
		\includegraphics[width=0.45\textwidth]{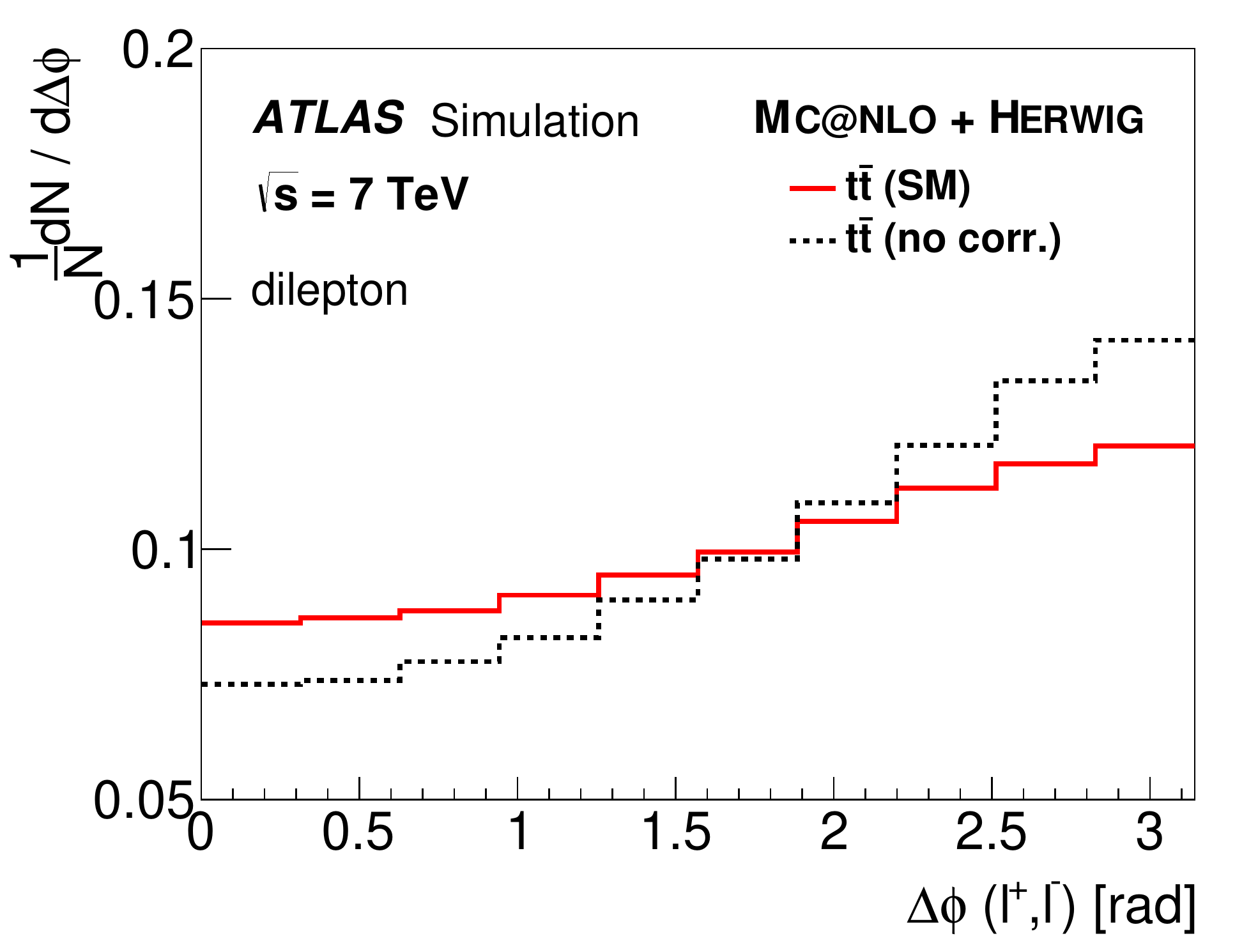}
			\label{fig:parton_dphi}
		}
		\caption{Distributions sensitive to \ttbar\ spin correlation \cite{ueberpaper}: \subref{fig:parton_sratio} S-Ratio. \subref{fig:parton_dphi} Azimuthal angle between the two analysers in the laboratory frame. The charged leptons from the dilepton channel served as spin analysers. \mcatnlo\ was used to generate events with SM spin correlation ($A=\text{SM}$) and uncorrelated \ttbar\ events ($A=0$).  The notation $A$ used in the figure corresponds to the spin correlation $C$ used in this thesis.}
		\label{fig:basis2}
\end{figure}

This observable makes use of the fact that at the LHC the like-helicity gluons dominate the production (see Figure \ref{fig:helicity_vs_mttbar}). The \newword{S-Ratio} of squared matrix elements for SM spin correlation and uncorrelated \ttbar\ spins, 
\begin{align}
  S &= \frac{(|{\cal M}|_{\rm RR}^2 +
    |{\cal M}|_{\rm LL}^2)_{\mathrm {corr}}}{(|{\cal M}|_{\rm RR}^2 +
    |{\cal M}|_{\rm LL}^2)_{\mathrm {uncorr}}}\\[0.2cm]
         &= \frac{m_t^2\{(t \cdot l^{+})  (t \cdot l^{-}) + (\tbar \cdot
       l^{+})  (\tbar \cdot l^{-}) - m_t^2  (l^{+} \cdot
       l^{-})\}}{(t \cdot l^{+})(\tbar \cdot l^{-})(t \cdot \tbar)
     },\,
     \label{eq:sratio}
\end{align}
 is calculated by the four-momentum vectors of the top ($t$) and the anti-top quark ($\bar{t}$) as well as two analysers, in this case charged leptons ($l^{\pm}$) \cite{Mahlon2010}. 
 A comparison between the distributions of $S$ for a spin correlation as predicted by the SM as well as for uncorrelated \ttbar\ pairs at parton level is shown in Figure \ref{fig:parton_sratio} \cite{ueberpaper}. 
 As the $\cos{\theta_i} \cos{\theta_j}$ distributions, the $S$-Ratio demands reconstruction of the full event kinematics. In particular in the case of the dilepton channel, this is challenging.

\subsubsection{$\Delta \phi$ in the Laboratory Frame}
\begin{figure}[ht]
	\centering
		\includegraphics[width=0.45\textwidth]{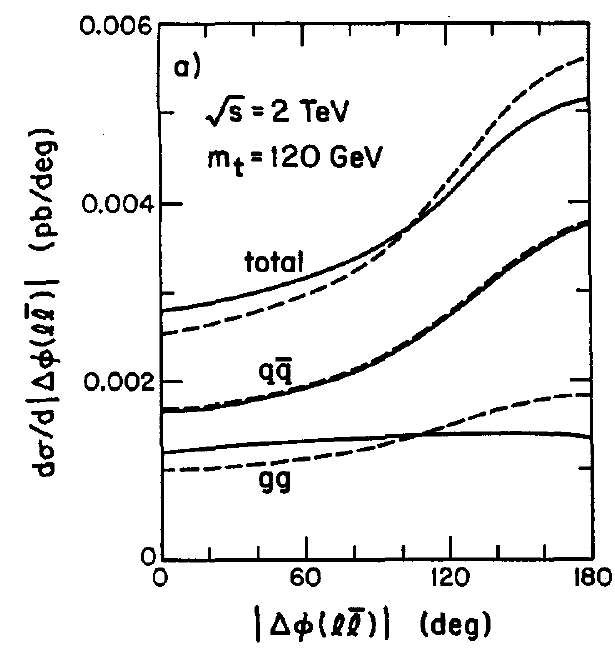}
		\caption{Azimuthal angle between the two charged leptons from \ttbar\ decays produced in $p\bar{p}$ collisions at $\sqrt{s} = 2\,\TeV$ \cite{Barger1989}. The SM prediction (solid line) is compared to the scenario of uncorrelated \ttbar\ pairs (dashed line). A top mass of 120 GeV was assumed.}
	\label{fig:dphi_theo_old}
\end{figure}

The $S$-Ratio as shown in Equation \ref{eq:sratio} can be expressed in the ZMF \cite{Mahlon2010} as
\begin{align}
  S &= \left( \frac{1-\beta^2}{1+\beta^2} \right)  
\left( \frac{(1+\beta^2) + (1-\beta^2)\cos{(l^{+}, l^{-})} 
- 2 \beta^2 \cos{(t,l^{+})} \cos{(\bar{t},l^{-}})}
{(1-\beta \cos{(t,l^{+})})(1-\beta \cos{(\bar{t},l^{-})})} \right).
\end{align}     
The angle between the lepton momenta suggest checking the angular separation of the two leptons in $\Delta \eta$, \dphi\ and $\Delta R = \sqrt{\left( \phi_i - \phi_j\right)^2 + \left( \eta_i - \eta_j \right)^2}$. While $\Delta \eta$ shows no separation power between the scenario of SM-like \ttbar\ spin correlation and uncorrelated \ttbar\ pairs \cite{Mahlon2010}, \dphi\ does. The \dphi\ distributions were studied in \cite{Mahlon2010} for an LHC setup with $\sqrt{s} = 14\,\TeV$ and showed impressive separation power. 
Before these studies were made, the sensitivity of \dphi\ was already known and checked in the context of Tevatron studies \cite{Barger1989, Arens1993}. In Figure \ref{fig:dphi_theo_old}, the \dphi\ distribution for a Tevatron setup and an underestimated top quark mass of $m_t = 120 \GeV$ is shown. It can be noticed that the $q\bar{q}$ production is insensitive. As $q\bar{q}$ production is dominating at the Tevatron, \dphi\ has no large sensitivity to the \ttbar\ spin correlation. The situation is reversed at the LHC where the $gg$ fusion is dominating, in particular in the like-helicity mode as shown in Figure \ref{fig:helicity_vs_mttbar_LHC}. This makes \dphi\ a very interesting observable, uniquely at the LHC.  
For the reasons given above, \dphi\ was not only utilized in the publication about the observation of \ttbar\ spin correlation at the LHC \cite{ATLAS_spin_paper}, it is also the observable used for the analysis presented in this thesis. Here, the additional complication with respect to the dilepton analyses performed by ATLAS and CMS \cite{ATLAS_spin_paper, ueberpaper, CMS_spin_paper, CMS_spin_prelim} is the identification of the hadronic spin analyser.

\subsubsection{Further Angular Variables}
\label{sec:further_spinvars}
While the \dphi\ distribution was an obvious candidate for the dilepton channel, an alternative for the \ljets\ channel was investigated in \cite{Mahlon2010} as well. As the \ljets\ channel basically allows for a full event reconstruction, boosts into other rest frames than the laboratory frame are allowed. A sensitive quantity is the $\cos{\theta}$ angle between two spin analysers in the ZMF. This variable provides access to the trace of the spin correlation matrix \cite{Baumgart2013}.

By using the down-type quark as analyser for the hadronically decaying top quark and by placing an additional cut on the invariant mass of the \ttbar\ pair ($m_{\ttbar} < 400 \GeV$),\footnote{The cut on the invariant mass was motivated by a larger separation between the sample with SM spin correlation and the uncorrelated sample. See also \cite{Bernreuther2010} for discussion on $m_{\ttbar}$.} a good separation power between SM-like and uncorrelated \ttbar\ pairs can be achieved, as shown in \cite{Mahlon2010}.  However, this method is limited: the boost requires a fully correct assignment of all involved decay products and furthermore a good energy resolution, in particular for the $m_{\ttbar}$ cut. Up to now, this variable has not yet been utilized in a spin correlation measurement.

In \cite{Baumgart2013} it is suggested to measure the sum and the difference of the two analysers' polar angles. As spin basis, the helicity basis should be used but with the top spin axis as common z-axis.\footnote{In contrast, separate z-axes are used for the top and anti-top quark in the definition of the helicity basis.} This provides access to different linear combinations of the spin matrix elements $\widehat{C}^{11}$, $\widehat{C}^{12}$, $\widehat{C}^{21}$ and $\widehat{C}^{22}$. Accessing the remaining elements of the spin matrix $\widehat{C}$ is possible by measuring distributions of the spin analysers azimuthal angle shifted by a phase depending on the spin analysers' polar angle \cite{Baumgart2013}.

\subsection{Measurement of \fsm}
\label{sec:fsm}
The degree of \ttbar\ spin correlation, $C$, can be predicted as shown in Section \ref{sec:exp_correlations}. $C$ and additional information of the full spin density matrix $\widehat{C}$ influence the shape of several angular distributions which were introduced in the last sections. 
For each of these observables, two distributions were shown: a SM distribution with an underlying \ttbar\ spin correlation as calculated in \ref{sec:exp_correlations} as well as a distribution with uncorrelated \ttbar\ pairs, $C=0$.
Equation \ref{eq:spincorr_C} shows that the spin correlation is a linear function of the amount of parallel and anti-parallel \ttbar\ spins. This allows to create templates $T$ corresponding to arbitrary values $X$ of the spin correlation $C$:
\begin{align}
T_X = \fsm \cdot T_{C = C_{\text{SM}}} + \left( 1 - \fsm \right) T_{C=0} 
\end{align}
By performing template fits of a distribution which is sensitive to the \ttbar\ spin correlation, it is possible to measure the mixing fraction \fsm. This immediately leads to a measured spin correlation $C$ via
\begin{align}
C = \fsm\ \cdot C_{\text{SM}}
\end{align}
Sometimes \fsm\ is referred to as ``fraction of SM spin correlation''. This is reasonable as it defines the amount of a SM sample mixed into a linear combination, but might be misleading. In particular, it does not have the properties of a fraction in the literal sense (e.g. being bound by 0 and 1).  
While $-1 \leq C \leq 1$ holds, \fsm\ is bound by the limits of $C$. The interpretation of \fsm\ is the following:
\begin{itemize}
\item {\bf $\fsm < 0$} Instead of a correlation, an anti-correlation was observed.\footnote{In case an anti-correlation was predicted, a correlation was observed.}
\item {\bf $\fsm = 0$} \ttbar\ pairs are uncorrelated.
\item {\bf $0 < \fsm\ < 1$} The \ttbar\ pairs are less correlated than predicted. 
\item {\bf $\fsm = 1$} The observed \ttbar\ spin correlation matches the SM prediction. 
\item {\bf $\fsm > 1$} The \ttbar\ spin correlation is higher than predicted. 
\end{itemize}
BSM physics can have different effects on different observables. Hence, it is possible to measure different values of \fsm\ for different observables. 

It is possible to quote \fsm\ and translate it into $C$ for a particular basis. However, such statements should only be made if the translated result corresponds to the measured distribution. For example, if \fsm\ was extracted via a fit of $\cos{\theta_i} \cos{\theta_j}$ using the helicity basis, it can be translated into $C_{\text{helicity}}$. A translation into $C_{\text{beamline}}$ will be misleading. 

The quantity \fsm\ was utilized in several measurements which will be presented in Section \ref{sec:spinresults}. It is also used in the analysis presented in this thesis to extract the amount of spin correlation from a distribution of \dphi.  
 
\section{Sensitivity of \texorpdfstring{$t \bar{t}$}{Top/Anti-Top Quark} Spin Correlation to Physics Beyond the Standard Model}
\label{sec:BSM}
The Standard Model allows to calculate the spin correlation $C$ of \ttbar\ pairs as shown in Table \ref{tab:exp_corr}. Next to the correlation itself, the SM also provides information about how the top quark spins are transferred to the decay products. This information is included in the spin analysing power $\alpha$, listed in Table \ref{tab:anapower}.

Possible new physics can affect both $C$ and $\alpha$. In the following, examples for BSM physics and their implication on \ttbar\ spin correlation are explained. The new physics processes are split into two classes: those affecting the spin correlation $C$ via a modified \ttbar\ production and those that modify the top quark decay and hence $\alpha$.

\subsection{New Physics in the \ttbar\ Production}
\begin{figure}[ht]
	\centering
			\subfigure[]{
		\includegraphics[width=0.47\textwidth]{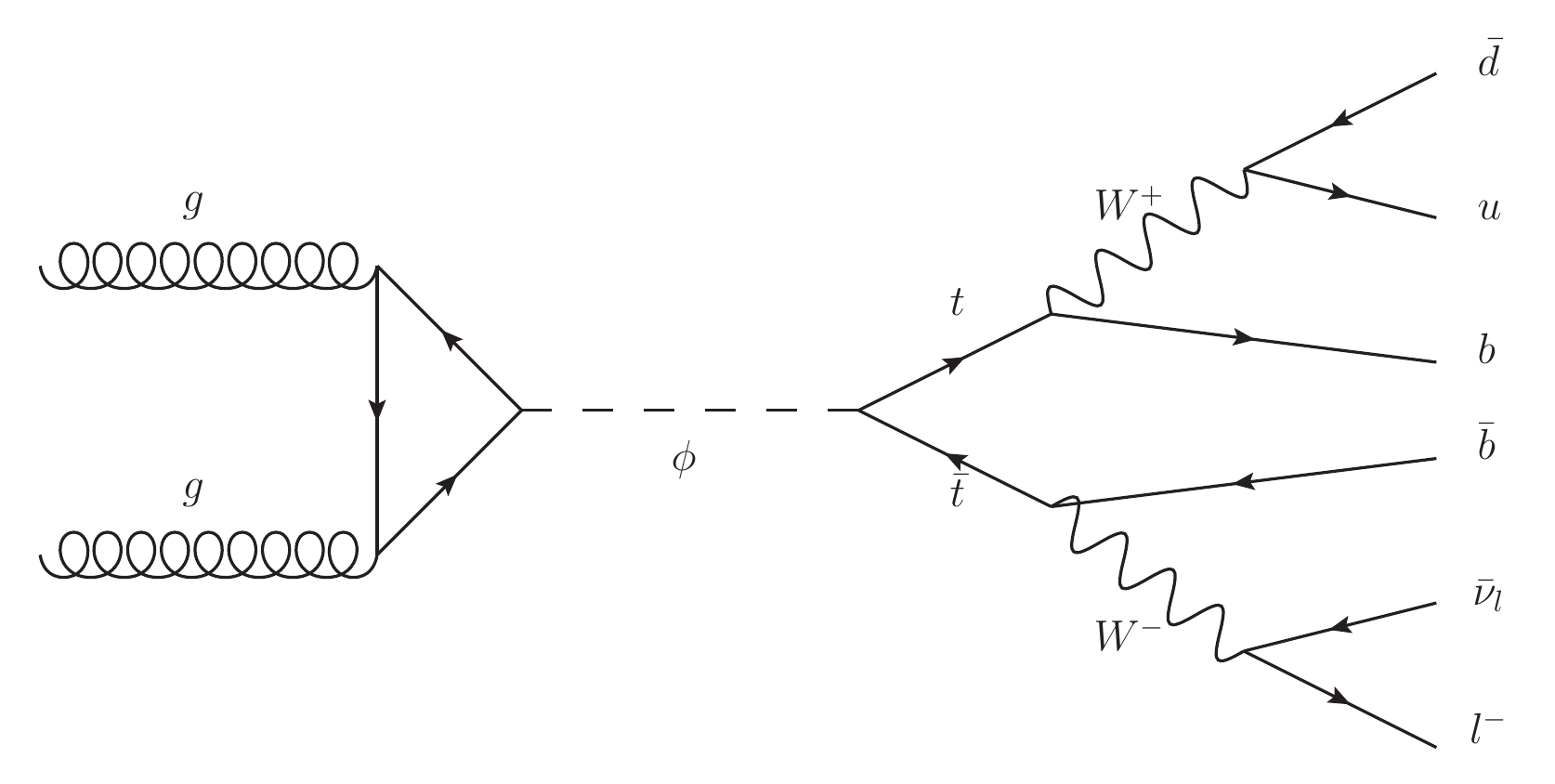}
			\label{fig:BSM_scalar_production}
		}
					\subfigure[]{
		\includegraphics[width=0.47\textwidth]{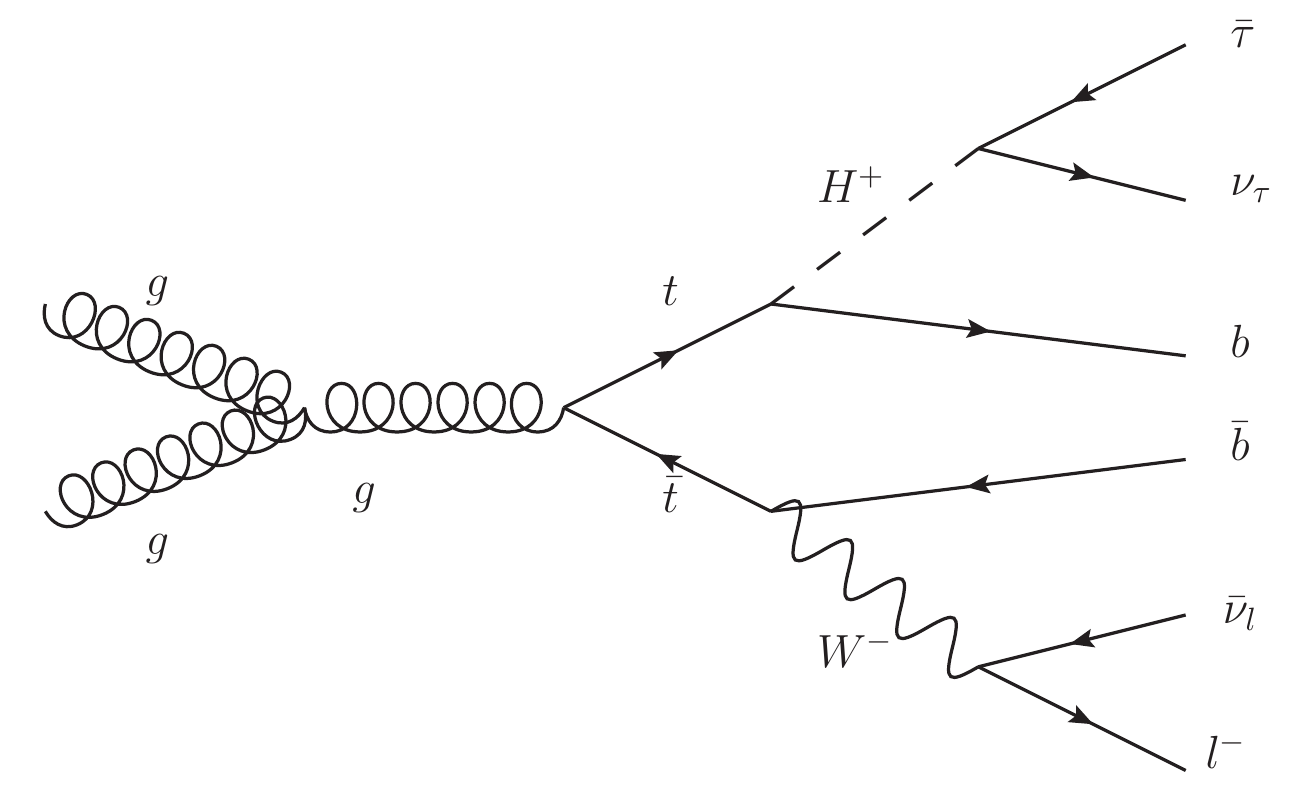}
			\label{fig:BSM_charged_higgs}
		}
		\caption{\subref{fig:BSM_scalar_production} An example for modified \ttbar\ production. A heavy scalar $\phi$ is replacing the virtual gluon. \subref{fig:BSM_charged_higgs} An example for a modified \ttbar\ decay. The $W$ boson is replaced by a scalar charged Higgs boson.}
		\label{fig:BSM_ttbar}
\end{figure}

An example for new physics in the \ttbar\ production is shown in Figure \ref{fig:BSM_scalar_production}: the virtual gluon in the \ttbar\ production is replaced by other particles like a heavy scalar $\phi$. 

There are several new physics models which include new particles that are able to replace the virtual gluon in the \ttbar\ production. One example which is in particular interesting to explain the deviation of $A_{\text{FB}}$ as measured by the CDF collaboration \cite{top_asymm_CDF, top_asymm_lep_CDF} is the existence of an axigluon \cite{Baumgart2011, Tavares2011}. It is part of theories which embed QCD into a more general $SU(3) \times SU(3)$ gauge group \cite{Frampton1987}.
 
Other theories embedding the full SM into a larger gauge group predict the existence of a heavy neutral gauge boson $Z'$, affecting the \ttbar\ spin correlation \cite{Arai2009}. 
Furthermore, \newword{Kaluza-Klein gravitons} $G$ \cite{Fitzpatrick2007} as part of the \newword{Randall-Sundrum model} \cite{Randall1999} lead to different spin correlation coefficients $C$ as calculated in \cite{Gao2010}. 
In \cite{Baumgart2011} a general overview of spin correlation modifications caused by Spin-0, Spin-1 and Spin-2 resonances is given. 

A rather model-independent approach to look for new physics in the production is the search for non-vanishing top quark chromomagnetic- and chromoelectric dipole moments of which the latter would lead to a CP violation in QCD and modify the \ttbar\ spin correlation \cite{Bernreuther2013, Baumgart2013}.  

Figure \ref{fig:BSM_summary} shows the modifications of $A_{\text{FB}}$ (as in Equation \ref{eq:AFB}) and $C_{\text{helicity}}$ by several BSM scenarios: axigluons $G'$, scalar colour triplets $\Delta$, scalar colour sextets $\Sigma$ and neutral components of a scalar isodoublet $\phi^0$ \cite{Fajfer2012}. The 68 and 95\,\% CL results from ATLAS \cite{ATLAS_spin_paper} are included as yellow band and grey dashed line. While the ATLAS results indicate the exclusion of a large parameter space of the BSM models, it should be kept in mind that updated ATLAS results \cite{ueberpaper} lower the yellow band and reduce the exclusion. This is due to the change of the central value of $C_{\text{helicity}} = 0.40$ \cite{ATLAS_spin_paper} to  $C_{\text{helicity}} = 0.37$ (via \dphi) and $C_{\text{helicity}} = 0.23$ (via $\cos{\theta_{l^+}} \cos{\theta_{l^-}}$) \cite{ueberpaper}. 

Such BSM interpretations need to be handled with care. The \dphi\ distribution and the  $\cos{\theta_{i}} \cos{\theta_{j}}$ distributions are sensitive to different elements of the \ttbar\ spin density matrix. Hence, BSM models can have a different impact on both. Treating different distributions as equivalent, translating the results via \fsm\ and interpreting one in terms of the other is at least delicate.

Another aspect in the context of \ttbar\ spin correlation is the search for a scalar partner of the top quark, \textit{stops}\index{Stop} or \textit{top squarks}\index{Top squark}. Such particles are predicted in the context of SUSY and can mimic the decay signature of top quarks. As these particles are scalar, they will have a large impact on the \dphi\ distribution. This was calculated in \cite{stop} and is shown in Figure \ref{fig:stop}.

\subsection{New Physics in the \ttbar\ Decay}
\label{sec:BSM_decay}
\begin{figure}[ht]
	\centering
			\subfigure[]{
		\includegraphics[width=0.45\textwidth]{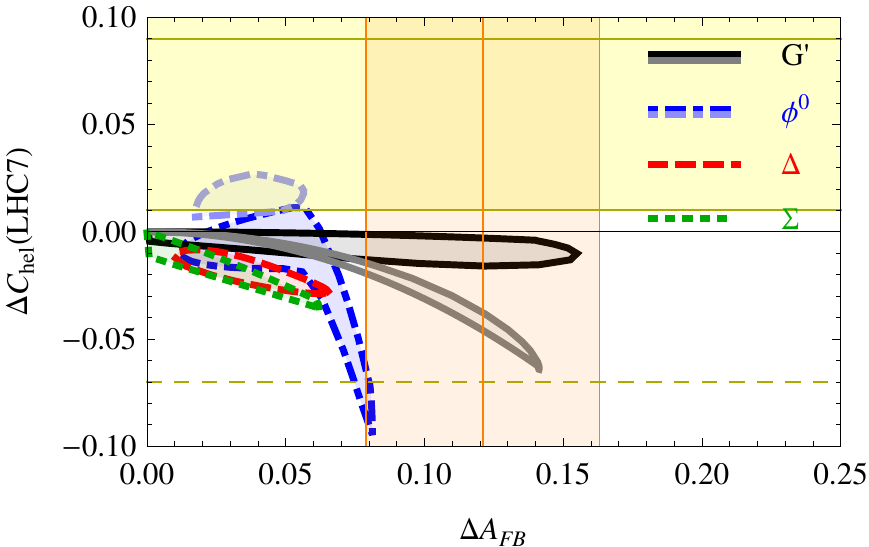}
			\label{fig:BSM_summary}
		}
													\subfigure[]{
		\includegraphics[width=0.45\textwidth]{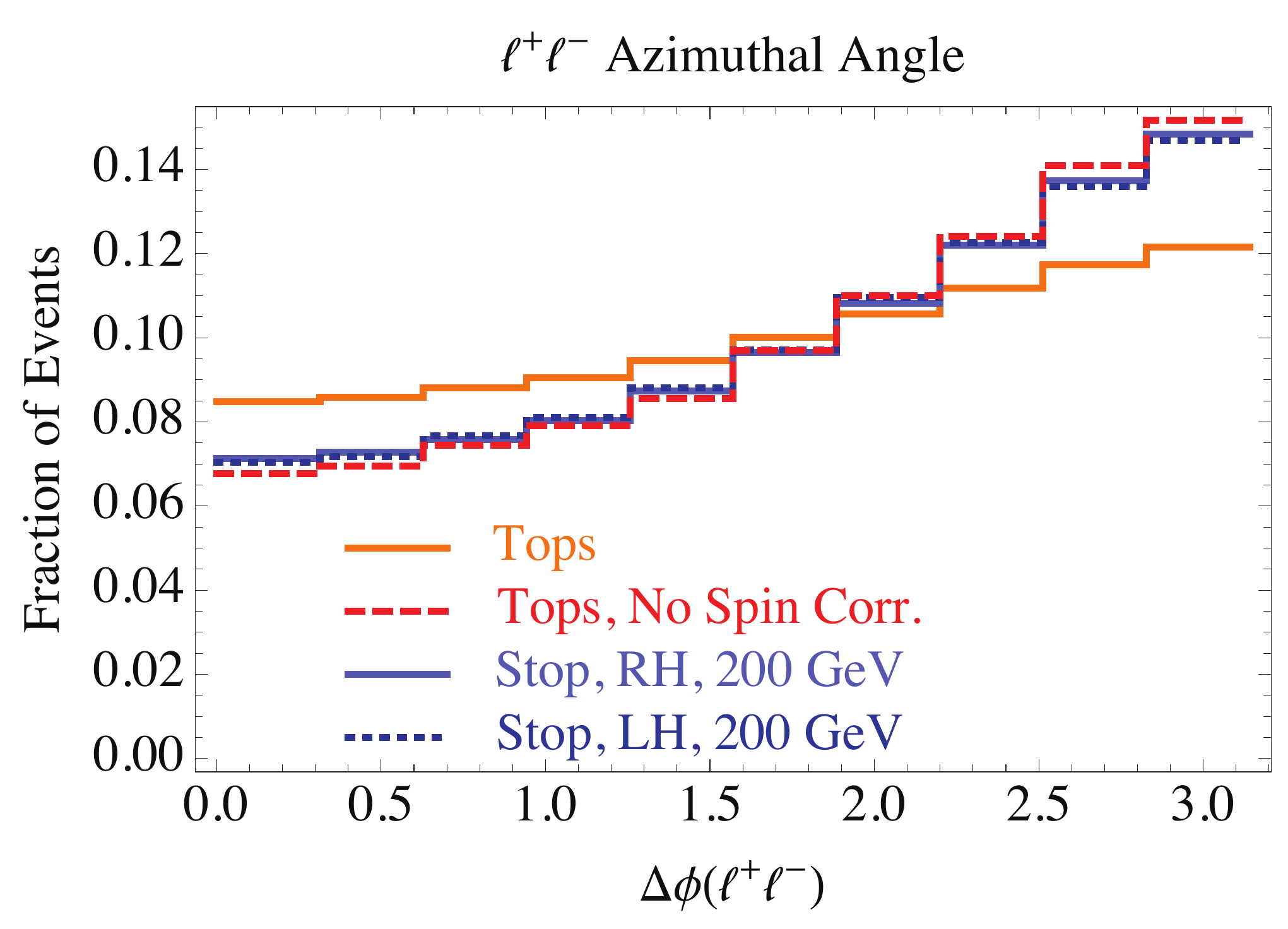}
			\label{fig:stop}
		}\\
							\subfigure[]{
		\includegraphics[width=0.45\textwidth]{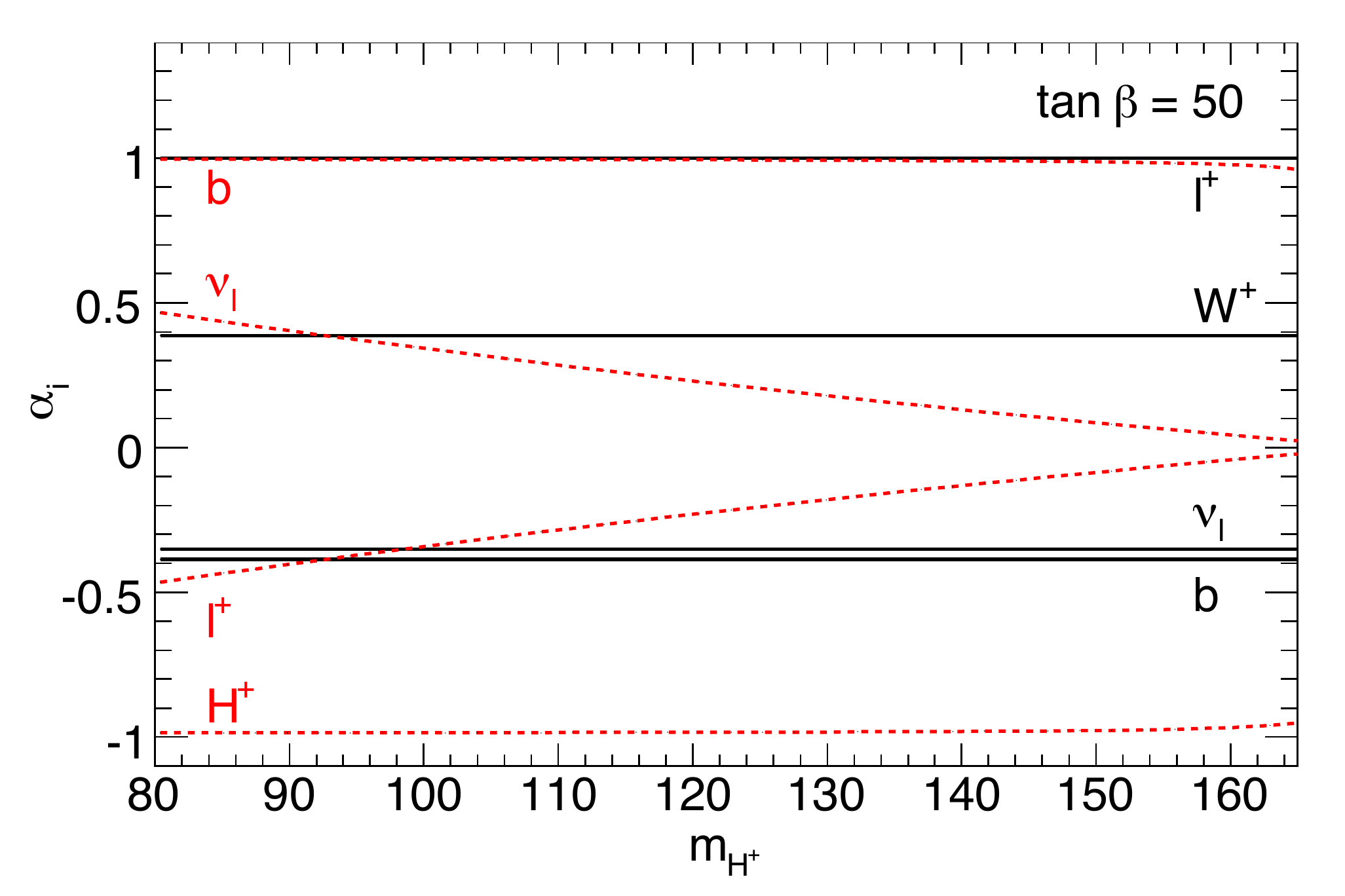}
			\label{fig:BSM_alpha}
		}
									\subfigure[]{
		\includegraphics[width=0.45\textwidth]{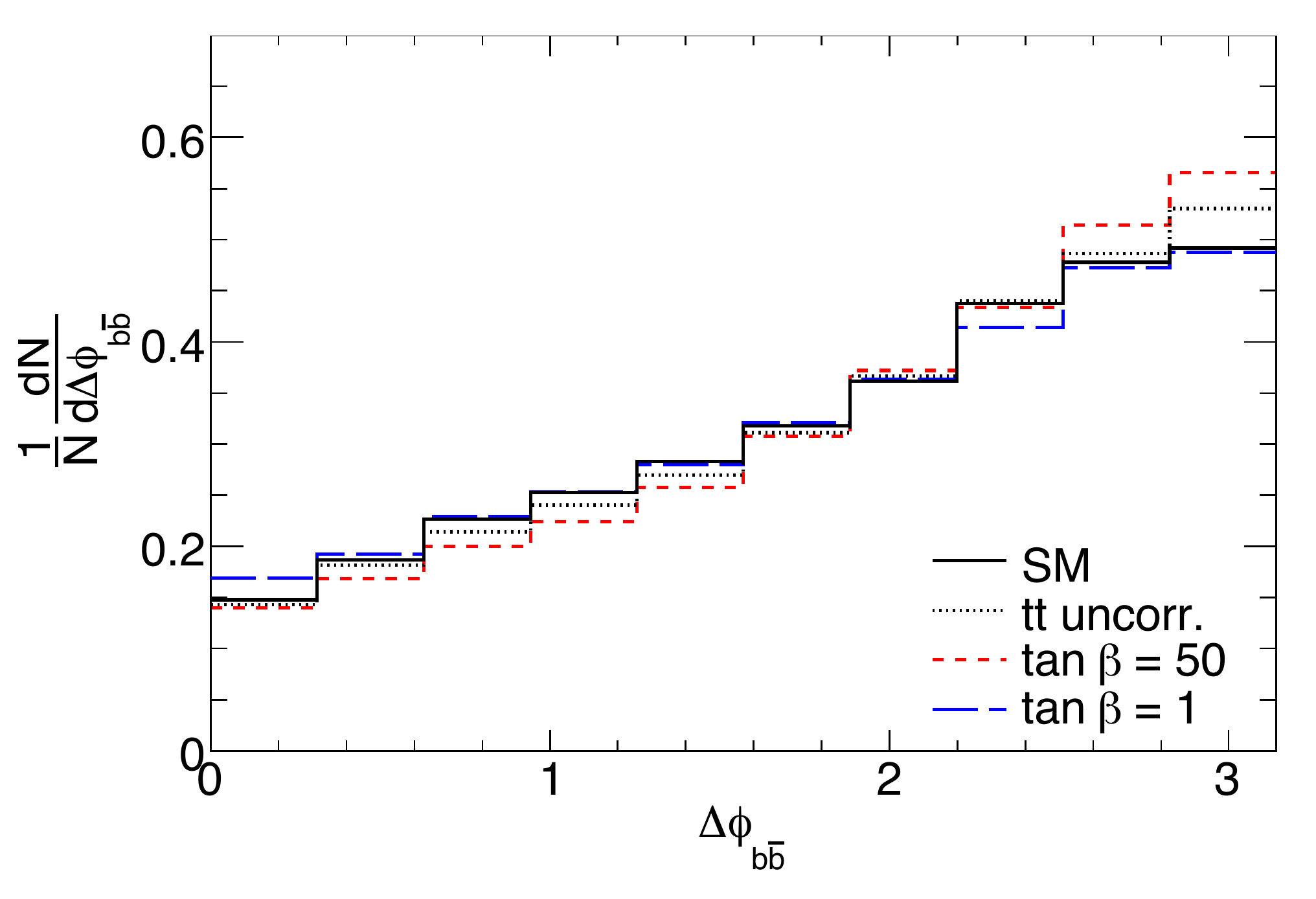}
			\label{fig:BSM_dphi_hplus}
		}
		\caption{\subref{fig:BSM_summary} Modifications of $A_{\text{FB}}$ and $C_{\text{helicity}}$ by several BSM scenarios \cite{Fajfer2012}. \mbox{\subref{fig:stop} $\Delta \phi(l^{+}, l^{-})$} distributions for top squarks in the dilepton channel, simulated at \rts = 8 TeV \cite{stop}. \subref{fig:BSM_alpha} Spin analysing powers in top decays via a charged Higgs $H^+$ \cite{Eriksson2007}. \subref{fig:BSM_dphi_hplus} Modifications of $\Delta \phi \left( b, \bar{b}\right)$ by the decay $t \rightarrow H^+ b$ \cite{Eriksson2007}.}
		\label{fig:BSM}
\end{figure}

Access to the \ttbar\ spin correlation is possible via the top quark decay products serving as analysers. It is explicitly assumed that the top decay vertex is of pure $V-A$ structure. The implications of a V+A mixture on the spin correlation was evaluated in \cite{Jezabek1994}. Next to this rather general test also specific models have been checked. The \newword{Two-Higgs-Doublet Model} \textit{2HDM}\index{2HDM | see {Two-Higgs-Doublet Model }}, required for example in many SUSY models, includes two additional charged Higgs bosons. If their mass is close to the one of the $W$ boson, it will be hard to identify them directly in top quark decays. But as a charged Higgs boson is a scalar particle, it will modify the \ttbar\ spin correlation \cite{Mahlon1996}. In \cite{Eriksson2007} the modifications of the spin analysing power $\alpha$ of the top decay products (Figure \ref{fig:BSM_alpha}) and the modifications of the \dphi\ shape (Figure \ref{fig:BSM_dphi_hplus}) were calculated. In these studies different ratios $\beta$ of the VEVs of the two Higgs doublets were evaluated. It should be mentioned that large values of $\tan{\beta}$ would also lead to significant changes in the $W$ helicity observables \cite{Eriksson2007}.

\section{Recent Measurements of \texorpdfstring{$t \bar{t}$}{Top/Anti-Top Quark} Spin Correlation}
\label{sec:recentresults}

In this section the most recent results concerning the measurement of top quark polarization and \ttbar\ spin correlation are presented. 
\subsection{Recent Top Polarization Measurements}
Top quarks produced in pairs via the strong interaction are predicted to be unpolarized \cite{Bernreuther2001}. This yields to vanishing coefficients $B$ in Equation \ref{eq:doublediff} and $P^{3}$ in Equation \ref{eq:analyser_angle}, respectively.
The D0 collaboration analysed $5.4~\text{fb}^{-1}$ of data, taken at $\sqrt{s} = 1.96\,\TeV$.
They found a good agreement between the $\cos{\theta}$ distributions and the SM expectation without quoting an explicit value $B$ for the polarization \cite{top_pol_D0}.
Instead, a Kolmogorov-Smirnov (KS) test of the SM prediction of $\cos{\theta}$, using the helicity basis, was performed. It lead to a  KS test probability of 14\,\% in the dilepton channel and 58\,\% in the \ljets\ channel. 
The charged leptons were used as analysers.

\begin{figure}[ht]
	\centering
			\subfigure[]{
		\includegraphics[width=0.6\textwidth]{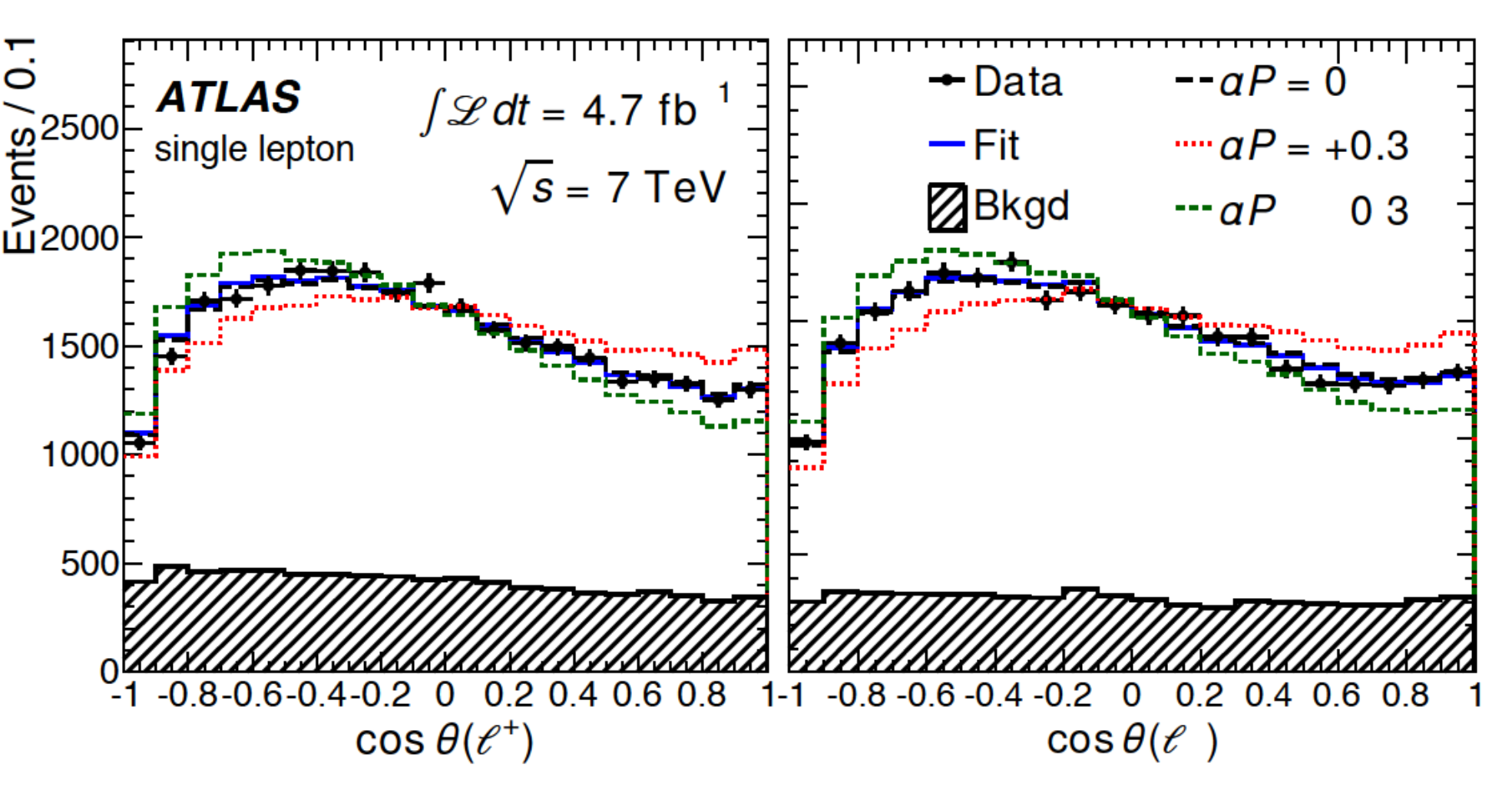}
			\label{fig:top_pol_ATLAS}
		}
					\subfigure[]{
		\includegraphics[width=0.33\textwidth]{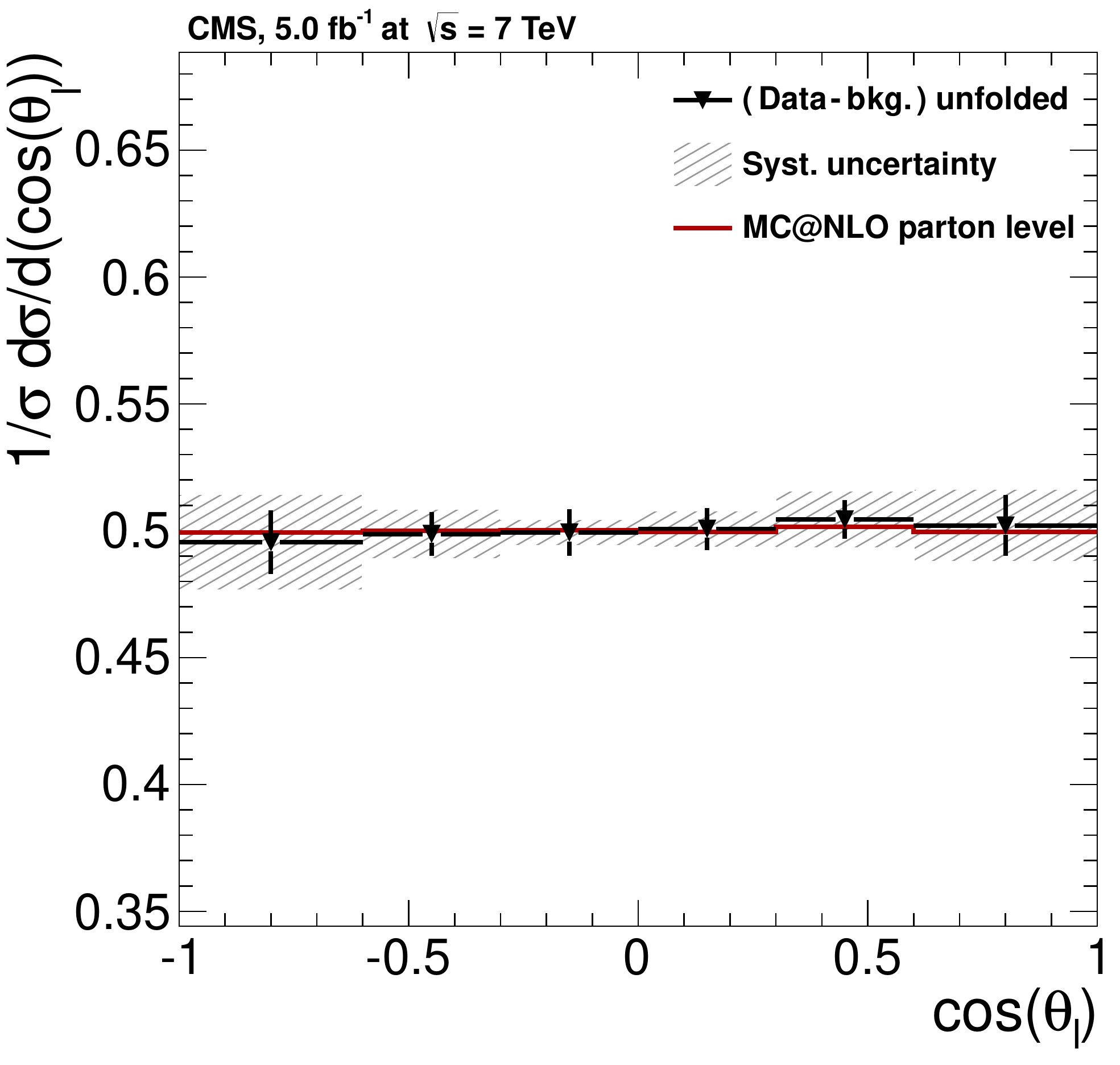}
			\label{fig:top_pol_CMS}
		}
		\caption{\subref{fig:top_pol_ATLAS} Distributions of the angles between the charged leptons and the helicity spin basis. The data was fit with a CP conserving polarization hypothesis \cite{ATLAS_top_pol}. \subref{fig:top_pol_CMS} Unfolded distribution of the angle between the charged lepton and the helicity spin basis \cite{CMS_spin_paper}.}
		\label{fig:top_pol_res}
\end{figure}

The strategy of fitting templates to the $\cos{\theta}$ distributions was also followed by ATLAS. As in the D0 measurement, the helicity basis was used. With $5.4~\text{fb}^{-1}$ of data taken at $\sqrt{s} = 7\,\TeV$ ATLAS measured $B$ in two scenarios \cite{ATLAS_top_pol}. In the first case, a non-vanishing polarization was assumed to stem from a CP conserving process. This lead to \mbox{$B_{CPc} = -0.035 \pm 0.014\,\text{(stat.)}\pm0.037\,\text{(syst.)}$}. In case of a maximally CP violating process causing the polarization it was measured as \mbox{$B_{CPv} = 0.020 \pm 0.016\,\text{(stat.)}{\,}^{+0.013}_{-0.017}\,\text{(syst.)}$}. Figure \ref{fig:top_pol_ATLAS} shows the measured and fitted distribution of $\cos{\theta}$ using the CP conserving hypothesis.

The CMS collaboration used $5.0~\text{fb}^{-1}$ of data taken at $\sqrt{s} = 7\,\TeV$ to measure the asymmetry
\begin{align}
A_P = \frac{N\left( \cos{\theta_l} > 0\right) - N\left( \cos{\theta_l} < 0\right)}{N\left( \cos{\theta_l} > 0\right) + N\left( \cos{\theta_l} < 0\right)}
\end{align}
of unfolded $\cos{\theta}$ distributions (see Figure \ref{fig:top_pol_CMS}). CP invariance was assumed and the charged leptons were used as analysers. 
This asymmetry is directly related to the polarization $B$ via $B=2 \cdot A_P$. \\CMS measured \mbox{$B = 0.010 \pm 0.026\,\text{(stat.)}\pm 0.040\,\text{(syst.)} \pm 0.016\,\text{(top \pt)}$}.

All top polarization results are in good agreement with the SM prediction of $B = 0$.

\subsection{Recent Spin Correlation Measurements}
\label{sec:spinresults}
At the Tevatron, the full dataset of 5.4 \ifb\ was analysed by the CDF and D0 collaborations. In the 
dilepton channel D0 performed a fit of \ttbar\ signal templates to the $\cos{\theta_{l^{+}}}\cos{\theta_{l^{-}}}$ distributions mixing uncorrelated events with events correlated as predicted by the SM \cite{D0_spin_dilep_angles}. 

Their result was a spin correlation \mbox{$C_{\text{beam}} = 0.10 {\,}^{+0.38}_{-0.40}\,\text{(stat.)} \pm 0.11\,\text{(syst.)}$} which agrees with the SM prediction $C_{\text{beam}}^{\text{SM}} = 0.78$. For the same dataset and \ttbar\ channel the spin correlation was measured via \fsm, explained in Section \ref{sec:fsm}, by using a matrix-element-based approach \cite{D0_spin_dilep_MEM}. This leads to $\fsm=0.74 {}^{+0.33}_{-0.35}\,\text{(stat.)} {\,}^{+0.15}_{-0.18}\,\text{(syst.)}$ which is in good agreement with the SM prediction of $\fsm=1.0$ and excludes the scenario of uncorrelated \ttbar\ spins at the 97.7\,\% CL. In the \ljets\ channel D0 measured with the same approach $\fsm=1.15{\,}^{+0.25}_{-0.26}\,\text{(stat.)} \pm{0.18}\,\text{(syst.)}$. Figure \ref{fig:D0_spin_ljets} shows the likelihood discriminant $R$ for the \ljets\ channel. The combination with the result in the dilepton channel leads to an evidence for \ttbar\ spin correlation by excluding the uncorrelated scenario at the $3\,\sigma$ level ($\fsm=0.85 \pm 0.29$) \cite{D0_spin_ljets}. 

The CDF collaboration fitted two-dimensional distributions of (\mbox{$\cos{\theta_{l^+}},\cos{\theta_{l^-}}$}) and (\mbox{$\cos{\theta_{b}},\cos{\theta_{\bar{b}}}$}) between the two charged leptons and the two $b$ jets in the dilepton channel using 5.1 \ifb\ of data and the beam line basis \cite{CDF_spin_dilep}. The templates were parameterized as a function of the spin correlation $C$. As a result, $C=0.042{\,}^{+0.563}_{-0.562}\,\text{(stat.} \oplus \text{syst.)}$ was obtained. In the \ljets\ channel CDF fitted two-dimensional distributions of  ($\cos{\theta_{l}} \cos{\theta_{d}},\cos{\theta_{l}} \cos{\theta_{b}}$) using both the beam line and the helicity basis \cite{CDF_spin_ljets} with 5.3 \ifb\ of data. In Figure \ref{fig:CDF_spin_ljets} the fitted result for the $\cos{\theta_{l}} \cdot \cos{\theta_{d}}$ distribution is shown. The two-dimensional fit lead to $C_{\text{beam}}=0.72 \pm 0.64\,\text{(stat.)} \pm{0.26}\,\text{(syst.)}$ and $C_{\text{helicity}}=0.48\pm 0.48\,\text{(stat.)} \pm{0.22}\,\text{(syst.)}$. 

\begin{figure}[ht]
	\centering
			\subfigure[]{
		\includegraphics[width=0.45\textwidth]{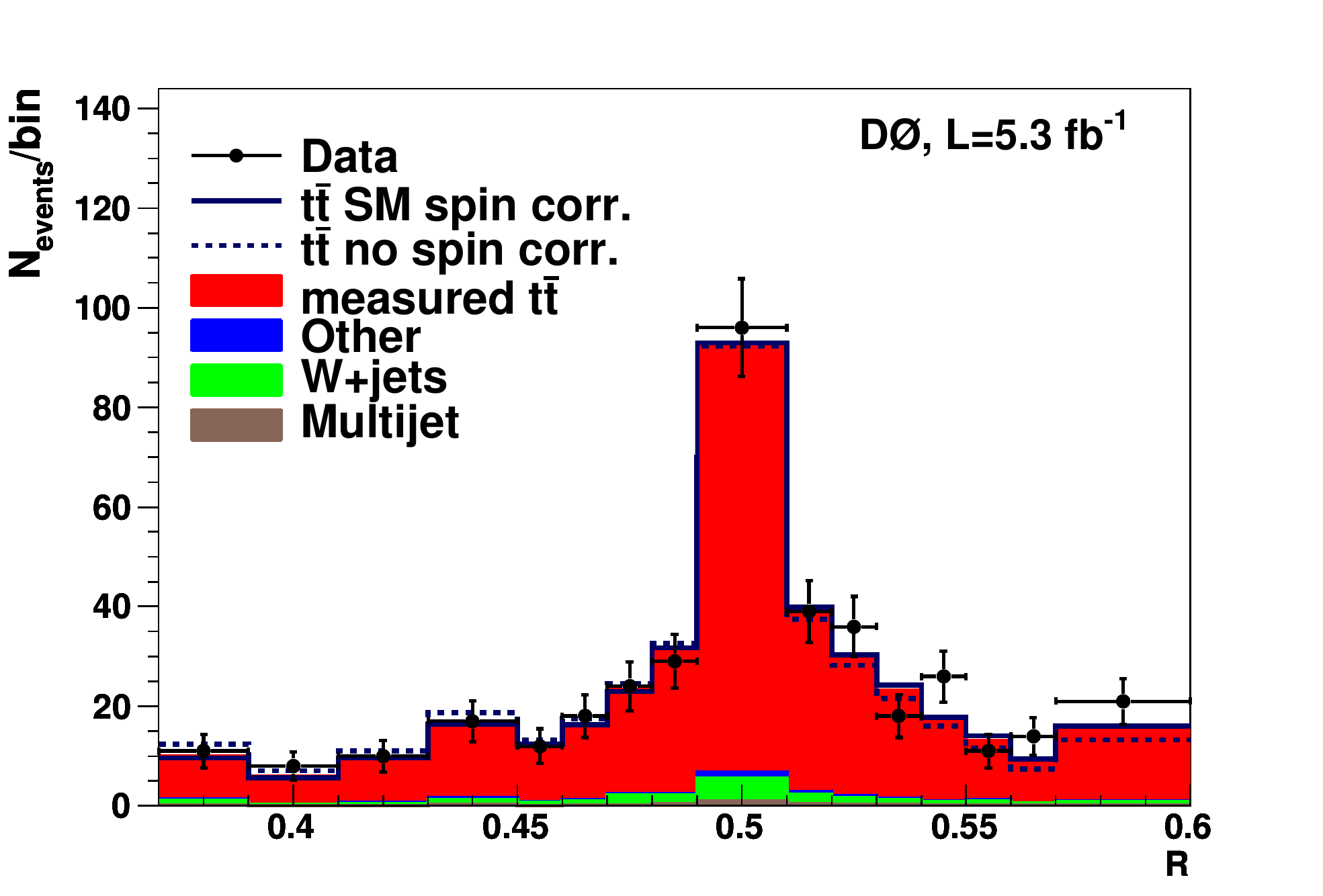}
			\label{fig:D0_spin_ljets}
		}
					\subfigure[]{
		\includegraphics[width=0.45\textwidth]{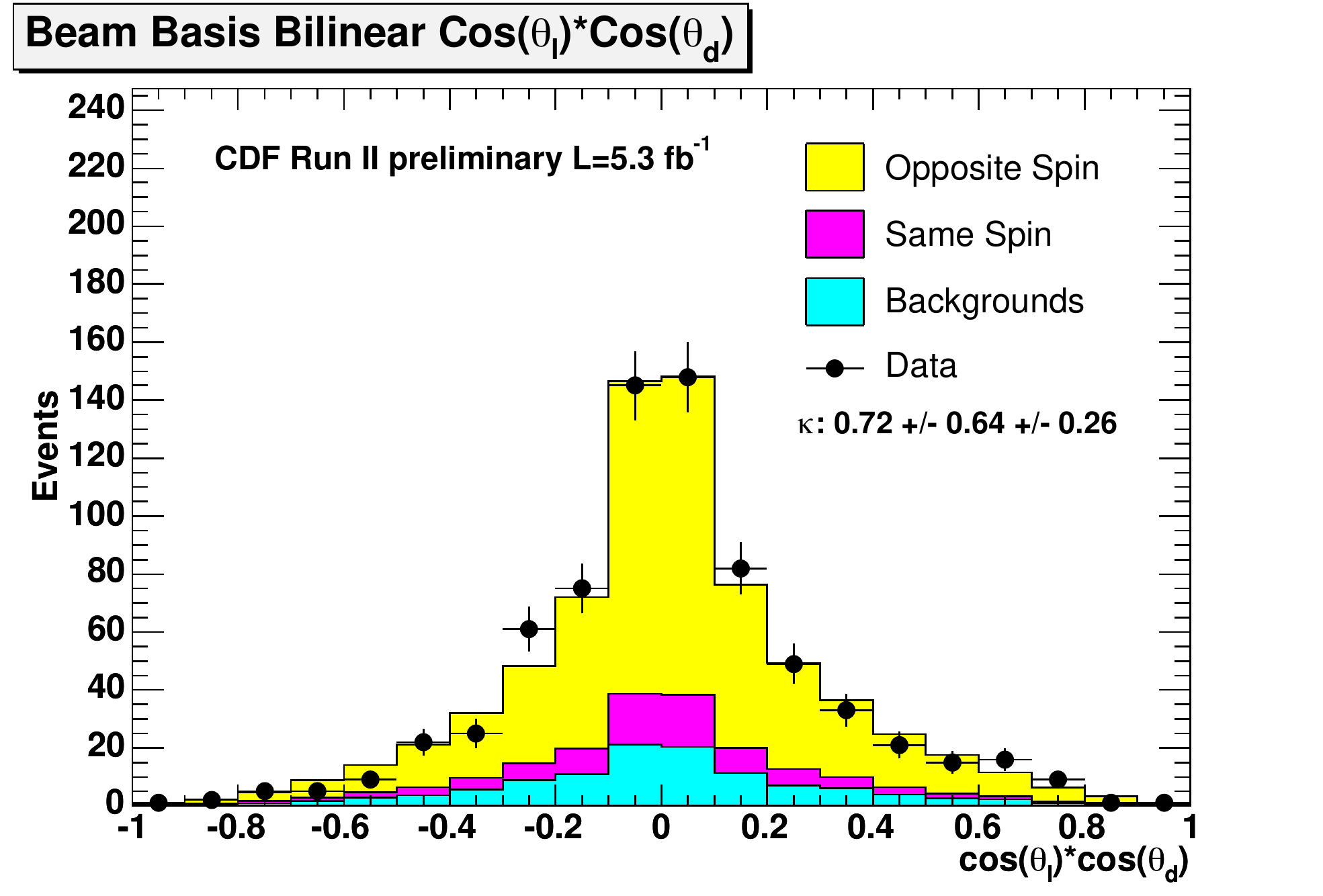}
			\label{fig:CDF_spin_ljets}
		}
		\caption{Results of measured \ttbar\ spin correlations at the Tevatron in the \ljets\ channel. \subref{fig:D0_spin_ljets} Likelihood discriminant $R$ for the analysis in the \ttbar\  \ljets\ channel of the D0 analysis. \subref{fig:CDF_spin_ljets} Result of the fitted $\cos{\theta_{l}} \cdot \cos{\theta_{d}}$ (beam line basis) distribution of the CDF analysis \cite{CDF_spin_ljets}.}
		\label{fig:spinresults_tevatron}
\end{figure}

Moving to the LHC shows that new approaches have to be chosen. On the one hand, the helicity basis offers a higher expected value of $C$ compared to the beam line basis. On the other hand the unique opportunity of using \dphi\ distributions is given. As this requires no full event reconstruction, except the identification of the spin analysers, the ATLAS and CMS experiments both measured $\Delta \phi \left( l^+, l^-\right)$ in the dilepton channel. Already 2.1 \ifb\ provided sufficient statistics, allowing to exclude the scenario of uncorrelated \ttbar\ pairs at the $5\,\sigma$ level by ATLAS \cite{ATLAS_spin_paper}. Two templates of $\Delta \phi \left( l^+, l^-\right)$ distributions were used: SM prediction of spin correlation and uncorrelated \ttbar\ events. A value of $\fsm=1.30 \pm 0.14\,\text{(stat)} {\,}^{+0.27}_{-0.22}\,\text{(syst.)}$ was measured and translated to $C_{\text{helicity}} = 0.40 \pm 0.04\,\text{(stat.)} {\,}^{+0.08}_{-0.07}\, \text{(syst.)}$ and $C_{\text{maximal}} = 0.57 \pm 0.06\, \text{(stat.)} {\,}^{+0.12}_{-0.10}\, \text{(syst.)}$, respectively. The sum of \dphi\ distributions for the $ee$, $\mu \mu$ and $e \mu$ channel is shown in Figure \ref{fig:ATLAS_spin_obs} for data and the two distributions of SM-like and uncorrelated \ttbar\ events. 
\begin{figure}[ht]
	\centering
			\subfigure[]{
		\includegraphics[width=0.47\textwidth]{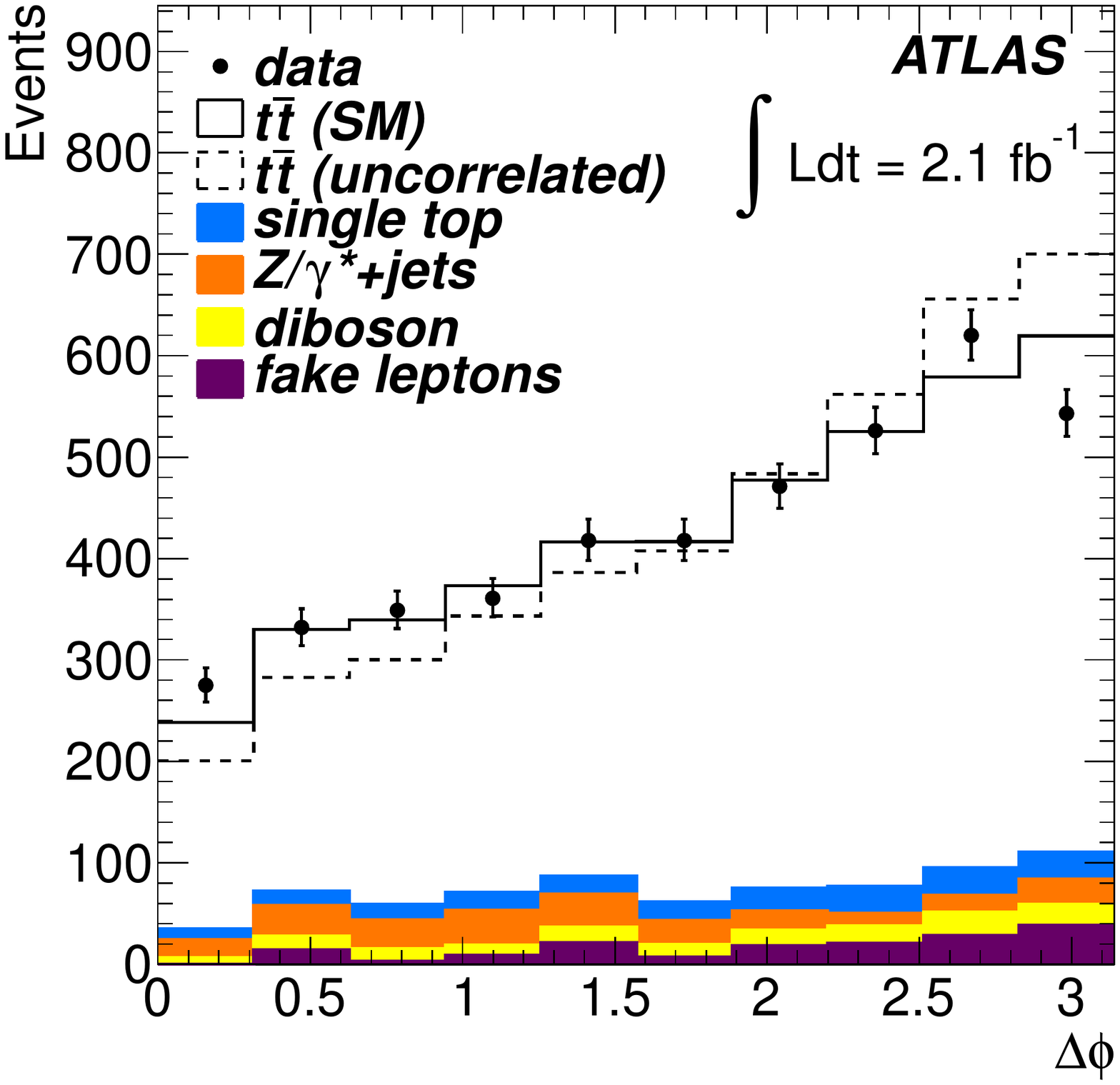}
			\label{fig:ATLAS_spin_obs}
		}
					\subfigure[]{
		\includegraphics[width=0.45\textwidth]{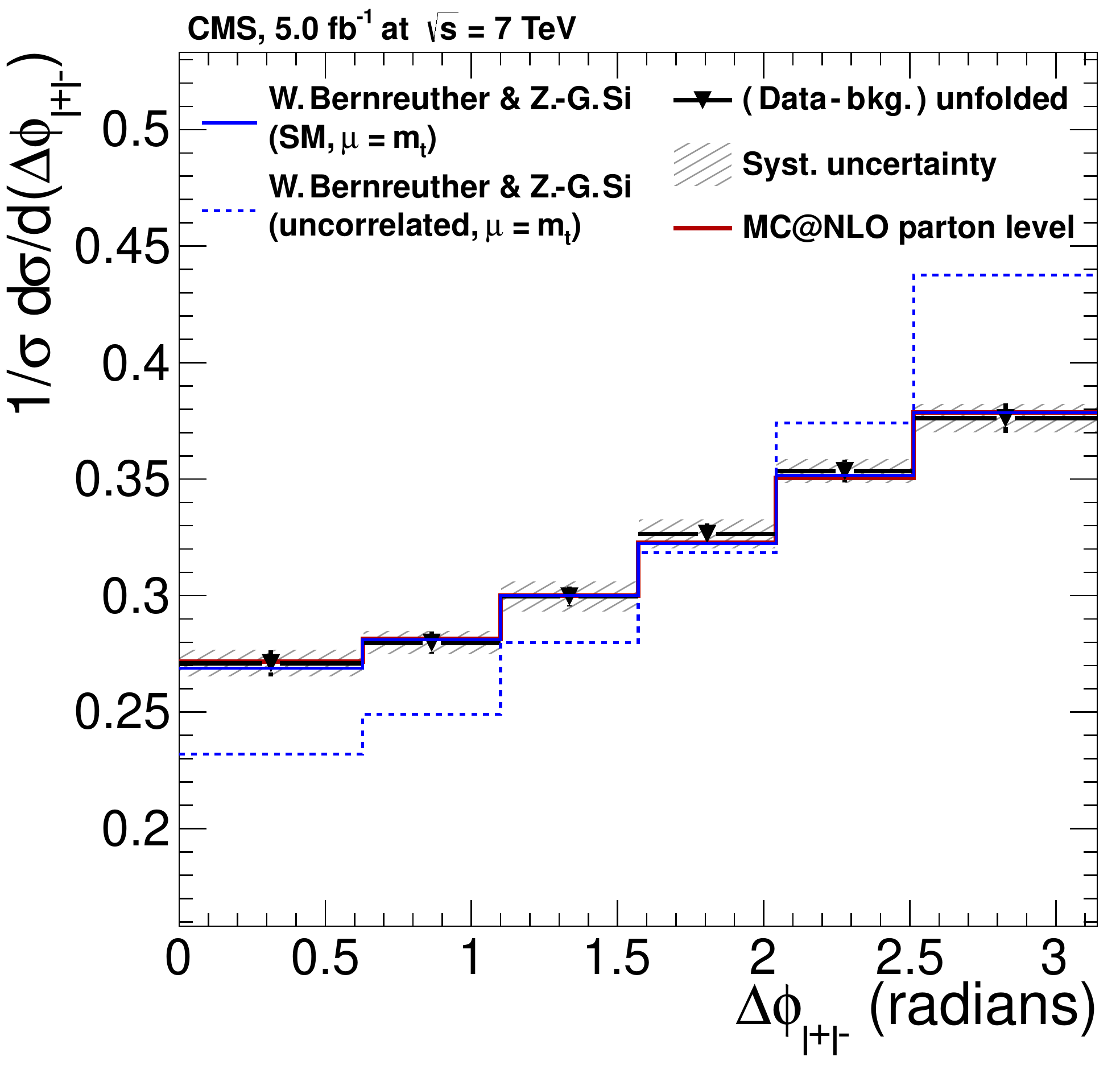}
			\label{fig:CMS_spin_paper}
		}
		\caption{LHC measurements of the $\Delta \phi \left( l^+, l^-\right)$ distributions in \ttbar\ events decaying in the dilepton channel. \subref{fig:ATLAS_spin_obs} ATLAS result leading to observation of spin correlation \cite{ATLAS_spin_paper}. \subref{fig:CMS_spin_paper} Unfolded distribution with subtracted background as measured by CMS together with NLO predictions \cite{CMS_spin_paper}.}
		\label{fig:spinresults_deltaphi}
\end{figure}
The full dataset taken at $\sqrt{s} = 7\,\TeV$ with an integrated luminosity of 4.6 \ifb\ was also analysed by ATLAS \cite{ueberpaper} fitting SM-like and uncorrelated \ttbar\ signal templates in the dilepton channel and extracting \fsm. The results for the \dphi, the $S$-Ratio (Figure \ref{fig:ATLAS_spin_Sratio}), and the $\cos{\theta_{l^+}}\cos{\theta_{l^-}}$ (Figure \ref{fig:ATLAS_spin_max} for the maximal basis) distributions are shown in Figure \ref{fig:ATLAS_spin_conf}. All results agree with the SM predictions. However, with the exception of \dphi, all other distributions result in slightly lower values of \fsm\ than predicted. The results of \cite{ueberpaper}, shown in Figure \ref{fig:ATLAS_spin_conf}, include measurements of the dilepton channel and the results of this thesis. The results of this thesis are the first LHC measurements in the \ljets\ channel. 
\begin{figure}[ht]
	\centering
		\includegraphics[width=0.7\textwidth]{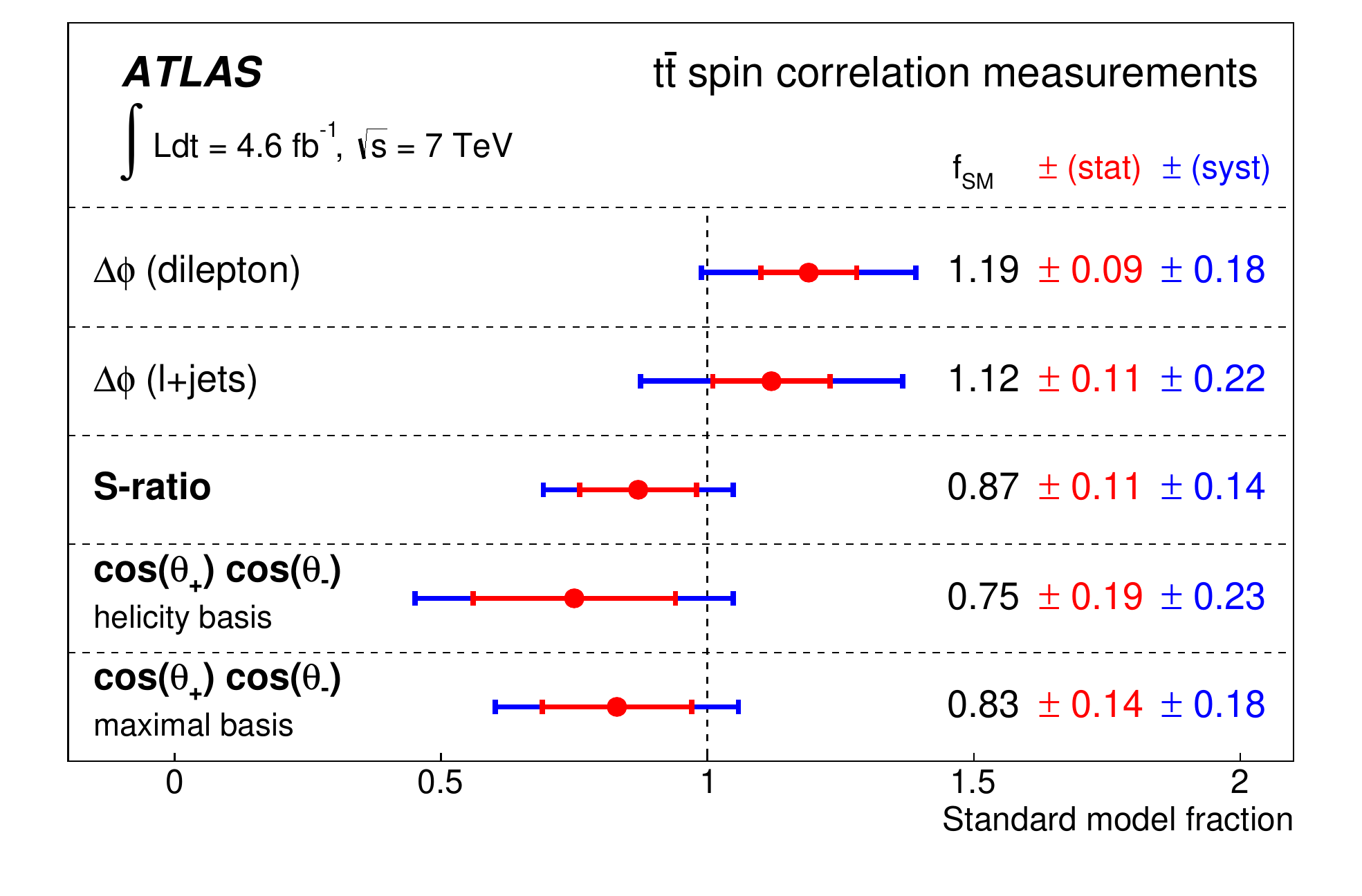}
		\caption{Overview of ATLAS results of measurements of the \ttbar\ spin correlation for the 4.6 \ifb\ 7 TeV dataset \cite{ueberpaper}.}
	\label{fig:ATLAS_spin_conf}
\end{figure}
CMS performed an unfolding of the  $\Delta \phi \left( l^+, l^-\right)$ and $\cos{\theta_{l^+}}\,\cos{\theta_{l^-}} \equiv c_1 \cdot c_2$ distributions using 5.0 \ifb\ of 7 TeV data. Asymmetries $A$, 
\begin{align}
A_{\Delta\phi} &= \frac{N (  \Delta \phi_{\ell^+\ell^-} > \pi/2 )-N ( \Delta \phi_{\ell^+\ell^-} < \pi/2 )}{N ( \Delta  \phi_{\ell^+\ell^-} > \pi/2 ) +N ( \Delta \phi_{\ell^+\ell^-} < \pi/2 )}, \\
A_{c_{1}c_{2}} &= \frac{N (c_{1} \cdot c_{2} > 0)-N (c_{1} \cdot c_{2} < 0)}{N (c_{1} \cdot c_{2} > 0)+N (c_{1} \cdot c_{2} < 0)},
\end{align}
which are related to the spin correlation, were measured and compared to the predictions of the SM. Table \ref{tab:CMS_spin} shows the results. 
\begin{table}[htbp]
\small
\begin{center}
\begin{tabular}{|c|c|c|c|c|}
\hline
{} &  {Data (unfolded)} & {\mcatnlo} & {NLO (SM)} & {NLO (uncorr.)} \\
\hline
\hline 
\rule{0pt}{0.5cm}$A_{\Delta\phi}$ &   ${\;\;\;}0.113 \pm 0.010  \pm 0.007 \pm 0.012$  &  ${\;\;\;}0.110 \pm 0.001$  &  $0.115{\,}^{+0.014}_{-0.016}$  &  $0.210{\,}^{+0.013}_{-0.008}$    \\ 
\rule{0pt}{0.5cm}$A_{c_{1}c_{2}}$ & $-0.021 \pm 0.023 \pm 0.027 \pm 0.010$  &  $ -0.078 \pm 0.001$  &  $-0.078 \pm 0.006$ &  $0$ \\ [0.5ex]
\hline
\end{tabular}
\end{center}
\caption{Asymmetries related to the \ttbar\ spin correlation measured by CMS \cite{CMS_spin_paper}. The uncertainties are statistical, systematic and an additional uncertainty from top \pt\ reweighting. For \mcatnlo\ the statistical uncertainty is quoted. The NLO calculation includes the uncertainty of a variation of the renormalization and factorization scale by a factor of two.}
\label{tab:CMS_spin}
\end{table}
Here, $A_{c_{1}c_{2}}$ is directly related to the spin correlation $C_{\text{helicity}}$ as defined in Equation \ref{eq:spincorr_C} \cite{Bernreuther2010} via $C_{\text{helicity}} = -4 A_{c_{1}c_{2}}$. It is remarkable that the \dphi\ distributions agree very well with the SM predictions as seen in Figure \ref{fig:CMS_spin_paper} while the translated value of $C_{\text{helicity}} = 0.08 \pm 0.15$ does not ($C_{\text{helicity}}^{\text{SM}} = 0.31$, see Table \ref{tab:exp_corr}). 

\begin{figure}[ht]
	\centering
			\subfigure[]{
		\includegraphics[width=0.45\textwidth]{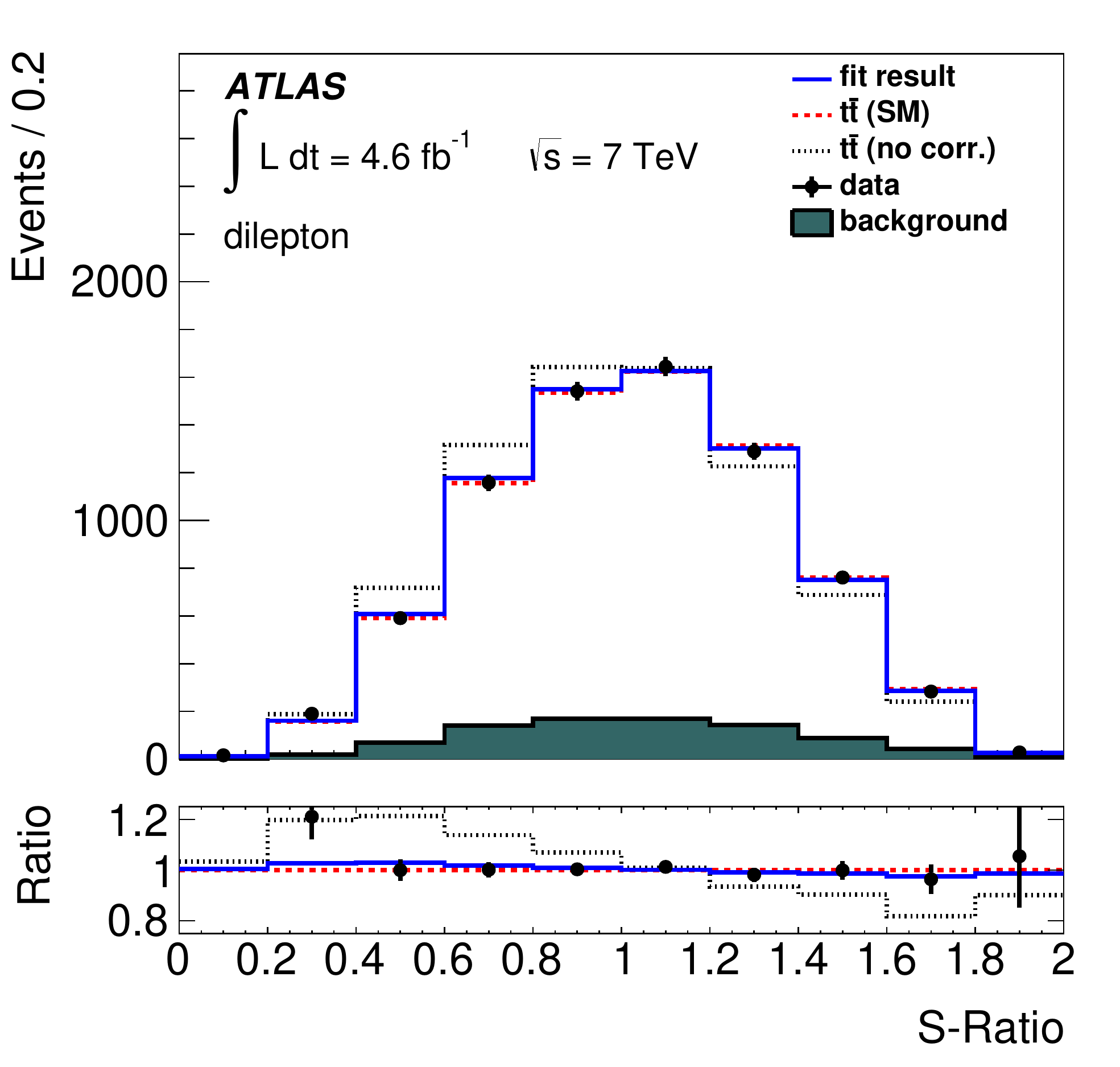}
			\label{fig:ATLAS_spin_Sratio}
		}
					\subfigure[]{
		\includegraphics[width=0.45\textwidth]{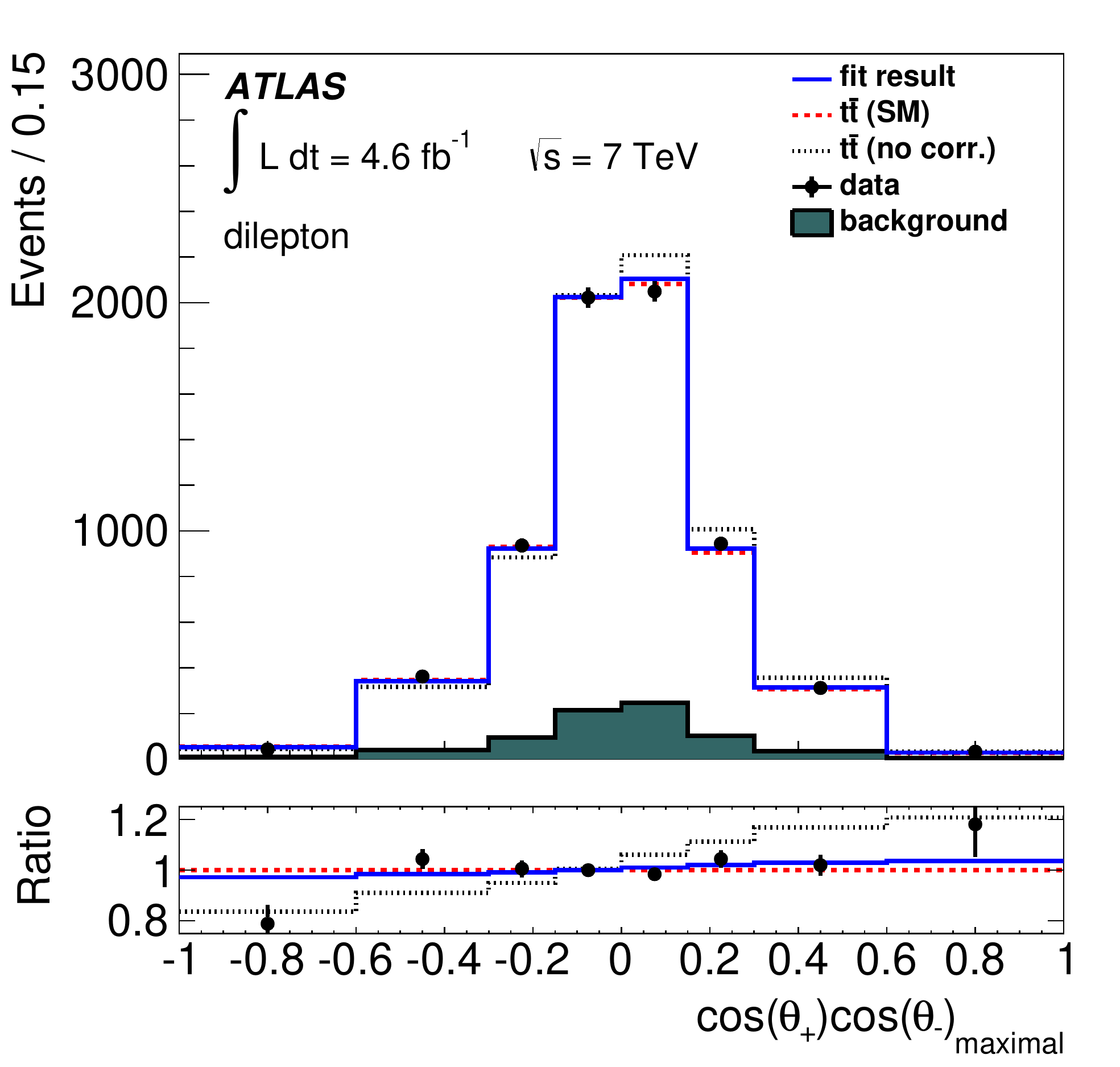}
			\label{fig:ATLAS_spin_max}
		}
		\caption{ATLAS measurements using 4.6 \ifb\ of data selecting \ttbar\ events in the dilepton channel and fitting \subref{fig:ATLAS_spin_Sratio} the $S$-Ratio and \subref{fig:ATLAS_spin_max} the $\cos{\theta_{l^+}}\,\cos{\theta_{l^-}}$ for the maximal basis \cite{ueberpaper}.}
		\label{fig:spinresults_atlasextra}
\end{figure}

To conclude, several bases and quantities sensitive to the \ttbar\ spin correlation have been measured to be consistent with the SM predictions. The uncertainties, however, are still too large to allow tight exclusion limits of BSM models. A small but not significant trend was observed, namely that the $\cos{\theta_i}\, \cos{\theta_j}$ distributions consistently lead to values below the SM prediction for all Tevatron and LHC measurements. This  was not the case using \dphi. 

Measurements of the top quark polarization \cite{ATLAS_top_pol, CMS_spin_paper} have not shown a deviation from zero, agreeing with the LO SM prediction.

With the exception of the results presented in this thesis, no measurement of \ttbar\ spin correlation at the LHC in the \ljets\ channel has been published. This motivated the choice of the \ljets\ channel for the analysis presented in this thesis. This analysis will investigate \ttbar\ spin correlation via hadronic analysers and test aspects that a measurement in the dilepton channel cannot access.

\chapter[Experimental Setup]{Experimental Setup}
Producing \ttbar\ pairs requires large production energies. As presently no lepton collider is able to deliver a \com\ energy of $\sqrt{s} = 2\,\mtop$, only the two most powerful hadron colliders are able to produce top quarks. The first one is the proton/anti-proton collider Tevatron located at Fermilab close to Chicago with $\sqrt{s} = 1.96\,\TeV$.\footnote{The Tevatron stopped operation in 2011.} The top quark's discovery was made at the two Tevatron experiments CDF and D0 in 1995 \cite{top_disc_CDF, top_disc_D0}. The second accelerator is the \textit{Large Hadron Collider (LHC)}\index{LHC} colliding protons with $\sqrt{s} = 7\,\TeV$ for the data taking period in 2011 before moving to $\sqrt{s} = 8\,\TeV$ for the 2012 dataset \cite{LHC}. Currently, the machine is being upgraded to operate with $\sqrt{s} = 13\,\TeV$ and $\sqrt{s} = 14\,\TeV$ for the time after the \textit{Long Shutdown 1 (LS1)}\cite{LS1}. 

Apart from the high \com\ energy, a high luminosity $\mathcal{L}$ is the key to maximize the total number of observed \ttbar\ events. With an integrated luminosity \intlum\ of \intlumi\ and a \ttbar\ production cross section of \ttbarxsec(7 TeV) = $173.3 \pm 2.3\,\text{(stat.)}\pm 9.8\,\text{(syst.)}$ pb \cite{top_xsec_lhccomb_atlas, top_xsec_lhccomb_cms}, about 800,000 \ttbar\ pairs were produced for the 2011 dataset. 

The \textit{ATLAS}\index{ATLAS} detector was used to take data analysed in this thesis. This chapter will introduce the accelerator and detector used.

\section{The LHC}
The Large Hadron Collider \cite{LHC} is the world's most powerful particle accelerator. It is based at the international particle research laboratory \textit{CERN}\index{CERN} (Organisation europ\'eenne pour la recherche nucl\'eaire) near Geneva, Switzerland. Based in a $26.7\,\text{km}$ long tunnel located $45$ to $170\,\text{m}$ below the surface\footnote{The LHC tunnel has an inclination of 1.4\,\%, leading to a variation of its altitude of about $\pm 60\,\text{m}$. See \cite{LHC2} for details.} \cite{LHC} between the Jura mountains and the Geneva Lake it can operate in three modes: proton/proton, proton/ion and ion/ion collision. The LHC uses protons for both beams in order to reach the design value of the instantaneous luminosity of $\mathcal{L} = 10^{34}\,\text{cm}^{-2}\text{s}^{-1}$. The design \com\ energy of $14\,\TeV$ will be reached after the LS1 will be completed. So far, the LHC was running with $\sqrt{s} = 7\,\TeV$ (2011 dataset) and $\sqrt{s} = 8\,\TeV$ (2012 dataset). Such high energies require the usage of superconductive magnets. 1232 superconducting dipole magnets use niobium-titanium coils operating at 1.9 K and producing a magnetic field of up to 8.3 Tesla. A sophisticated magnet design allows housing both beam pipes in the same cryostat. Additional 392 superconducting quadrupole magnets are used to focus and stabilize the beam, supported by further multipole magnets of higher order. Reaching the design values, the LHC will contain 2808 bunches per beam, each consisting of $1.15 \cdot 10^{11}$ protons and spaced with a distance of 25 ns. Details about the LHC machine parameters for the design values and the dataset analysed in this thesis are given in Section \ref{sec:dataset}.

The LHC is fed with protons from the CERN accelerator chain: Hydrogen atoms are ionized and accelerated up to 50 MeV at the \textit{LINAC II}\index{LINAC II} before they are injected to the \textit{Booster}\index{Booster} (1.4 GeV) which then fills the \textit{Proton Synchroton}\index{Proton Synchroton} (PS\index{PS | see {Proton Synchrotron }}, 25 GeV). From here the protons are lead to the \textit{Super Proton Synchrotron}\index{Super Proton Synchrotron} (SPS, 450 GeV) before they reach their final destination, the LHC. Ions start being accelerated at the \textit{LINAC III}\index{LINAC III} and the \textit{LEIR}\index{LEIR} (Low Energy Ion Ring) before being filled to the PS (450 GeV). The CERN accelerator complex is illustrated in Figure \ref{fig:acc_map}.

\begin{figure}[ht]
	\centering
		\includegraphics[width=\textwidth]{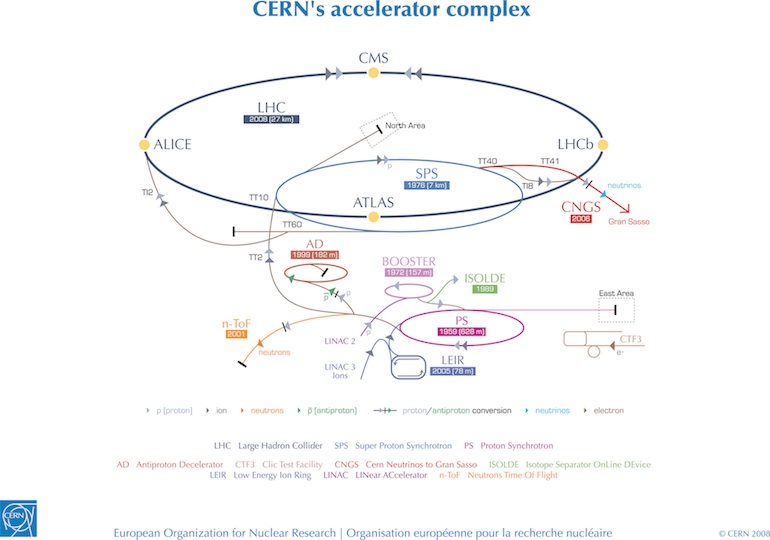}
		\caption{The CERN accelerator complex \copyright CERN.}
	\label{fig:acc_map}
\end{figure}
The two proton beams are crossed and brought to collision at four interaction points. Each of them is surrounded by an experiment: \textit{ATLAS}\index{ATLAS} \cite{ATLAS} and  \textit{CMS}\index{CMS} \cite{CMS} are two general purpose detectors. They both cover almost the full solid angle and aim for high luminosities and low $\beta^{*}$ \footnote{Beam shapes can be modelled by  Gaussian distributions in the transverse plane. The $\beta$ function describes how the constant beam emittance is reduced to the beam width $\sigma$ during collimation via $\sigma = \sqrt{\varepsilon \beta}$. The value of $\beta$ at the interaction point is indicated with $\beta^{*}$.} to discover rare events. The main physics goals of these experiments are the search for a Higgs boson, Dark Matter candidates and signatures for supersymmetry. The \textit{ALICE}\index{ALICE} \cite{ALICE} experiment focuses on the analysis of heavy ion collisions searching for signatures of the \textit{quark gluon plasma}\index{Quark Gluon Plasma} (\textit{QGP}\index{QGP | see {Quark Gluon Plasma }}) and analysing the behaviour of hadronic matter at high densities and temperatures. The \textit{LHCb}\index{LHCb} \cite{LHCb} experiment focuses on B-physics and physics at low scattering angles. It is asymmetric and covers only a part of the phase space. Its physics goals are the study of CP violation\index{CP violation} and BSM physics involving heavy flavours. 

Next to the four big experiments several smaller ones are located close to the interaction point, such as \textit{MOEDAL}\index{MOEDAL} \cite{MOEDAL} searching for magnetic monopoles, \textit{LHCf}\index{LHCf} \cite{LHCf} that studies hadron interaction models used in cosmic ray analyses and \textit{TOTEM}\index{TOTEM} \cite{TOTEM} for elastic and diffractive cross section measurements.

\section{The ATLAS Detector}
The ATLAS detector is a general purpose detector covering almost the full solid angle. It consists of several layers of tracking, calorimetry and muon chamber devices. ATLAS is capable of dealing with event rates of up to 40 million events per second resulting from the high luminosities provided by the LHC. As up to 50-140 (for the design values of the LHC, depending on the chosen filling scheme \cite{LS1}) hard scattering events can pile up during a bunch crossing, an excellent tracking system is required in order to associate the reconstructed physics objects with different interaction processes. The tracking devices are also used for tagging jets as \textit{$b$-jets}\index{b@$b$-jet}. Such jets emerged from an initial B-meson leading to a secondary vertex within the tracking system with a probability large enough to be utilized for \bjet\ tagging. 

The calorimeters are needed to determine the energy of electrons, photons and jets precisely. With a good spatial resolution the calorimeter system is able to provide a high mass resolution. 

An additional muon system combined with a high magnetic field is the basis for muon reconstruction, triggering and high precision measurement.

Before describing individual components of ATLAS in the next sections, some conventions about the coordinate system will be explained as they are used throughout the whole thesis. ATLAS uses a right-handed coordinate system with the beam direction defining the z-axis. The x/y-plane is transverse to the beam axis with the x-axis pointing from the interaction point in the centre of ATLAS to the centre of the LHC ring. The y-axis points upwards. The azimuthal angle $\phi$ is used in the x/y-plane. The polar angle $\theta$ is measured from the beam axis. The rapidity $y \equiv \frac{1}{2}\ln\left( \frac{E - p_L}{E+p_L}\right)$ (using the longitudinal momentum component $p_L$) is preferred to $\theta$ as its intervals and corresponding differential cross sections are invariant under Lorentz boosts along the z-axis \cite{PDG}. Instead of the rapidity $y$ the \textit{pseudo-rapidity}\index{Pseudo-rapidity} $\eta \equiv - \ln \left( \tan\left( \frac{\theta}{2}\right) \right)$ is often used as an approximation for $p \gg m$ \cite{PDG}. For massless objects both expressions are equivalent.  

Many parts of the ATLAS detector are split into a central part with a barrel structure and a forward part with an end-cap structure. An overview of the ATLAS detector with its components is shown in Figure \ref{fig:ATLAS_overview}.
\begin{figure}[ht]
	\centering
		\includegraphics[width=0.95\textwidth]{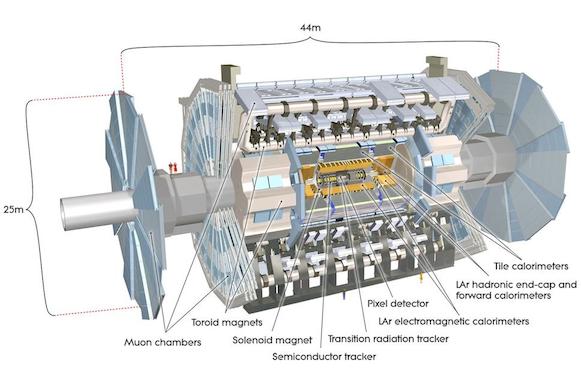}
		\caption{The ATLAS detector with its components \cite{ATLAS}.}
	\label{fig:ATLAS_overview}
\end{figure}

\subsection{Inner Detector}
\label{sec:ID}
Close to the interaction region the particle flux is quite high as up to 1,000 particles are expected to be created in each bunch crossing \cite{ATLAS}, depending on the luminosity delivered by the LHC. Those particles, which are charged, will leave tracks in the inner detector. The number of tracks depends on the instantaneous luminosity and the average number of interactions per bunch crossing, $\langle \mu \rangle$. In \cite{pixelperf} this number of tracks was measured as $N_{\text{tracks}} \approx 10 / \langle \mu \rangle$ at $\sqrt{s} = 7\,\TeV$ for tracks with $\pt > 400 \mev$.

There are several requirements on the devices measuring the tracks of these particles: As the particle density and the production rates are very high, the measurement needs to be made with very high granularity for two reasons. On the one hand, only a high granularity enables a separation of all the particle's tracks and a reconstruction of the corresponding vertices. On the other hand, the high granularity implies a high number of readout channels. The higher that number is, the lower the rate per channel gets. Lowering this rate per channel is mandatory for the high event rates. Next to the requirement of performing the measurement as precisely as possible, the detector must also minimize the disturbance of the particle's trajectory. The material -- quoted in terms of radiation lengths\footnote{The radiation length \radlength\ is defined as the average distance after which an electron loses its energy down to the fraction of $\frac{1}{e}$ via Bremsstrahlung \cite{PDG}.}  \radlength\ -- needs to be minimal. This reduces the possibility of track deflections and photon conversions. The latter effect is leading to a misidentification of photons as charged particles. 

The \textit{Inner Detector}\index{Inner Detector} (\textit{ID}\index{ID | see {Inner Detector }}) of ATLAS consists of tracking systems using three different techniques. These are all enclosed in a 2 T solenoidal field for momentum determination and charge separation. They all make use of the fact that charged particles ionize material and leave charges that can be kept as signals. The highest resolution is provided by the \textit{Silicon Pixel Detector} having the smallest distance of $R \geq 4.55\,\text{cm}$ to the interaction point. Three layers in the barrel and three discs on each end-cap provide about 80.3 million readout channels with an accuracy of $10\times115\,\mu\text{m}^2$ ($R/\phi \times z$ for the barrel, $R/\phi \times R$ for discs).

The Pixel Detector is surrounded by the \textit{Silicon Microstrip Tracker}\index{Silicon Microstrip Tracker} (\textit{SCT}\index{SCT | see {Silicon Microstrip Tracker }}). Instead of pixels it uses small-angle (40 mrad) stereo silicon strips with an intrinsic resolution of $17 \times 580\,\mu\text{m}^2$ ($R/\phi \times z$ for barrel, $R/\phi \times R$ for discs) for about 6.3 million readout channels. The SCT is also split into a barrel and a disc part.

The outermost part of the ID system is the \textit{Transition Radiation Tracker}\index{Transition Radiation Tracker} (\textit{TRT}\index{TRT | see {Transition Radiation Tracker }}). Straw tubes covering the range up to $\abseta = 2.0$ provide $R/\phi$ information only, using about 351,000 channels with an intrinsic accuracy of $130\,\mu$m per straw. The straws are filled with a $\text{Xe/CO}_2\text{/O}_2$ mixture. Transition radiation is emitted when charged particles pass through the material with different dielectric constants \cite{transrad}. The intensity of the emitted transition radiation depends on the relativistic $\gamma$ factor of the particle passing through the TRT. For a given momentum, this allows separating heavy from light particles, so for example electrons and pions.

The whole tracking system as shown in Figure \ref{fig:ID} covers a range of $\abseta < 2.5$ and provides a momentum resolution of $\sigma_{p_T} / p_T = 0.05\,\text{\%} \cdot p_T~\text{[GeV]} \oplus 1\,\text{\%}$ \cite{ATLAS}.
\begin{figure}[ht]
	\centering
		\includegraphics[width=0.8\textwidth]{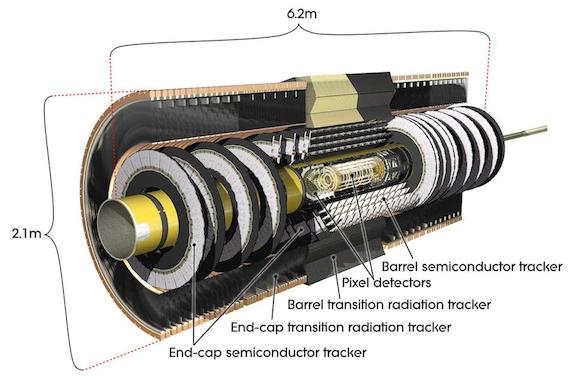}
		\caption{The ATLAS Inner Detector \cite{ATLAS}.}
	\label{fig:ID}
\end{figure}
In order to cope with the increasing number of events per bunch crossing and to increase the spatial resolution, an additional layer was added to the silicon pixel detector during the LS1 phase. This new detector, the \newword{Insertable B-Layer} (\textit{IBL}\index{IBL | see {Insertable B-Layer }}) \cite{IBL} will be the new innermost component of the ATLAS detector.

\subsection{Calorimeters}
\label{sec:calorimeters}
By inducing electromagnetic and hadronic showers and measuring their electromagnetic components, the calorimeter system is able to determine the particles' energies. These electromagnetic and hadronic showers must be fully contained in the calorimeter system. 
Hence, it needs to provide enough material in terms of the radiation length \radlength\ or the nuclear interaction length $\lambda_0$ \footnote{The nuclear interaction length $\lambda_0$ is defined analogously to \radlength, but for hadronic interactions \cite{PDG}.} to stop particles up to energies of several hundreds of \GeV. As muons are too heavy to radiate a sufficient amount of energy via Bremsstrahlung, they do not induce electromagnetic showers. Hence, they propagate through the calorimeter system, leaving traces of ionized particles. Even though this is insufficient for a reliable estimate of the muon energy, it still allows adding information for muon tracking.
 
The ATLAS calorimeter system has a sampling structure including active material for the readout of the signal and also passive material for the shower induction. The calorimeter system shown in Figure \ref{fig:calo} is divided into the \textit{Electromagnetic Calorimeter}\index{Electromagnetic Calorimeter} (\textit{ECal}\index{ECal | see {Electromagnetic Calorimeter }}), starting right after the solenoid magnet surrounding the ID system, and the \textit{Hadronic Calorimeter}\index{Hadronic Calorimeter} (\textit{HCal}\index{HCal | see {Hadronic Calorimeter }}) behind the ECal. While the ECal has a sufficient size to stop most electrons and photons via electromagnetic showers, the HCal is needed in addition to stop hadronically showering particles. 
\begin{figure}[ht]
	\centering
		\includegraphics[width=0.8\textwidth]{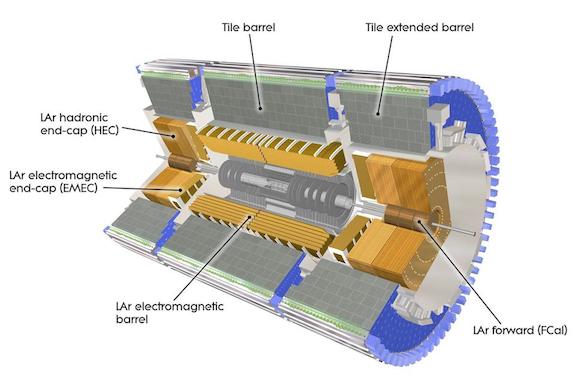}
		\caption{The ATLAS calorimeter system \cite{ATLAS}.}
	\label{fig:calo}
\end{figure}
The measured energy needs to be determined with a high precision. Next to a good energy resolution, analyses involving photons need another feature provided by the calorimetry. As photons leave no track in the Inner Detector, their direction can only be determined by the point of impact in the calorimeter. Thus, the ECal provides a very high granularity in particular in its first layer. It also contains a presampler to determine the energy loss in the parts in front of the calorimetry. The ECal uses liquid argon (LAr) as active material and lead/stainless steel as passive material. It has an accordion shape to ensure full $\phi$ coverage at high granularity. The $\eta$ coverage for the ECal is $\abseta < 1.475$ for the barrel part and $1.375 < \abseta < 3.2$ for the two end-caps. The granularity of $\Delta \eta \times \Delta \phi$ varies as a function of \abseta\ between $0.025 \times 0.025$ and $0.1 \times 0.1$ and uses about 180,000 readout channels \cite{ATLAS}. 

In contrast to the ECal, the HCal uses two different techniques in the barrel and the end-cap part. Steel is used as absorber for the barrel and scintillating tiles as active material. It covers the region up to $\abseta < 1.7$. In contrast, the \textit{Hadronic End Cap}\index{Hadronic End Cap} (\textit{HEC}\index{HEC | see {Hadronic End Cap }}) uses a LAr/Copper combination and extends the HCal to $\abseta < 3.2$. The forward part with $3.1 < \abseta < 4.9$ is covered by the \textit{Forward Calorimeter}\index{Forward Calorimeter} (\textit{FCal}\index{FCal | see {Forward Calorimeter }}) using LAr as active and copper (EM part) and tungsten (hadronic part) as absorbers. 

The total resolution of the calorimetry is  

\begin{tabular}{ll}
$\sigma_{E} / E = 10\,\text{\%}~/\sqrt{E}~\text{[GeV]} \oplus 0.7\,\text{\%}$&{ECal}\\
 $\sigma_{E} / E = 50\,\text{\%}~/\sqrt{E}~\text{[GeV]} \oplus 3\,\text{\%}$ &{HCal (barrel and end-cap)} \\
 $\sigma_{E} / E = 100\,\text{\%}~/\sqrt{E}~\text{[GeV]} \oplus 10\,\text{\%}$ & {FCal}
\end{tabular}

The total thickness of the calorimeter system is $\approx20~\radlength$ ($\abseta < 1.4$) and $\approx30~\radlength$ ($1.4 < \abseta < 3.2$) for the ECal and $\approx 10 ~\lambda_0$ for the combined ECal, HCal and FCal \cite{ATLAS}.

\subsection{Muon Chambers}
As the calorimeter system stops all detectable particles except muons, the \textit{muon spectrometer}\index{Muon spectrometer} (\textit{MS}\index{MS | see {Muon spectrometer }}) is placed in the outermost region of ATLAS. A toroidal magnetic field, described in Section \ref{sec:magnets}, is placed outside the calorimeters. This additional magnetic field, and the caused curvature of the muon tracks, allow for an additional momentum measurement for muons. The information of the outer muon spectrometer shown in Figure \ref{fig:MS} is combined with the track information provided by the ID to a combined muon track. Different techniques are used in the MS. \textit{Monitored drift tubes}\index{Monitored drift tubes} (\textit{MDTs}\index{MDT | see {Monitored drift tubes }}) are used for precision tracking in both the barrel part of the MS ($\abseta < 1.4$) as well as the end-cap part ($1.6 < \abseta < 2.7$). The end-cap tracking is supported by additional \textit{Cathode Strip Chambers}\index{Cathode Strip Chamber} (\textit{CSCs}\index{CSC | see {Cathode Strip Chamber }}) with a high granularity in the region of $2.0 < \abseta < 2.7$ to cope with the high event rates. 
\begin{figure}[ht]
	\centering
		\includegraphics[width=0.75\textwidth]{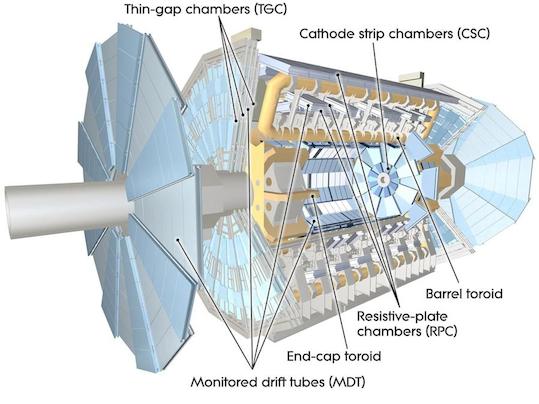}
		\caption{The ATLAS muon system \cite{ATLAS}.}
	\label{fig:MS}
\end{figure}
For triggering, \textit{Resistive plate chambers}\index{Resistive place chambers} (\textit{RPCs}\index{RPC | see {Resistive place chambers }}) are used for the barrel and \textit{Thin-gap chambers}\index{Thin-gap chamber} (\textit{TGCs}\index{TGC | see {Thin-gap chamber }}) are used for the end-cap part. Both systems offer a fast readout. Next to triggering, the RPCs and TGCs are also used to provide secondary tracking information. The whole MS provides about one million channels and a total resolution of $\sigma_{p_T} / p_T = 10\,\text{\% at 1 TeV}$ \cite{ATLAS}.   

\subsection{Magnet System}
\label{sec:magnets}
The ATLAS magnet system consists of four components. They create a magnetic field deflecting the particles in order to allow for momentum measurements of the tracking devices. All of them use superconducting NbTi conductors (+Cu for the toroid) which are stabilized with Al. The first one, a solenoid providing a 2 T magnetic field at the centre of the detector, surrounds the ID and is aligned parallel to the beam axis. The material budget of the solenoid is kept low with $\approx 0.66~\radlength$ \cite{solenoid} in order to avoid particle interactions which disturb the calorimeter measurements.

A second set of magnets provides the toroidal field for the MS. It consists of an air-core barrel magnet with eight racetrack shaped coils \cite{barrel_toroid} and two air-core end-cap magnets with eight squared coils each. The magnetic field varies between 0.15 T and 2.5 T for the barrel, with an average of 0.5 T. The field of the end-cap part varies between 0.2 and 3.5 T (1 T average) \cite{ATLAS}.
 
\subsection{Trigger System}

The event rate of about 40 MHz (at design value) is far too high to allow for the storage of all the corresponding collision data. The \textit{Trigger and Data Acquisition}\index{Trigger and Data Acquisition} (\textit{TDAQ}\index{TDAQ | see {Trigger and Data Acquisition }}) system needs to filter interesting events at a rate of 200 Hz. This is done via a three level trigger system, divided into the L1, L2 and Event Filter (EF) trigger. The system of L2 and HL triggers is also referred to as \newword{High Level Trigger} (\textit{HLT}\index{HLT | see {High Level Trigger }}).

The first one, L1, reduces the rate from 40 MHz to 75 kHz. In contrast to L2 and the EF, which are software based, it is hardware based as it needs to be extremely fast. By using low granularity information from the calorimeters and the MS it defines \textit{Regions of Interest}\index{Region Of Interest} (\textit{ROI}\index{ROI | see {Region Of Interest }}) within the detector. These ROIs contain objects which are defined in the \textit{trigger menus}\index{Trigger menu}. Such objects can be muons, jets, electrons, photons or $\tau$-leptons with a high transverse momentum. Also, events with a high amount of totally deposited transverse momentum or missing transverse momentum ($\etmiss$, defined in Section \ref{sec:etmiss}) can be triggered. 

Based on the L1 information, the L2 trigger reads out the full detector information. Only a certain part of the detector, the ROI, is read out at this stage. In contrast to the L1 trigger, the L2 is also able to add information from the ID. 
After the L2 decision the event rate is reduced further below 3.5 kHz. 

If an event is stored or not is decided by the EF trigger which uses the full detector information available for each event. It reduces the event rate below 200 Hz. 

\subsection{Luminosity Measurement}
In many analyses the data is compared to a prediction. In order to predict the expected number of events the luminosity of the dataset under study must be known. 

By knowing the number of average inelastic interactions per bunch crossing, $\mu$ (or the number of visible ones $\mu^{\text{vis}}$), the number of bunches $n_b$, the revolution frequency $f_r$ and the production cross section of inelastic proton/proton reactions $\sigma_{\text{inel}}$ (and the efficiency $\varepsilon$ to actually observe them) the luminosity can be calculated as
\begin{equation}
\mathcal{L} = \frac{\mu n_b f_r}{\sigma_{\text{inel}}} = \frac{\mu^{\text{vis}} n_b f_r}{\varepsilon \sigma_{\text{inel}}}
\end{equation}
according to \cite{lumi_7tev}. ATLAS uses several detectors for an online measurement of the luminosity during data taking. The most important one is \textit{LUCID}\index{LUCID} \cite{LUCID}, a Cherenkov detector placed at $\pm 17\,\text{m}$ from the interaction point, $10\,\text{cm}$ away from the beam line. It consists of 16 Al tubes filled with $\text{C}_4\text{F}_{10}$ and attached photomultipliers that are used to collect the Cherenkov light. 

At $\pm 140\,\text{m}$ from the interaction point the \textit{Zero-Degree Calorimeters}\index{Zero-Degree Calorimeters} (\textit{ZDC}\index{ZDC | see {Zero-Degree Calorimeters }}) \cite{ZDC} are located right behind the place where the common beam line is split into two. The final-triplet quadrupoles of the LHC deflect all charged particles out of the acceptance of the ZDC \cite{lumi_7tev}. It measures events with mesons decaying into photons and neutrons emitted at very forward angles. Such events play an important role in centrality measurements of heavy ion collisions \cite{ZDC}. 

The main device used for beam loss monitoring is the \textit{Beam Conditions Monitor}\index{Beam Conditions Monitor} (\textit{BCM}\index{BCM | see {Beam Conditions Monitor }}) \cite{BCM}. It consists of radiation hard diamond sensors located at $\pm 184\,\text{cm}$ from the interaction point. For low luminosity runs before the 2011 dataset, the \textit{Minimum Bias Trigger Scintillators}\index{Minimum Bias Trigger Scintillators} (\textit{MBTS}\index{MBTS | see {Minimum Bias Trigger Scintillators }}) \cite{ATLAS_lumi, improved_lumi} located at $\pm 365\,\text{cm}$ from the interaction point have also been used.  For offline luminosity measurements the ATLAS ID and parts of the EMCal (inner wheel of the EMEC and first layer of FCal \cite{lumi_7tev}) are also used. All ways of measuring the luminosity described above are relative measurements with a need for an absolute calibration. 

In the future, an absolute calibration of the luminosity will be possible with the \textit{ALFA}\index{ALFA} \cite{ALFA} detector. Located at $\pm 240\,\text{m}$ from the interaction point in one of the \textit{Roman Pots}\index{Roman Pot} \cite{ALFA} it can be moved as close as $1\,\text{mm}$ to the proton beam. For runs with special beam settings (low $\beta^{*}$, low emittance) the measurement of elastic proton/proton scattering at low angles can be used to calculate the luminosity. This is possible as the total proton/proton cross section is proportional to the imaginary part of the elastic scattering amplitude in the limit of zero momentum transfer, as stated by the \textit{optical theorem}\index{Optical theorem} \cite{ALFA}.

Until ALFA is fully operational, \textit{van der Meer (vdM) scans}\index{Van der Meer scan} \cite{vdmscan} are used to measure the horizontal and vertical beam profiles $\Sigma_x$ and $\Sigma_y$ by scanning the two proton beams across each other horizontally and vertically \cite{beamshape}. These profiles are translated to the absolute luminosity via
\begin{equation}
\mathcal{L} = \frac{n_b f_r n_1 n_2}{2 \pi \Sigma_x \Sigma_y}.
\end{equation}

Based on these vdM scans the luminosity for the dataset used in this analysis is measured with an uncertainty of 1.8\,\% \cite{improved_lumi}.

\chapter[Analysis Objects]{Analysis Objects}
\label{sec:objects}
Within the field of particles physics, the laws of nature are studied at the fundamental level. At this level, elementary forces interact with elementary particles. Most objects of interest are, however, not accessible by experiments at that level. All quarks except the top quark are immediately bound within confinement and only observable as composite objects. Others such as the top quark, $\tau$ leptons, $W$, $Z$ and Higgs bosons will immediately decay before they can be observed by a detector. Physics objects can be described at different stages, illustrated in Figure \ref{fig:levels}. 
\begin{figure}[ht]
	\centering
		\includegraphics[width=0.95\textwidth]{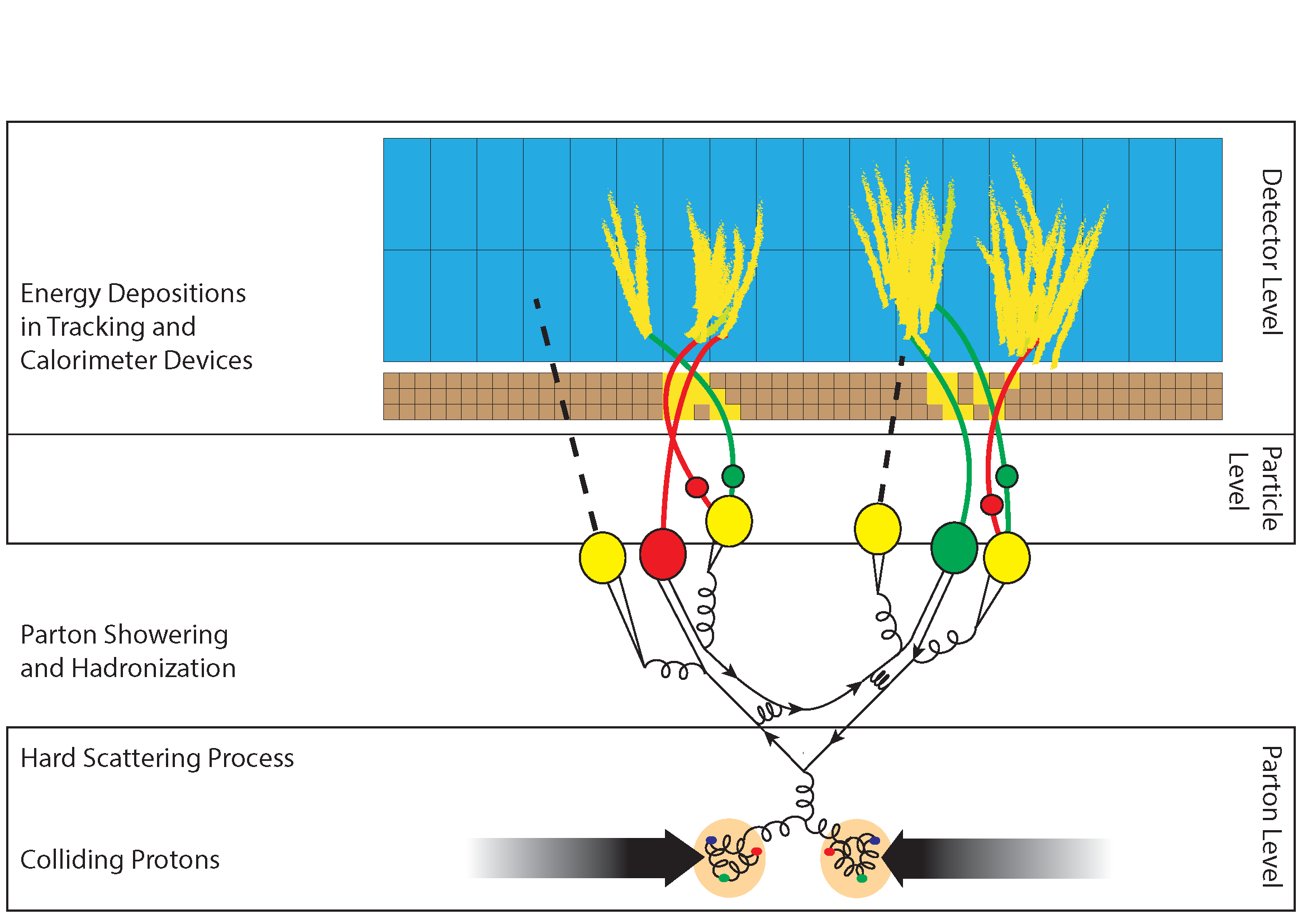}
		\caption{Illustration of a particle detection process and the different levels of object descriptions.}
	\label{fig:levels}
\end{figure}
The hard interaction process, described at leading order, is often referred to as the \textit{parton level}\index{Parton level}.\footnote{Even though the class of partons only includes quarks and gluons, also leptons and bosons can be described at the same level.} After the hard scattering the process of parton showering takes place, forming bound states. These are observable as particles which are in principle detectable. This level is called the \textit{particle level}\index{Particle level}. Reaching the detector, the particles will interact and leave signatures. The detection of these signatures takes place at the \textit{detector level}\index{Detector level}. Each object at parton level has a distinct type of signature that it leaves in the detector. This allows reconstructing objects from these signatures. \textit{Reconstruction level}\index{Reconstruction level} is a synonym for the detector level.
  
The goal of the event reconstruction, described in Chapter \ref{sec:selandreco}, is to map the objects at the detector level to the initial objects at parton level. This links the measurement to the analysis of the physics process of interest. 

In this chapter the objects recorded in the detector are described. Figure \ref{fig:objects} shows a \ttbar\ candidate event as it is measured in the detector. The detector objects are highlighted. This example of a \ttbar\ decaying in the dilepton channel includes all objects of interest for the analysis presented in this thesis.

All of the objects described in the following will be called candidates, as their detector signature does not necessarily need to be produced by the expected particle. \textit{Jets}\index{Jet}, bunches of particles stemming from a hadronisation process, for instance might also be mis-reconstructed as electrons in case they deposit a large fraction of their energy in the EM calorimeter. Section \ref{sec:fakes} discusses the derivation of a data-driven estimate of such jets mimicking leptons (referred to as \textit{fake leptons}\index{Fake leptons}).
\begin{figure}[ht]
	\centering
		\includegraphics[width=0.65\textwidth]{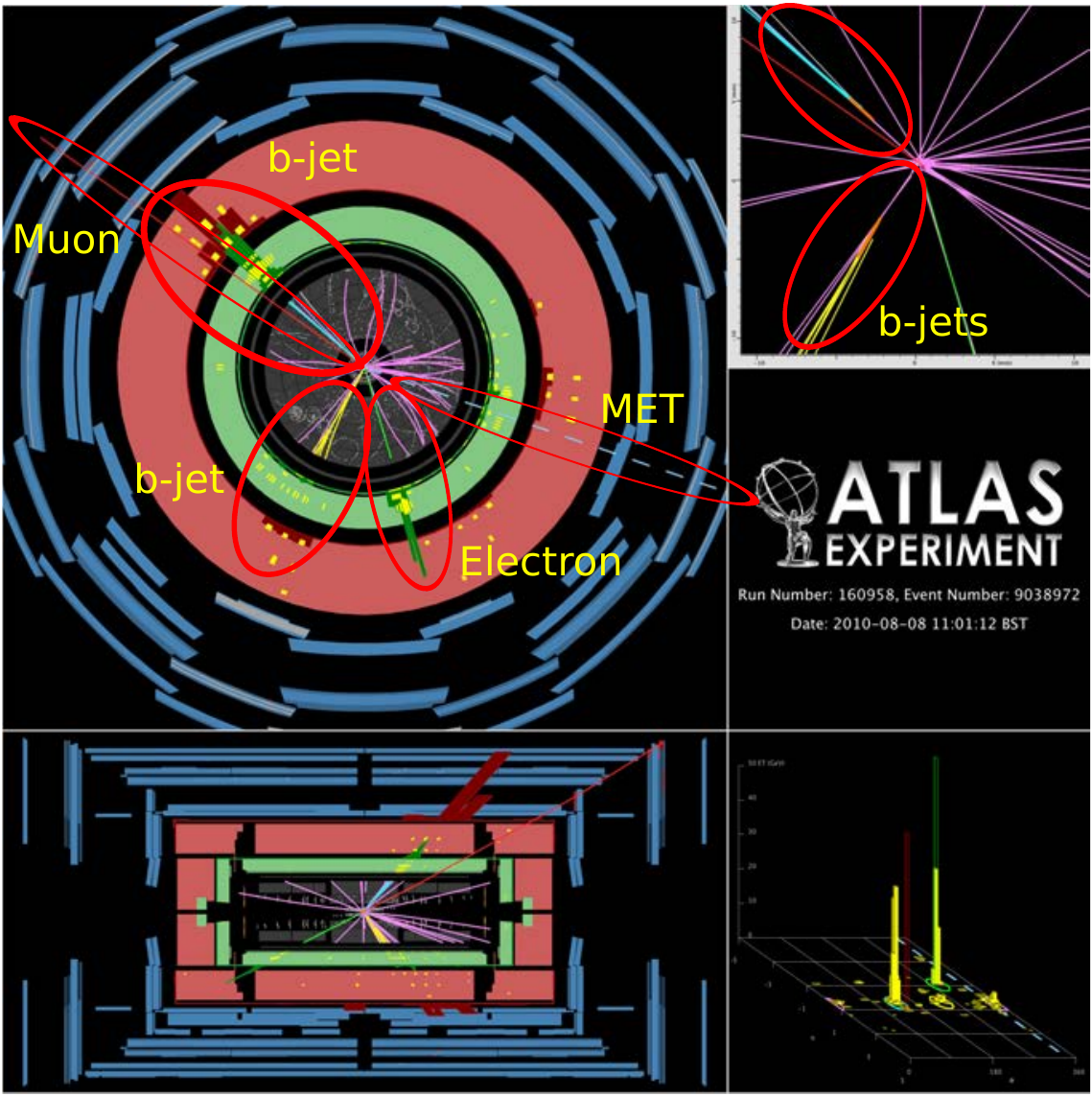}
		\caption{An ATLAS event display showing a \ttbar\ candidate event with both top quarks decaying leptonically, one into an electron, the other one into a muon. The event is shown in the $r/\phi$ plane (upper left) and the $r/z$ plane (lower left). The electron is represented by the green track pointing to the corresponding clusters in the ECal. The signature of the muon is visible as red track passing the muon spectrometer. The two jets in the event were both tagged as $b$-jets. Their secondary vertices are visible in the zoomed view of the primary vertex (upper right). Picture translated from \cite{lemmerbuch}, original version from \cite{atlas_public_eventdisplays}.}
	\label{fig:objects}
\end{figure}
From now on antiparticles will not be mentioned explicitly but being included in the name of their corresponding particles.

\section{Electrons}
\label{sec:electrons}
Being charged, electrons are supposed to leave a track in the ID and to be stopped in the ECal by the process of an electromagnetic shower \cite{PDG}. The reconstruction starts with the search of tracks pointing at clusters of energy deposition in the ECal with $\et > 2.5$ GeV.\footnote{The transverse energy \et\ is defined as $\et = E_{\text{cluster}} / \cosh{\left(\eta_{\text{track}}\right)}$.} Such seed clusters, consisting of $3 \times 5$ cells in $\eta \times \phi$ in the central ECal layer, are matched to tracks with $\pt > 0.5$ GeV from the ID. After track matching the clusters are fully reconstructed in clusters of $3 \times 7$ cells for the barrel and $5 \times 5$ for the end-cap of the ECal.
The cluster energy consists of the following components: the energy deposited in front of the ECal, the actual cluster energy in the ECal as well as added energies from lateral (energy out of the cluster) and longitudinal (deposition after the ECal) energy leakage.
The four-momentum vector of the electron is built by taking the energy from the cluster and the momentum direction from the track. This choice is motivated by the higher precision with which the quantities can be measured: tracking ensures a precise determination of the direction and the calorimeter is more precise in the determination of the energy.\footnote{This is not the case for low-energetic electrons, which will not be studied in this thesis.}
Due to the acceptance of the ID only electron candidates with $\left| \eta_{\text{cluster}}\right| < 2.47$ are considered. 
In case the energy cluster of an electron candidate is affected by a malfunctioning front-end board or high voltage supply, it is rejected. 

The selection of electrons and the rejection of jets mis-identified as electrons is made by a cut-based classification. These cuts take information from both the calorimeter and the ID into account. The sets used for the present analysis base on the original classification into \textit{loose}, \textit{medium} and \textit{tight} electrons in the order of decreasing efficiency and increasing purity \cite{reco_electrons, electron_reco}. 
The original \textit{tight} definition has been updated and optimized for the data taking period in 2011 to the \textit{tight++}. While the \textit{tight++} electrons are used for the event selection in this analysis the \textit{medium++} (looser identification cuts as tight++ \cite{reco_electrons, electron_reco}) electrons are needed for the estimation of the fake lepton background (see Section \ref{sec:fakes}). 
The electron identification efficiency for the different classifications are shown as a function of pile-up in Figure \ref{fig:el_reco}.
\begin{figure}[ht]
	\centering
		\includegraphics[width=0.5\textwidth]{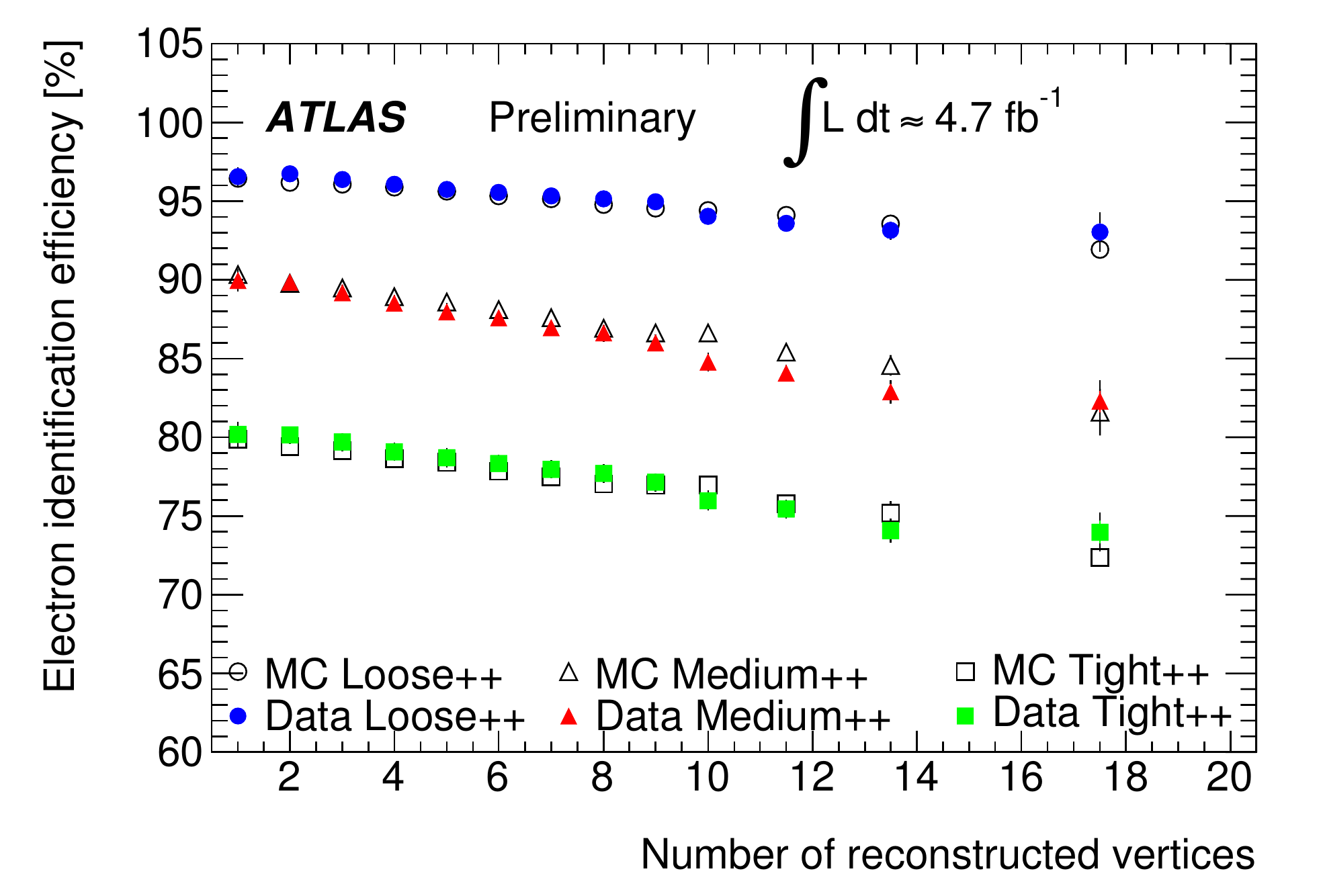}
		\caption{Identification efficiency for different electron types as a function of pile-up \cite{reco_el_eff}.}
	\label{fig:el_reco}
\end{figure}

Selected electrons need to pass additional isolation requirements to further suppress contributions from misidentified jets. The amount of energy (momentum) in a cone with a size of  $R = 0.2$ ($ R = 0.3$) around the electron cluster (track) that does not belong to the actual electron object is summed. The values, referred to as \etcone\ \index{EtCone20} and \ptcone\ \index{PtCone30} may not pass a certain value, depending on the \et\ and the $\eta$ of the object. These values are corrected for energy leakage and additional depositions from pile-up events to ensure a constant isolation efficiency of 90\,\%. A \textit{tag-and-probe method}\index{Tag-and-probe method} (T\&P method) \cite{reco_electrons, electron_reco} is used to determine the efficiency.
To avoid a double counting of a detector signature as both electron and jet, the reconstructed jet (see Section \ref{sec:jets}) which is closest to the electron is removed if it is within a distance $\Delta R = 0.2$ to the electron. The electron candidate is rejected in case that after this removal there is still a jet present within $\Delta R = 0.4$. This removal also ensures a proper isolation of the electron. 
In order to reach the efficiency plateau of the electron triggers (see Section \ref{sec:eventselection}) a cut of $\et > 20$ GeV is used. 
To ensure that an electron is being produced at the primary vertex, the distance on the $z$-axis between the reconstructed primary vertex and the electron needs to be less than $2\,\text{mm}$. 

As the simulated events do not mimic the real performance of objects perfectly, correction factors have to be applied. 
A \textit{scale factor}\index{Scale factor} (SF) is defined as the ratio of the efficiency in data ($\varepsilon_{\text{data}}$) and in the simulation ($\varepsilon_{\text{MC}}$). Such scale factors are applied to the simulation to properly model the performance. Table \ref{tab:electron_corrections} shows an overview of corrections for electron objects. 
\begin{table}[htbp]
\begin{center}
\begin{tabular}{|c|c|c|}
\hline
Corrected Quantity  & Applied to & Parameterization  \\

\hline
\hline
Energy correction for crack region& Data, MC & \eta, \et\\
Electron energy scale & Data & $\eta_{\text{cluster}}$, $\phi_{\text{cluster}}$, $\et$\\
Electron energy resolution  & MC & \eta, $E$\\
Trigger efficiency  & MC & $\eta$, \et, data taking period\\
Selection and isolation efficiency &MC & \eta, \et \\
Reconstruction efficiency & MC & $\eta$\\

\hline

\end{tabular}
\end{center}
\caption{Corrections and scale factors for electrons.}
\label{tab:electron_corrections}
\end{table}
These corrections were mostly studied by evaluating $Z \rightarrow ee$, $J/\psi \rightarrow ee$ and $W \rightarrow e\nu$ events using a T\&P method \cite{reco_electrons, electron_reco}. For the latter type of events the ratio of electron energy in the calorimeter to electron momentum in the tracker was studied. Details about the electron energy calibration methods are provided in \cite{electron_calib}.
Due to the large amount of passive material in front of the transition region between barrel and end-cap calorimeters (\textit{crack region}\index{Crack region}) with $1.42 < \left|  \eta_{\text{cluster}}\right| < 1.55$, electrons in this region suffer from a poorer reconstruction efficiency. On top of the default energy calibration, an additional calibration for electrons in this region is applied to both data and MC simulations. Despite the additional correction, electron candidates in the crack region with $1.37 < \left|  \eta_{\text{cluster}}\right| < 1.52$  are fully rejected for this analysis due to the low reconstruction efficiency  and low resolution.

\section{Muons}
\label{sec:muons}
Muons leave tracks in the ID and the MS system. The \textit{MuId}\index{MuId algorithm} algorithm \cite{reco_muons} starts from the outermost layer with track segments of the MS. It matches tracks to the ID track segments, takes energy losses in the calorimeter into account and refits the full muon track. 
The muon momentum is required to be larger than 20 GeV. The acceptances of the ID and MS require $\abseta < 2.5$. As for the electrons, the muon's distance on the $z$-axis to the reconstructed primary vertex needs to be less than $2\,\text{mm}$. 
Similar to the electron classification, muons are classified as loose and tight. The former classification is used for the signal selection and the latter for the fake lepton estimation. 
Loose muons demand a certain track quality: A hit in the B-layer of the ID is required if expected. At least one hit in the pixel detector must be recorded. Non-operational pixel sensors are automatically counted. The sum of hits in the SCT plus the number of crossed non-operational sensors must be at least six. The total number of non-operational pixel and SCT sensors must not be larger than two. With $n_{\text{TRT}} =n_{\text{TRT hits}}  + n_{\text{TRT outliers}} $ as the sum of the TRT track hits and the TRT outliers (as defined in \cite{reco_muons}) it is required that $n_{\text{TRT}} > 5$ and $n_{\text{TRT outliers}} / n_{\text{TRT}} < 0.9$ for $\abseta < 1.9$ and $n_{\text{TRT outliers}} / n_{\text{TRT}} < 0.9$ in the case of  $n_{\text{TRT}} > 5$ for $\abseta \geq 1.9$.

Tight muons need to fulfill the loose criteria as well as further isolation requirements. As defined for the electrons, the \ptcone\ and \etcone\ definitions have been used to ensure tracking and calorimeter isolation. \etcone\ $< 4~\GeV$ and \ptcone\ $< 2.5~\GeV$ are required.
 
Muon candidates with a distance $\Delta R < 0.4$ to a jet with $\pt > 25 $ GeV and \mbox{$\left| \text{JVF} \right| > 0.75$} (see Section \ref{sec:jets} for a definition) are removed from the event to suppress contributions from heavy flavour decays within jets.

As for electrons, the momentum and the efficiencies of selected muons are not perfectly modelled and need to be corrected. These corrections are summarized in Table \ref{tab:muon_corrections}. A summary of the muon reconstruction performance in ATLAS is given in \cite{muon_performance}.

\begin{table}[htbp]
\begin{center}
\begin{tabular}{|c|c|c|}
\hline
Corrected Quantity  & Applied to & Parameterization  \\

\hline
\hline
Momentum scale (both ID and MS)& MC & $\eta$, $\pt$\\
Momentum resolution (both ID and MS) & MC & \eta, $\pt$\\
Trigger efficiency  & MC & $\eta$, $\phi$, data taking period\\
Isolation efficiency &MC & data taking period \\
Reconstruction efficiency & MC & \pt, $\eta$, $\phi$\\

\hline

\end{tabular}
\end{center}
\caption{Corrections and scale factors for muons.}
\label{tab:muon_corrections}
\end{table}

\section{Jets}
\label{sec:jets}
After production or scattering, quarks will form hadrons or (in case of the top quark) decay, making it impossible to detect them as free partons. Particles being produced in this process of hadronisation will deposit energy in the ECal and HCal and leave tracks in the ID. These energy depositions are collimated and can be reconstructed as jets. Only the energy depositions induced by electromagnetic processes will be visible. This includes both electromagnetic showers as well as ionization processes from hadronic showers. As no method of \textit{hadronic compensation}\index{Hadronic compensation} \cite{PDG} is used at ATLAS and certain jet components (e.g. neutrinos and slow neutrons) don not leave electromagnetic signatures, the total jet energy must be deduced from the electromagnetic component.

Energy depositions in \textit{topological clusters}\index{Topological Clusters} \cite{topo_clusters} are reconstructed as jets using the \textit{anti-$k_t$ algorithm}\index{Anti-$k_t$ algorithm} \cite{antikt} with a distance parameter $R=0.4$ via the \newword{FASTJET} \cite{fastjet} software.

The first level of energy calibration is called \newword{EM scale}, as only detectable electromagnetic energy depositions are taken into account. At this stage, the energy is corrected for contributions arising from the in-time and out-of-time pile-up.\footnote{In-time pile-up refers to energy depositions from objects not belonging to the hard scattering process, but to the same bunch crossing. Out-of-time pile-up describes energy-depositions from up to 12 (2011 dataset) preceding and one following bunch crossings \cite{JES}.} These corrections depend on the number $n_\text{PV}$ of reconstructed primary vertices, the average number of interactions per bunch crossing, $\left< \mu \right>$,  for the specific luminosity block and the pseudorapidity $\eta$ of the jet \cite{reco_jets}. 
The four-momentum of the jet is corrected for the position of the primary vertex \cite{reco_jets}.
Jets suffer from a bias of their reconstruction direction towards better instrumented regions with respect to poorer detected regions. Thus, an additional $\eta$ correction is applied. It is significant  in the calorimeter transition regions only (see Figure \ref{fig:jet_etacorr}) \cite{reco_jets}. 

After the reconstruction on the level of the calibrated EM scale, jets need to be calibrated to the hadronic scale. 
A variety of calibration methods is described in \cite{reco_jets}. For this analysis jets are calibrated using the \newword{EM+JES calibration}. For this calibration Monte Carlo simulations are used to calculate the jet response $R^{\text{jet}}_{\text{EM}} = E^{\text{jet}}_{\text{EM}} / E^{\text{jet}}_{\text{truth}}$. It is defined as the ratio of the jet energy visible via EM depositions divided by the total (true) jet energy.
The inverse of $R$, parameterized in \pt\ and $\eta$, is then used to scale the jet energy on the EM scale up to the calibrated jet energy. This level of calibrated jet energy is referred to as the EM+JES scale.
Figure \ref{fig:jetresponse} shows the jet response as a function of pseudo-rapidity for different bins of jet energy calibrated to the EM+JES scale.

\begin{figure}[ht]
	\centering
			\subfigure[]{
		\includegraphics[width=0.45\textwidth]{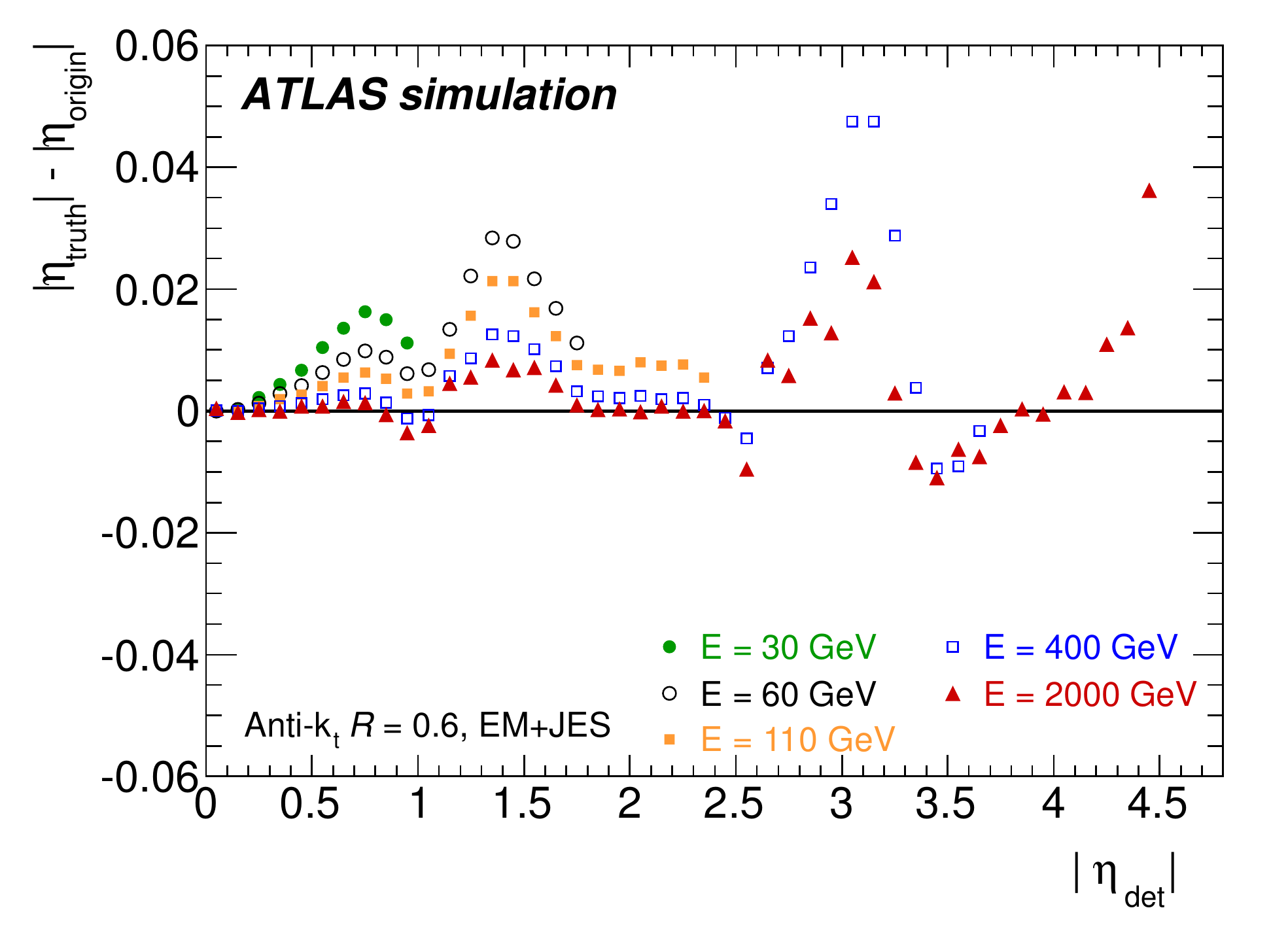}
			\label{fig:jet_etacorr}
		}
	\subfigure[]{
		\includegraphics[width=0.45\textwidth]{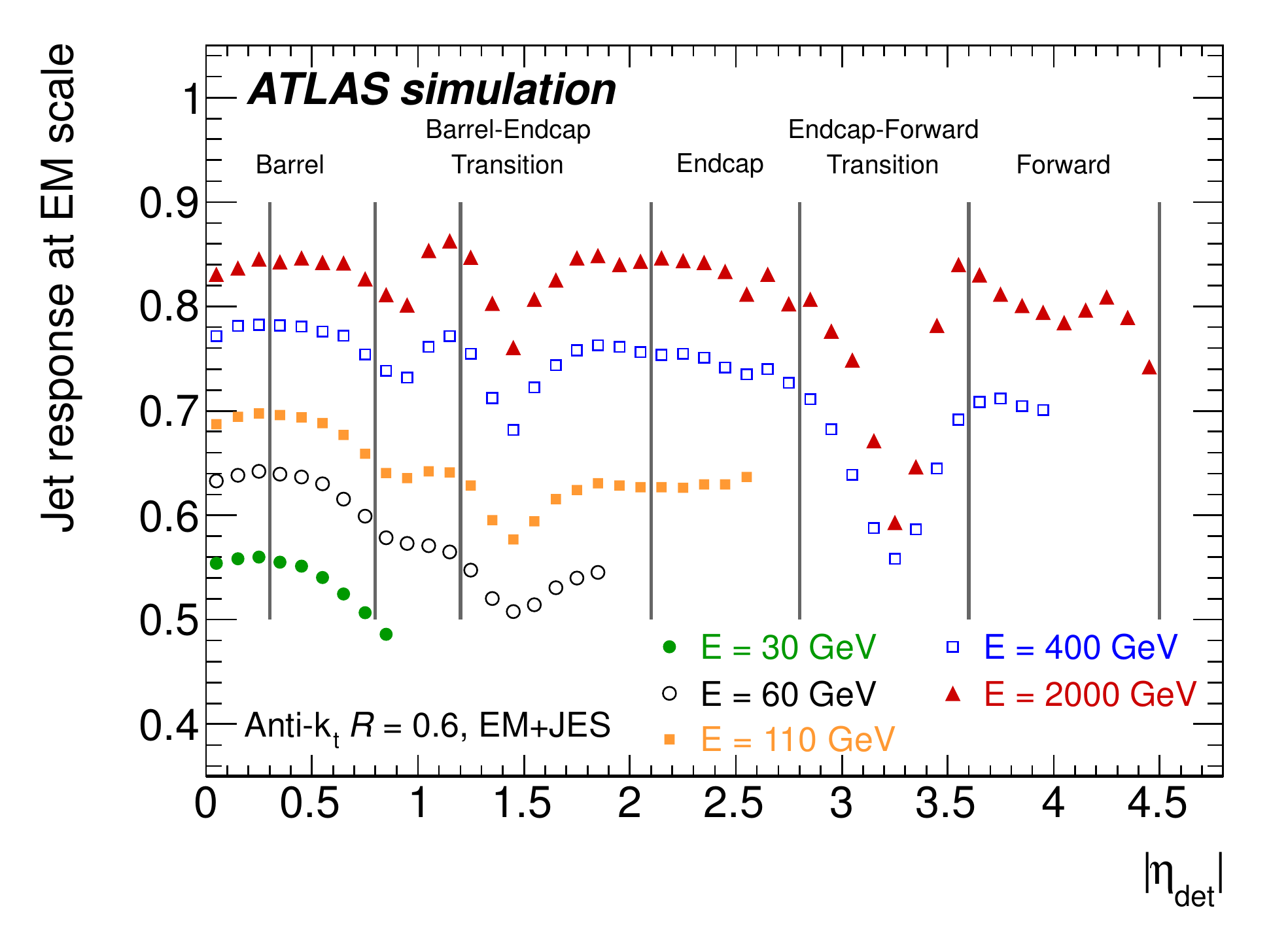}
			\label{fig:jetresponse}
		}

		\caption{\subref{fig:jet_etacorr} Difference between the
origin corrected $\eta_{\text{jet}}$ and the true  $\eta_{\text{jet}}$ in bins of the calorimeter
jet energy calibrated with the EM+JES scheme as a function of $\eta_{\text{detector}}$ \cite{reco_jets}. \subref{fig:jetresponse} Jet response $R^{\text{jet}}_{\text{EM}} = E^{\text{jet}}_{\text{EM}} / E^{\text{jet}}_{\text{truth}}$ \cite{reco_jets}. }
\end{figure}

On top of the EM+JES calibration, which is only based on MC simulations, in-situ calibration techniques have been applied \cite{JES}. In the central detector region with $\abseta < 1.2$ the momentum balance between jets and reference objects is used to derive additional corrections to the MC driven JES calibration. Photons and $Z$ bosons serve as reference objects for jets with $\pt < 800~\GeV$. For larger transverse momenta, multijet systems are also used for balancing. The forward region of the detector is calibrated with respect to the central part using di-jet \pt\ balancing.

Jets are required to have $\abseta < 2.5$ and $\pt > 25$ GeV. If a jet has a distance of $\Delta R < 0.2$ to a reconstructed electron, it is removed to avoid jet/electron double counting. Additional background such as beam-gas events, beam-halo events, showers induced by cosmic rays as well as calorimeter noise is suppressed be applying additional quality cuts on the jets \cite{reco_jets}. An event will be rejected in case it contains a jet with $\pt > 20$ GeV which did not pass the quality criteria.

With increasing instantaneous luminosity the average number of interactions per bunch-crossing rises. Hence, the contribution of jets which are reconstructed but do not belong to the main interaction of interest increases  as well. 
To cope with this problem, a discriminating variable called \newword{Jet Vertex Fraction} (\textit{JVF}\index{JVF | see {Jet Vertex Fraction }}) is introduced which separates jets belonging to the primary vertex of interest from other background jets. This variable is introduced in the following section. To be accepted, a jet needs to pass the criterion of $\left| \text{JVF} \right| \geq 0.75$.

\subsection{Jet Vertex Fraction}
\label{sec:jvf}
Jets taken into account for the event reconstruction are all supposed to stem from the same primary vertex. However, also jets from pile-up events are selected and can pass the quality criteria. Hence, an additional discrimination between primary and pile-up jets is needed. 
The discrimination bases on a procedure introduced by the D0 collaboration \cite{jvf_d0}. It is realized by checking all tracks corresponding to the jet for their primary vertex. Figure \ref{fig:jvf} illustrates the principle. 
\begin{figure}[ht]
	\centering
		\includegraphics[width=0.95\textwidth]{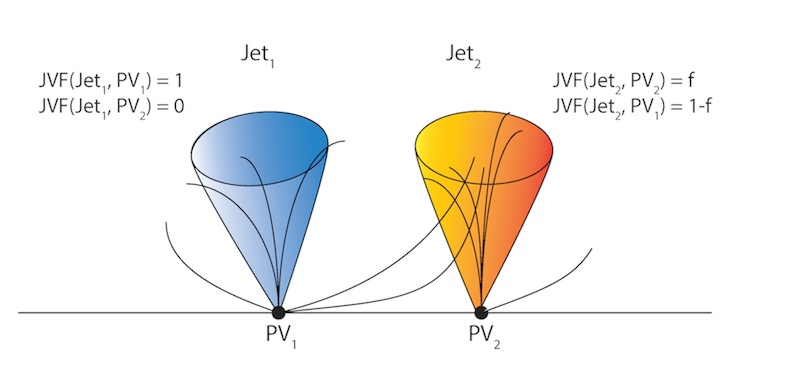}
		\caption{Illustration of the principle of the Jet Vertex Fraction variable. While for the left jet all tracks originate from $\text{PV}_1$ and none from $\text{PV}_2$, the right jet has only a certain fraction $f$ of tracks from $\text{PV}_2$ and another fraction $1-f$ from $\text{PV}_1$.}
	\label{fig:jvf}
\end{figure}
Each track is associated with at least one primary vertex. In Figure \ref{fig:jvf} all tracks of $\text{jet}_1$ originate from the primary vertex $\text{PV}_1$, and none from $\text{PV}_2$. In the case of $\text{jet}_2$, only the fraction $f$ of tracks originates from $\text{PV}_2$. Such a fraction can be calculated for each pair of jet and primary vertex. The higher this fraction $f$ is, the more likely it is that the jet-to-vertex assignment was done correctly. Cutting on this fraction, the jet vertex fraction (JVF), allows discriminating primary jets from pile-up jets.

Figure \ref{fig:jvf} is just a simplification of the real JVF determination. In fact, a track might be allocated to several vertices depending on the size of a variable distance window. The JVF is computed as shown in Equation \ref{eq:jvf}:
\begin{align}
\text{JVF}(\text{jet}, \text{vertex}) = \frac{\sum_{\text{tracks of }\text{jet}}  (\text{vertex} \in \text{track.vertexlist}) \cdot \pt^{\text{track, }\text{jet}}}  {\sum_{\text{tracks of }\text{jet}} \pt^{\text{track,}\text{ jet}}}
\label{eq:jvf}
\end{align}
The expression $(\text{vertex} \in \text{track.vertexlist})$ is 1 if the vertex is contained in the list of all vertices assigned to the track and 0 if not. 

Figure \ref{fig:jvf_stab} shows the effect of the pile-up suppression in $Z+\text{jets}$ events. As in case no track could be matched to a jet a JVF of $-1$ is assigned, the absolute value of the JVF is taken for the cut not to exclude such jets.
\begin{figure}[ht]
	\centering
	\subfigure[]{
		\includegraphics[width=0.45\textwidth]{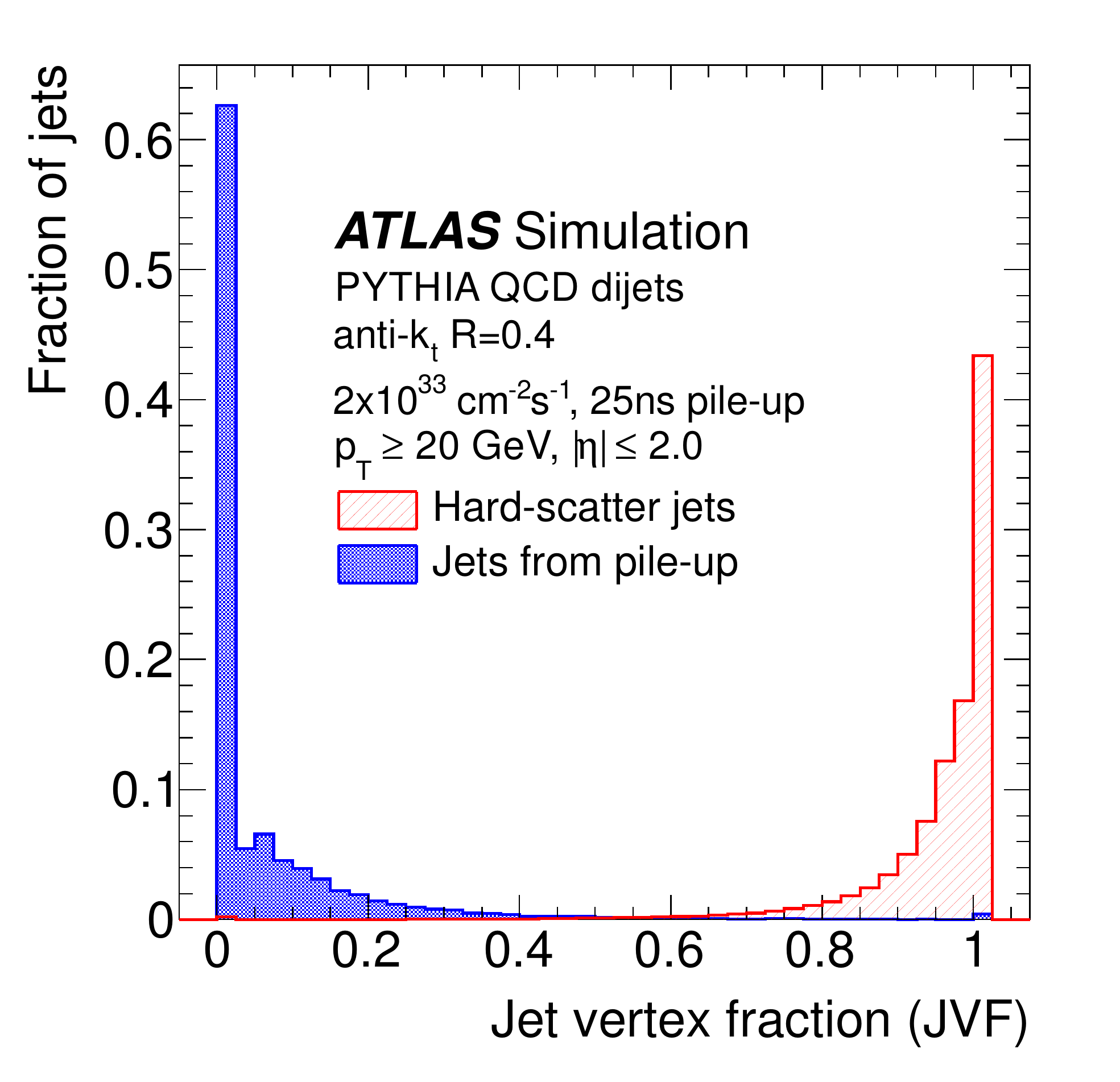}
			\label{fig:jvf_dist}
		}
		\subfigure[]{
		\includegraphics[width=0.45\textwidth]{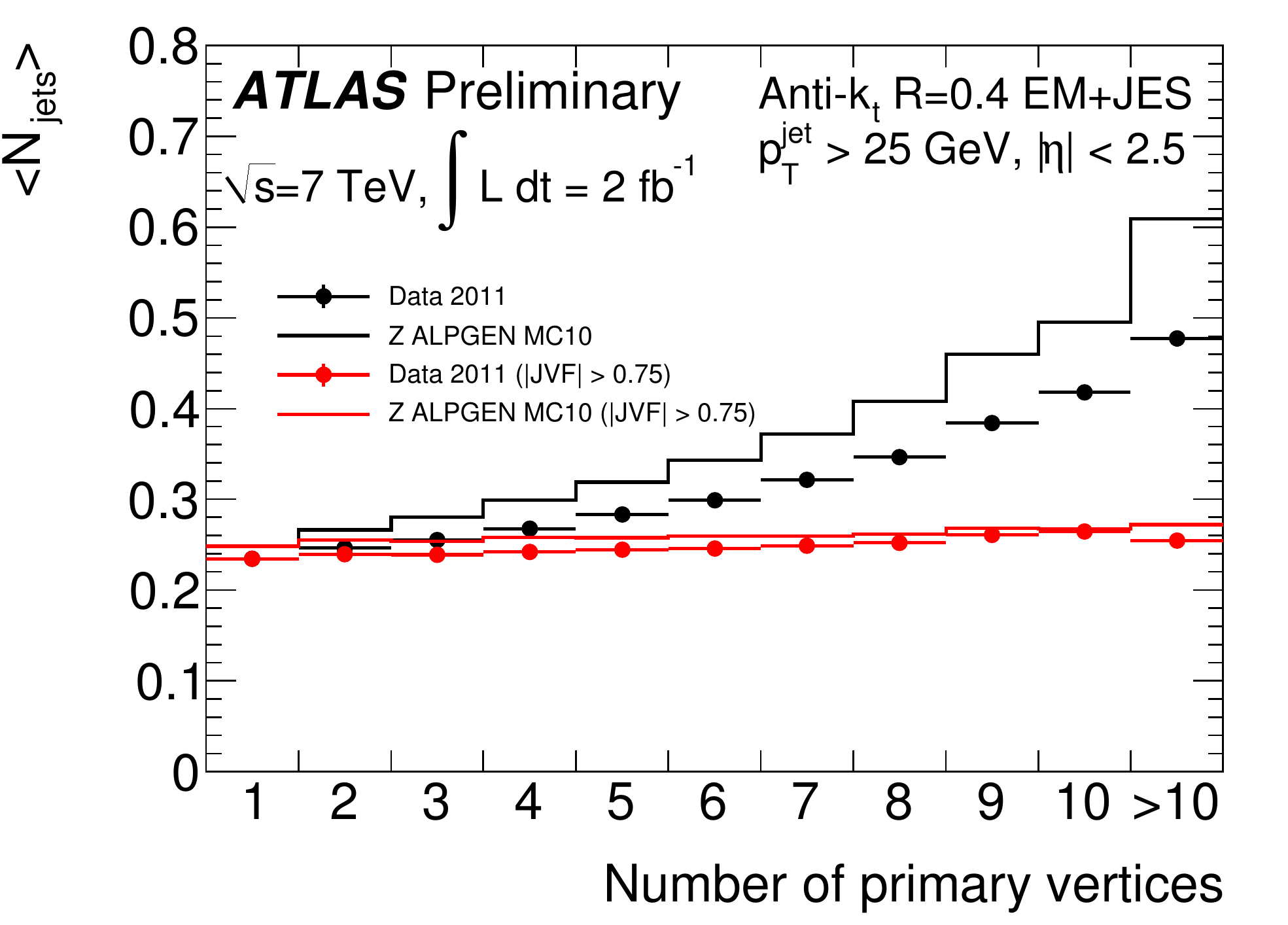}
			\label{fig:jvf_stab}
		}
		\caption{\subref{fig:jvf_dist} Distribution of the JVF variable for jets from the hard scattering process and for pile-up jets \cite{jvf_website}. \subref{fig:jvf_stab} Pile-up suppression by the application of the JVF cut, shown as stability of the number of reconstructed jets against the number of primary vertices for $Z+\text{jets}$ events \cite{jvf_website}.}
	\label{fig:jvf_data}
\end{figure}
As the distribution of the JVF variable is not perfectly modelled by the Monte Carlo simulation, scale factors have to be applied. These are derived using a T\&P method on a sample of $Z \rightarrow ee$ and $Z \rightarrow \mu \mu$ events using specific selections to get samples of hard-scattering jets and pile-up jets.
Further details of the JVF studied with data from the 2012 dataset can be found in \cite{jvf_2012}.

\subsection{B-Tagging}
\label{sec:btagging}
Hadronising \bQ s offer the unique opportunity to tag their jets experimentally. As top quarks decay to almost 100\,\% into a \bQ\ \cite{PDG}, their identification is of special interest. 
B-hadrons inside a $b$-jet are massive ($m_B \approx 5$ GeV) and have a relatively long lifetime ($c \tau \approx 0.5\,\text{mm}$). Hence, displaced secondary decay vertices of these B-hadrons can be reconstructed as shown in Figure \ref{fig:objects} due to the high spatial resolution of the ATLAS ID. Furthermore, B-hadrons have a significant branching fraction of $\approx 10$\,\% for $B \rightarrow l \nu X$ \cite{PDG}, where $l$ represents either an electron or a muon (10\,\% per lepton, so 20\,\% for both). 
The high \pt\ of the B-hadron's decay products, caused by the B-hadron's high mass, leads to a high track multiplicity inside the $b$-jet. The \btag ging algorithms IP3D, SV1 and JetFitterCombNN \cite{b-tagging}, which are used at ATLAS,  make use of these discriminative jet properties and provide a quantity called \textit{$b$-tag weight}\index{b@$b$-tag weight}. Cutting on this weight allows tagging \bjet s with a certain purity and efficiency. The fraction of real \bjet s contained in a \btag ged sample is given as purity. The efficiency gives the probability to tag a \bjet\ as such. 

In this analysis the \mvone\ tagger is used. It is based on a neural network using the output weights of the {\tt JetFitter}, {\tt IP3D} and {\tt SV1} algorithms \cite{b-tagging} as input. This tagger was set up at a working point with a  \btag ging efficiency of 70\,\% and a purity of 91\,\% for simulated \ttbar\ events by requiring the \btag\ weight $w > 0.6017$. Scale factors need to be applied to tagged jets in MC simulations to properly model the data. 
Using the default calibration based on di-jet events \cite{b-tagging_calib} leads to a disagreement of the \btag ged jet multiplicity (Figure \ref{fig:btagmismodel_before}). 
\begin{figure}[ht]
	\centering
	\subfigure[]{
		\includegraphics[width=0.45\textwidth]{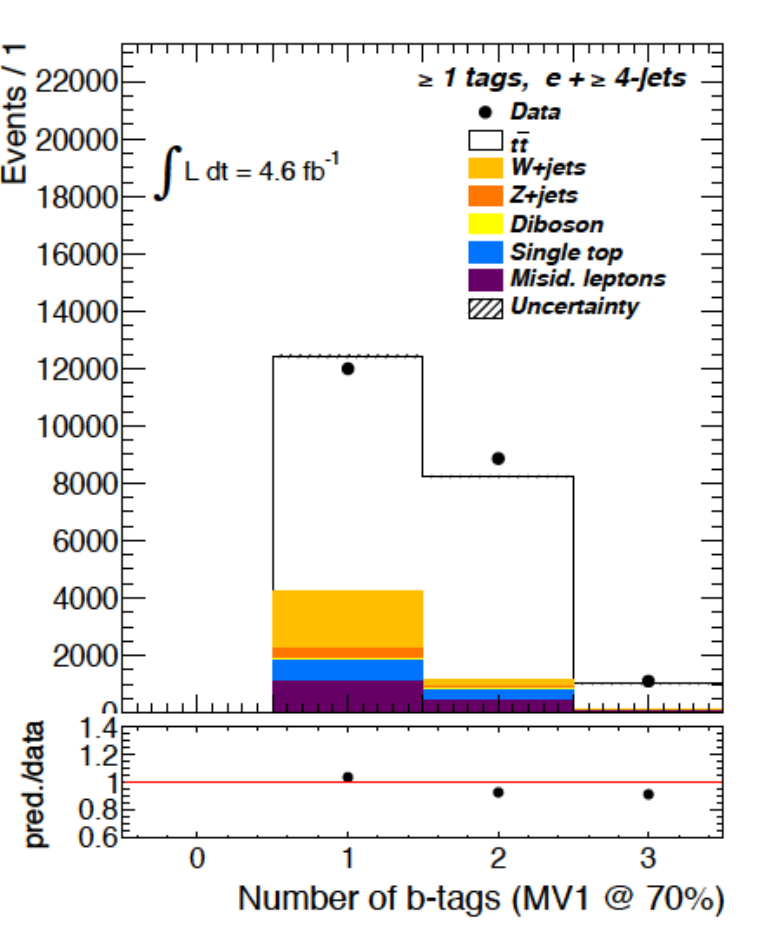}
			\label{fig:btagmismodel_before}
		}
		\subfigure[]{
		\includegraphics[width=0.45\textwidth]{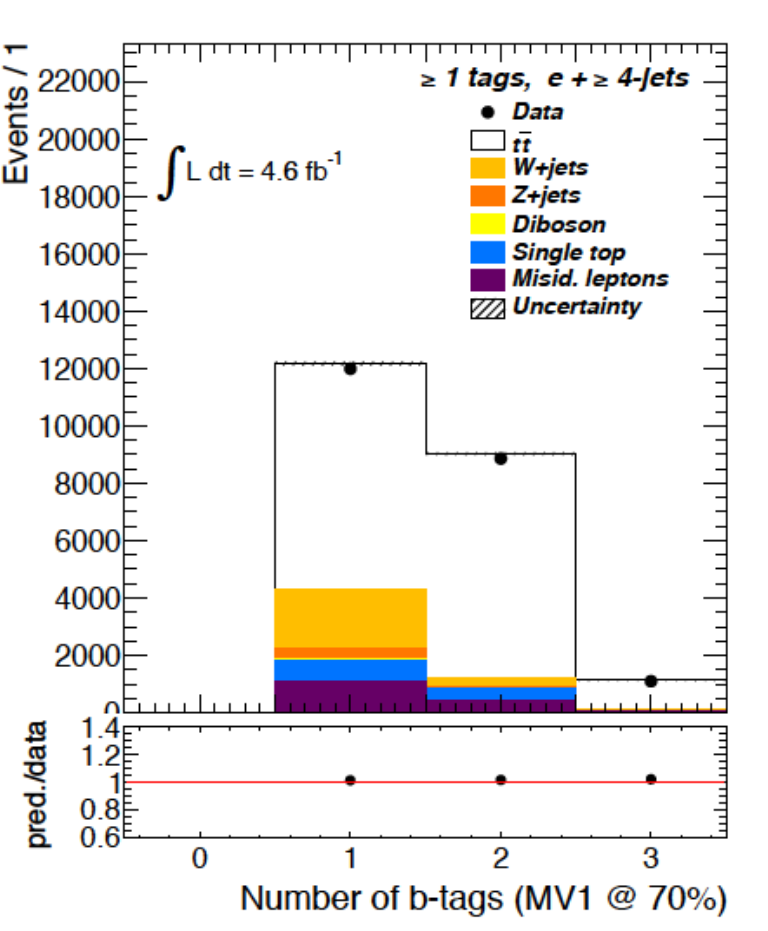}
			\label{fig:btagmismodel_after}
		}
		\caption{Distribution of the \bjet\ multiplicity using \subref{fig:btagmismodel_before} the default calibration based on di-jet events ({\tt pTrel}+{\tt System8}) \cite{b-tagging_calib} and  \subref{fig:btagmismodel_after}  the combined calibration ({\tt pTrel}+{\tt System8}+dileptonic \ttbar) \cite{b-tagging_dilepcal} for \ttbar\ pairs decaying into an electron and at least four jets.}
	\label{fig:btagmismodel}
\end{figure}

The default calibration was known to become unreliable for \bjet s with a high \pt\ \cite{b-tagging_calib}. A modified calibration is needed to perform the spin correlation analysis, as a proper \btag\ multiplicity modelling is essential. The default calibration is combined with a \ttbar\ calibration derived from dileptonic \ttbar\ events \cite{b-tagging_dilepcal}. Taking the dileptonic calibration ensures statistical independence of the calibration and the analysed data in the \ljets\ channel. 
The default calibration scale factors ({\tt pTrel}+{\tt System8}) are compared to the scale factors using a combined default and \ttbar\ calibration (KinSelDL) in Figure \ref{fig:btag_calib}. All scale factors agree within uncertainties.
\begin{figure}[ht]
	\centering
		\includegraphics[width=0.95\textwidth]{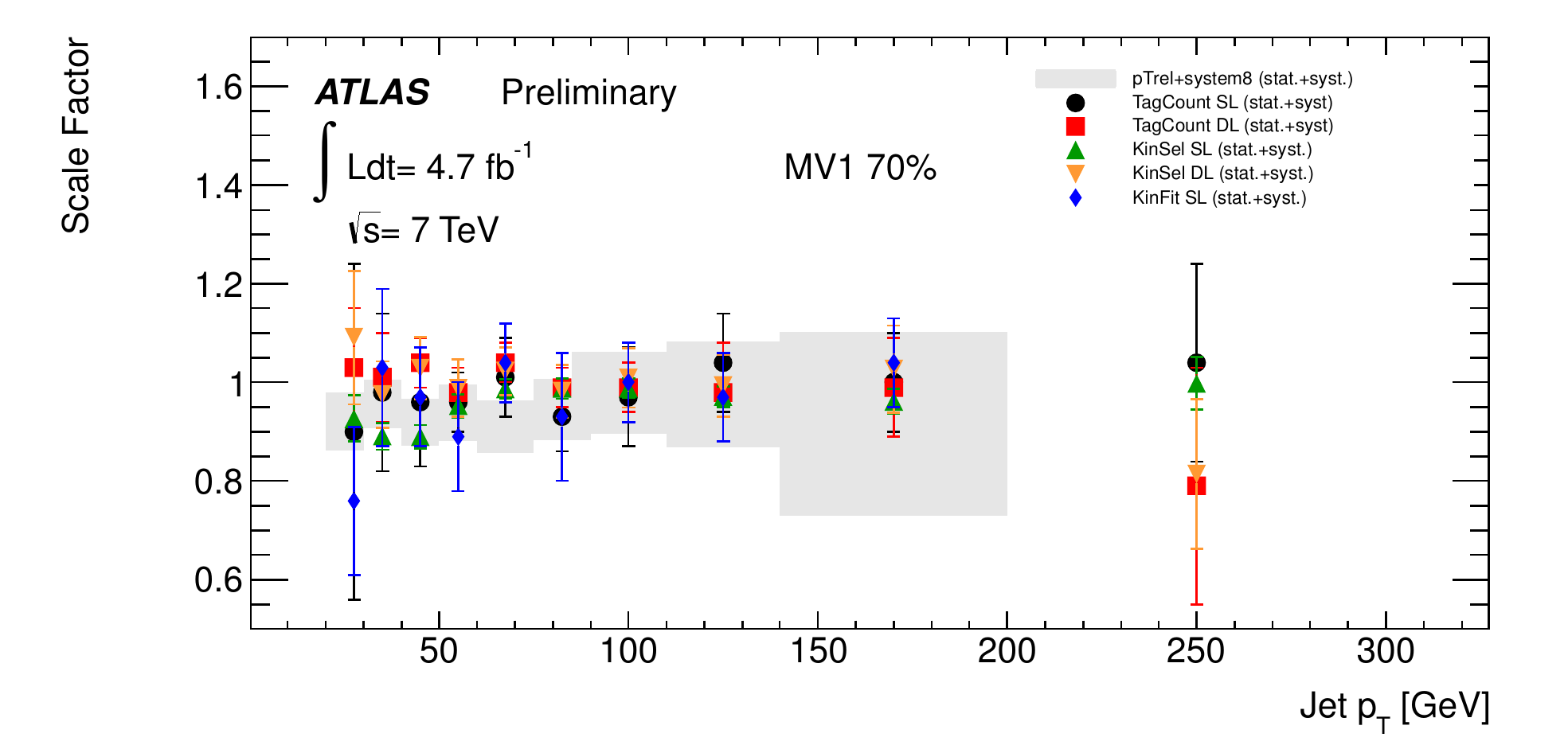}
		\caption{Overview of \btag\ scale factors using different calibration methods \cite{b-tagging_dilepcal}.}
	\label{fig:btag_calib}
\end{figure}
By comparing the \bjet\ multiplicity using default calibration (Figure \ref{fig:btagmismodel_before}) to the one calibrated with the improved combined calibration (\ref{fig:btagmismodel_after}) one can clearly see the improvement.

\section{Missing Transverse Momentum}
\label{sec:etmiss}
The disadvantage of hadron colliders compared to lepton colliders is the unknown initial state. While the momenta of the colliding protons at the LHC are known, their colliding constituents' momenta are only known at the level of probability distributions according to the PDFs.  

Particles not leaving a signature in the ATLAS detector, such as neutrinos, can be reconstructed indirectly by the application of the laws of momentum conservation. As the initial state of the $pp$ collisions is not known, full momentum conversation cannot be used. Nevertheless, momentum conservation can be applied to the transverse plane. Here the total momentum before and after the collision is zero. This allows reconstructing the sum of momenta for particles not leaving a signature. It is referred to as \newword{Missing Transverse Momentum}.  \etmiss\ expresses its magnitude.\footnote{The term \etmiss\ is misleading and sometimes also referred to as \textit{Missing Transverse Energy}. Using the argument of momentum conservation, the directly missing quantity must be a vector -- the momentum.} It is calculated from the energy depositions in the calorimeters and the tracks in the muon spectrometer which are grouped to the objects to which they are assigned and calibrated accordingly. The remaining energy depositions are grouped into the \newword{CellOut term}:
\begin{align}
       E_{x(y)}^{\text{miss}} =
             E_{x(y)}^{\text{miss},electron} +
              E_{x(y)}^{\text{miss,jets}}  +
               E_{x(y)}^{\text{miss,softjets}}   + 
              E_{x(y)}^{\text{miss,muon},\mu} +
             E_{x(y)}^{\text{miss,CellOut}}
\end{align}
The magnitude of the \etmiss\ is given by
\begin{align}
 E_{\mathrm{T}}^{\text{miss}}=\sqrt{\left(E_{x}^{\text{miss}}\right)^{2} +\left(E_{y}^{\text{miss}}\right)^{2}}.
\end{align}
As the object definitions may vary from analysis to analysis so may the definition of the \etmiss. In this analysis the following \etmiss\  object definition was used:
\begin{itemize}
\item Electrons as defined in Section \ref{sec:electrons} are required to have $\pt > 10$ GeV. All energy scale correction factors except the out-of-cluster correction are applied.
\item Jets as defined in Section \ref{sec:jets} with $\pt > 20$ GeV are used at the EM+JES calibration level.
\item Jets with $7~\text{GeV}< \pt < 20$ GeV are included by the softjets term. They are being calibrated  to the EM scale only.
\item Combined muons (see Section \ref{sec:muons}) with $\abseta < 2.5$ GeV are included. Out of the acceptance of the ID, for $2.5 < \abseta <2.7$, no combined muons can be reconstructed. Still, muons reconstructed using the MS only are included for this $\eta$ region. The muon energy deposited in the calorimeters is added to the cellout term in case the muon is isolated ($\Delta R (\text{muon, jet}) > 0.3$) and added to the jets term in case it is not. 
\item All further energy cells are added to the CellOut term.
\end{itemize} 
A detailed description of the MET performance on the 2011 dataset is given in \cite{reco_MET}.

\section{$\tau$ Leptons}
\label{sec:taus}
Both top quarks of a produced \ttbar\ pair decay into a \bQ\ and a $W$ boson of which about 11\,\% decay into a $\tau$ lepton \cite{PDG}. These can contribute to the selected signal sample depending on their decay channel (and the decay channel of the other $W$ boson). As $\approx 35$\,\% of all $\tau$ leptons directly decay into an electron or muon via $\tau \rightarrow e \bar{\nu}_e \nu_\tau$ or  $\tau \rightarrow \mu \bar{\nu}_\mu \nu_\tau$ \cite{PDG} the detector signature will look very similar to a prompt $W \rightarrow e \bar{\nu}_e$ or $W \rightarrow \mu \bar{\nu}_\mu$ decay. A larger \etmiss\ and changed kinematics of the electrons and muons stemming from decayed $\tau$ leptons make the difference.  
This analysis does not make use of an explicit $\tau$ lepton reconstruction. Events including $\tau$ leptons might pass the selection and the $\tau$ decay products will be reconstructed as (non-prompt) electrons, muons or jets.

\chapter[Dataset, Signal and Background Modelling]{Signal and Background Modelling}
Physics results in high energy physics experiments can often only be obtained by the comparison of simulated to measured events. This includes the signal and background events of which some are estimated by data-driven methods. The used dataset as well as the modelled signal and background events are described in this chapter. For a comparison of simulated and measured event properties a well-defined event selection is needed. Hence, this comparison is shown in the following Chapter \ref{sec:selandreco}.

The simulated events include PDFs, the hard scattering process, parton showering as well as a simulation of the energy depositions of the particles in the ATLAS detector. The latter was achieved using the GEANT4 \cite{Geant4} toolkit.

\section{Dataset}
\label{sec:dataset}
Data taken with the ATLAS detector at $\sqrt{s} = 7\,\TeV$ are analysed for the measurement of \ttbar\ spin correlation. This dataset will be called ``the 2011 dataset'' or ``the 7 \TeV\ dataset''. Each time the ATLAS TDAQ records data\footnote{A block of data taking usually starts with the beginning of collisions and end with a beam dump.}, a \textit{run number}\index{Run number} is assigned to the data taken. Each run is then separated into \textit{luminosity blocks}\index{Luminosity block} (\textit{LBs}\index{LB | see {Luminosity block }}). Further, only data passing certain data quality criteria are taken into account. The data quality can be degraded if certain components of the detector are not available or are providing bad data due to misconfiguration or malfunction. Physics analyses use a \textit{Good Runs List}\index{Good Runs List} (\textit{GRL}\index{GRL | see {Good Runs List }}) including the run numbers and LBs of the data to be analysed. The GRL of the 2011 dataset includes the run numbers 178044-191933. These runs were taken in the time between the 22nd of March 2011 and the 30th of October 2011. After a technical stop in September 2011
the $\beta^{*}$ of the proton beams was reduced from 1.5 m to 1.0 m, leading to an increased instantaneous luminosity. The evolution of the instantaneous luminosity is shown in Figure \ref{fig:lumi}. Figure \ref{fig:intlumi} shows the evolution of the total integrated luminosity of the 2011 dataset, separated in luminosity delivered by the LHC, recorded by ATLAS and accepted by the GRLs. 
\begin{figure}[ht]
	\centering
		\includegraphics[width=0.95\textwidth]{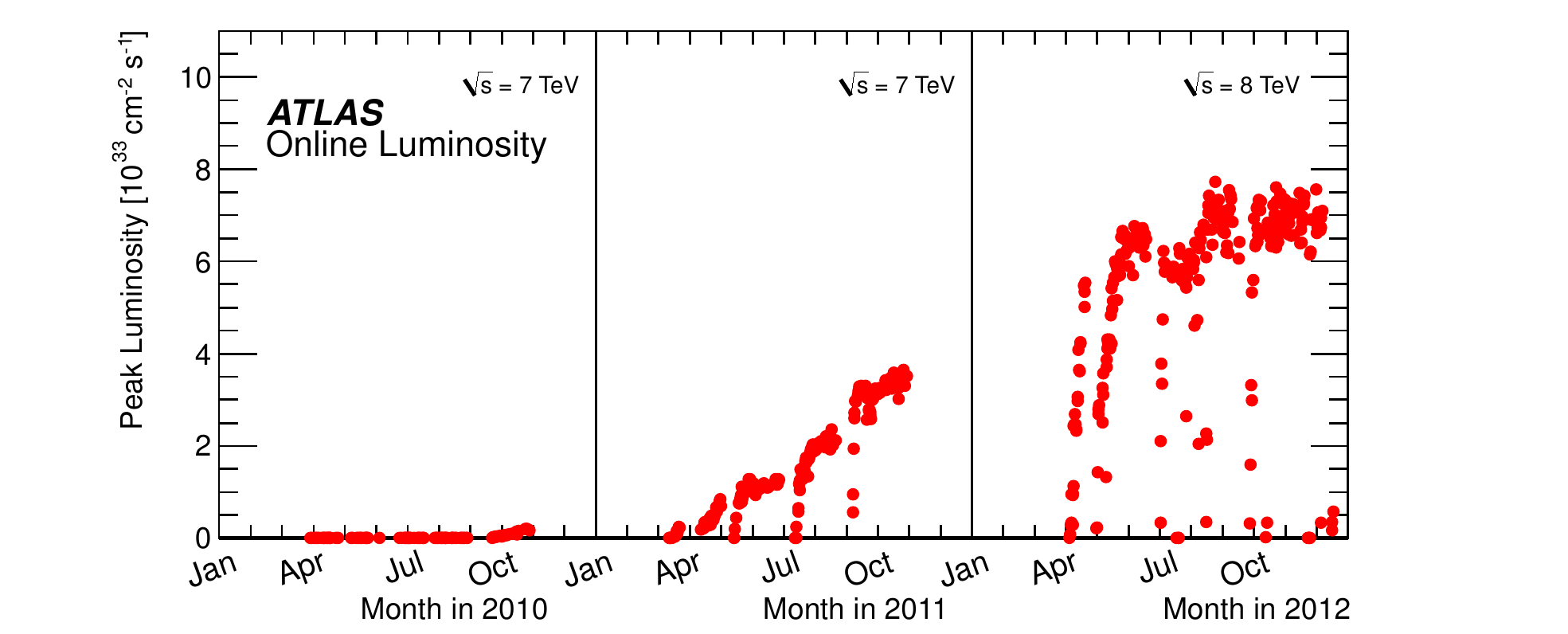}
		\caption{Evolution of the instantaneous luminosity delivered to ATLAS during data taking in 2010-2012 \cite{atlas_public_lumi}.}
	\label{fig:lumi}
\end{figure}The average number of interactions per bunch crossing, $\mu$, also increased due to the lowering of $\beta^{*}$ (see Figure \ref{fig:mu_2011}). With increasing $\left< \mu \right>$ the so-called $\textit{pile-up}\index{Pile-up}$ rises: energy depositions reconstructed as objects not belonging to the hard scattering process are also included in the event and have to be vetoed. The procedure to veto such pile-up objects during the event selection was described in Section \ref{sec:jvf}.
\begin{figure}[h]
	\centering
\begin{tabular}{>{\centering}b{0.49\textwidth}<{\centering} >{\centering}b{0.49\textwidth}<{\centering} }
	\subfigure[]{
			\includegraphics[width=0.48\textwidth]{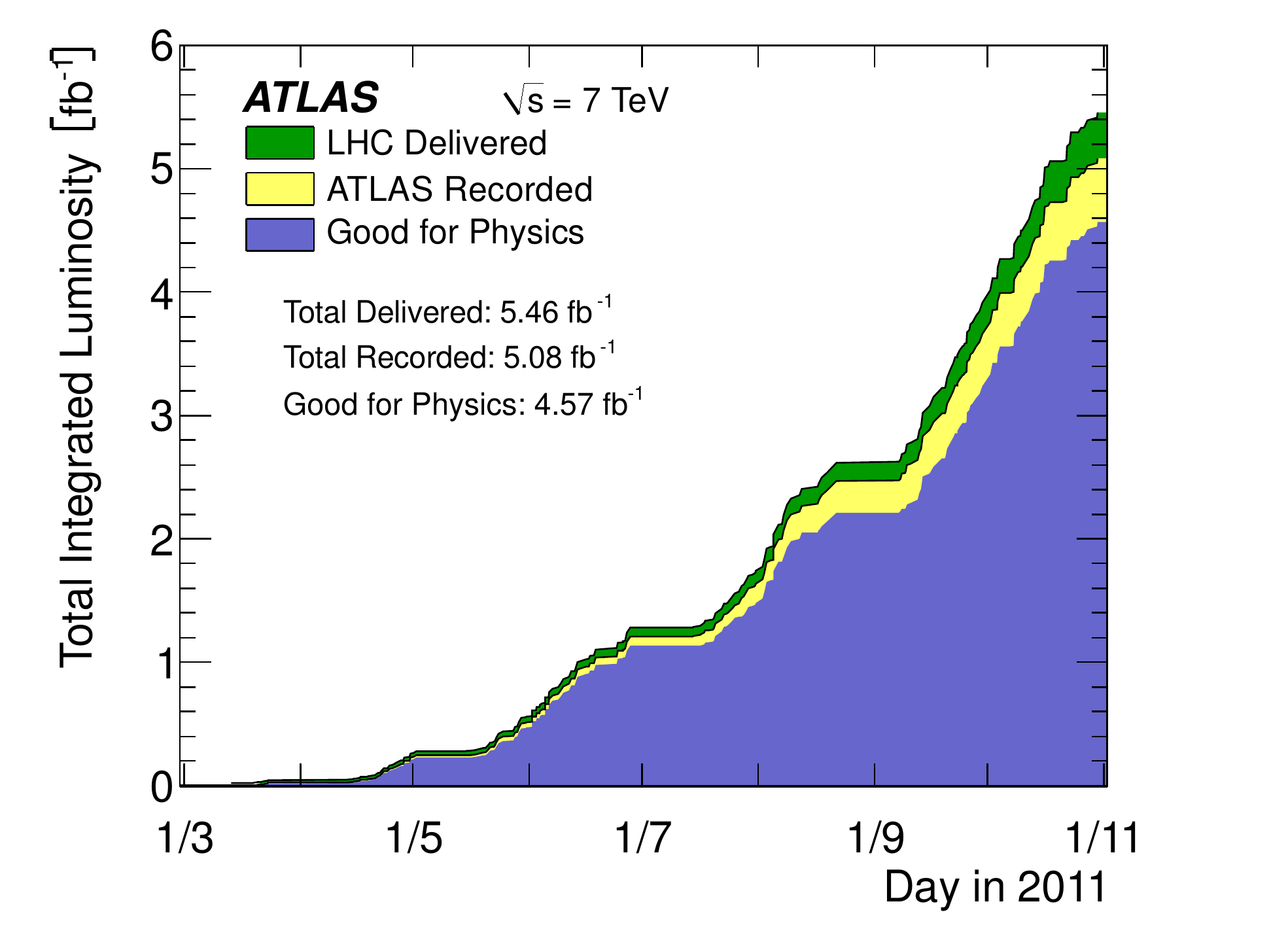}
			\label{fig:intlumi}
		}&
	\subfigure[]{

		\includegraphics[width=0.48\textwidth]{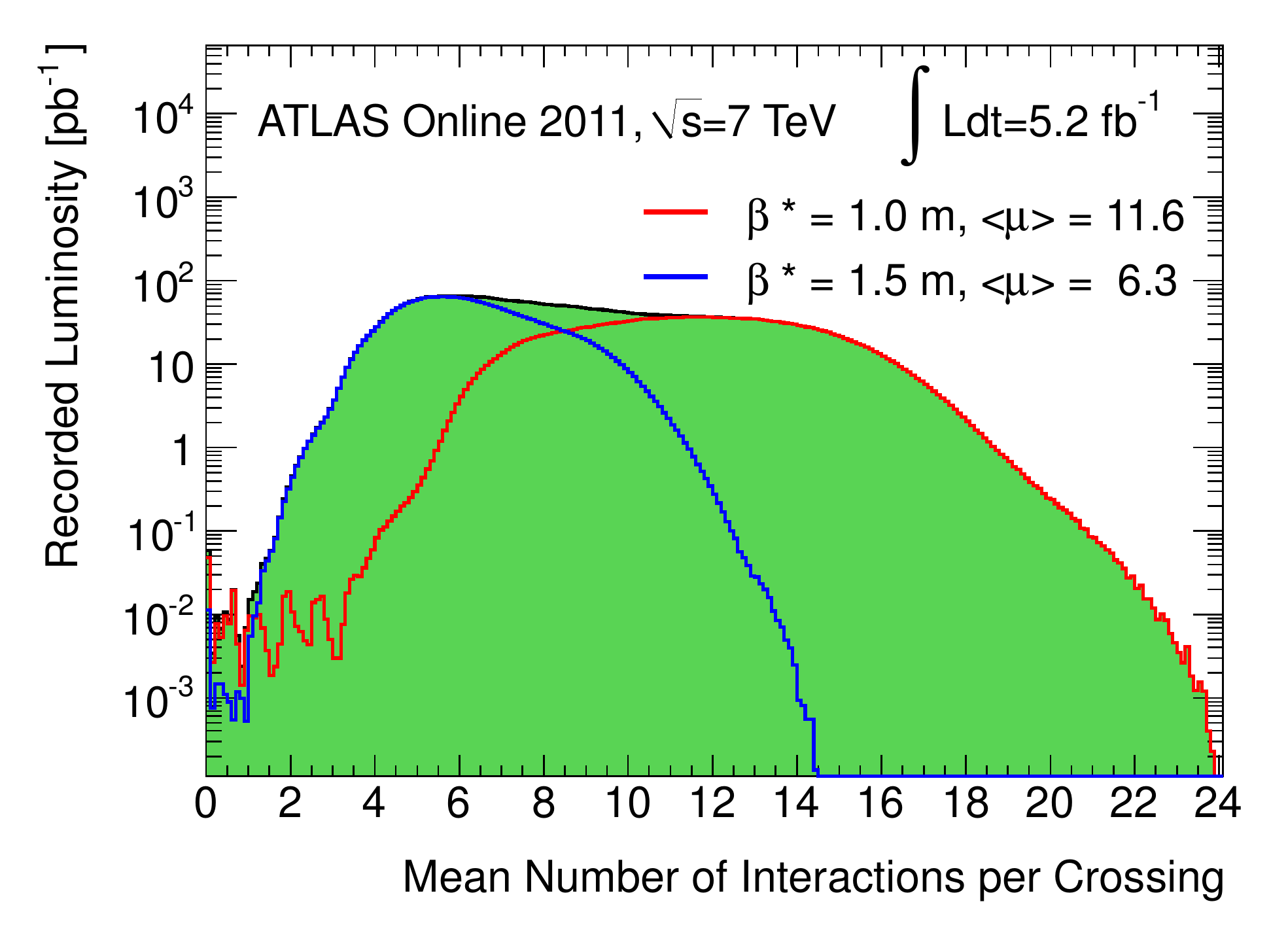}
			\label{fig:mu_2011}
		}
\end{tabular}
		\caption{\subref{fig:intlumi} Evolution of the integrated luminosity \intlum\ delivered, recorded and accepted by the GRL for the 2011 dataset \cite{atlas_public_lumi}. \subref{fig:mu_2011} The mean number of interactions per bunch crossing for the 2011 dataset \cite{atlas_public_lumi}. }
\end{figure}
During the data taking in 2011 the LHC beam parameters had not yet reached their design values. Table \ref{tab:beampar_2011} lists the beam parameters at the beginning and the end of the 7 \TeV\ run and compares them to the LHC design values. The total integrated luminosity accepted by the GRLs  corresponds to an integrated luminosity of \intlum\ = \intlumi. The datasets for both the taken data and the used simulations are listed in Appendix \ref{app:datasets}.

\begin{table}[htbp]

\begin{center}
\begin{tabular}{|c|c|c|c|c|c|}
\hline
& Protons/Bunch & Bunches & Bunch Spacing& $\beta^{*}$ & Peak Luminosity\\
\hline
\hline
2011 		& $1.45 \cdot 10^{11}$ & 1380 & 50 ns & 1.0 m &${\;\;}3.7 \cdot 10^{33} \text{ cm} ^{-2}\text{ s}^{-1}$\\
Nominal	& $1.15 \cdot 10^{11} $& 2808 & 25 ns & 0.55 m&$10.0 \cdot 10^{33} \text{ cm} ^{-2}\text{ s}^{-1}$\\
\hline

\end{tabular}
\end{center}
\caption{LHC beam parameters at the end of the 7 \TeV\ run as well as the design values \cite{IPAC2013}.}
\label{tab:beampar_2011}
\end{table}

\section{\texorpdfstring{$t\bar{t}$}{Top/Anti-Top Quark} Signal Samples}
\label{sec:signal}
Top quark pair production is simulated using the NLO generator \mcatnlo\ v4.01 \cite{MCatNLO, MCatNLO2, MCatNLO3, MCatNLO4} assuming a 
top quark mass of 172.5~GeV and using the NLO PDF CT10~\cite{CT10}. For parton showering and hadronisation, \herwig\ 6.520 \cite{herwig} is used. \jimmy\ 4.31 \cite{jimmy} simulates multi-parton interactions. The event generator is tuned according to the \newword{ATLAS Underlying Event Tune} AUET2-CT10 \cite{ATLAS_gentune}. 

Samples without spin correlation are generated by setting the \mcatnlo\ parameters 
$\mathrm{IL}_1 = \mathrm{IL}_2 = 7$ \cite{MCatNLO4}.  In this case the top and anti-top quark decay is not performed by \mcatnlo, but by \herwig. As the top spin information is not propagated to \herwig, top quark spins are effectively decorrelated.  As a side effect, the sample with uncorrelated \ttbar\ pairs has a top width of $\Gamma_t = 0$. This effect was studied to have no impact on the actual analysis. Samples including Standard Model spin correlation
are generated by setting the variables $\mathrm{IL}_1 = \mathrm{IL}_2 = 0$.

The $t\bar{t}$ cross section for $pp$ collisions at a \com\ energy of $\sqrt{s} = 7\,\tev$ is $\sigma_{t\bar{t}}= 177{\,}^{+10}_{-11}$~pb for a top quark mass of $172.5 \gev/c^2$. It has been calculated at next-to-next-to leading order (NNLO) in QCD including resummation of next-to-next-to-leading logarithmic (NNLL) soft gluon terms with top++2.0 \cite{Cacciari:2011hy, Baernreuther:2012ws, Czakon:2012zr, Czakon:2012pz, Czakon:2013goa, Czakon:2011xx}. The PDF and $\alpha_S$ uncertainties were calculated using the PDF4LHC prescription \cite{Botje:2011sn} with the MSTW2008 68\,\% CL NNLO \cite{MSTW, Martin:2009bu}, CT10 NNLO \cite{CT10, Gao:2013xoa} and NNPDF2.3 5f FFN \cite{NNPDF} PDF sets, added in quadrature to the scale uncertainty. The NNLO+NNLL value, as implemented in Hathor 1.5 \cite{Aliev:2010zk}, is about 3\,\% larger than the exact NNLO prediction.

All \ttbar\ final states except the full hadronic ones are included in these two samples. The full hadronic \ttbar\ final states are included in the fake lepton estimation, as described in Section \ref{sec:fakes}. 

It should be mentioned that despite the fact that a NLO generator, \mcatnlo, was used to model the signal, the \ttbar\ spin correlation is not implemented at full NLO \cite{NLO_generators}. The implementation of the spin correlation treatment in \mcatnlo, as described in \cite{MCatNLO3}, does not include spin dependent virtual corrections \cite{NLO_generators}. The actual effects on the analysis can be neglected. The described missing NLO contributions affect only part of the difference between LO and NLO. Even the total difference between NLO and LO spin correlation is low ($\approx 2\,\%$) \cite{Bernreuther2004}.

\section{MC Driven Backgrounds}

\subsection{Single Top}
For the background arising from single-top 
production in the $s$- and the $Wt$-channel, \mcatnlo+\herwig\ is used with NLO PDF CT10, invoking the \textit{diagram removal scheme}\index{Diagram removal scheme} \cite{diagramremoval} to remove overlaps between the single-top and \ttbar\ final states. For the $t$-channel, \acermc\ \cite{acermc} interfaced to \pythia\ \cite{pythia} 6.452 with modified LO PDFs (MRST LO$*$$*$, LHAPDF 20651) \cite{mrstcal, mrstcal2} is used.

\subsection{Diboson}
Background contributions arising from $WW$, $ZZ$ and $WZ$ production (\textit{diboson background}\index{Diboson background}) are simulated by the \herwig\ generator with modified LO PDFs (MRST LO$*$$*$, LHAPDF 20651).\footnote{In most of the other cases \herwig\ served as parton shower and hadronisation generator and JIMMY for the underlying event model.} As \herwig\ is a LO generator, the cross sections of the diboson processes are scaled to match the NLO prediction. 

\subsection{$W$+Jets}
$W$ boson production in association with multiple jets is the dominating source of background events. To simulate these, \alpgen\ v2.13~\cite{alpgen} is used. It implements the exact LO matrix elements for final states with up to five partons using the LO PDF set CTEQ6L1~\cite{cteq6l}. To simulate parton showering, hadronisation and multi-parton interactions, the \herwig\ and \jimmy\ generators are used as for the simulation of the \ttbar\ signal.  Dedicated samples are used for the production of heavy flavour samples ($W$+$c$+jets, $W$+$c\bar{c}$+jets and $W$+$b\bar{b}$+jets).
The MLM~\cite{Mangano:2006rw} matching scheme of the \alpgen\ generator is used to remove overlaps 
between the $n$ and $n+1$ parton samples with parameters {\tt RCLUS}=0.7 and 
{\tt ETCLUS}=20~GeV\@. Also, phase space overlaps across the different flavour samples are removed. 
While MC simulations are used to determine the kinematic shapes of the $W+\text{jets}$ background, its normalization and flavour composition is derived by a data-driven approach, described in Sections \ref{sec:wjets_norm} and \ref{sec:wjets_flav}.

\subsection{$Z$+Jets}
For the estimation of the background contribution caused by the Drell-Yan production $Z/\gamma^{*} \rightarrow \ll$ plus additional jets, \alpgen+\herwig\ with the LO PDF set CTEQ6L1 is used as for the $W$+jets background. Additional jets are simulated with up to five additional partons on matrix element level. Even though this simulation takes into account interferences between $Z$ and $\gamma^{*}$ bosons, it is briefly called $Z$+jets background. Two sets of samples were used for $Z$+jets background: 
inclusive  $Z$+jets samples and in addition dedicated $Z+b\bar{b}$ samples. 
Overlapping phase spaces are removed. The cross sections are scaled to match the NNLO predictions.

\section{Data Driven Backgrounds}
\subsection{Fake Lepton Estimation}
\label{sec:fakes}
The objects reconstructed as isolated leptons can in fact also be either jets with a high electromagnetic component or hadrons from jets decaying into leptons that seem to be isolated.  
This \textit{fake lepton background}\index{Fake lepton background} is caused by QCD induced multijet events.\footnote{Hence it is also called multijet background or misidentified lepton background.} Such events would demand a large amount of MC statistics and suffer from a limited ability of proper MC modelling. Thus, a data-driven approach is chosen to estimate the fake lepton background. It is based on the \textit{matrix method}\index{Matrix method} which is introduced before the channel specific estimates are explained. 

\subsubsection{Matrix Method}
For the matrix method \cite{matrixmethod} two dedicated samples are produced by applying different selection criteria to the taken data. Two different lepton isolation criteria, loose and tight (see sections \ref{sec:electrons} and \ref{sec:muons}), are used. Of these two, the tight definition has more stringent isolation requirements. The total number of events passing each of the criteria is a sum of real and fake leptons:
\begin{align}
N^{\text{loose}}& = N^{\text{loose}}_{\text{real}} + N^{\text{loose}}_{\text{fake}},\\
N^{\text{tight}} &= N^{\text{tight}}_{\text{real}} + N^{\text{tight}}_{\text{fake}}.
\end{align}
As the tight sample is a subset of the loose, selection efficiencies for real and fake leptons can be defined as 
\begin{align}
\varepsilon_{\text{real}} = \frac{N^{\text{tight}}_{\text{real}}}{N^{\text{loose}}_{\text{real}}}, \quad \varepsilon_{\text{fake}} = \frac{N^{\text{tight}}_{\text{fake}}}{N^{\text{loose}}_{\text{fake}}}.
\end{align}
The number of fake leptons within the tight sample can be expressed as
 \begin{align}
 N^{\text{tight}}_{\text{fake}} = \frac{ \varepsilon_{\text{fake}}}{ \varepsilon_{\text{real}} -  \varepsilon_{\text{fake}}} \left(\varepsilon_{\text{real}} N^{\text{loose}} - N^{\text{tight}} \right)
 \end{align}
Hence, by knowing the real and fake efficiencies the selected loose and tight samples can be used to obtain a sample of tight fake leptons. Technically this is done by applying the weights
\begin{align}
w_{\text{loose}} = \frac{\varepsilon_{\text{fake}} \varepsilon_{\text{real}}}{\varepsilon_{\text{real}} - \varepsilon_{\text{fake}}}, \quad w_{\text{tight}} = \frac{\varepsilon_{\text{fake}} \left(  \varepsilon_{\text{real}} -1 \right)}{\varepsilon_{\text{real}} - \varepsilon_{\text{fake}}},
\end{align}
to the events, leading to positive weights for loose and negative weights for tight events, and merging them.
In the following the determination of $\varepsilon_{\text{real}}$ and $\varepsilon_{\text{take}}$ for both the \mujets\ and the \ejets\  channel are described. 

\subsubsection{\mujets\ Channel}

For the \mujets\ channel, two different approaches (A and B) are used and the resulting fake lepton background estimation is averaged. For both methods loose muons are defined the same way as tight ones but without the requirement on the isolation (\ptcone\ and \etcone). In method A, both $\varepsilon_{real}$ and $\varepsilon_{fake}$ are parameterized in $|\eta|$ and $\pt$ of the muon. In method B, $\varepsilon_{real}$ is found to be constant as a function of both \pt\ and \eta\ while $\varepsilon_{fake}$ is parameterized in $|\eta_{\mu}|$. The fake dominated control region for method A is defined by cutting on the transverse $W$ boson mass\footnote{The transverse $W$ boson mass is defined as $ \Wmt = \sqrt{2 \pt^l \pt^\nu ( 1 - \cos (\phi^l - \phi^\nu))}$.} $\Wmt < 20$~GeV and \etmiss\
$+\Wmt < 60$~GeV. Method B uses $\etmiss\ < 20$~GeV and \etmiss\
$+\Wmt < 60$~GeV instead. 

Further, the signal efficiencies are obtained from Monte Carlo simulation (which agrees within 1\,\% with the values derived by the T\&P method on data \cite{muon_reco_eff}). The fake efficiencies are obtained by an extrapolation of an impact parameter significance $d_0 / \sigma(d_0)$ \cite{b-tagging} dependent tight to loose ratio. This method makes use of the fact that fake muons originate from heavy flavour jets and hence have a larger impact parameter significance. 

\subsubsection{\ejets\ Channel}
In the \ejets\ channel, $\varepsilon_{real}$ is derived via a tag-and-probe method for $Z \rightarrow ee$ events. As the topologies of \ttbar\ and $Z \rightarrow ee$ events are different and also affect $\varepsilon_{real}$, a correction factor derived from Monte Carlo samples is applied to account for that. The fake efficiencies are derived in a control region with \etmiss\ $<$ 20 GeV in which the contribution of further backgrounds, estimated via Monte Carlo samples, is subtracted.
Both efficiencies are parameterized as functions of $|\eta_{e}|$ and $\Delta R(e, \text{closest jet})$. The analysis described in this thesis demands in particular a good angular distribution of the background.  For this reason, the $\Delta R$ parameterized fake lepton estimate was validated and implemented.
\begin{figure}[ht]
	\centering
			\subfigure[]{
		\includegraphics[width=0.45\textwidth]{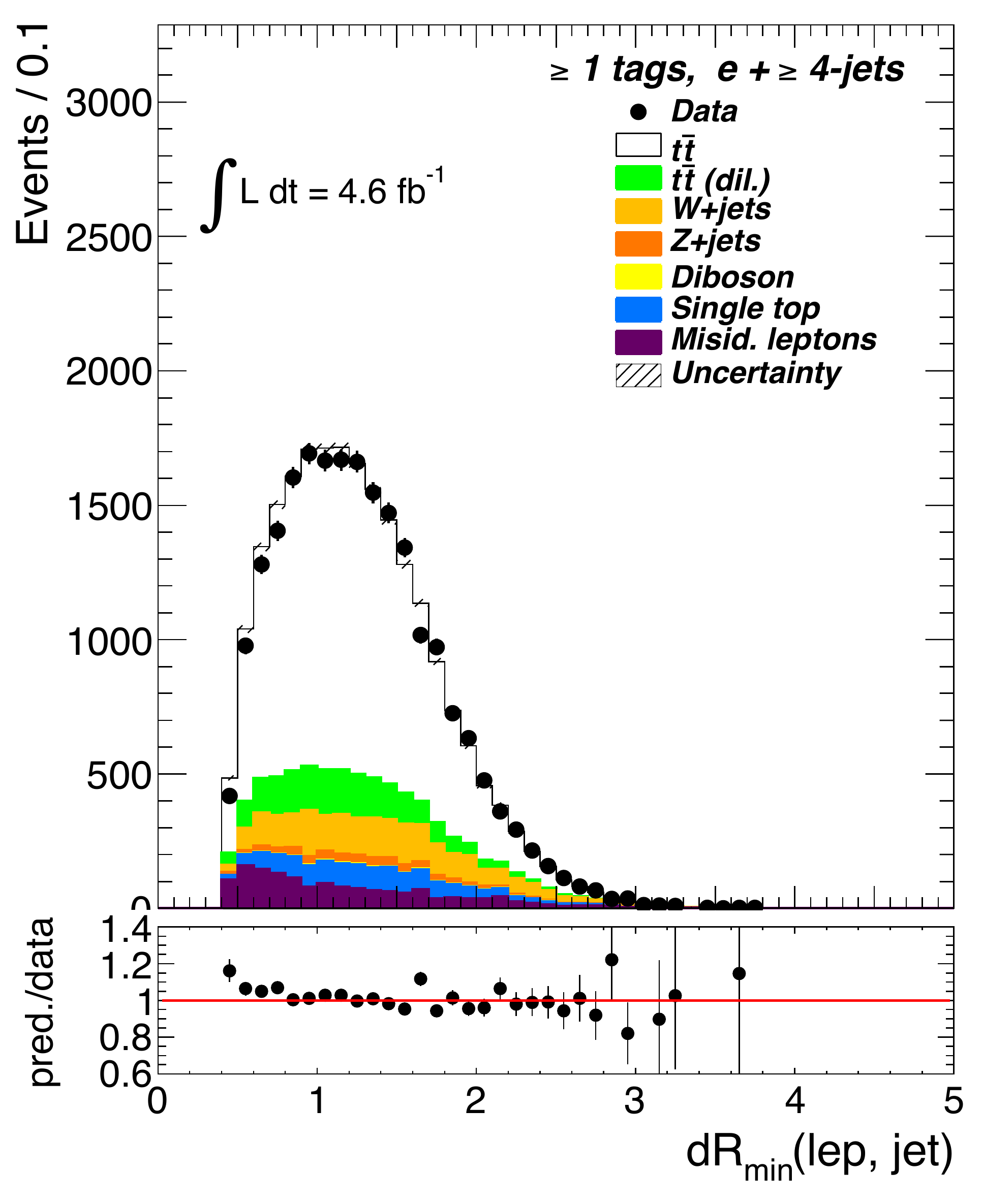}
			\label{fig:QCD_before}
		}
					\subfigure[]{
		\includegraphics[width=0.45\textwidth]{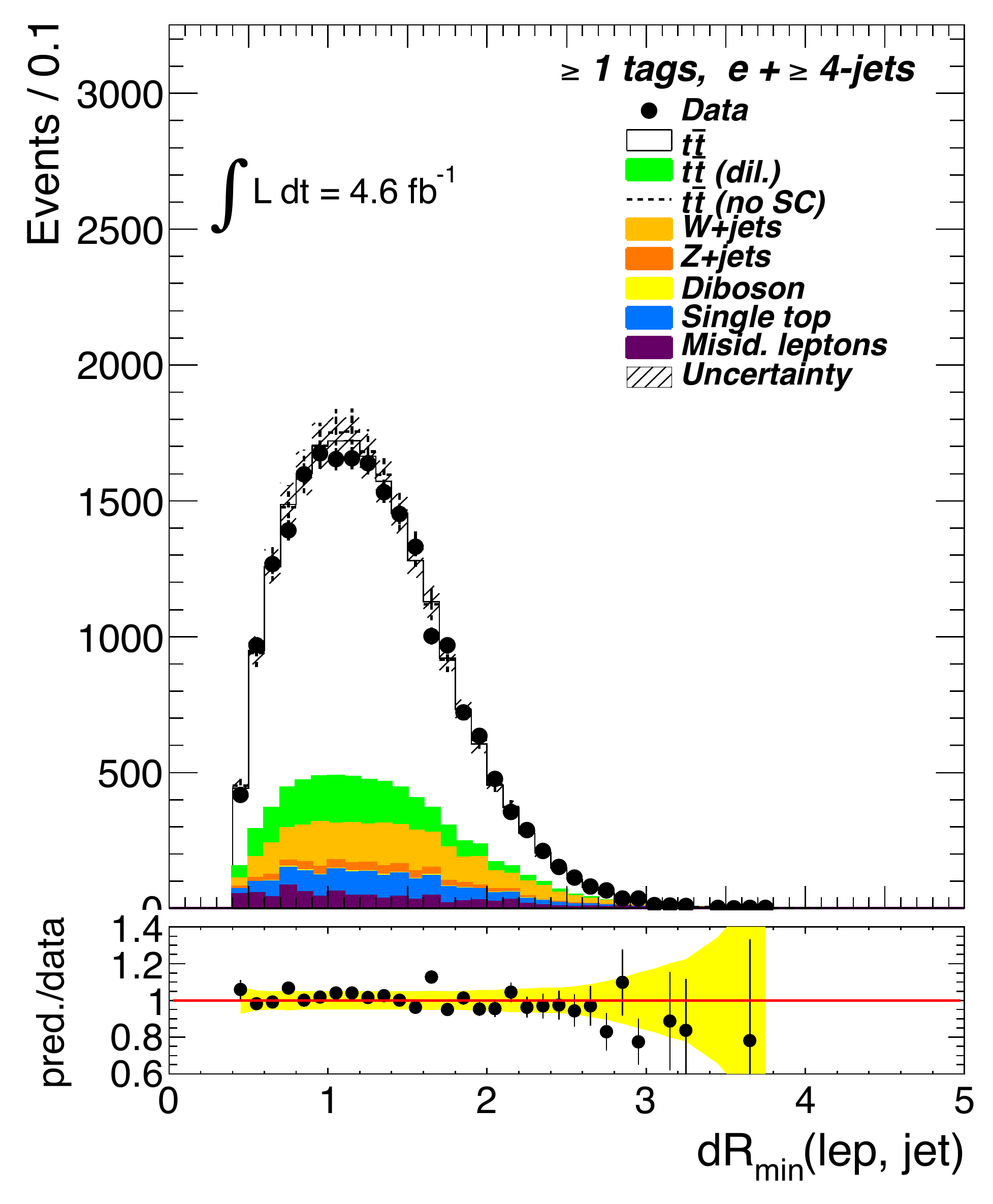}
			\label{fig:QCD_after}
		}
		\caption{Improvement by the choice of a new parameterization for the fake lepton estimate. Mismodelling is visible particularly in the $\Delta R (\text{lepton, jet})$ distributions: \subref{fig:QCD_before} Old parameterization in \pt\ and $\eta$ of the lepton. The template statistics uncertainty is shown for the prediction. \subref{fig:QCD_after} New parameterization with an additional $\Delta R (\text{lepton}, \text{jet})$ dependence. The template statistics uncertainty, the normalization uncertainty for the data driven yields and the theory uncertainty on the cross sections are shown for the prediction and propagated to the error band in the ratio.}
\end{figure}
The improvement gained by the new parameterization for the fake lepton background can be seen by comparing Figure \ref{fig:QCD_before} to \ref{fig:QCD_after}. In particular, events with leptons and jets having a close distance are more accurately modelled.

\subsection{$W$+jets Normalization}
\label{sec:wjets_norm}
While MC simulations are used to estimate the $W$+jets background, a data-driven approach estimates its normalization with lower uncertainty than that from the MC prediction. The approach makes use of the charge asymmetry (CA) of $W^{\pm}$ production at the LHC with its $pp$ collision mode. It leads to $r = \sigma(pp \rightarrow W^+) / \sigma(pp \rightarrow W^-) \approx 1.5$ due to different parton densities for $u$- and $d$-quarks (see Figure \ref{fig:PDF}). While the total normalization of $W$+jets is not well modelled by the MC, the ratio $r$ is. It can be used to determine the total yield of $W$+jets events $N_W$ by measuring the number of events including a positive ($D^+$) or negative ($D^-$) lepton in data.
\begin{align}
N_W = N_{W^+} + N_{W^-} &= \left( \frac{N_{W^+}^{\text{MC}} + N_{W^-}^{\text{MC}}}{N_{W^+}^{\text{MC}} - N_{W^-}^{\text{MC}}}\right) \left( D^+ - D^- \right)\\
&= \left( \frac{r_{\text{MC}} +1 }{r_{\text{MC}} -1}\right)  \left( D^+ - D^- \right).
\label{eq:wjetsnorm}
\end{align}
The common event selection as described in Section \ref{sec:eventselection}, except cuts on the \btag ging, is used to measure the difference $\left( D^+ - D^- \right)$. \ttbar\ events, fake lepton background and $Z$-jets events are produced symmetrically (with respect to the lepton charge) to a good approximation. Hence, the assumption that the difference $\left( D^+ - D^- \right)$ is caused by the $W$-asymmetry is valid after the background from single top production is subtracted.

To obtain the normalization for the selection of $n$ jets of which at least one jet is \btag ged the equation
\begin{align}
W^n_{\geq 1 \text{tag}} =W^n_{\text{pretag}} f^{2\text{jet}}_{\text{tag}} f^{2\rightarrow n}_{\text{tag}} 
\end{align}
is used. It includes the fraction $ f^{2\text{jet}}_{\text{tag}}$ of tagged to untagged jets (\textit{tagging fraction}\index{Tagging fraction}) for the 2-jet selection and the ratio  $f^{2\text{jet}}_{\text{tag}}$ of tagging fractions between the 2-jet and the $n$-jet selection, which is derived by MC simulation.

\subsection{$W$+jets Flavour Composition}
\label{sec:wjets_flav}
Next to the total normalization of $W$+jets events, the flavour composition of the sample needs to be determined by a data-driven approach. The total number of $W$+jets events can be divided into the jet flavour types $bb$, $cc$, $c$ and light with corresponding fractions $F$ (summing up to one). An event of each subclass has a probability $P$ to be \btag ged.
Using this classification, the total number of $W$+jets events in a \btag ged sample  with $i$ jets  is given by 
\begin{align}
N^{W^{\pm},\text{tag}} &= N^{W^{\pm},\text{pretag}} \left( F_{bb, i}P_{bb, i} + F_{cc, i} P_{cc, i} + F_{c,i}P_{c,i} + F_{\text{light}, i}  P_{\text{light}, i}\right) \\
&= N^{W^{\pm},\text{pretag}} \left( F_{bb, i}P_{bb, i} + k_{ccbb}F_{bb, i} P_{cc, i} + F_{c,i}P_{c,i} + F_{\text{light}, i}  P_{\text{light}, i}\right) .
\end{align}
Instead of determining $F_{cc, i}$ individually it is expressed by $F_{bb, i} $ and the ratio of $F_{cc, i} / F_{bb, i} $, which is derived from MC simulation. Also, the tagging probabilities $P$ are estimated from MC. 
The flavour compositions are determined for subsamples with positive and negative leptons separately. By using $N^{W^{\pm},\text{pretag}}$ from the CA normalization and requiring that $N^{W^{\pm},\text{tag}}$ matches between simulation and data, the flavour fractions are rescaled by factors $K$ accordingly. The rescaled flavour fractions are then put back into the CA normalization method iteratively until convergence is reached. As a baseline, events with two jets are used. Applying the scale factors $K$ to other jet multiplicity samples can cause the sum of fractions to deviate from one:\begin{align}
K_{bb, 2} F_{bb, i}^\text{MC} + K_{cc, 2} F_{cc, i}^\text{MC} + K_{c, 2} F_{c, i}^\text{MC} + K_{\text{light}, 2} F_{\text{light}, i}^\text{MC}=A.
\end{align} 
A correction is applied by rescaling each fraction $K$ with $A$:
\begin{align}
K_{xx, i} = \frac{K_{xx, 2}}{A}.
\end{align}

\chapter[Event Selection and Reconstruction]{Event Selection and Reconstruction}
\label{sec:selandreco}
With the \ljets\ channel as the \ttbar\ final state under study, advantages and challenges come along. Every event provides the necessary information for full reconstruction of all the objects of interest. The dilepton channel suffers from the ambiguity of two undetected neutrinos, making it hard to reconstruct both of them correctly. In contrast, using assumptions such as the masses of the top quark and the $W$ boson, the whole \ttbar\ event can be reconstructed in the \ljets\ channel. This includes the single neutrino from the leptonic top decay. The challenge of the \ljets\ channel reconstruction is the separation of the two $T_3 = \pm \frac{1}{2}$ non-$b$-jets. 

In this chapter the applied selection cuts are explained before the data is compared to the prediction. The agreement of the MC generator to the measured data is discussed before the reconstruction of the events is explained in detail, in particular the separation of the two light jets from the $t\rightarrow Wb \rightarrow bqq'$ decay. The correct selection of top spin analysers is validated. Finally, reconstruction efficiencies are presented and compared to other reconstruction methods.

\section{\texorpdfstring{$t\bar{t}$}{Top/Anti-Top Quark} Selection in the Lepton+Jets Channel}
\label{sec:eventselection}
As no dedicated $\tau$ lepton reconstruction is used, the chosen \ljets\ channel splits into an \ejets\ and a \mujets\ channel. Still, $\tau$+jets events with leptonically decaying $\tau$ leptons  are part of the signal as well. Also dileptonic events including a hadronically decaying $\tau$ are selected. In Section \ref{sec:topdecay} it was explained how the ideal \ljets\ decays compare to the actual ones. 

Events from the \ejets\ and \mujets\ channel are selected from single lepton trigger streams. 
For each period the selected trigger is the unprescaled\footnote{A trigger prescale of $p$ randomly drops a fraction $(1-\frac{1}{p})$ of  events that had passed the trigger chain to reduce the data rate. Unprescaled triggers have $p = 1$, so no events are dropped.} trigger with the lowest \pt\ threshold. To avoid a loss of trigger efficiency, the \pt\ cut of the selected lepton is chosen such that it is well above the \pt\ thresholds of the trigger leptons.\footnote{$\pt^{\mu} > 20 \GeV$ and $\et^{e} > 25 \GeV$.} With this choice the reconstructed objects are said to be `within the trigger efficiency plateau'. The chosen trigger streams are listed in Table \ref{tab:triggers} and depend on the data taking period. 

\begin{table}[htbp]
\begin{center}
\small
\begin{tabular}{|c|c|}
\hline
Electron Trigger & Muon Trigger \\
\hline\hline
\begin{tabular}{c}EF\_e20\_medium (before period K) \\ EF\_e22\_medium (period K) \\ EF\_e22vh\_medium1 or EF\_e45\_medium1 (period L-M) \end{tabular}& \begin{tabular}{c}EF\_mu18 (before period J) \\ EF\_mu18\_medium (starting period J)\end{tabular}\\
\hline
\end{tabular}
\end{center}
\caption{Used trigger streams depending on the data taking period.}
\label{tab:triggers}
\end{table}
In the following sections, the expression 'good' refers to objects passing the quality criteria as defined in Chapter \ref{sec:objects}. 

\subsection{\ejets\ Selection}
\begin{enumerate}
 \item The electron trigger must have fired.
 \item The event must contain at least one primary vertex with at least five tracks.
 \item Exactly one good electron is found.
 \item No good muon is found.
 \item The good electron must match the object that fired the trigger.
 \item No jet with $\pt\ > 20~\GeV$ failing the quality cuts may be included in the event.
 \item At least four good jets with \pt~$> 25$~GeV, $\abseta < 2.5$, and a jet vertex fraction $|\textrm{JVF}|>0.75$ are found. 
 \item \etmiss~$>  30$~GeV is required.
 \item The transverse $W$ mass \Wmt\ must be larger than 30 GeV. 
 \item At least one \bjet\ must be identified using the  \mvone\ tagger at the 70\,\% efficiency working point.
\end{enumerate}
The cuts on \etmiss\ and \Wmt\ suppress the multijet background containing fake leptons.

\subsection{\mujets\ Selection}
\begin{enumerate}
 \item The muon trigger must have fired.
 \item The event must contain at least one primary vertex with at least five tracks.
 \item Exactly one good muon is found.
 \item No good electron is found.
 \item The good muon must match the object that fired the trigger.
 \item Electrons and muons must not share a track. 
 \item No jet with $\pt\ > 20~\GeV$ failing the quality cuts may be included in the event.
 \item At least four good jets with \pt~$> 25$~GeV, $\abseta < 2.5$, and a jet vertex fraction $|\textrm{JVF}|>0.75$ are found. 
 \item \etmiss~$> 20$ GeV is required.
 \item \Wmt + \etmiss~must be larger than 60 GeV. 
 \item At least one \bjet\ must be identified using the  \mvone\ tagger at the 70\,\% efficiency working point.
\end{enumerate}
The cuts on \etmiss\ and \Wmt\ suppress the multijet background containing fake leptons. As this particular background contamination is lower in the \mujets\ channel compared to the \ejets\ channel, these cuts were chosen less stringent.

\subsection{Yields after Selection}
The number of events after the event selection and after the application of scale factors 
for the signal and background MC, scaled to the integrated 
luminosity of the data, is shown in Table~\ref{tab:cutflow}.  The number of observed events in the data is also shown. The uncertainties come from the uncertainties on the cross sections for the MC driven backgrounds and by the variation of the real and fake efficiencies according to their uncertainties for the fake lepton background. The numbers of selected events before the application of the \btag ging cut are listed in Table \ref{tab:cutflow_pretag} in the Appendix.
\begin{table}[htbp]
\begin{center}
\begin{tabular}{
 |   l  |
    S[table-format=5.1]@{\,\( \pm \)\,}
    S[table-format=5.1] |
    S[table-format=-5.1]@{\,\( \pm \)\,}
    S[table-format=5.1]|
    } 
\hline
$n_{\text{jets}} \geq 4$, $n_{\text{b-tags}} \geq 1$ & \multicolumn{2}{c|}{\ejets} & \multicolumn{2}{c|}{\mujets} \\
\hline
\hline
$W+$jets (DD/MC) & 2320 & 390 & 4840 & 770\\
$Z+$jets (MC) & 450 & 210 & 480 & 230\\
Misident. leptons (DD) & 840 & 420 & 1830 & 340\\
Single top (MC) & 1190 & 60 & 1980 & 80\\
Diboson (MC) & 46 & 2& 73 & 4\\
\hline
Total (non-\ttbar) & 4830 & 620& 9200 & 890\\
\hline
\ttbar\ (MC, l+jets) & 15130 & 900 & 25200 & 1500\\

\ttbar\ (MC, dilepton) & 2090 & 120 & 3130 & 190\\
\hline
Expected & 22100 & 1100 & 37500 & 1800\\
Observed & \multicolumn{2}{c|}{21770} & \multicolumn{2}{c|}{37650}\\
\hline
\end{tabular}
\end{center}
\caption{Number of selected data events and background composition for $n_{\text{jets}} \geq 4$. The given uncertainties are statistical for data driven backgrounds and given by the uncertainties on the cross sections for the Monte Carlo based backgrounds.}
\label{tab:cutflow}
\end{table}

The yield agreement as a function of the run period is shown in Figure \ref{fig:periods} for a selection with at least four jets.  
\begin{figure}[htbp]
\begin{center}
\subfigure[]{
\includegraphics[width=0.4\textwidth]{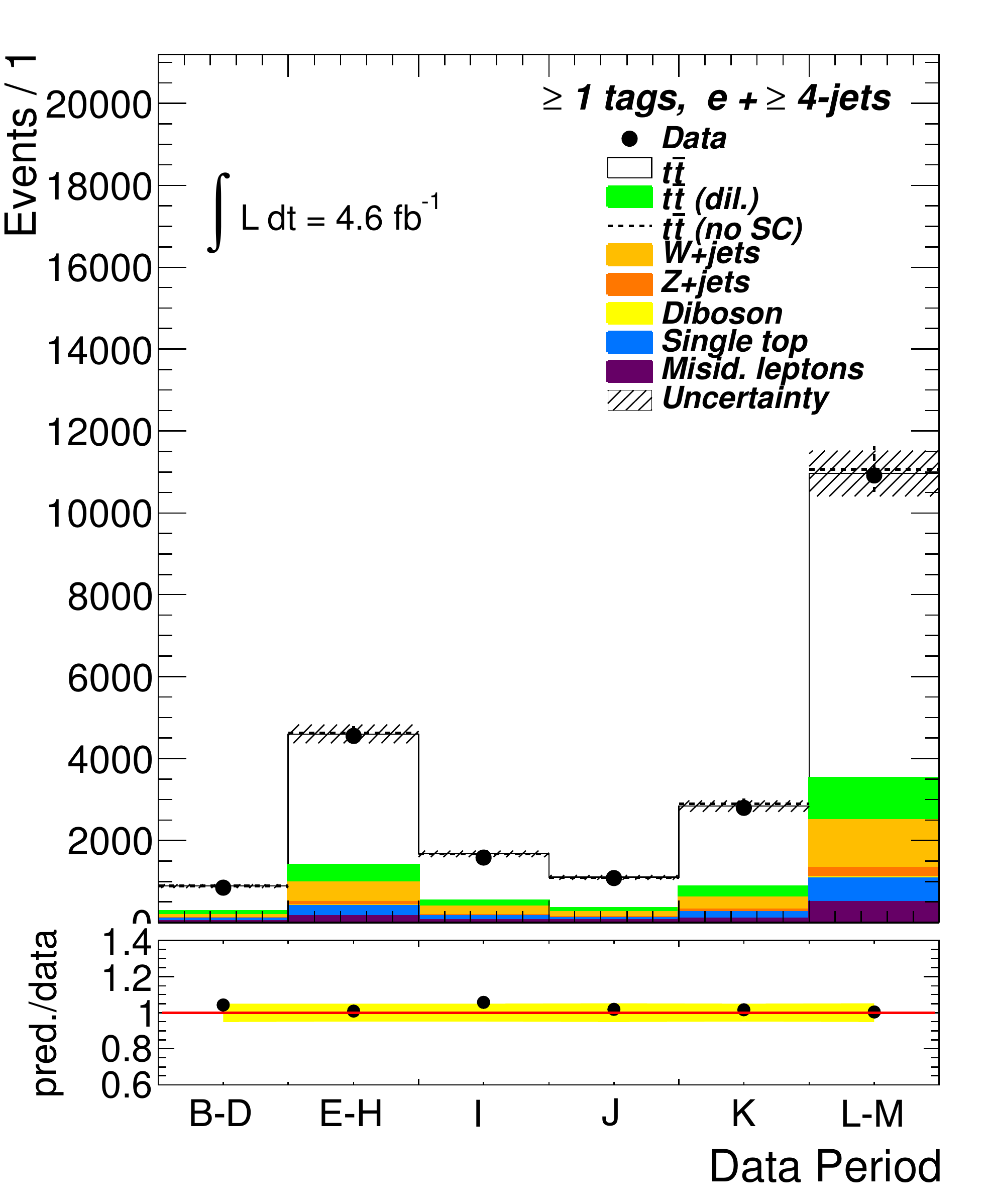}
\label{fig:periods_el}
}
\subfigure[]{
\includegraphics[width=0.4\textwidth]{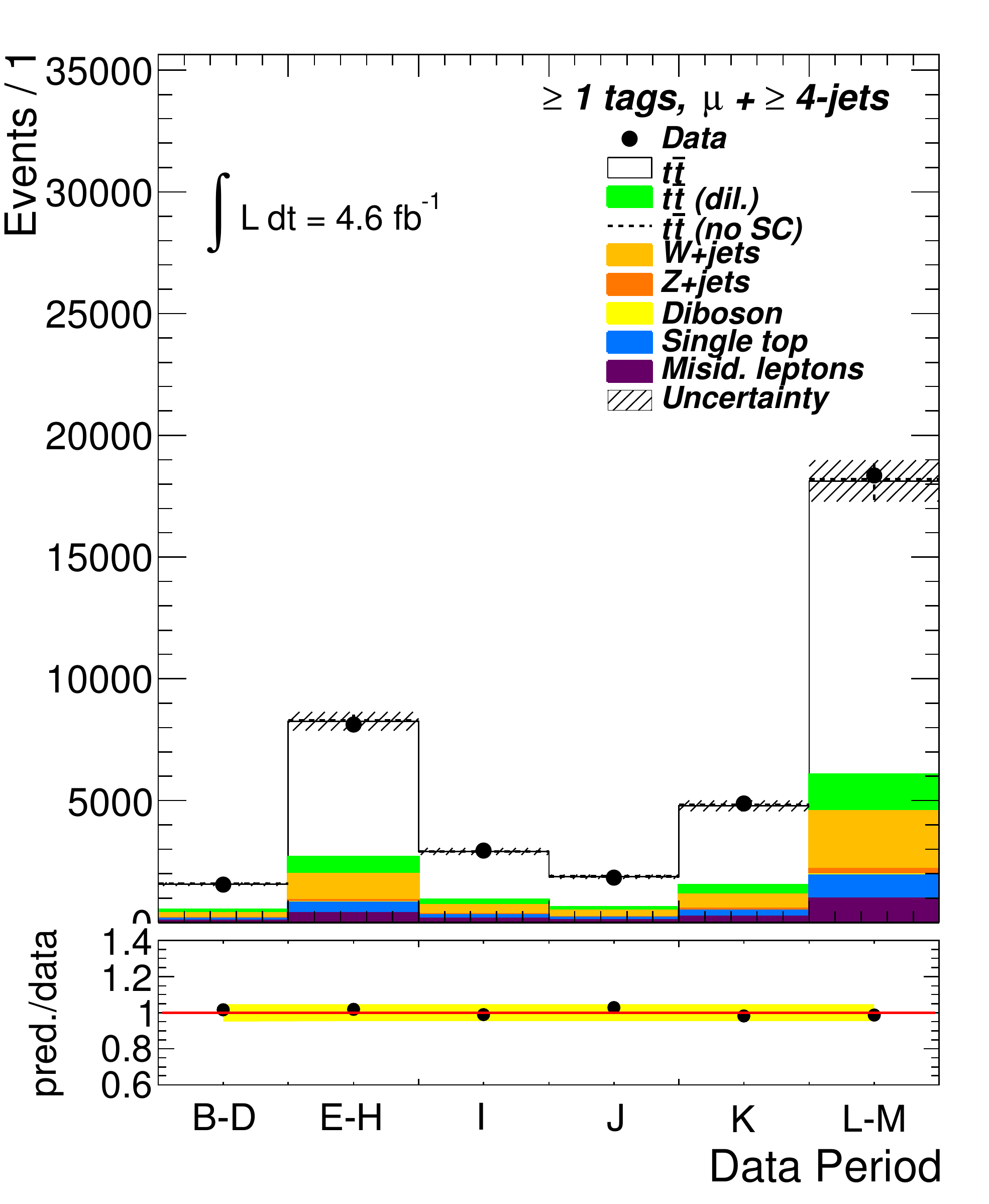}
\label{fig:periods_mu}
}
\end{center}
\caption{Yield for data (points) and the different Monte
  Carlo contributions (filled histograms) split into the different run
  periods for \subref{fig:periods_el} the \ejets\ channel and \subref{fig:periods_mu} the \mujets\ channel. The default selection of at least four jets with at least one \btag ged jet was used. 
}
\label{fig:periods}
\end{figure}

\section{Data/MC Agreement}
In this section, the agreement between data and prediction is shown. They are in good agreement. The uncertainties on the prediction (propagated into the yellow uncertainty band in the ratio) are given by the uncertainties on the calculated cross sections for the MC driven backgrounds and the normalization uncertainty on the lepton fake background. These uncertainties are treated as uncorrelated.

All of the following plots require the default selection to be passed. The inclusive $n_{\text{jets}} \geq 4$ selection is shown. Control distributions before the application of the \btag\ requirement as well as a split into subsamples of $n_{\text{jets}} = 4$, $n_{\text{jets}} \geq 5$, $n_{\text{b-tags}} = 1$ and $n_{\text{b-tags}} \geq2$ can be found in Appendix \ref{sec:app_CPs}.
\begin{figure}[htbp]
\begin{center}
\includegraphics[width=0.3\textwidth]{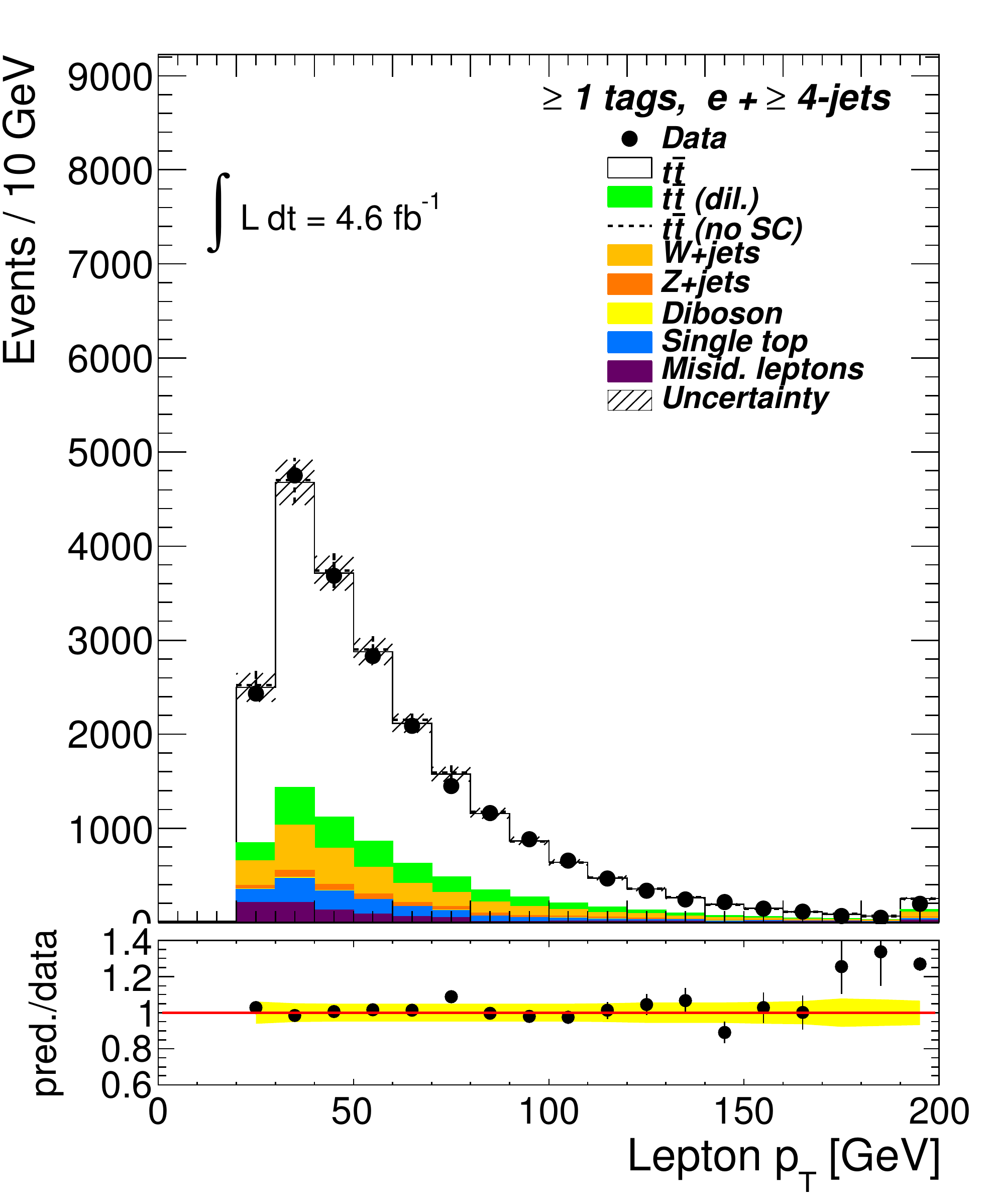}
\includegraphics[width=0.3\textwidth]{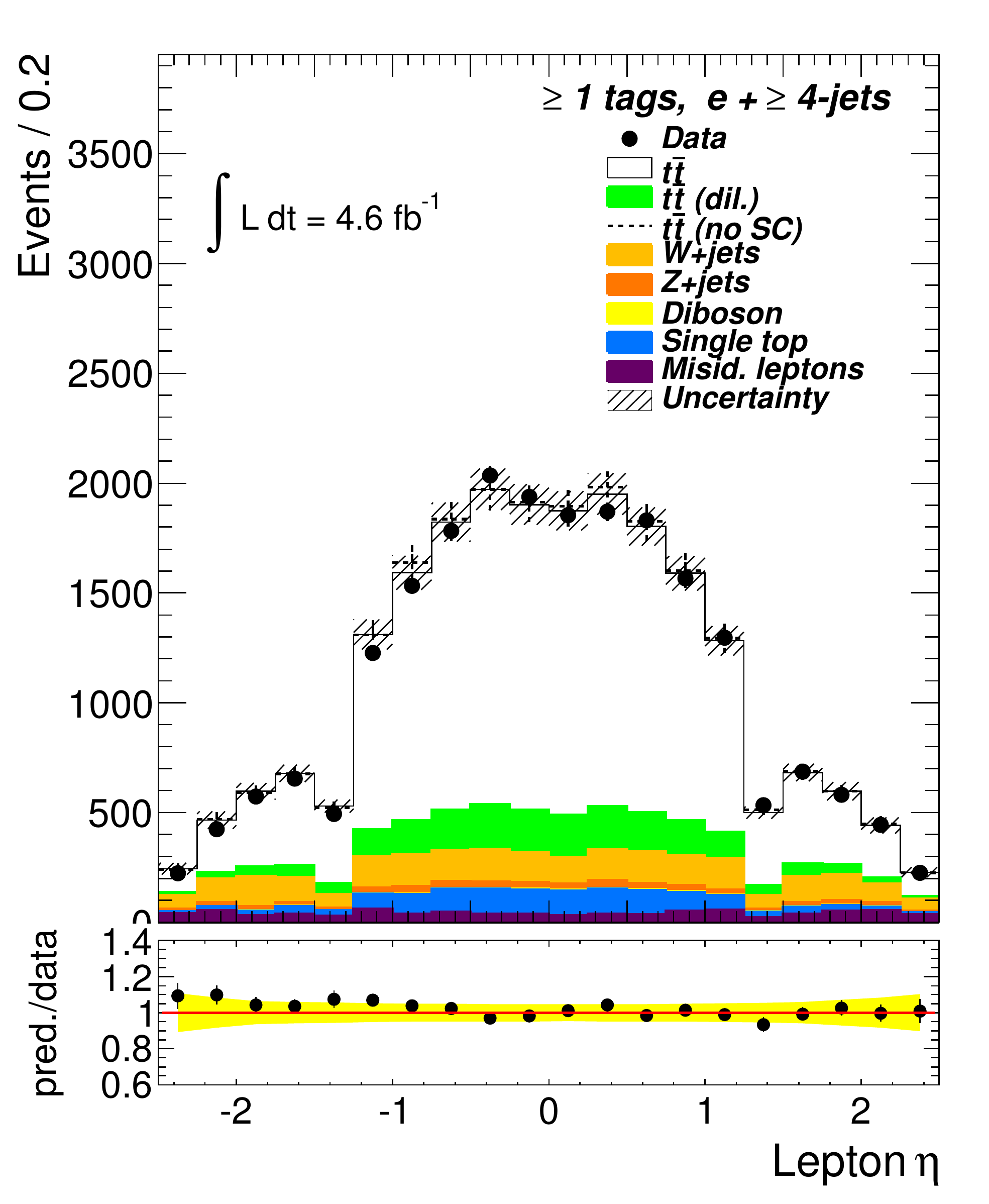}
\includegraphics[width=0.3\textwidth]{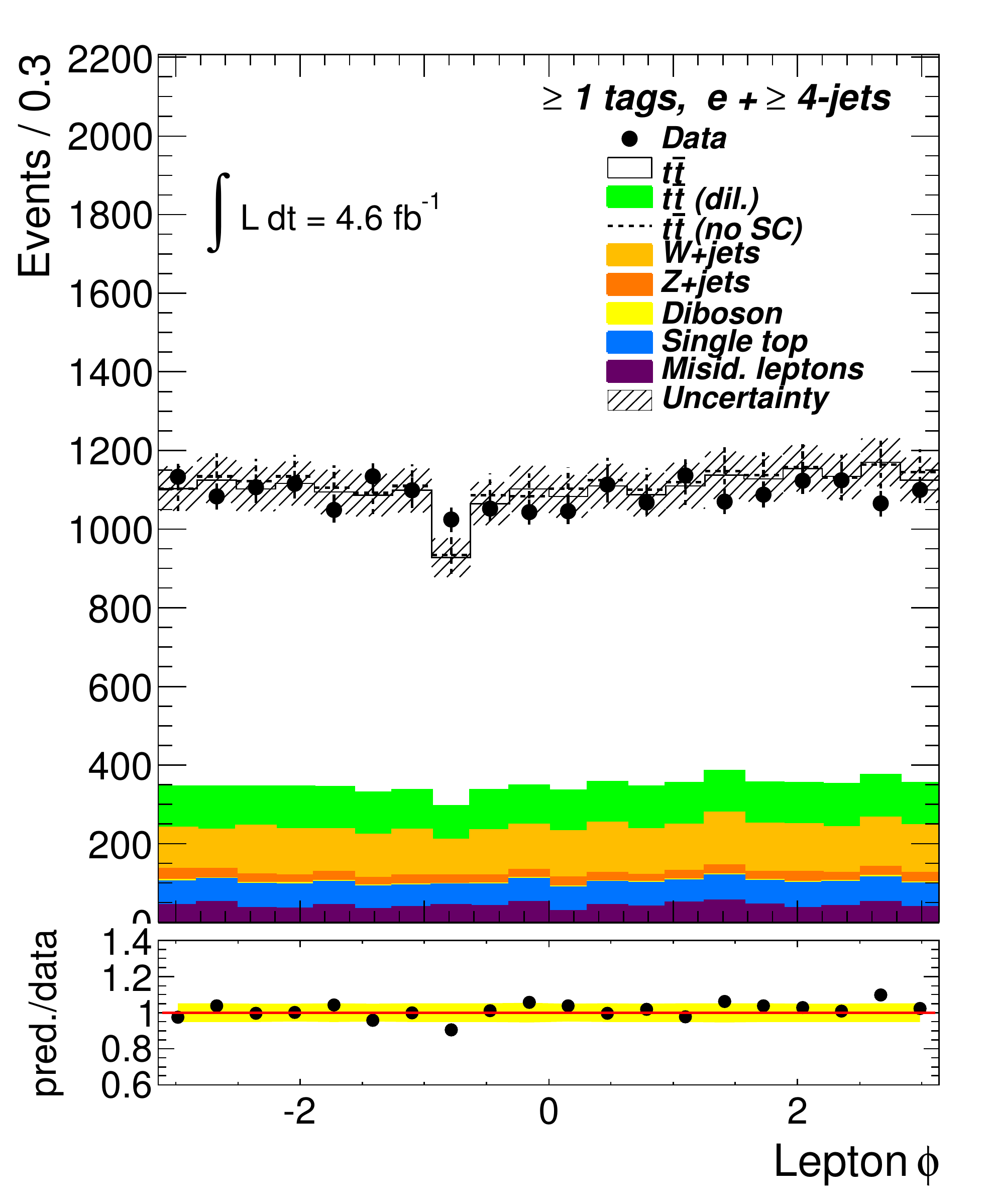}
\includegraphics[width=0.3\textwidth]{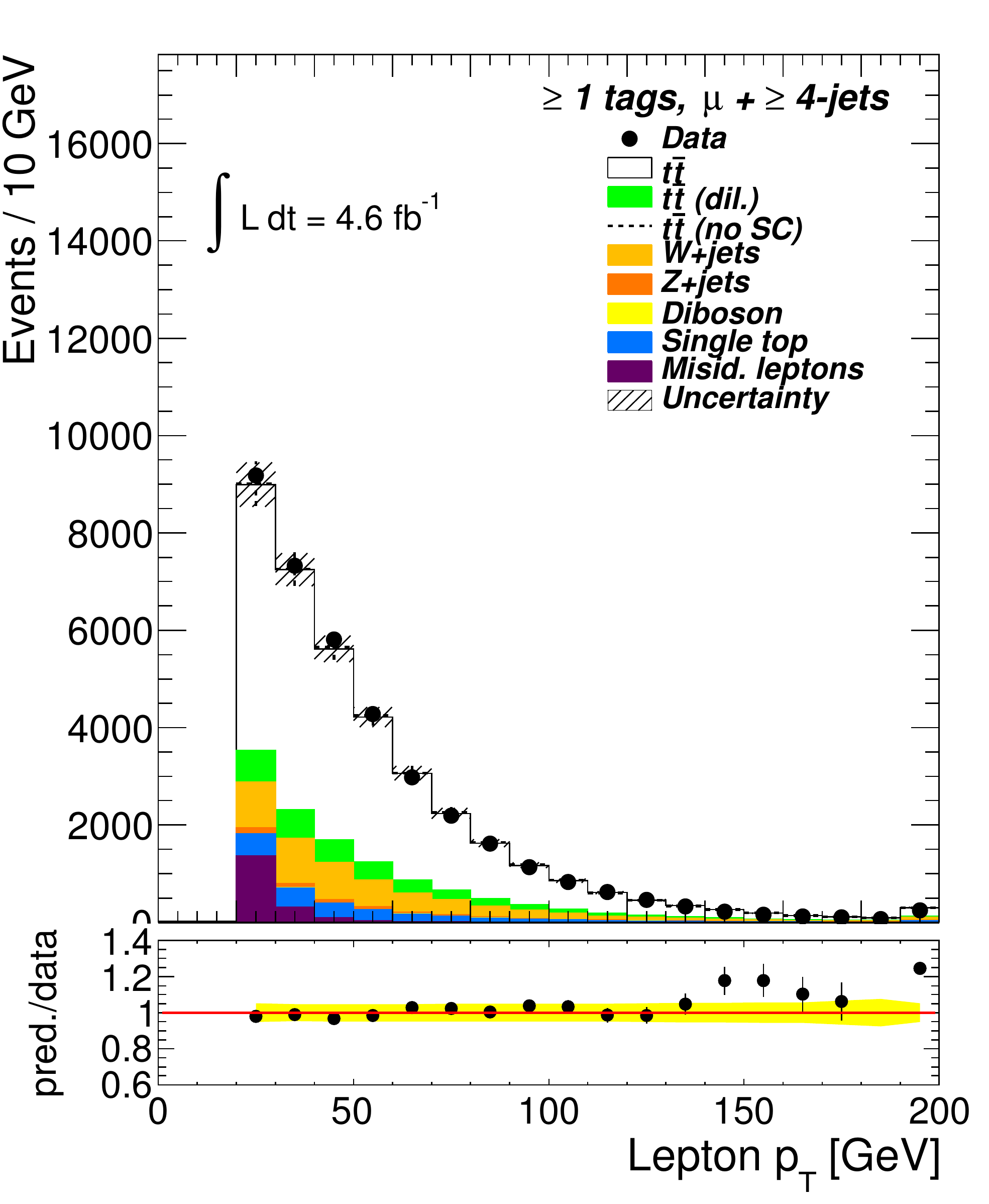}
\includegraphics[width=0.3\textwidth]{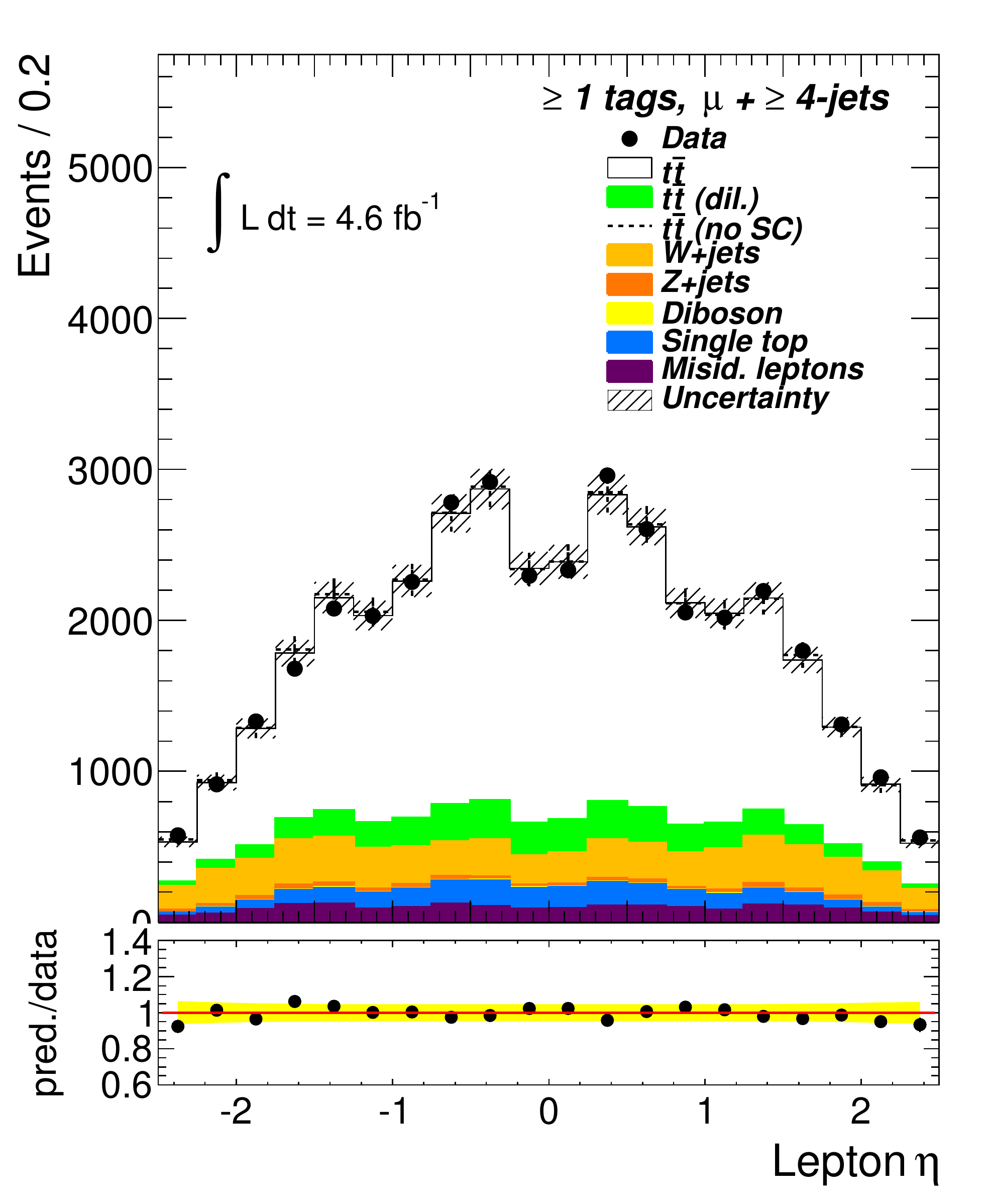}
\includegraphics[width=0.3\textwidth]{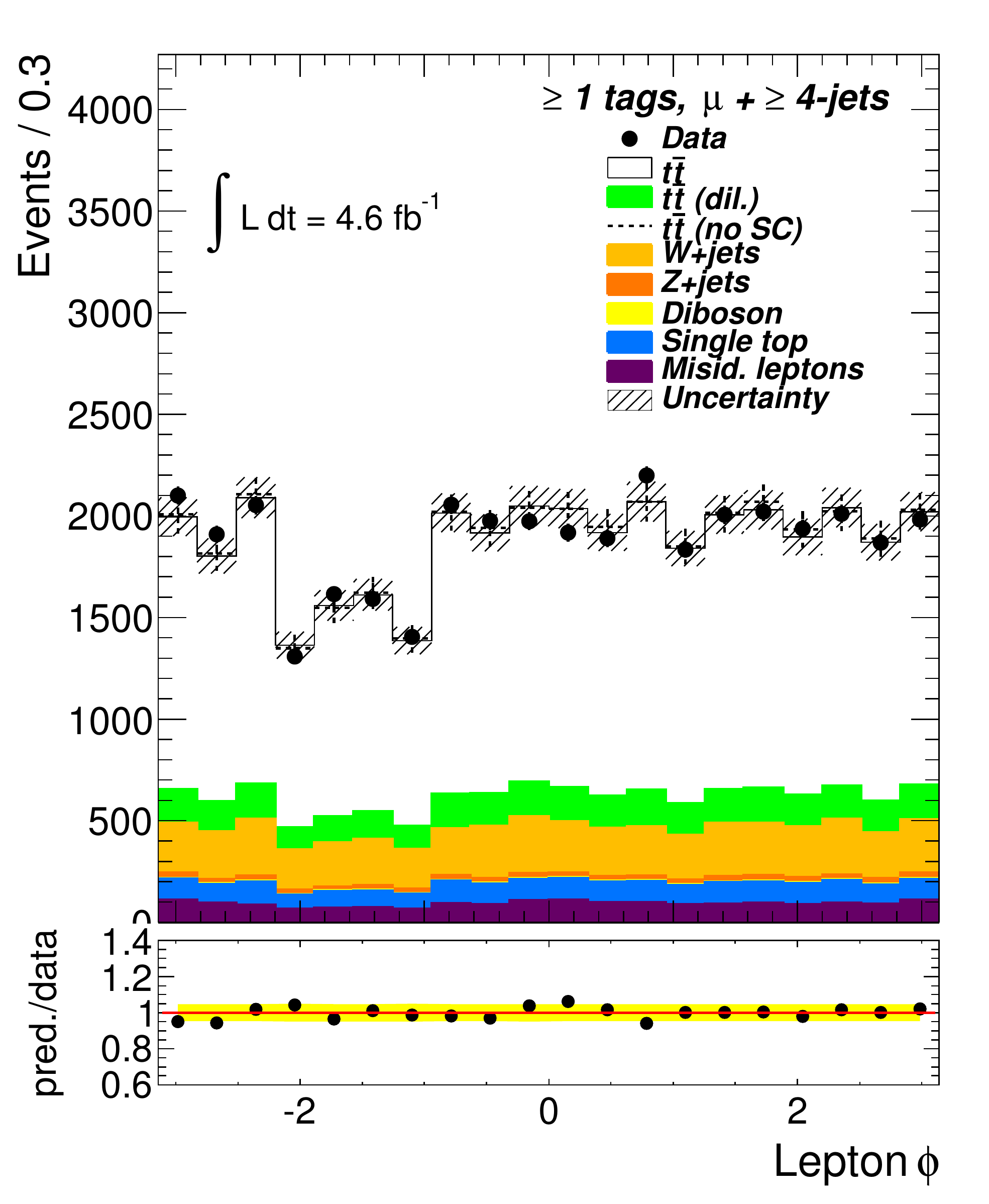}
\end{center}
\caption{Control distributions for the lepton \pt, $\eta$
  and $\phi$ distribution of the \ejets\ (top) and \mujets\ (bottom) channel ($n_{\text{jets}} \geq 4, n_{\text{b-tags}} \geq 1$). 
}
\label{fig:lepton_4incl_1tags}
\end{figure}

\begin{figure}[htbp]
\begin{center}
\includegraphics[width=0.3\textwidth]{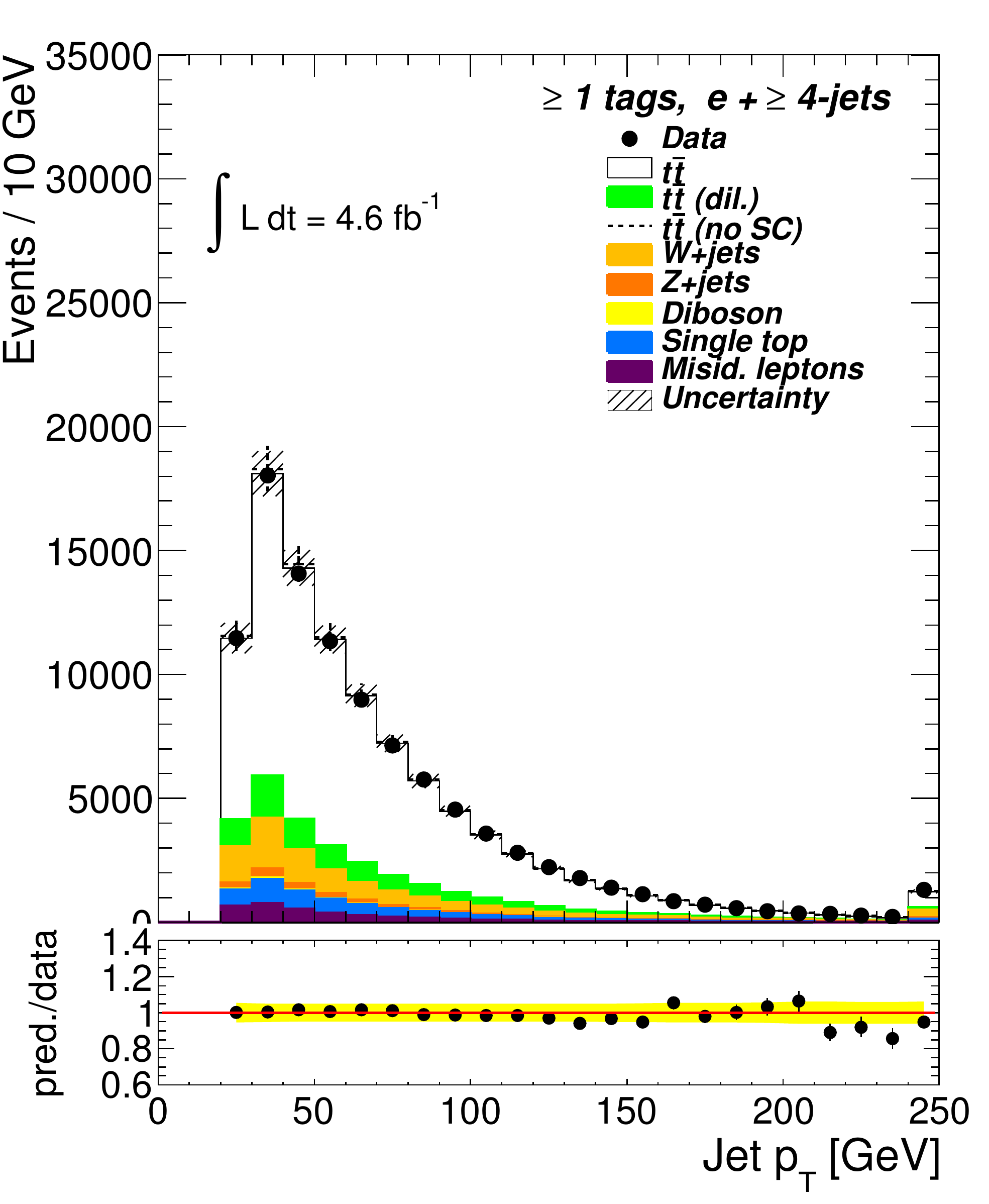}
\includegraphics[width=0.3\textwidth]{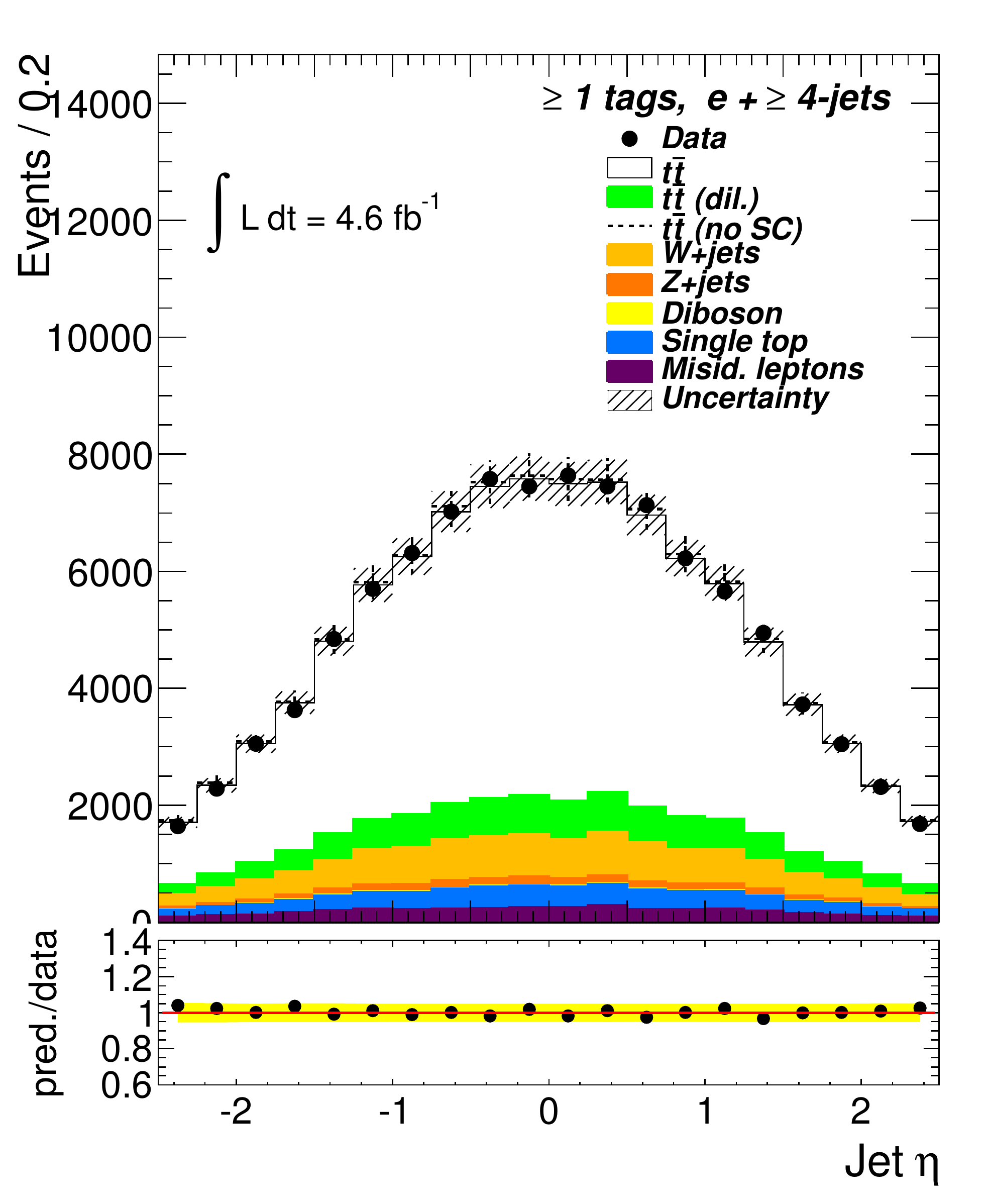}
\includegraphics[width=0.3\textwidth]{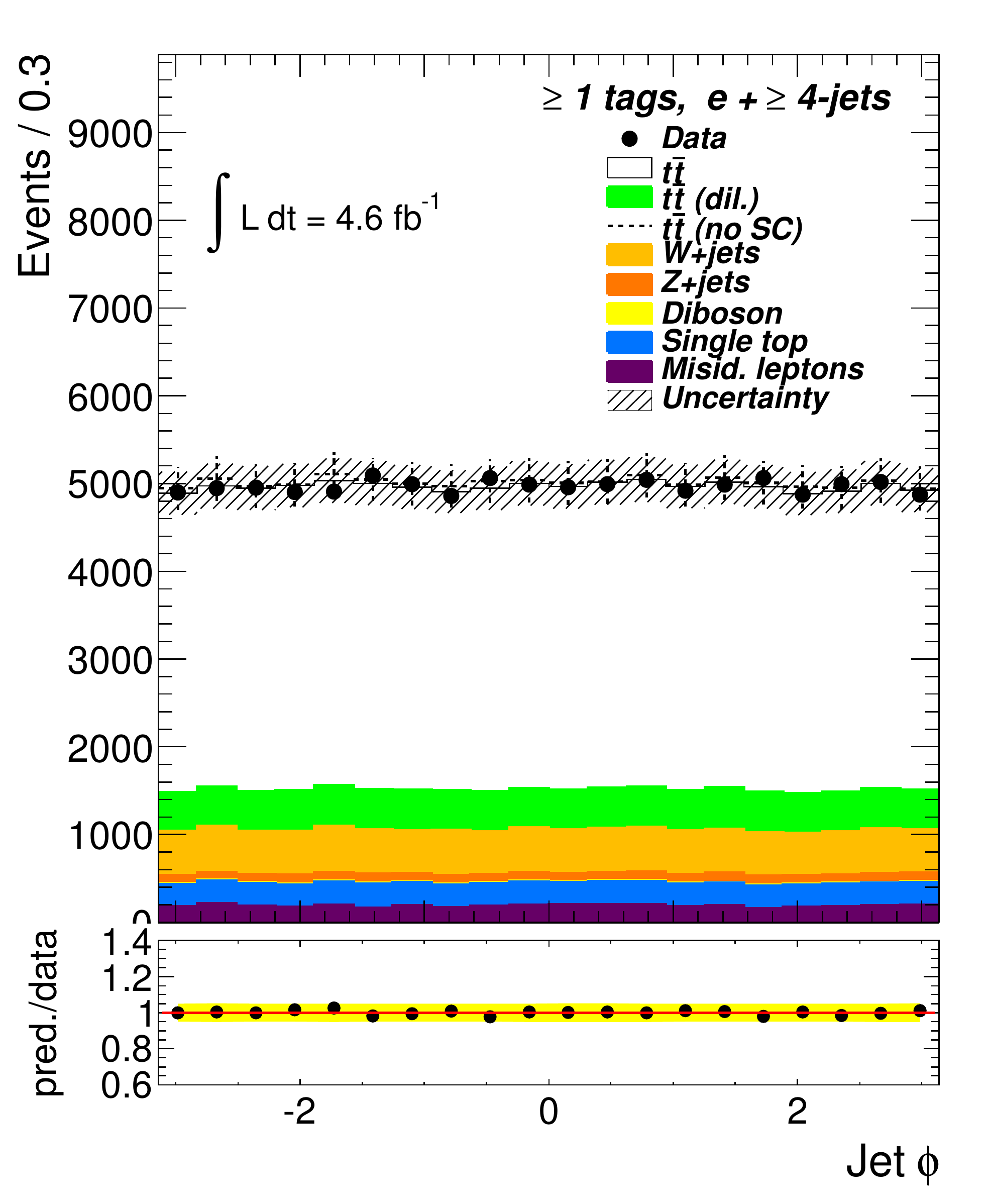}
\includegraphics[width=0.3\textwidth]{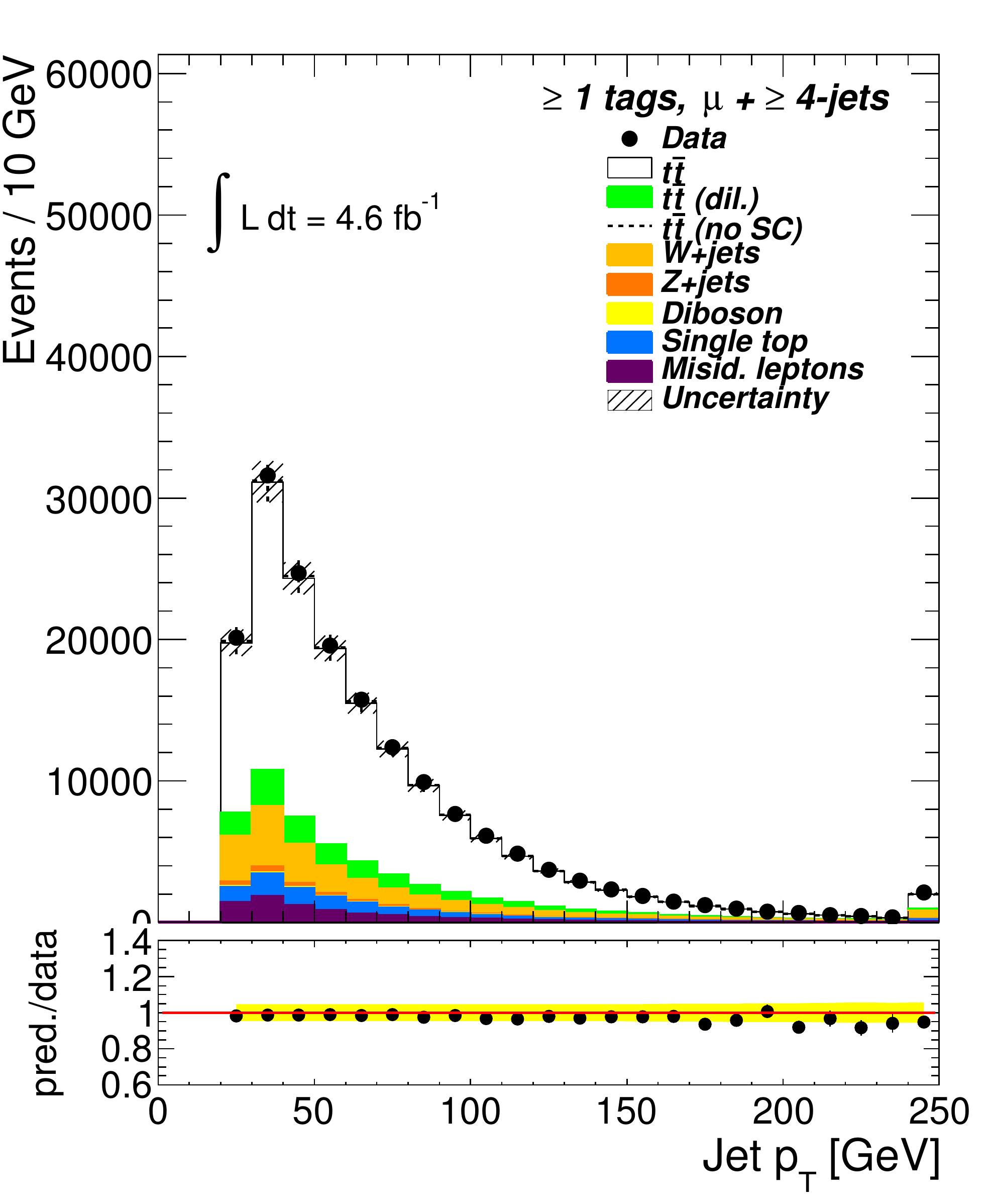}
\includegraphics[width=0.3\textwidth]{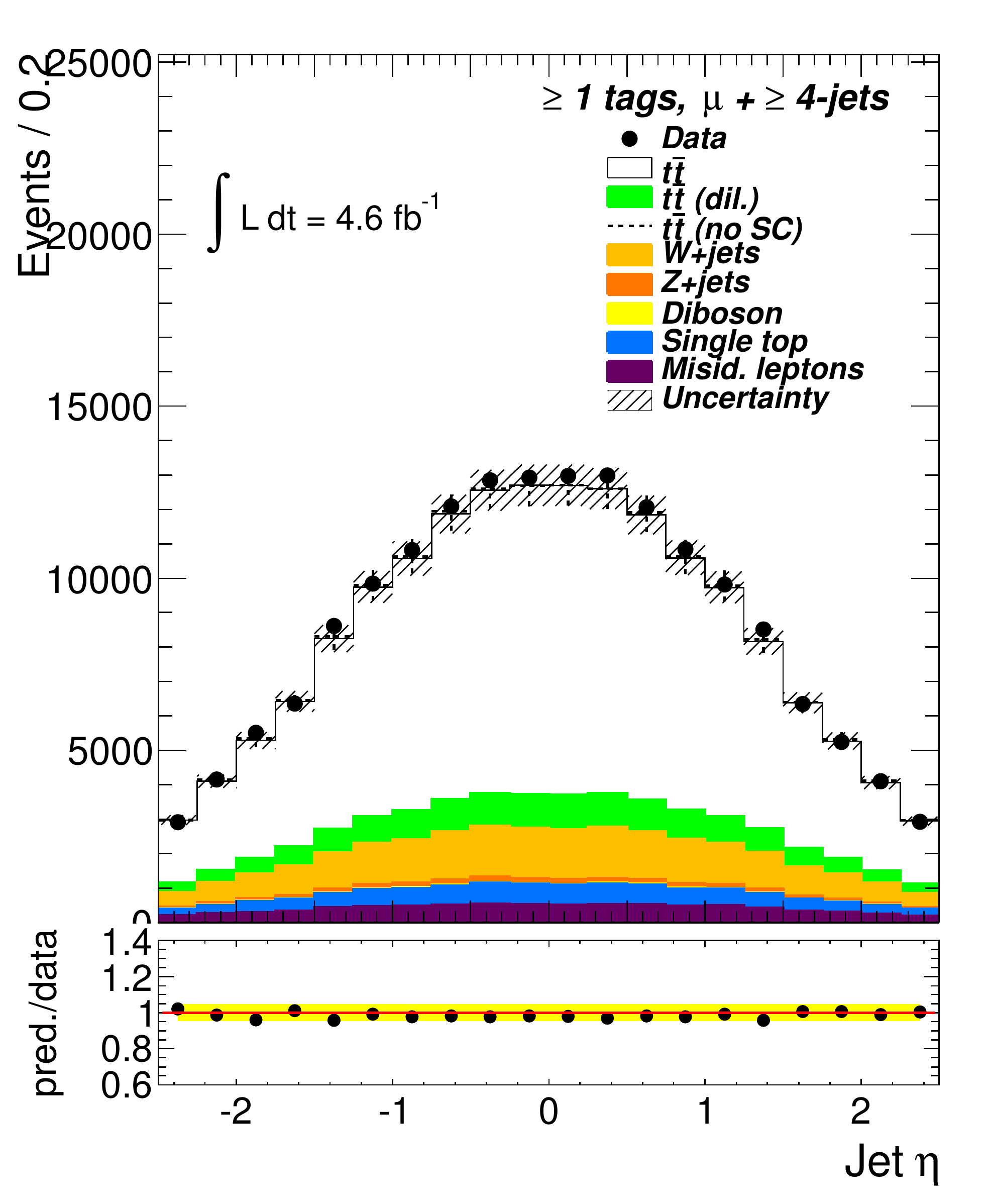}
\includegraphics[width=0.3\textwidth]{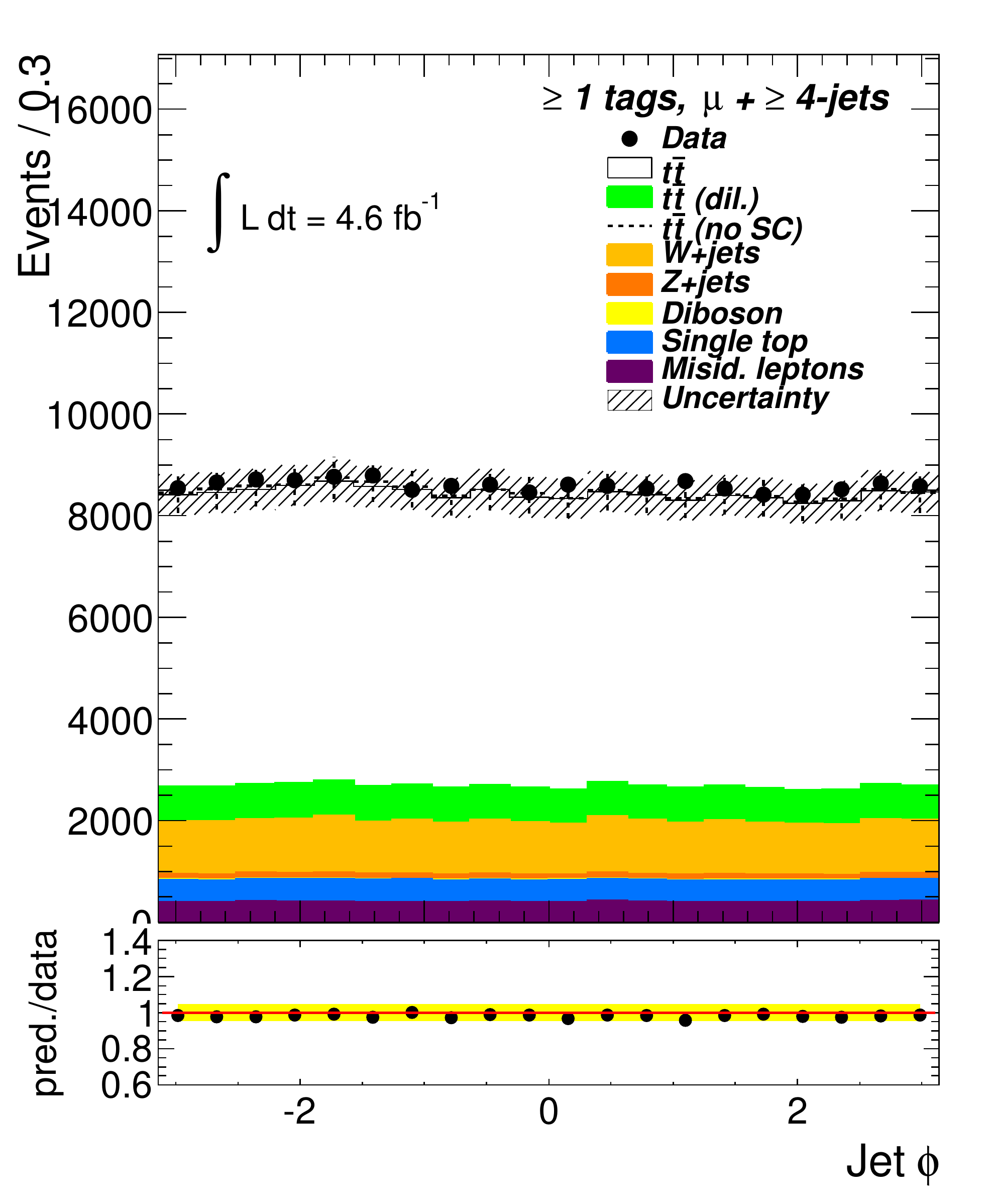}
\end{center}
\caption{Control distributions for the jet \pt, $\eta$
  and $\phi$ distribution of the \ejets\ (top) and \mujets\ (bottom) channel (all selected jets, $n_{\text{jets}} \geq 4, n_{\text{b-tags}} \geq 1$). 
}
\label{fig:jet_4incl_1tags}
\end{figure}

\begin{figure}[htbp]
\begin{center}
\includegraphics[width=0.3\textwidth]{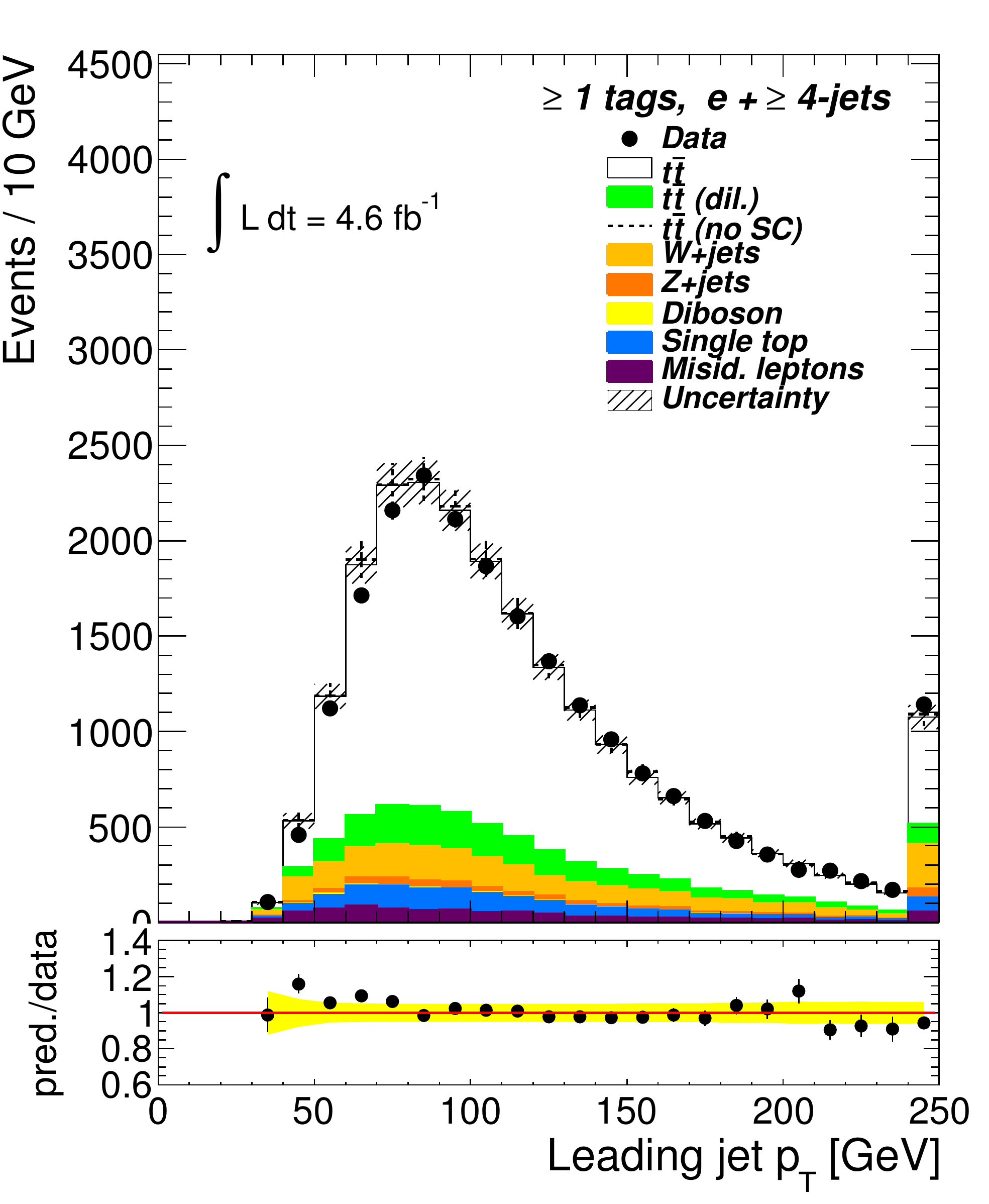}
\includegraphics[width=0.3\textwidth]{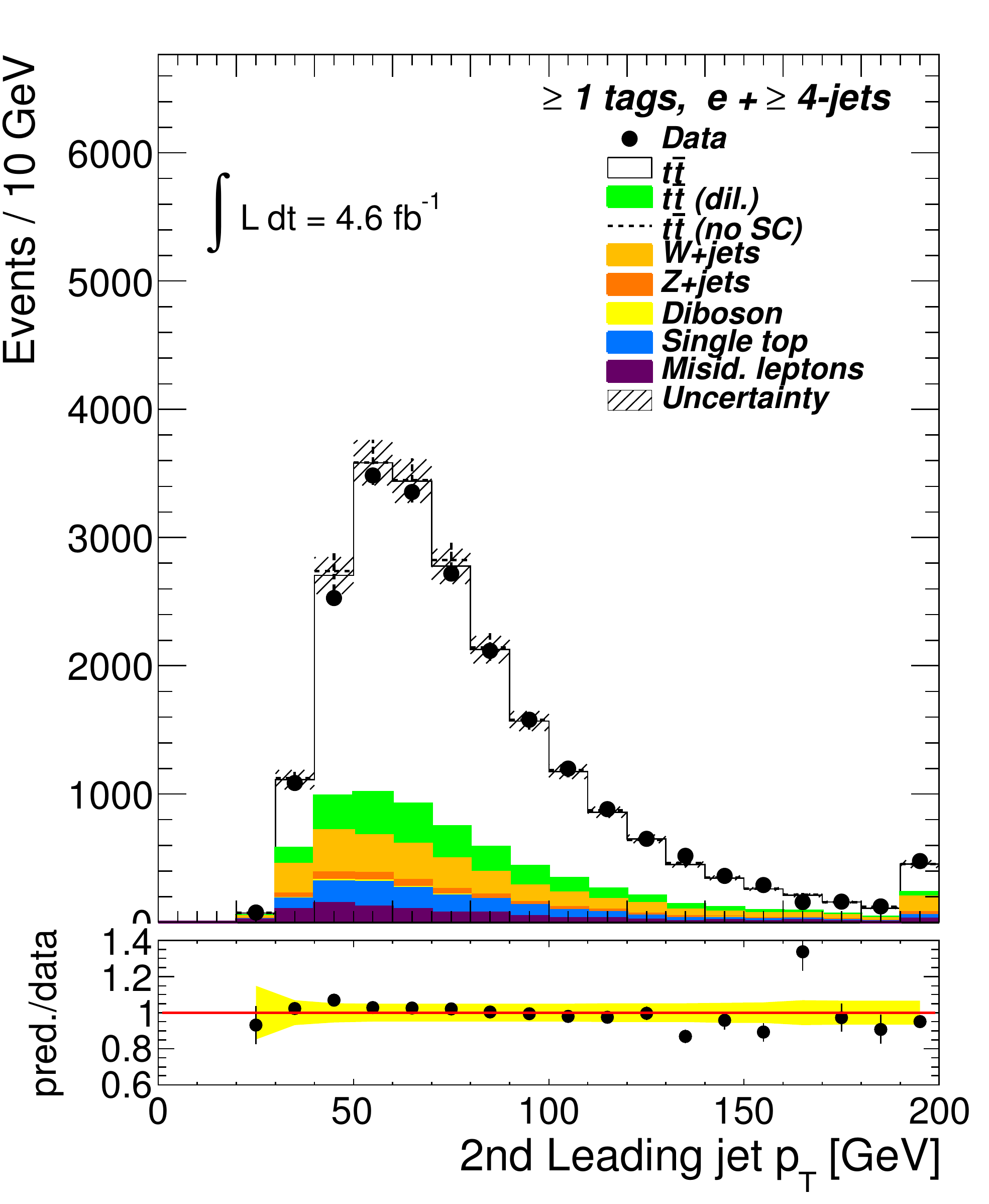}
\includegraphics[width=0.3\textwidth]{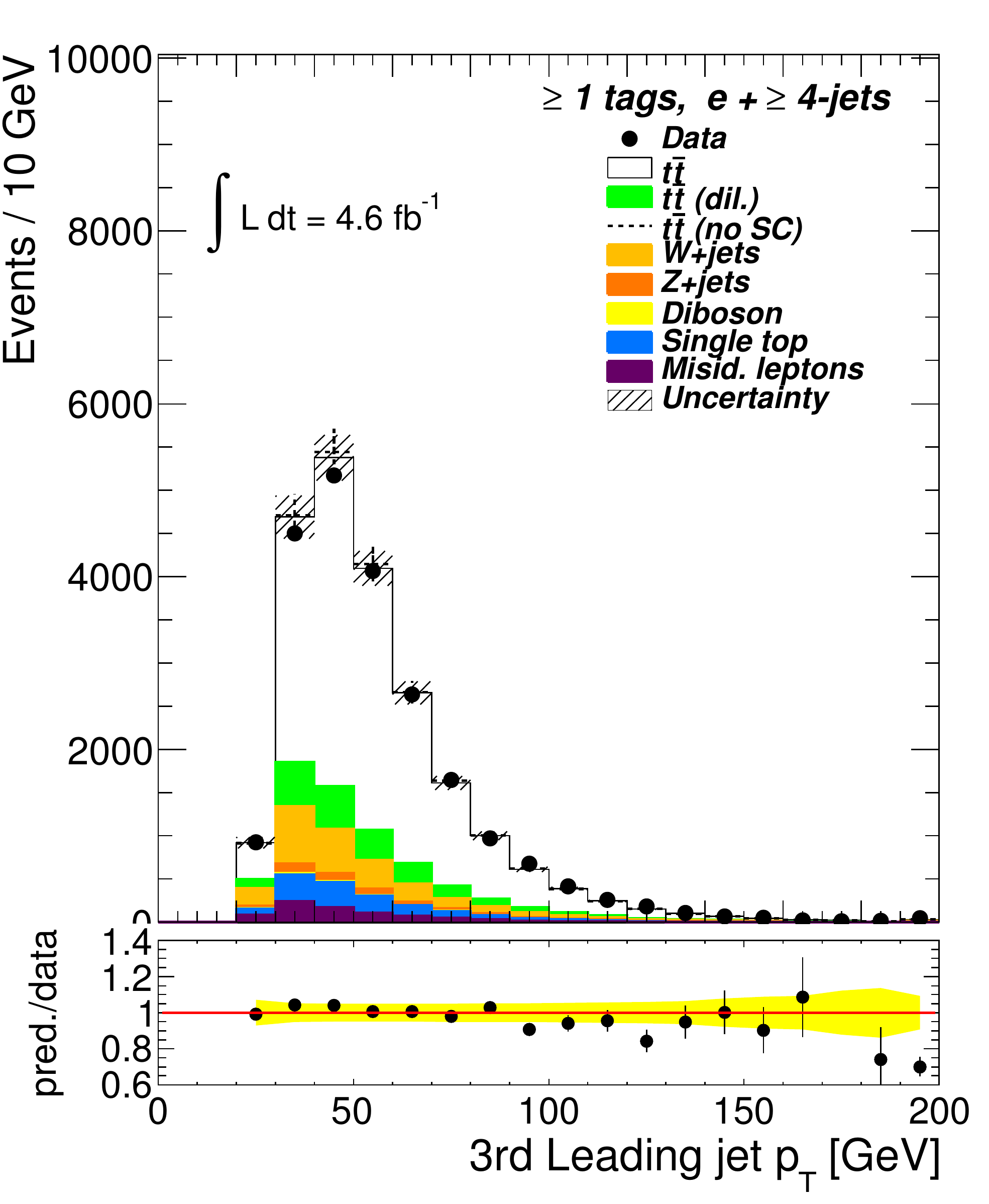}
\includegraphics[width=0.3\textwidth]{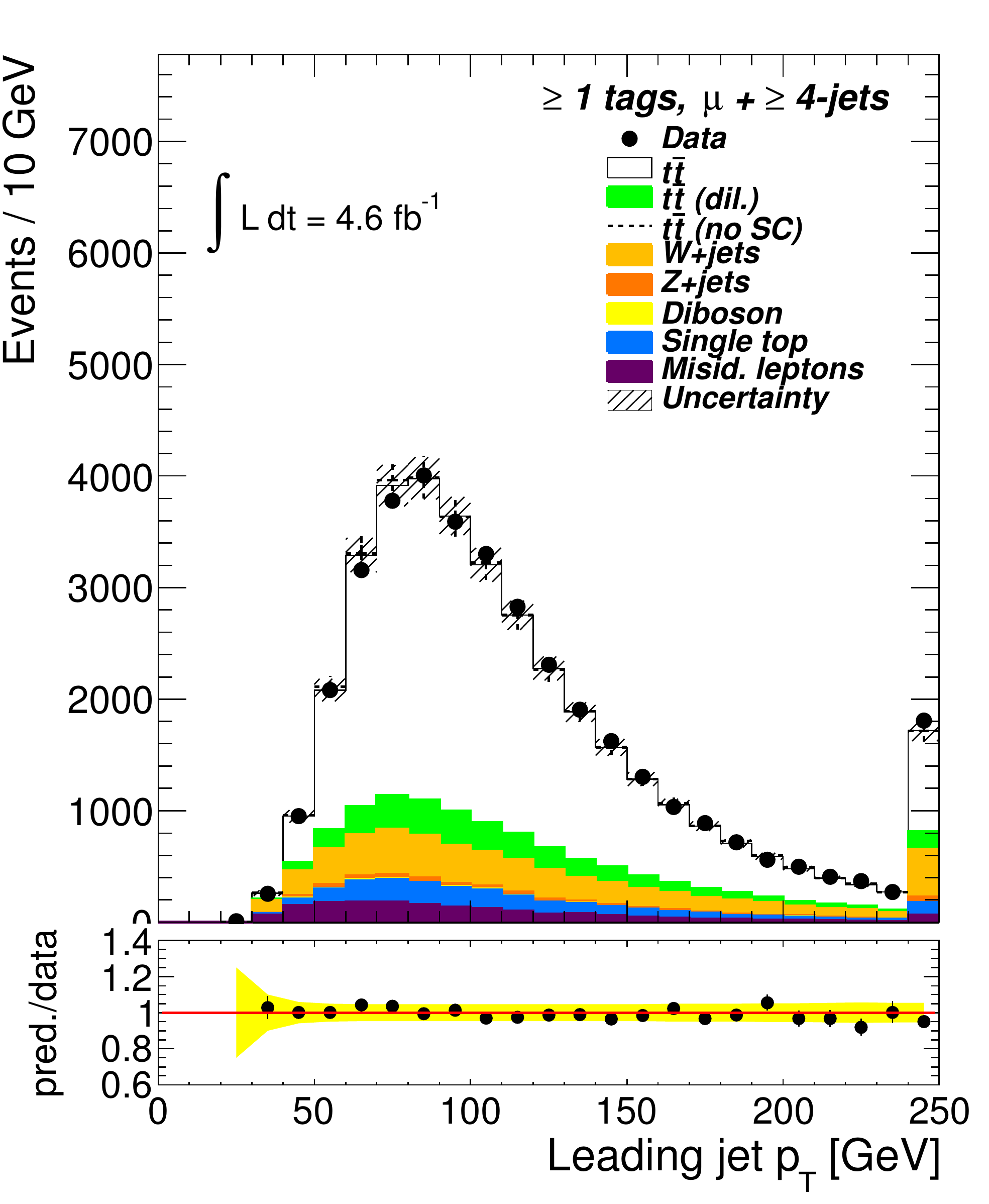}
\includegraphics[width=0.3\textwidth]{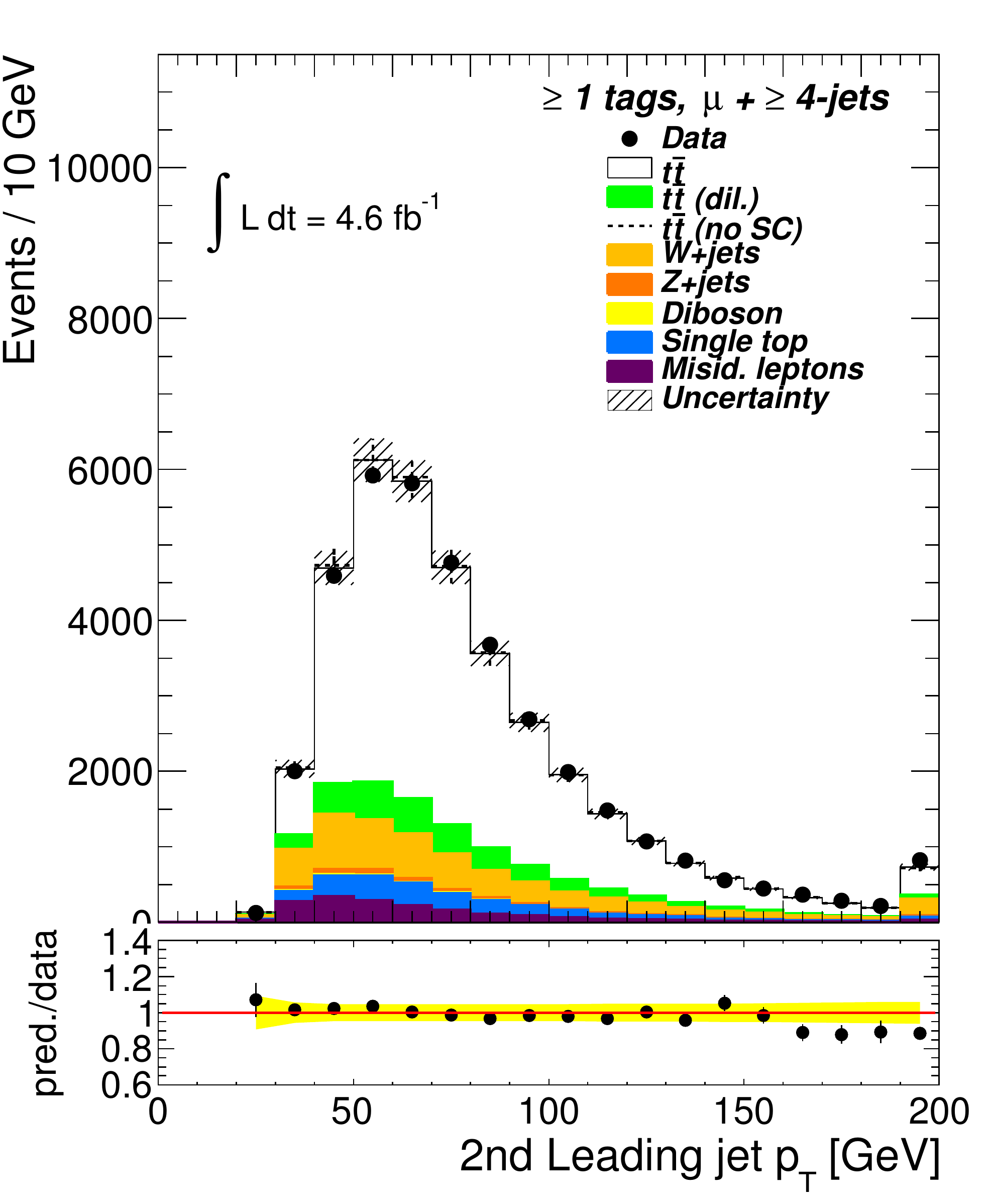}
\includegraphics[width=0.3\textwidth]{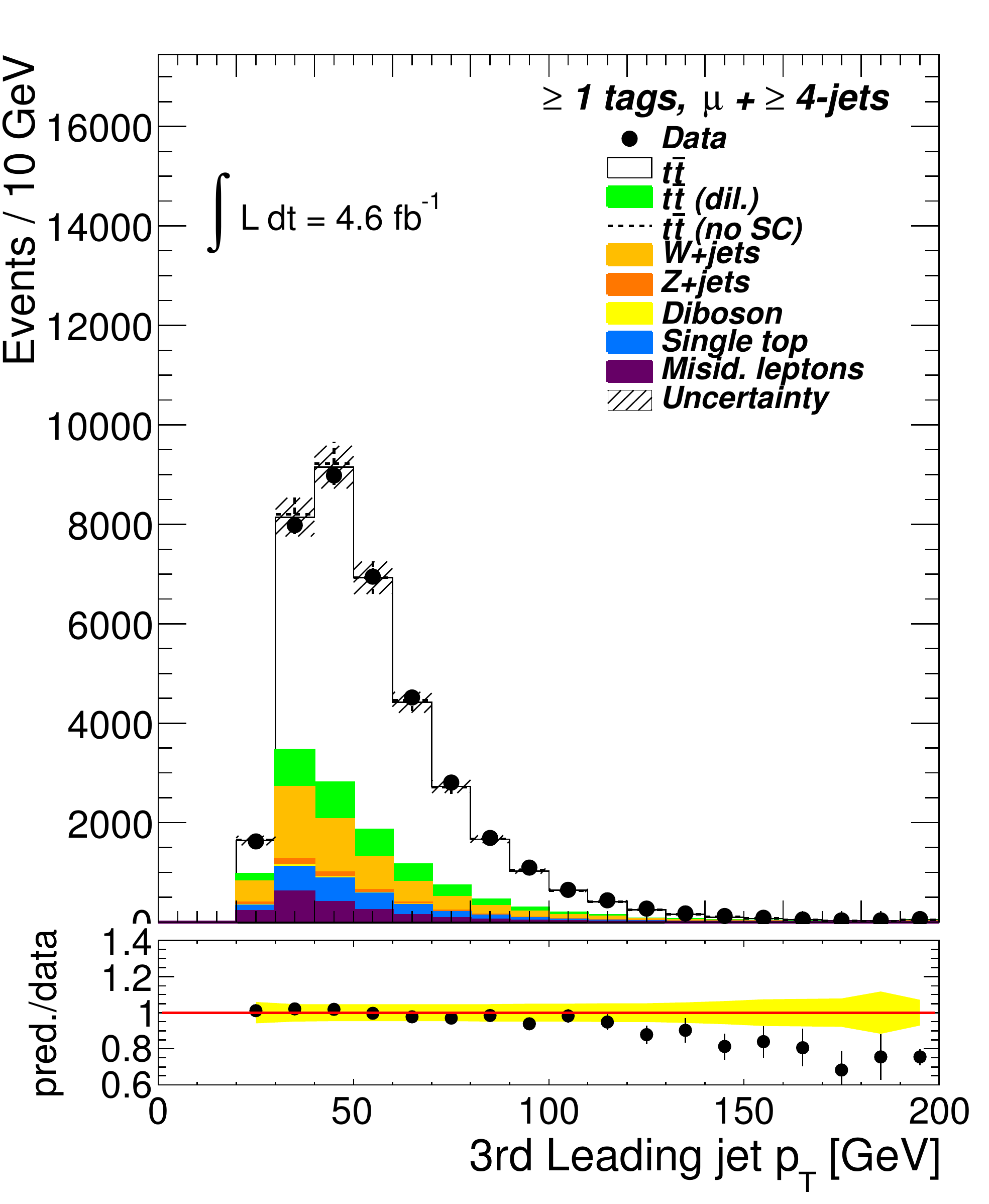}
\end{center}
\caption{Control distributions for the jet \pt\ of the three
  highest \pt\ jets of the \ejets\ (top) and \mujets\ (bottom) channel ($n_{\text{jets}} \geq 4, n_{\text{b-tags}} \geq 1$).
}
\label{fig:subjets_4incl_1tags}
\end{figure}

\begin{figure}[htbp]
\begin{center}
\includegraphics[width=0.3\textwidth]{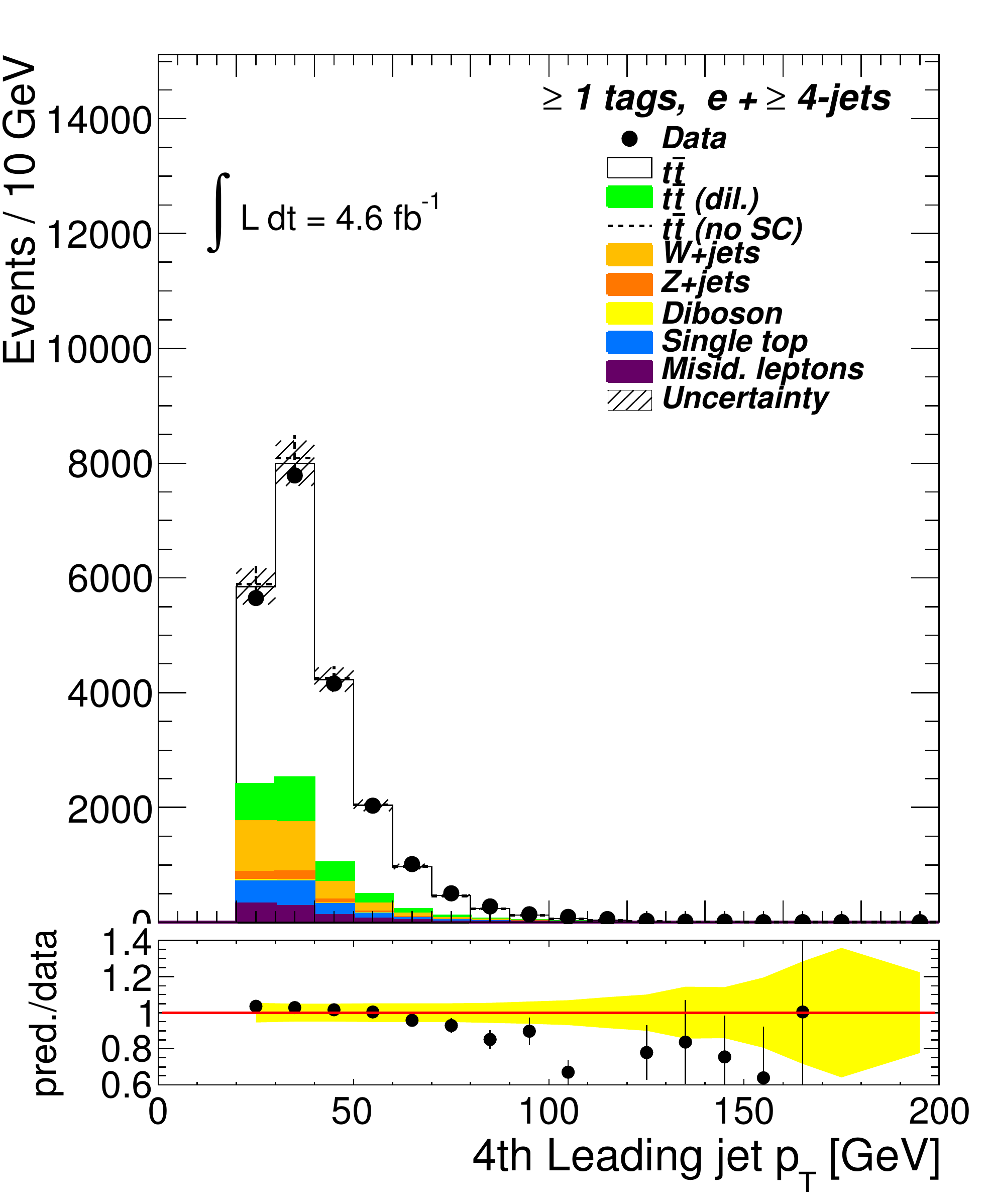}
\includegraphics[width=0.3\textwidth]{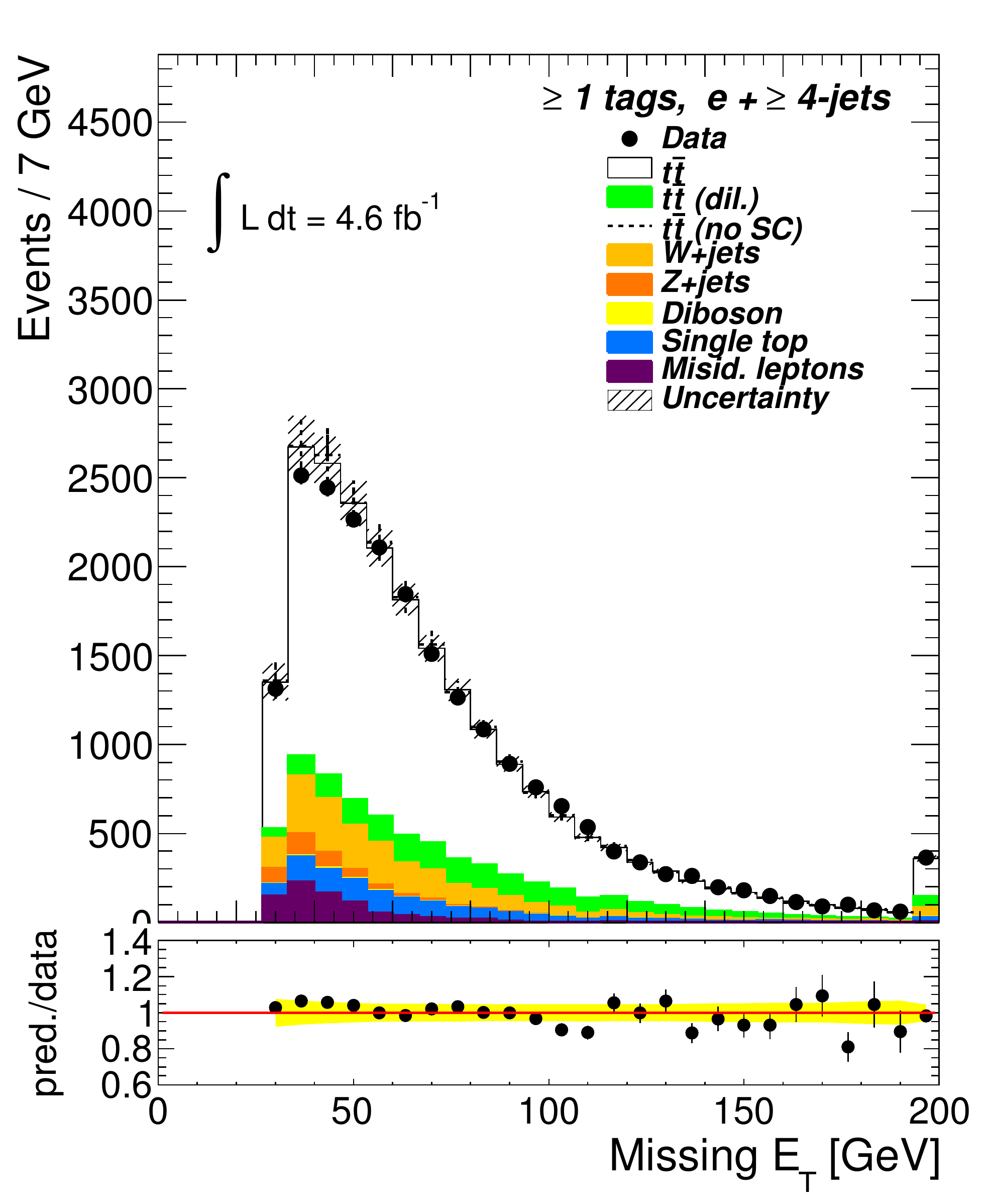}
\includegraphics[width=0.3\textwidth]{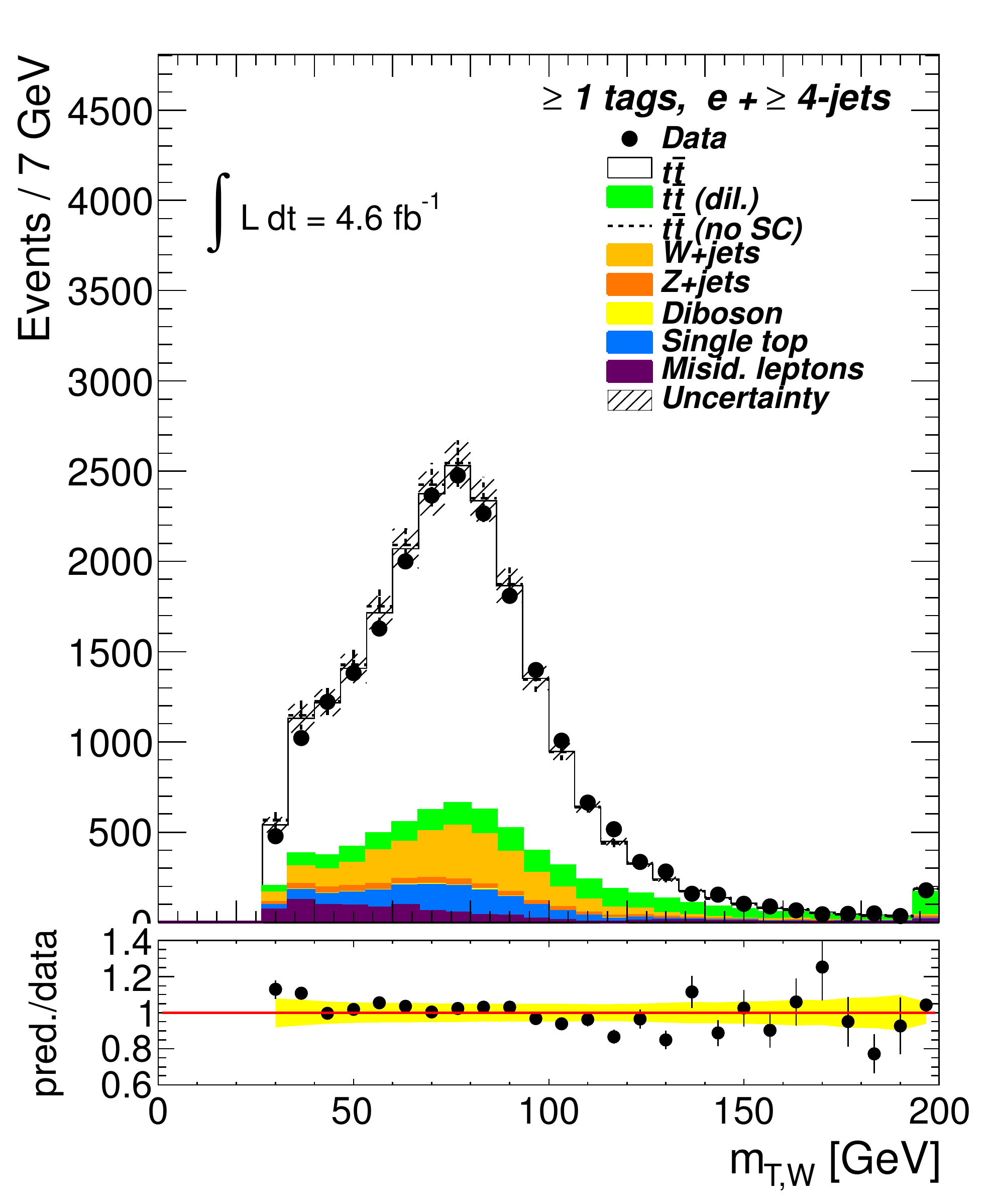}\\
\includegraphics[width=0.3\textwidth]{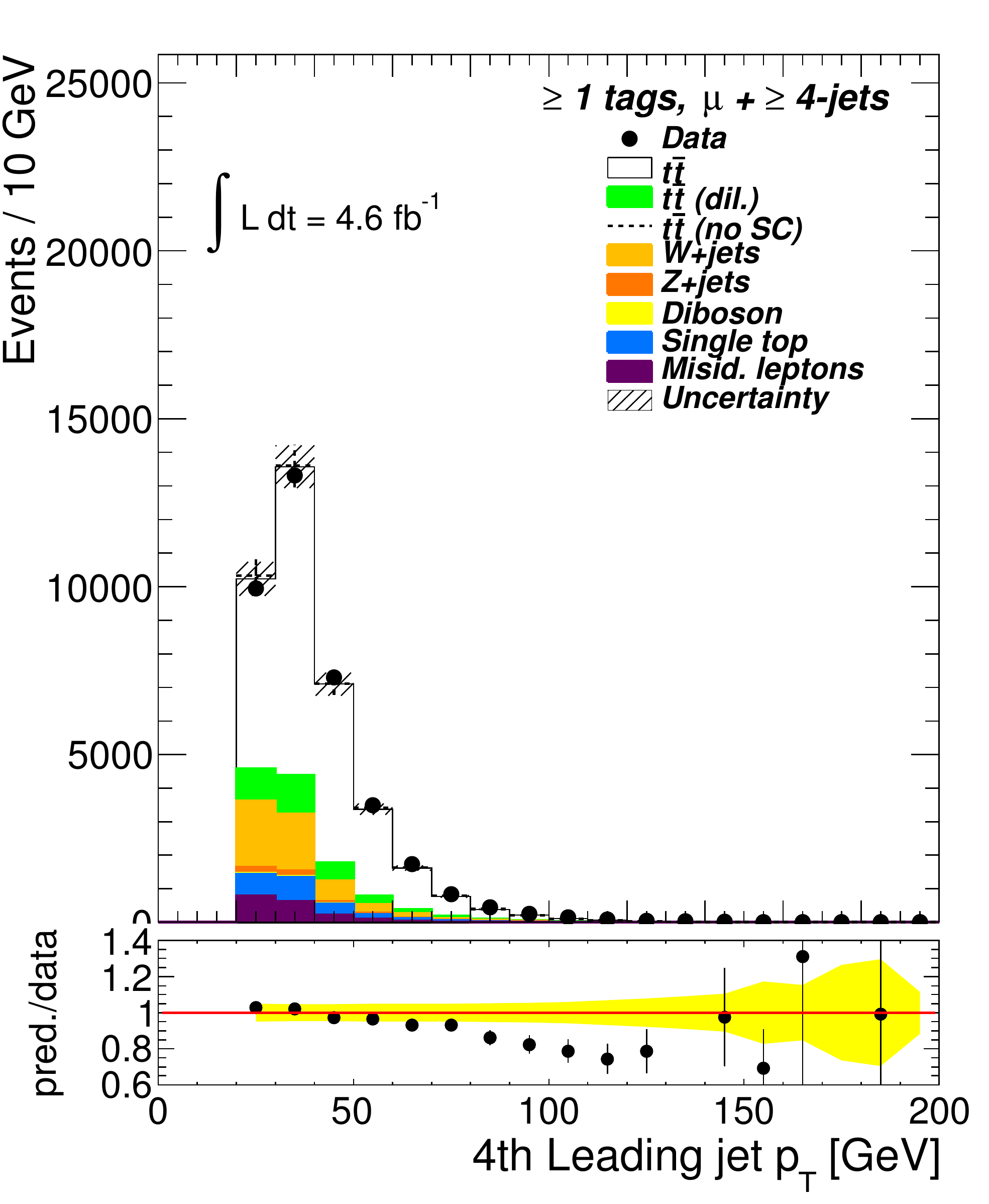}
\includegraphics[width=0.3\textwidth]{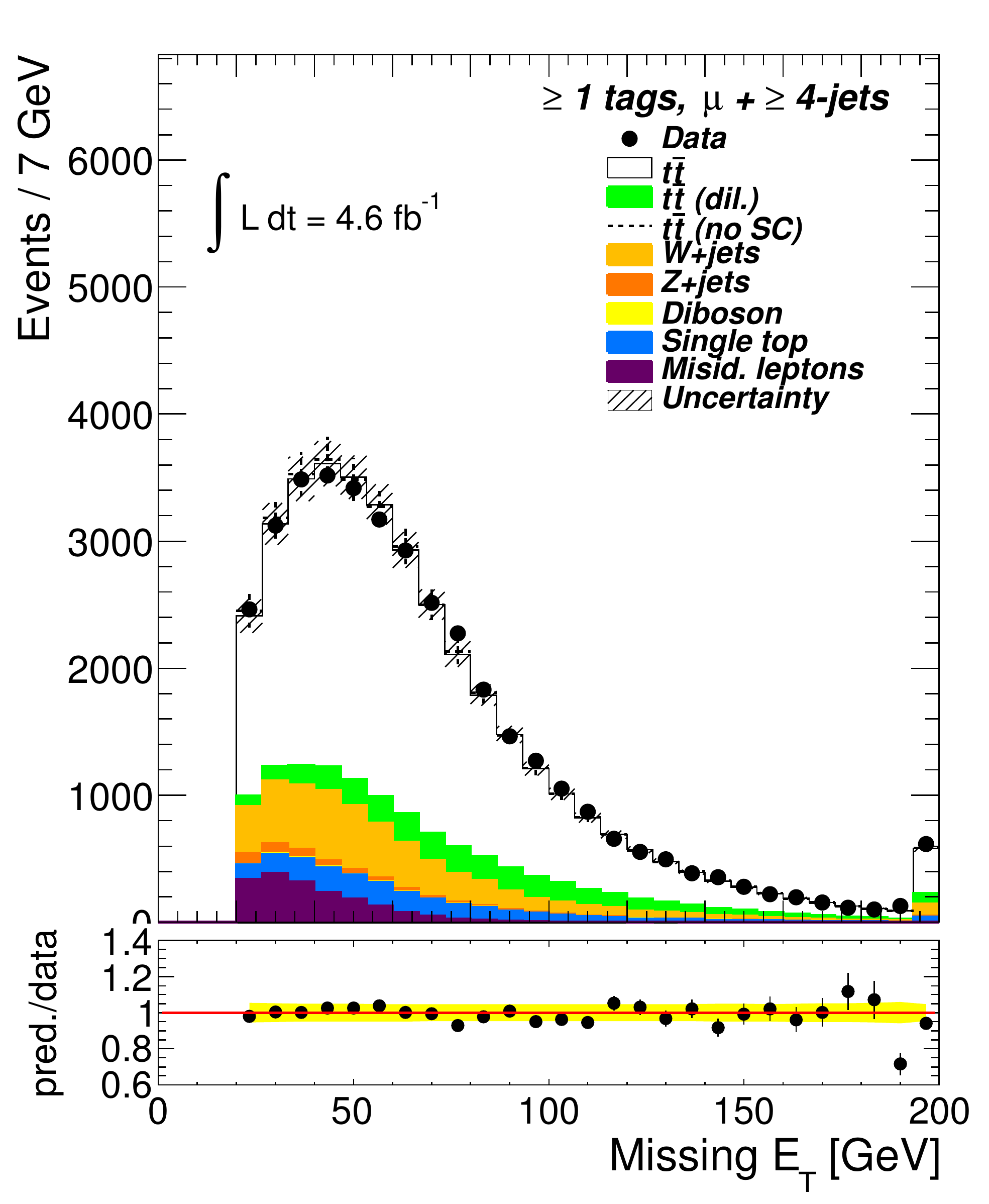}
\includegraphics[width=0.3\textwidth]{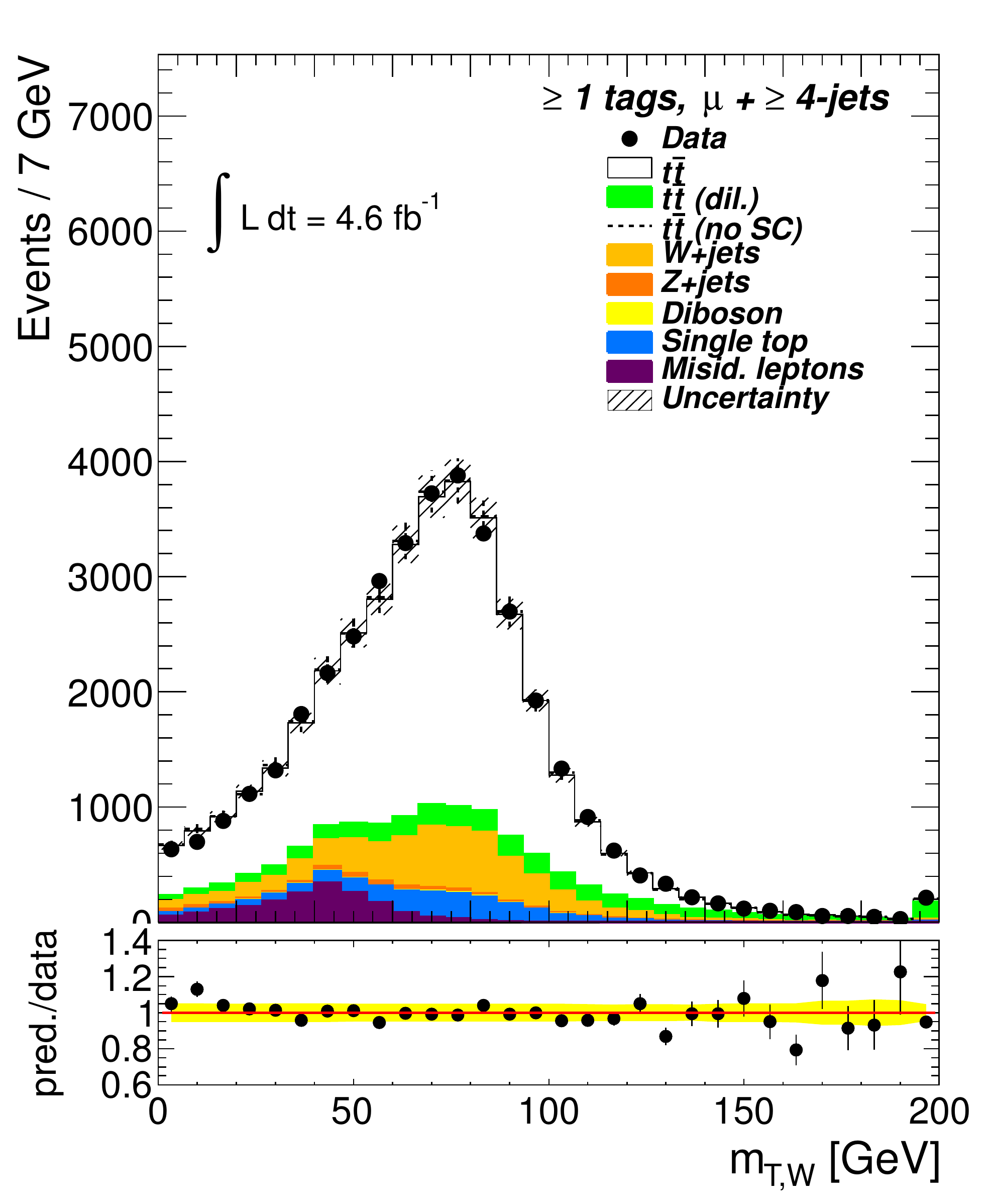}

\end{center}
\caption{Control distributions for the \pt\ of jet with the 4th highest \pt, the missing transverse momentum and \Wmt\ of the \ejets\ (top) and \mujets\ (bottom) channel ($n_{\text{jets}} \geq 4, n_{\text{b-tags}} \geq 1$).
}
\label{fig:kinematic_4incl_1tags}
\end{figure}

\section{Mismodelling of the Jet Multiplicity}
\label{sec:jetmismodeling}
The jet multiplicity it not well described by the \mcatnlo\ generator. 
As shown in Figure \ref{fig:jetmult_prob}, the \mcatnlo\ generator predicts less jets than observed in data. Furthermore, the jet \pt\ spectrum is modelled softer (see figures \ref{fig:subjets_4incl_1tags} and \ref{fig:kinematic_4incl_1tags}). 
This fact was extensively studied, also independently of this thesis. An extensive study of the jet multiplicity in \ttbar\ events can be found in \cite{jetmult}. 

Figure \ref{fig:jetmult_prob} shows the jet multiplicity mismodelling of \mcatnlo.

 \begin{figure}[htbp]
 	\centering
			\subfigure[]{
		\includegraphics[width=0.3\textwidth]{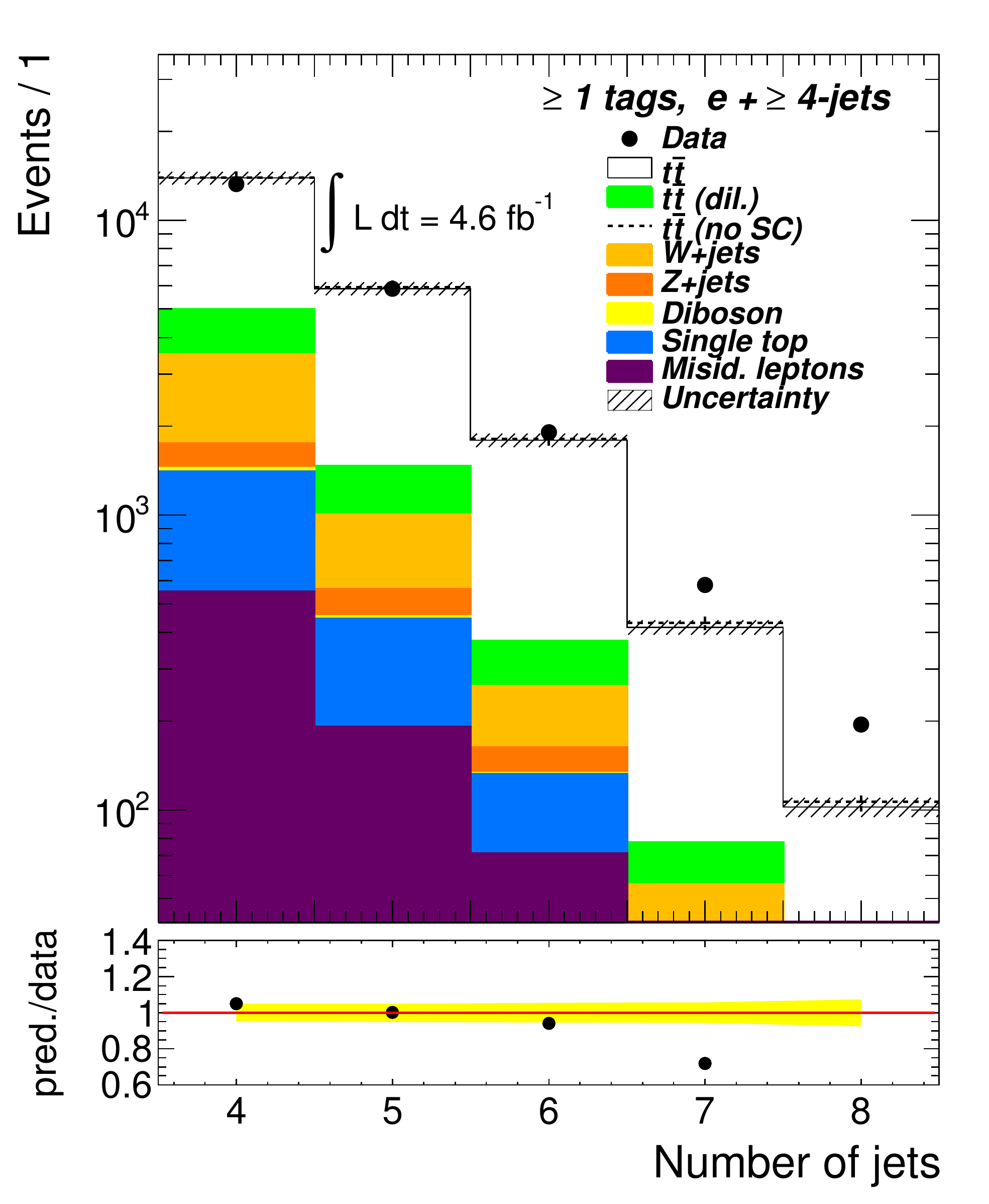}
			\label{fig:jetmult_prob_el}
		}
					\subfigure[]{
		\includegraphics[width=0.3\textwidth]{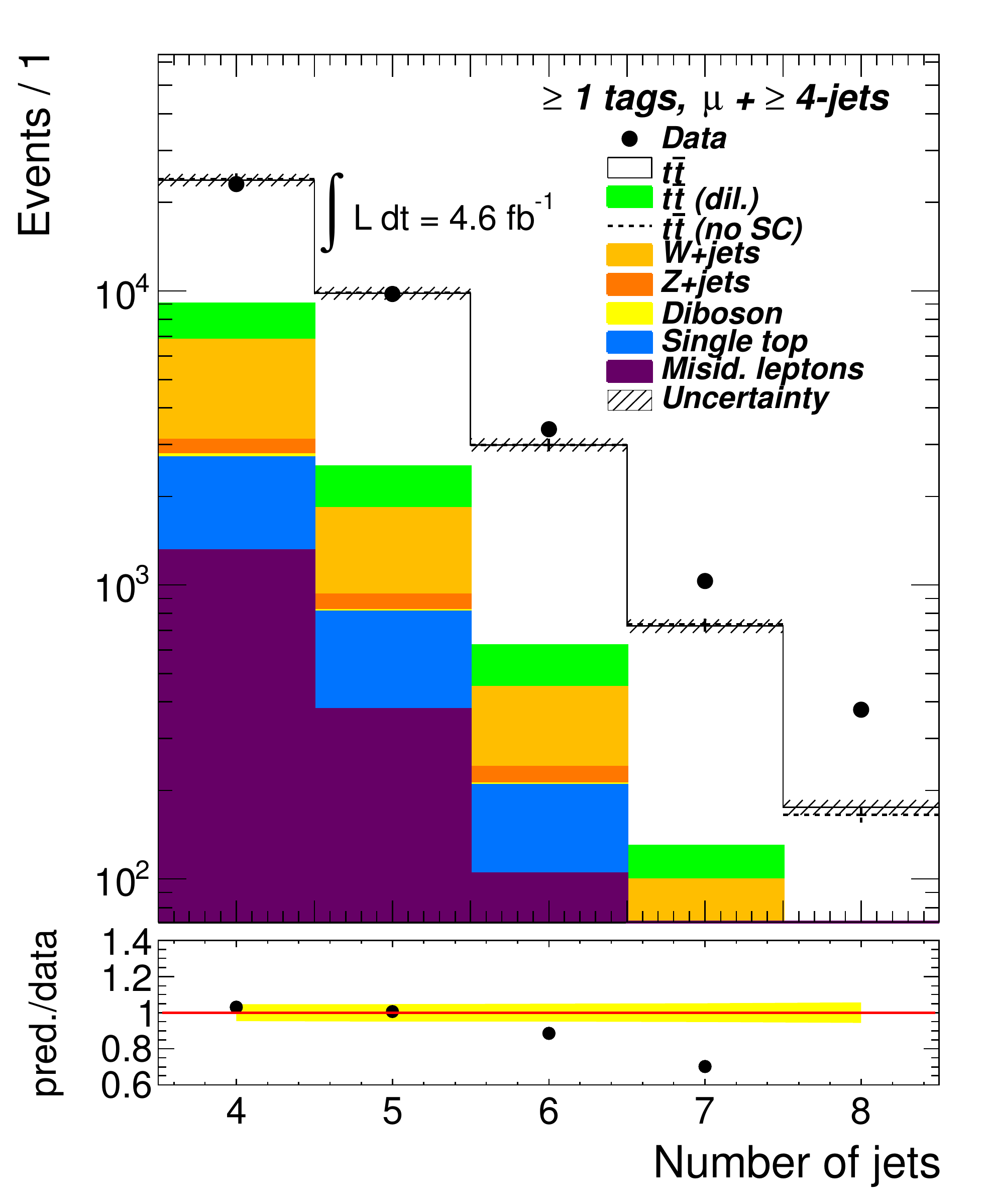}
			\label{fig:jetmult_prob_mu}
		}		

\caption{Jet multiplicity using \mcatnlo\ as \ttbar\ signal generator. \subref{fig:jetmult_prob_el} \ejets\ channel. \subref{fig:jetmult_prob_mu}  \mujets\ channel.}
\label{fig:jetmult_prob}
\end{figure}
A better modelling of the jet multiplicity is provided by other generators, such as \powheg. This is shown in Figure \ref{fig:jetmult_ok}, where \powheg+\pythia\ was chosen as MC generator for the \ttbar\ signal. 
 \begin{figure}[htbp]
 	\centering
			\subfigure[]{
		\includegraphics[width=0.3\textwidth]{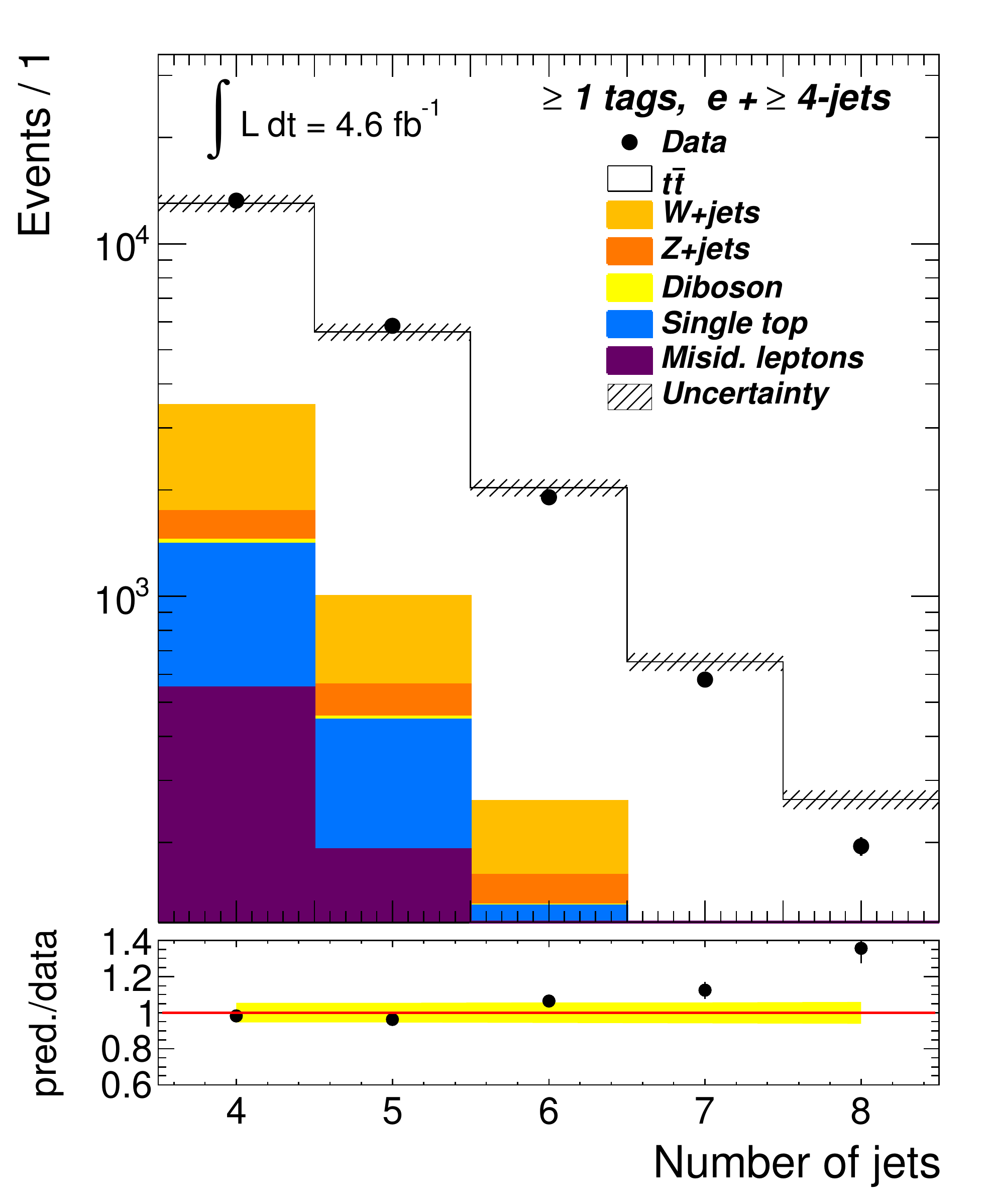}
			\label{fig:jetmult_ok_el}
		}
					\subfigure[]{
		\includegraphics[width=0.3\textwidth]{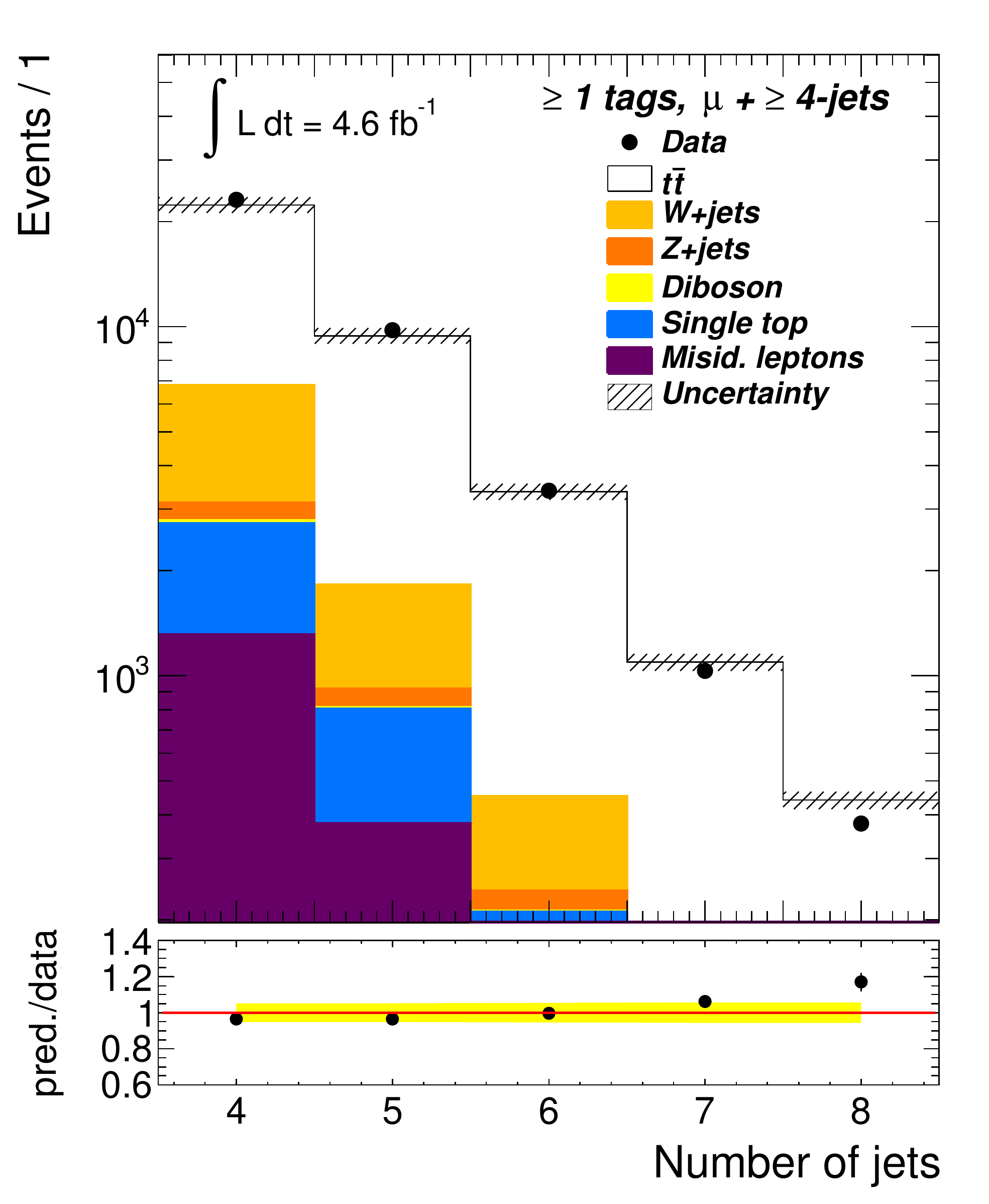}
			\label{fig:jetmult_ok_mu}
		}		

\caption{Jet multiplicity using \powheg+\pythia\ as \ttbar\ signal generator. \subref{fig:jetmult_ok_el} \ejets\ channel. \subref{fig:jetmult_ok_mu}  \mujets\ channel.}
\label{fig:jetmult_ok}
\end{figure}

Up to the time that this analysis was finalized, only \mcatnlo\ samples with both SM spin correlation and uncorrelated \ttbar\ pairs were available. Hence, no alternative generator could be used. Therefore, the effect of mismodelling was included as a source of uncertainty which will be discussed in Section \ref{sec:modeling_uncertainties}. Furthermore, an additional parameter was introduced in the analysis that allowed for an in-situ correction of the jet multiplicity (explained in Section \ref{sec:jetmultcorr}).

\section{Reconstruction of \texorpdfstring{$t\bar{t}$}{Top/Anti-Top Quark} Events with a Kinematic Likelihood Fit}
\label{sec:KLFitter}
The challenging part of studying the \ttbar\ spin correlation in the \ljets\ channel is the correct assignment of the reconstructed quantities to the spin analysers, in particular the different jets. To reconstruct the selected events, a kinematic likelihood fitter (\mbox{\KLFitter}\index{KLFitter}) \cite{klfitter} based on the \newword{Bayesian Analysis Toolkit} (\textit{\bat}\index{BAT | see {Bayesian Analysis Toolkit }}) \cite{BAT} is used. \KLFitter\ is a well-established tool used in several analyses, for example \cite{WHel_ATLAS, mtop_ATLAS_template, top_asymm_ATLAS, ATLAS_top_pol, top_xsec_diff_ATLAS, ttH_ATLAS}.
It allows reconstructing events back to leading order parton level. It is based on a certain model of the final state (\ttbar\ in this case) and uses known constraints to map reconstructed objects (jets, charged leptons and \etmiss) to the corresponding LO parton objects.
These constraints are e.g. the masses of the $W$ bosons and the (anti)top quarks.
Whereas the angles of reconstructed objects are assumed to be correctly reconstructed, their energies and momenta are allowed to vary within their detector resolutions. The possible variations of the energies and momenta are given by \textit{transfer functions}\index{Transfer function} (TF) $W\left(\tilde{E}_{\mathrm{object}} \; | \; E_{\text{LO parton}}\right)$, describing the probability for a LO parton (electron) with energy $E$ to be reconstructed as jet (electron) with an energy $\tilde{E}$. For muons, energy is replaced by transverse momentum. 
Different sets of TFs were derived for different acceptance regions of the detector using MC truth information. The likelihood $\mathcal{L}$ used in \KLFitter\ can be written as: 
\begin{eqnarray}
\mathcal{L} & = &
B\left\{m(\lqone \lqtwo) \; | \; \mw, \gammaw\right\} \cdot
B\left\{m(\lep \nul) \; | \; \mw, \gammaw\right\} \cdot \nonumber \\
&&
B\left\{ m(\lqone \lqtwo \bhad) \; | \; \mtop, \gammatop \right\} \cdot
B\left\{ m(\lep \nul \blep) \; | \; \mtop, \gammatop\right\} \cdot \nonumber \\
&&
W\left(\tilde{E}_{\mathrm{jet}_1} \; | \; E_{\bhad}\right) \cdot
W\left(\tilde{E}_{\mathrm{jet}_2} \; | \; E_{\blep}\right) \cdot
W\left(\tilde{E}_{\mathrm{jet}_3} \; | \; E_{\lqone}\right) \cdot
W\left(\tilde{E}_{\mathrm{jet}_4} \; | \; E_{\lqtwo}\right) \cdot \nonumber \\
& &
W\left(\tilde{E}_{\mathrm x}^{\mathrm{miss}} \; | \; p_{x, \nul}\right) \cdot
W\left(\tilde{E}_{\mathrm y}^{\mathrm{miss}} \; | \; p_{y, \nul}\right) \cdot
\left\{ \begin{matrix} W\left(\tilde{E}_{\lep} \; | \; E_{\lep}\right) \; , \; \mbox{\ejets\ channel} \\ W\left(\tilde{p}_{\mathrm{T,} \lep} \; | \; p_{\mathrm{T,} \lep}\right) \; , \; \mbox{\mujets\ channel} \end{matrix} \right.
\label{eq:klfitter_LH}
\end{eqnarray}

$W$ are the transfer functions and $B$ Breit-Wigner distributions of the top and $W$ mass which were fixed to 172.5 \GeV\ and 80.4 \GeV, respectively. All functions are normalized to give an integral of 1.0.
The maximization\footnote{Technically, the negative logarithm of the likelihood is minimized.}
 of the likelihood is called kinematic fitting. It is performed for all possible combinations within the permutation table mapping reconstructed particles to a model of truth particles. 
The number of permutations depends on the number of reconstructed jets which are passed to \KLFitter. In the ideal case the four reconstructed jets represent the number of model partons. Then, only the correct jet-to-parton assignment needs to be found. 
In case of \ttbar\ events the number of permutations is 4!=24 (two \bQ\ and two light quark jets are mapped to the corresponding quarks) with 4!/2=12 maximized likelihood values. The number of different likelihood values is reduced by a factor 2 as the kinematics are invariant under exchange of the jets from the hadronically decaying $W$ boson. 

One possible setup of \KLFitter\ is to use the permutation with the highest likelihood value as best permutation. A more sophisticated approach is calculating a quantity called \textit{event probability} \index{Event probability} $p_i$ for each permutation $i$. It is based on a normalized\footnote{Normalized with respect to all permutations.} likelihood and additional extensions $\Delta p_{i,j}$:
\begin{align}
p_i = \frac{\mathcal{L}_i \prod_j \Delta p_{i, j}}{\sum_i \mathcal{L}_i \prod_j \Delta p_{i, j}}
\end{align}
One possible extension is the usage of \btag ging information. The simplest approach assigns a weight of zero to a permutation in case a \btag ged jet is on the position of a light quark model particle:

\begin{align}
\Delta p_{i, \text{veto}} =
  \begin{cases}
   0 & \text{if model parton is light and jet is tagged}, \\
   1 & \text{else}.
  \end{cases}\end{align}
Another way of using a veto method is to veto permutations where the jet assigned to a \bQ\ is not tagged or a combination of both veto methods.
A more sophisticated extension takes the efficiencies $\varepsilon_b$ of \btag ging and the mistag rate $\varepsilon_l$ \cite{mistag} of light jets into account:\footnote{The mistag rate $\varepsilon_l$ is the probability to tag a non-$b$-jet as $b$-jet.}
\begin{align}
\Delta p_{i, \text{tag}} = & \left \{ \begin{matrix} \varepsilon_b \; , \; \mbox{$b_{\mathrm{had}}$ was \btag ged} \\ (1-\varepsilon_b) \; , \; \mbox{$b_{\mathrm{had}}$ was not \btag ged} \end{matrix} \right \} \cdot
\left \{ \begin{matrix} \varepsilon_b \; , \; \mbox{$b_{\mathrm{lep}}$ was \btag ged} \\ (1-\varepsilon_b) \; , \; \mbox{$b_{\mathrm{lep}}$ was not \btag ged} \end{matrix} \right \} \cdot \nonumber \\
& \left \{ \begin{matrix}\varepsilon_l \; , \; \mbox{$q_1$ was \btag ged} \\ (1-\varepsilon_l) \; , \; \mbox{$q_1$ was not \btag ged} \end{matrix} \right \} \cdot
\left \{ \begin{matrix} \varepsilon_l \; , \; \mbox{$q_2$ was \btag ged} \\ (1-\varepsilon_l) \; , \; \mbox{$q_2$ was not \btag ged} \end{matrix} \right \}
\end{align}
A \btag\ algorithm can be used with a certain cut on the \btag\ weight, required to tag a jet. This cut defines a working point that comes along with certain values of $\varepsilon_b$ and $\varepsilon_l$. 
As both $\varepsilon_b$ and $\varepsilon_l$ depend on the working point used in an analysis, the method is also referred to as \textit{working point method}\index{Working point method}.
In Section \ref{sec:udsep} an extension is introduced that allows separating light up- and down-type jets. This is necessary in order to successfully map all spin analysers to their reconstructed objects.

\section{Transfer Functions}
\label{sec:TFs}
Transfer functions $W$ model the relation between energies and momenta of reconstructed objects and the partons of the LO decay signature (labelled as 'truth'). Such a mapping is needed to account for detector resolution effects.  The resolution is modelled by a double-Gaussian function:
\begin{align}
W(E_{\mathrm{reco}}, E_{\mathrm{truth}}) = \frac{1}{2 \pi (p_2 + p_3\, p_5)} \left( e^{- \frac{(\Delta E - p_1)^2}{2 p_2^2} } + p_3 e^{- \frac{(\Delta E - p_4)^2}{2 p_5^2} } \right)
\label{eq:TF}
\end{align}
where $\Delta E = \frac{E_{\mathrm{truth}} -
  E_{\mathrm{reco}}}{E_{\mathrm{truth}}}$. The double-Gaussian functions account for the detector resolution and higher order effects.
  Transfer functions exist for all reconstructed quantities: Jets, Muons, Electrons and \etmiss. 
For objects mainly based on calorimeter information (jets, electrons), the parameters $p_i$ are
functions of $E_{\mathrm{truth}}$. For muons the energies $E$ are
replaced with the transverse momentum \pt. From now on all references to an energy include the corresponding formulation for transverse momentum. 

For the determination of the transfer functions a dedicated tool, {\tt TFTool}\index{TFTool}, was developed. 
{\tt TFTool} uses a sample of selected \ttbar\ events, produced with the \mcatnlo\ generator. 
For each event every model object (two \bQ s, two light jets and a charged lepton) is tried to be matched to a reconstructed object. 
Objects $i$ and $j$ are matching in case the distance is $\Delta R < 0.3$. Furthermore, the matching was to be bi-uniquely, meaning that within a distance $\Delta R = 0.3$ exactly one parton must match a reconstructed object and also vice versa. 

Objects are classified as $b$-jets, light jets, electrons or muons. The $b$-jet requirement is checked on truth level, meaning that the jet must emerge from a \bQ. No explicit requirement on the tagging was made. This is of course just one specific classification choice. Instead of classifying the model, also the reconstructed object could be classified in \btag ged and untagged jets. Furthermore, it is possible to separate jets containing a soft muon and those that do not. For the used dataset this option is excluded as muons are removed from the jets during the event selection process as described in Section \ref{sec:eventselection}. 

The resolution of the reconstructed objects varies for different parts of the detector and is non-uniform in $\left| \eta \right|$. Hence, different transfer functions are derived for individual object types and $\left| \eta \right|$ regions. 
{\tt TFTool} assumes individual parameterizations of $p_i$ for each object type and keeps the type of parameterization fixed across $\left| \eta \right|$.\footnote{Only the parameterization type is kept fixed, not the numerical parameter values.}

The parameterization as a function of $E_\text{truth}$ is motivated by the underlying physics effects which determine the detector resolution. In the case of calorimeter energy, the resolution for higher energies is $\frac{\sigma_E}{E} \sim \frac{1}{\sqrt{E}}$ (see Section \ref{sec:calorimeters}). In contrast to this, the muon resolution decreases linearly: $\frac{\sigma_{\pt}}{\pt} \sim \pt$. Other parameters which are not related to the resolution are estimated as a linear function of $E_\text{truth}$: $p_i = a_i + b_i E_{\text{truth}}$. 
For bins of $E_{\text{truth}}$ and $\left| \eta \right|$, histograms of $\frac{E_{\mathrm{truth}} -
  E_{\mathrm{reco}}}{E_{\mathrm{truth}}}$ are created and filled with the values of matched objects. The binning in \abseta\ is fixed and follows the detector structure by using a dedicated bin for the calorimeter transition region, for instance (see Table \ref{tab:TF_eta}). The binning in $E_{\text{truth}}$ is variable to allow for sufficient statistics in each bin. 

\begin{table}[htbp]
\begin{center}
\small
\begin{tabular}{l l l l l l }
Light Jets & $\left[0.0,  0.8\right]$ &$\left[0.8, 1.37\right]$ &$\left[1.37,  1.52\right]$ &$\left[1.52,  2.5\right]$ &$\left[2.5,  4.5\right]$\\
B-Jets & $\left[0.0,  0.8\right]$ &$\left[0.8, 1.37\right]$ &$\left[1.37,  1.52\right]$ &$\left[1.52,  2.5\right]$ &$\left[2.5,  4.5\right]$\\
Electrons & $\left[0.0,  0.8\right]$ &$\left[0.8,  1.37\right]$ &$\left[1.37,  1.52\right]$ &$\left[1.52,  2.5\right]$ &{}\\
Muons & $\left[0.0,  1.11\right]$ &$\left[1.11,  1.25\right]$ &$\left[1.25,  2.5\right]$ &{} &{}\\
\end{tabular}
\end{center}
\caption{\abseta\ binning used for the transfer functions.}
\label{tab:TF_eta}
\end{table}
For each of the filled histograms, a double Gaussian function as defined in Equation \ref{eq:TF} is fitted. During this first fit, referred to as local fit, the transfer function parameters $p_i$ are not parameterized globally as function of $E_{\text{truth}}$. In Figure \ref{fig:TF_localfit} local fits for truth energies of about 100 GeV are shown.

 \begin{figure}[htbp]
 	\centering
			\subfigure[]{
		\includegraphics[width=0.45\textwidth]{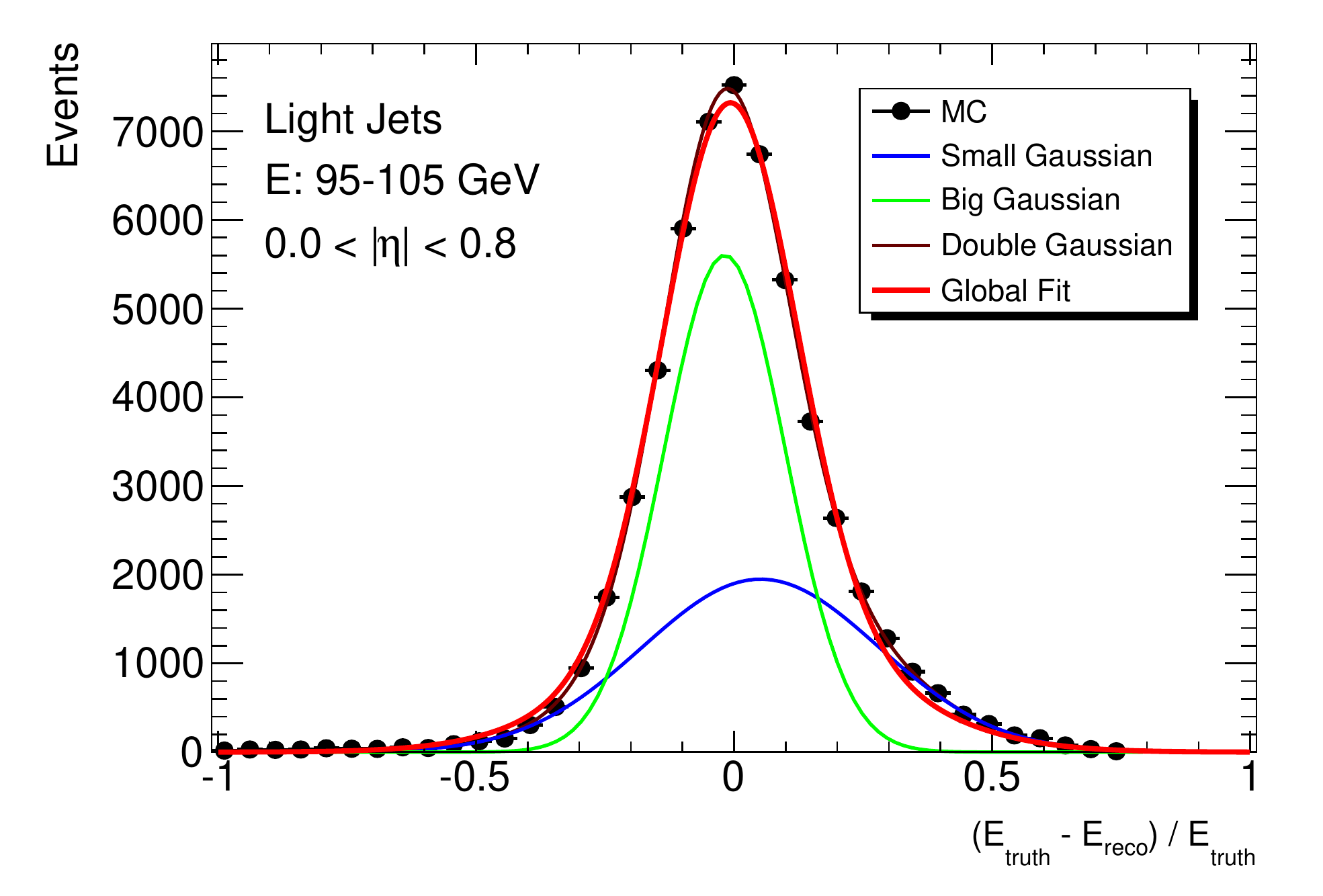}
			\label{fig:localfit_lJets}
		}
					\subfigure[]{
		\includegraphics[width=0.45\textwidth]{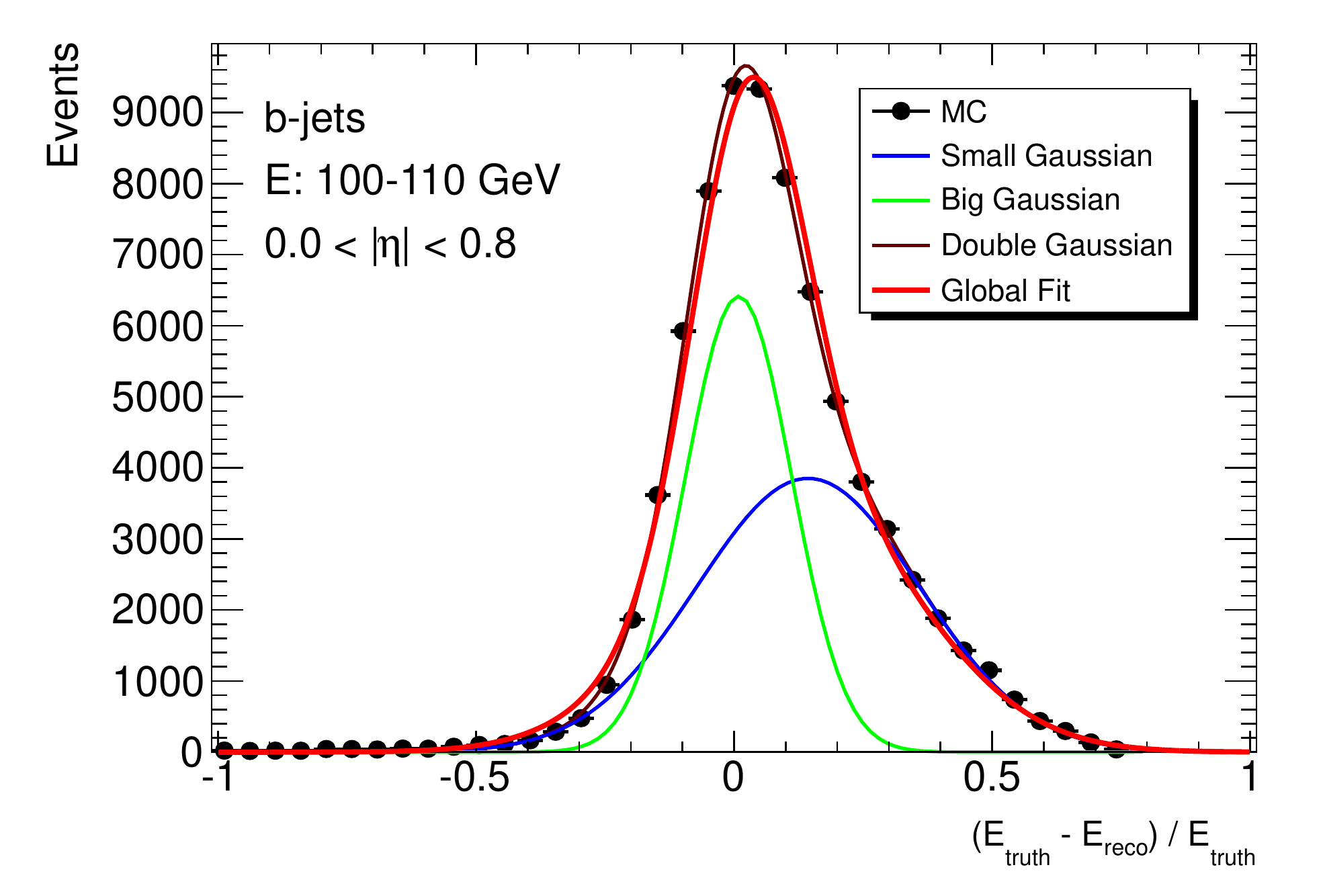}
			\label{fig:localfit_bJets}
		}\\	
							\subfigure[]{
		\includegraphics[width=0.45\textwidth]{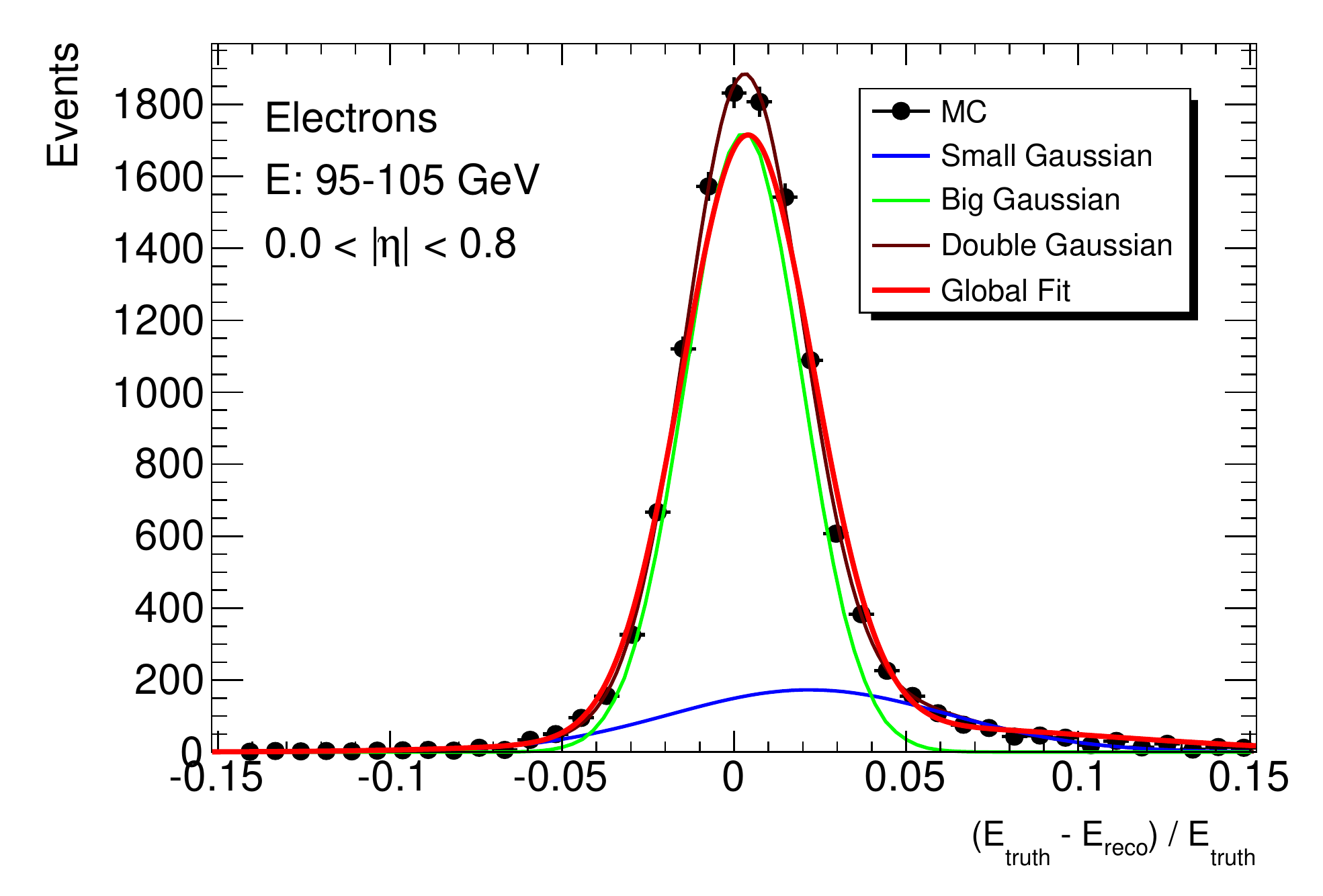}
			\label{fig:localfit_electrons}
		}		
							\subfigure[]{
		\includegraphics[width=0.45\textwidth]{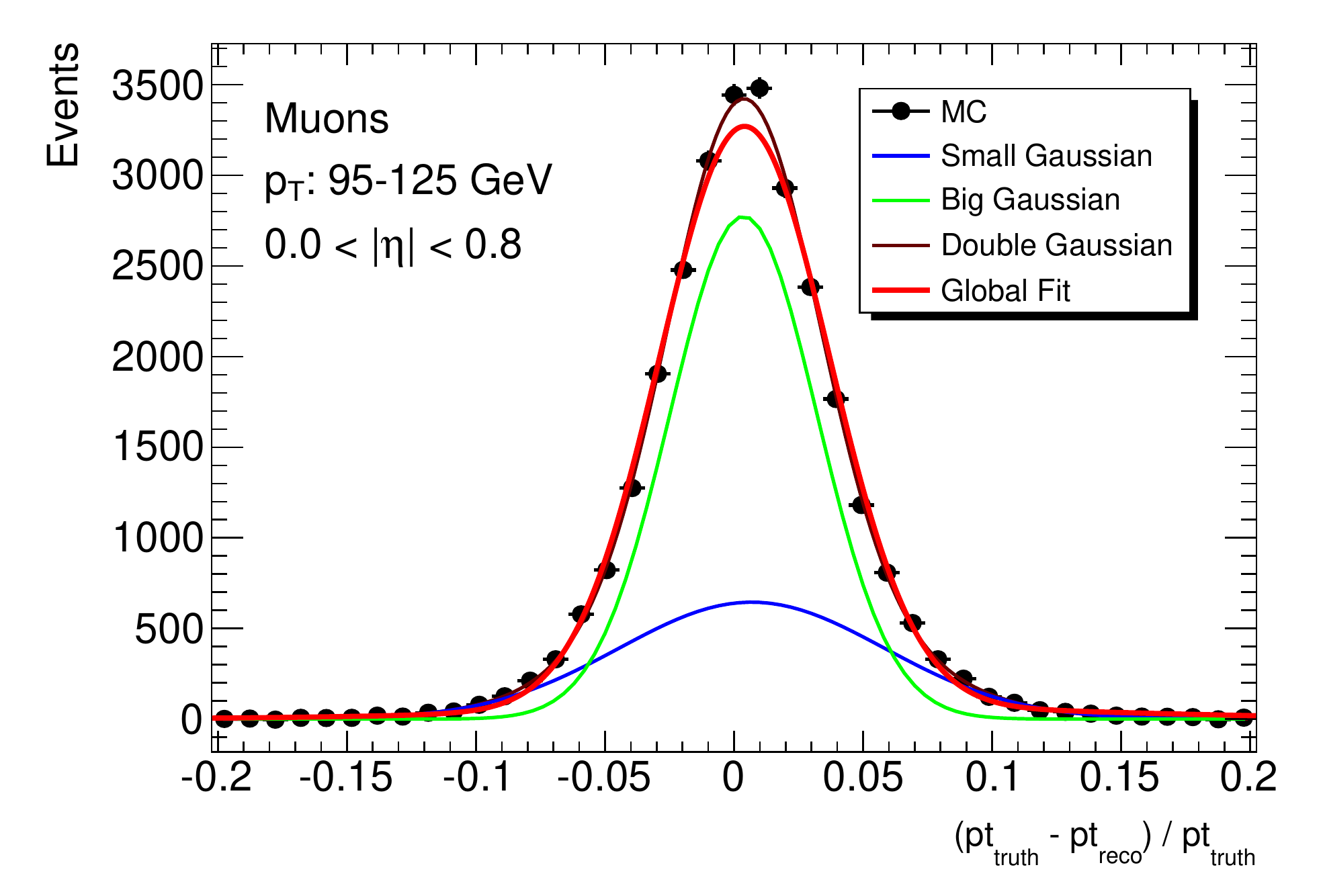}
			\label{fig:localfit_muons}
		}			

\caption{Transfer function fits for \subref{fig:localfit_lJets} light jets, \subref{fig:localfit_bJets} b-jets, \subref{fig:localfit_electrons} electrons and \subref{fig:localfit_muons} muons for truth energies / transverse momenta of about 100 GeV. The entries derived from a signal MC sample are fitted locally with a double Gaussian function, composed of a small and a big Gaussian function. The global TF fit is also shown.}
\label{fig:TF_localfit}
\end{figure}
The local fits  are a sum of a small and a big Gaussian function. Jets have on average significantly worse resolution than leptons. Furthermore, jets have a trend to be reconstructed with less energy than the original parton has, while the lepton transfer function is symmetric. This can be explained by final state radiation and out-of-cone contributions. For \bjet s this effect is even larger, visible in the more dominant tail on the right side of Figure \ref{fig:localfit_bJets} compared to Figure \ref{fig:localfit_lJets}. In the next step, all values of $p_i$ are plotted against $E_{\text{truth}}$ for each of the \abseta\ bins and object types. These parameters are then fitted with an approximated dependence on $E_{\text{truth}}$. This stage is called parameter estimation. As an example for light jets in the region $0.0 \leq \abseta < 0.8$ the result of such a fit is shown in Figure \ref{fig:ParEst_lJets}. The uncertainties on the entries are given by the uncertainties on the local fits. They are increased by a scaling factor to stabilize the parameter estimation fit.
\begin{figure}[htbp]
\begin{center}
\includegraphics[width=\textwidth]{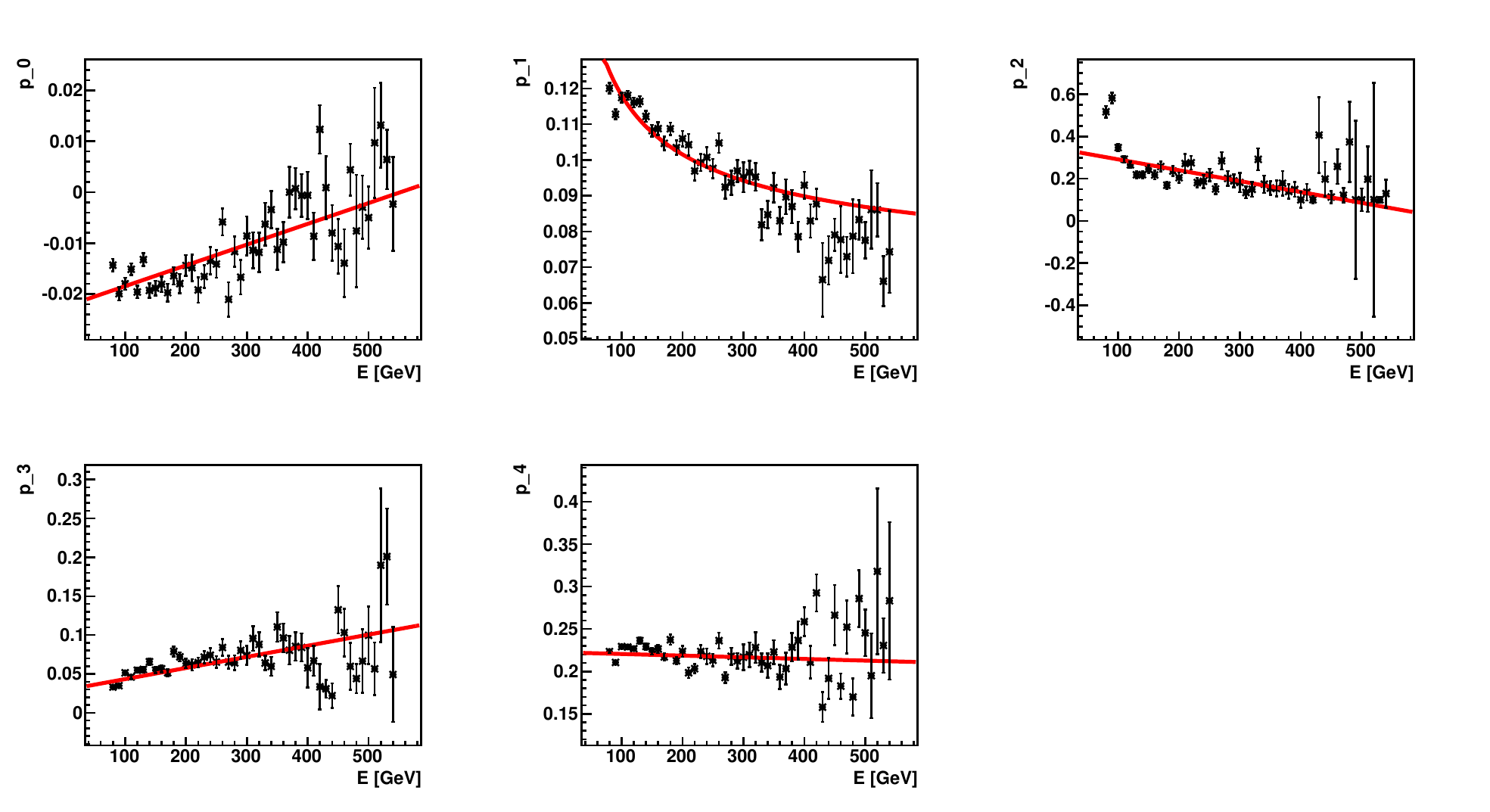}
\end{center}
\caption{Estimation of parameters $p_i$ as a function of $E_{\text{truth}}$ for light jets in the region $0.0 \leq \abseta < 0.8$. The parameters were estimated as functions of $E_{\text{truth}}$ according to Table \ref{tab:TF_parameterizations}.
}
\label{fig:ParEst_lJets}
\end{figure}
One can see that the approximations work reasonably well. The jet energies of $E_{\text{truth}} < 100 \GeV$ are cut off for the approximation and are extrapolated for the parameterization. This decision was made to keep the fit stable. Removing this cutoff would cause the parameter estimation fits to fail. Correlations among the parameters cause jumps between several local fit options. These need to be kept under control to find a common and global parameterization. 

Table \ref{tab:TF_parameterizations} lists the chosen parameterizations for the TF parameters $p_i$. 
\begin{table}[htbp]
\begin{center}
\small
\begin{tabular}{|c|c|c|c|c| }
\hline
{}& Light Jets & B-Jets & Electrons & Muons\\
\hline
\hline
$p_1$ & $a_1 + b_1 E_{\text{truth}}$ & $a_1 + b_1E_{\text{truth}}$& $a_1 + b_1 E_{\text{truth}}$ & $a_1 + b_1 E_{\text{truth}}$ \\
$p_2$ & $a_2 / \sqrt{E_{\text{truth}}} + b_2 $ & $a_2 / \sqrt{E_{\text{truth}}} + b_2 $ & $a_2 / \sqrt{E_{\text{truth}}} + b_2 $ & $a_2+ b_2 E_{\text{truth}}$ \\
$p_3$ & $a_3 + b_3 E_{\text{truth}}$ & $a_3 + b_3 E_{\text{truth}}$ & $a_3 + b_3 E_{\text{truth}}$ & $a_3 + b_3 E_{\text{truth}}$\\
$p_4$ & $a_4 + b_4 E_{\text{truth}}$ & $a_4+ b_4 E_{\text{truth}}$ & $a_4 + b_4 E_{\text{truth}}$ & $a_4 + b_4 E_{\text{truth}}$ \\
$p_5$ & $a_5+ b_5 E_{\text{truth}}$ & $a_5 + b_5 E_{\text{truth}}$ & $a_5 + b_5 E_{\text{truth}}$ & $a_5 + b_5 E_{\text{truth}}$ \\
\hline
\end{tabular}
\end{center}
\caption{Parameterizations of the transfer function parameters $p_i$ as a function of the LO parton energy $E_{\text{truth}}$.}
\label{tab:TF_parameterizations}
\end{table}
The transfer functions have been continuously adapted to data taking conditions and object definitions. New running conditions, calibrations and event selections made the changes necessary. Since then, the parameterizations have changed. Table \ref{tab:TF_parameterizations} reflects the status at the time this analysis was performed. 
 
Using the parameter estimation fit results for $a_i$ and $b_i$ as starting values, a global fit of the transfer functions $W$ for all bins of $E_{\text{truth}}$ is performed. The results of this global fit, shown in Figure \ref{fig:TF_localfit}, is then taken as set of transfer function parameters and implemented to \KLFitter. 

One remark on the transfer functions should be made: It is not expected that the $\Delta E$ distributions follow a Gaussian distribution. The reason is that the resolution $\Delta E = \frac{E_{\mathrm{truth}} -
  E_{\mathrm{reco}}}{E_{\mathrm{truth}}} \sim \frac{1}{\sqrt{E_{\text{truth}}}}$. Hence, within a bin of $E_{\text{truth}}$ the resolution is only constant in the limit of a vanishing bin width.

  In Figure \ref{fig:ljets_evolve} the implemented set of TFs for light jets in the central \abseta\ region is shown. The illustration was chosen such that the distribution of possible reconstructed jet energies is shown for a given value of $E_{\text{truth}} = E_{\text{parton}}$. The vertical lines indicate the latter value.
\begin{figure}[htbp]
\begin{center}
\includegraphics[width=\textwidth]{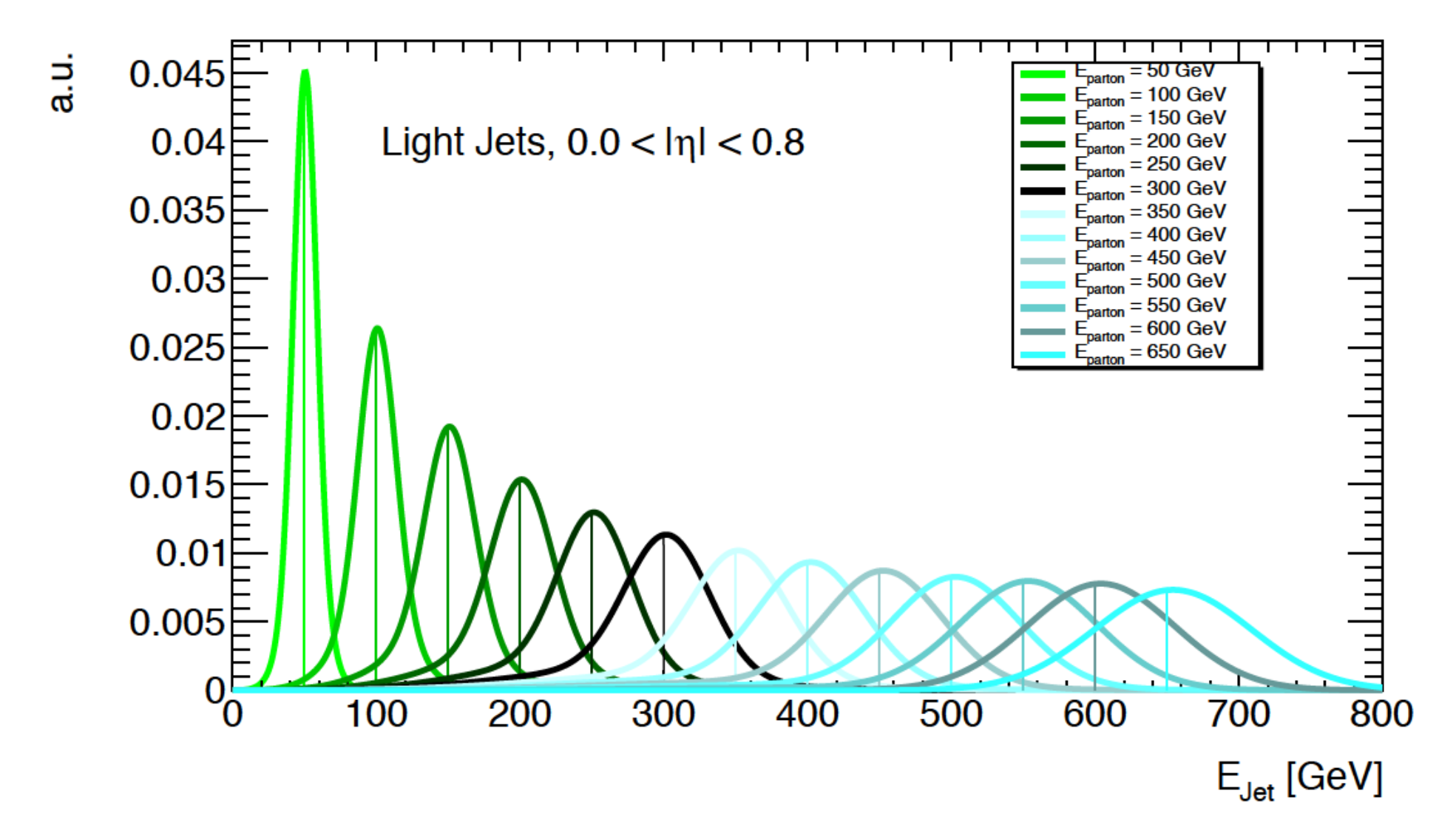}
\end{center}
\caption{Transfer functions for light jets with $0.0 \leq \abseta\ < 0.8$. Vertical lines indicate the parton energies.
}
\label{fig:ljets_evolve}
\end{figure}
While for ($b$-)jets, electrons and muons {\tt TFTool} was used to derive the transfer functions, the \etmiss\ transfer functions had a dedicated procedure and tool. For early studies the TF was set as constant for all events by fitting the distributions of $E_{x,y}^{\text{miss}} - p_{x,y}^{\nu}$ with a Gaussian function. It has been optimized at a later stage as follows. 
The basic idea is that the TFs map the $x$ and $y$ components of the neutrino momentum to the $x$ and $y$ components of the measured missing transverse momentum. Since it is known that the \etmiss\ depends on the scalar sum of deposited energy in the calorimeters \cite{etmiss}, \sumet, this quantity has been used to parameterize the width of  $E_{x,y}^{\text{miss}} - p_{x,y}^{\nu}$. 
The parameterization function was heuristically chosen to be of the Sigmoid type
\begin{align}
\sigma \left( \sumet\ \right)= p_0 + \frac{p_1}{1 + e^{-p_2 \left(\sumet - p_3 \right)}}
\end{align}
Figure \ref{fig:TF_etmiss} shows that the dependence on \sumet\ is not negligible. 
\begin{figure}[htbp]
\begin{center}
\includegraphics[width=0.75\textwidth]{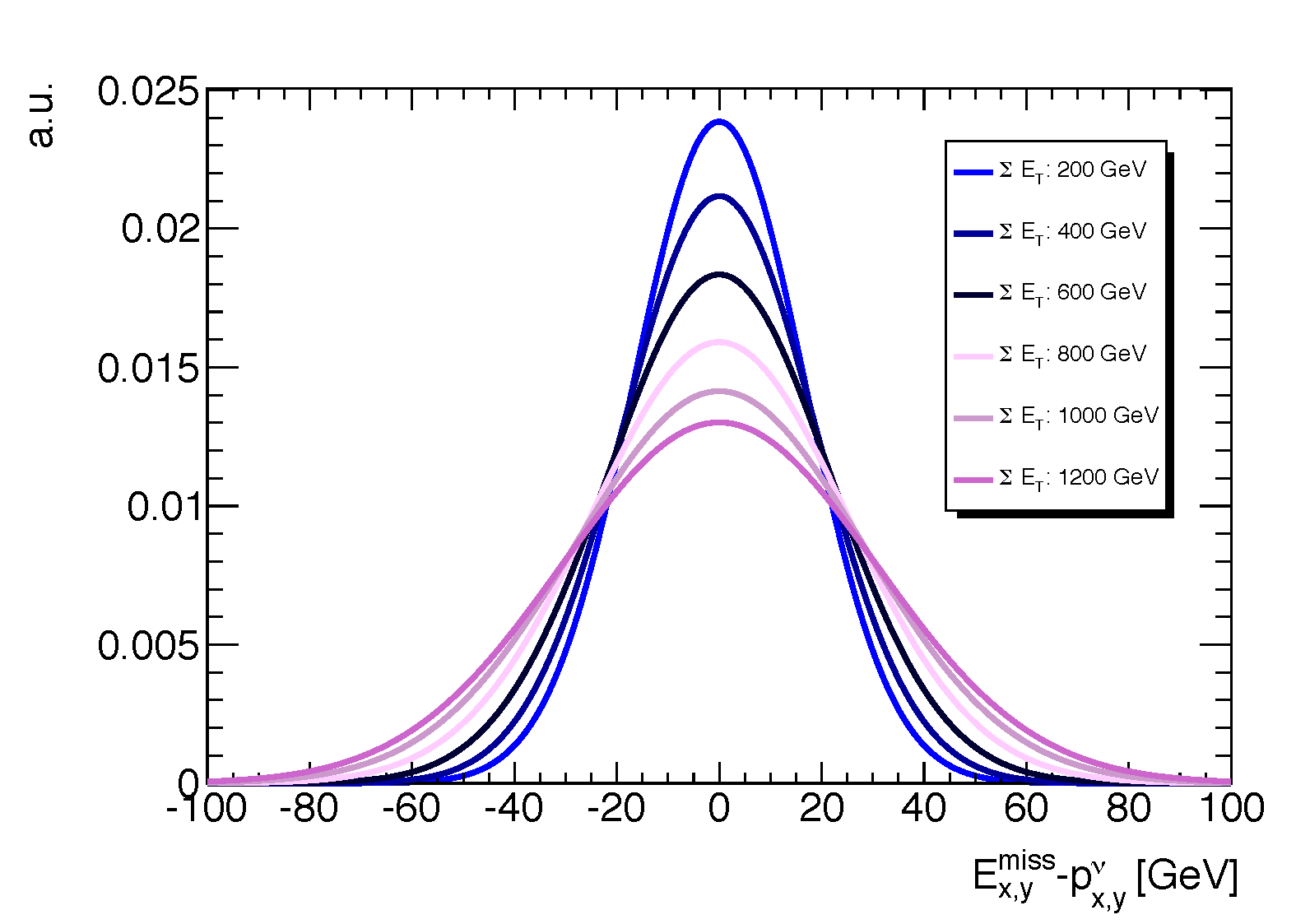}
\end{center}
\caption{Transfer functions for Neutrinos / \etmiss, parameterized as a function of \sumet.
}
\label{fig:TF_etmiss}
\end{figure}

The \etmiss\ TF leads to a width of the $E_{x,y}^{\text{miss}} - p_{x,y}^{\nu}$ distribution of $\approx 18$ GeV for $\sumet = 400 \GeV$. This is higher than the value of about $10$ GeV quoted as \etmiss\ resolution in \cite{etmiss}. As the \ttbar\ topology is more complex than the di-jet events used in \cite{etmiss}, this is not surprising. It was checked that the TFs for the $x$ and $y$ components of the neutrinos are equivalent, as expected.

Studies on uncertainties of the TFs can be found in \cite{fritz_BA}. Concerning the evaluation of uncertainties of the spin correlation analysis, no dedicated TF uncertainties have been derived.  The TFs assume a true model and deviations from this in the simulated events are evaluated via the common systematic uncertainties evaluation, such as detector modelling (see Chapter \ref{sec:systematics}). Furthermore, the fit uncertainties of the TF parameters were found to be small. 

However, the used TFs are not a perfect model of the mapping of measured to partonic energies. This can be observed in particular at low jet energies where pile-up effects play a crucial role and disturb the expected proportionality of the jet energy resolution to $\frac{1}{\sqrt{E}}$. This can be seen in Figure \ref{fig:TF_lowE} and leads to the cutoff for the parameter estimation fit. 
 \begin{figure}[htbp]
 	\centering
			\subfigure[]{
		\includegraphics[width=0.45\textwidth]{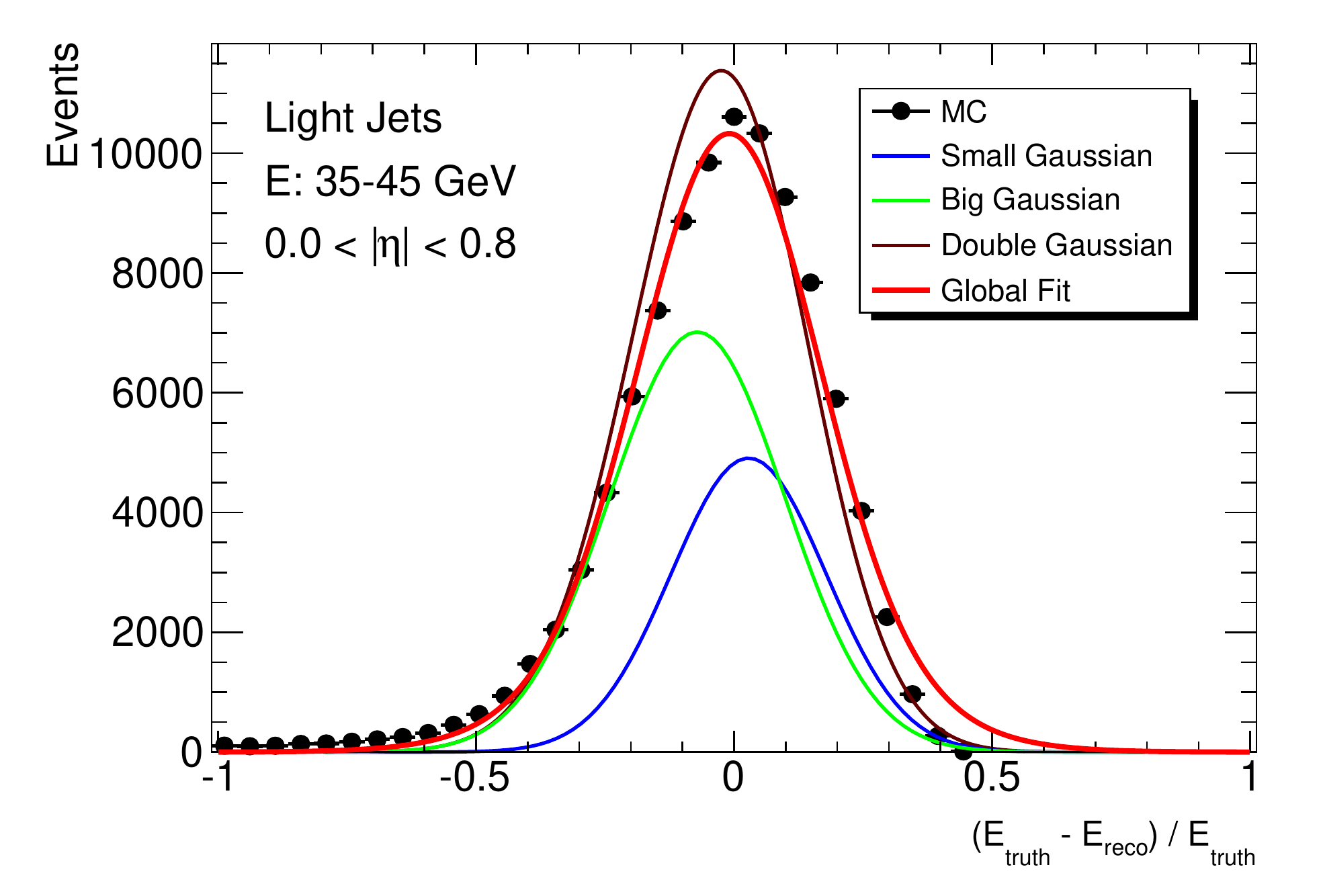}
			\label{fig:TF_lJets_40}
		}
					\subfigure[]{
		\includegraphics[width=0.45\textwidth]{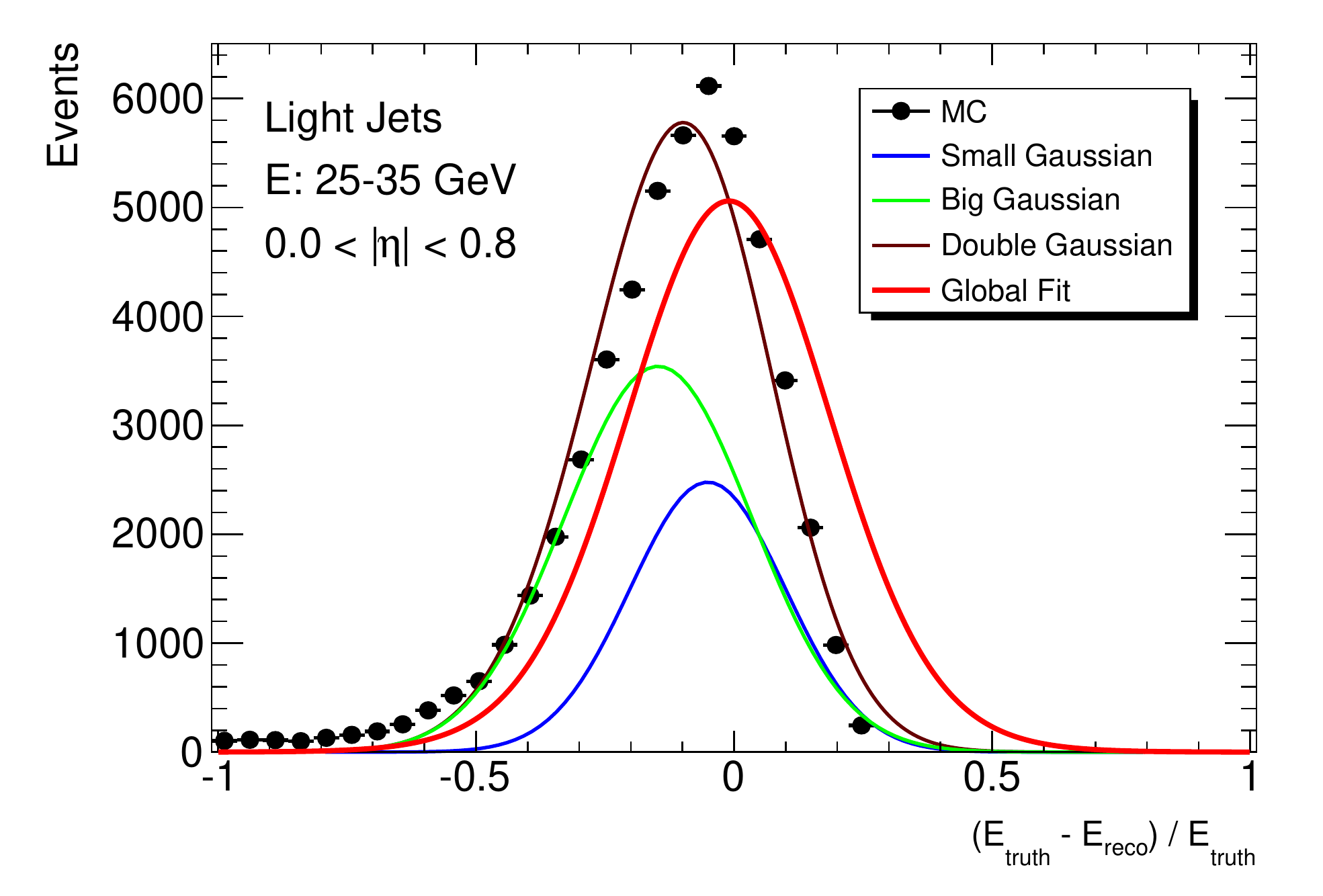}
			\label{fig:TF_lJets_30}
		}		
\caption{Description of the detector resolution effects for light jets at low energies. \subref{fig:TF_lJets_40} $35 \leq E_{\text{jet}} < 45~\GeV$, \subref{fig:TF_lJets_30}  $25 \leq E_{\text{jet}} < 35~\GeV$. }
\label{fig:TF_lowE}
\end{figure}
At low energies two things are observed: The reconstructed energy is not lower, but larger than the corresponding parton energy. Pile-up contributions can lead to such overestimations. Furthermore, the global fit is no longer able to properly model the TFs at low energies. This imperfect modelling is of no concern for the presented analysis since \KLFitter\ provides both fitted jet energies as well as a mapping of reconstructed to model objects. In this analysis only the latter property of \KLFitter\ was used, since for the chosen spin correlation observable only angular distributions, which are not modified by \KLFitter\ anyway, are used.

For analyses using the fitted outputs from \KLFitter\ possible biases have to be considered. They  can be eliminated either by a more advanced parameterization of the transfer function parameters $p_i$ or by a more advanced calibration of the jets. Another approach would be to map the reconstructed energies to particle level instead of parton level for which the description of objects (particles) is better justified than for LO partons. 
A detailed discussion goes beyond the scope of the thesis but it is worth to be studied in the future.

\section{KLFitter Extension for Up/Down-Type Quark Separation}
\label{sec:udsep}
By default, \KLFitter\ is not able to separate the two light jets from the $W$ boson decay as permuting them keeps the likelihood $\mathcal{L}$ invariant. The same holds true for the \btag ging extensions presented so far where only $b$-jets and non-$b$-jets are separated.

A dedicated extension to the likelihood has been developed, taking into account quantities with separation power between the two light jets. There are two facts that help to separate these: their flavour and the V-A structure of the weak decay vertex. 

The V-A structure of the $W$ decay vertex predicts differences in the energies between the two light jets \cite{Jezabek1994, Brandenburg2002}. While the suggested frame for the energy determination is the top quark rest frame, a difference in \pt\ is also visible in the laboratory frame which is easier to determine. Figure \ref{fig:sep_pt} shows the different \pt\ distributions for the prompt \bQ\ from the top decay and the light up- and down-type jets from the $W$ decay.
 \begin{figure}[htbp]
 	\centering
			\subfigure[]{
		\includegraphics[width=0.45\textwidth]{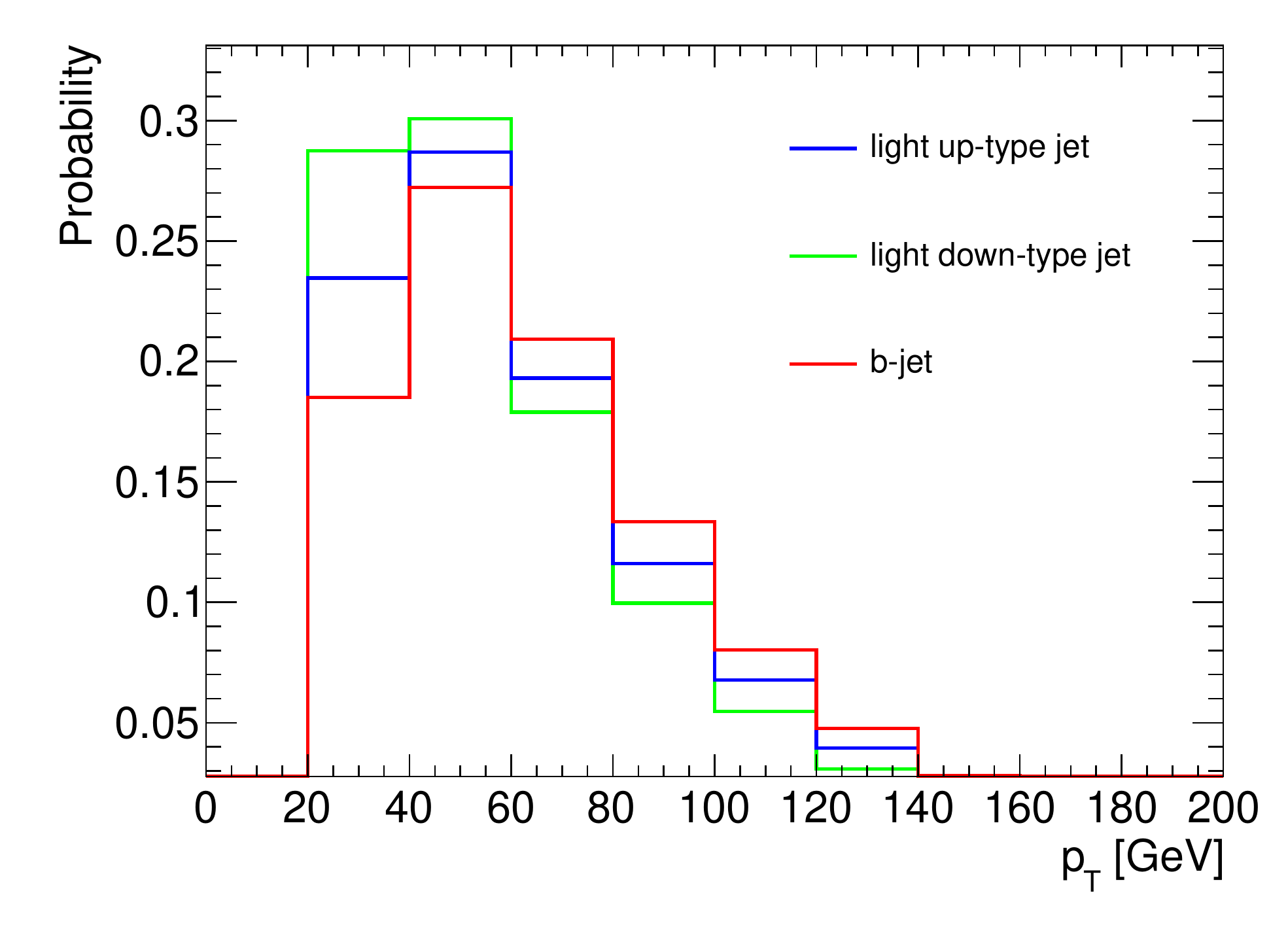}
			\label{fig:sep_pt}
		}
							\subfigure[]{
		\includegraphics[width=0.45\textwidth]{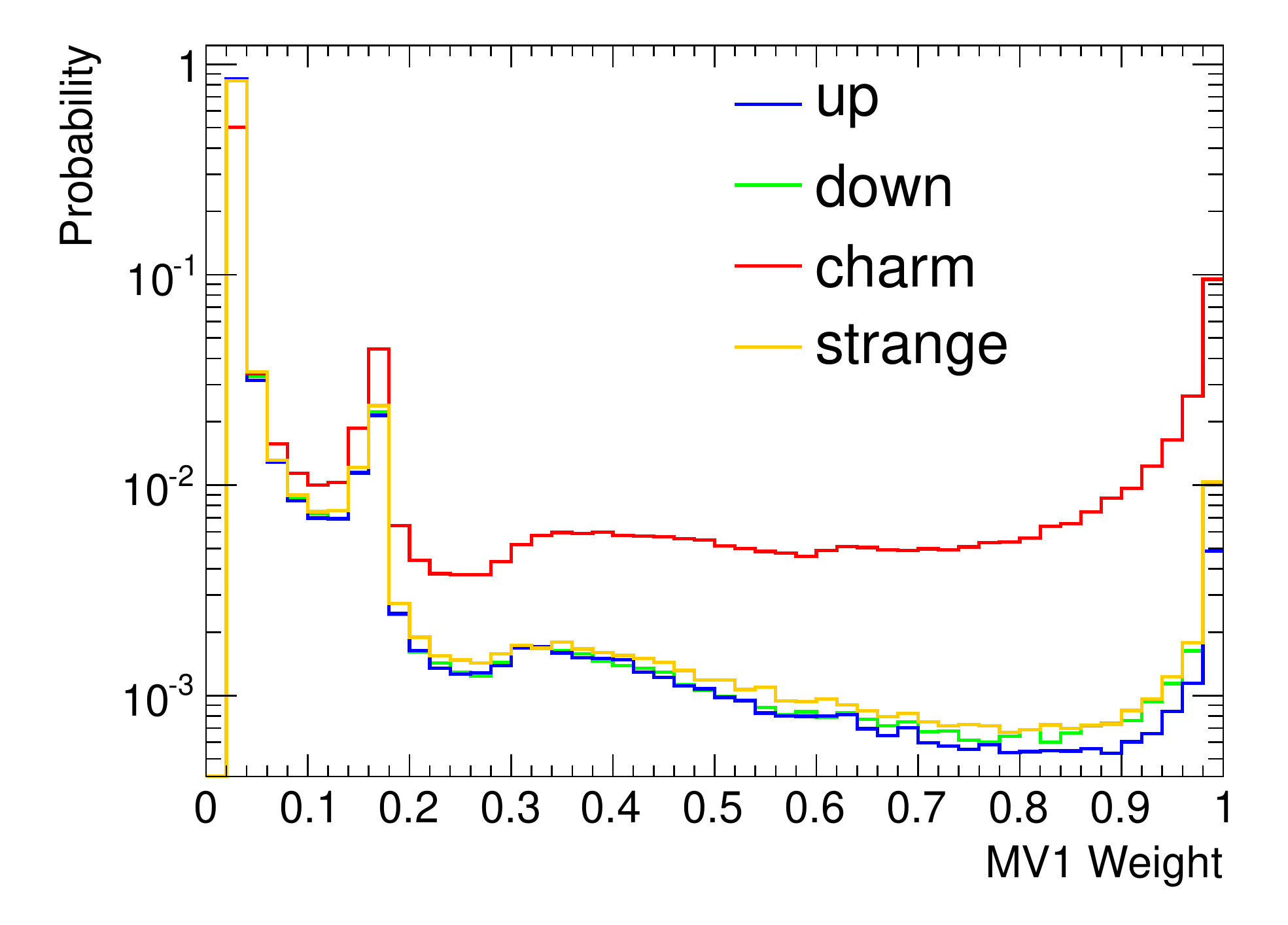}
			\label{fig:sep_weight_light}
		}	
					\subfigure[]{
		\includegraphics[width=0.45\textwidth]{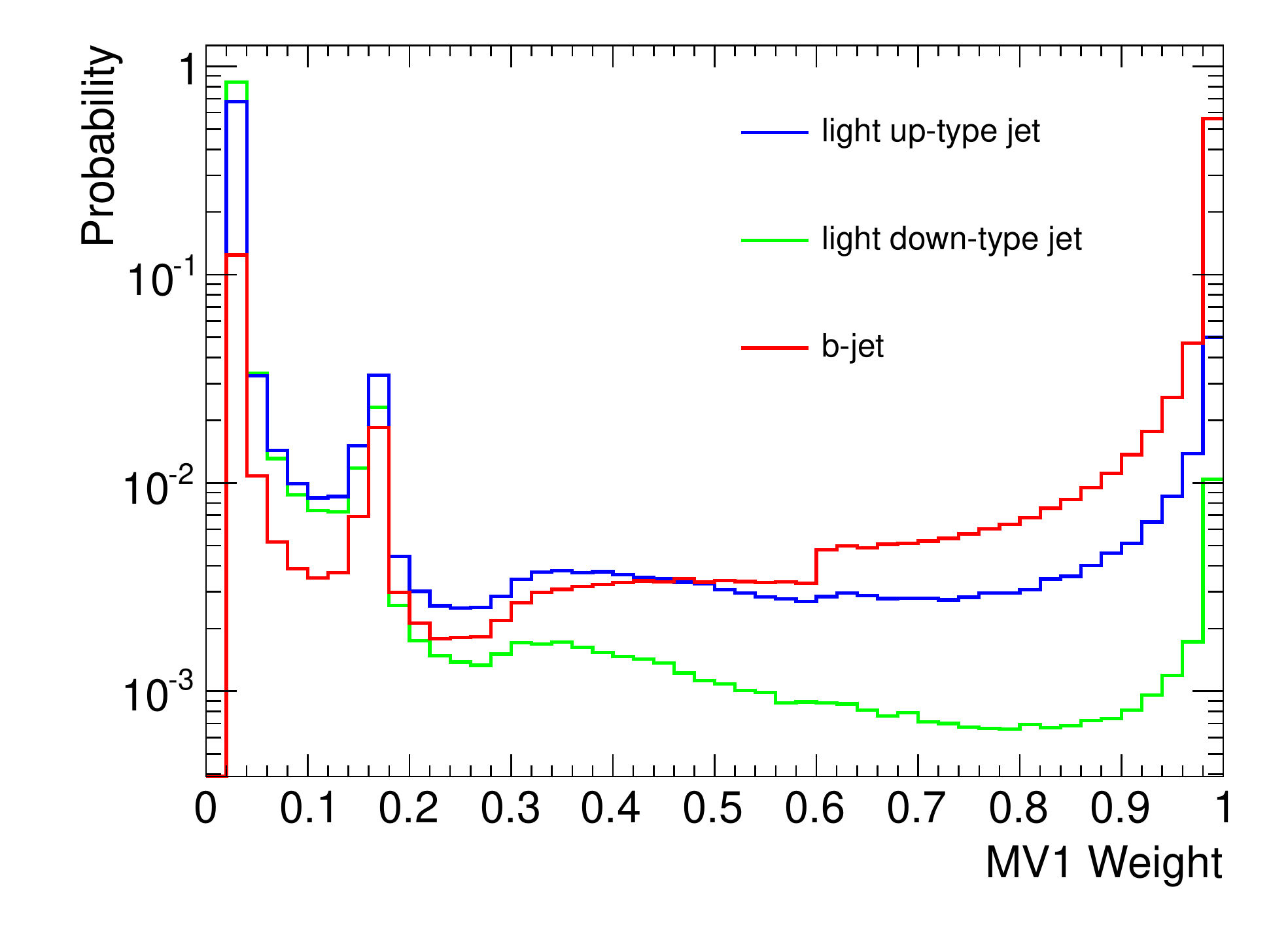}
			\label{fig:sep_weight}
		}		
\caption{\subref{fig:sep_pt} \pt\ spectra for the three jet types from the hadronic top quark decay. 
\subref{fig:sep_weight_light} Weight of the \mvone\ \btag ger for different light quark types of the $W$ decay. \subref{fig:sep_weight} Weight of the \mvone\ \btag ger for the three jet types from the hadronic top quark decay.}
\label{fig:sep_1d}
\end{figure}

Another possibility is to access the flavour of the two light jets. At first sight both are of 'non-$b$-type'. But while for a decay of $W^{+} \rightarrow u\bar{d}$ both quarks are clearly light, the decay $W^{+} \rightarrow c\bar{s}$ contains a charm quark, heavier than all other light quarks. Techniques used for \btag ging can in principle be used to develop dedicated $c$-taggers. As on the one hand these are not yet well established and on the other hand a dedicated tagging is not desired, the output of the \mvone\ \btag ger is sufficient to see differences between the different light jet types. Figure \ref{fig:sep_weight_light} confirms the expectation that down and strange quarks have almost identical \mvone\ tag weight distributions while up and charm quarks do not. The minimal difference for the former pair arises from the different \pt\ distributions which are correlated to the tagger weight. 

As the $W^{+}$ boson decays in 50\,\% of the cases into a $c\bar{s}$ pair, the \mvone\ tagger weight distributions can be grouped into (prompt) $b$-, up- and down-type quarks. The up- and down-type separation power is shown in Figure \ref{fig:sep_weight}. 

\begin{figure}[htbp]
\begin{center}
\subfigure[]{
\includegraphics[width=0.31\textwidth]{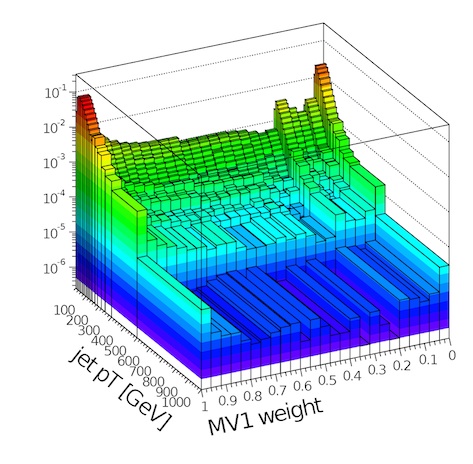}
\label{fig:klfitterweight_uncal_bQ}
}
\subfigure[]{
\includegraphics[width=0.31\textwidth]{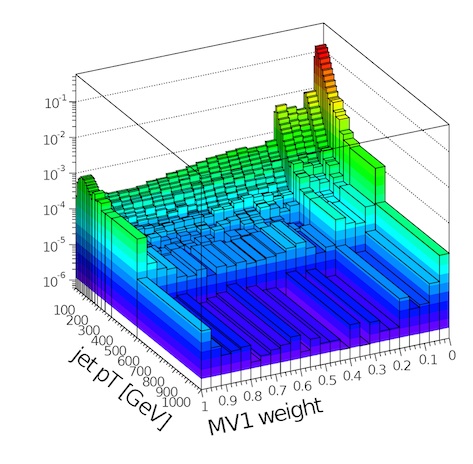}
\label{fig:klfitterweight_uncal_dQ}
}
\subfigure[]{
\includegraphics[width=0.31\textwidth]{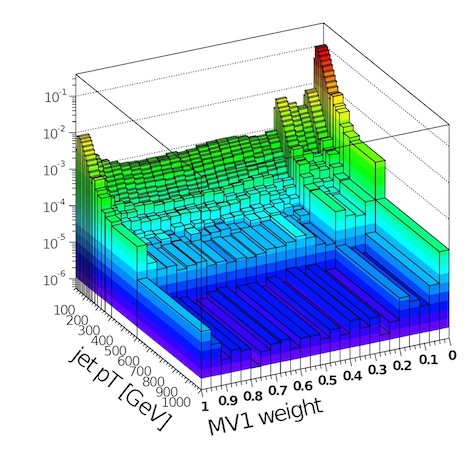}
\label{fig:klfitterweight_uncal_uQ}
}
\end{center}
\caption{Distribution of the \btag ging weights of the \mvone\ tagger vs. the \pt\ of a jet for three types of jets: \subref{fig:klfitterweight_uncal_bQ} $b$-jets coming directly from the top quark decay, \subref{fig:klfitterweight_uncal_dQ} down-type quarks coming from the $W$ boson and \subref{fig:klfitterweight_uncal_uQ} up-type quarks coming from the $W$ boson. }
\label{fig:klfitter_pT_and_tagweight_distrib}
\end{figure}

In order to consider both the differences in \pt\ and in the \btag ging weight $w$ as well as the correlation between the quantities, two-dimensional, normalized, distributions for the three jet types ($b$, light up-type and light down-type) are created and used to calculate event probability extensions according to 
\begin{eqnarray}
\Delta p_{i, \text{u/d sep.}}  = &P_{\text{2D}}^{\text{b-type}}(\pt(\text{jet}_{\text{blep}}), w_{\text{jet}_{\text{blep}}}) \cdot P_{\text{2D}}^{\text{b-type}}(\pt(\text{jet}_{\text{bhad}}), w_{\text{jet}_{\text{bhad}}}) \cdot \nonumber \\ &P_{\text{2D}}^{\text{u-type}}(\pt(\text{jet}_{\text{uQ}}), w_{\text{jet}_{\text{uQ}}}) \cdot P_{\text{2D}}^{\text{d-type}}(\pt(\text{jet}_{\text{dQ}}), w_{\text{jet}_{\text{dQ}}})
\end{eqnarray}

By using the extension of the \KLFitter\ event probability, up- and down-type quarks can be separated to a larger extent than with more simple approaches. Such a simple alternative could be the choice of the least energetic jet as down-type quark jet \cite{Jezabek1994b} as it is also used in other reconstruction methods (see Section \ref{sec:recocomp}). Table \ref{tab:simpleeff} compares the simple approach used on top of a \KLFitter\ reconstruction to the event probability extension with dedicated up- and down-type separation. 
\begin{table}[htbp]
\begin{center}
\begin{tabular}{|c||c|c|}
\hline
{}&   default & $\Delta p$ extension \\
\hline
\dQ\ & 31\,\% & 35\,\%\\
\hline
\bQ\ & 50\,\% & 55\,\%\\
\hline
\end{tabular}
\end{center}
\caption{Reconstruction efficiencies for the \dQ\ and \bQ\ obtained via selection and jet-to-parton mapping via \KLFitter. For the default \KLFitter\ setup the jet with the lower energy is taken as \dQ\ candidate.}
\label{tab:simpleeff}
\end{table}

The reconstruction efficiency is defined as the fraction of real \ljets\ events passing the selection for which the model parton matches the jet it is assigned to within a distance of $\Delta R < 0.3$:
\begin{align}
\varepsilon_{\text{reco}} = \frac{N_{\text{match}}}{N_{\text{selected}}}.
\end{align}

A detailed discussion about the reconstruction efficiencies and possible optimizations via cuts on certain event variables is discussed in the next section.

\section{Reconstruction Efficiencies and Optimizations}
\label{sec:optimizations}
A good reconstruction algorithm always tries to map the reconstructed detector objects to physics objects with the highest efficiency. What exactly defines a high efficiency is defined by the individual analyses. In the case of \ttbar\ analyses it is desirable to know which of the reconstructed objects stems from which parton of the \ttbar\ decay. One part of the reconstruction concerns the mapping of partons to reconstructed objects. Sometimes an emphasis on a certain type of parton is made while another type might be even ignored. 
\KLFitter\ reconstructs the full \ttbar\ topology and even fits the reconstructed kinematic quantities to best estimates of the underlying parton properties. Only the mapping capabilities of \KLFitter\ are used in this thesis, not the actual fitting. The overall performance of \KLFitter\ is illustrated in Figure \ref{fig:reco_eff_global}.
\begin{figure}[ht]
	\centering
		\includegraphics[width=0.95\textwidth]{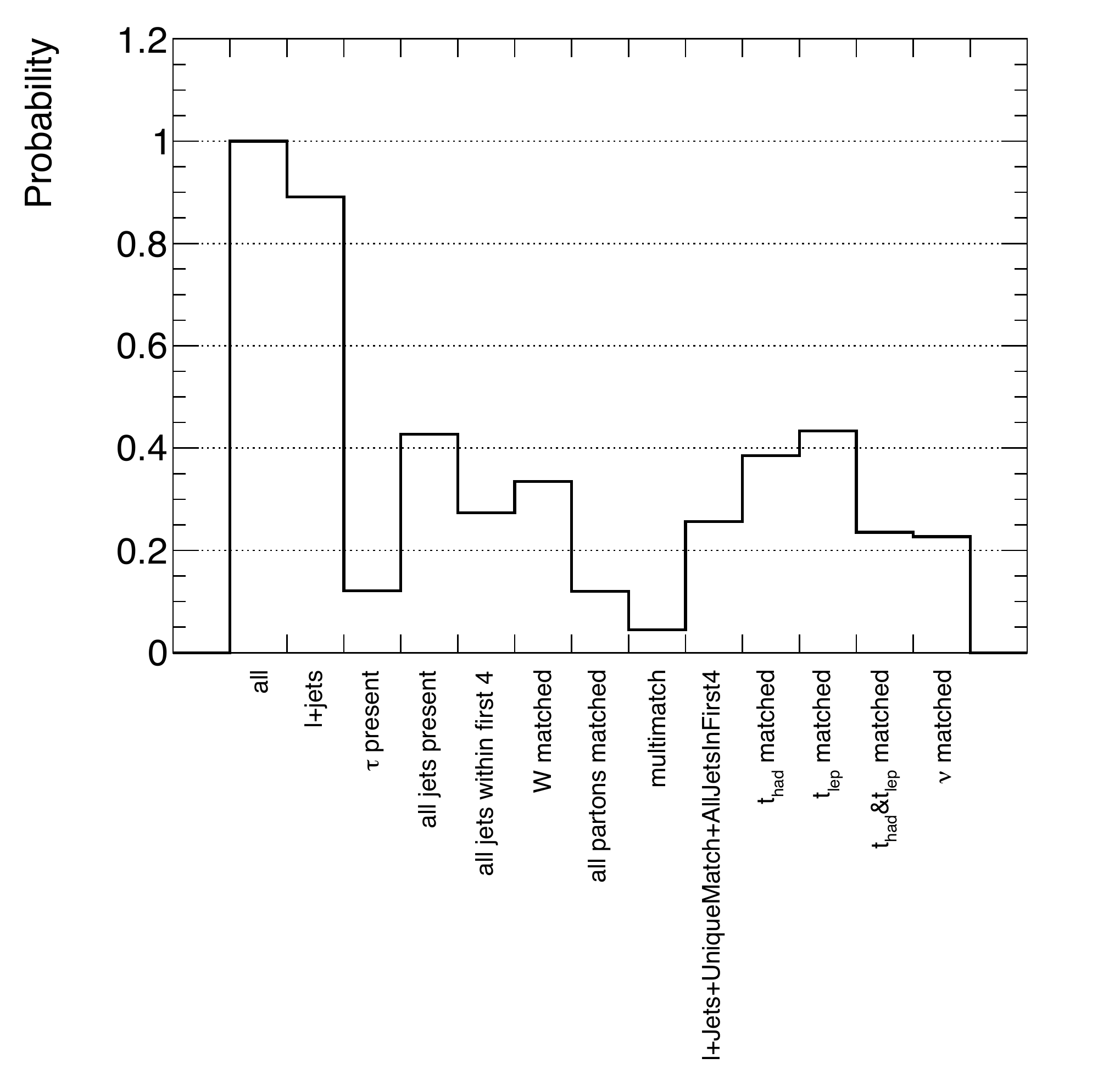}
		\caption{Event properties and reconstruction efficiencies using \KLFitter\ in the \mbox{\mujets} channel. Numbers are given with respect to all events passing the event selection. }
	\label{fig:reco_eff_global}
\end{figure}
An \mcatnlo\ sample of \ttbar\ events passing the \mujets\ events selection is used. The bin label should be read as follows:
\begin{itemize}
\item \textbf{ all} All events passing the event selection. 
\item\textbf{ l+jets} Real \ljets\ events. The rest of about 10\,\% is given by \ttbar\ events decaying into the dilepton channel. 
\item {\textbf{$\tau$ present}} A $\tau$ lepton is present.
\item \textbf{all jets present} All model partons match a jet which was successfully reconstructed. 
\item\textbf {all jets within first 4} The jets matching the four partons are also the highest four in \pt\ and hence given to \KLFitter.
\item\textbf {W matched} The two jets assigned to the hadronically decaying top quark match but might be exchanged. 
\item\textbf {all partons matched} Each jet was correctly assigned to the corresponding parton.
\item\textbf {multimatch} In at least one case more than one jet matched a parton. 
\item\textbf {l+Jets+UniqueMatch+AllJetsInFirst4} The event is a real \ljets\ event, each parton matches a jet and there is no multiple matching. The four matched jets must be the jets with the highest \pt\ and hence passed to \KLFitter.
\item\textbf {$t_{\text{had}}$ matched} The reconstructed hadronically decaying top quark matches the partonic top quark within $\Delta R < 0.4$. 
\item\textbf {$t_{\text{lep}}$ matched} The reconstructed leptonically decaying top quark matches the partonic top quark within $\Delta R < 0.4$. 
\item \textbf {$t_{\text{had}}$~\&~$t_{\text{lep}}$ matched} Both the reconstructed hadronically and leptonically decaying top quark match the corresponding partonic top quark within $\Delta R < 0.4$. 
\item\textbf {$\nu$ matched} The reconstructed neutrino matches the partonic real neutrino. 
\end{itemize}

It is important to understand that the relatively low reconstruction efficiencies of 35\,\% (\dQ) and 55\,\% (\bQ) are mainly a problem of acceptance, not of the internal reconstruction performance. To disentangle the two effects, Figure \ref{fig:reco_eff_details} lists reconstruction efficiencies normalized to two different references.
\begin{figure}[ht]
	\centering
		\includegraphics[width=0.95\textwidth]{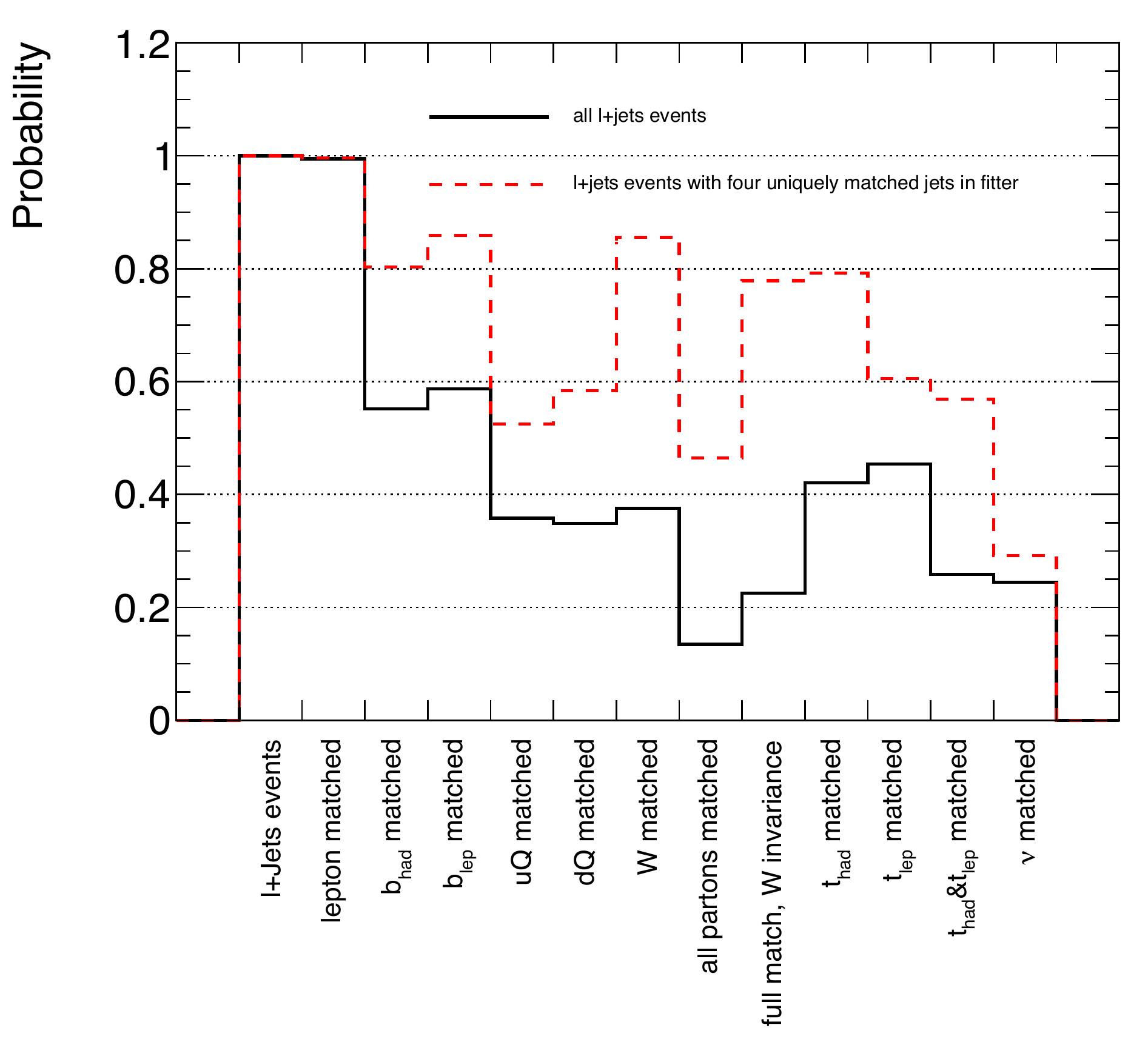}
		\caption{Reconstruction efficiencies using \KLFitter\ in the \mujets\ channel. Numbers are given with respect to all events passing the event selection (solid line) and to such events where all four jets passed to \KLFitter\ were bi-uniquely matched to partons (dashed line).}
	\label{fig:reco_eff_details}
\end{figure}
 In the first case efficiencies with respect to all selected \ljets\ events are shown. In the second case matched events are normalized to only those where all four jets passed to \KLFitter\ were bi-uniquely matched to partons. Hence, the second case reflects the internal performance of the reconstruction algorithm as it considers only those events for which the algorithm had a chance to correctly reconstruct all jets. 

The most powerful spin analysers for hadronically decaying top quarks are the \dQ\ and the \bQ. While the \dQ\ has the same full analysing power of $\left| \alpha \right|\approx 1$ as the charged lepton, the \bQ\ still has a lower but still reasonable spin analysing power of $\left| \alpha \right|\approx 0.4$ and -- in contrast to the \dQ\ -- a high reconstruction efficiency. As being the most interesting spin analysers, the reconstruction efficiencies of the \dQ\ and the \bQ\ will be studied in detail in this section. 

It is obvious that the reconstruction efficiencies of 35\,\% for the \dQ\ and 55 \,\% for the \bQ\ are inclusive quantities and depend on the event properties. The object definition and event selection gained stability against pile-up, especially by the introduction of the JVF cut (see Section \ref{sec:jets}). While for early studies the efficiency was dropping significantly for an increasing number of primary vertices, it could be stabilized with the current object definition and event selection as shown in Figure \ref{fig:eff_npvtx}. This is important as the analysis needs to be independent of the running conditions. In particular, future spin analyses for which the average number of interactions will increase with the higher luminosity setups need to be stable with respect to pile-up. 
 \begin{figure}[htbp]
 	\centering
						\subfigure[]{
		\includegraphics[width=0.45\textwidth]{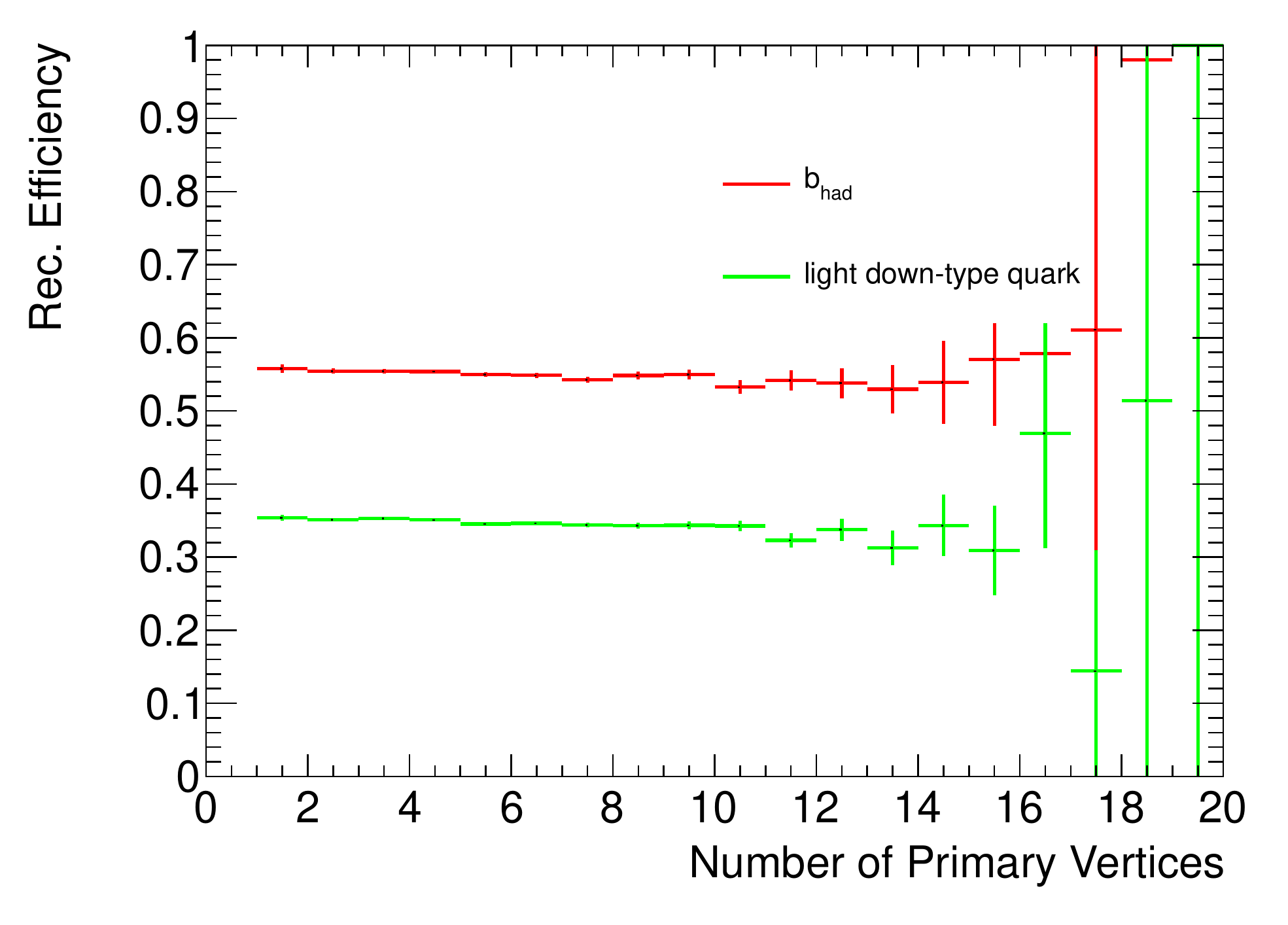}
			\label{fig:eff_npvtx}
		}	
					\subfigure[]{
		\includegraphics[width=0.45\textwidth]{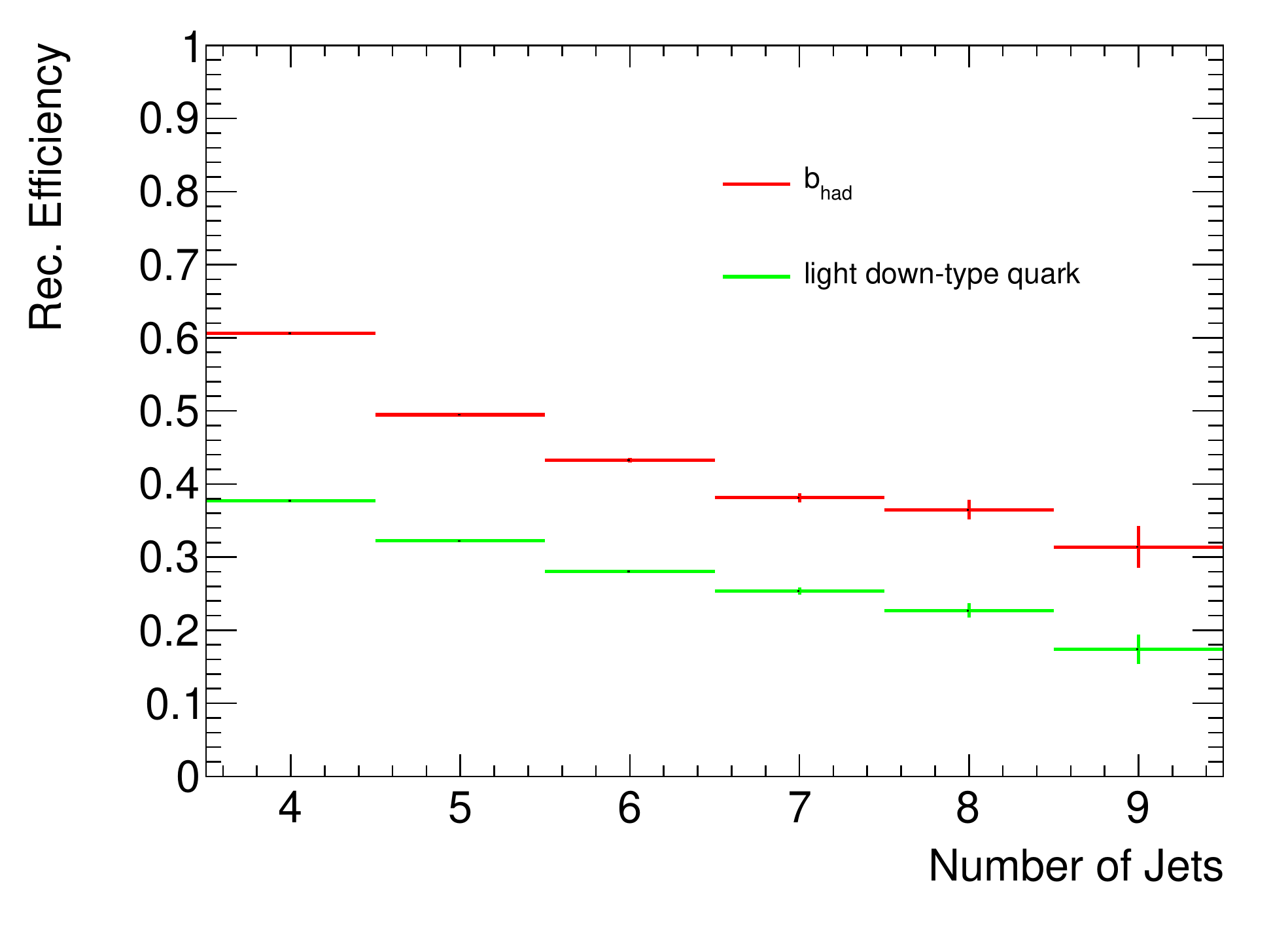}
			\label{fig:eff_njets}
		}\\
									\subfigure[]{
		\includegraphics[width=0.45\textwidth]{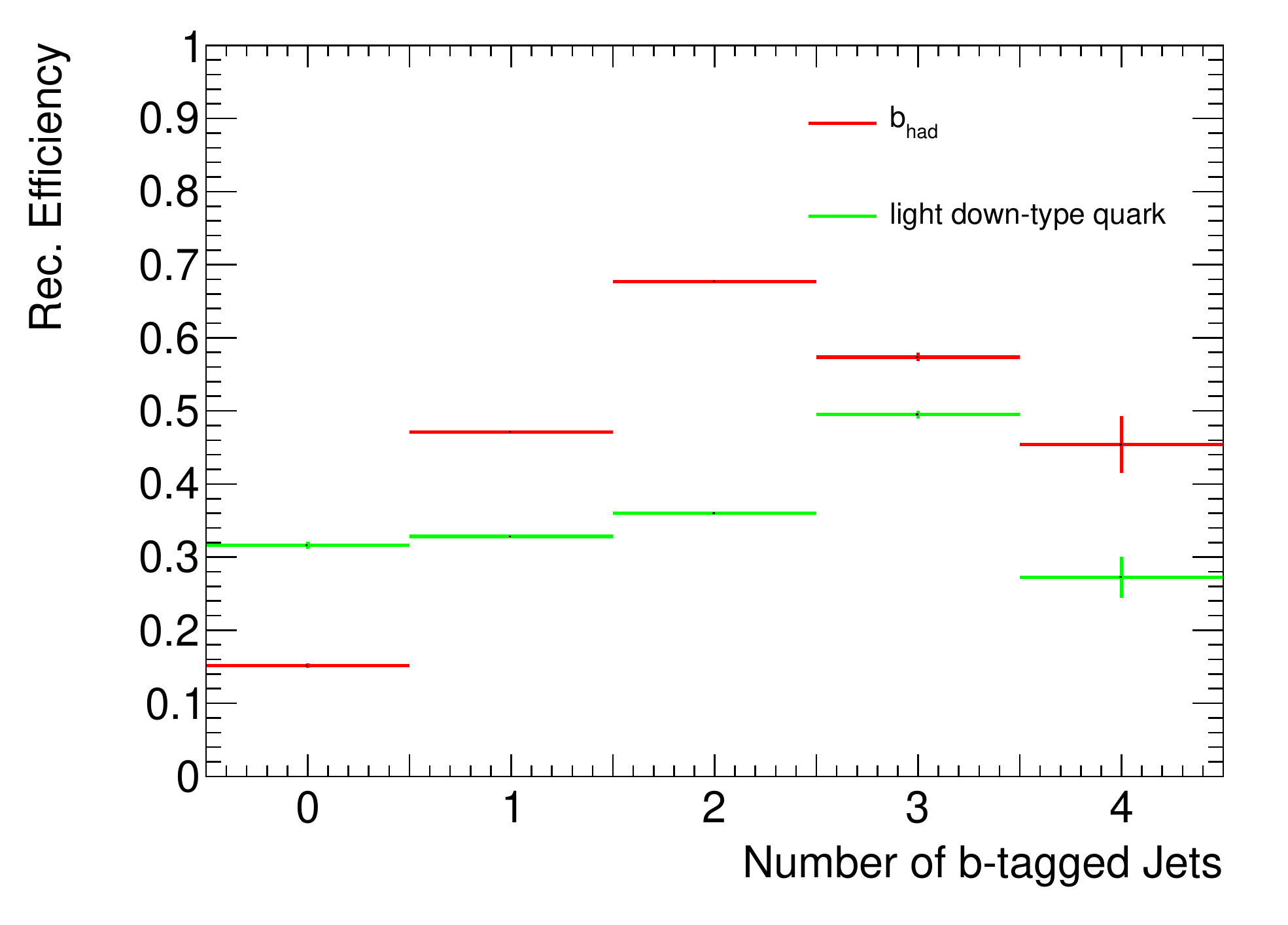}
			\label{fig:eff_ntags}
		}	
					\subfigure[]{
		\includegraphics[width=0.45\textwidth]{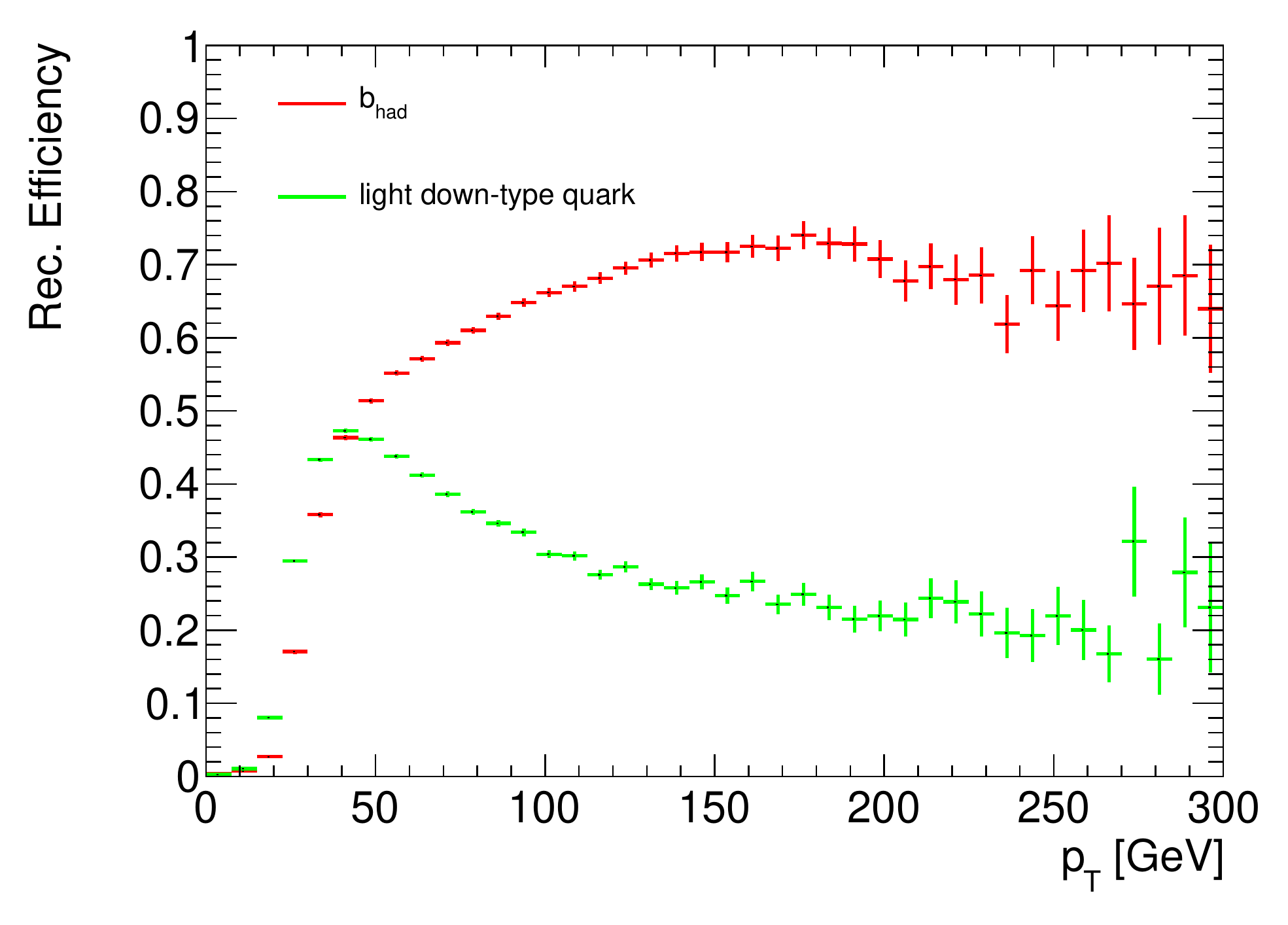}
			\label{fig:eff_pt}
		}\\
									\subfigure[]{
		\includegraphics[width=0.45\textwidth]{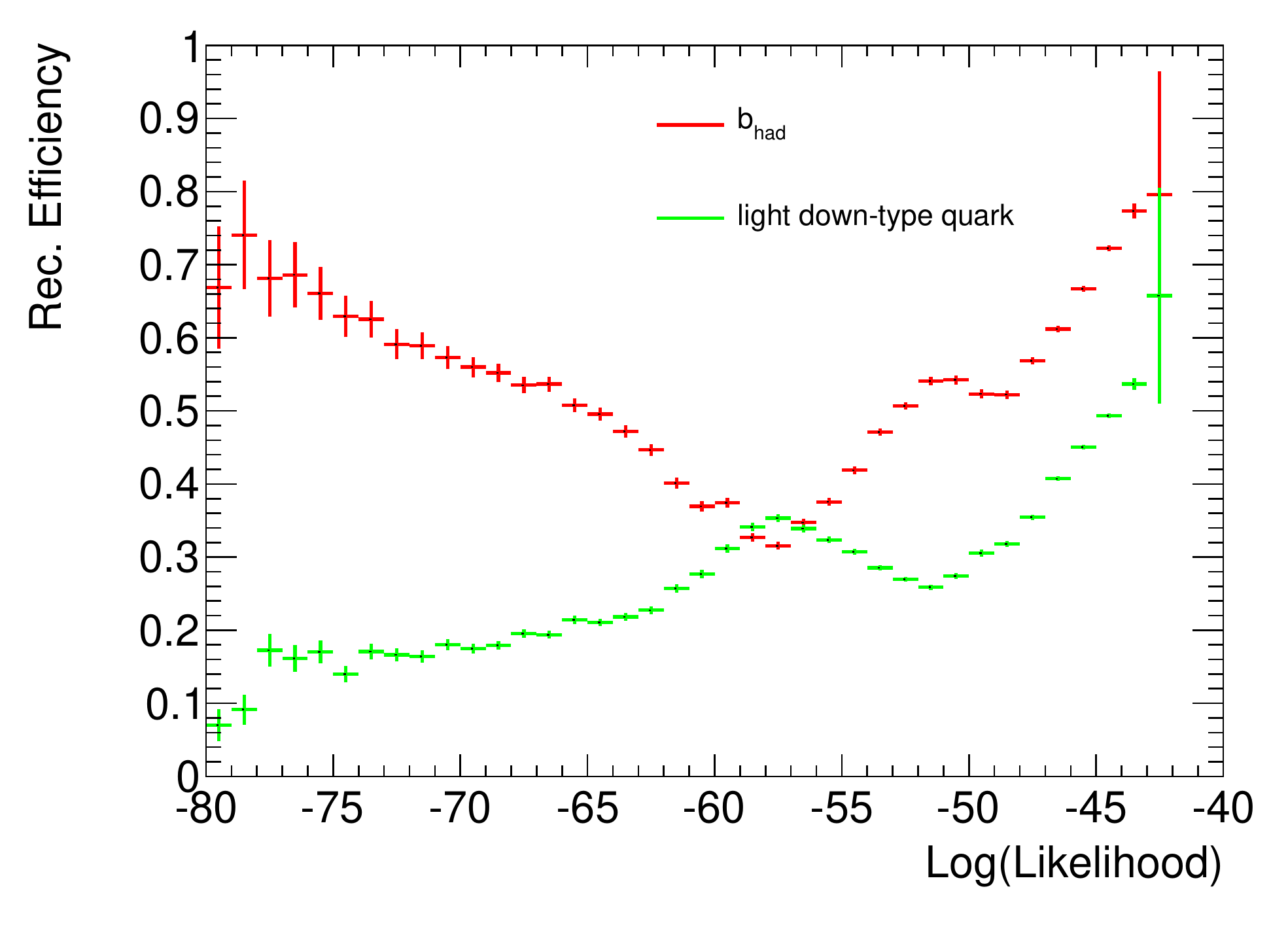}
			\label{fig:eff_LH}
		}
							\subfigure[]{
		\includegraphics[width=0.45\textwidth]{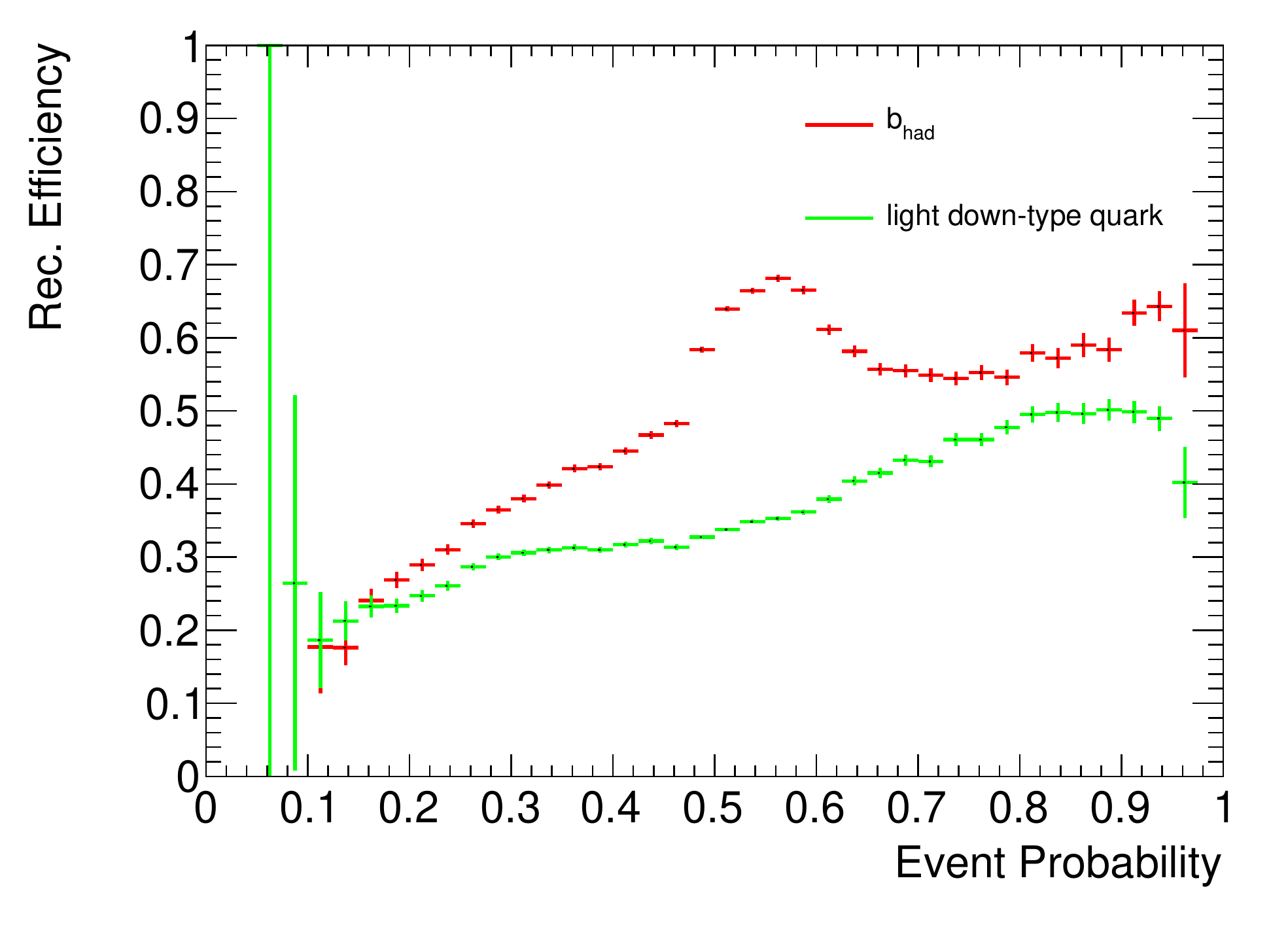}
			\label{fig:eff_evprob}
		}	

\caption{Reconstruction efficiencies of \bQ s and down-type quarks as a function of \subref{fig:eff_npvtx} the number of reconstructed primary vertices, \subref{fig:eff_njets} the number of reconstructed jets, \subref{fig:eff_ntags} the number of \btag ged jets, \subref{fig:eff_pt} the transverse momentum of the jets assigned to the \dQ\ and the \bQ, \subref{fig:eff_LH} the logarithm of the \KLFitter\ likelihood and \subref{fig:eff_evprob} the \KLFitter\ event probability. }
\label{fig:eff_diff}
\end{figure}
An ideal situation would be that each of the four LO partons hadronises into a single, non-overlapping jet (``model jet''), which is within the detector acceptance and the selection criteria. As the selection requires at least four jets and initial and final state radiation even increase this number, additional jets lead to additional combinatorial background. It is obvious that a larger number of jets increases the probability that a pile-up jet mimics the kinematics of a jet stemming from a model parton. In particular, the probability that the four model jets are the ones with the highest \pt\ -- and thus the ones which are passed to \KLFitter\ -- decreases. Figure \ref{fig:eff_njets} illustrates this. 

In Figure \ref{fig:eff_ntags} the reconstruction efficiency as a function of the number of jets passed to \KLFitter, which were \btag ged, is shown. The result reflects the reconstruction method: Maximal \bQ\ reconstruction efficiency can be reached for exactly two \btag ged jets as this clearly separates the $b$-jets from the light jets. Higher \btag\ multiplicities increase the chance of interchanging light and b-jets. For the \dQ, two \btag s separate light from \bjet s. Three tags are likely in the case of a $W \rightarrow cs$ decay for which the $c$-quark was mistagged. Its high \btag\ weight helps \KLFitter\ to separate it from the up-type quark. Four tags increase the similarity of the up- and down-type quark properties and decreases the reconstruction efficiency. 

The \pt\ dependence of the reconstruction efficiency is opposite for down-type quarks and \bQ s. In general, a high \pt\ reduces the probability of a jet to be interchanged with a low-\pt\ pile-up jet. This is observed for the \bQ\ as shown in Figure \ref{fig:eff_pt}. For the \dQ, the trend of increasing reconstruction efficiency reverses for high \pt. This fact is due to the dedicated \dQ\ reconstruction, in particular the up-/down-type separation. It is based on the -- on average -- lower \pT\ of the \dQ\ (due to the V-A structure of the weak decay vertex). In case the \pt\ of the \dQ\ is too high, the kinematics of the parent $W$ boson force the \pt\ of the up-type quark to be low. The assumption that the \pt\ of the \dQ\ is lower than the \pt\ of the up-type quark is wrong in these cases and leads to a wrong assignment.

The main quantities describing the quality of the reconstruction are the \KLFitter\ output values for the likelihood (Figure \ref{fig:eff_LH}) and the event probability (Figure \ref{fig:eff_evprob}). 
As the likelihood is a complex quantity, the reconstruction efficiency dependence on it is not intuitive. In general, the argument holds that an event that has a proper \ttbar\ topology leads to high values of the likelihood. This implies a high reconstruction efficiency. What is surprising at first sight is the fact that for low values of a likelihood ($\log(\text{likelihood}) < -55$) the reconstruction efficiency for the \bQ\ increases. 

The likelihood has several components as shown in Equation \ref{eq:klfitter_LH}. During the fitting process the jet and lepton energies are varied within the transfer functions to allow the jet combinations to match the masses of the $W$ boson and the top quark. By plotting the individual components of the likelihood (see Appendix \ref{sec:app_lhcomp} for details), one can see that the tail of low likelihood values is caused by low values of the $W$ boson and top quark mass Breit-Wigner functions. As the fit maximizes the total likelihood, the fitting process is always a trade-off between varying the transfer function value or the Breit-Wigner value. While the width of the Breit-Wigner functions is fixed and given by the width of the decaying $W$ boson and top quark, the width of the transfer functions varies. It is in particular a function of the jet energy. As described in Section \ref{sec:TFs}, the jet resolution increases with higher jet energies due to the intrinsic calorimeter resolution. Hence, high energetic jets cause less flexibility of the transfer function. In case the event has no good \ttbar\ topology, the Breit-Wigner functions will then have low values. This is the first conclusion about the low values of the likelihood. 

The second conclusion is: If low values of the likelihood are dominated by high energetic jets, this leads to a high reconstruction efficiency for the \bQ\ and a low reconstruction efficiency for the \dQ, as it was shown in Figure \ref{fig:eff_pt}. 

What remains is a discussion of the dependence of the reconstruction efficiencies on the event probability. It is shown in Figure \ref{fig:eff_evprob}. In case of a high event probability one particular permutation exists for which -- and only for this permutation -- the assumptions of a \ttbar\ event topology match the observations. This is demonstrated by high correlation of event probability and reconstruction efficiency. The peak of the \bQ\ reconstruction efficiency around 0.5 is also plausible. Assuming the case of indistinguishable up- and down-type quarks, the two permutations for which the up- and down-type jets are permuted have equal event probabilities. In the best case where these permutations are clearly favoured, the event probability is about 0.5. In the case of indistinguishable up- and down-type quarks these are likely up and down quarks, not charm and strange quarks.\footnote{In the latter case the quark jets could be separated by the \btag\ weight.} If the pair of $W$ jets does not contain a heavy flavour charm quark, there is less risk that the charm quark is interchanged with the hadronic \bQ. This fact connects the high reconstruction efficiency with the similarity of the up- and down-type quark signatures and the peak of the event probability at 0.5. 

Following this thought, a high event probability implies a clear separation between the light up- and down-type quark. This can only be reached if it is a charm/strange quark pair and the \btag\ weight can be utilized. The consequence of the third jet carrying a high \btag\ weight is a higher probability of interchanging it with the \bQ. Hence, the \bQ\ reconstruction efficiency will drop. This is exactly what is observed. 

In terms of future improvements and for a proper understanding of the reconstruction details it is useful to know what went wrong in case of a misreconstruction of the \dQ\ or the \bQ. Figure \ref{fig:matchpartner} shows to which parton a reconstructed model jet of a certain type ($b_{\text{had.}}$, $b_{\text{lep.}}$, light up-type, light down-type) could be matched.
 \begin{figure}[htbp]
 	\centering
						\subfigure[]{
		\includegraphics[width=0.7\textwidth]{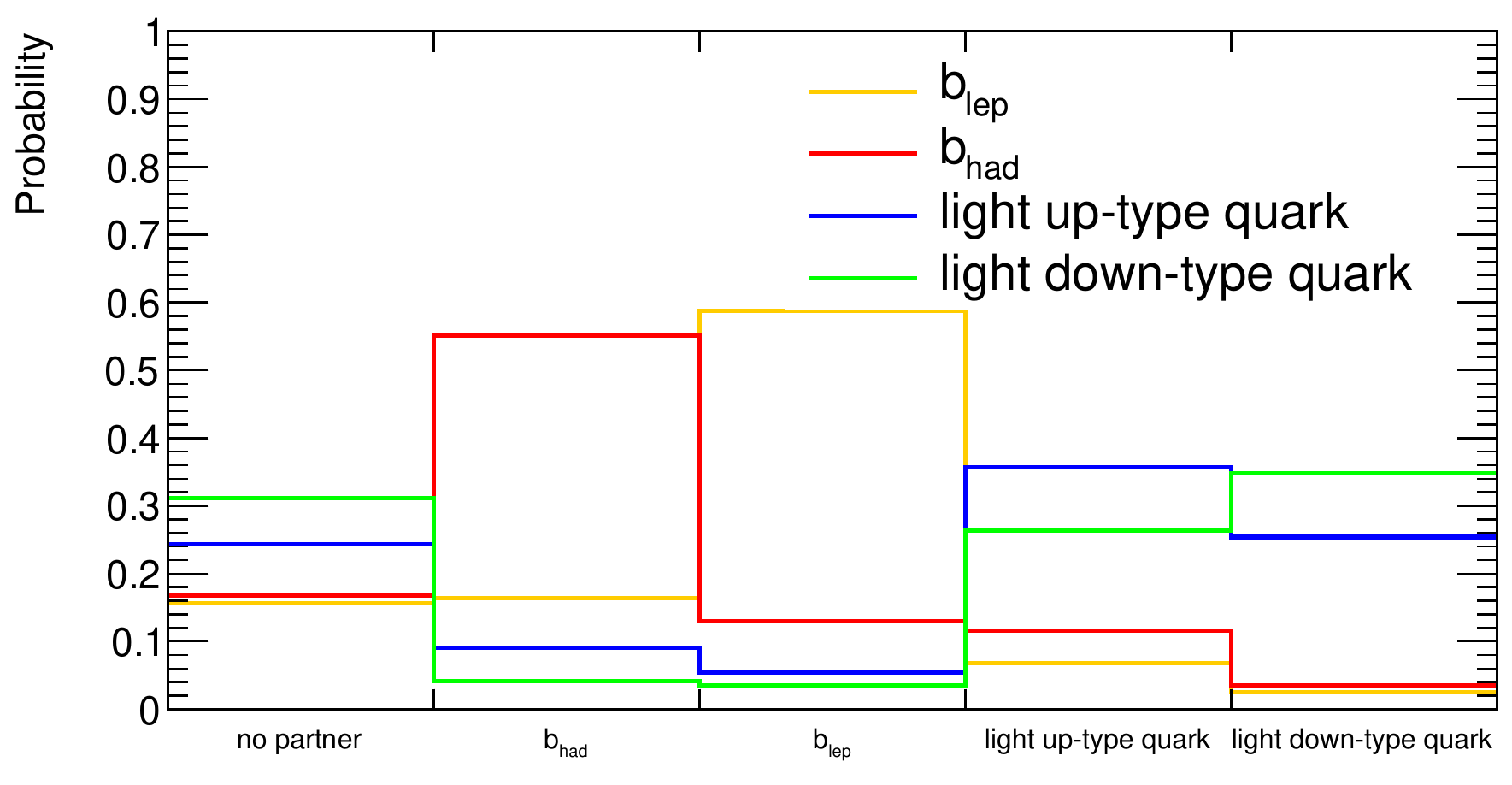}
			\label{fig:matchpartner}
		}	
					\subfigure[]{
		\includegraphics[width=0.7\textwidth]{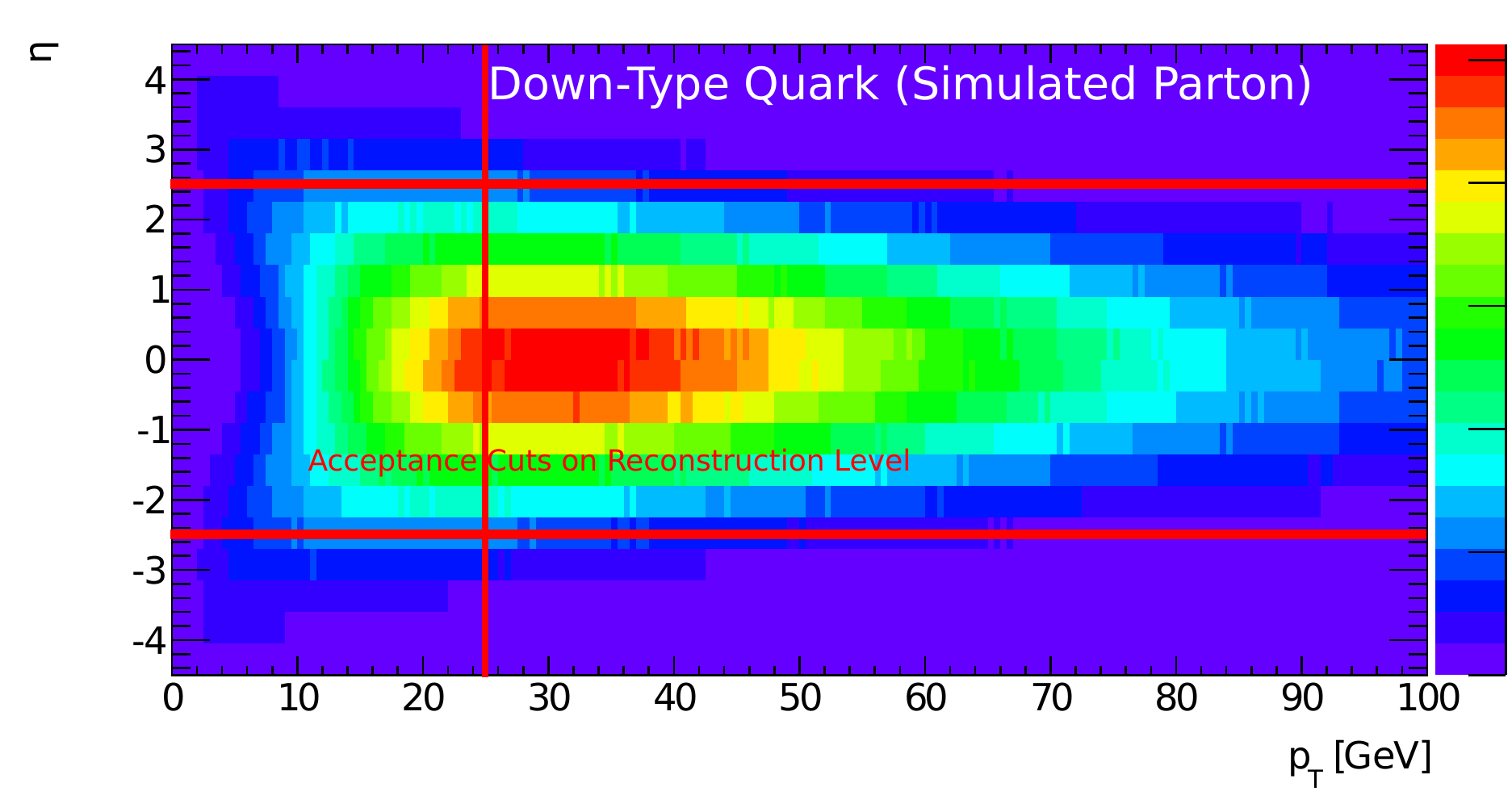}
			\label{fig:truth_dQ_acc}
		}

\caption{\subref{fig:matchpartner} Jet allocated to model partons by \KLFitter\ (coloured lines) and the partons to which they were matched (x-axis). \subref{fig:truth_dQ_acc} $\eta$ and \pt\ of down-type quarks for the full phase space, before event selection. The red lines indicate the cuts applied on reconstruction/detector level.}
\label{fig:whatgoeswrong}
\end{figure}
 Each jet can be matched to one of the model partons or to none at all. The success of the reconstruction is reflected in the fact that the most probable option is that a jet was assigned to the corresponding model parton. It is also visible that the light jets with their \pt\ being lower than the ones of the $b$-jets suffer most from acceptance and selection cuts. As expected the two light quarks are interchanged quite often. But it is remarkable that it is more likely for a \dQ\ to be not reconstructed at all than to be interchanged with the up-type quark. The fact that the \dQ\ has on average a lower \pt\ explains the higher probability that its jet is not within the four jets which are used for reconstruction (as these have the highest \pt). Figure \ref{fig:truth_dQ_acc} shows the $\eta$ and \pt\ distribution of the down-type quark before a selection is applied. The horizontal and vertical lines indicate the cuts applied on detector level ($\pt > 25~\GeV$ and $\abseta < 2.5$).

The dependences of the reconstruction efficiency suggest to place cuts on certain output quantities to create a sub-sample with higher reconstruction efficiency. Higher efficiencies always come along with a reduction in statistics. Furthermore, certain cuts have different effects on the reconstruction efficiencies of the \dQ\ and the \bQ. The following cuts have been evaluated in terms of efficiency gain and statistics loss:
\begin{itemize}
\item {\bf Top Quark Mass Cut} The invariant mass of the three jet system allocated to the decay products of the hadronically decaying top quark is required to satisfy $\left| m_{t} - m_{jjj}\right| < 35~\GeV$.
\item {\bf $W$ Boson Mass Cut} The invariant mass of the two jet system allocated to the decay products of the $W$ boson of the hadronically decaying top quark are required to satisfy \mbox{$\left| m_{W} - m_{jj}\right| < 25~\GeV$}.
\item {\bf Top Quark and $W$ Boson Mass Cut} Both of the former two requirements have to be satisfied.
\item {\bf Likelihood Cut} The logarithm of the likelihood output of \KLFitter\ must be larger than -50.
\item {\bf Event Probability Cut} The event probability of \KLFitter\ has to be larger than 0.5.
\item {\bf Jet Multiplicity Cut} Exactly four jets must be present in the selected event.
\item {\bf B-Tag Multiplicity Cut} At least two of the reconstructed jets have to be tagged as $b$-jets.
\end{itemize}
The results are visualized in Figure \ref{fig:opticuts}.  The distributions of the likelihood and the event probability is shown in Figure \ref{fig:klfitter_control3} in the context of the validation of the reconstructions process. 

\begin{figure}[htbp]
 	\centering
						\subfigure[]{
		\includegraphics[width=0.7\textwidth]{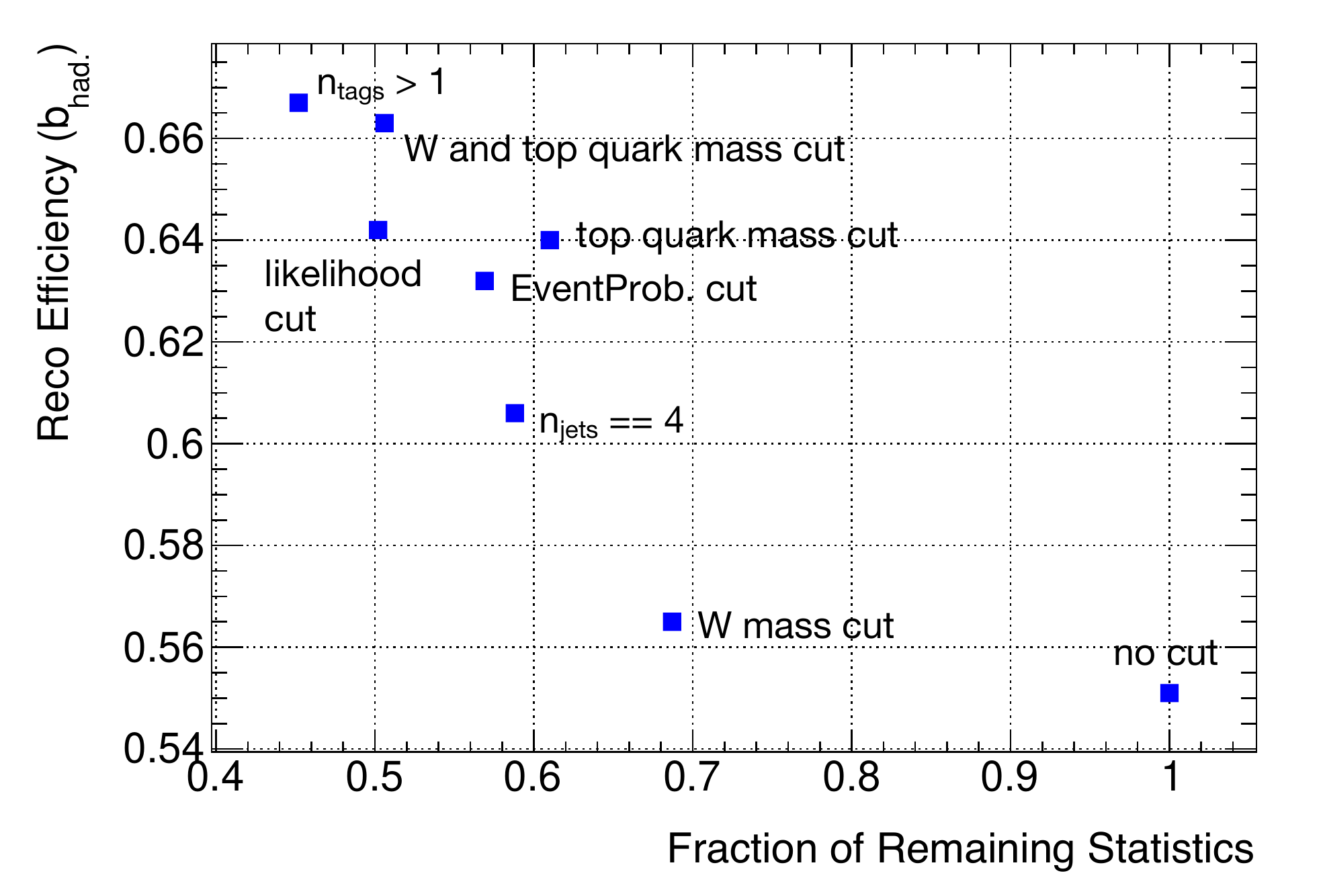}
			\label{fig:opticuts_bQ}
		}	
					\subfigure[]{
		\includegraphics[width=0.7\textwidth]{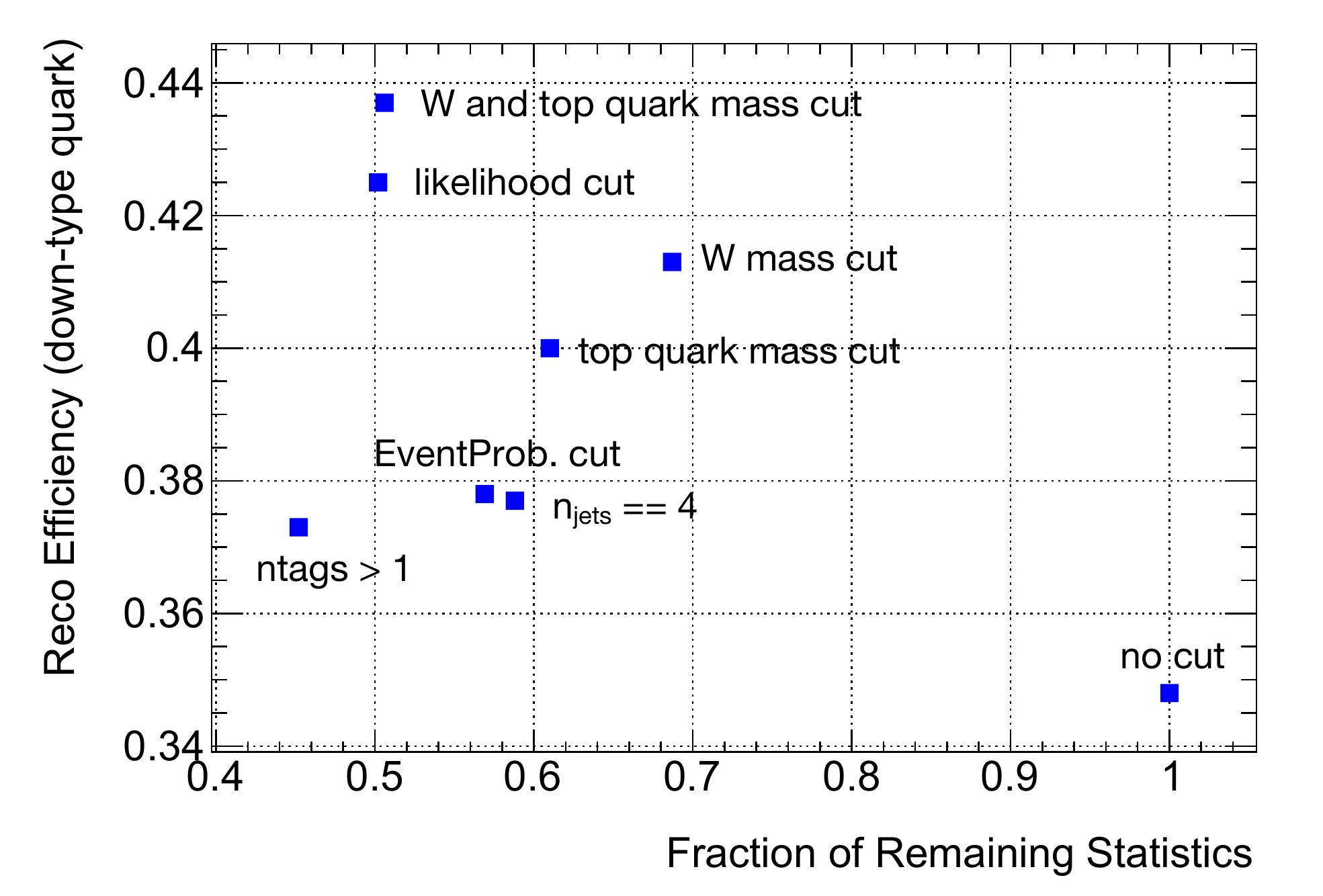}
			\label{fig:opticuts_dQ}
		}

\caption{Cuts on event properties and their effects on the statistics and reconstruction efficiencies for the \subref{fig:opticuts_bQ} \bQ\ and \subref{fig:opticuts_dQ} \dQ.}
\label{fig:opticuts}
\end{figure}

A combination of the $W$ boson and top quark mass cut is found to be the most powerful cut for both spin analysers. Other cuts can be considered if an individual analyser is going to be optimized. In this case also the specific cuts can be further tuned. 

In this analysis no cut was placed. Instead, the selected samples were split into subsamples according to their jet multiplicity and \btag\ multiplicity. This choice is further discussed in Section \ref{sec:channels}.

\section{KLFitter Setup}
\KLFitter\ can be used with several options, which were presented in the last sections. The actual setup which was used to perform the analysis presented in this thesis is briefly presented. 

The number of jets passed to \KLFitter\ was set to four. These four jets were selected by a \pt\ ordering of all reconstructed jets. The jets with the highest \pt\ were used. This was tested to lead to the highest reconstruction efficiency. Even though this excludes a proper event reconstruction where the jets matching the four model partons are not within the four jets with the highest \pt, it also decreases combinatorial background. The latter effect was found to be dominating.

No $b$-jet veto method was used. Instead, the \btag ging information was taken into account via the working point mode and the additional up/down-type quark separation. 

The top quark mass in used in the \KLFitter\ likelihood was fixed to 172.5 GeV. An alternative option is to leave it as a free parameter. The former option was chosen as it leads to a higher reconstruction efficiency. Using the free parameter leads to a benefit only if the fitted top quark mass is utilized in an analysis and needs to be unbiased. This was not the case in the present studies. 

\section{Comparison to Other Reconstruction Methods}
\label{sec:recocomp}
It was shown that the \KLFitter\ with the up-/down-type quark separation extension provides good reconstruction efficiencies. As several other \ttbar\ reconstruction methods exist, it is worth to check their reconstruction efficiencies. Two alternatives to \KLFitter\ will be introduced and compared in the following section. The introduction will be limited to the description of the reconstruction of the hadronically decaying top quark. It has to be stressed that several variations of these methods exist and that the descriptions provided here only represent one specific choice. Furthermore, none of these methods was optimized for \bQ\ or \dQ\ reconstruction. 

\subsection{\ptmax}
The basic idea of the \ptmax\ method is that the three-jet system with the highest transverse momentum corresponds to the three jets from the hadronically decaying top. All reconstructed jets are considered. Out of the three selected top quark jets, the jet pair with an invariant mass closest to the $W$ mass is selected as $W$ boson jet pair. A veto on \btag ged jets is set which can cause the reconstruction method to provide no solution. 
Both of the $W$ boson jets are then boosted into the hadronic top quark rest frame. The jet with the lower energy is used as \dQ\ jet and the top quark jet not associated with a $W$ boson as \bQ\ jet. 

\subsection{Topological Method}
At least two \btag ged jets are required for the topological method. Out of all non-tagged jets the pair with an invariant mass closest to the $W$ mass is selected as $W$ boson jet pair. The invariant mass of all three-jet systems consisting of the two $W$ boson jets and a third tagged jet is calculated. The three-jet system with the invariant mass closest to the top quark mass defines the \bQ\ jet as non-$W$ boson jet. Within the two $W$ boson jets the \dQ\ jet is chosen the same way as for the \ptmax\ method. 

\subsection{Performance Comparison}
Both the \ptmax\ and the topological method were compared to \KLFitter\ under the following conditions: The event passes the selection as described in Section \ref{sec:eventselection} and must be a real $\ttbar \rightarrow \ljets$ event. The results of matched \dQ\ and \bQ\ jets are provided in Table \ref{tab:efficomp}. As some reconstruction methods have additional constraints on the event, they only provide solutions for a certain fraction of the selected events (last column). The reconstruction efficiencies are provided both with respect to the number of events for which a solution exists and with respect to all selected events.
\begin{table}[htbp]
\begin{center}
\small
\begin{tabular}{|c|c|c|c|}
\hline
{}&\multicolumn{2}{c|}{Rec. Efficiency (w.r.t. all Events)} & Remaining Stat.\\
{}&\bQ\ & \dQ\ &{}\\
\hline
{\KLFitter} & 55\,\%  (55 \,\%) & 35\,\%  (35\,\%)& 100\,\%\\
{\ptmax} & 44\,\% (36\,\%)  & 32\,\% (26\,\%)& 82\,\%\\
{Topological Method} & 42\,\% (21\,\%)&29\,\% (15\,\%)  & 51\,\%\\
\hline
\end{tabular}
\end{center}
\caption{ \ttbar\ reconstruction efficiency for different algorithms. The efficiencies are given with respect to all selected \ttbar\ events (in parentheses) and to events passing additional criteria given by the algorithm. The loss of statistics due to these criteria is also quoted.}
\label{tab:efficomp}
\end{table}

The comparison shows that the used setup of \KLFitter\ does not only perform best but does furthermore not cause a reduction of statistics.

\chapter[Analysis Strategy]{Analysis Strategy}
Many ways to measure the spin correlation in \ttbar\ events exist. They access different components of the \ttbar\ spin matrix $\widehat{C}$, explained in Section \ref{sec:SC}. For dileptonic decays of \ttbar\ pairs there is a clear preference at the LHC. Charged leptons have the best spin analysing power and can be measured with a high precision. As further the measurement of the azimuthal angle between the two charged leptons in the laboratory frame does not need any event reconstruction, $\Delta \phi (l^{+}, l^{-})$ was the observable of choice for the first LHC results \cite{ATLAS_spin_paper, CMS_spin_paper} and also lead to the observation of spin correlation in \ttbar\ events \cite{ATLAS_spin_paper}. 

In the \ljets\ channel, however, the choice is not straightforward. The selection of both, the observable and the spin analyser are motivated in the next session, followed by cross checks of the correct reconstruction of the spin analyser as explained in sections \ref{sec:KLFitter} and \ref{sec:udsep}. After that, the measurement of the observable is explained, including the usage of nuisance parameters for systematic uncertainties as well as a correction dedicated to mismodelling of the jet multiplicity in the MC signal generator.
The chapter concludes with a validation of the linearity of the method, a discussion about the correlation of \dQ\ and \bQ, and the effects of the correlation on the measured results.

\section{Choice of Observable} 
The fact that the measurement of the azimuthal angle between two spin analysers in the laboratory frame is suitable for \ttbar\ production at the LHC is also valid for the \ljets\ channel. In principle, there exists also a hadronic analyser with the same spin analysing power as a charged lepton, namely the \dQ\ (see Table \ref{tab:anapower}). Even though its measurement has slightly worse energy and angular resolution compared to a charged lepton, the true challenge lies in its correct identification. 
In Chapter \ref{sec:selandreco}, an advanced reconstruction technique was introduced, which is able to reconstruct the \dQ\ analyser. As this method reconstructs the full \ttbar\ event, even the more complex observables could in principle be studied. However, as these require the correct reconstruction of all spin analysers, the separation power between signal with spin correlation according to the SM and with uncorrelated spins are significantly diluted. 

By using simulated \ttbar\ events, the separation power of a SM spin correlation sample and a sample of uncorrelated \ttbar\ pairs was tested at the detector level for different spin correlation observables. It was found to be significant only for the \dphi\ distributions. Thus, \dphi\ was chosen as observable for the \ljets\ channel. 

Next to the most powerful hadronic analyser, the \dQ, the second most powerful analyser, the \bQ, was evaluated at parton level. The respective distributions for the full phase space, without any object cuts on \pt\ and $\eta$, are shown in Figure \ref{fig:dphi_truth}.
 \begin{figure}[htbp]
 	\centering
			\subfigure[]{
		\includegraphics[width=0.45\textwidth]{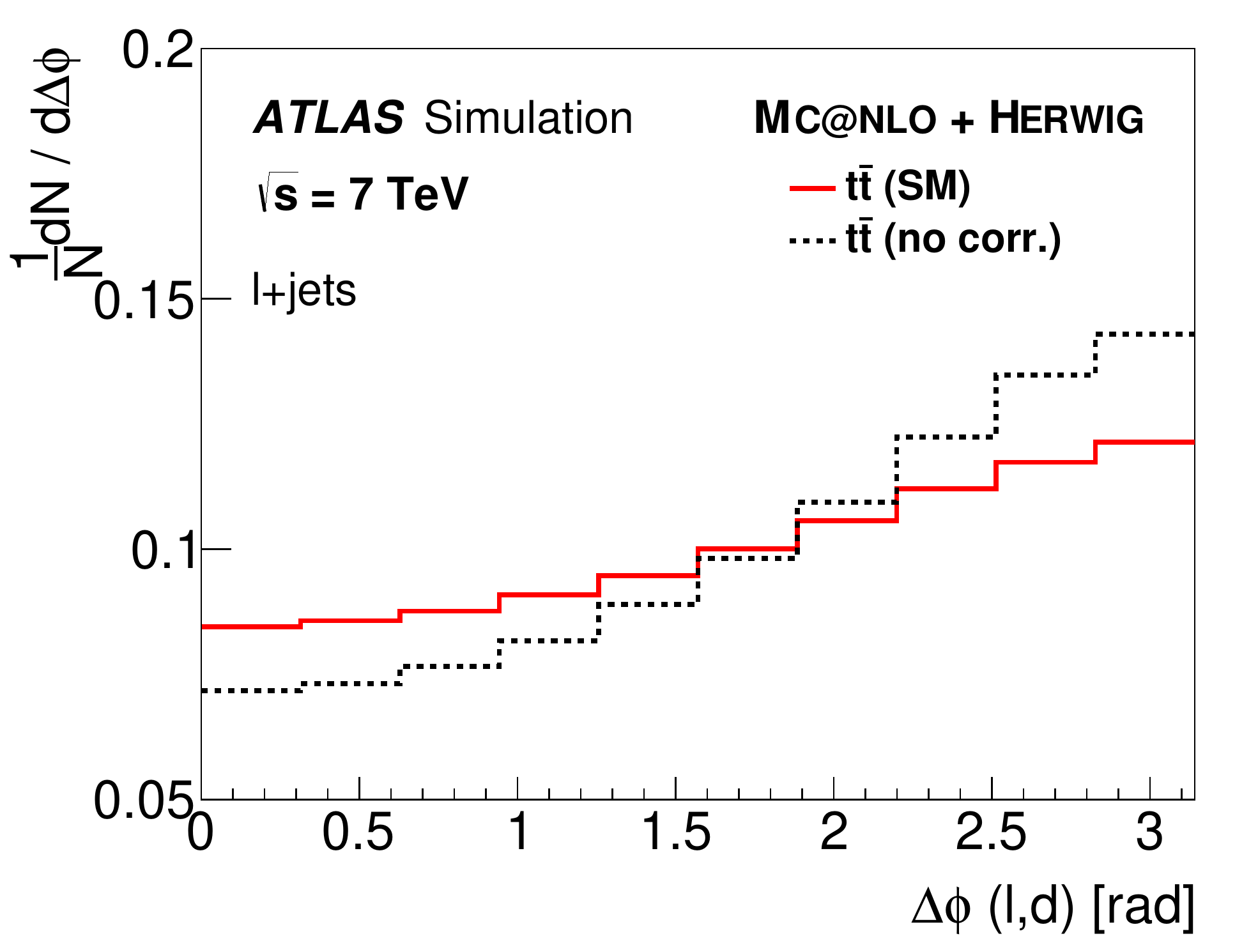}
			\label{fig:truth_dQ}
		}
					\subfigure[]{
		\includegraphics[width=0.45\textwidth]{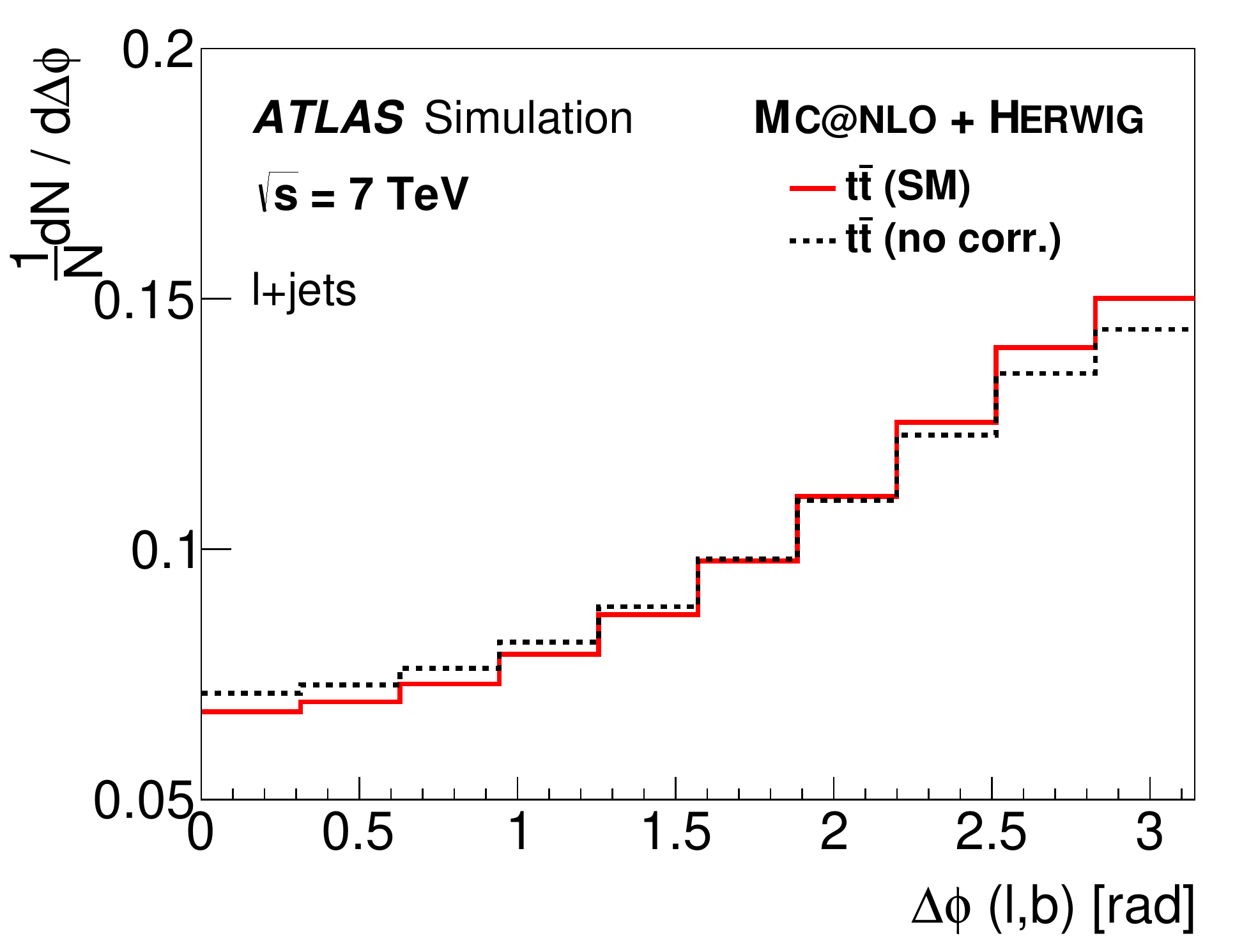}
			\label{fig:truth_bQ}
		}		
\caption{Distribution of the azimuthal angle between the charged lepton and the \subref{fig:truth_dQ} \dQ\ and the \subref{fig:truth_bQ} \bQ\ in the laboratory frame \cite{ueberpaper}. The distributions show the full phase space on parton level without usage of cuts on \pt\ and \eta\ of the objects.}
\label{fig:dphi_truth}
\end{figure}
The larger spin analysing power of the \dQ\ in \ref{fig:truth_dQ} with respect to the \bQ\ in \ref{fig:truth_bQ} is clearly visible. Also, the reversed sign on the analysing power is reflected in the fact that the deviations of SM expectation and uncorrelated \ttbar\ pairs go in opposite directions. This fact plays an important role in this analysis. A mismodelling of \ttbar\ kinematics in a MC generator needs to be distinguished from a deviation in the \ttbar\ spin correlation. 
Such a distinction is possible due to the opposite shape changing effects\footnote{While for the \dQ\ the \dphi\ distribution is flatter for correlated \ttbar\ pairs than for uncorrelated ones, it is steeper for the \bQ.} of spin correlation on the two analysers.

The quantities of interest are the \dphi\ distributions on the detector level. Due to the limitations of the reconstruction efficiency and the acceptance, the separation power between the scenarios with SM spin correlation and uncorrelated \ttbar\ pairs are diluted. The \dphi\ distributions at the detector level are shown in Figure \ref{fig:dphi_reco}.
 \begin{figure}[htbp]
 	\centering
			\subfigure[]{
		\includegraphics[width=0.45\textwidth]{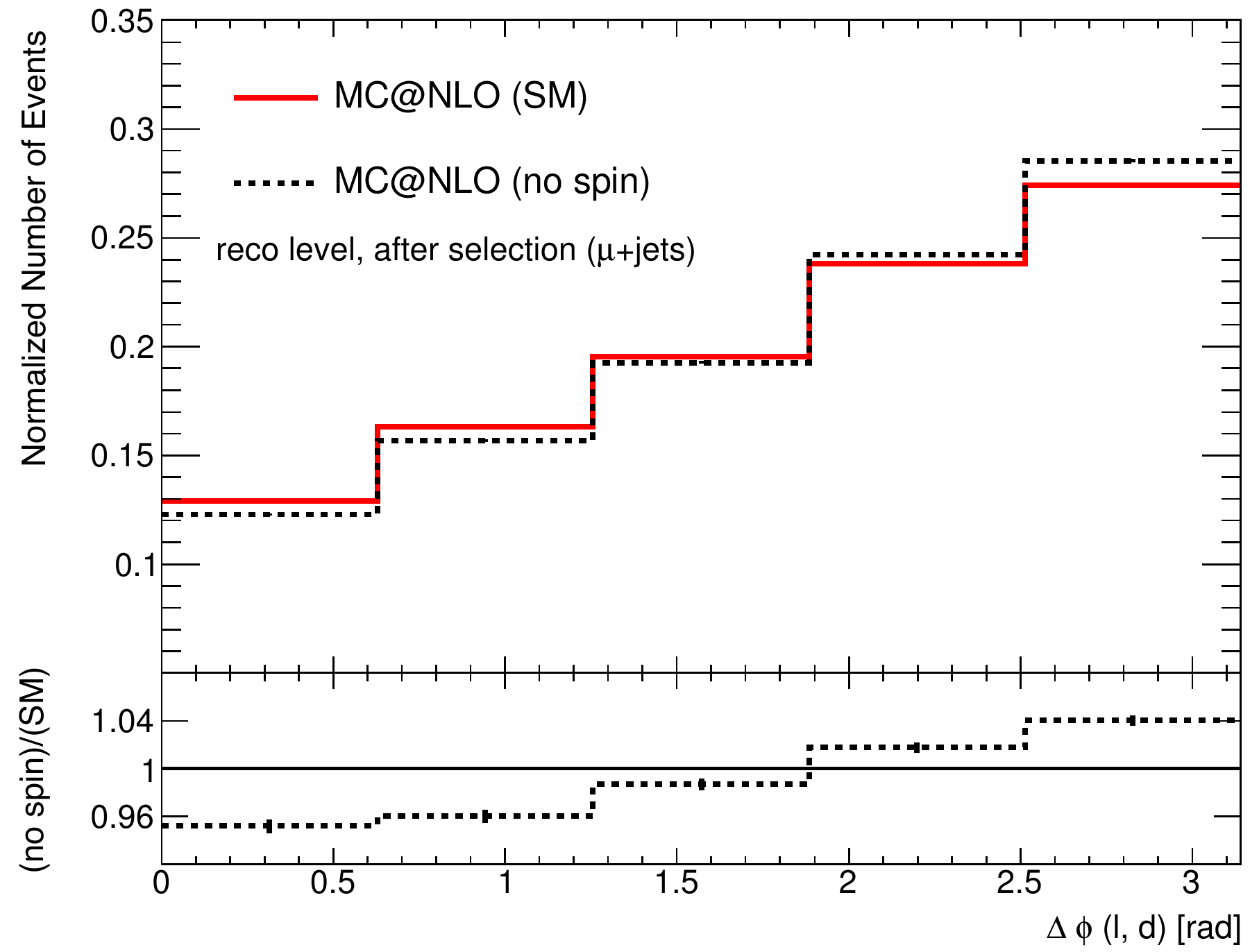}
			\label{fig:reco_dQ}
		}
					\subfigure[]{
		\includegraphics[width=0.45\textwidth]{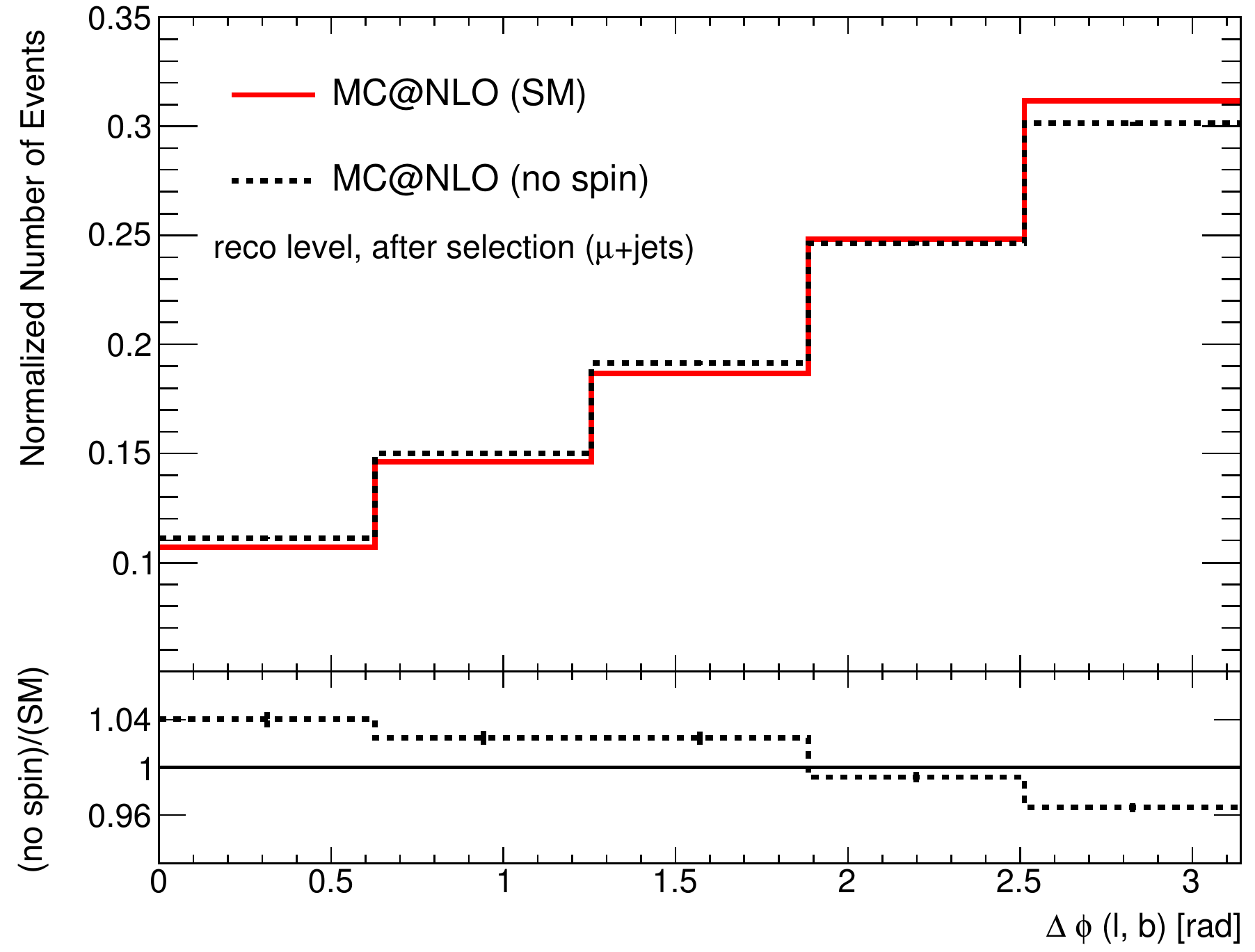}
			\label{fig:reco_bQ}
		}		
\caption{Distribution of the azimuthal angle between the charged lepton and the jet assigned to the \subref{fig:reco_dQ} \dQ\ and the \subref{fig:reco_bQ} \bQ\ in the laboratory frame. The distributions use reconstructed quantities in the \mujets\ channel after the event selection is applied. The shown uncertainties (barely visible) represent the MC statistics uncertainty.}
\label{fig:dphi_reco}
\end{figure}
It is remarkable that both quantities have comparable separation power. This shows that the reduced spin analysing power of the \bQ\ (as shown in Figure \ref{fig:dphi_truth}) is at least partially compensated by the larger reconstruction efficiency as summarized in Table \ref{tab:simpleeff}.

As a consequence, both the \dQ\ and the \bQ\ are used as analysers. It is shown in Section \ref{sec:correlation} that  treating both analysers as uncorrelated quantities is valid. From now on, the expression ``\dQ'' will refer to the jet associated to the \dQ\ and accordingly for the \bQ, if not stated otherwise.

\section{Spin Analyser Validation} 
\label{sec:anaval}
The reconstruction of the hadronic analysers as explained in Section \ref{sec:udsep} is an advanced reconstruction technique and based on several input quantities. The performance of the reconstruction must be the same on data and on simulated events. Otherwise, a reconstruction efficiency, which is different in data and simulation, can mimic a higher or lower spin correlation. 
The validation of the reconstruction procedure via a comparison of the input and output quantities of the reconstruction algorithm is the subject of this section. 

As \btag ging is important for both the explicit tagging of jets as $b$-jets as well as for the continuous distribution of the weights for up/down-type separation, these quantities need to be checked for a good agreement of prediction and data. In the first two columns of Figure \ref{fig:klfitter_control1}, the data/MC agreement of the \btag\ weight (logarithmic y-axis) and the number of \btag ged jets is shown. Here and in the following validation plots, the upper row shows results of the \ejets\ channel and the lower row those of the \mujets\ channel. No significant deviation of the data from the prediction can be observed. The slight slope in the \btag\ multiplicity is corrected during the fit by using the \btag\ efficiency uncertainties as nuisance parameters. 
Another quantity of interest is the \pt\ of the \dQ\ jet. In the third column of Figure \ref{fig:klfitter_control1}, the \pt\ spectrum of the reconstructed \dQ\ jet is shown. The kinematic mismodelling of \mcatnlo\ as seen in Figures \ref{fig:subjets_4incl_1tags} and \ref{fig:kinematic_4incl_1tags} propagates to the \dQ\ jet. No additional mismodelling or artificial correction by the reconstruction is observed. A good agreement of the jet \pt\ spectrum is observed using \powheg+\pythia\ (see Figure \ref{fig:app_jetpt} in the Appendix). 
\begin{figure}[htbp]
\begin{center}
\includegraphics[width=0.3\textwidth]{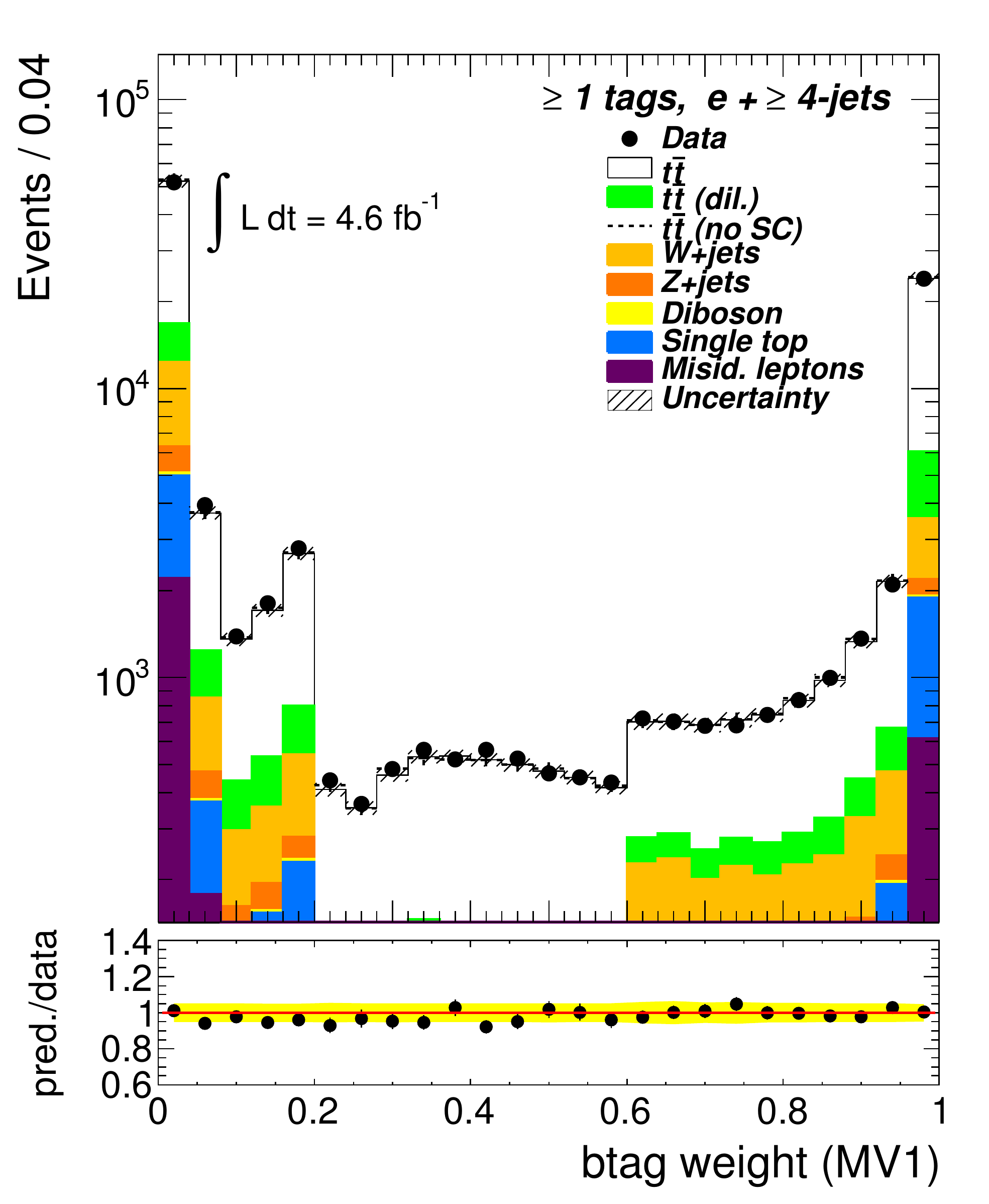}
\includegraphics[width=0.3\textwidth]{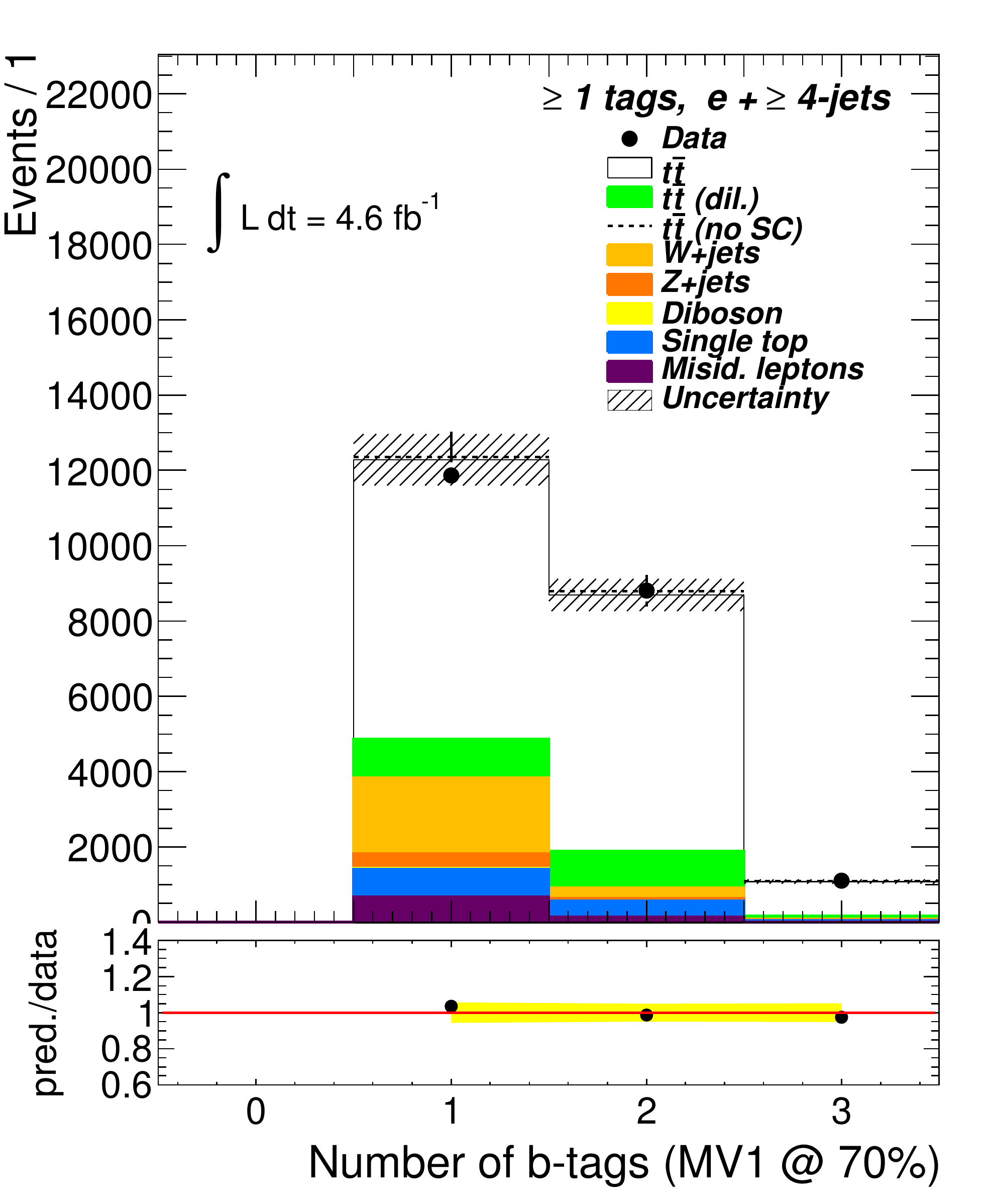}
\includegraphics[width=0.3\textwidth]{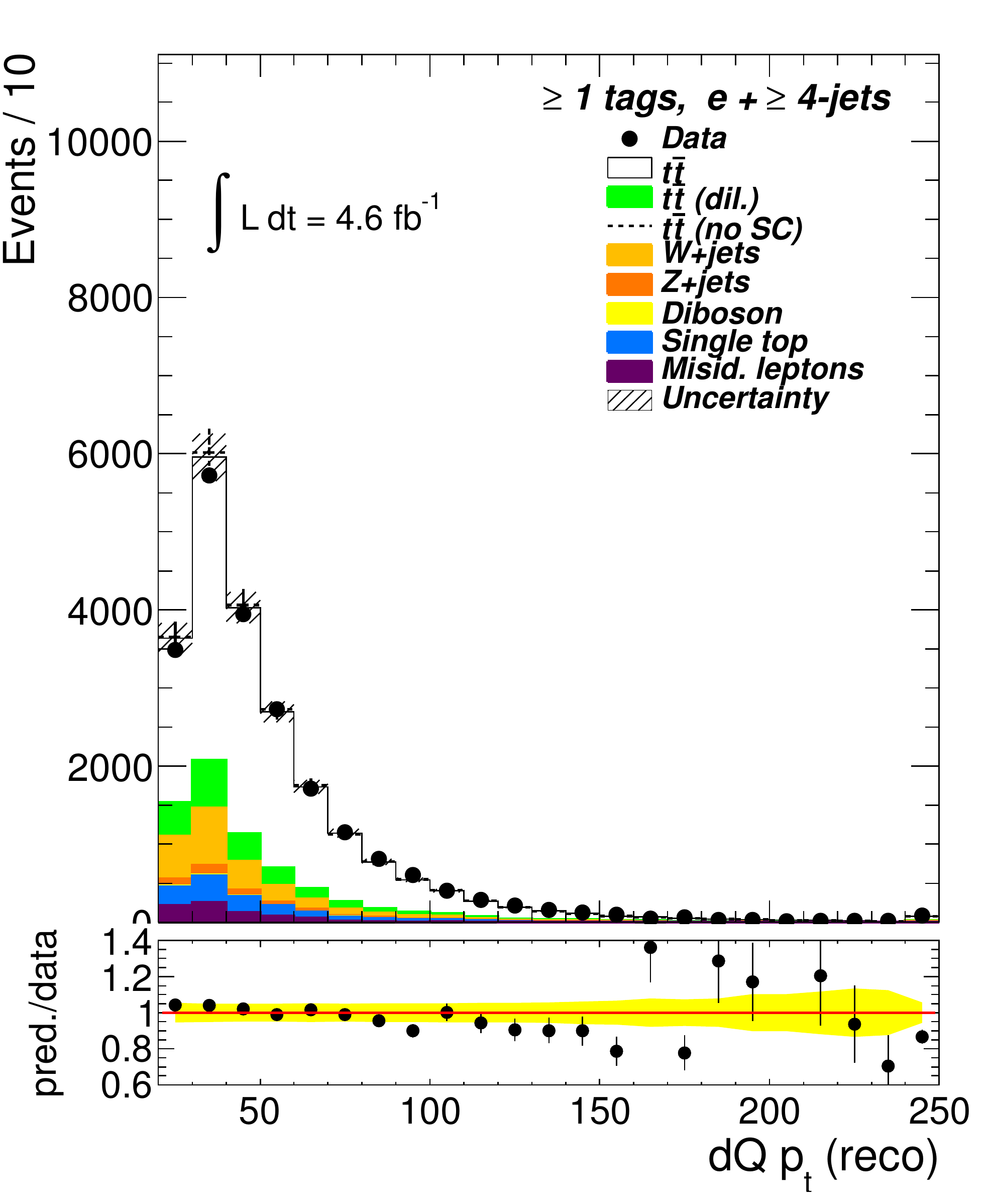}\\
\includegraphics[width=0.3\textwidth]{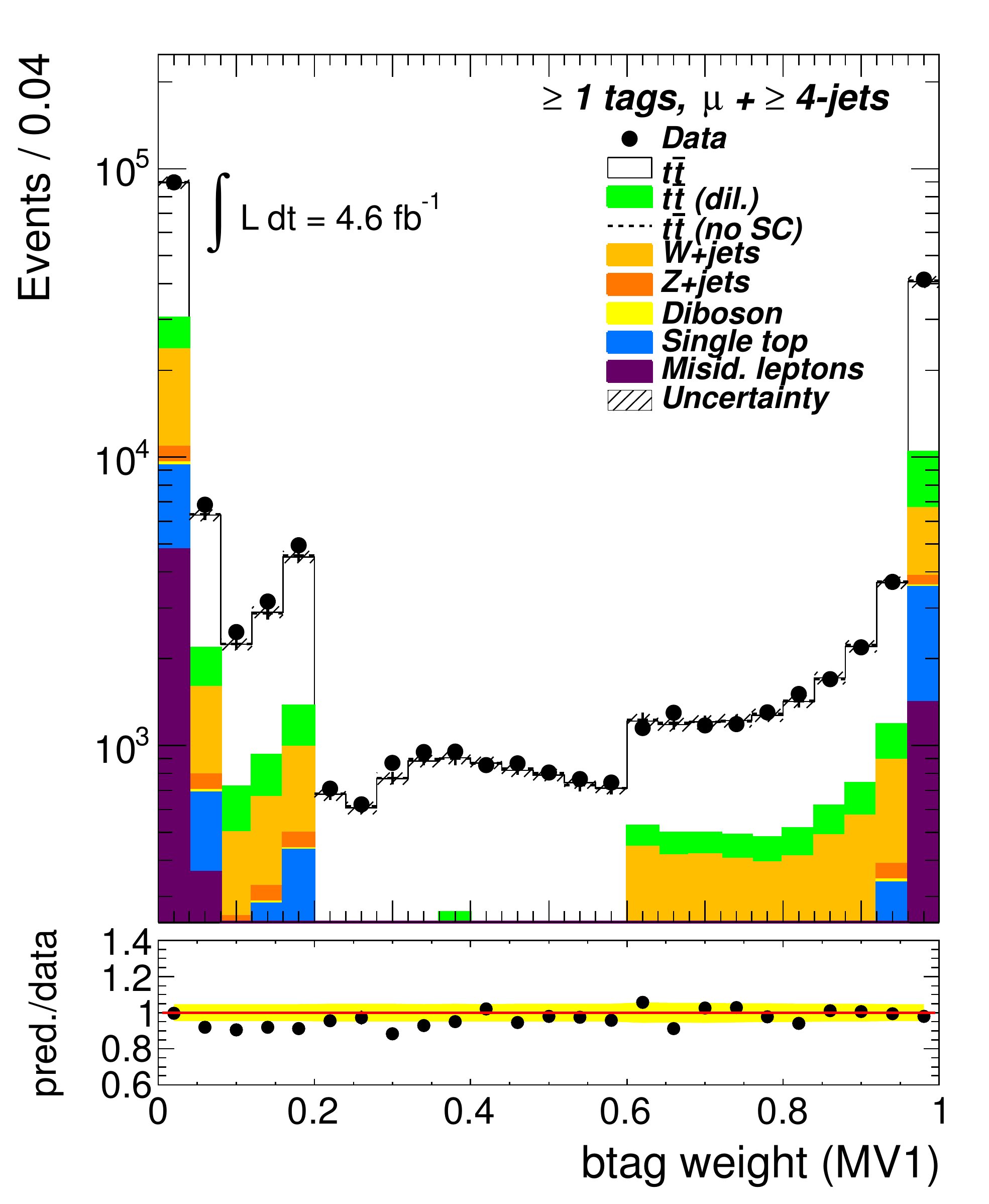}
\includegraphics[width=0.3\textwidth]{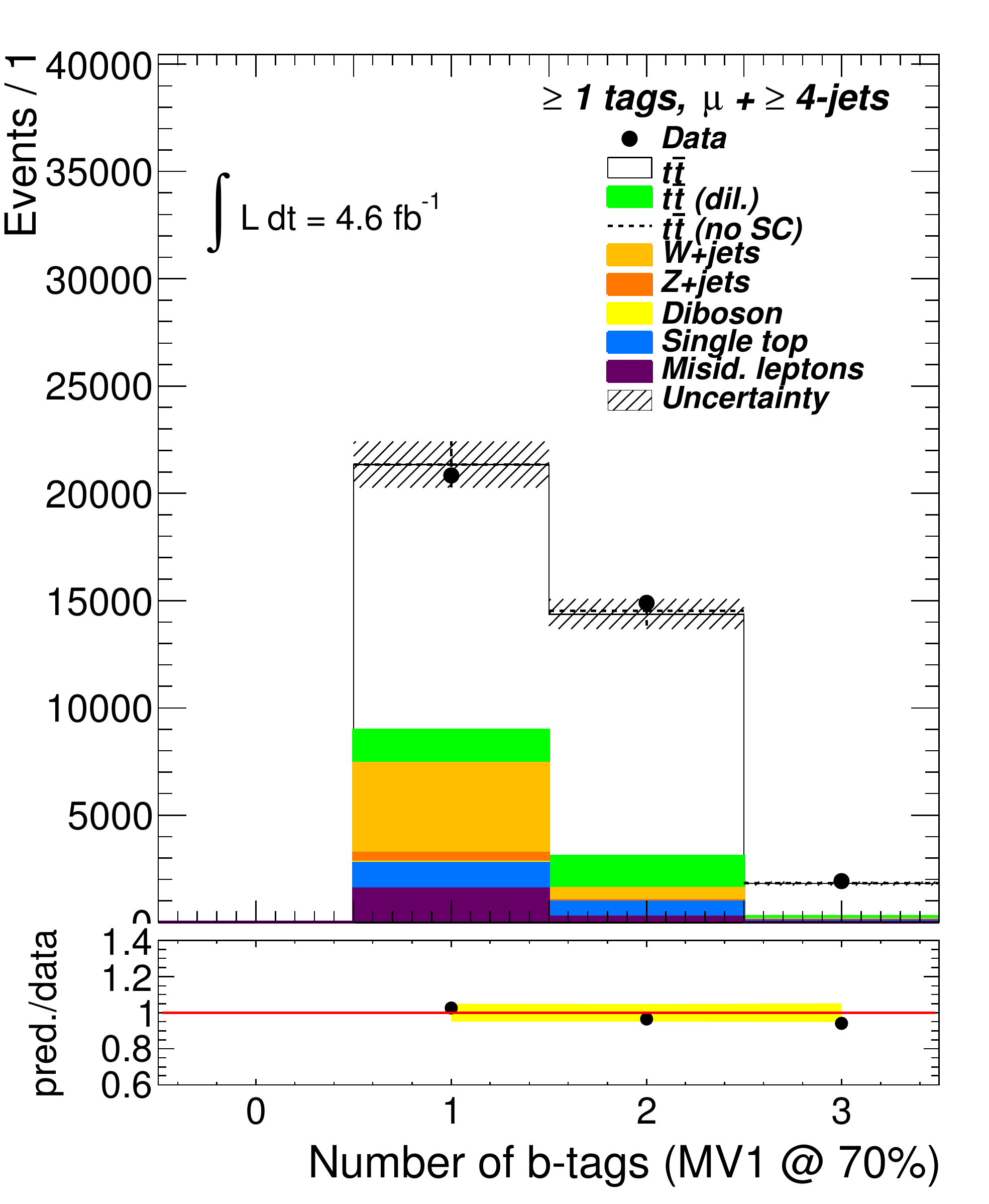}
\includegraphics[width=0.3\textwidth]{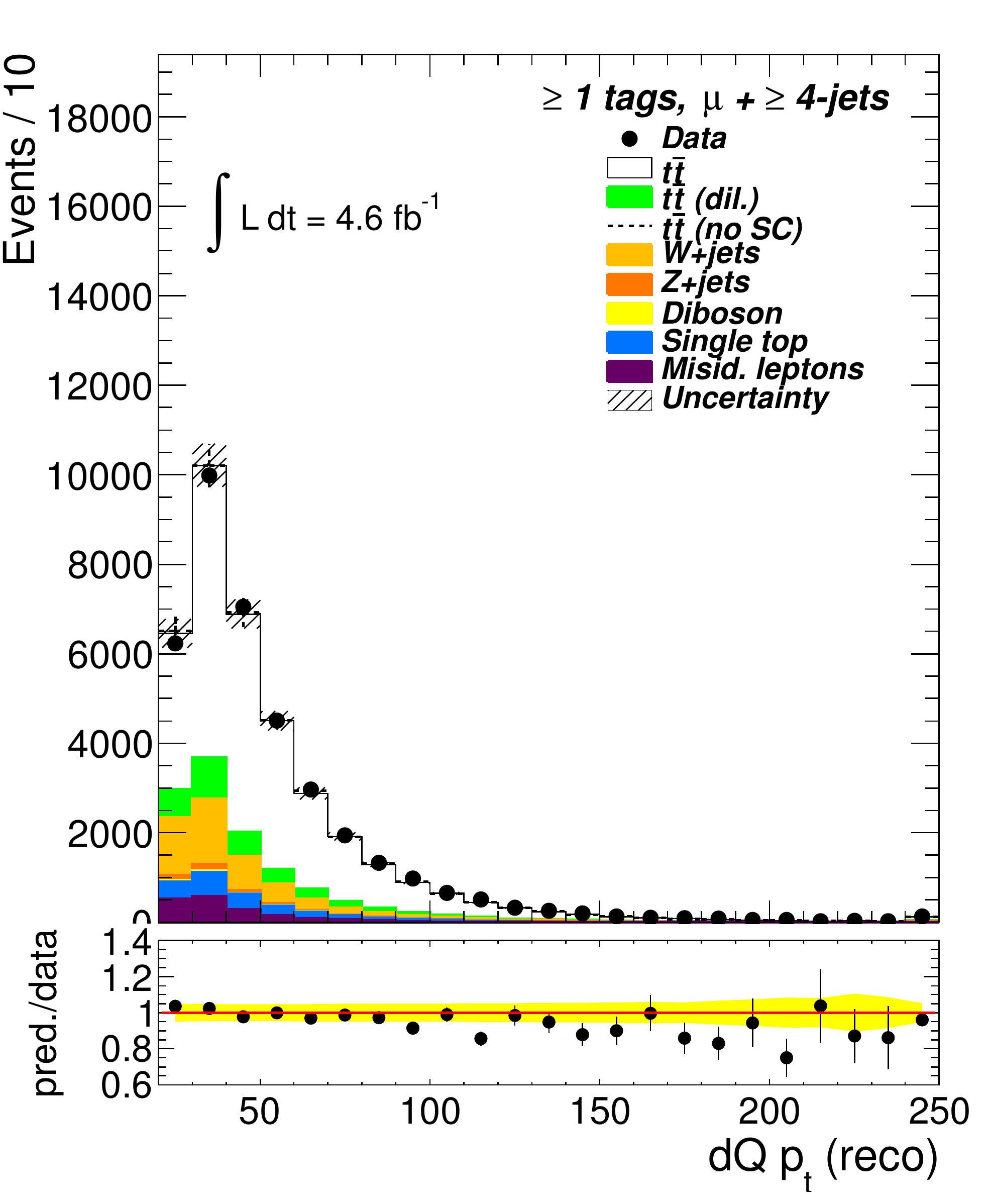}
\end{center}
\caption{The weight of the \mvone\ \btag ger (left), the number of tagged jets (centre) and the \pt\ of the down-type quark candidate (right) for the \ejets\ channel (upper row) and the \mujets\ channel (lower row). 
}
\label{fig:klfitter_control1}
\end{figure}
Closely related to the \pt\ distribution is the jet index corresponding to the \bQ\ and \dQ. This index is assigned during the \pt\ ordering of the jets, starting with zero for the jet with the highest \pt. The first two columns of Figure \ref{fig:klfitter_control2} show good agreement for the \dQ\ and the \bQ, respectively.
\begin{figure}[htbp]
\begin{center}
\includegraphics[width=0.3\textwidth]{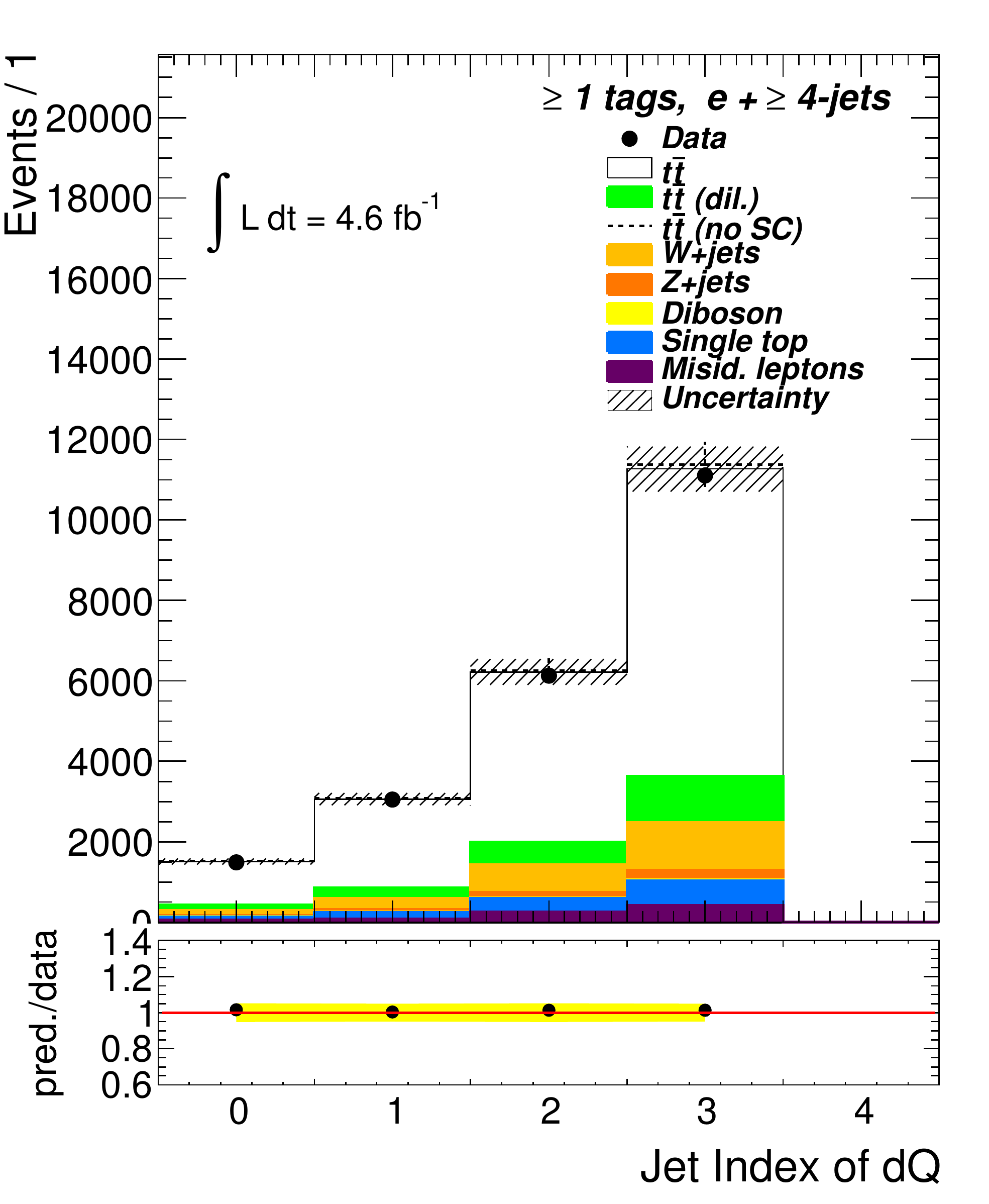}
\includegraphics[width=0.3\textwidth]{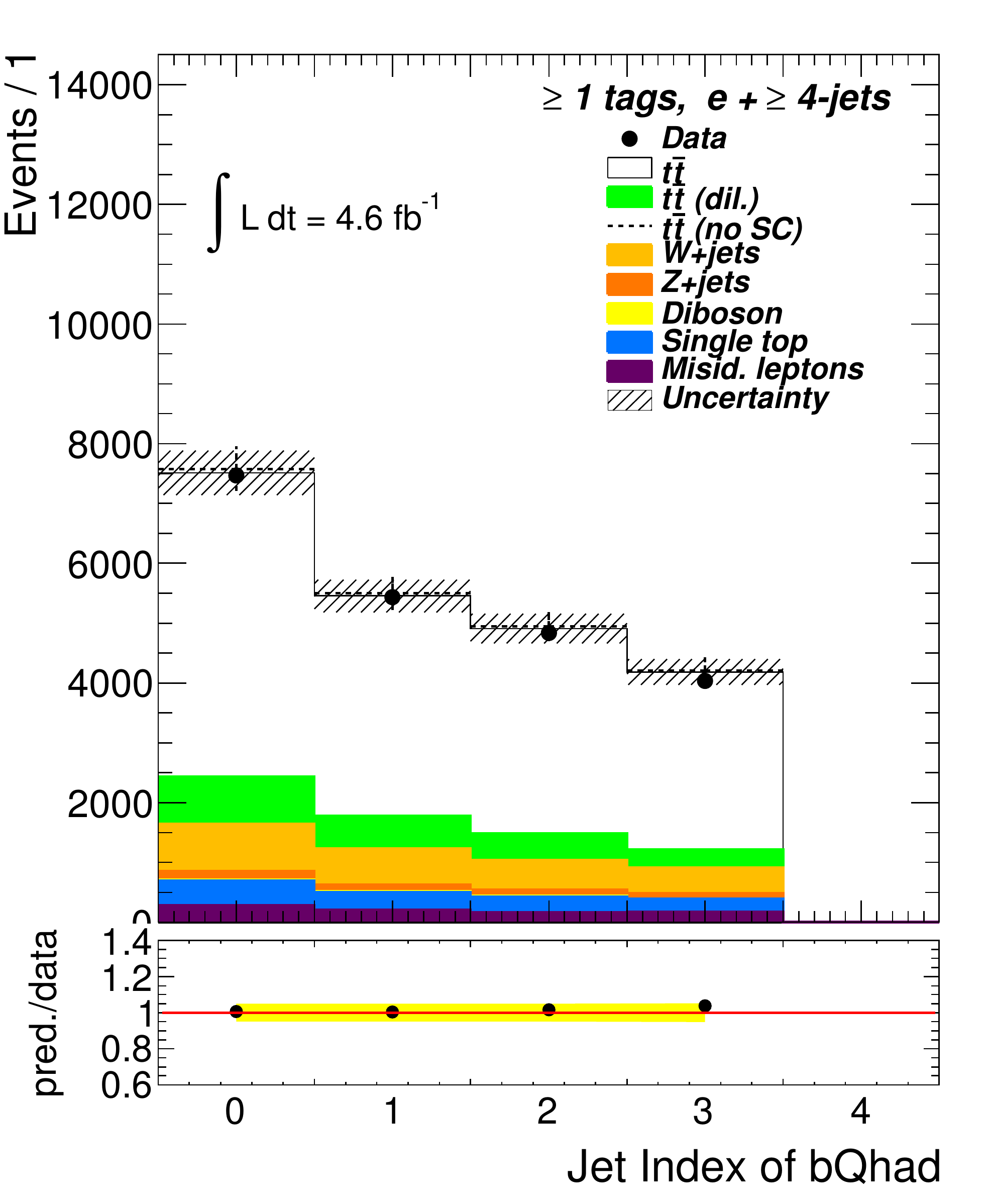}
\includegraphics[width=0.3\textwidth]{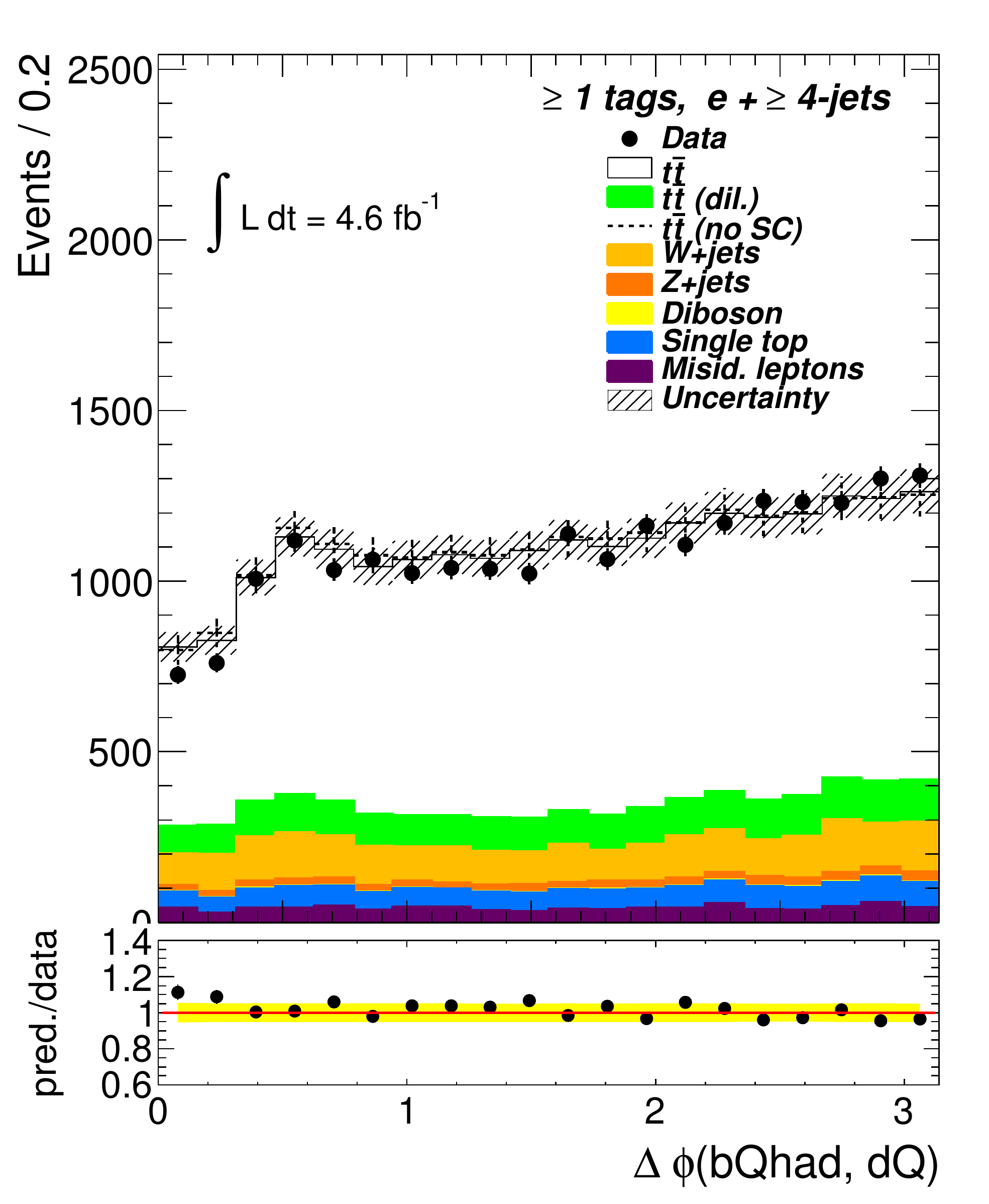}\\
\includegraphics[width=0.3\textwidth]{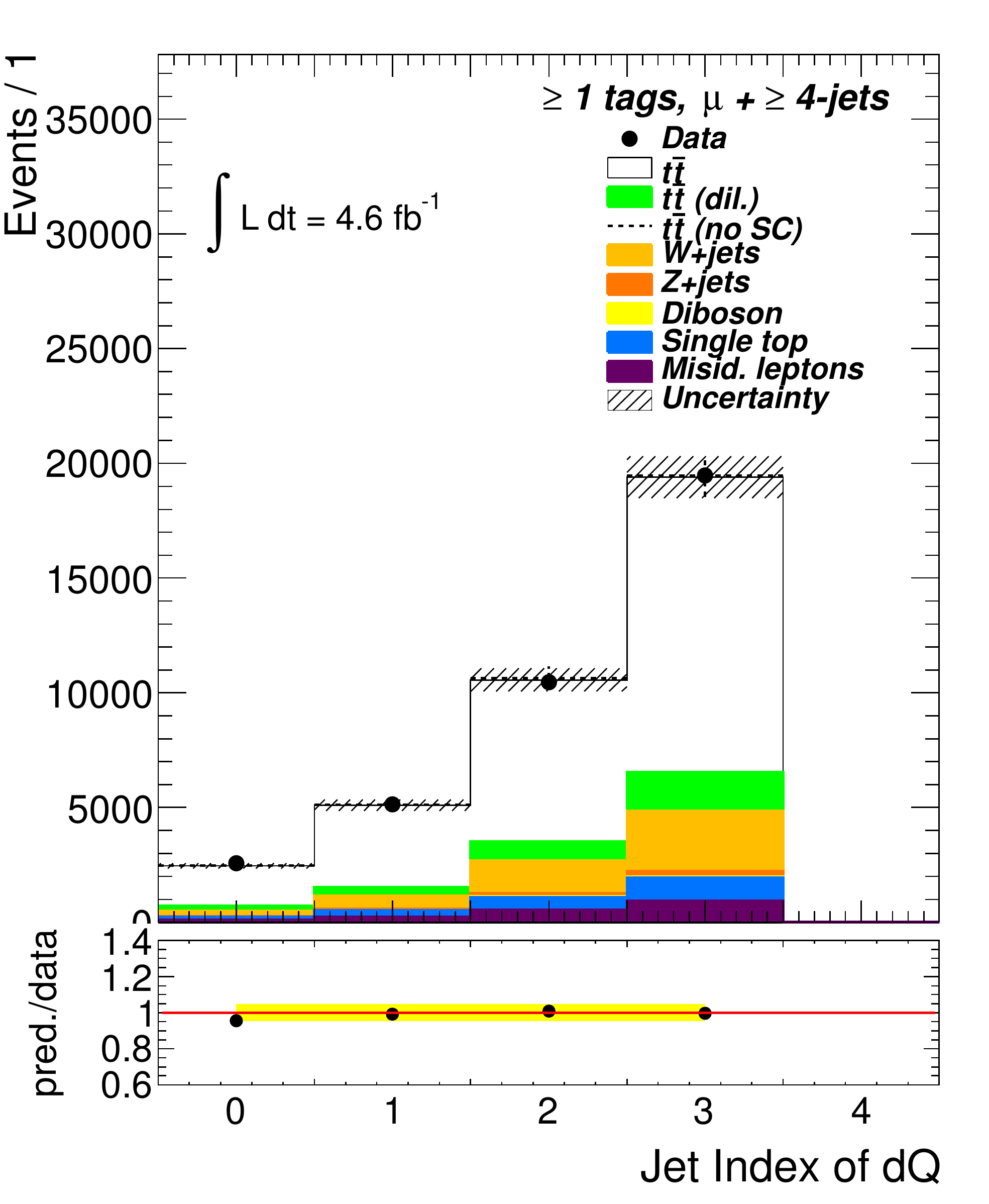}
\includegraphics[width=0.3\textwidth]{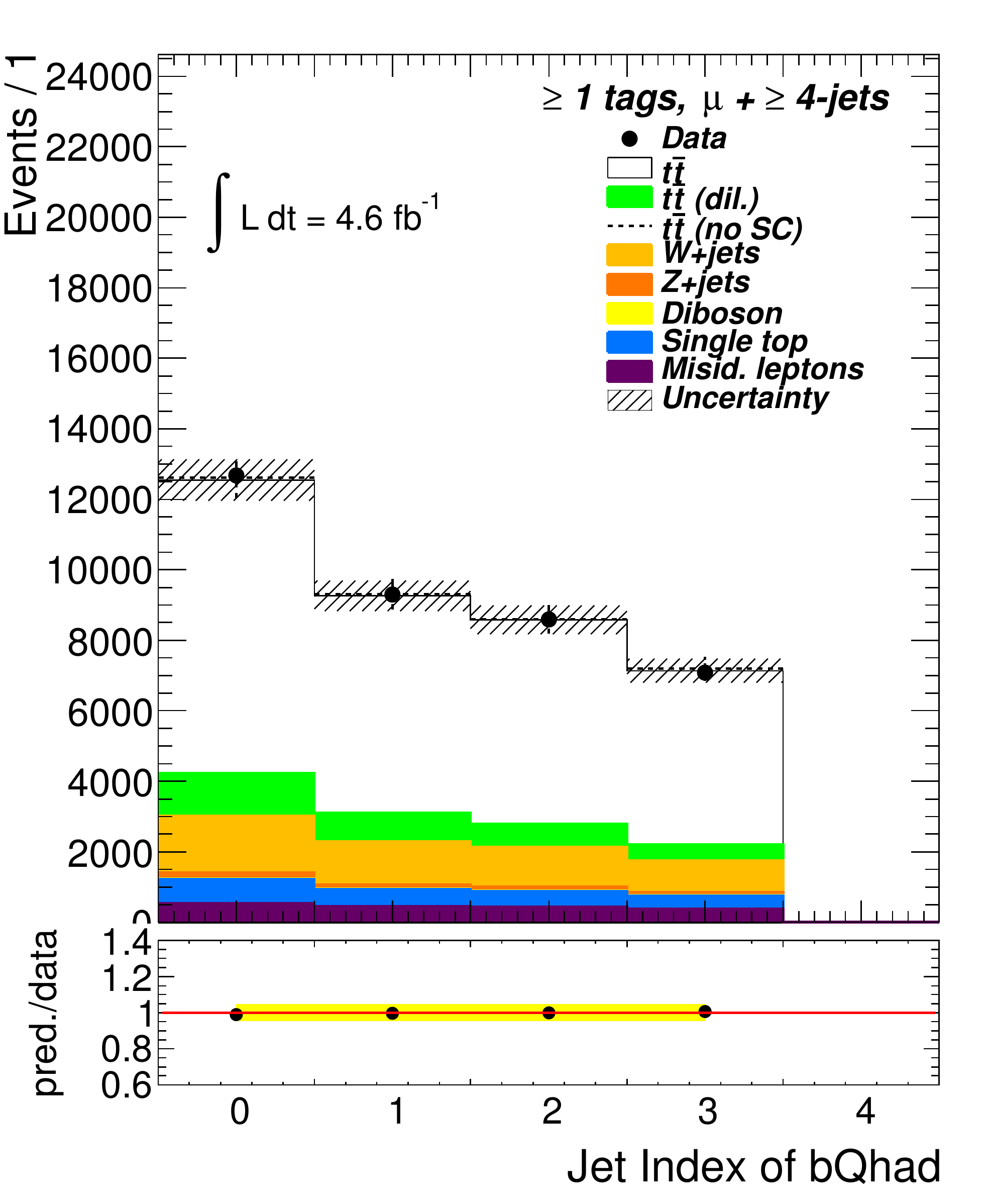}
\includegraphics[width=0.3\textwidth]{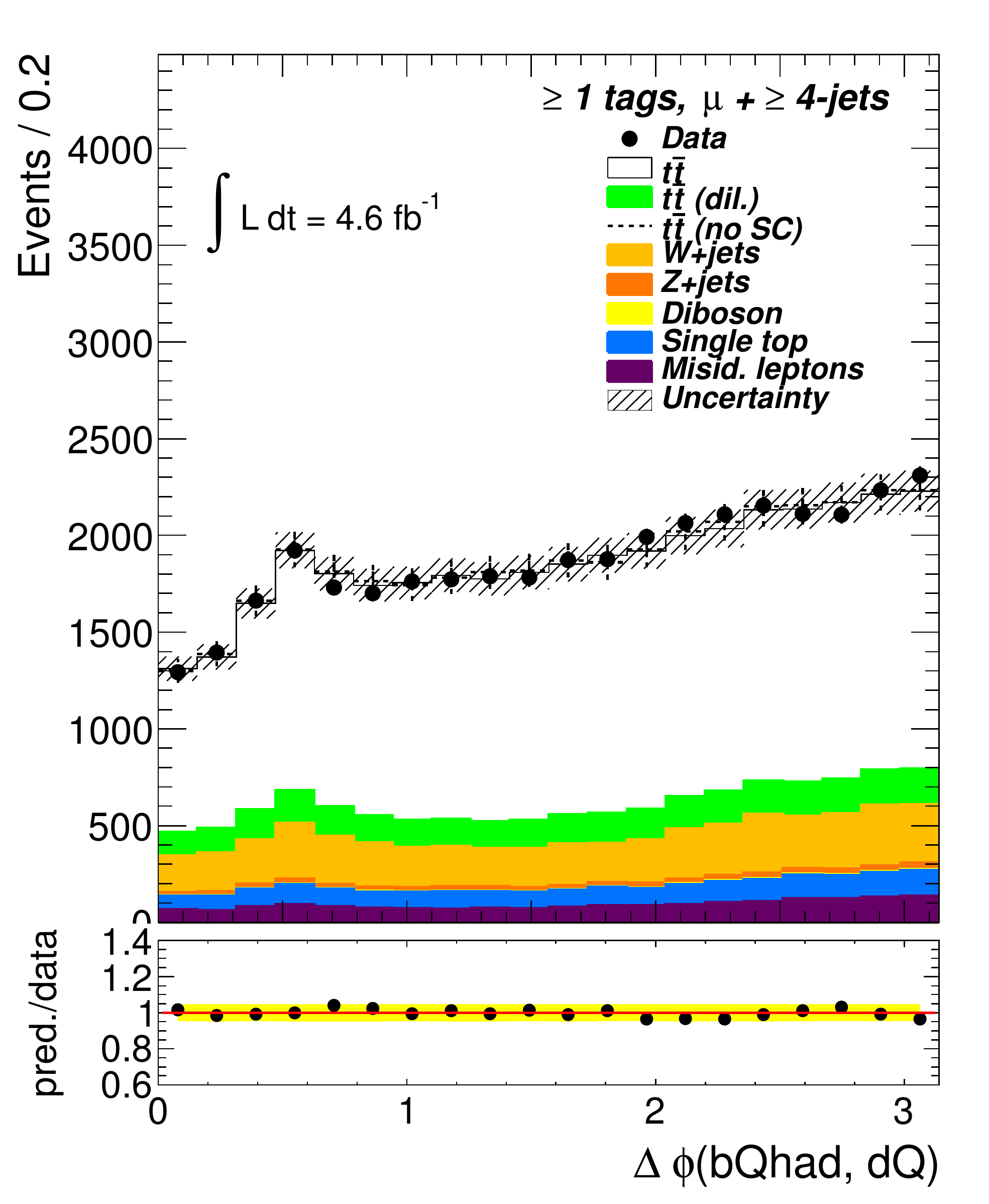}
\end{center}
\caption{The \pt\ ranking index of the down-type quark jet (left), the index of \bQ\ jet from the hadronic top (middle) and the $\Delta \phi$ angle between the reconstructed hadronic \bQ\ and the down-type-quark (left) for the \mbox{\ejets} channel (upper row) and the \mbox{\mujets} channel (lower row). 
}
\label{fig:klfitter_control2}
\end{figure}
As this analysis studies angles between spin analysers, angular variables of the reconstructed quantities also need to be investigated. The azimuthal angle between the \bQ\ and the \dQ\ was chosen as a control distribution and is shown in the third column of Figure \ref{fig:klfitter_control2}.

An overall summary of the reconstruction via \KLFitter\ is given by the likelihood and the event probability. Both are shown in Figure \ref{fig:klfitter_control3}. 

\begin{figure}[htbp]
\begin{center}
\includegraphics[width=0.3\textwidth]{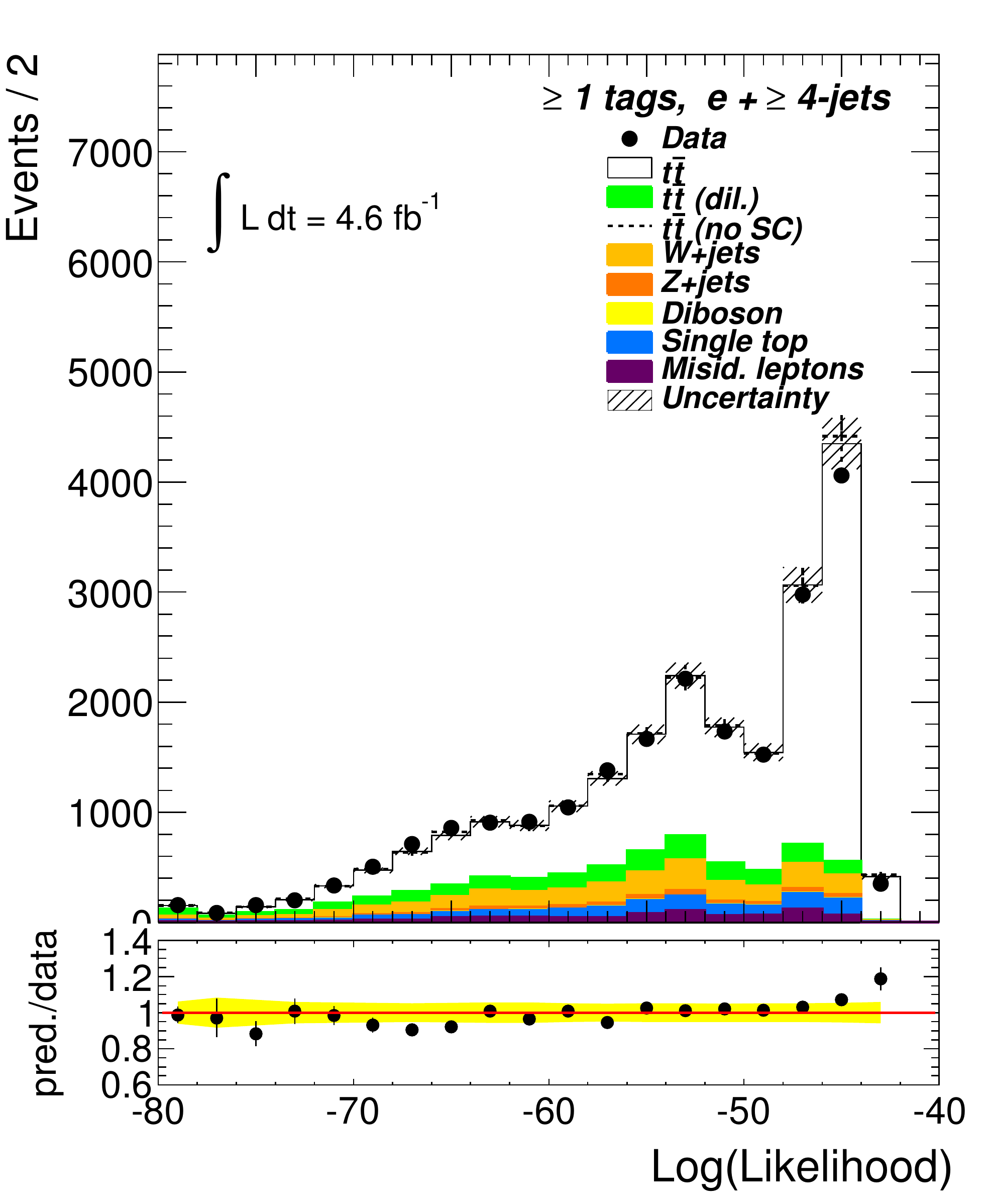}
\includegraphics[width=0.3\textwidth]{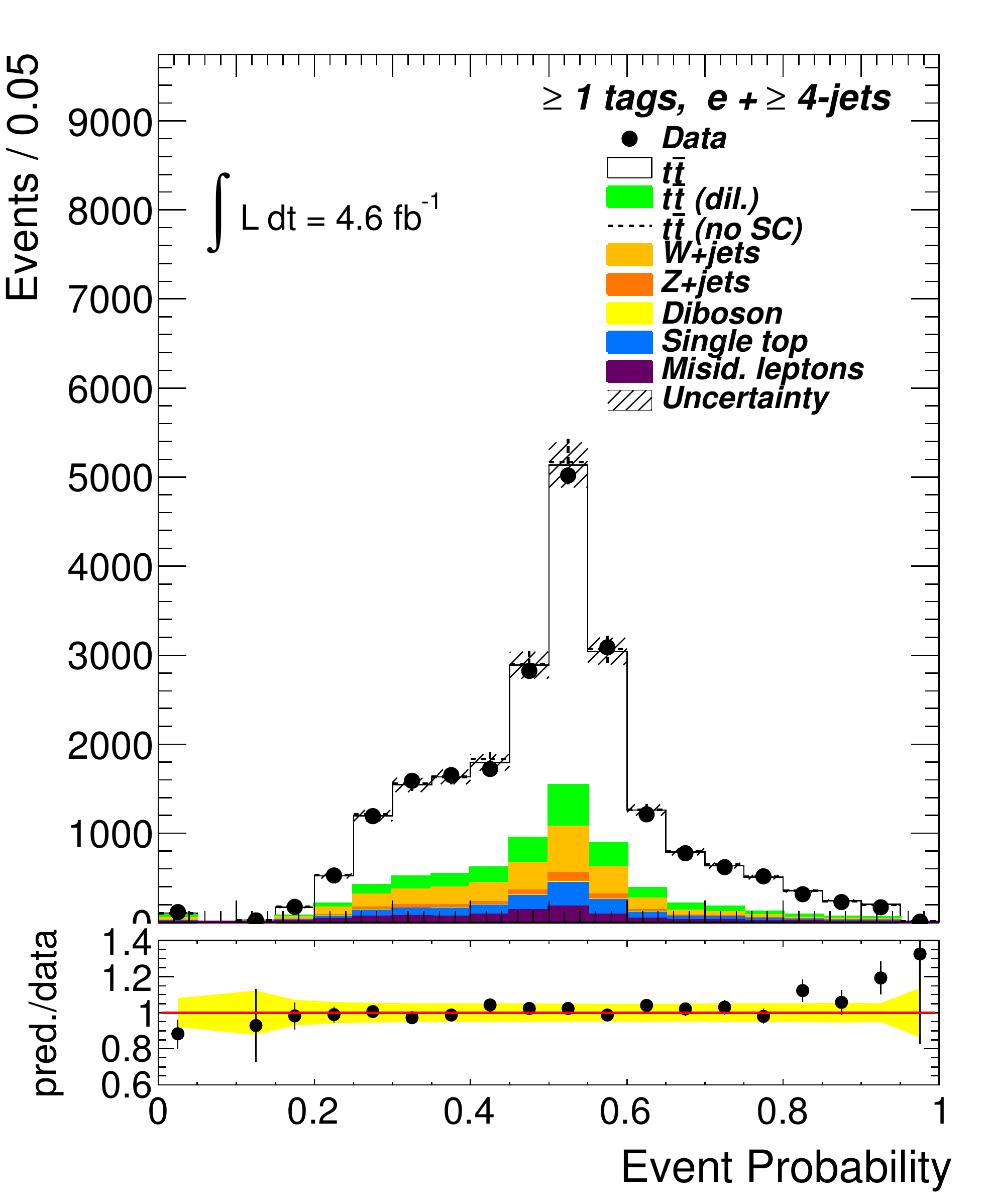}\\

\includegraphics[width=0.3\textwidth]{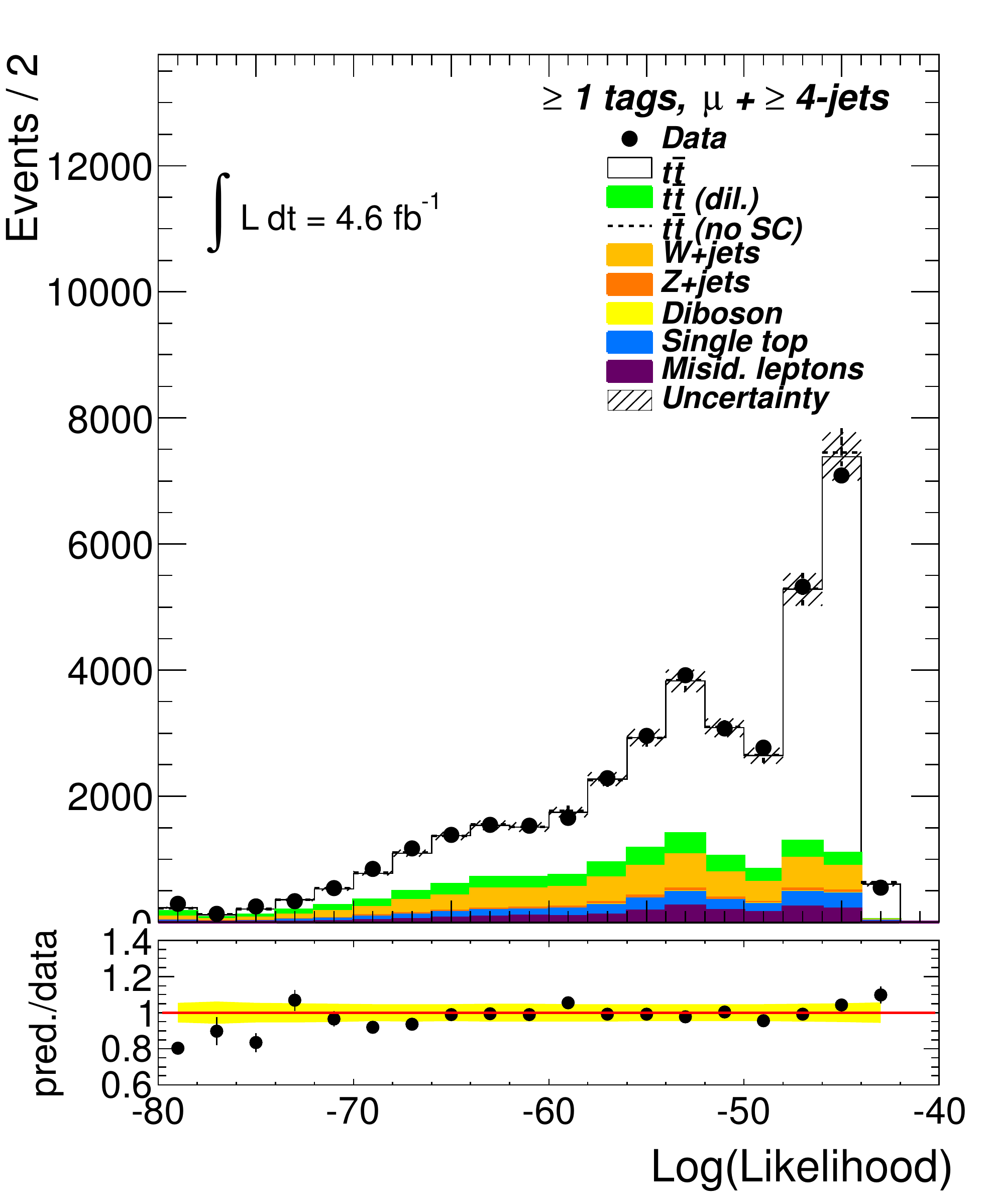}
\includegraphics[width=0.3\textwidth]{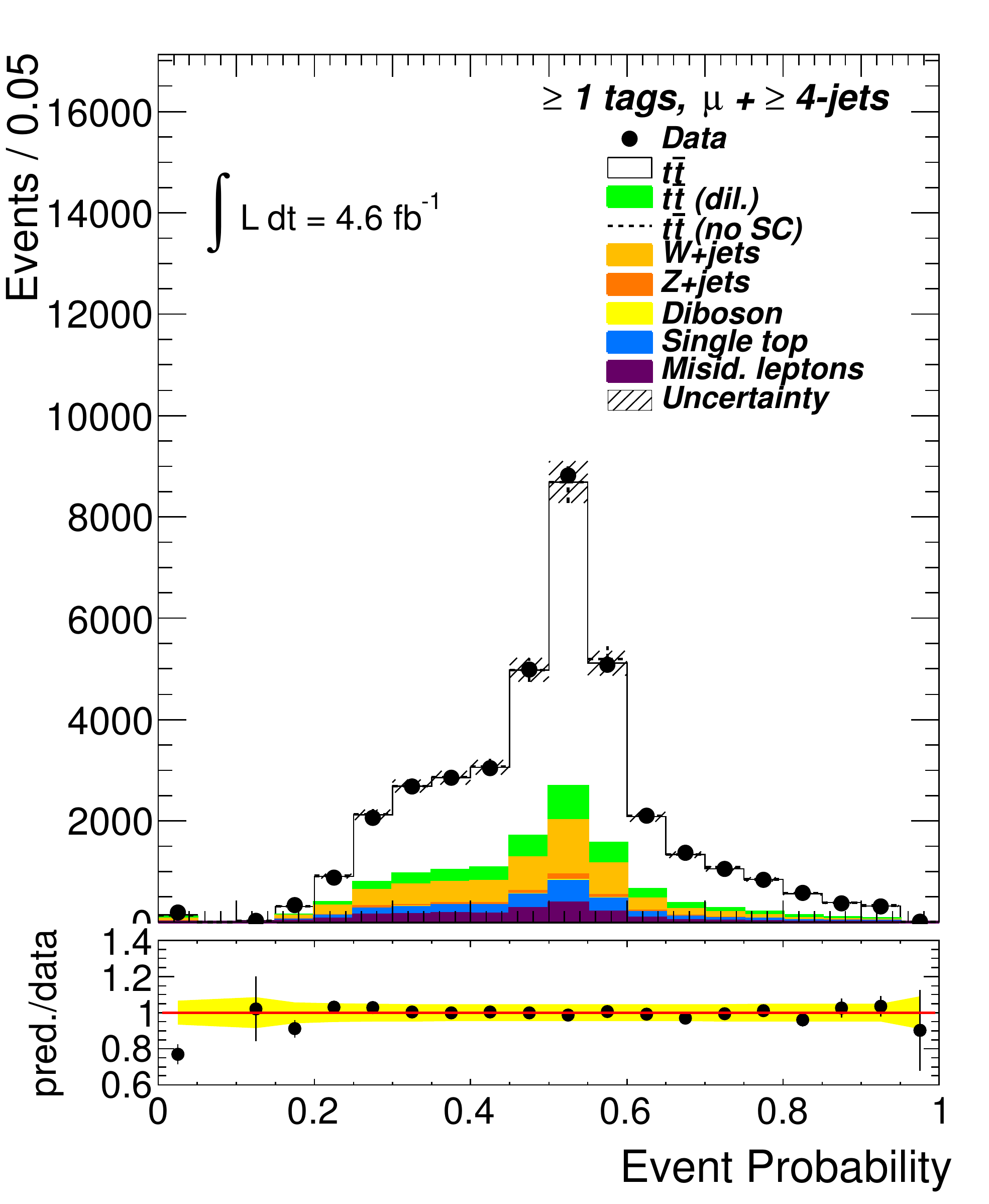}

\end{center}
\caption{The logarithm of the KLFitter likelihood (left) and the event probability (right) for the \ejets\ channel (upper row) and the \mujets\ channel (lower row). 
}
\label{fig:klfitter_control3}
\end{figure}
In general, no mismodelling is observed in the reconstruction of the \dQ\ and \bQ\ spin analysers. 
The data is considered validated and the measurement is performed as described in the next sections. 

\section{Binned Likelihood Fit}
\label{sec:fit}
The spin correlation measurement must deduce the \ttbar\ signal -- as well as its spin properties --  and the background contribution.  This is realized with a template fit, based on the principle of a binned likelihood fit. Templates are created for each signal and background composition. MC simulation is used for all templates except the fake lepton background, which is derived from data. The measured dataset is split into several channels, which are explained in the next section. 
By using the templates from the prediction for signal and background events as well as the measured data distribution, it is possible to define a likelihood
\begin{equation}
\mathcal{L}=\prod^{C}_{j=1}\prod^{N}_{i=1}\frac{e^{-(s_{ij}+b_{ij})}(s_{ij}+b_{ij})^{n_{ij}}}{n_{ij}!},
\end{equation}
where $C$ is the number of channels, $N$ is the number of bins per template, and $n_{ij}$, $s_{ij}$ and $b_{ij}$ are
the number of events in the i'th bin and the j'th channel in the data, signal and
background distribution, respectively.
The signal distribution is a linear combination of the two available signal samples:  \ttbar\ pairs with spins correlated as predicted by the SM and \ttbar\ pairs with uncorrelated spins. The fraction \fsm\ of SM like spin correlation defines the mixing:
\begin{equation}
s_{ij}=\varepsilon_{j}^{\text{signal}} N_{\ttbar} \cdot (\fsm \cdot p^{\text{SM \ttbar}}_{ij} + (1-\fsm) \cdot p^{\text{unc. \ttbar}}_{ij})
\label{eq:fit_signal}
\end{equation}
where $p^{\text{SM \ttbar}}_{ij}$ and $p^{\text{unc. \ttbar}}_{ij}$ are the entries
in bin $i$ of the normalized template for the SM and the sample with uncorrelated spins,
respectively, in channel $j$. The total \ttbar\ yield is given by the parameter $N_{\text{\ttbar}}$. It can also be reformulated as the expected \ttbar\ yield $N_{\text{exp. \ttbar}}$ scaled by a factor $c$, $N_{\text{\ttbar}} = c \cdot N_{\text{exp. \ttbar}}$. The efficiency $\varepsilon_{j}$ denotes the fraction of the total \ttbar\ yield that is reconstructed in channel $j$. 

The background contribution for bin $i$ and channel $j$ breaks down to:
\begin{equation}
b_{ij}=\sum^{3}_{k=1} N_{k} \cdot \varepsilon_{kj} \cdot p_{ijk} 
\label{eq:fit_BG}
\end{equation}
summed over the different background contributions $k$, each having its own efficiency $\varepsilon$: $W$+jets, fake lepton background and remaining backgrounds. $N_{k}$ represents the total number of events from the background type $k$. $\varepsilon_{kj}$ is the relative contribution of the background events of type $k$ in the channel $j$ to the total number of background events $N_{k}$. $p_{ijk}$ is the entry of bin $i$ of the normalized background template for the background type $k$ in the channel $j$.

Technically, this fitting is implemented by transforming the two normalized signal templates $p^{\text{SM \ttbar}}$ and $p^{\text{unc. \ttbar}}$ into $p^{\text{sum \ttbar}}$ and $p^{\text{diff \ttbar}}$ via
\begin{align}
p^{\text{sum \ttbar}} &= \frac{1}{2} \left( p^{\text{SM \ttbar}} + p^{\text{unc. \ttbar}} \right),\\
p^\text{{diff \ttbar}} &= \frac{1}{2} \left( p^{\text{SM \ttbar}} - p^{\text{unc. \ttbar}} \right).
\end{align}
The signal contribution is a linear combination of $p_{\text{sum \ttbar}}$ and $p_{\text{diff. \ttbar}}$ :
\begin{equation}
s_{ij}=\varepsilon_{j}^{\text{signal}} (N_{\text{sum \ttbar}} \cdot p_{\text{sum \ttbar}, ij} + N_{\text{diff \ttbar}} \cdot p_{\text{diff \ttbar}, ij} )
\end{equation}
with the parameter values for the total yield $N_{\text{sum \ttbar}}$ and the parameter for the scaling of the difference of SM and uncorrelated events $N_{\text{diff \ttbar}}$. The values of  $p_{\text{sum \ttbar}, ij}$ and $p_{\text{diff \ttbar}, ij}$ represent the entries of bin $i$ of the templates $p_{\text{sum \ttbar}}$ and $p_{\text{diff \ttbar}}$ in the channel $j$. 

As these are just linear transformations, the fitted parameter values $N_{\text{sum \ttbar}}$ and $N_{\text{diff \ttbar}}$ can be easily translated into the parameters of interest, namely the cross section scale factor $c$ and the spin correlation fraction $\fsm$.

The linear transformation is introduced to add numerical stability to the fit. Without the transformation the two signal parameters would be fully anti-correlated. The transformation resolves this issue. 

In addition to the basic fit parameters, further parameters are added. One set of parameters accounts for the systematic uncertainties, taking them into account as additional \textit{nuisance parameters}\index{Nuisance parameters} (NPs) by fitting their effects on the template to the data and thus constraining their impact. This procedure is described in Section \ref{sec:nuipars}.

After the choice of channels is explained in the following section, the treatment of fit parameters is described in Section \ref{sec:fitpars}.

\subsection{Analysis Channels} 
\label{sec:channels}
Each channel is a subset of the whole data available and has distinct properties: signal to background ratio, reconstruction efficiency, impact of systematic uncertainties and statistical uncertainty. 
Some channels are pre-defined as the \ejets\ and the \mujets\ channel are reconstructed in different, orthogonal data streams. Others are defined by the analysis strategy. In the combination of \dQ\ and \bQ\ results, the different analysers are treated as separate channels. Section \ref{sec:correlation} is dedicated to the question if the treatment of the \dQ\ and the \bQ\ as independent variables is justified. 

Further splitting of the \ejets\ and the \mujets\ data is possible and reasonable. The first splitting divides the data into a channel with exactly four jets and one with at least five jets. As shown in Section \ref{sec:optimizations}, this creates a subsample with a higher reconstruction efficiency for both the \dQ\ and the \bQ, namely for $n_{\text{jet}} = 4$. Another motivation for this splitting is the jet multiplicity mismodelling of the \mcatnlo\ generator. It is possible to introduce an additional parameter to the fit correcting the efficiencies of the  $n_{\text{jet}} = 4$ with respect to the  $n_{\text{jet}} > 5$ channel, allowing an in-situ correction of the mismodelling (see Section \ref{sec:jetmultcorr}).

The number of \btag ged jets is also a criterion to split the data sample into subsets of higher and lower reconstruction efficiency. As the \btag\ multiplicity is also not perfectly modelled (see Section \ref{sec:anaval}), the introduced nuisance parameters dedicated to the \btag ging uncertainties can correct this mismodelling in-situ. 

All channels used in the analysis are defined and listed in Table \ref{tab:channels}. Results are obtained for the individual channels, combinations with the same analyser and a full combination.

\begin{table}[htbp]
\begin{center}
\begin{tabular}{|c|c|c|c|c|}
\hline
Channel & Analyser & Lepton Flavour & Jet Multiplicity & B-Tags \\
\hline
\hline
1&\multirow{8}{*}{\dQ} & \multirow{4}{*}{electron} & \multirow{2}{*}{$=4$} & {=1} \\
2&{} &{} & {} &{$>1$} \\
3&{} &{} & \multirow{2}{*}{$>4$} &{$=1$} \\
4&{} &{} & {} &{$>1$} \\
5&{} &\multirow{4}{*}{muon} & \multirow{2}{*}{$=4$} &{$=1$} \\
6&{} &{} & {} &{$>1$} \\
7&{} &{} & \multirow{2}{*}{$>4$} &{$=1$} \\
8&{} &{} & {} &{$>1$} \\
\hline
9&\multirow{8}{*}{\bQ} & \multirow{4}{*}{electron} & \multirow{2}{*}{$=4$} & {=1} \\
10&{} &{} & {} &{$>1$} \\
11&{} &{} & \multirow{2}{*}{$>4$} &{$=1$} \\
12&{} &{} & {} &{$>1$} \\
13&{} &\multirow{4}{*}{muon} & \multirow{2}{*}{$=4$} &{$=1$} \\
14&{} &{} & {} &{$>1$} \\
15&{} &{} & \multirow{2}{*}{$>4$} &{$=1$}\\
16&{} &{} & {} &{$>1$}\\
\hline
\end{tabular}
\end{center}
\caption{Analysis channels. }
\label{tab:channels}
\end{table}

\subsection{Usage of Priors}
\label{sec:priors}
As the fitting framework \bat\ is using the Bayesian approach, a-priori information about the parameters can be included in the fit. This is realized by the addition of \textit{priors}\index{Prior} to each parameter $p_i$. These are multiplied to the likelihood. Different types of priors exist: Delta priors fix a parameter to one certain value. Constant priors have no effect at all as they are constant in the parameter space. Gaussian priors are normalized Gaussian functions that take an expected value as central value and an uncertainty on the expectation as width.\footnote{The expression 'width' of Gaussian distributions refers to the standard deviation $\sigma$ and may not be confused with the 'Full Width at Half Maximum', FWHM.} This kind of prior is used to constrain the background yields according to their normalization uncertainties. 
The Gaussian priors used in this analysis are explained in Section \ref{sec:fitpars}.

\subsection{Systematic Uncertainties as Nuisance Parameters}
\label{sec:nuipars}
Systematic uncertainties affect the results by changing both the shape and the yield of the measured distributions. Next to the more traditional way of evaluating their effect via ensemble tests, the option of including them already in the fit can be a good alternative if certain requirements are fulfilled. If for an uncertainty both the $\pm 1\,\sigma$ variations lead to a well-defined\footnote{'Well-defined' in a sense that the variations are not simply random fluctuations.} template the effects on each bin can be quantified and linearly interpolated. This allows to assign an additional fit parameter (nuisance parameter) $\beta_i$ to the uncertainty $i$. Within the fit, the effect is considered via modified efficiencies used in Equations \ref{eq:fit_signal} and \ref{eq:fit_BG}:
\begin{align}
\tilde{\varepsilon} = \varepsilon + \sum_{\text{unc.}} \beta_i \Delta \varepsilon_i.
\end{align}
Here, $\Delta \varepsilon_i$ is the relative change of yield per bin caused by the systematic effect $i$. Values of $\beta = \pm 1$ correspond to the effects of $\pm 1\,\sigma$ deviations.
By including systematic effects as nuisance parameters they can hence improve the data/MC agreement caused by miscalibration covered by systematic uncertainties. Systematic uncertainties included as nuisance parameters propagate their uncertainties into the fit uncertainty.

A complete list of nuisance parameters used in this analysis is provided in Section \ref{sec:NP_test}.

\subsection{Jet Multiplicity Correction}
\label{sec:jetmultcorr}
The division into subsets of $n_{\text{jets}}=4$ and $n_{\text{jets}}>4$ allows for a correction of the mismodelling of jet multiplicity by \mcatnlo. Such a correction needs to be applied to the signal samples as the predicted signal yield in the $n_{\text{jets}}>4$ subset is too low. No such correction was applied to the background samples.

As the efficiencies for each type of signal are set and fixed before the fit, there are two possibilities for the fit to fill the gap between prediction and data yield in the $n_{\text{jets}}>4$ channels:
\begin{itemize}
\item{The backgrounds will fill the gap and will thus be overestimated.}
\item{As the efficiencies are deduced from the MC including deviating jet multiplicities, they will be incorrect. By filling the gap in the  $n_{\text{jets}} \geq 5$ channels, the signal will also increase in the $n_{\text{jets}}=4$ channels. The fit will end up in a compromise between an overestimation of the $n_{\text{jets}}=4$ channels and an underestimation of the $n_{\text{jets}}>4$ channels. }
\end{itemize}
Both possibilities are inconvenient. Thus, an additional nuisance parameter is introduced. It modifies the efficiency of the $n_{\text{jets}}>4$ channels by $\pm 10$\,\% per integer value of the parameter. The method was tested first by performing a combined fit of all \dQ\ and all \bQ\ channels for which the effect of jet multiplicity mismodelling is the same but the \dphi\ shapes are different. With a starting value of 0 and a constant prior, the correction parameters were fitted to be $0.88 \pm 0.16$ (\dQ) and $0.86 \pm 0.17$ (\bQ). This means the efficiencies are corrected by 1.088 and 1.086, respectively.
As a cross check, a second procedure was tested: Before fitting, the \ttbar\ yield in the $n_{\text{jets}} >4$ channel was manually scaled up until the best possible data/MC agreement was reached. This scale factor was then fixed and manually applied to the fit. The result is consistent with that of the more flexible nuisance parameter approach.

\subsection{Fit Parameters}
\label{sec:fitpars}

Table \ref{tab:priors} lists all parameters used in the fit. Used priors are set according to the uncertainty on the estimation (see Chapter \ref{sec:systematics}). 

\begin{table}[htbp]
\begin{center}
{\footnotesize
\begin{tabular}{|c|c|c|c|}
\hline
Parameter & Description & {Gaussian Prior} \\
{} & {}  &Width / Mean\\
\hline
\hline
$p_0$ &  $\frac{1}{2}(N_{SM} + N_{UC})$ & {---}\\
$p_1$ &  $\frac{1}{2}(N_{SM} - N_{UC})$ & {---}\\
$p_2$ &  $N_{\text{rem. BG}, e+\text{jets}} + N_{\text{rem. BG}, \mu+\text{jets}}$  &  0.20 \\
$p_3$ &  $N_{W+\text{jets}, n_{\text{jets}} = 4, e+\text{jets}}$  & 0.18 \\
$p_4$ &  $N_{W+\text{jets}, n_{\text{jets}} \geq 5, e+\text{jets}}$  & 0.27 \\
$p_5$ &  $N_{QCD, n_{\text{jets}} = 4, e+\text{jets}}$  & 0.50\\
$p_6$ &  $N_{QCD, n_{\text{jets}} \geq 5, e+\text{jets}}$  & 0.50\\
$p_7$ &  $N_{W+\text{jets}, n_{\text{jets}} = 4, \mu+\text{jets}}$  &  0.16 \\
$p_8$ &  $N_{W+\text{jets}, n_{\text{jets}} \geq 5, \mu+\text{jets}}$ & 0.23 \\
$p_9$ &  $N_{QCD, n_{\text{jets}} = 4, \mu+\text{jets}}$  & 0.20\\
$p_{10}$ &  $N_{QCD, n_{\text{jets}} \geq 5, \mu+\text{jets}}$ & 0.20\\
$p_{11}$ &  $\varepsilon\left(n_{\text{jet}} \geq 5\right)$ correction & {---}\\
\hline

\end{tabular}
}
\end{center}
\caption{The parameters used for the fit, priors used and their corresponding widths divided by the mean values. The mean value of each Gaussian prior is the expectation value of the corresponding fit parameter. The widths are representing the normalization uncertainties as described in Chapter \ref{sec:systematics}.} 
\label{tab:priors}
\end{table}
Further parameters are added for systematic uncertainties (when applicable, see Section \ref{sec:NP_test}).

\section{Method Validation}
\label{sec:linchecks}
A fit works linearly if pseudo data with a given \fsm\ is also fitted as such. For the evaluation of the linearity, pseudo data are created for 16 values of \fsm\ between $-1.0$ and $2.0$. 100,000 ensembles are created by applying Poissonian fluctuations to each bin of signal templates mixed to a given \fsm. The mean of the fit output for each \fsm\ is then fitted with a linear function. It is required that the slope is unity and the offset zero. The linearity test is successfully passed (see Figure \ref{fig:lincheck_comb}). No deviations of the expected slopes and offsets are observed up to two decimal places.

The pull $p$ is defined as the difference between the fitted value of \fsm\ and the value of \fsm\ used to create the pseudo data which was fitted, divided by the uncertainty on \fsm\ of a fit:
\begin{align}
p \equiv \frac{f^{out}_{\text{SM}} - f^{in}_{\text{SM}}}{\Delta f^{out}_{\text{SM}}}.
\end{align}
For each value of \fsm\ the pull distribution is plotted and the mean and RMS values are compared to the expectations of zero and unity, respectively. No deviations from the expectations are observed. The pull mean and RMS values are checked over the full range of $-1.0 < \fsm\ < 2.0$. 
The pull mean values are distributed around zero within 0.01 for the \dQ, the \bQ, and the full combination (see Figure \ref{fig:pull_mean_comb}). The pull RMS values were distributed around unity within 0.02 for the \dQ, the \bQ, and the full combination. 
 \begin{figure}[htbp]
 	\centering
			\subfigure[]{
		\includegraphics[width=0.45\textwidth]{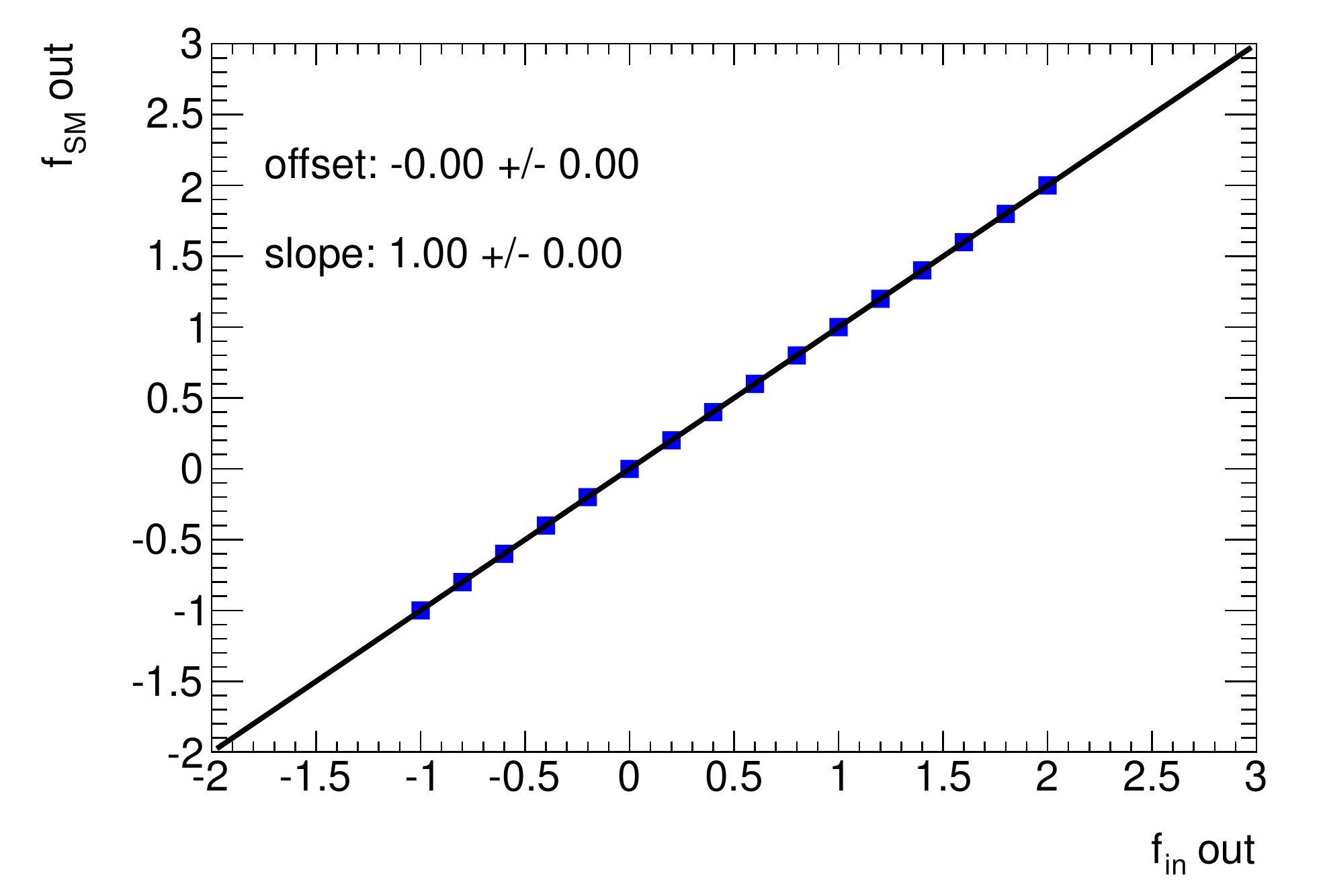}
			\label{fig:lincheck_comb}
		}
					\subfigure[]{
		\includegraphics[width=0.45\textwidth]{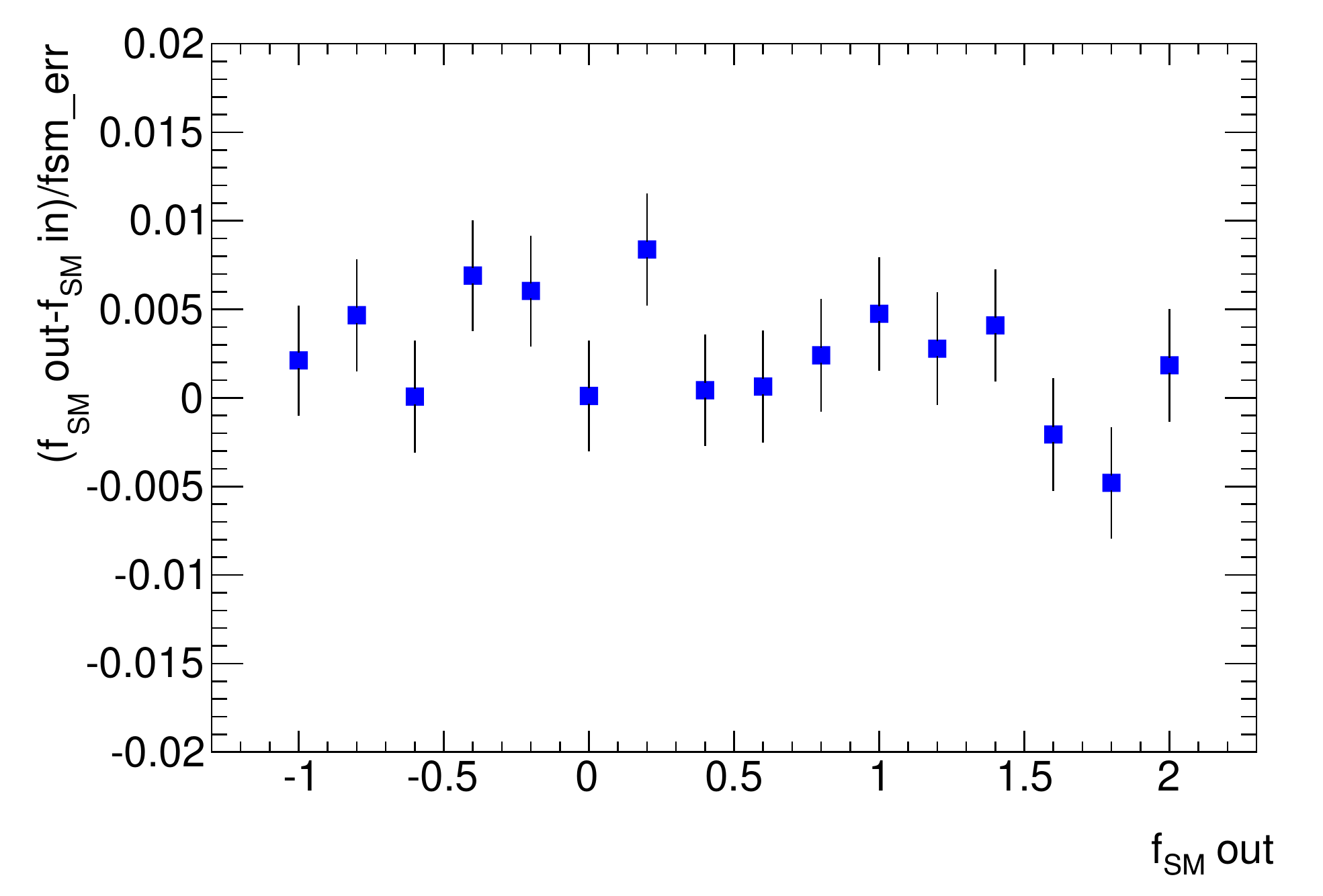}
			\label{fig:pull_mean_comb}
		}		
\caption{\subref{fig:lincheck_comb} Linearity test for the combination of \dQ\ and \bQ\ analysers. \subref{fig:pull_mean_comb} Pull mean distribution for the combination of \dQ\ and \bQ\ analysers.}
\label{fig:linchecks}
\end{figure}

\section{Expected Statistical Uncertainty}
\label{sec:exp_stat_unc}
The expected statistical uncertainty is evaluated via ensemble tests. Pseudo data are created from the signal templates for \ttbar\ (at $\fsm\ = 1.0$) and background events according to the integrated luminosity of \intlumi. A Poissonian fluctuation is applied to the expected distribution before a fit is applied. For this test, 100,000 ensembles are created and fitted. A distribution of the fit output value for \fsm\ was fitted with a Gaussian distribution and its width is taken as expected statistical uncertainty. \\
Table \ref{tab:exp_stat_unc} shows the expected statistical uncertainty for each of the eight channels and two analysers as well as for the combinations of all \dQ\ and \bQ\ channels.

\begin{table}[htbp]
\begin{center}
\begin{tabular}{|c|c|c|c||c|}
\hline
Analyser & Lepton Flavour & Jet Multiplicity & B-Tags & Expected $\Delta \fsm$ (stat.) \\
\hline
\hline
\multirow{8}{*}{\dQ} & \multirow{4}{*}{electron} & \multirow{2}{*}{$=4$} & {=1} & 0.37\\
{} &{} & {} &{$>1$} & 0.34\\
{} &{} & \multirow{2}{*}{$>4$} &{$=1$} & 0.58\\
{} &{} & {} &{$>1$} & 0.49\\
{} &\multirow{4}{*}{muon} & \multirow{2}{*}{$=4$} &{$=1$} & 0.52\\
{} &{} & {} &{$>1$} & 0.31\\
{} &{} & \multirow{2}{*}{$>4$} &{$=1$} & 0.46\\
{} &{} & {} &{$>1$} & 0.35\\
\hline
\multirow{8}{*}{\bQ} & \multirow{4}{*}{electron} & \multirow{2}{*}{$=4$} & {=1} & 0.70\\
{} &{} & {} &{$>1$} & 0.39\\
{} &{} & \multirow{2}{*}{$>4$} &{$=1$} & 0.77\\
{} &{} & {} &{$>1$} & 0.68\\
{} &\multirow{4}{*}{muon} & \multirow{2}{*}{$=4$} &{$=1$} & 0.50\\
{} &{} & {} &{$>1$} & 0.32\\
{} &{} & \multirow{2}{*}{$>4$} &{$=1$} & 0.75\\
{} &{} & {} &{$>1$} & 0.53\\
\hline
\hline
{\dQ} &{all} & {all} &{all} & 0.14\\
{\bQ} &{all} & {all} &{all} & 0.18\\
\hline
{Combination} &{all} & {all} &{all} & 0.11\\
\hline

\end{tabular}
\end{center}
\caption{Expected statistical uncertainty for the eight individual channels for both analysers as well as for the combination of analysers. }
\label{tab:exp_stat_unc}
\end{table}

The gain of sensitivity due to the combination of channels is visible. Even though the separation of the \dQ\ and \bQ\ is comparable, it is still on average lower for the \dphi\ distributions utilizing the \bQ. 
Two \btag s increase the purity (less background) and the reconstruction efficiency. The latter is also increased in the case of four jets, due to a decreased combinatorial background.

\section{Analyser Correlation}
\label{sec:correlation}

The azimuthal angles \dphi\ between the lepton and the \dQ\ and the lepton and the \bQ\ are used in this analysis as two independent variables for the full combination fit. Their independence is cross-checked by three main points:
\begin{itemize}
\item The two observables must obtain their spin analysing power from different effects. This is true as the analysing power of the \bQ\ arises from longitudinally polarized $W$ bosons and is degraded by transversely polarized $W$ bosons. The  analysing power of the \dQ\ arises from both the longitudinally and the transversally polarized $W$ bosons. Independent treatment and combination is suggested in \cite{Stelzer1996}.
\item{At parton level, the two observables need to be uncorrelated. }
\item{At detector level, the two observables need to be uncorrelated. }
\end{itemize}
These last two points are addressed in this section. The correlation of the two observables \dphidQ\ and \dphibQ\ is evaluated by plotting the two quantities in two-dimensional histograms. Table \ref{tab:indep_corr} lists the correlation coefficients on the detector level for the signal, the backgrounds and the data. For the signal, also parton level results are shown.

\begin{table}[htbp]
\begin{center}
\begin{tabular}{|c|c|c|c|c|c|c|}
\hline
Correlation  & \multicolumn{2}{c|}{ SM Sample} & \multicolumn{2}{c|}{ Uncorr. Sample}   & BG & Data\\
{}  &Parton Level& Rec. Level&Parton Level& Rec. Level& \multicolumn{2}{c|}{Rec. Level }\\
\hline
\hline
\ejets\ & \multirow{2}{*}{-0.04}&-0.12& \multirow{2}{*}{-0.05} & -0.12 & -0.10 &-0.12\\
\mujets\ & {} & {-0.11}& {} & -0.12 & -0.11 & -0.12\\
\hline
\end{tabular}
\end{center}
\caption{Linear correlation coefficients between $\Delta \phi(l,d)$ and  $\Delta \phi(l,b)$ on the detector level in the \ejets\ and the \mujets\ channel for the SM sample, the sample without spin correlation, the background (BG) and in data.}
\label{tab:indep_corr}
\end{table}
The observed correlation is small and consistent between data and prediction.  
In case the correlation affects the measurement it is expected to show up in a wrong estimation of the expected statistical uncertainty and deviations in the linearity check. Hence, the linearity checks from Section \ref{sec:linchecks} are repeated taking the correlation between \dQ\ and \bQ\ into account. This is realized by drawing real ensembles as subsets of the full MC sample instead of applying Poissonian fluctuations. 

For each generated ensemble, both the \dphidQ\ and \dphibQ\ values are filled into histograms to create pseudo data. This procedure replaces the application of Poissonian fluctuation, which is applied independently to each bin, leading to a vanishing correlation between the analysers. For each value of \fsm, 50 ensembles are created. 

The new linearity check leads to a slope of $0.99 \pm 0.01$ and an offset of $-0.01 \pm 0.01$. Within uncertainties, no deviation is observed. An equivalent cross check of the pull distribution for correlated ensembles is not meaningful. Using the procedure of real ensembles creates a bias in the estimation of the statistical uncertainty. This is a consequence of the way the pseudo data templates are created. To obtain a distribution of a certain \fsm\ requires a linear combination of events from the SM \ttbar\ sample and the sample of uncorrelated \ttbar\ pairs. Thus, the number of drawn events is larger than the expected \ttbar\ events. Hence, only for the cases of $\fsm=1.0$ and $\fsm=0.0$ the pull behaves as expected as only events from the SM and the uncorrelated sample are needed. The number of ensembles also matches the number of expected events in this and only this case. No difference in the expected statistical uncertainty was observed between the assumption of uncorrelated analysers and the correlated treatment.

\chapter[Systematic Uncertainties]{Systematic Uncertainties}
\label{sec:systematics}
No simulation including modelling of the underlying physics is expected to be perfect. Limited knowledge and simplified models cause systematic uncertainties, which affect the precision of the measurement. This concerns both the physics processes under study as well as the modelling of the detector. The systematic uncertainties relevant for this analysis are introduced in the next section. Two different approaches for the evaluation of the uncertainties will be discussed.

The classical way to evaluate systematic uncertainties is to perform ensemble tests: systematic templates are created by varying the default templates according to the systematic uncertainty by $1\,\sigma$ up and down respectively. Poissonian fluctuations of the bins of the modified templates are applied to create a set of ensembles. The fit output distribution of this pseudo data follows a Gaussian distribution. Comparing the mean of this distribution to the nominal fit result gives the size of the systematic uncertainty. This is done for all those systematic uncertainties where either a special prescription is needed (as ``taking the largest of effects A, B and C and symmetrize'''), where no systematic up and down variation is available (as switching a setting off which is on by default) or where a continuous interpolation between a default and a systematic variation makes no sense (as for the comparison of two different MC generators). \\
Another approach is the introduction of nuisance parameters. Instead of performing ensemble tests with the templates varied by $1\,\sigma$ up and down, these templates are used to calculate the modification of the signal template as a linear function of the parameter value of the respective uncertainty. This procedure was described in Section \ref{sec:nuipars}\\

Both procedures have been used depending on the type of systematic uncertainty and the results are provided after a detailed list of all uncertainties in the following section. The discussion of the effects on the measurement is presented at the end of this chapter. 

\section{List of Systematic Uncertainties}
\label{sec:syslist}
In this section all evaluated uncertainties are described in detail. They are grouped into detector related uncertainties affecting jet and lepton reconstruction, \etmiss\ and luminosity, background and signal modelling as well as method specific uncertainties that are caused by the limited template statistics.
\subsection{Jet Uncertainties}
\subsubsection{Jet Energy Scale}
Uncertainties on the different in-situ JES calibration techniques, as discussed in Section \ref{sec:jets}, are combined and assigned to categories, depending on their source \cite{JES}. The total number of 54 uncertainties are further reduced via combination into groups. In the end, the following numbers of uncertainties remain:
\begin{itemize}
\item Detector description (2)
\item Physics modelling (4)
\item Statistics and method (3)
\item Mixed category (2)
\end{itemize}
On top of these eleven in-situ JES uncertainties additional sources of uncertainties are determined \cite{JES}:
\begin{itemize}
\item \textbf{$\eta$-Intercalibration (2)} Statistical and modelling (dominated by \pythia\ vs. \herwig\ difference in forward region) uncertainties. 
\item \textbf{Pile-Up (2)} Effects of the number of primary vertices (in-time pile-up) and average number of interactions per bunch crossing (out-of-time pile-up) on the JES.
\item \textbf{High \pt\ Jets} Difference between the high \pt\ single hadron response in-situ and in test beam measurements.
\item \textbf{MC Non-Closure} Difference between the MC generators used in the calibration and in the analysis. 
\item \textbf{Close-By Jets} Uncertainty on the effect of varied jet energy response due to close-by jets.
\item \textbf{Flavour Composition} Uncertainty on the fraction of gluon jets leading to a different jet response.
\item \textbf{Flavour Response} Uncertainty on the particular gluon and light quark jet responses.
\item \textbf{$b$-JES} Uncertainty on the jet response difference for $b$-jets. It replaces uncertainties on flavour composition and response in case a jet is tagged as $b$-jet.
\end{itemize}
In total, 21 components of the JES uncertainty are available and evaluated. An overview of the total JES uncertainty as a function of the jet \pt\ is shown in Figure \ref{fig:JES_insitu} and Figure \ref{fig:JES_total}. 

 \begin{figure}[htbp]
 	\centering
			\subfigure[]{
		\includegraphics[width=0.45\textwidth]{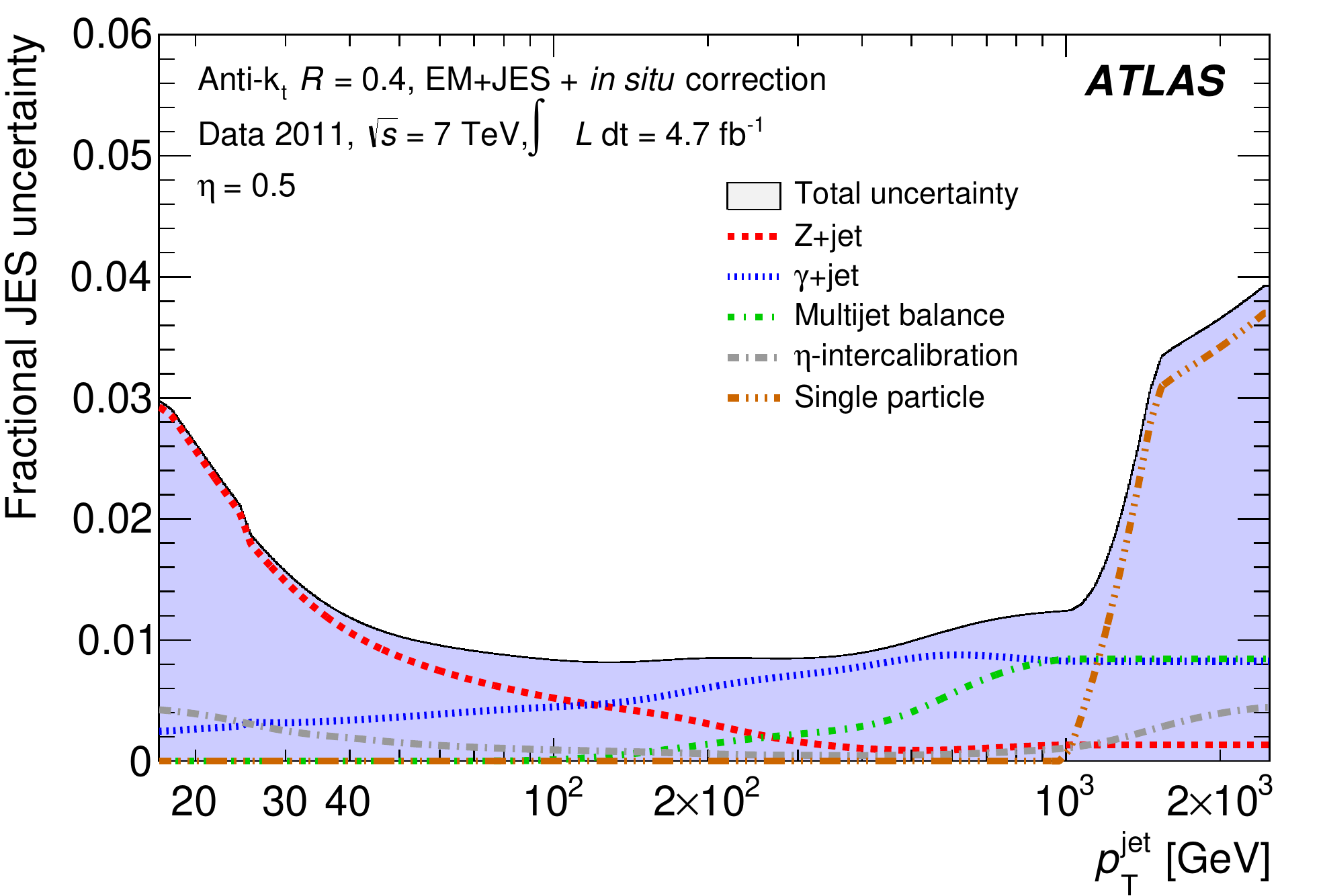}
			\label{fig:JES_insitu}
		}
					\subfigure[]{
		\includegraphics[width=0.45\textwidth]{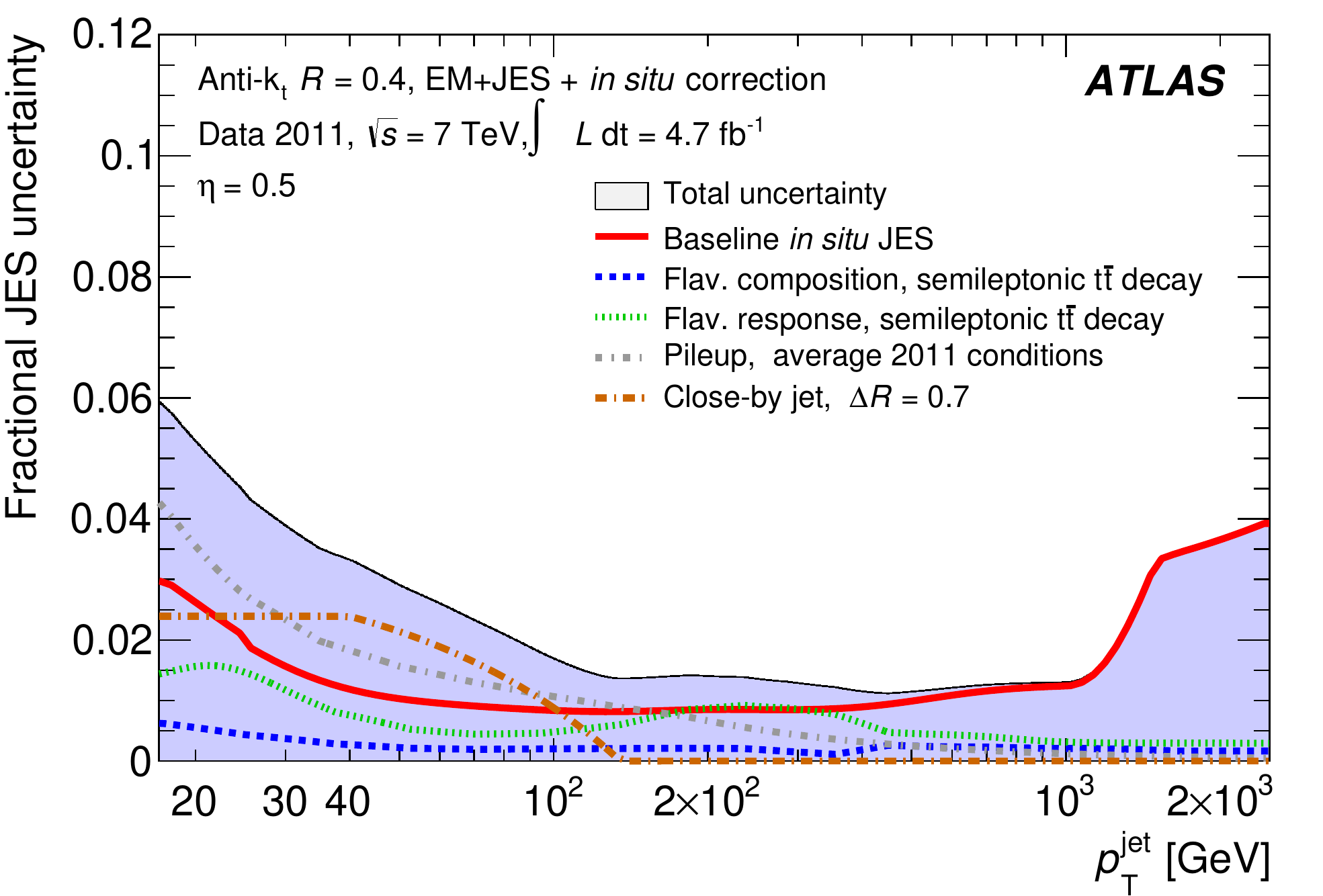}
			\label{fig:JES_total}
		}		
\caption{JES systematic uncertainty as a function of jet \pt. \subref{fig:JES_insitu} Total in-situ contribution and components \cite{JES}. \subref{fig:JES_total} Total JES uncertainty (without $b$-jet JES uncertainty) with \ttbar\ (\ljets\ channel) specific components \cite{JES}.}
\label{fig:JES}
\end{figure}

\subsubsection{Jet Energy Resolution}
The jet energy resolution was measured using the bisector method \cite{bisector} and di-jet \pt\ balance \cite{JER}. The energy resolution determined in data and MC agree within 10\,\%. This difference is covered by the uncertainties of the resolution measurement. Hence, no correction of the MC resolution is applied. The jet energy resolution uncertainty on the analysis is evaluated by smearing the jets in the MC according to the uncertainties of the resolution measurement in an updated version of \cite{JER} using the full 2011 dataset.

\subsubsection{Jet Reconstruction Efficiency}
By comparing track jets to calorimeter jets, a difference in the jet reconstruction efficiency between data and Monte Carlo simulation is found \cite{jetreco}. The efficiency in data is slightly smaller. For the evaluation of the jet reconstruction efficiency, jets were randomly rejected according to the mismatch in efficiency. The jet reconstruction efficiency in MC is lowered by 0.23\,\% for jets with a \pt\ between 20 and 30 GeV. Jets with a higher \pt\ are not affected.

\subsubsection{Jet Vertex Fraction}
As described in Section \ref{sec:jets}, scale factors are applied to the jet selection efficiency and inefficiency for both jets emerging from the hard scattering process as well as pile-up jets. 
Scale factors are applied to the hard scatter jet selection efficiency $\varepsilon_{\text{HS}}$ and the mistag rate $I_{\text{HS}}$. 
The \pt\ dependence of the JVF SFs is parameterized and fitted. The uncertainties on these fits are taken as one contribution to the JVF SF uncertainty. Another contribution comes from effects of varied selection cuts applied to the $Z$+jet sample, which is used to determine the JVF SF. 
The total JVF uncertainty is obtained by propagating the uncertainties of the JVF scale factors ($\Delta \varepsilon_{\text{HS}} \approx 0.5\,\%$, $\Delta I_{\text{HS}} \approx$ 5-10\,\%) to the total event weight. 

\subsubsection{B-Tagging Scale Factors}
The scale factors for tagged $b$-jets, $c$-jets and mistagged jets are derived by combining several calibration methods as described in Section \ref{sec:btagging}. The uncertainties on the corresponding scale factors are indicated in Figure \ref{fig:btag_calib}.
For this analysis the uncertainties are accessed using the \textit{eigenvector method}\index{Eigenvector method}. The covariance matrices of all uncertainties are summed. The square roots of the corresponding eigenvalues are then used as components of the total uncertainty.
These components are available for the efficiencies of $b$-jets (9), $c$-jets (5) and the mistag rate (1). A similar approach of the eigenvector method is used in the context of PDF uncertainties, described for example in \cite{lhcprimer}. 

\subsection{Lepton Uncertainties}
\subsubsection{Lepton Trigger Scale factor}
Uncertainties on the trigger scale factors are derived for both electrons ($\approx$ 0.5-1.0\,\%) and muons (1-2\,\%). They depend on the data taking period as well as on the $\eta$ and $E_T$ (electrons) or $\eta$ and $\phi$ (muons). The uncertainties contain components from limited $Z$ boson sample statistics and systematic uncertainties for different T\&P selections. 

\subsubsection{Lepton ID and Isolation Efficiency}
The uncertainties on electron ID and isolation efficiency scale factors (2-3\,\%) depend on the $\eta$ and $E_T$ of the electron. Statistical limitations, a pile-up dependence, the modelling of the underlying events as well as a difference between the isolation efficiency in $W/Z$ and top quark events contribute to the uncertainty.

For muons, the isolation efficiency uncertainty ($\approx 0.7\,\%$) depends on the data period and is composed of a statistical and a systematic component. 

\subsubsection{Lepton Reconstruction Efficiency}
The uncertainty on the electron reconstruction efficiency (0.6-1.2\,\%) depends only on \abseta\ of the electron, while uncertainties on the muon reconstruction efficiency ($\approx 0.3\,\%$) depend on the data taking period as well as on $\eta$ and $\phi$ of the muon. For the muon, the statistical and systematic uncertainty components are added linearly. 

\subsubsection{Electron Energy Resolution}
The electron energy resolution is smeared in the Monte Carlo simulation in order to match the resolution in data. Each smearing factor has a relative uncertainty  of $\leq$ 10\,\% for electron energies up to 50\,\GeV\ and up to 60\,\% for high energetic electrons. For the evaluation of the uncertainty of the electron resolution, the smearing is performed with the systematic variation of the smearing factors. 

\subsubsection{Electron Energy Scale}
Before the energy resolution smearing is applied, the energy of the electron is scaled up and down by the corresponding uncertainty. The uncertainties of up to 1.5\,\% depend on the $E_T$ of the electron as well as on the $\eta$ of $\phi$ of the corresponding energy cluster. Dominating contributions result from the modelling of the detector material and the presampler energy scale.

\subsubsection{Muon Momentum Scale}
A muon momentum scale correction (up to 1.5\,\%) is applied to the MC simulation by default. For the evaluation of the corresponding uncertainty, it is completely switched off. The caused effect is quoted as symmetrized uncertainty. 

\subsubsection{Muon Momentum Resolution}
The muon momentum resolution is varied separately for the ID and the muon spectrometer components according to their uncertainties. Uncertainties on the resolution smearing factors vary between 2-12\,\% (muon spectrometer) and 4-27\,\% (ID), respectively. The largest difference of the two up and the two down variations is taken as uncertainty. 

\subsection{Missing Transverse Momentum}
Two different types of uncertainties affect the \etmiss. On the one hand, the uncertainties of the objects used to calculate the \etmiss\ are propagated. On the other hand, dedicated \etmiss\ uncertainties exist:
 The \textit{pile-up uncertainty} takes into account effects of additional energy in the calorimeter coming from pile-up events. The uncertainties on the \textit{CellOut} term (11-14\,\%) for calorimeter energy outside reconstructed objects and the \textit{SoftJets} term (9-11\,\%) for soft underlying events are 100\,\% correlated and evaluated together. The effects of both the \textit{pile-up uncertainty} (6.6\,\% effect on both the CellOut and the SoftJets term) and the combined \textit{CellOut/SoftJets} uncertainty are added in quadrature to obtain the total \etmiss\ uncertainty.

\subsection{Luminosity}
The total luminosity of \intlumi\ for the full 2011 dataset has an uncertainty of 1.8\,\%, measured via van der Meer scans \cite{improved_lumi}. To account for this, the expected yields were changed in the priors accordingly and the fit was repeated with the priors modified up and down by 1.8\,\%.

\subsection{Uncertainties on the Background}

\subsubsection{Fake Lepton Normalization}
The uncertainty of the QCD fake estimation is evaluated by varying the real and fake efficiencies according to their uncertainties and adding their effects in quadrature. This yields to an normalization uncertainty of 50\,\% in the \mbox{\ejets} channels and 20\,\% in the \mbox{\mujets} channels. These uncertainties are then used for the prior widths on the background yields as described in Section \ref{sec:fitpars}. The same priors were used for the $ n_{\text{jets}} = 4$ and the  $n_{\text{jets}} \geq 5$ channels.

\subsubsection{Fake Lepton Shape}
For the \ejets\ channel the effects on the shape arising from the efficiency uncertainties for real and fake electrons are added in quadrature and taken as systematic uncertainty. In the \mujets\ channel, two different methods were used and averaged. Their difference is taken as systematic uncertainty.

\subsubsection{$W$+Jets Normalization}
The $W$+jets background was determined using MC samples. A data-driven approach, described in Section \ref{sec:wjets_norm}, is used to correct the normalization and the heavy flavour composition. 

The factor $r_{MC}$ (Equation \ref{eq:wjetsnorm}), used to determine the normalization of the $W$+jets background, will vary with modifications of the chosen MC generator parameters, the JES, the PDF, lepton ID misidentification and \btag ging scale factor uncertainties. The resulting $W$+jets normalization uncertainties are used as width for the $W$+jets priors in the fit, as described in Section \ref{sec:fitpars}. Different priors were used for the $ n_{\text{jets}} = 4$ and the  $n_{\text{jets}} \geq 5$ channels.

\subsubsection{$W$+Jets Shape}
Uncertainties on the $W$+jets shape are assigned to the flavour and jet multiplicity dependent scale factors as described in Sections \ref{sec:wjets_norm} and \ref{sec:wjets_flav}. The jet multiplicity bins were treated as uncorrelated.  The uncertainties contain components addressing the modelling, reconstruction and dedicated $W$+jets generator settings for the factorization and parton matching scales.

Details about the $W$+jets shape and normalization uncertainties can be found in \cite{top_xsec_diff_ATLAS}.

\subsubsection{Remaining Background Sources}
$Z$+jets, diboson and single top backgrounds are varied according to the uncertainties on the theoretical prediction. For $Z$+jets events the uncertainty is determined using Berends-Giele scaling \cite{Berends1991} to be 48\,\% for events with exactly four jets. For each additional jet, 24\,\% additional uncertainty is added in quadrature.
The uncertainties on the single top cross section are 3\,\% for the $t$-channel \cite{st_xsec_theo_t}, 4\,\% for the $s$-channel \cite{st_xsec_theo_s} and 8\,\% for the $Wt$-channel \cite{st_xsec_theo_Wt}.

The uncertainty on the diboson background is 5\,\% plus 24\,\% per additional jet not originating from a hadronically decaying boson.
 
The total effect on the remaining background sums up to 19\,\% for the \ejets\ channel and 15\,\% for the \mujets\ channel, conservatively covered by 20\,\% on the total remaining background. As in the other cases of normalization uncertainties, the uncertainty is propagated to the prior used in the fit.  

\subsection{\ttbar\ Modelling Uncertainties}
\label{sec:modeling_uncertainties}
A good modelling of the signal is necessary to correctly interpret the results. As the spin correlation is measured via kinematic distributions, all sources of uncertainties affecting them are of particular interest. This section is dedicated to uncertainties on the modelling of the \ttbar\ signal.

\subsubsection{Parton Distribution Functions}
\label{sec:unc_PDF}
The PDF used for the \ttbar\ signal generator \mcatnlo\ is the CT10 NNLO set \cite{CT10, Gao:2013xoa}. By using the LHAPDF framework \cite{lhapdf} weights depending on the initial partons' proton momentum fractions $x_i$ and the scale $Q^2$ can be obtained to rescale the samples to different PDF sets. For the evaluation of the PDF uncertainty, three different PDF sets including their nominal and error sets are compared: MSTW2008nlo68cl  \cite{MSTW, Martin:2009bu}, CT10 and NNPDF2.3 \cite{NNPDF}.

Pseudo data is generated from the reweighted samples and used for ensemble testing. The fit output values are plotted in Figure \ref{fig:pdf_unc}.
\begin{figure}[htbp]
\begin{center}
			\subfigure[]{
\includegraphics[width=0.45\textwidth]{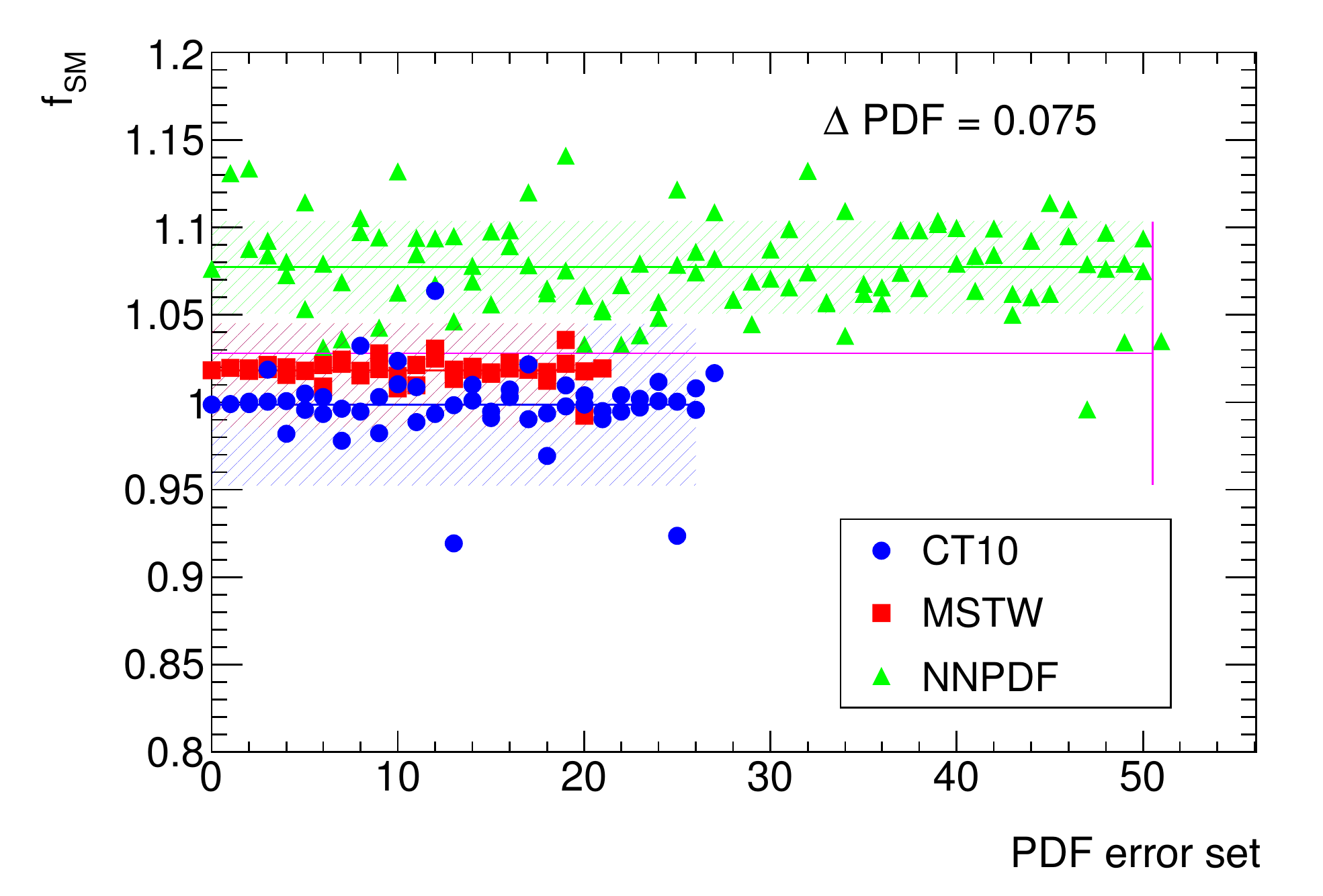}
\label{fig:PDF_dQ}
}
			\subfigure[]{
\includegraphics[width=0.45\textwidth]{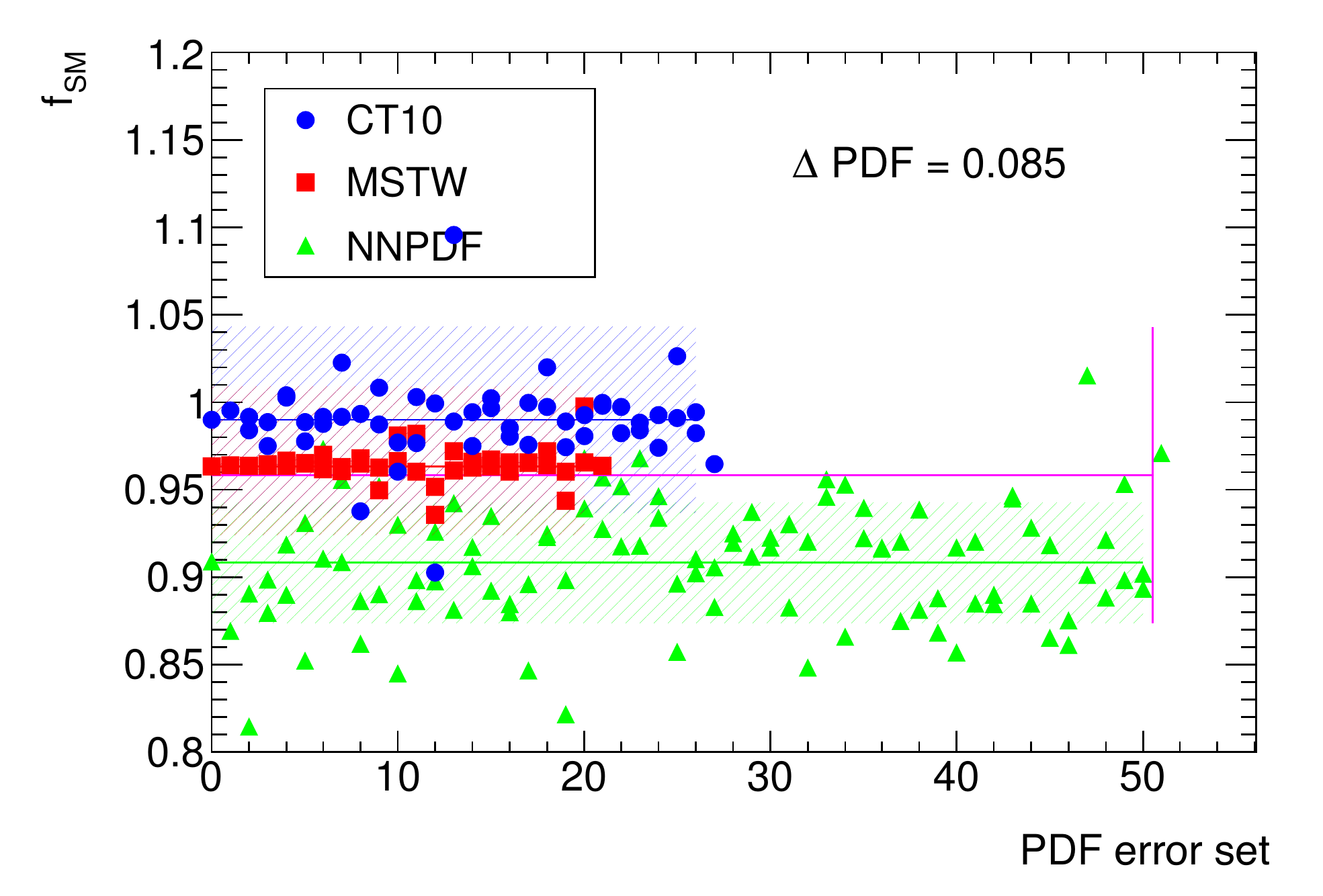}
\label{fig:PDF_bQ}
}
			\subfigure[]{
\includegraphics[width=0.45\textwidth]{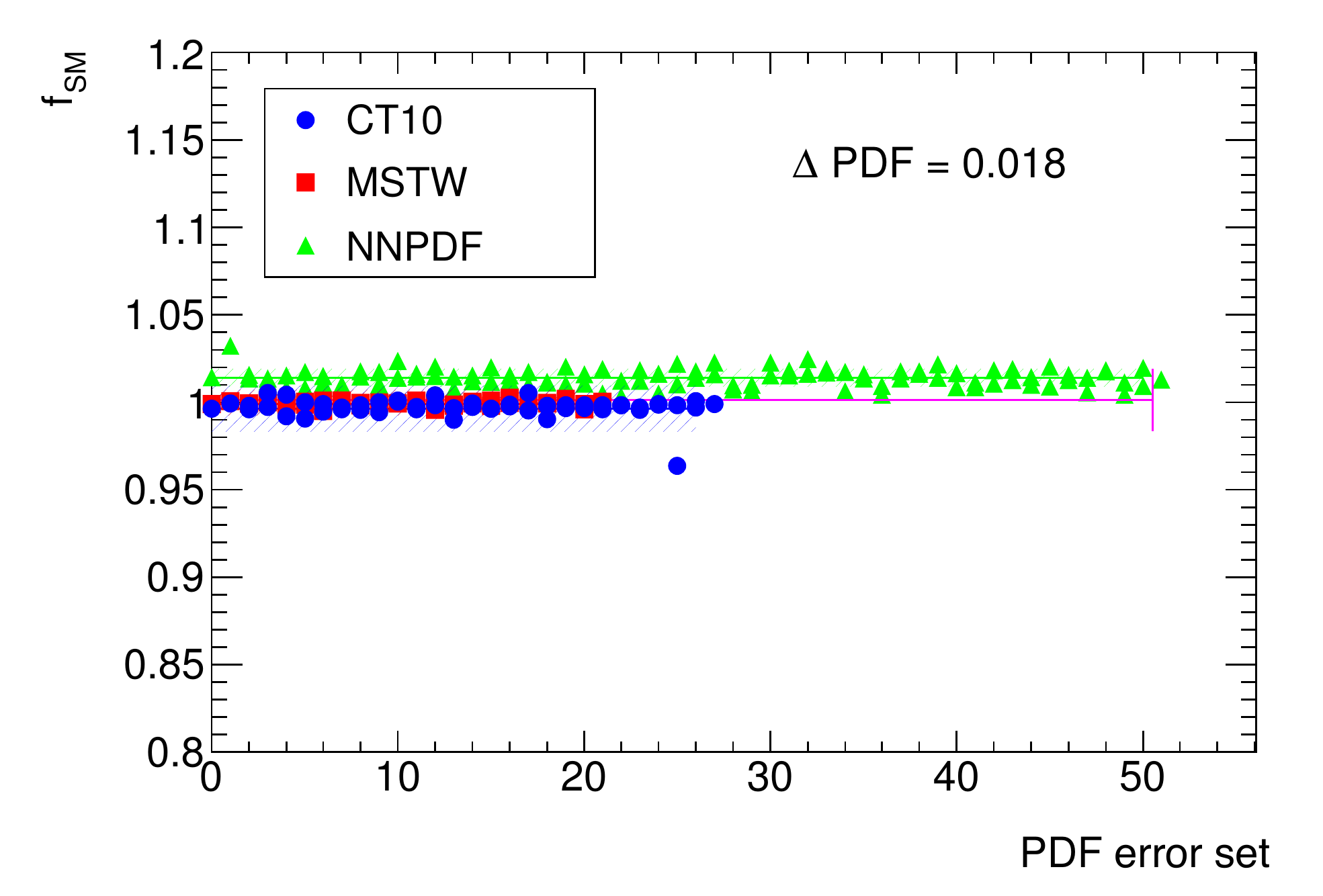}
\label{fig:PDF_dbQ}
}
\end{center}
\caption{Fitted \fsm\ values for the MSTW2008nlo68cl, CT10 and NNPDF2.3 PDF set and their corresponding error sets. The results are shown for the \subref{fig:PDF_dQ} \dQ, the \subref{fig:PDF_bQ} \bQ\  and \subref{fig:PDF_dbQ} the full combination of the fit.}
\label{fig:pdf_unc}
\end{figure}  Each bin contains one up- and down variation of the error set, except the nominal one in the first bin. For each PDF set, an error band is drawn. According to the definition of the PDF errors, these bands are the RMS (NNPDF), the asymmetric Hessian (MSTW) and the symmetric Hessian (CT10). The outer edges of the error bands define the total PDF uncertainty, as indicated in the plots. One can see that the modifications due to different PDF sets have effects going in opposite directions for the two spin analysers. Thus, the combination of the two analysers can reduce the PDF uncertainty significantly.

\subsubsection{Top Quark Mass}
Samples with varied masses for the top quark are used for ensemble testing. Figure \ref{fig:mass_unc} shows the mean fit output values for both analysers and the combination. 
\begin{figure}[htbp]
\begin{center}
\includegraphics[width=0.75\textwidth]{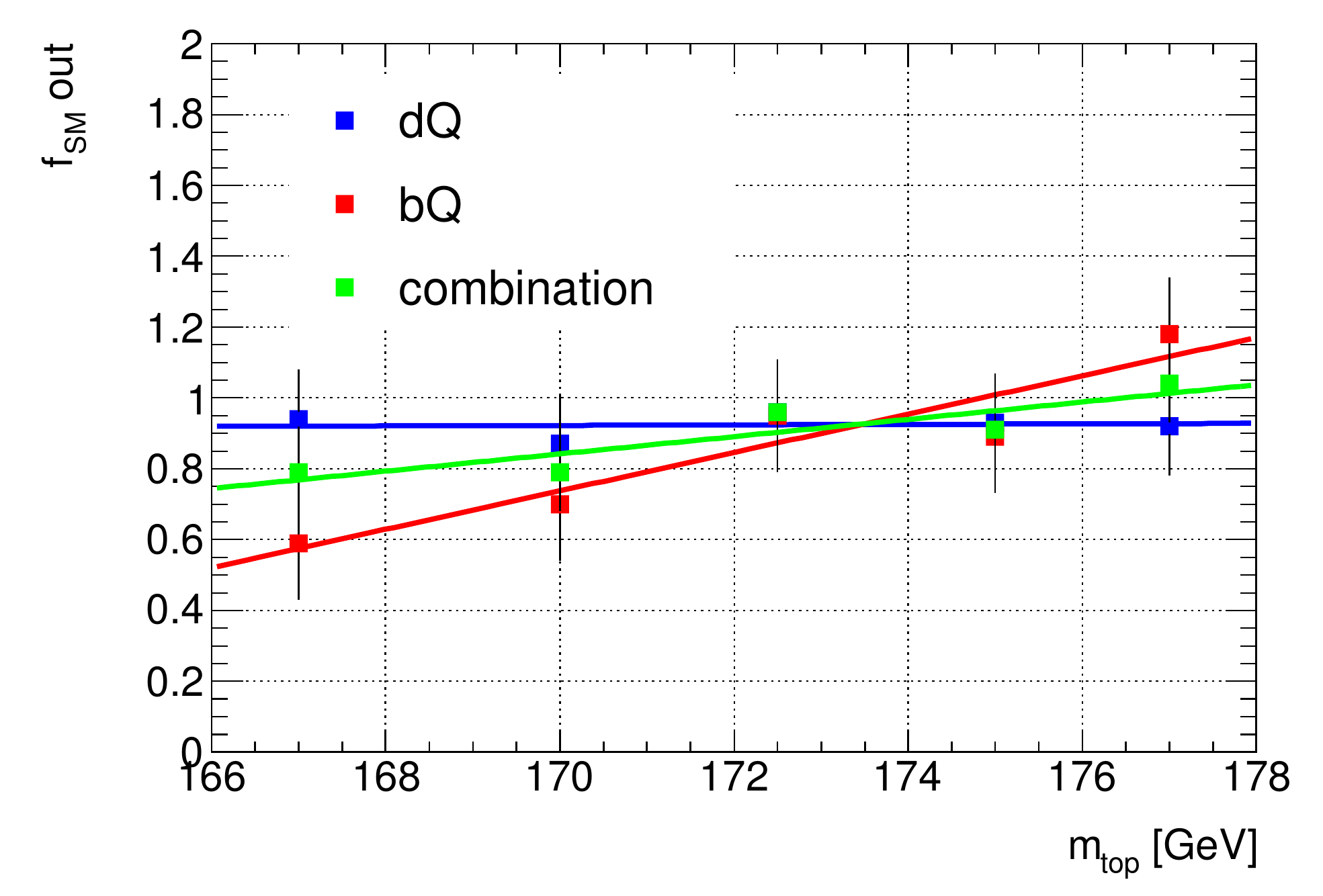}
\end{center}
\caption{Fit values for \fsm\ for different values of the top quark mass. For each analyser, a linear fit is performed.}
\label{fig:mass_unc}
\end{figure} 
To evaluate the dependence on the top quark mass, linear fits are performed. The slope $s$ is used to calculate the uncertainty of the fitted \fsm\ values due to limited knowledge of the top quark mass:
\begin{align}
\Delta \fsm = s \cdot \Delta m_t
\end{align} 
Several options for choosing $\Delta m_t$ exist. Examples are the uncertainty on the world combination ($\Delta m_t = 0.76 \GeV$ \cite{m_top_world}) or the LHC combination ($\Delta m_t = 0.95 \GeV$ \cite{m_top_LHC}) as well as the deviation between the mass used in the generator and the world combination ($\left| m_t^{\text{MC}} - m_t^{\text{world}}\right| = \left| 172.5 \GeV - 173.34 \GeV\right| = 0.84 \GeV$). The uncertainty on the LHC combination was used in order to be conservative and cover the deviation of $m_t$ used in the generator.

While the \dQ\ is relatively stable against variations of the top mass, the \bQ\ is not. This comes from the fact that the spin analysing power for the \dQ\ is always 1, independent of the kinematics of the top decay. As the spin analysing power of \bQ\ depends on the $W$ boson polarization state, which itself depends on the top mass (see Equation \ref{eq:alpha_b} or Figure 7 in \cite{Mahlon1996}), a dependence of the \bQ\ as analyser is expected.
The obtained values for the slopes $s$ are $s_{dQ} < 0.01 \GeV^{-1}$, $s_{bQ} = 0.05 \GeV^{-1}$ and $s_{\text{comb.}} = 0.02 \GeV^{-1}$.

\subsubsection{Top \pt\ uncertainty}
\label{sec:toppt_unc}
Recent measurements of the differential top quark cross section \cite{top_xsec_diff_ATLAS, top_xsec_diff_ATLAS_conf} showed that the top \pt\ spectrum of \mcatnlo\ and the unfolded measurement in data agree within uncertainties. But especially for the high values of the top \pt\ the agreement is at the edge of the uncertainties. Furthermore, a slope in the ratio is visible in Figure \ref{fig:toppt_xsec_vs_pt}. It is not only the MC generator itself, which causes the top \pt\ differences. Also the used PDF set plays an important role, as shown in Figure \ref{fig:toppt_PDF}.

\begin{figure}[htbp]
\begin{center}
			\subfigure[]{
\includegraphics[width=0.5\textwidth]{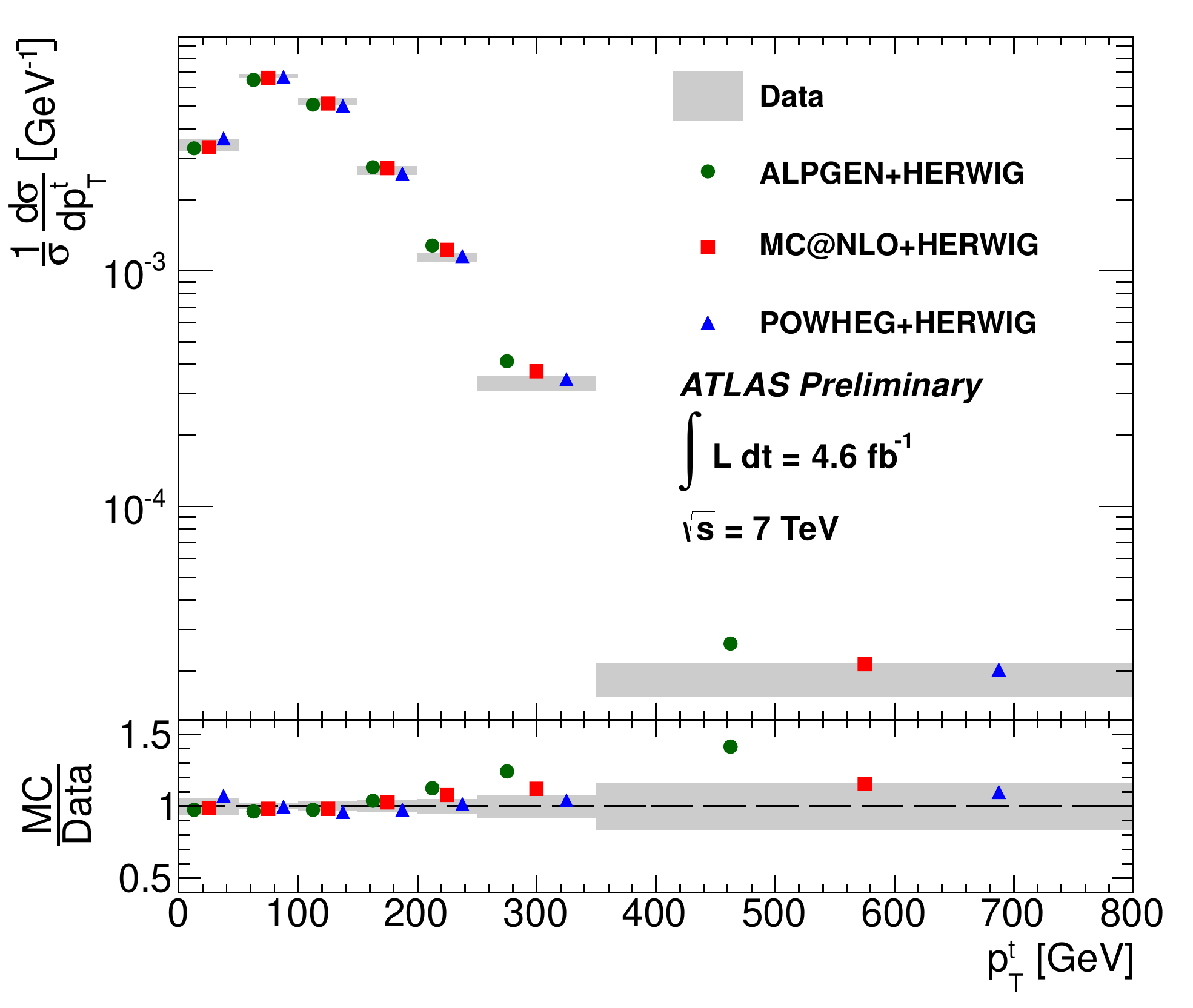}
\label{fig:toppt_xsec_vs_pt}
}
			\subfigure[]{
\includegraphics[width=0.44\textwidth]{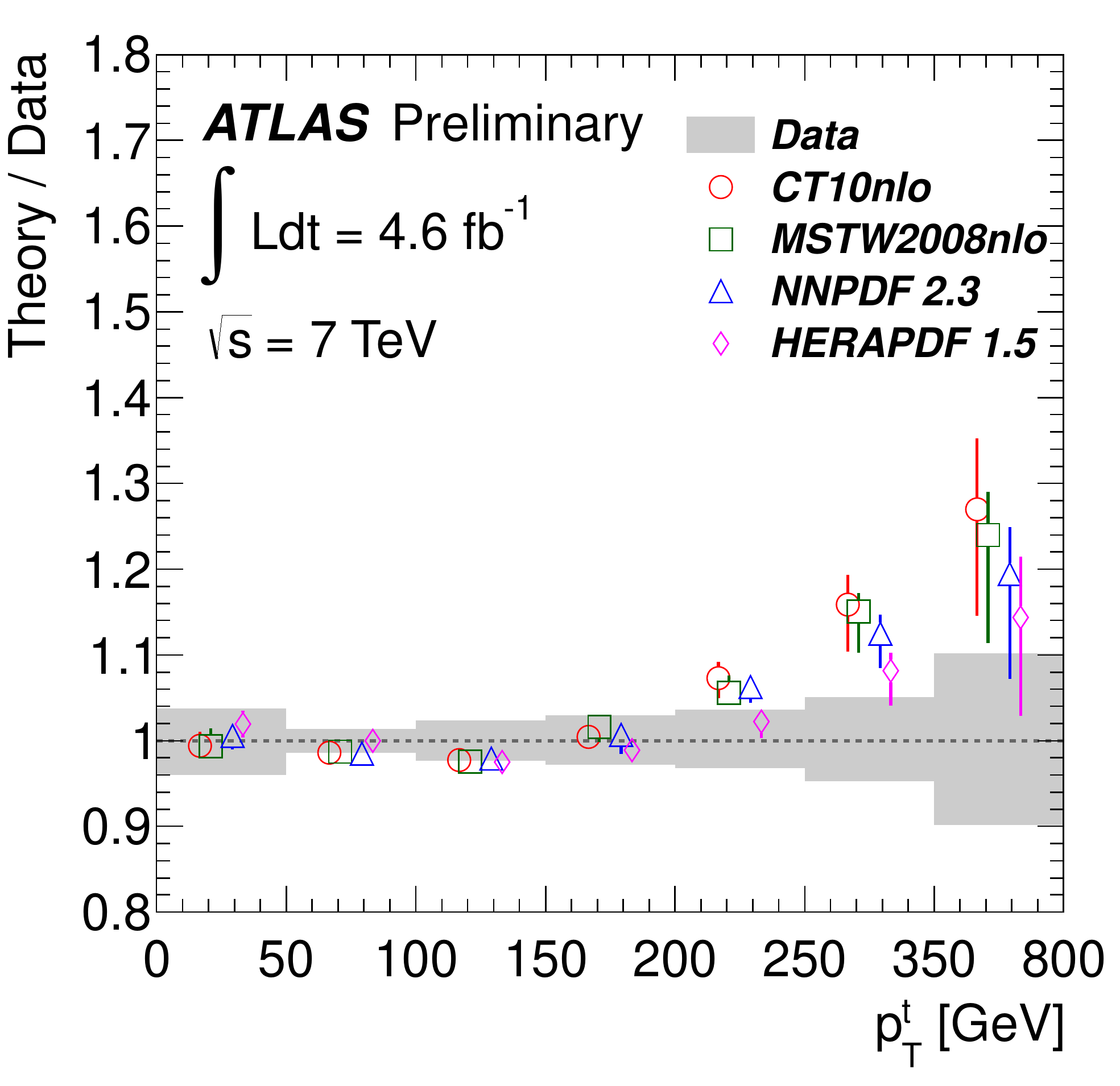}
\label{fig:toppt_PDF}
}
\end{center}
\caption{\subref{fig:toppt_xsec_vs_pt} Measured differential top quark cross section as a function of the top quark transverse momentum \cite{top_xsec_diff_ATLAS_conf}. The unfolded data is compared to different generators. \subref{fig:toppt_PDF} Ratios of the NLO QCD predictions to the measured normalized differential cross sections for different PDF sets \cite{top_xsec_diff_ATLAS_conf}.}
\label{fig:toppt_conf}
\end{figure} 

The effect of a mismodelled \pt\ spectrum of the top quark is investigated. First, the results of\cite{top_xsec_diff_ATLAS_conf} had to be reproduced. For that, top quark\footnote{The \herwig\ status code 155 was used to access the top quarks.} \pt\ distributions for the full phase space were compared. \mcatnlo\ and \powheg+\herwig\ are used as generators and are compared to the unfolded measurement of \cite{top_xsec_diff_ATLAS_conf}. 

\begin{figure}[htbp]
\begin{center}
			\subfigure[]{
\includegraphics[width=0.45\textwidth]{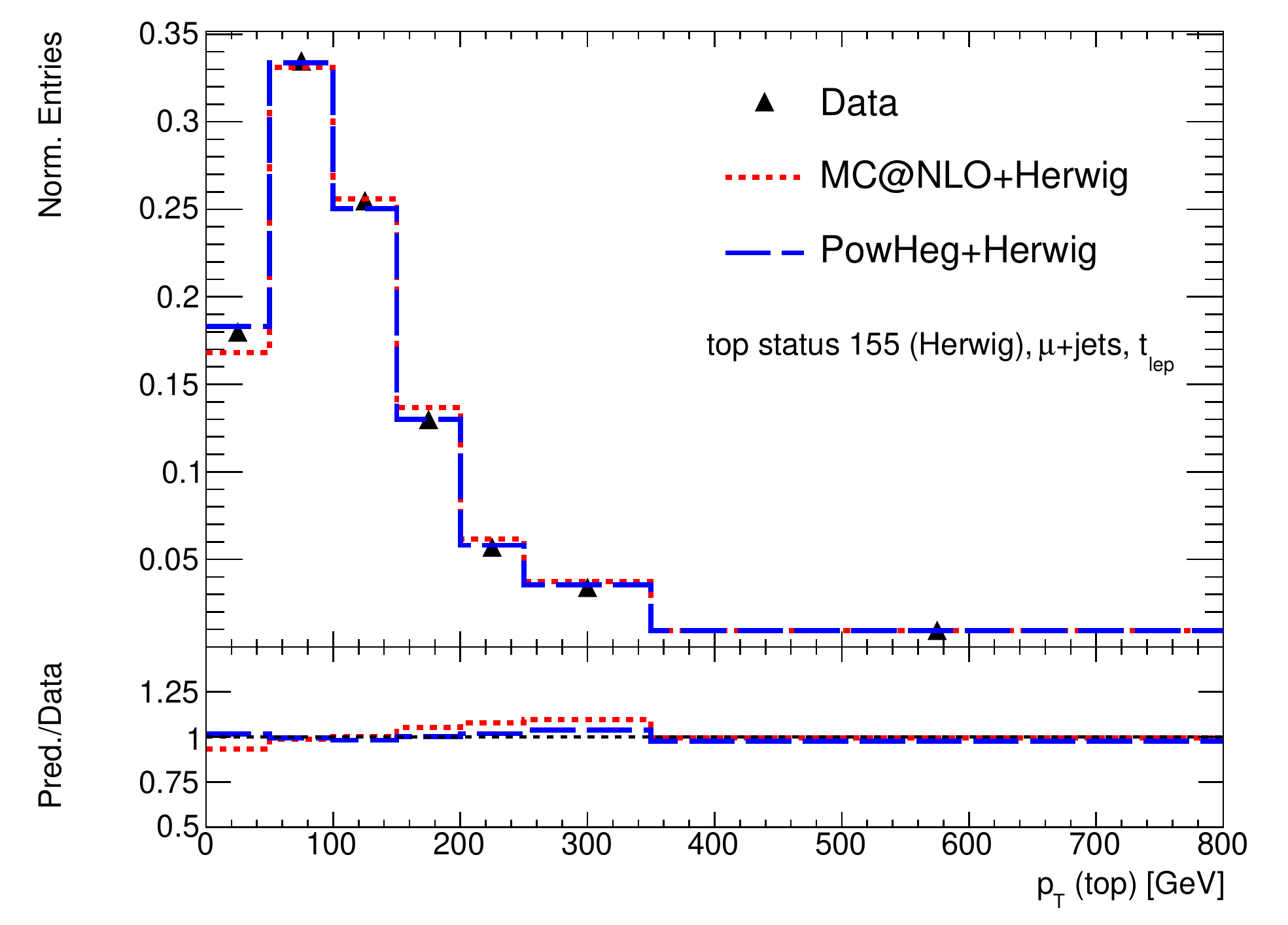}
\label{fig:toppt_meas_tlep_mujets}
}
			\subfigure[]{
\includegraphics[width=0.45\textwidth]{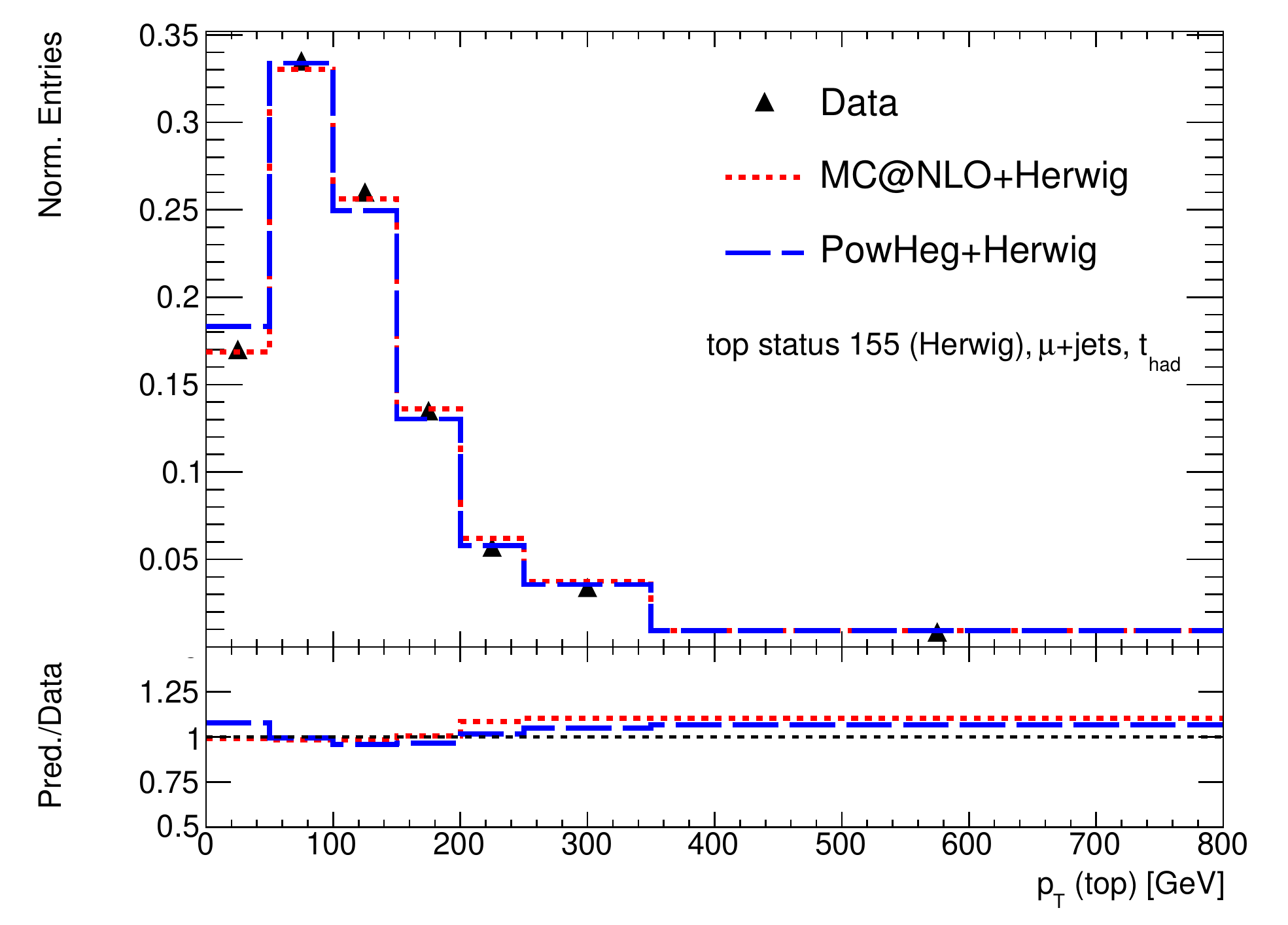}
\label{fig:toppt_meas_thad_mujets}
}
\end{center}
\caption{The \pt\ spectrum of \subref{fig:toppt_meas_tlep_mujets} the leptonic  and \subref{fig:toppt_meas_thad_mujets} the hadronic  top quark. The unfolded data measurement of \cite{top_xsec_diff_ATLAS_conf} is compared to \mcatnlo\ and \powheg+\herwig. Here, results from the \mujets\ channel are shown.}
\label{fig:toppt_ratios}
\end{figure} 
The distributions, shown in Figure \ref{fig:toppt_ratios}, reproduce the results from \cite{top_xsec_diff_ATLAS_conf}. The ratio of the hadronic top \pt\ spectrum measured in data and the one of \mcatnlo\ for the full phase space is used to reweight the signal sample. The provided combined top \pt\ spectrum of the \ejets\ and \mujets\ channels is used to calculate the scale factors as it had the smallest uncertainties and no discrepancies between the \ejets\ and \mujets\ numbers are observed. Table \ref{tab:topptSFs} shows the scale factors. The reweighted sample is then used to perform ensemble tests. The difference between the fitted value of \fsm\ and \fsm\  = 1.0 is then quoted as uncertainty for the top \pt. 

\begin{table}[htbp]
\begin{center}
\begin{tabular}{|c||c|c|c|c|c|c|c|}
\hline
top \pt\ [GeV]& 0-50 & 50-100 & 100-150 & 150-200 & 200-250 & 250-350 & 350-800 \\
\hline

Scale Factor & 1.01 & 1.02 & 1.02 & 0.99 & 0.92 & 0.88 & 0.86\\
\hline
\end{tabular}
\end{center}
\caption{Scale factors used to reweight the  top \pt\ spectrum of \mcatnlo\ to the one measured in data.}
\label{tab:topptSFs}
\end{table}

As the top \pt\ uncertainty is found to be large it was also split into a low and high top \pt\ part to check the effect of each region. Table \ref{tab:topptresults} shows the top \pt\ uncertainties from reweighting of the full top \pt\ spectrum (default, quoted as final uncertainty), the ``low top \pt\ region'' ($\leq 200$ GeV) and the ``high top \pt\ region'' ($>200$ GeV). 

\begin{table}[htbp]
\begin{center}
\begin{tabular}{|c|c|c|c|}
\hline
Top \pt\ Region & \dQ\ & \bQ\ & Combination \\
 \hline
 \hline
Full & 0.17& 0.24 & 0.01\\
Low & 0.03 & 0.05 &$<$ 0.01\\
High & 0.14 & 0.20 &0.01\\
\hline
\end{tabular}
\end{center}
\caption{Uncertainty on \fsm\ by reweighting the top \pt\ spectrum of \mcatnlo\ to the one measured in data. The uncertainty was evaluated by reweighting the full spectrum, the ``low top \pt\ region'' ($\leq $ 200 GeV) and the ``high top \pt\ region'' ($>$ 200 GeV).     }
\label{tab:topptresults}
\end{table} 
The high top \pt\ region has a higher influence on the total uncertainty. This fact is not trivial as the bulk of events is at the low top \pt\ spectrum. 
Figure \ref{fig:toppt_SM_normtosum} shows the \dphi\ distribution for the Standard Model expectation normalized to one. The blue and red line represent the fraction of the total spectrum for the low and high top \pt\ region. In Figure \ref{fig:toppt_SM_normto1}, the \dphi\ distributions for the low and high top \pt\ regions are normalized to one, which helps to illustrate the different \dphi\ shapes. When pseudo data is constructed from the samples reweighted in top \pt, the following effects can be observed: The steep contribution (high top \pt) is scaled down and the total distribution gets flatter. This will be interpreted as a higher spin correlation if the templates used for fitting stay unrescaled. Figure \ref{fig:toppt_reco_example} compares the SM and the uncorrelated \ttbar\ spin sample to the SM sample, which is reweighted in top \pt. Figure \ref{fig:topptexpl2} shows the same plots but for the \bQ\ as analyser. One can see the same effect by the top \pt\ reweighting but the opposite interpretations in terms of \fsm.
 \begin{figure}[htbp]
\begin{center}
			\subfigure[]{
\includegraphics[width=0.45\textwidth]{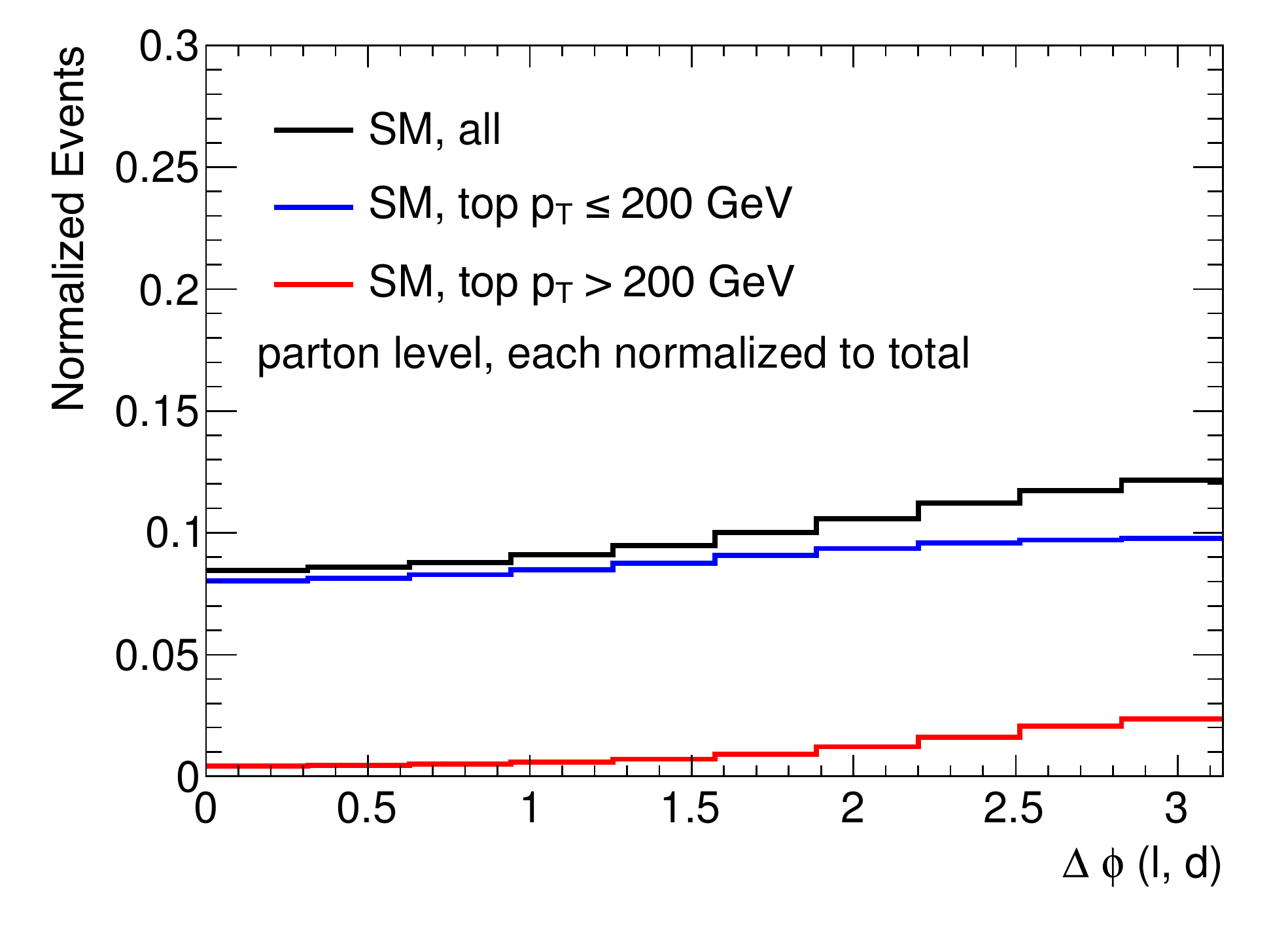}
\label{fig:toppt_SM_normtosum}
}
			\subfigure[]{
\includegraphics[width=0.45\textwidth]{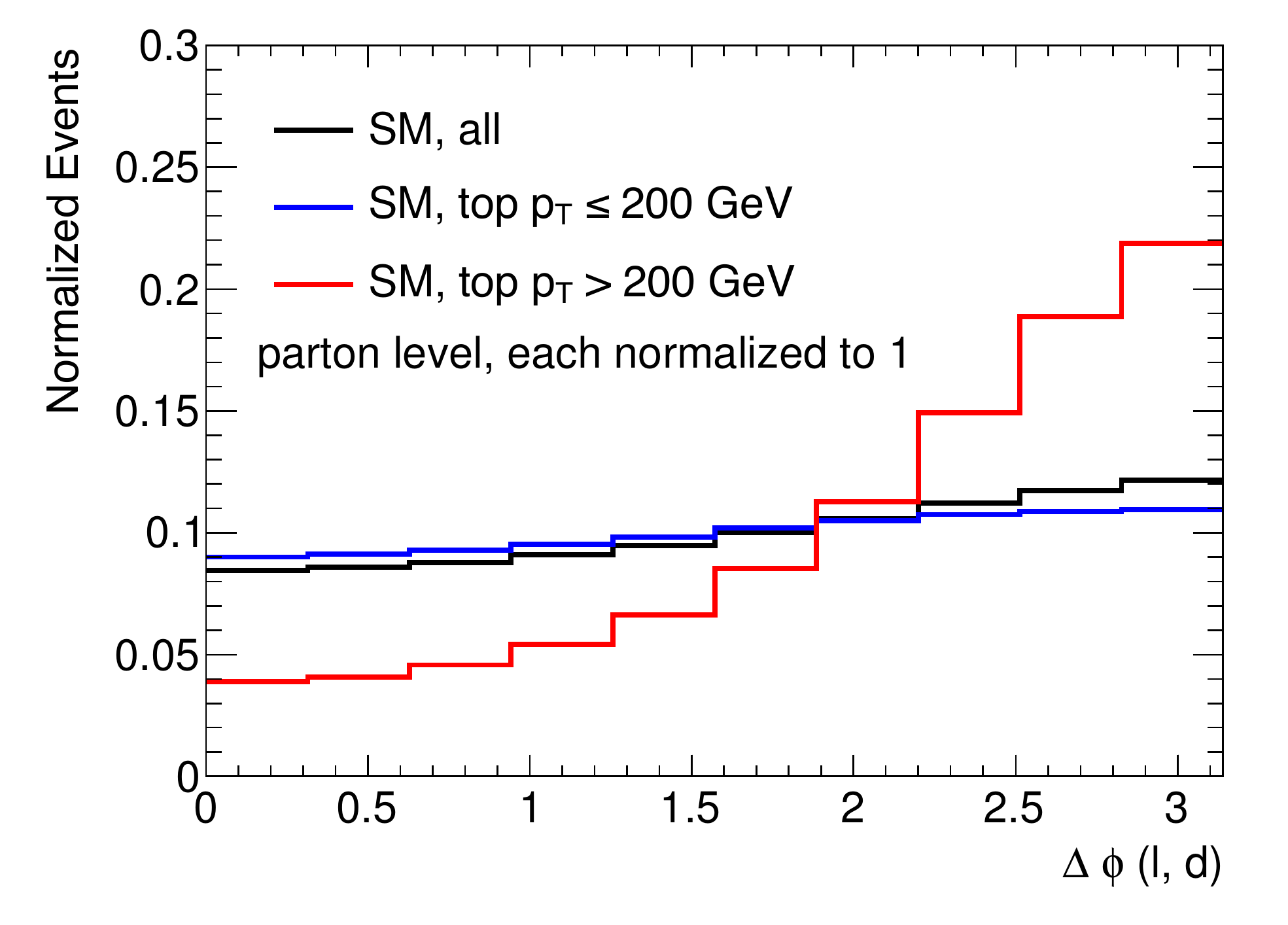}
\label{fig:toppt_SM_normto1}
}
			\subfigure[]{
\includegraphics[width=0.6\textwidth]{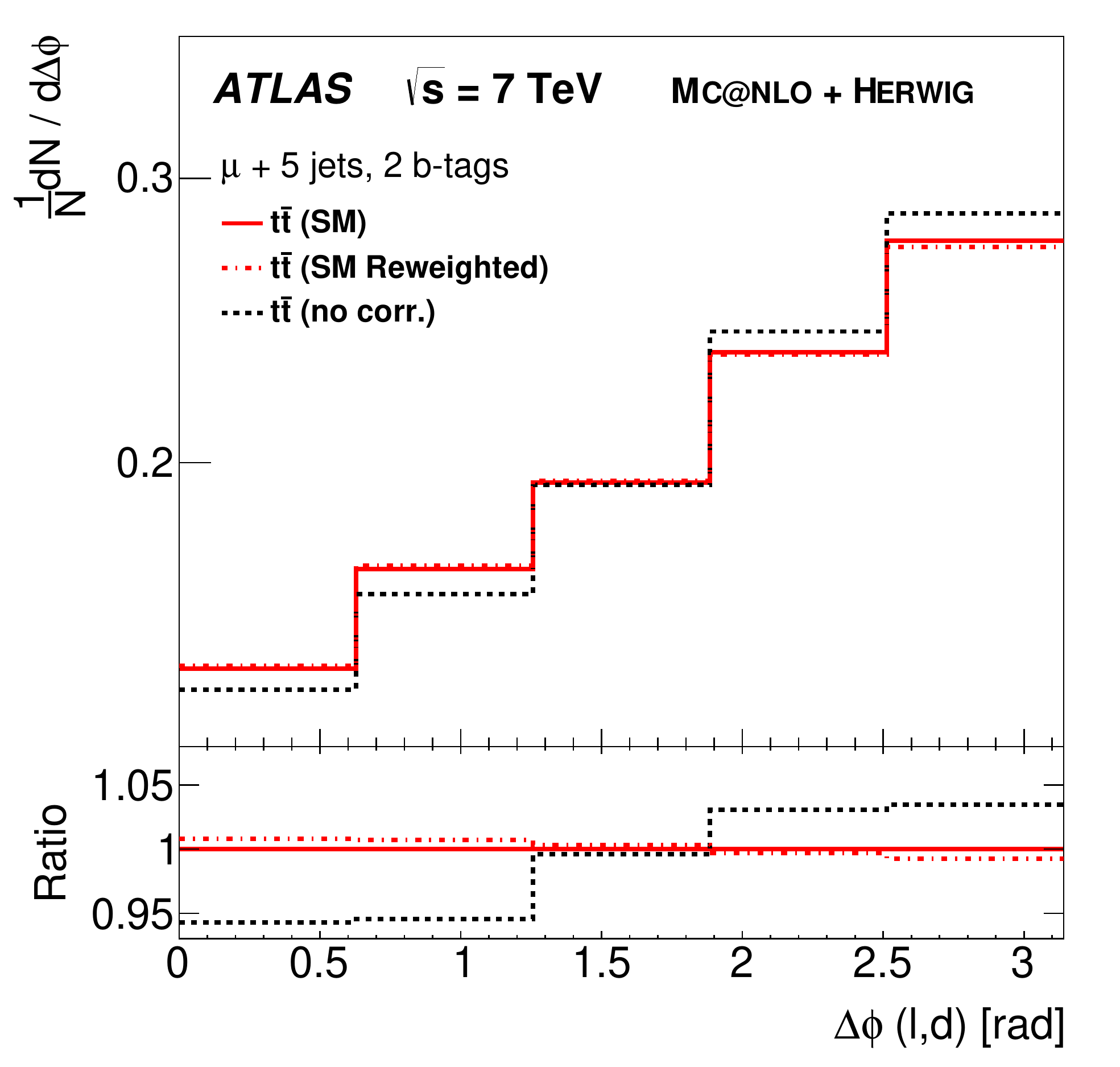}
\label{fig:toppt_reco_example}
}
\end{center}
\caption{\dphi\ distributions using the \dQ\ as analyser. \subref{fig:toppt_SM_normtosum} The \dphi\ distribution for the Standard Model expectation normalized to one. The contributions from the low and high top \pt\ region are indicated. \subref{fig:toppt_SM_normto1} Same distributions as \subref{fig:toppt_SM_normtosum}, but with the low and high top \pt\ region also normalized to one. As an example, the effect of reweighting is shown in \subref{fig:toppt_reco_example}: The SM and the uncorrelated \ttbar\ sample are compared to the SM prediction, which is reweighted in top \pt \cite{ueberpaper}. The ratio plot shows the ratios uncorrelated over SM (dashed) and reweighted over SM (red dash-dotted).}
\label{fig:topptexpl}
\end{figure} 

 \begin{figure}[htbp]
\begin{center}
			\subfigure[]{
\includegraphics[width=0.45\textwidth]{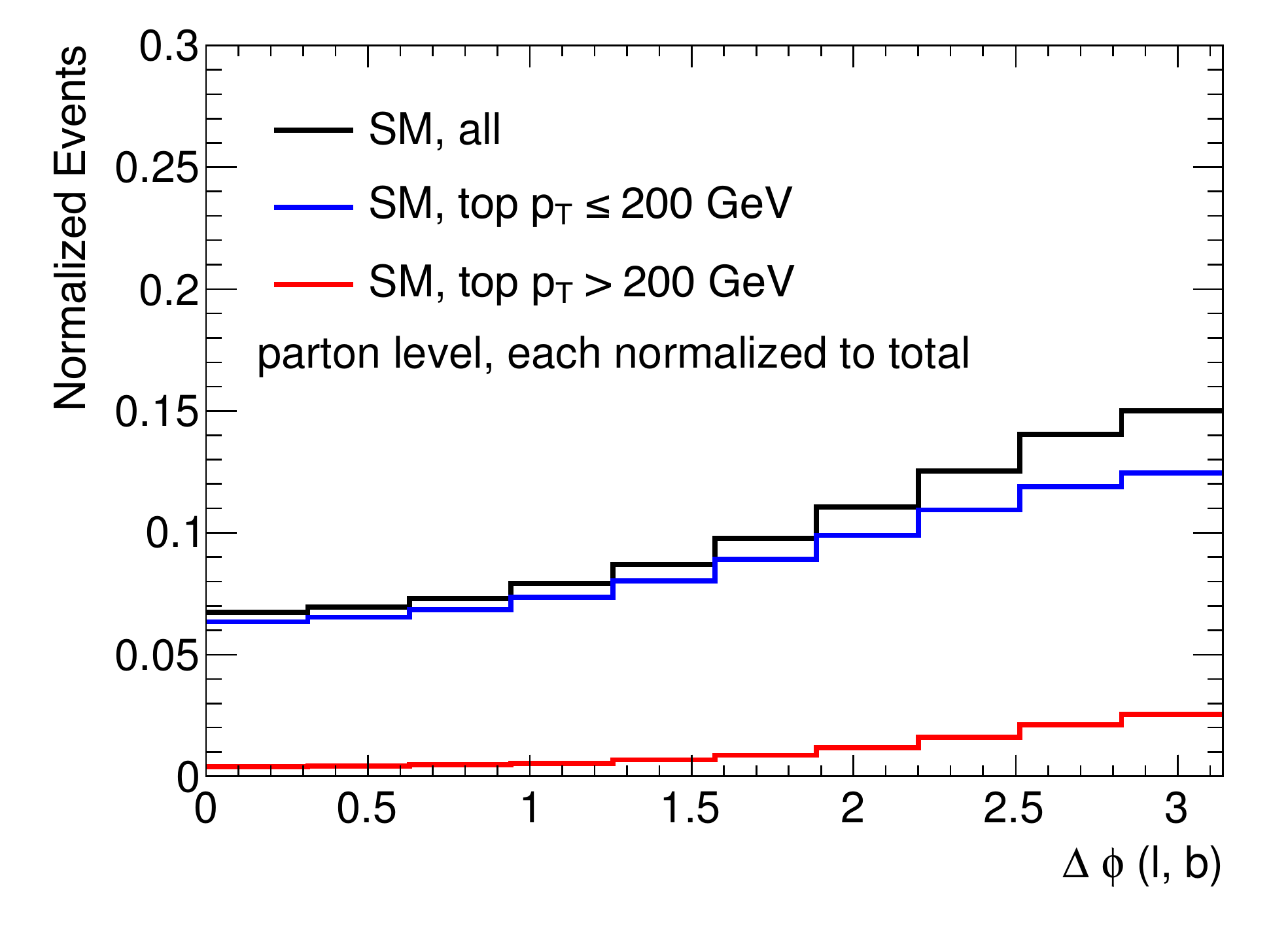}
\label{fig:toppt_SM_normtosum_b}
}
			\subfigure[]{
\includegraphics[width=0.45\textwidth]{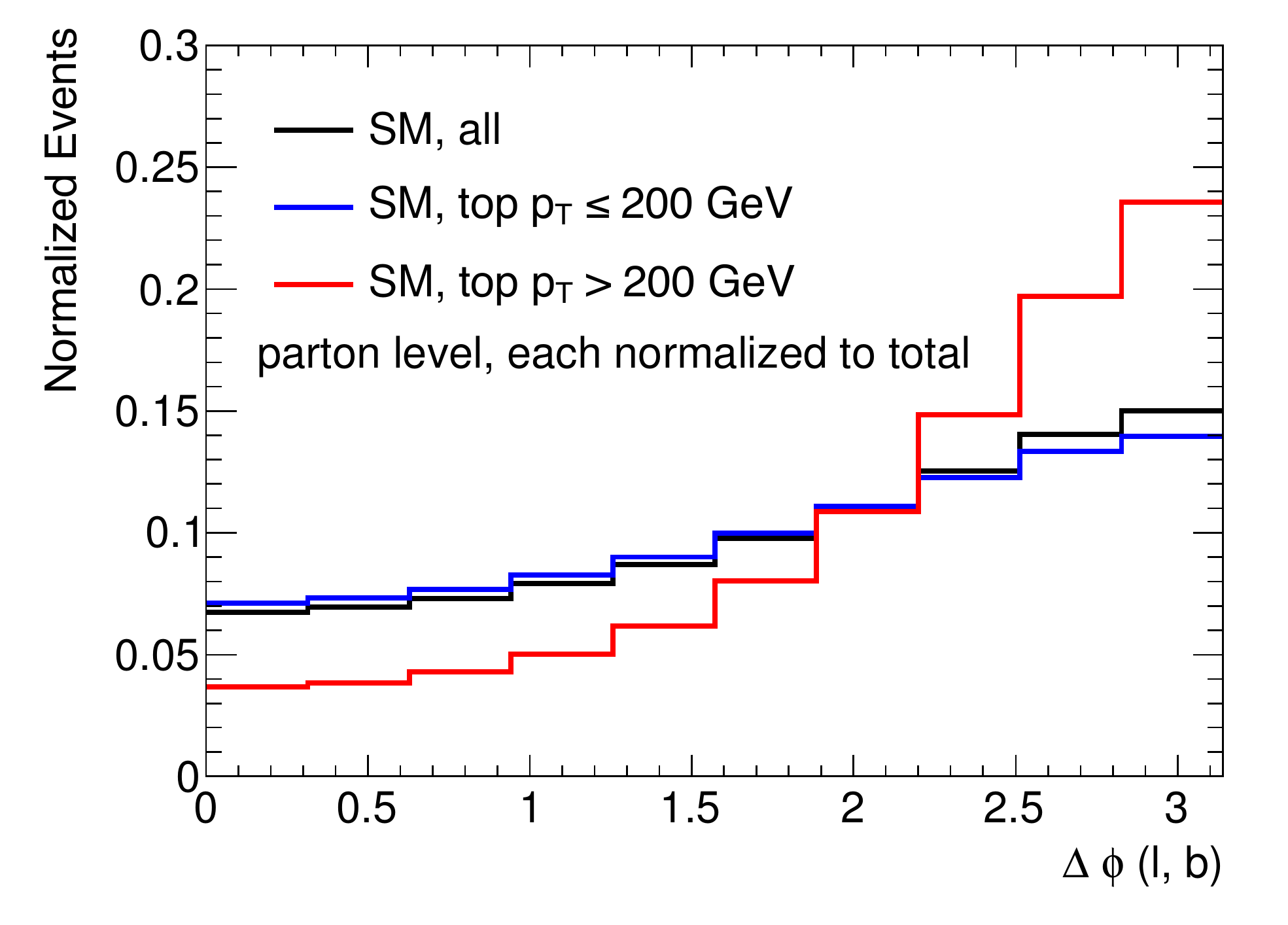}
\label{fig:toppt_SM_normto1_b}
}
			\subfigure[]{
\includegraphics[width=0.6\textwidth]{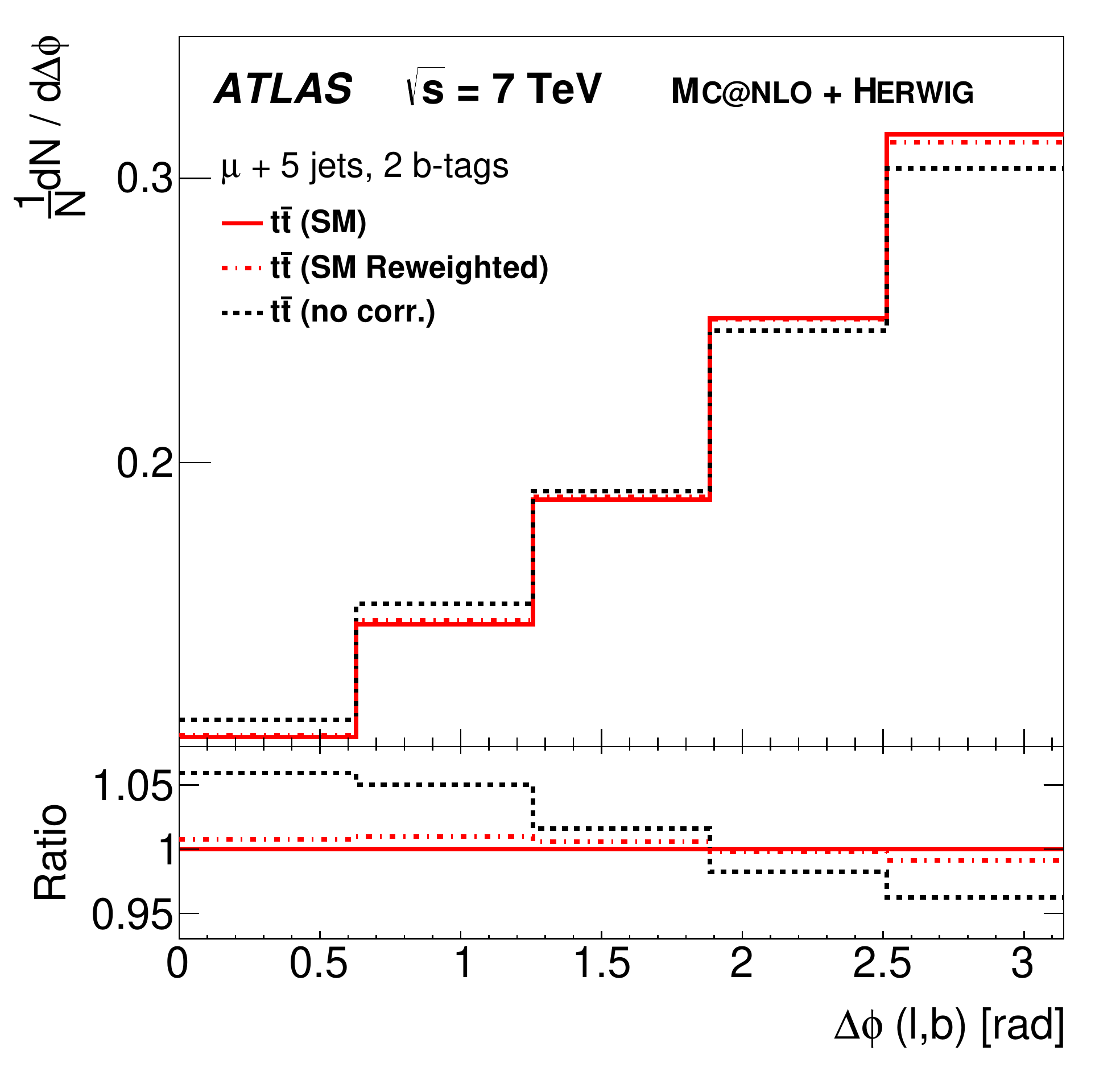}
\label{fig:toppt_reco_example_b}
}
\end{center}
\caption{\dphi\ distributions using the \bQ\ as analyser. \subref{fig:toppt_SM_normtosum_b} The \dphi\ distribution for the Standard Model expectation normalized to one. The contributions from the low and high top \pt\ region are indicated.\subref{fig:toppt_SM_normto1_b} Same distributions as \subref{fig:toppt_SM_normtosum}, but with the low and high top \pt\ region also normalized to one. As an example, the effect of reweighting is shown in \subref{fig:toppt_reco_example_b}: The SM and the uncorrelated \ttbar\ sample are compared to the SM prediction, which is reweighted in top \pt \cite{ueberpaper}. The ratio plot shows the ratios uncorrelated over SM (dashed) and reweighted over SM (red dash-dotted).}
\label{fig:topptexpl2}
\end{figure} 

\subsubsection{Colour Reconnection}
The colour of quarks and gluons is conserved. Colour strings, propagating from the initial state to the final hadronisation products can break and need to be reconnected after factorization.  
In order to estimate the effect due to the imperfect modelling of the colour reconnection \cite{color_reconnection}, templates are created with \powheg+\pythia\ using the P2011 CTEQ5L \pythia\ tune \cite{perugia_tune}. Another sample, using the same tune but with the colour reconnection completely turned off (using the \textit{NOCR} setting), is used for a second set of templates. Via ensemble tests, the difference between the fitted \fsm\ results is evaluated, symmetrized and taken as the uncertainty for colour reconnection. 

\subsubsection{Underlying Event}
To study variations of the underlying event, the P2011 CTEQ5L \pythia\ tune is set to a mode (\textit{mpiHi}) where the production rate of semi-hard jets coming from multiple parton-parton reactions is increased. It is compared to the nominal P2011 CTEQ5L \pythia\ tune sample \cite{perugia_tune}. Via ensemble tests, the difference between the fitted \fsm\ results is evaluated, symmetrized and taken as uncertainty for underlying events.

\subsubsection{Parton Showering / Hadronisation}
In contrast to spin correlation analyses in the dilepton channel, the measurement in the \ljets\ channel depends significantly on the jet properties. On the one hand these are used to map the jets to the corresponding partons. On the other hand the jet kinematics are interpreted as representations of the \ttbar\ spin correlation.

To study the effect due to uncertainties of parton showering and hadronisation, a common generator, \powheg, is interfaced to \pythia\ and to \herwig. The \powheg+\pythia\ and \powheg+\herwig\ samples are both used for ensemble testing. The difference between the fit results using \powheg+\pythia\ and \powheg+\herwig\ is taken as uncertainty on the parton showering. 

\pythia\ and \herwig\ represent two different physics models for the calculation of parton showering. While \pythia\ is based on the Lund string model \cite{pythia}, \herwig\ uses the cluster fragmentation model \cite{herwig}.

In order to only account for the showering differences, \ttbar\ events including a $W \rightarrow \tau \nu$ decay needed to be vetoed. The reason is that $\tau$ polarization information is mistakenly neglected by ${\tt TAUOLA}$ \cite{tauola}, responsible for $\tau$ decays, in \powheg+\herwig. For the evaluation of the parton showering, this $\tau$ veto is applied to all samples used for creating pseudo data and fitting. Furthermore, both samples suffered from a bug in the top spin correlation for antiquark-gluon and gluon-antiquark production channels. 
As both generators suffer from the same error and only the difference between them is studied, there is no residual effect on the determined uncertainty of the measured \ttbar\ spin correlation.

Another large difference between the generators is the top \pt\ modelling. As this quantity is a large uncertainty itself (influencing the jet kinematics and the \dphi\ shape), the \powheg+\pythia\ sample is reweighted to match the top \pt\ spectrum of \powheg+\herwig. This is necessary for two reasons: As the top \pt\ uncertainty is quoted explicitly, it should not be double-counted in the showering uncertainty. Furthermore, it must be avoided that the difference in top \pt\ cancels a showering uncertainty effect. Indeed, this is observed when evaluating the uncertainty without reweighting: While the uncertainty for the \dQ\ combination is very large, there is no effect on the \bQ\ combination. The reweighting of the top \pt\ removes that effect. The difference between the top \pt\ spectra of the generators is shown in Figure \ref{fig:truth_top_pt}. As a cross check the top \pt\ reweighting is also applied the other way around (\powheg+\herwig\ to \powheg+\pythia). The resulting uncertainty is the same. 
\begin{figure}[htbp]
\begin{center}
\includegraphics[width=0.7\textwidth]{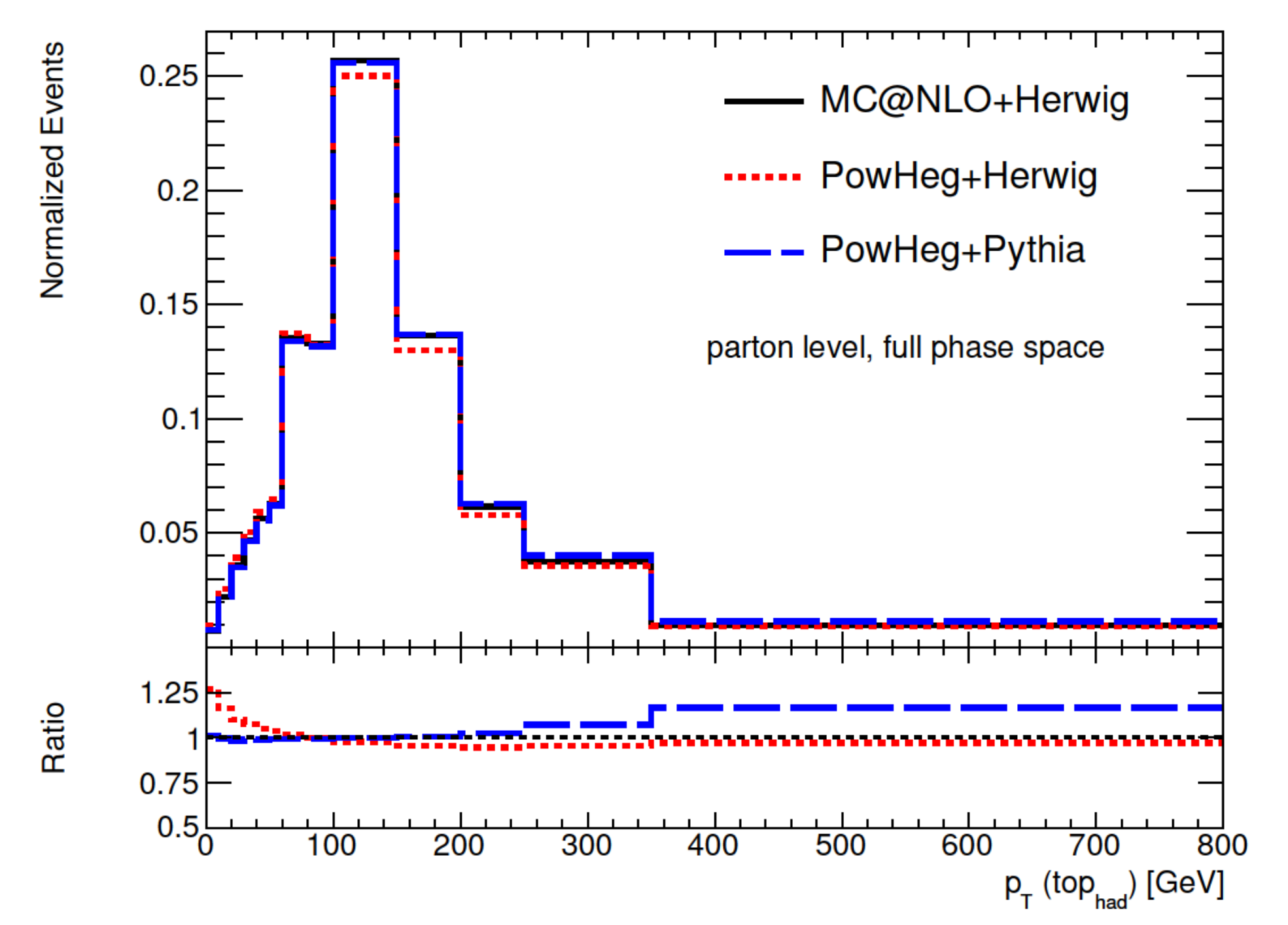}
\end{center}
\caption{The top \pt\ spectrum on parton level for different generators. The used top status codes were 155 for \herwig\ and 3 for \pythia. The ratios are with respect to \mcatnlo.}
\label{fig:truth_top_pt}
\end{figure} 
The uncertainties on parton showering and hadronisation as shown are listed in Table \ref{tab:psunc}. Two sets of uncertainties are provided: The default values without reweighting and the results using \powheg+\pythia\ reweighted to the top \pt\ spectrum of \powheg+\herwig. For a comparison to the default determination of the parton shower/hadronisation uncertainty (without top \pt\ reweighting), the top \pt\ uncertainty and the top \pt\ reweighted PS/hadronisation uncertainty, added in quadrature, are also listed. 
\begin{table}[htbp]
\begin{center}
\begin{tabular}{|c|c|c|c|}
\hline
Procedure & \dQ\ & \bQ\ & Combination \\
 \hline
 \hline
PS (no reweighting) & 0.33& 0.02 & 0.20\\
PS (top \pt\ reweighted) & 0.08 & 0.29 &0.16\\
PS (top \pt\ reweighted) $\oplus$ top \pt\ unc.& 0.19 & 0.38 &0.16\\
\hline
\end{tabular}
\end{center}
\caption{The uncertainties on \fsm\ originating from the difference between \powheg+\pythia\ and \powheg+\herwig. The first row provides the uncertainties using the default \powheg+X samples without any reweighting. The second row provides the uncertainties using \powheg+\pythia\ reweighted to the top \pt\ spectrum of \powheg+\herwig. The third row shows results with adding the dedicated top \pt\ uncertainty as it is evaluated in Section \ref{sec:toppt_unc} to the uncertainties in the second row in quadrature.}
\label{tab:psunc}
\end{table}

\subsubsection{Renormalization/Factorization Scale Variation}
Effects of a varied renormalization/factorization scale on the predicted value of the spin correlation $C$ were studied in \cite{Bernreuther2004}. The \mcatnlo\ sample used for the \ttbar\ signal is also available with varied values for the renormalization and the factorization scale $\mu$. The default value $\mu_0$ is varied by a factor 0.5 and 2.0, as in \cite{Bernreuther2004}. Ensemble tests are performed for both the up and down variation of $\mu_0$. The difference is quoted as uncertainty. 

\subsubsection{Initial and Final State Radiation}
\alpgen+\pythia\ samples with dedicated modifications of the P2011 CTEQ5L \pythia\ tune \cite{perugia_tune} are used to increase (\textit{radHi} setting) and decrease (\textit{radLo} setting) the amount of initial and final state radiation. They are compared with ensemble tests. Half of the difference between the two samples is symmetrized and taken as uncertainty.

\subsubsection{Template Statistics}
The precision of the measurement is limited by the available template statistics. To account for this limitation, a corresponding uncertainty contribution is evaluated.

 A common procedure to evaluate the effect is applying Poissonian fluctuations to the bins of the templates used for fitting. The fluctuations are based on the MC statistics. The width of the output distributions of \fsm\ is supposed to be taken as uncertainty. 

Evaluated uncertainties $\Delta \fsm$ are expected to be independent of the input value of \fsm, $f^{\text{SM, in}}$, and the average deviation between fit input and output is expected to be zero:
\begin{align}
\frac{\partial \left(\Delta \fsm \right)}{\partial f^{\text{SM, in}}} = 0 \\
\langle f^{\text{SM, in}} - f^{\text{SM, out}} \rangle = 0
\end{align}

Both expectations are found to be violated, leading to a close investigation of the observed effect. Several tests are performed:
\begin{itemize}
\item Only the signal templates are fluctuated.
\item The MC statistics of the SM spin correlation sample are reduced to the one without spin correlation to check for an effect due to different MC statistics of two templates used for mixing.\footnote{The sample with uncorrelated \ttbar\ pairs contains only $\frac{2}{3}$ of the statistics of the SM correlation sample.} 
\item Both signal templates are fluctuated individually.
\end{itemize}
While the first tests did not change the situation, the last one helped to understand the effect. It is illustrated in Figure \ref{fig:tempstat}.

\begin{figure}[htbp]
\begin{center}
			\subfigure[]{
\includegraphics[width=0.47\textwidth]{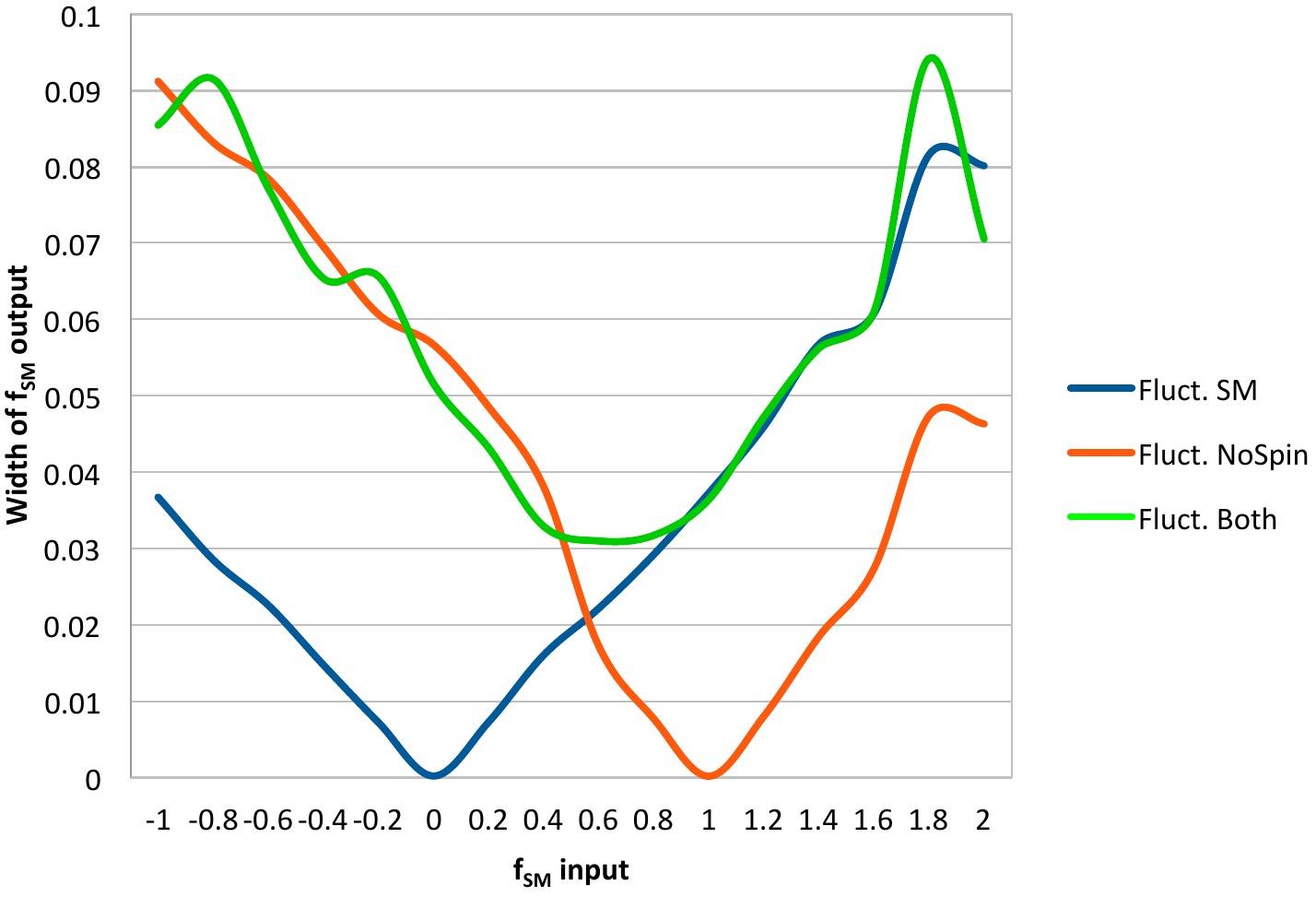}
\label{fig:tempstat_width}
}
			\subfigure[]{
\includegraphics[width=0.47\textwidth]{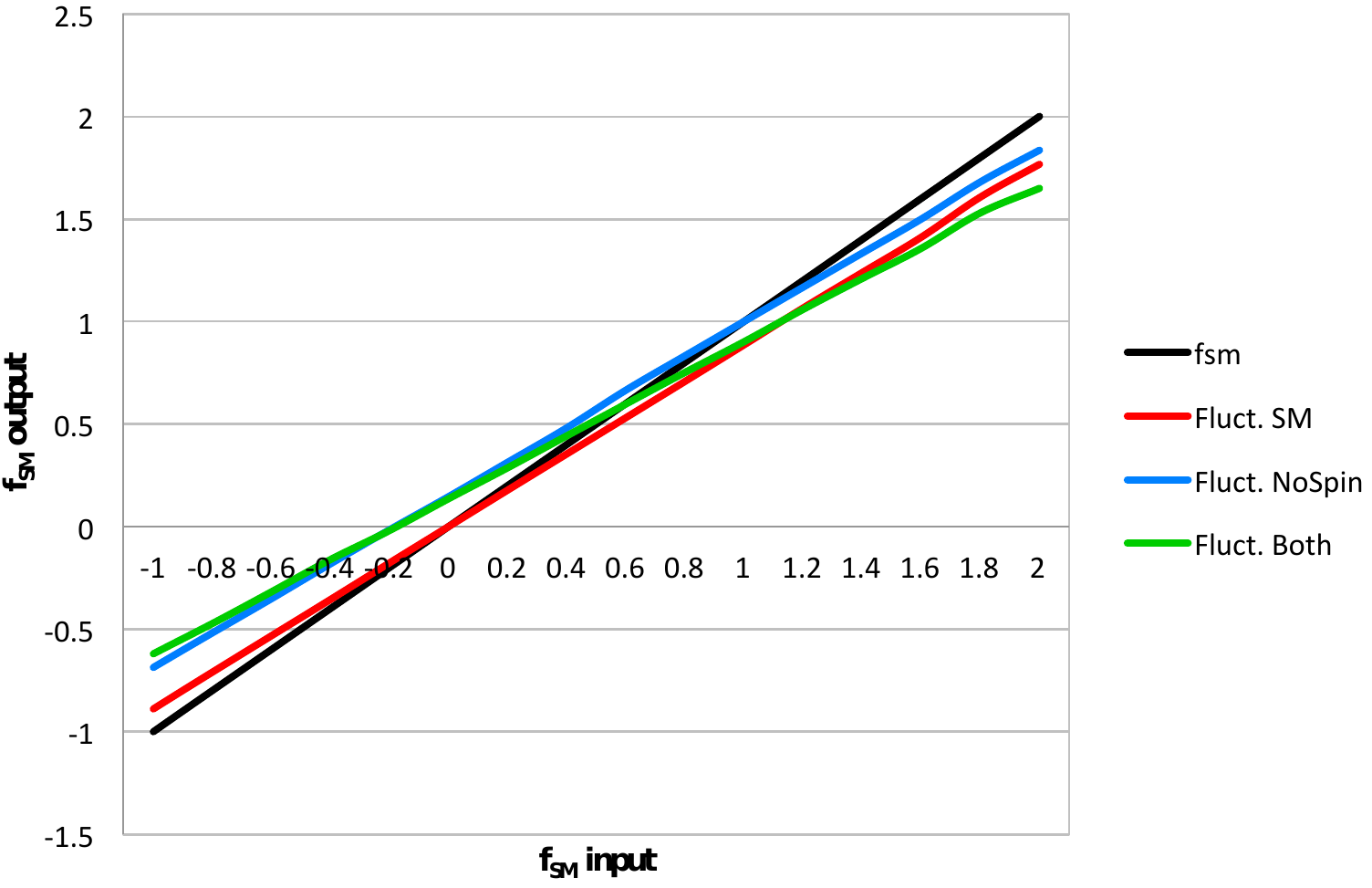}
\label{fig:tempstat_mean}
}
\end{center}
\caption{\subref{fig:tempstat_width} Width and \subref{fig:tempstat_mean} mean of the output \fsm\ distributions after fluctuation of the \ttbar\ signal templates bins according to their MC statistics uncertainty. Poissonian fluctuations were applied to either only the SM \ttbar\ spin correlation sample, the uncorrelated \ttbar\ spin sample or both. }
\label{fig:tempstat}
\end{figure} 

Figure \ref{fig:tempstat_width} shows the dependence of the MC statistics uncertainty on $f^{\text{SM, in}}$. Three scenarios are evaluated: Poissonian fluctuations are applied to the template with SM \ttbar\ spin correlation, to the uncorrelated \ttbar\ sample and to both.
No effect is expected for pseudo data with \fsm\ = 1.0 if only the template without spin correlation is modified. The signal has no contribution of the uncorrelated template and the non-fluctuated sample can fully describe the pseudo data. Vice versa, modifying only the Standard Model spin correlation template leaves the fit output unchanged for pseudo data with \fsm\ = 0.0. Using pseudo data with $\fsm \neq 0$ ($\fsm \neq 1$) adds a contribution of the $\fsm = 1$ ($\fsm = 0$) linearly with \fsm.

In Figure \ref{fig:tempstat_mean} the mean values of the \fsm\ output distributions is shown. This effect only shows up when the MC statistics is low with respect to the total separation power of the signals. The effect vanished for artificially increased separation powers and reduced MC statistics uncertainties.

As the fit loses separation power for fluctuated signal templates, it has problems to assign the data to one of the two scenarios and the fit tends towards values for which the sum of both MC statistics uncertainties is the smallest ($\fsm \approx 0.5$, as seen in Figure \ref{fig:tempstat_width}).
For the template statistics uncertainty, both effects are taken into account: the deviation of the mean and the width of the \fsm\ distribution were added in quadrature. All templates are included in the uncertainty, not just the \ttbar\ templates.

\section{Test for NP Inclusion}
\label{sec:NP_test}
Only such uncertainties for which a well-defined template for both a $\pm 1\,\sigma$ variation of a systematic effect exist are taken as candidates for using a nuisance parameter. Such a template is called ``well-defined'' if an uncertainty has a continuous and symmetric effect on a template. This excludes two-point uncertainties (e.g. smearing on/off or comparison of two generators) and uncertainties which need a dedicated evaluation (e.g. checking several effects and quoting the largest).

Uncertainty candidates passing these criteria, listed in Tables \ref{tab:NP_check1} and \ref{tab:NP_check2}, are tested for statistical relevance. 
This means that their systematic effect has to be larger than the Monte Carlo statistical uncertainty of the corresponding template. The testing procedure is as follows:
\begin{itemize}
\item If at least for two bins for either the up or the down variation the difference to the nominal is larger than the statistical uncertainty of that particular bin, the systematic uncertainty is defined as significant.
\item If the total deviation from the nominal sample is larger than the total statistical uncertainty, it is called significant.
\end{itemize}
This test is done for all systematic uncertainties, all eight channels, both analysers and all template types. It is included as a nuisance parameter only where it is relevant. For example, uncertainties only affecting the signal are only used as nuisance parameters for the signal templates. The used templates are the $W$+jets background, the fake lepton background, the remaining background as well as two signal templates: the sum and the difference of the SM \ttbar\ spin correlation template and the uncorrelated \ttbar\ signal template. An exception is made for the signal templates. If either the sum or the difference of the templates is significantly affected, both were linked to the nuisance parameter.

Figure \ref{fig:NP_check_example} shows an example NP which was tested for all channels. Three different channels with different outcome are shown:  
In the first channel (Figure \ref{fig:NP_example3a}), the systematic variations (red and blue lines) show clear significance with respect to the MC statistics (green) for all bins. In the second channel (Figure \ref{fig:NP_example3b}) no bin itself is statistically significant, but the sum of differences is and in the third case (Figure \ref{fig:NP_example3c}), the uncertainty is clearly not significant.

\begin{figure}[htbp]
\begin{center}
			\subfigure[]{
\includegraphics[width=0.45\textwidth]{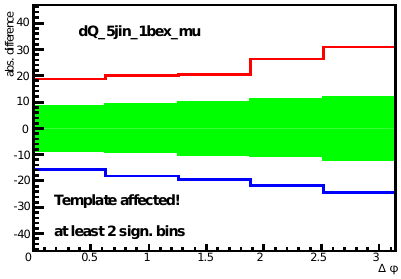}
\label{fig:NP_example3a}
}
			\subfigure[]{
\includegraphics[width=0.45\textwidth]{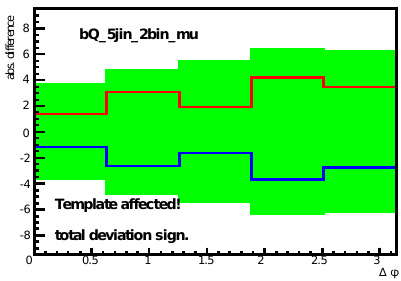}
\label{fig:NP_example3b}
}
			\subfigure[]{
\includegraphics[width=0.45\textwidth]{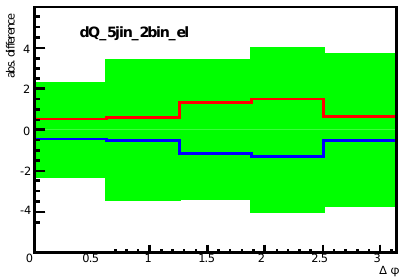}
\label{fig:NP_example3c}
}
\end{center}
\caption{An example nuisance parameter, tested on one template type and for several channels. Three channels are shown as example. \subref{fig:NP_example3a} The systematic variations (red and blue) show clear significance with respect to the MC statistics (green) for all bins. \subref{fig:NP_example3b} No bin itself is statistically significant, but the sum of differences is. \subref{fig:NP_example3c} The uncertainty is clearly not significant.}
\label{fig:NP_check_example}
\end{figure}

Tables \ref{tab:NP_check1} and \ref{tab:NP_check2} show a list of all systematic uncertainties that are tested and the channels on which they have an effect. 
\begin{table}[htbp]
\begin{center}
\begin{tabular}{|c||c|c|c|c|}
\hline
Uncertainty & \multicolumn{4}{c|}{Has effect on template with...} \\
{} & (SM + Unc.)  & (SM - Unc.)  & Rem. BG & $W$+Jets\\
\hline
\hline
JES/EffectiveNP\_Stat1 & yes & no & yes &---\\
JES/EffectiveNP\_Stat2 & no & no & no &---\\
JES/EffectiveNP\_Stat3 & no & no & no &---\\
JES/EffectiveNP\_Model1 & yes & no & yes &---\\
JES/EffectiveNP\_Model2 & no & no & no &---\\
JES/EffectiveNP\_Model3 & no & no & no &---\\
JES/EffectiveNP\_Model4 & no & no & no &---\\
JES/EffectiveNP\_Det1 & yes & no & yes &---\\
JES/EffectiveNP\_Det2 & no & no & no &---\\
JES/EffectiveNP\_Mixed1 & no & no & no &---\\
JES/EffectiveNP\_Mixed2 & yes & no & no &---\\
JES/Intercal\_TotalStat & yes & no & yes &---\\
JES/Intercal\_Theory & yes & yes & yes &---\\
JES/SingleParticleHighPt & no & no & no &---\\
JES/RelativeNonClosureMC11b & yes & no & no &---\\
JES/PileUpOffsetMu & yes & no & yes &---\\
JES/PileUpOffsetNPV & yes & no & no &---\\
JES/Closeby & yes & yes & yes &---\\
JES/FlavorComp & yes & yes & yes &---\\
JES/FlavorResponse & yes & yes & yes &---\\
JES/BJES & yes & no & yes &---\\
\hline
btag/break0 & no & no & no &---\\
btag/break1 & no & no & no &---\\
btag/break2& no & no & no &---\\
btag/break3 & no & no & no &---\\
btag/break4 & no & no & no &---\\
btag/break5 & yes & no & no &---\\
btag/break6 & yes & no & no &---\\
btag/break7 & no & no & no &---\\
btag/break8 & yes & no & yes &---\\
ctag/break0 & yes & no & no &---\\
ctag/break1 & yes & no & no &---\\
ctag/break2 & no & no & no &---\\
ctag/break3 & yes & no & no &---\\
ctag/break4 & yes & no & yes &---\\
mistag 	& yes & no & yes &---\\
\hline
\end{tabular}
\end{center}
\caption{List of candidates for the nuisance parameter fit (part 1). }
\label{tab:NP_check1}
\end{table}

\begin{table}[htbp]
\begin{center}
\begin{tabular}{|c||c|c|c|c|}
\hline
Uncertainty & \multicolumn{4}{c|}{Has effect on template with...} \\
{} & (SM + Unc.)  & (SM - Unc.)  & Rem. BG & $W$+Jets\\
\hline
\hline
JVF & yes & no & yes &---\\
MET/CellOut+SoftJet & no & no & no &---\\
MET/PileUp & no & no & no &---\\
\hline
el/Trigger 	& yes & no & no &---\\
el/Reco 	& yes & no & no &---\\
el/ID 		& yes & no & yes &---\\
el/E\_scale & yes & no & no &---\\
el/E\_resolution & no & no & no &---\\
mu/Trigger & yes & no & yes &---\\
mu/Reco & yes & no & no &---\\
mu/ID & yes & no & no &---\\
\hline
WJets/bb4 & --- & --- & --- &yes\\
WJets/bb5 & --- & --- & --- &yes\\
WJets/bbcc & --- & --- & --- &yes\\
WJets/c4 & --- & --- & --- &yes\\
WJets/c5 & --- & --- & --- &yes\\
\hline
\end{tabular}
\end{center}
\caption{List of candidates for the nuisance parameter fit (part 2).}
\label{tab:NP_check2}
\end{table}

\section{Evaluation of Non-Profilable Uncertainties}
The uncertainties that cannot be treated as nuisance parameters in the fit are evaluated with ensemble tests. 100,000 ensembles are drawn. Either the difference of the output \fsm\ to the expected \fsm\ = 1.0 is quoted or the special procedures are applied as explained in Section \ref{sec:syslist}.
Table \ref{tab:sys_ensembles} shows the list of these additional uncertainties for the combined \dQ\ and \bQ\ fits as well as for the full combination. 
\begin{table}[htbp]
\begin{center}
\begin{tabular}{|c||c|c|c|}
\hline
Uncertainty & \multicolumn{3}{c|}{$\Delta \fsm $} \\
{} & \dQ\ & \bQ\ & Combination\\
\hline
\hline
\rule{0pt}{0.5cm} QCD shape (\ejets) &${}^{+ 0.009}_{- 0.010}$ & ${}^{+ 0.019}_{- 0.028}$ &  ${}^{+ 0.009}_{- 0.007}$\\
\rule{0pt}{0.5cm} QCD shape (\mujets) &${}^{+ 0.001}_{- 0.004}$ &${}^{+ 0.019}_{- 0.006}$ &${}^{+ 0.006}_{- 0.005}$\\
\rule{0pt}{0.5cm} PDF & $\pm 0.075$ & $\pm 0.085$ & $\pm 0.018$\\
Jet Energy Resolution & $\pm 0.026$ & $\pm 0.017$ & $\pm 0.019$\\
Jet Reconstruction Efficiency & $\pm 0.001$ & $\pm 0.016$ & $\pm 0.008$\\
Muon Momentum Scale & $\pm 0.001$ & $\pm 0.008$ & $\pm 0.003$\\
Muon Momentum Resolution & $\pm 0.002$ & $\pm 0.004$ & $\pm 0.002$\\
Luminosity & $\pm 0.002$ & $\pm 0.000$ & $\pm 0.000$\\
Renormalization/Factorization Scale & $\pm 0.186$ & $\pm 0.132$ & $\pm 0.062$\\
Parton Showering & $\pm 0.084$ & $\pm 0.296$ & $\pm 0.160$\\
Underlying Event & $\pm 0.058$ & $\pm 0.029$ & $\pm 0.046$\\
Colour Reconnection & $\pm 0.009$ & $\pm 0.021$ & $\pm 0.012$\\
Initial/Final State Radiation & $\pm 0.015$ & $\pm 0.212$ & $\pm 0.073$\\
Top \pt\ & $\pm 0.169$ & $\pm 0.238$ & $\pm 0.016$\\
Template Statistics & $\pm 0.053$ & $\pm 0.069$ & $\pm 0.036$\\
\hline
\end{tabular}
\end{center}
\caption{List of systematic uncertainties evaluated with ensemble tests.}
\label{tab:sys_ensembles}
\end{table}

\section{Important Aspects of Systematic Uncertainties}
This section is dedicated to explain effects observed for certain sources of uncertainties and their implications on spin correlation measurements. In particular, cancellation effects are explained.

One important point in the discussion of the systematic uncertainties is the correlation between the two analysers. In most of the cases, the effect of a systematic variation is antisymmetric in terms of the resulting value of \fsm: A systematic variation leading to a higher result for \fsm\ with the \dQ\ as analyser leads to a lower result of \fsm\ using the \bQ\ as analyser. The reason is quite illustrative - after switching off spin correlation in \ttbar\ events, the \dphi\ distribution becomes steeper for the \dQ\ and flatter for the \bQ\ as shown in Figure \ref{fig:dphi_truth}. 

A higher top quark \pt\ serves as good example for such an effect. The decay products of both top quarks get more collimated due to the additional boost. This tends to a back-to-back topology of the two \ttbar\ spin analysers and hence to a larger azimuthal distance. The result is a shift to higher values of \dphi, independent of the analyser. To illustrate this effect, which is consistent for \dQ\ and \bQ\ analysers, a sample of uncorrelated \ttbar\ pairs is used. This decouples the influences of kinematics and spin configurations. Figure \ref{fig:dphi_topptslices} shows the shift to higher values of \dphi\ for higher top quark \pt. Parton level quantities are used and no phase space cuts are applied.

\begin{figure}[htbp]
\begin{center}
			\subfigure[]{
\includegraphics[width=0.45\textwidth]{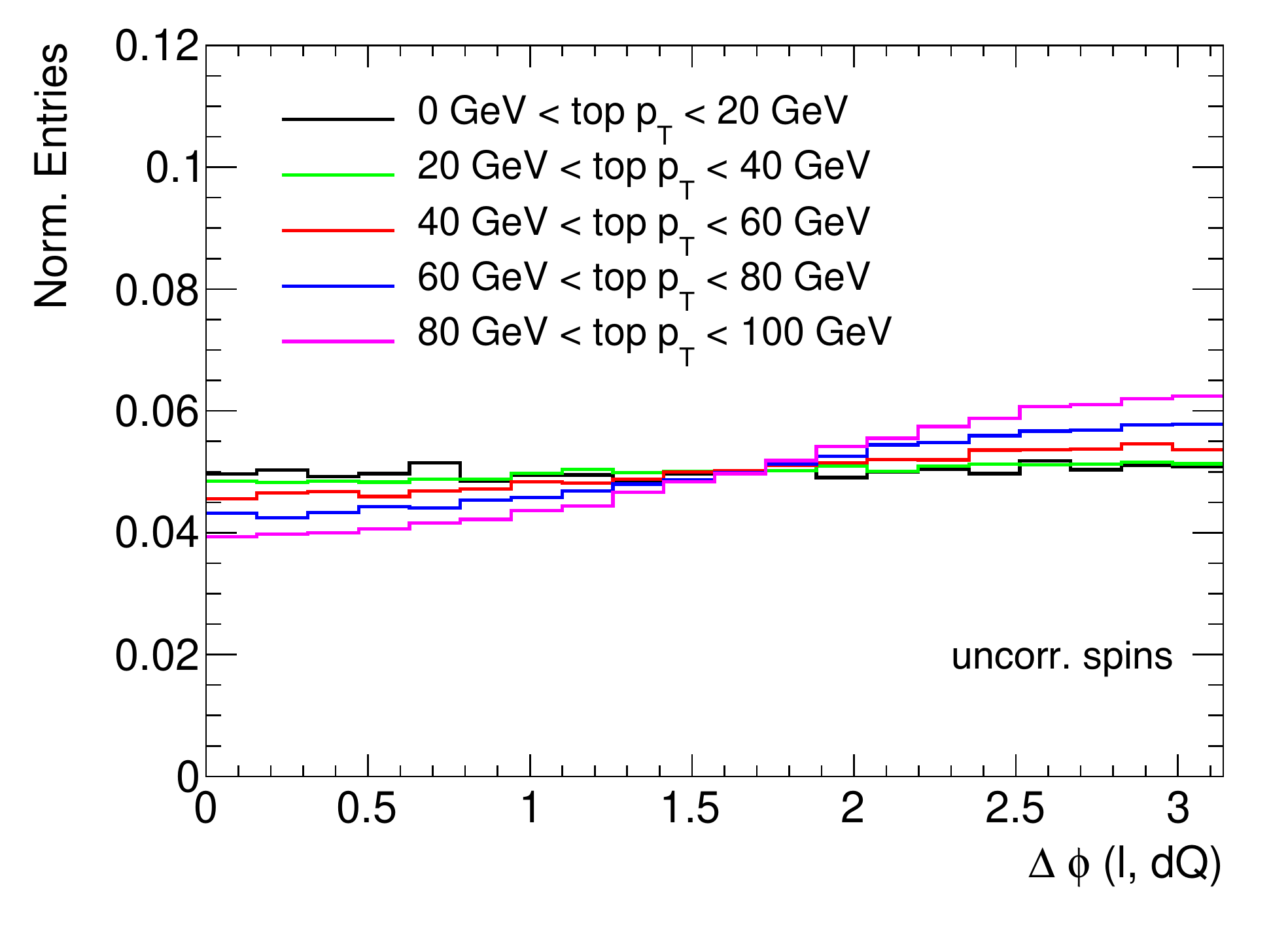}
\label{fig:dphi_dQ_topptslices}
}
			\subfigure[]{
\includegraphics[width=0.45\textwidth]{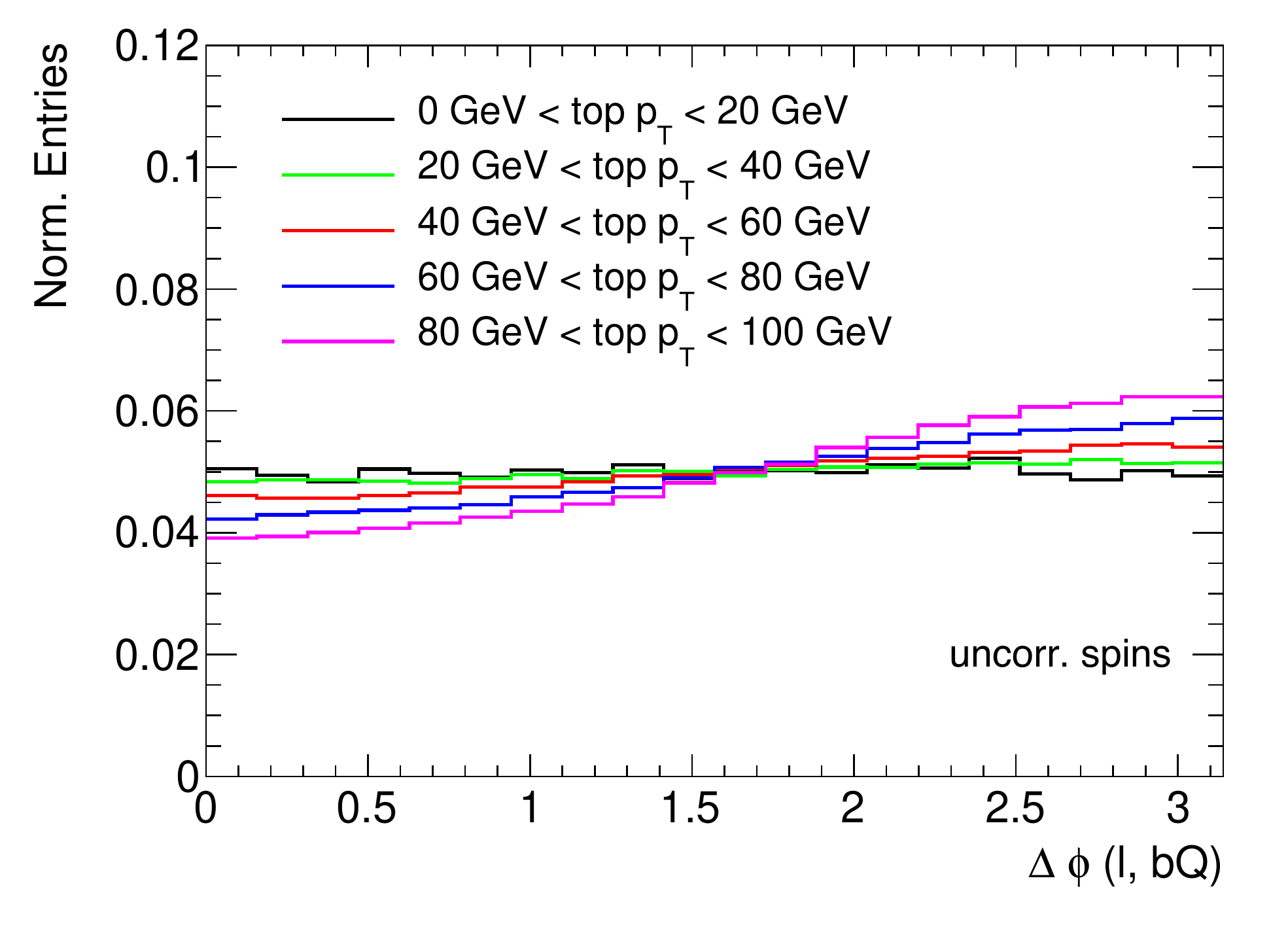}
\label{fig:dphi_bQ_topptslices}
}
\end{center}
\caption{Effects of increased top quark \pt\ on the azimuthal angle between the lepton and the \subref{fig:dphi_dQ_topptslices} \dQ\ and \subref{fig:dphi_bQ_topptslices} \bQ. The \mcatnlo\ sample with uncorrelated \ttbar\ pairs is used to decouple spin and kinematic effects on the \dphi\ shape. Parton level quantities are used.}
\label{fig:dphi_topptslices}
\end{figure} 

While for higher top quark \pt\ the \dphi\ distribution becomes steeper, it becomes flatter for higher \pt\ of the \ttbar\ pair. In this case the transverse boost of the \ttbar\ system collimates both top quarks and their decay products, preferring lower values of \dphi. Figure \ref{fig:dphi_ttbarptslices} shows the effect.

\begin{figure}[htbp]
\begin{center}
			\subfigure[]{
\includegraphics[width=0.45\textwidth]{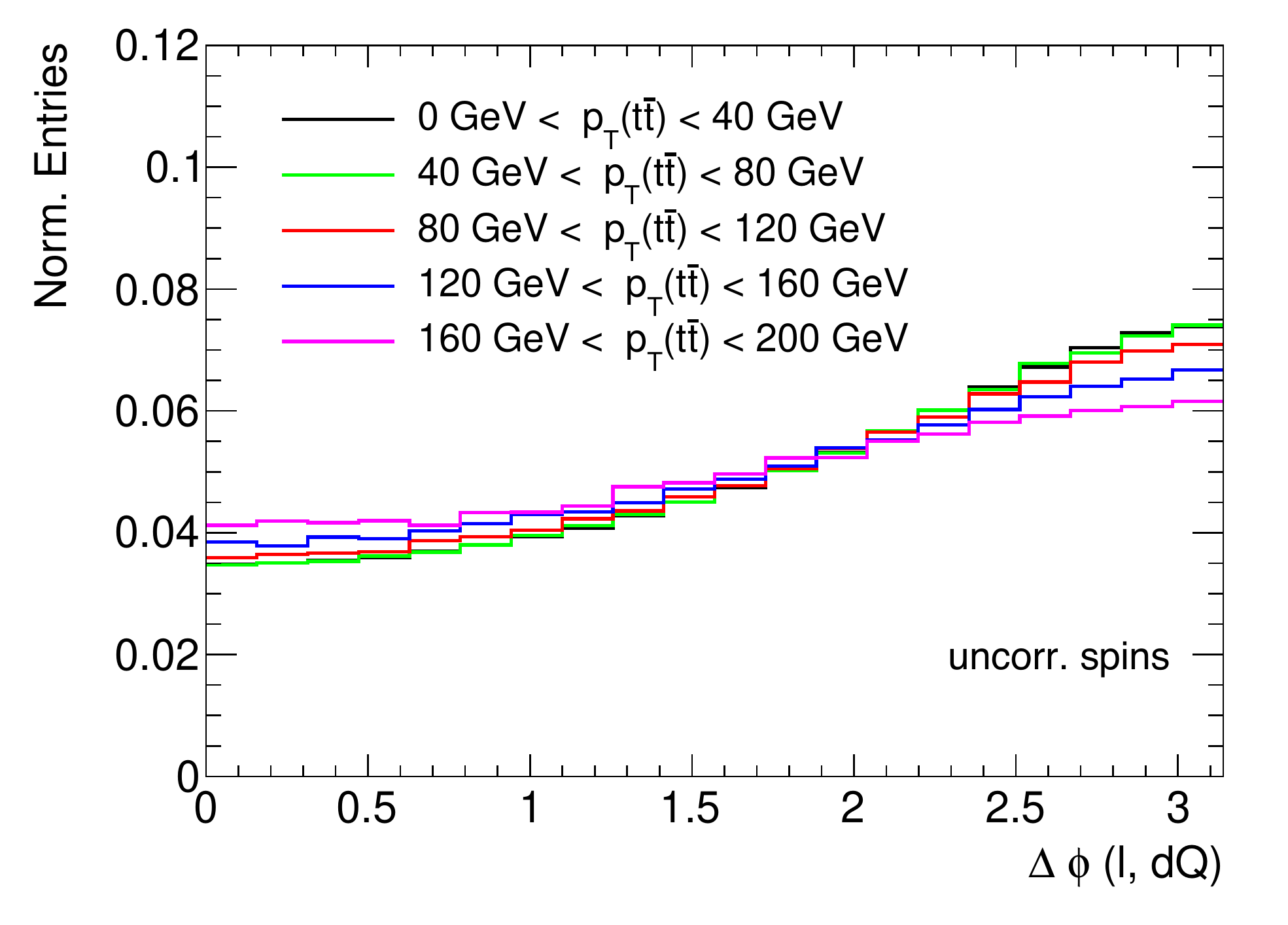}
\label{fig:dphi_dQ_ttbarptslices}
}
			\subfigure[]{
\includegraphics[width=0.45\textwidth]{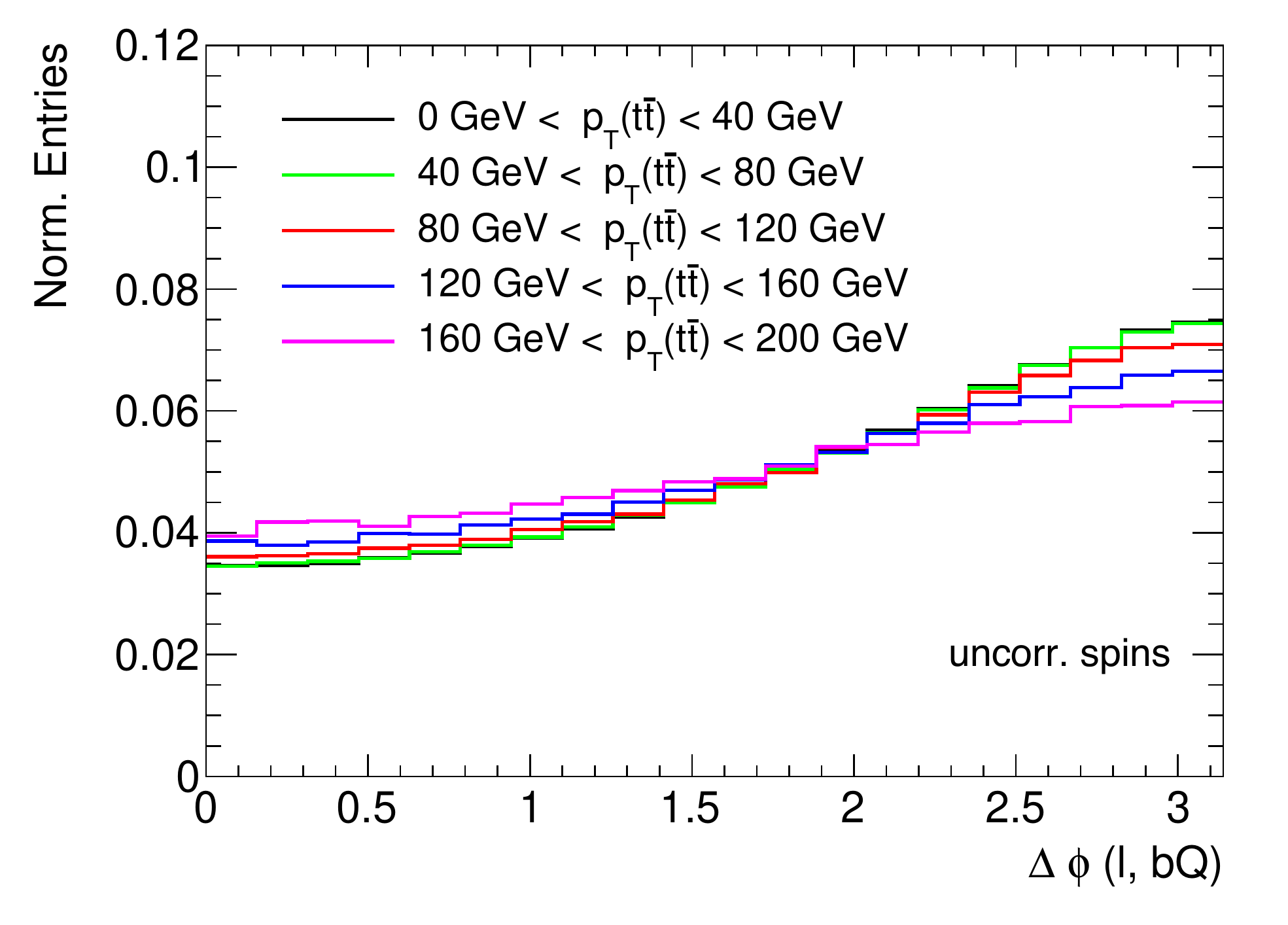}
\label{fig:dphi_bQ_ttbarptslices}
}
\end{center}
\caption{Effects of increased \pt\ of the \ttbar\ system on the azimuthal angle between the lepton and the \subref{fig:dphi_dQ_ttbarptslices} \dQ\ and \subref{fig:dphi_bQ_ttbarptslices} \bQ. The \mcatnlo\ sample with uncorrelated \ttbar\ pairs was used to decouple spin and kinematic effects on the \dphi\ shape. Parton level quantities were used.}
\label{fig:dphi_ttbarptslices}
\end{figure} 

Interpreting such an effect in terms of spin correlation is opposite for \dQ\ and \bQ\ spin analysers. 
Corresponding effects were shown in the context of the evaluation of the PDF uncertainty. The same holds true for the renormalization/factorization scale variation, which is also one of the most significant uncertainties.

In case a mismodelling is observed in data it will be reflected in deviations of \fsm\ into different directions.

\chapter[Results]{Results and Discussion}
This chapter presents the results of the \ttbar\ spin correlation analysis.
First, the fit results for the eight individual channels are presented for the \dQ\ as well as for the \bQ\ analyser. The individual channels are fitted without nuisance parameters. As a next step, the channels are combined for both spin analysers and a full combination is performed. Nuisance parameters are added to check the effect on the fitted \fsm\ results. Finally, the results are presented with their full uncertainties. 

After the presentation of the results the chapter concludes with a consistency check of the fit and a discussion of the effects due to systematic uncertainties.

\section{Single Channel Results}
Individual fit channels were neither for the \dQ\ nor for the \bQ\ spin analyser expected to show a significant difference between the scenarios with SM \ttbar\ spin correlation and uncorrelated \ttbar\ pairs. 
As furthermore deviations between the eight channels are not expected to be motivated by spin correlation effects, the analysis aims for a combination. For a cross check, the individual results are still listed in this section. Figure \ref{fig:statonly} shows the results for the fitted values of \fsm\ using the \dphi\ distribution between the charged lepton and the \dQ\ and \bQ, respectively. For this fitting setup no nuisance parameters are used. The quoted uncertainties are purely statistical. 
\begin{figure}[htbp]
\begin{center}
			\subfigure[]{
\includegraphics[width=0.7\textwidth]{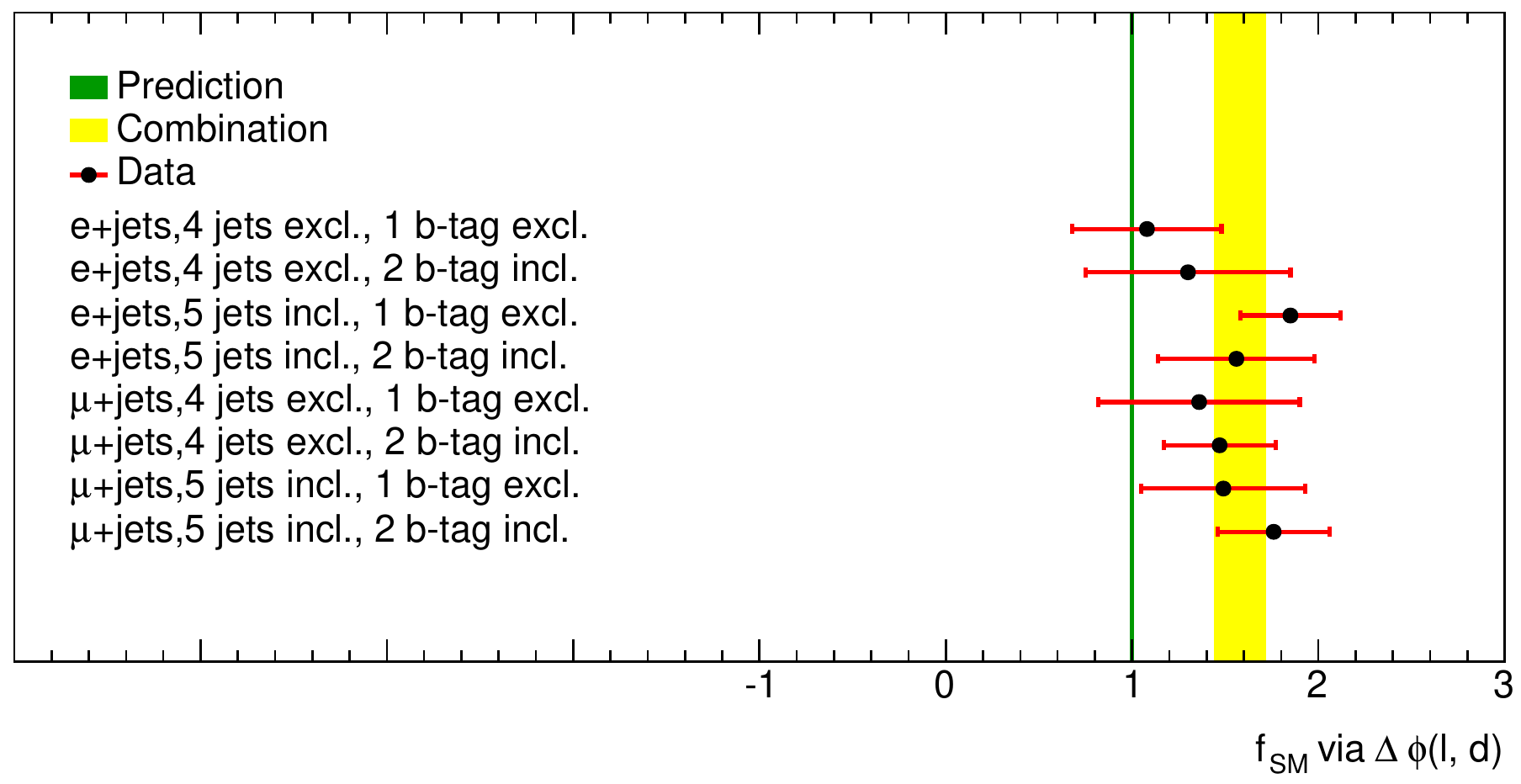}
\label{fig:dQ_statonly}
}
			\subfigure[]{
\includegraphics[width=0.7\textwidth]{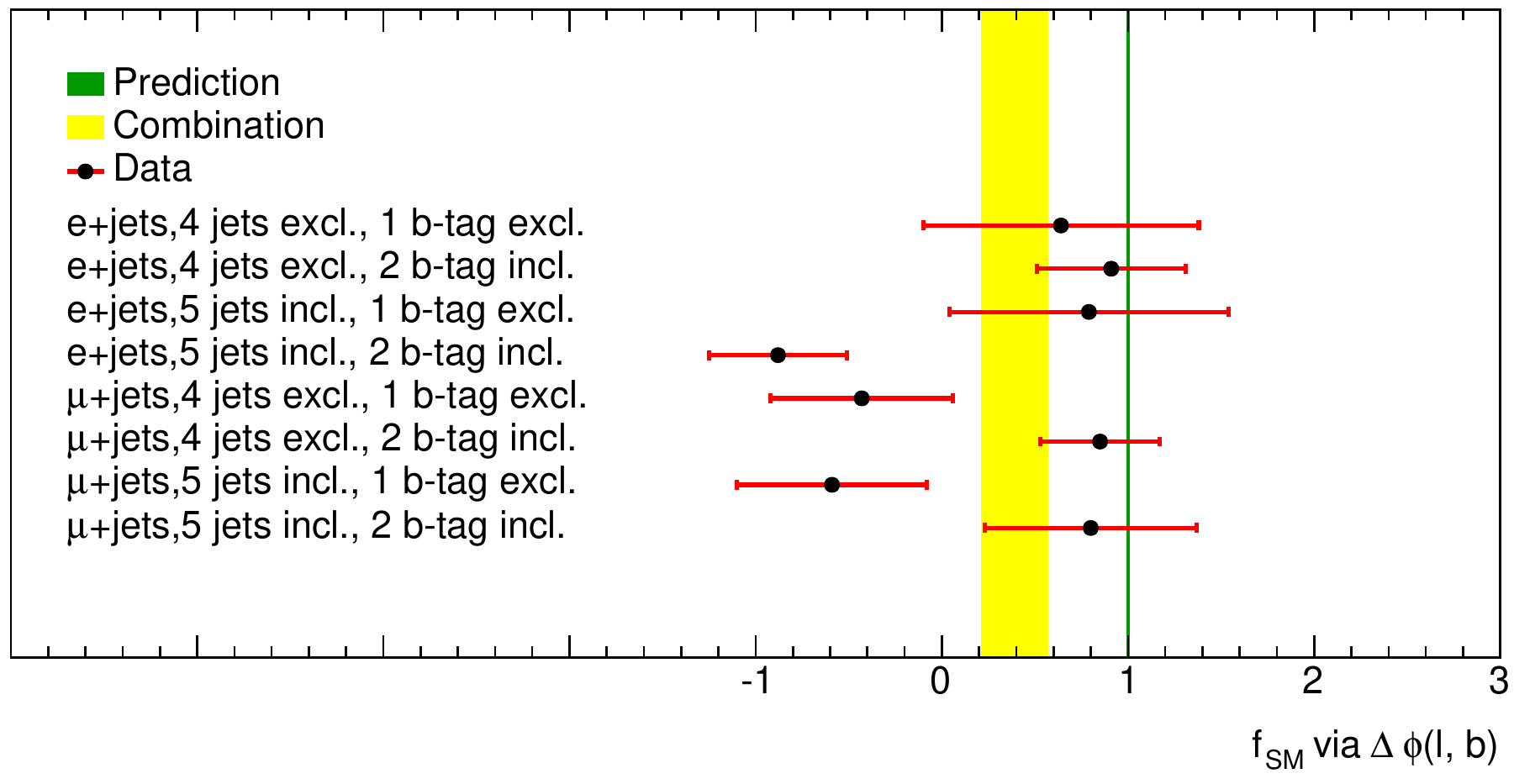}
\label{fig:bQ_statonly}
}
\end{center}
\caption{Comparison of single channel fit results using \subref{fig:dQ_statonly} the \dQ\ and the \subref{fig:bQ_statonly} \bQ\ as spin analyser. The fits were performed without nuisance parameters. The quoted uncertainties are statistical. Next to the individual fit results the SM expectation (green line) and the result of the combined fit (yellow band) are shown.}
\label{fig:statonly}
\end{figure} 

The following observations are made:
\begin{itemize}
\item While for the \dQ\ the individual results lie consistently above the SM expectation of $\fsm = 1.0$, the opposite is observed for the \bQ. 
\item The jet multiplicity mismodelling of \mcatnlo\ (see Section \ref{sec:jetmismodeling}) does not significantly disturb the measurement. Otherwise, a tension between the $n_{\text{jet}} = 4$ and the $n_{\text{jet}} \geq 5$ bins would have been observed. This was not the case.
\item A systematic effect \bQ\ results in the \mujets\ channels can be observed: The results of \fsm\ are higher in the $n_{\text{b-tags}} \geq 2$ channels than in the $n_{\text{b-tags}} = 1$ channels.
\end{itemize}

\section{Combined Fits without Nuisance Parameters}
For both the \dQ\ and the \bQ, the eight channels were combined. The combinations have a statistically significant tension. It is indicated by the yellow bands in Figure \ref{fig:statonly}. 
A full combination of both spin analysers is also performed. It leads to a good agreement with the SM prediction. Table \ref{tab:results_stat_fit} shows the results of the individual combinations including the statistical uncertainty from the data as well as the normalization uncertainties from the background sources. The latter is accounted for by the usage of priors in the fit.
\begin{table}[htbp]
\begin{center}
\begin{tabular}{|c|c|}
\hline
Combination &$\fsm \pm \text{statistical uncertainty}$\\
\hline
\dQ\ &$1.58 \pm 0.14$ \\
\bQ\ &$0.39 \pm 0.18$\\
Full Combination & $1.14 \pm 0.11$\\
\hline
\end{tabular}
\end{center}
\caption{Fit results for \fsm\ with statistical uncertainties and background normalization uncertainties. }
\label{tab:results_stat_fit}
\end{table}

\section{Combined Fits Using Nuisance Parameters}
Adding nuisance parameters to the fit has two effects: The fit uncertainty includes the component arising from the systematic uncertainties evaluated via NPs. Furthermore, the central values might change. This is because changes in the shape of the \dphi\ distributions can be fitted with either a modified signal composition -- and hence a modified \fsm\ -- or with systematic variations. In the second case the corresponding nuisance parameters are fitted to non-zero values.
The addition of the nuisance parameters leads to the fit results shown in Table \ref{tab:results_NP_fit}.
\begin{table}[htbp]
\begin{center}
\begin{tabular}{|c|c|}
\hline
Combination &$\fsm \pm \text{(statistical $\oplus$ NP) uncertainty}$\\
\hline
\dQ\ &$1.53 \pm 0.20$\\
\bQ\ & $0.53 \pm 0.22$\\
Full Combination & $1.12 \pm 0.15$\\
\hline
\end{tabular}
\end{center}
\caption{Final fit results for \fsm\ including statistical uncertainties and uncertainties due to nuisance parameters.}
\label{tab:results_NP_fit}
\end{table}
\section{Final Fit Results}
The final results also include those systematic uncertainties that are evaluated via ensemble tests. This evaluation has no effect on the central values. For quoting the final result the uncertainty due to NPs is separated from the statistical component via $\Delta f^{\text{SM, NP}} = \sqrt{\left(\Delta f^{\text{SM, stat.+NP}}\right)^2 - \left(\Delta f^{\text{SM, stat}}\right)^2}$. The results are shown in Table \ref{tab:results_total}.
\begin{table}[htbp]
\begin{center}
\begin{tabular}{|c|c|}
\hline
Combination & $\fsm \pm \text{stat.} \pm \text{NP syst.} \pm \text{add. syst.}$\\
\hline
\hline
{\dQ} & $1.53  \pm 0.14 \pm 0.14 \pm 0.29$\\
{\bQ} & $0.53  \pm 0.18 \pm 0.13 \pm 0.47$\\
{Full Combination}& $1.12 \pm 0.11 \pm 0.09 \pm 0.20$\\
\hline
\end{tabular}
\end{center}
\caption{Final fit results for \fsm\ including statistical uncertainties, uncertainties due to nuisance parameters as well as additional systematic uncertainties.}
\label{tab:results_total}
\end{table}
What can be noticed, in particular when comparing the final fit result to the result without NPs, is that
\begin{itemize}
\item both the \dQ\ and the \bQ\ combinations are now consistent with the SM expectation of $\fsm=1.0$.
\item the nuisance parameters affect the central value of the \bQ\ combination to a large extent.
\item the combination significantly reduces the uncertainties. 
\end{itemize}

In the following section the fit output is investigated closely and checked for consistency.

\section{Fit Consistency Checks}
Several quantities need to be checked to evaluate the consistency of the fit output. This is the purpose of this section. 

The \dphi\ distributions after fitting allow checking for a proper modelling of the data. The distributions of posterior probability density functions of the fit give a hint if \mbox{significant} changes of the assumed background yields were fitted. The measured \fsm\ should not result from significant changes in the background normalization but rather in a mixing of the two signal samples. As a last check, the fit values of the nuisance parameters need to be investigated. Significant deviations and constraints of their expected uncertainties would need to be well justified. 

\subsection{Post-fit Plots}
The following plots show the prediction of each of the eight channels for both the SM spin correlation and uncorrelated \ttbar\ pairs. This is compared to the data and the best-fit results including the uncertainties from the fit. The best-fit results are from the full combination fit.

\begin{figure}[htbp]
\begin{center}
\includegraphics[width=0.45\textwidth]{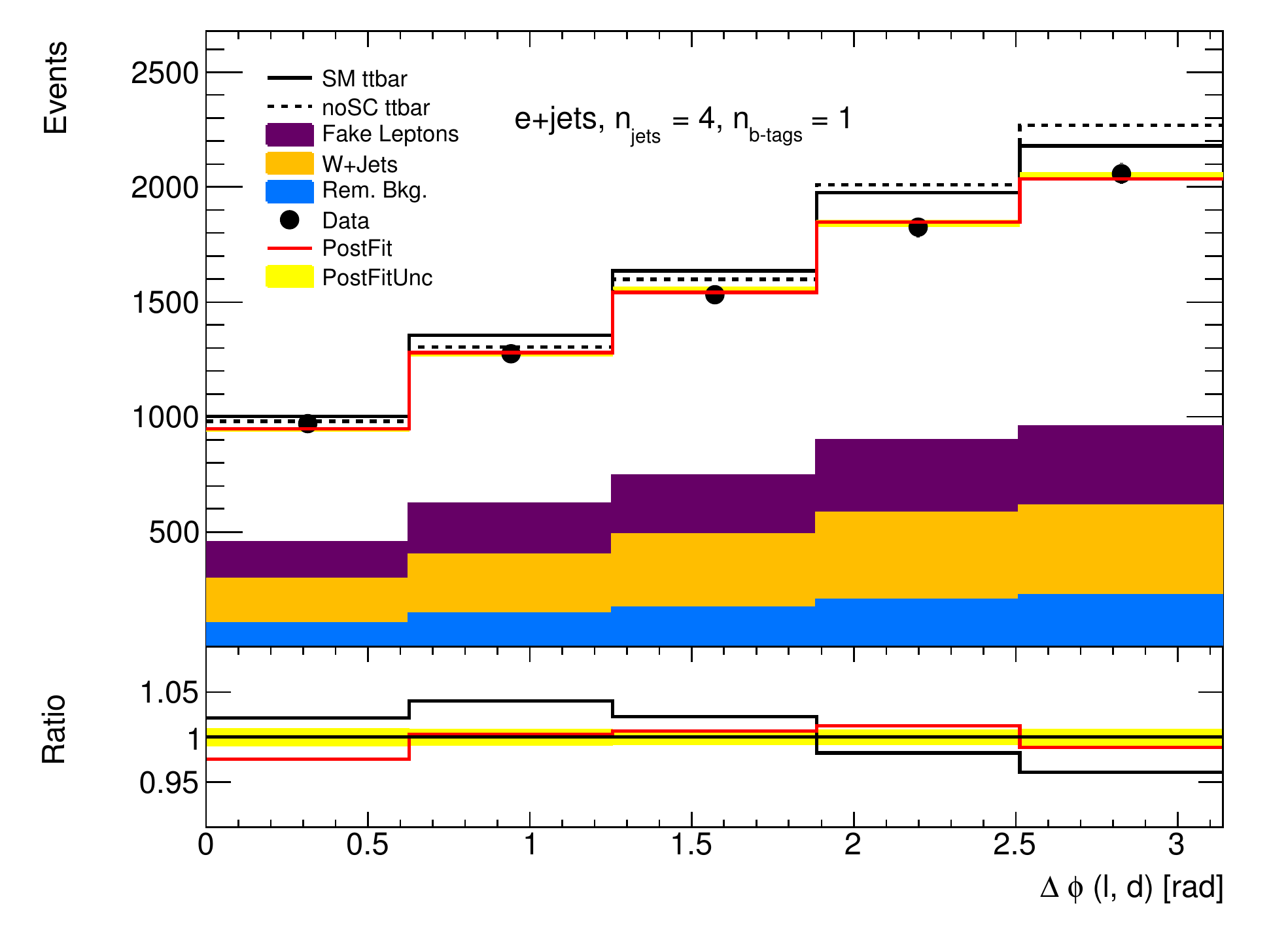} 
\includegraphics[width=0.45\textwidth]{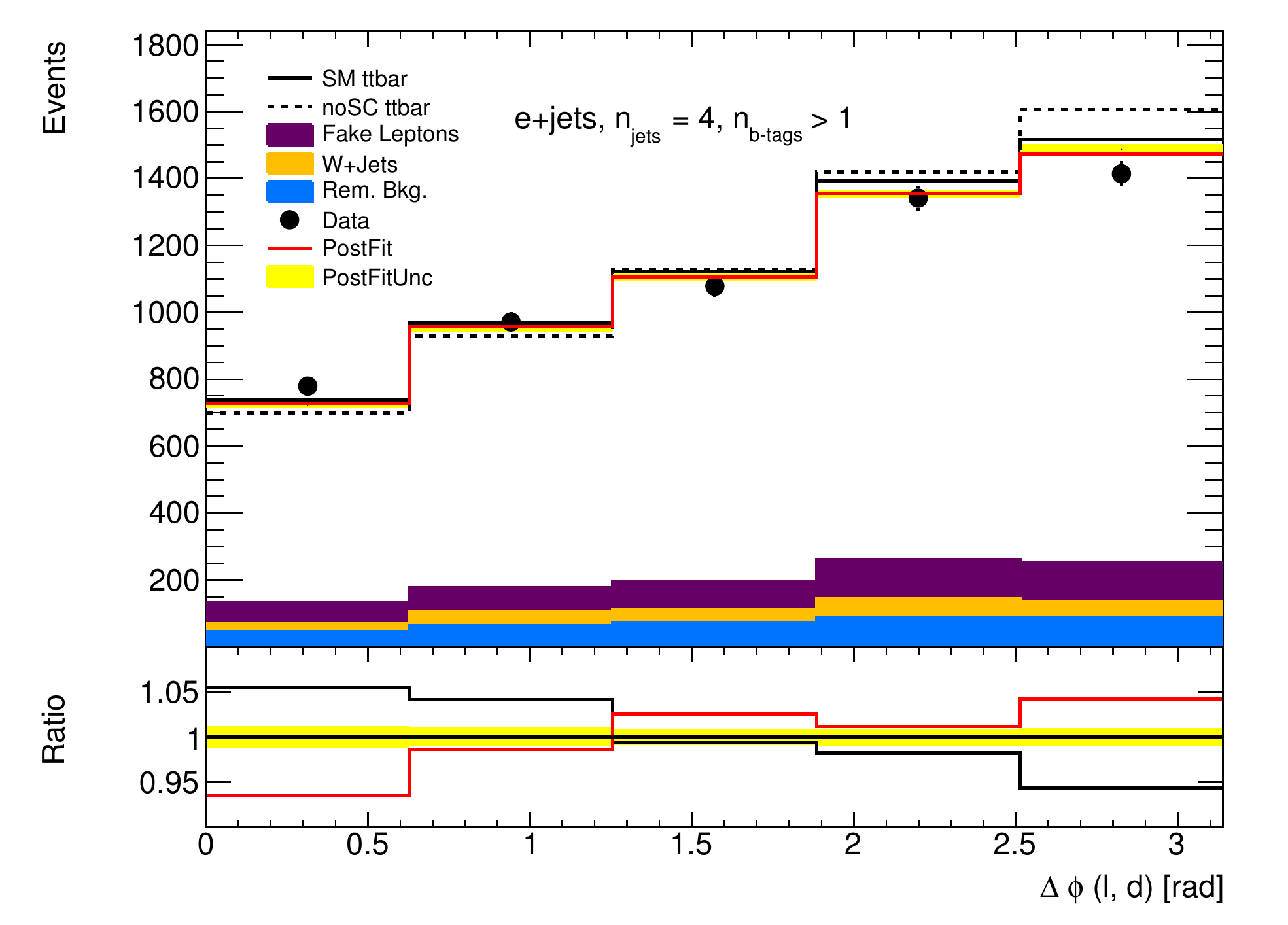} \\
\includegraphics[width=0.45\textwidth]{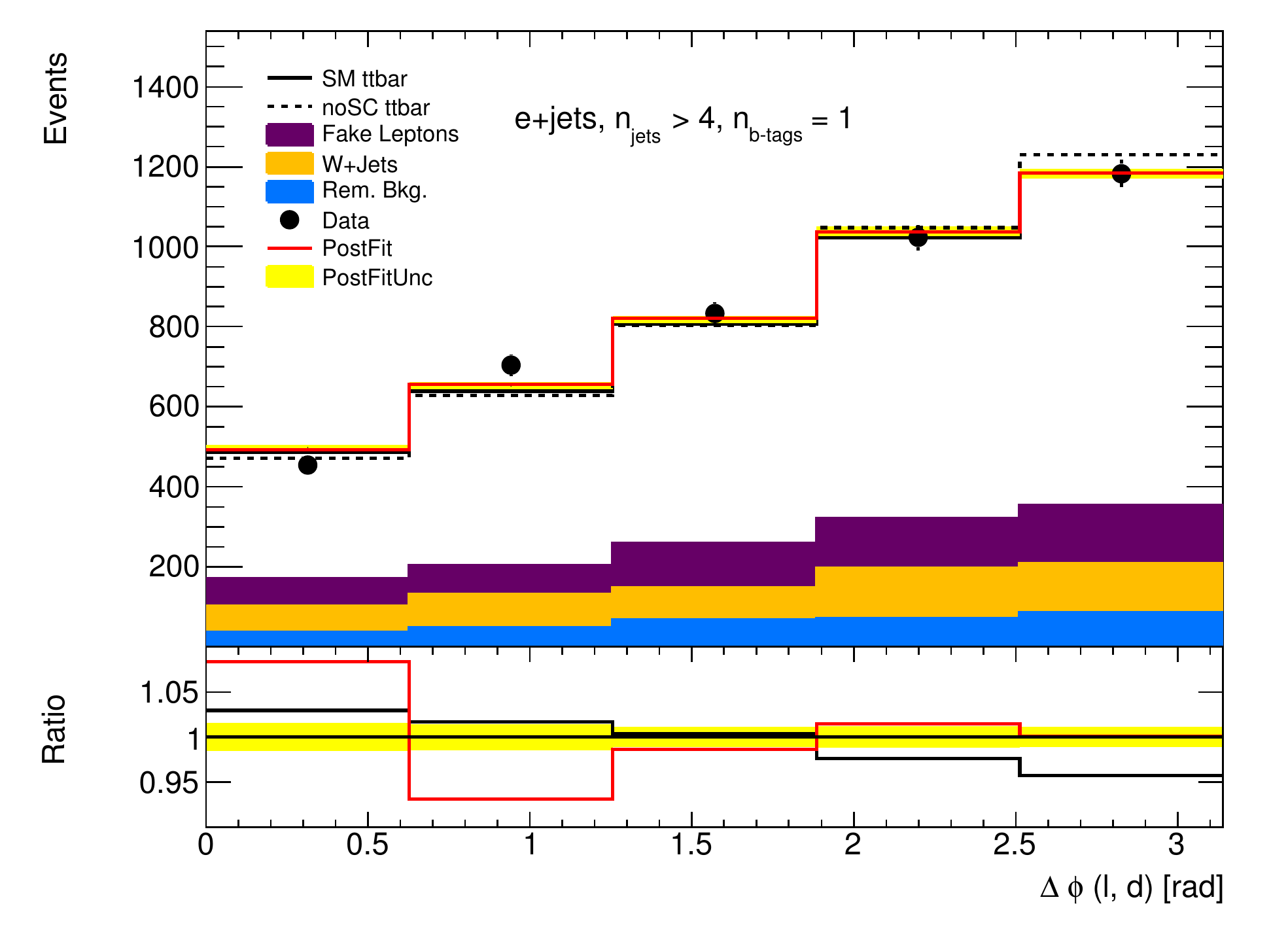} 
\includegraphics[width=0.45\textwidth]{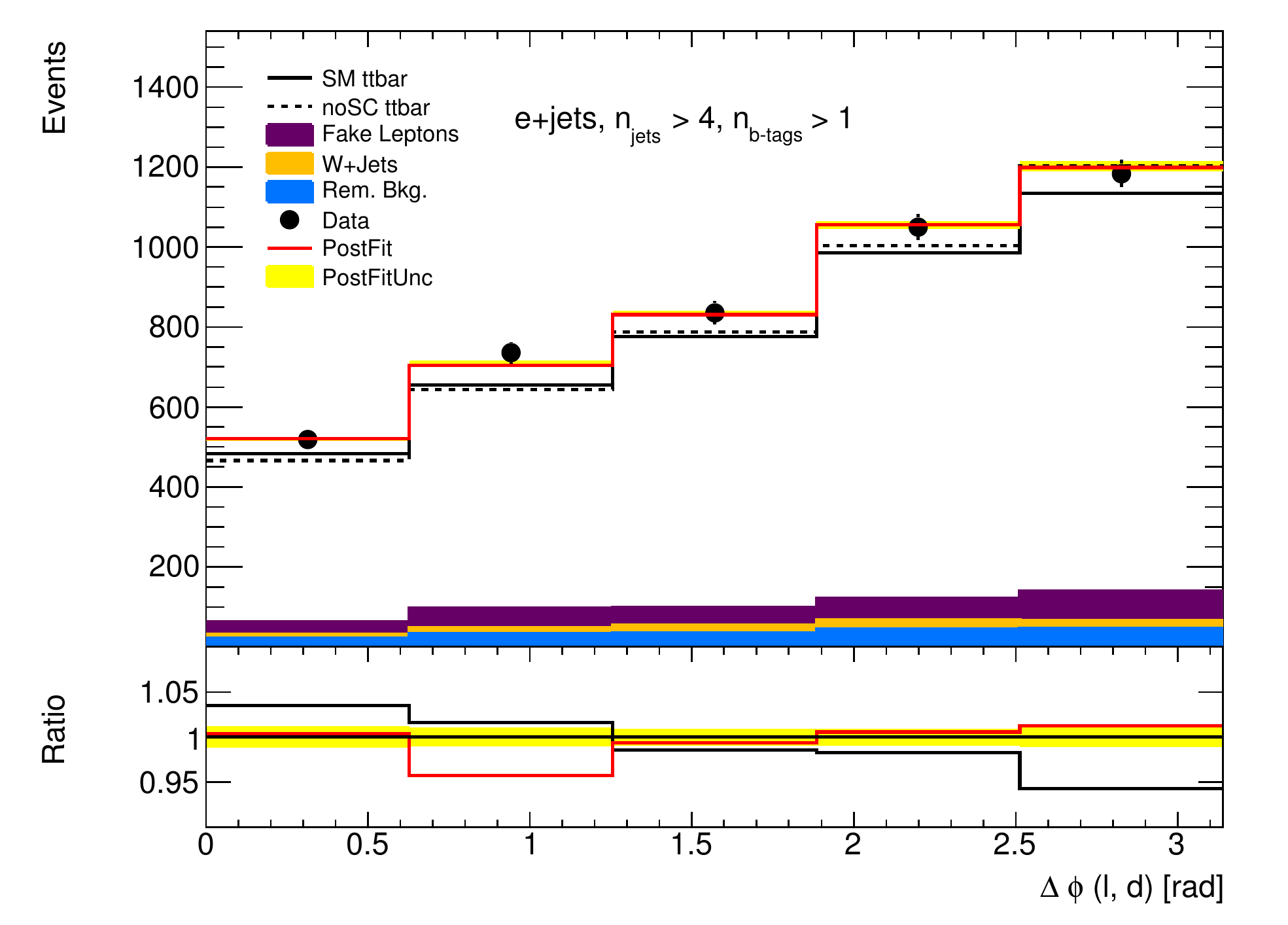} 
\end{center}
\caption{Prediction of the SM spin correlation and uncorrelated \ttbar\ pairs (black dashed and dotted) compared to data (black dots) and the best-fit result (red line) including uncertainties (yellow area). The ratios of SM and uncorrelated prediction (black line) as well as best-fit to data (red line) are shown. These plots show the four \ejets\ channels using the \dQ\ as analyser. 
}
\label{fig:sensitivity_el_dQ}
\end{figure} 

\begin{figure}[htbp]
\begin{center}
\includegraphics[width=0.45\textwidth]{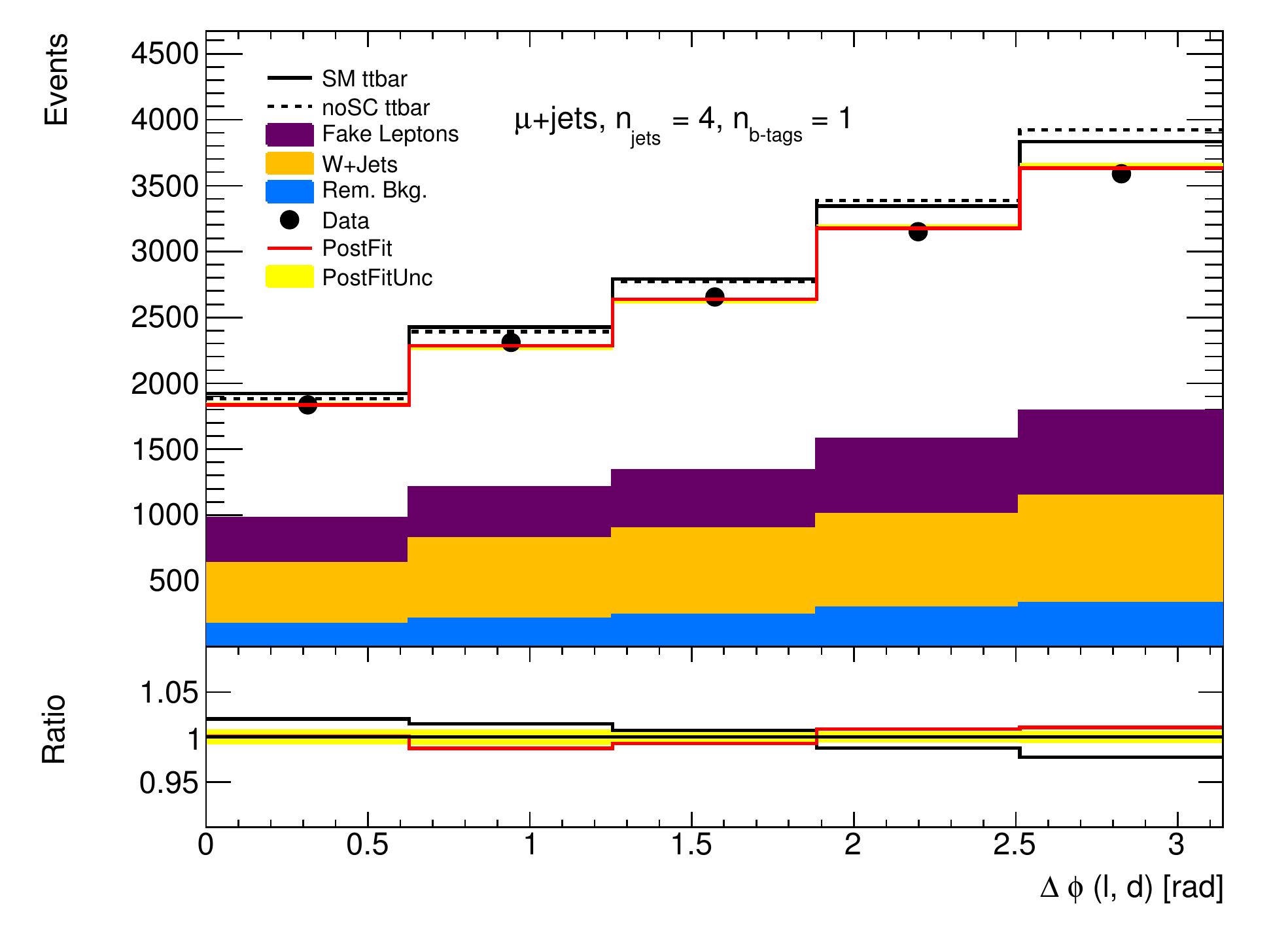} 
\includegraphics[width=0.45\textwidth]{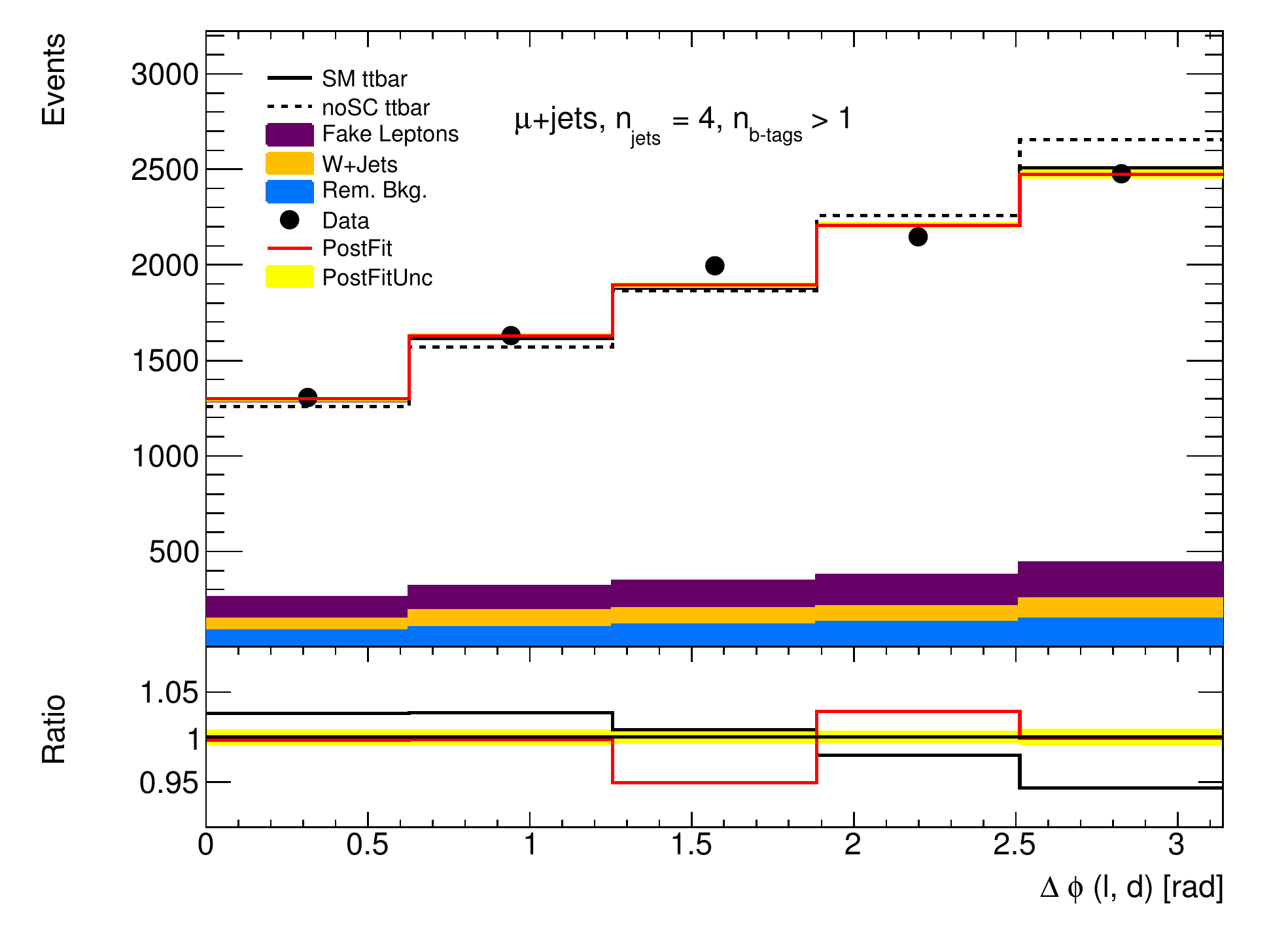} \\
\includegraphics[width=0.45\textwidth]{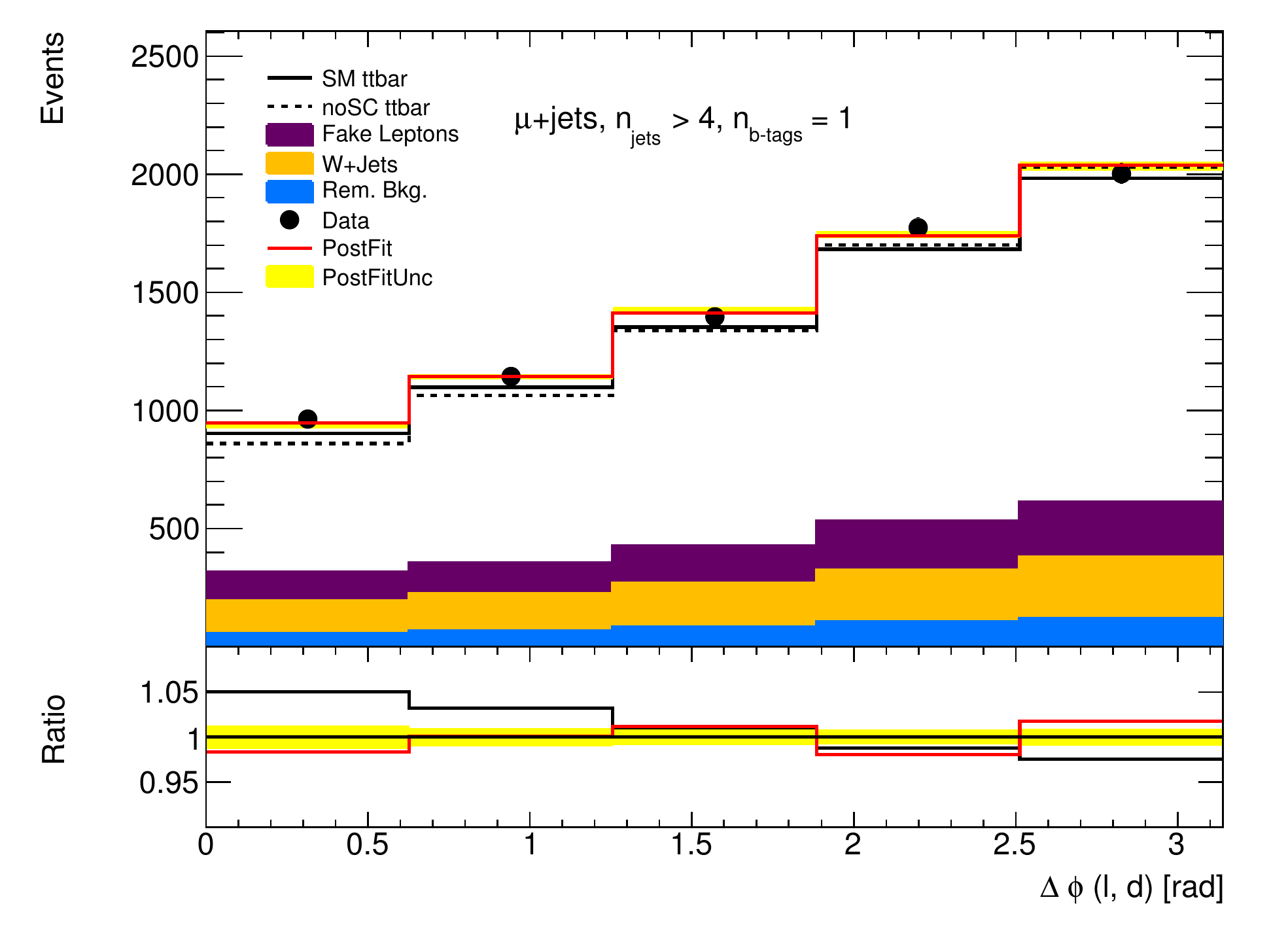} 
\includegraphics[width=0.45\textwidth]{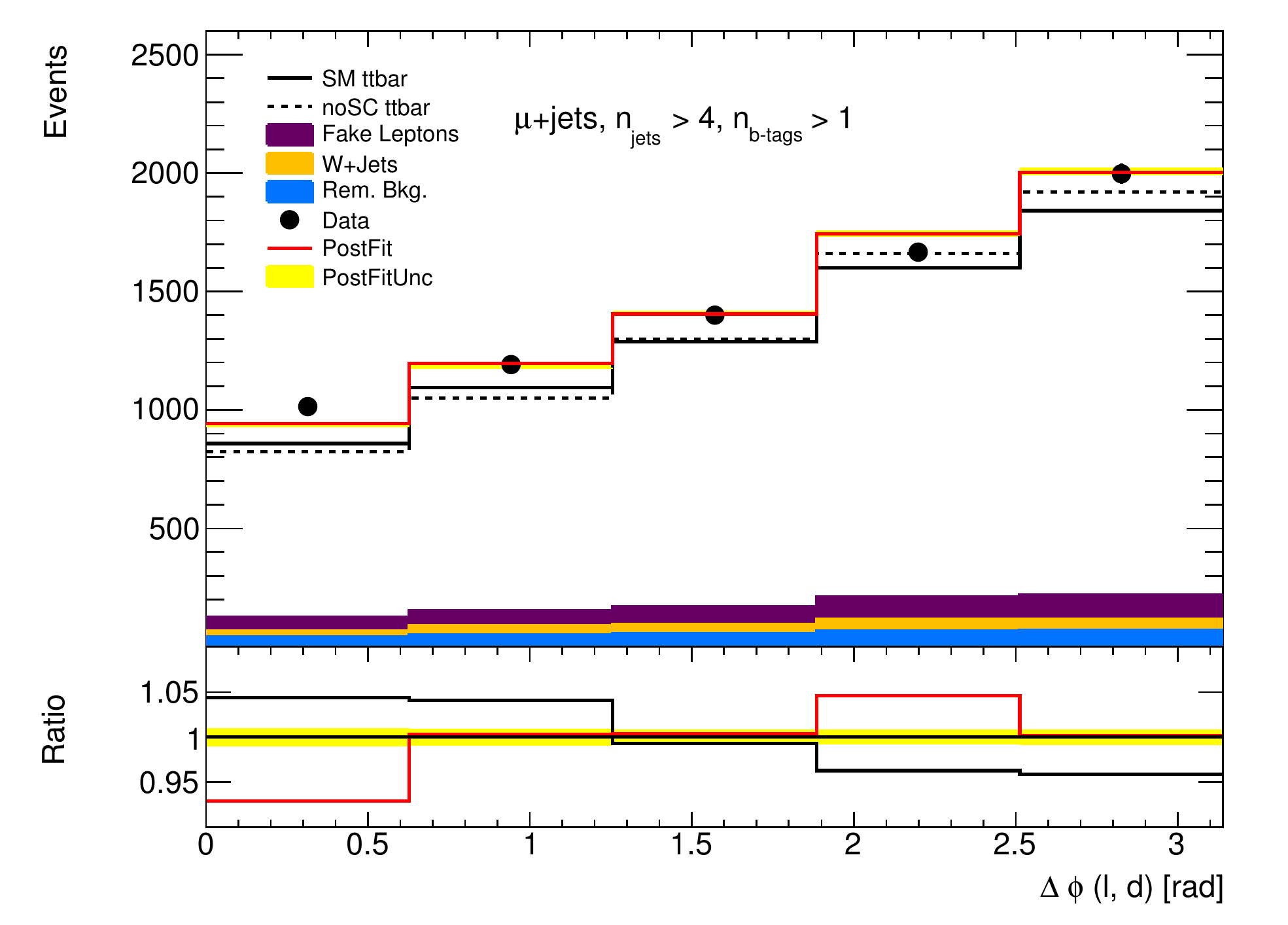} 
\end{center}
\caption{Prediction of the SM spin correlation and uncorrelated \ttbar\ pairs (black dashed and dotted) compared to data (black dots) and the best-fit result (red line) including uncertainties (yellow area). The ratios of SM and uncorrelated prediction (black line) as well as best-fit to data (red line) are shown. These plots show the four \mujets\ channels using the \dQ\ as analyser. 
}
\label{fig:sensitivity_mu_dQ}
\end{figure} 

\begin{figure}[htbp]
\begin{center}
\includegraphics[width=0.45\textwidth]{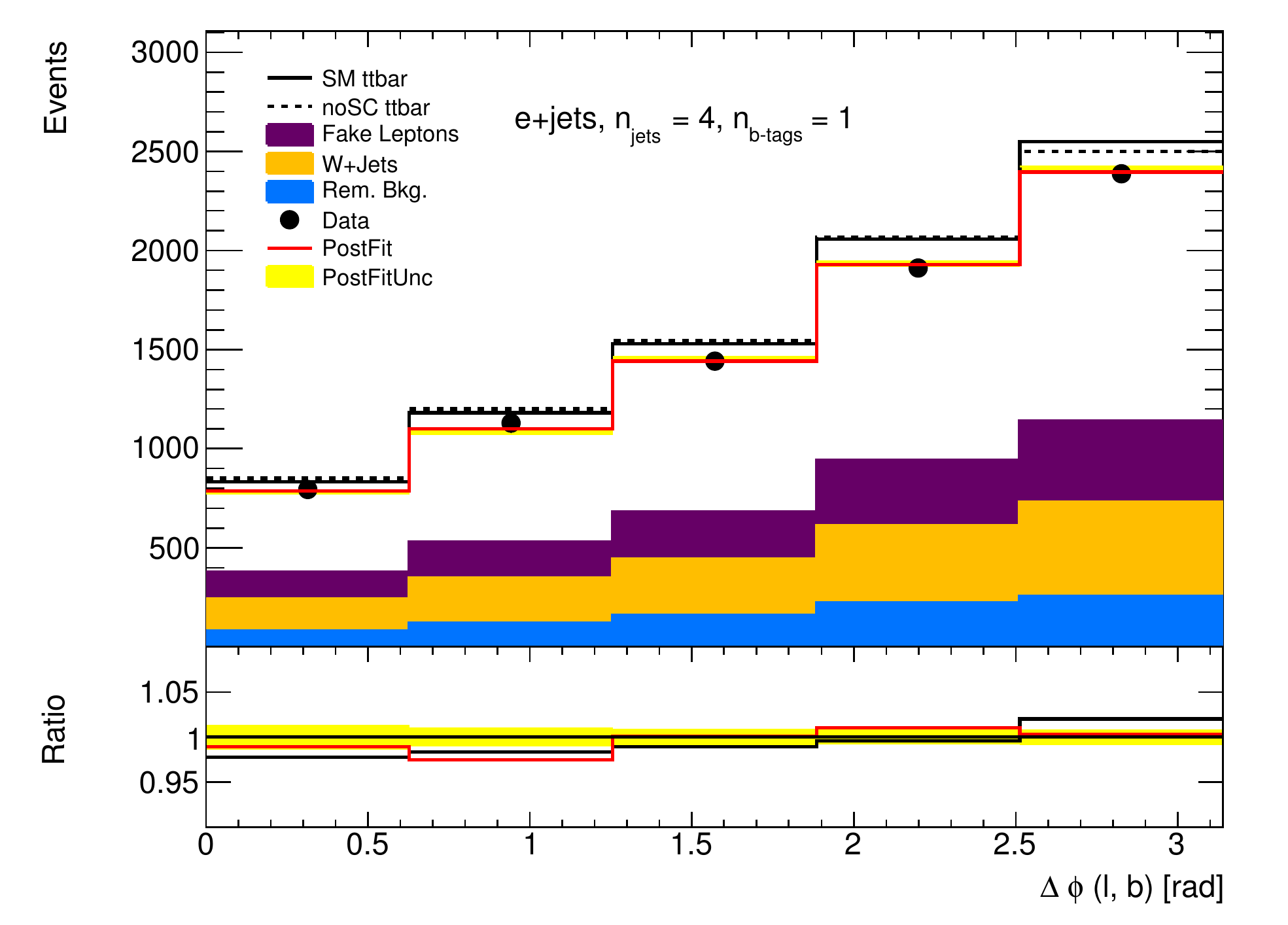} 
\includegraphics[width=0.45\textwidth]{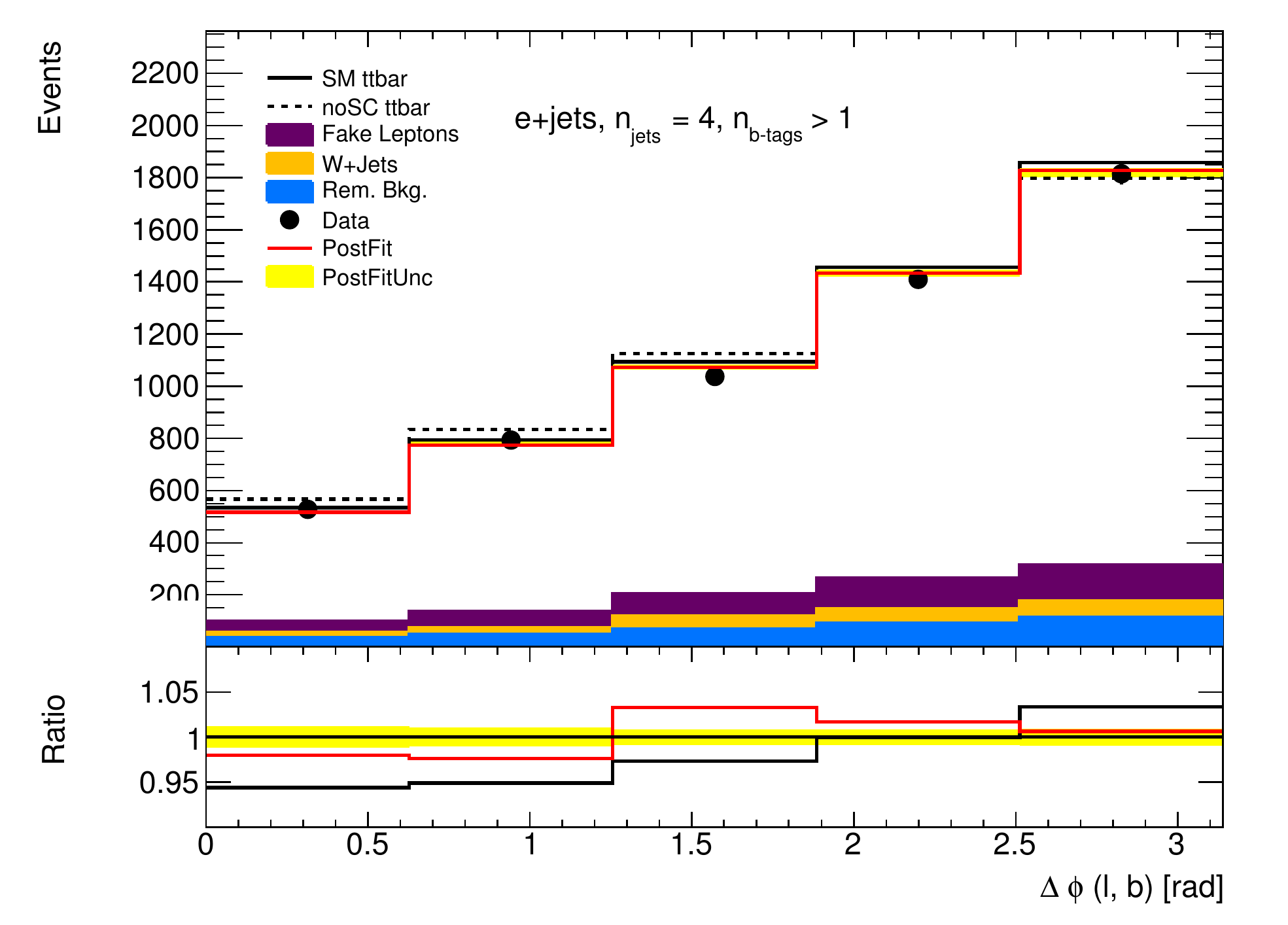} \\
\includegraphics[width=0.45\textwidth]{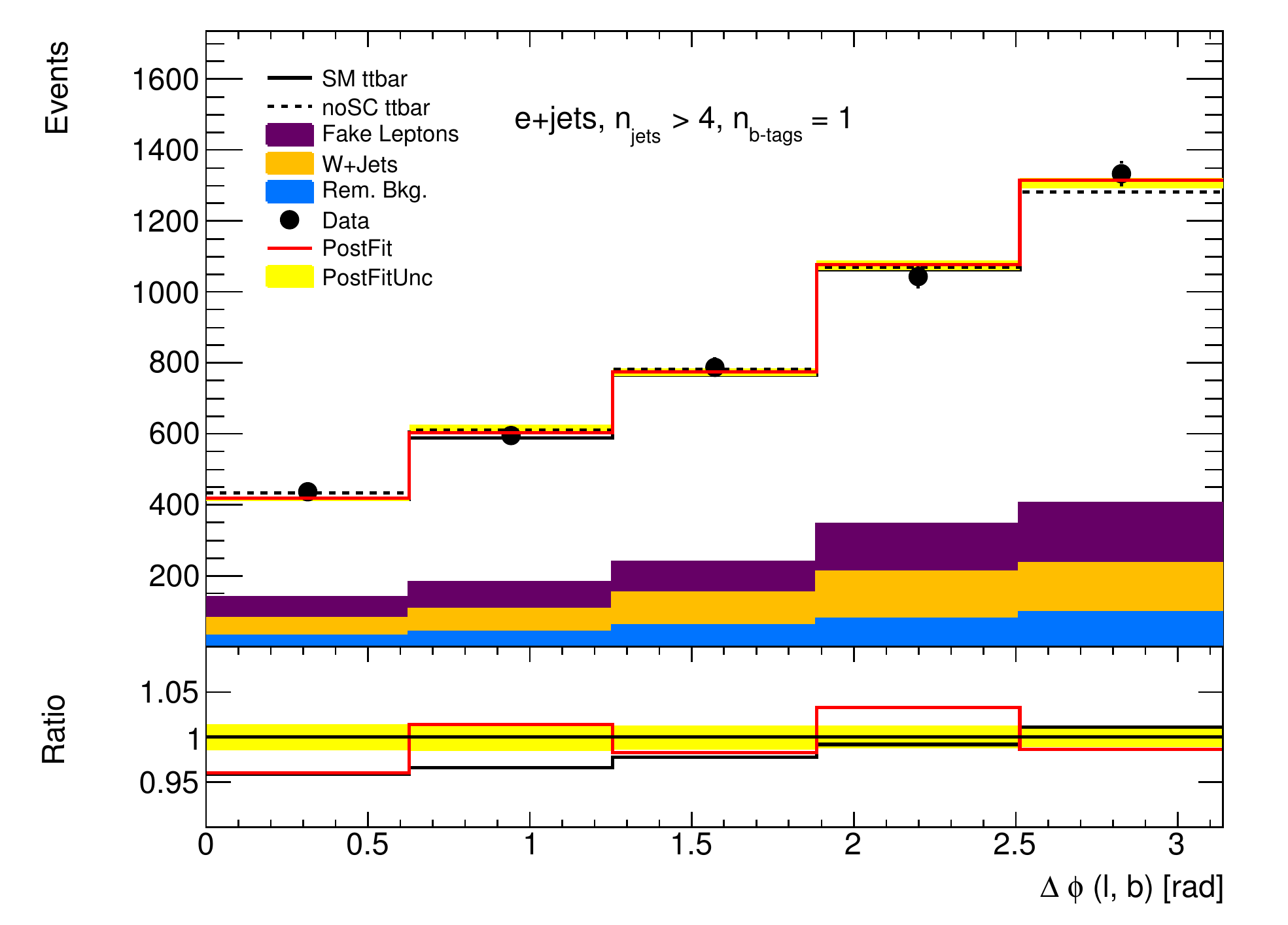} 
\includegraphics[width=0.45\textwidth]{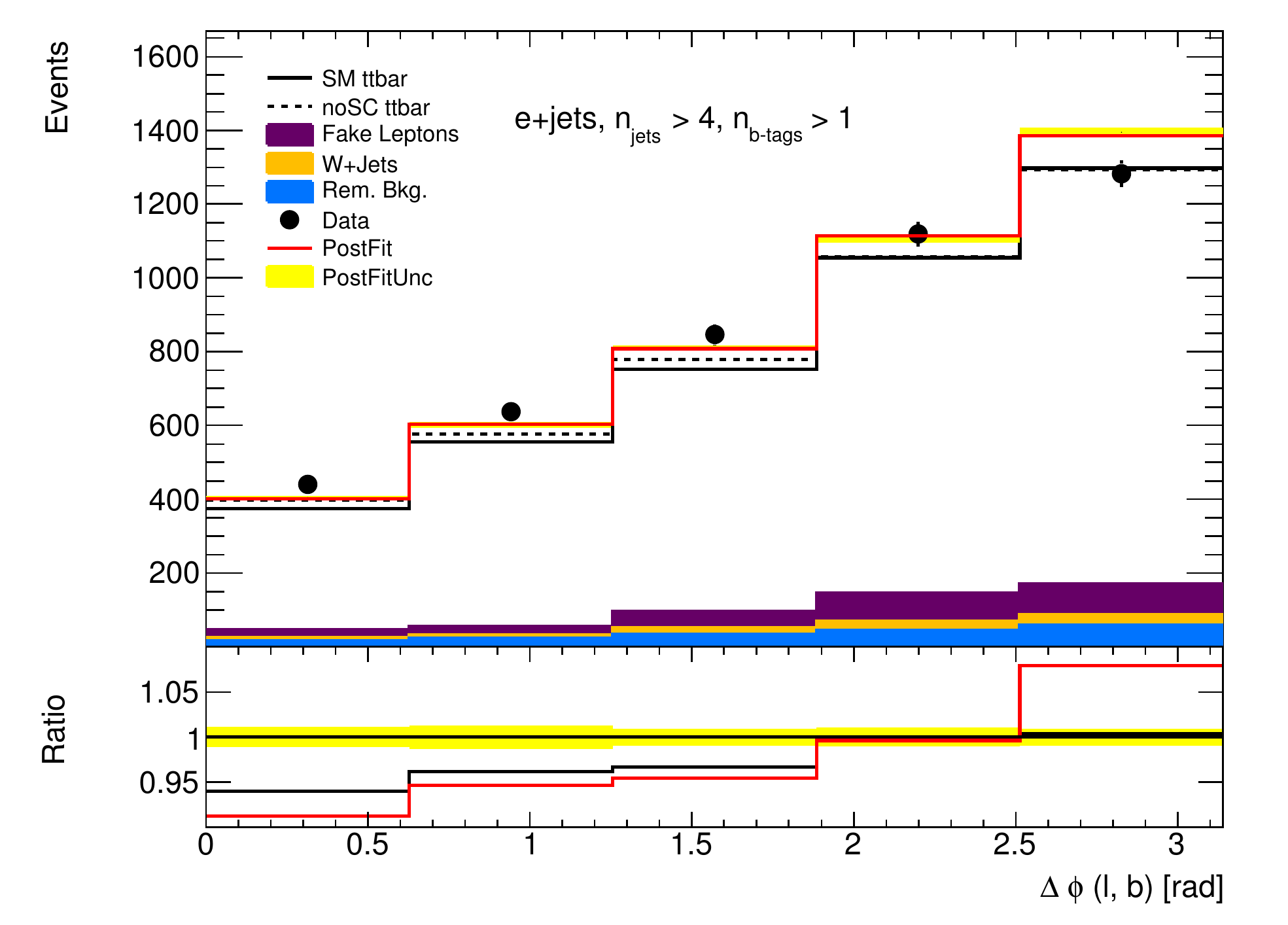} 
\end{center}
\caption{Prediction of the SM spin correlation and uncorrelated \ttbar\ pairs (black dashed and dotted) compared to data (black dots) and the best-fit result (red line) including uncertainties (yellow area). The ratios of SM and uncorrelated prediction (black line) as well as best-fit to data (red line) are shown. These plots show the four \ejets\ channels using the \bQ\ as analyser. 
}
\label{fig:sensitivity_el_bQ}
\end{figure} 

\begin{figure}[htbp]
\begin{center}
\includegraphics[width=0.45\textwidth]{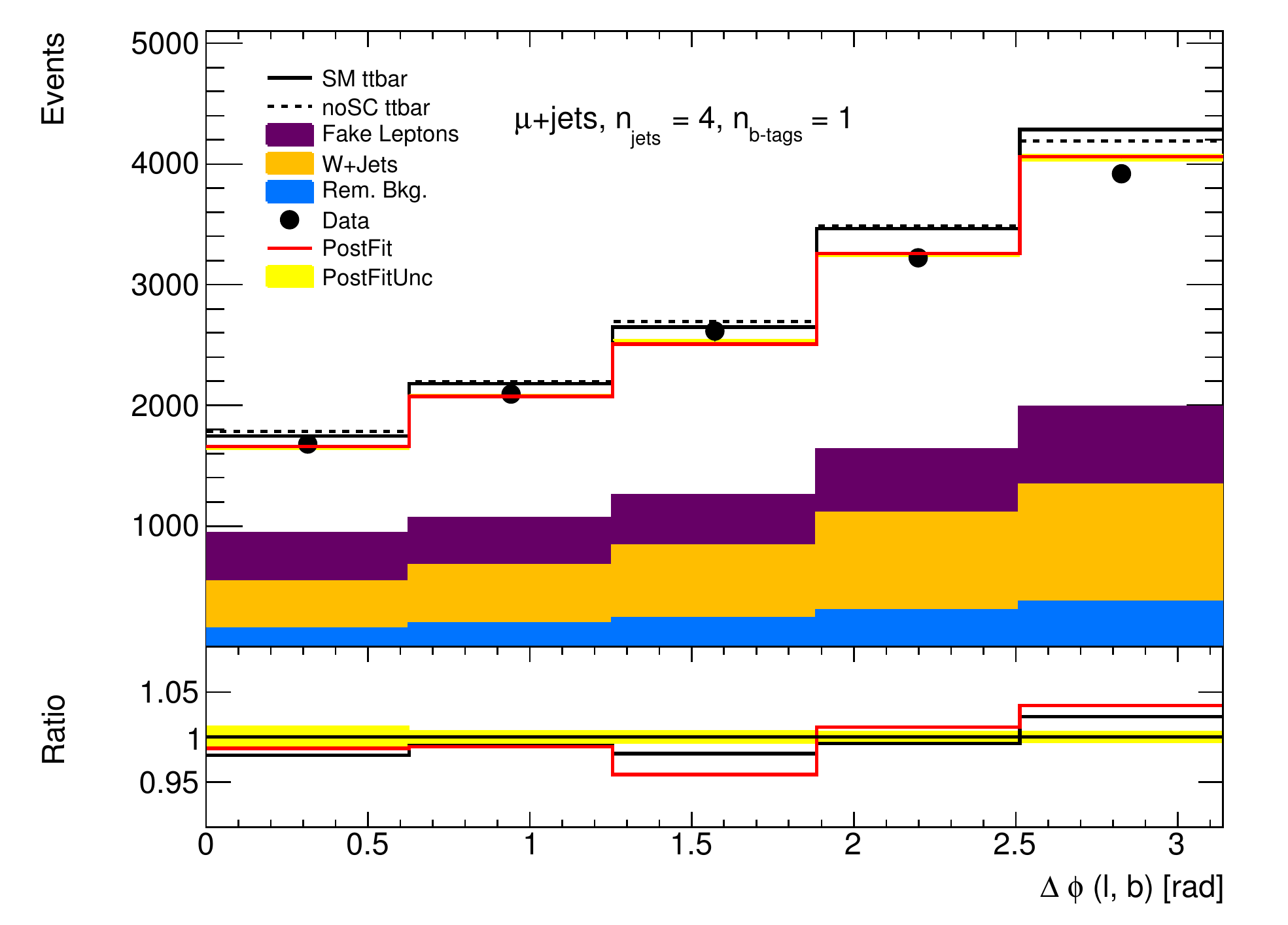} 
\includegraphics[width=0.45\textwidth]{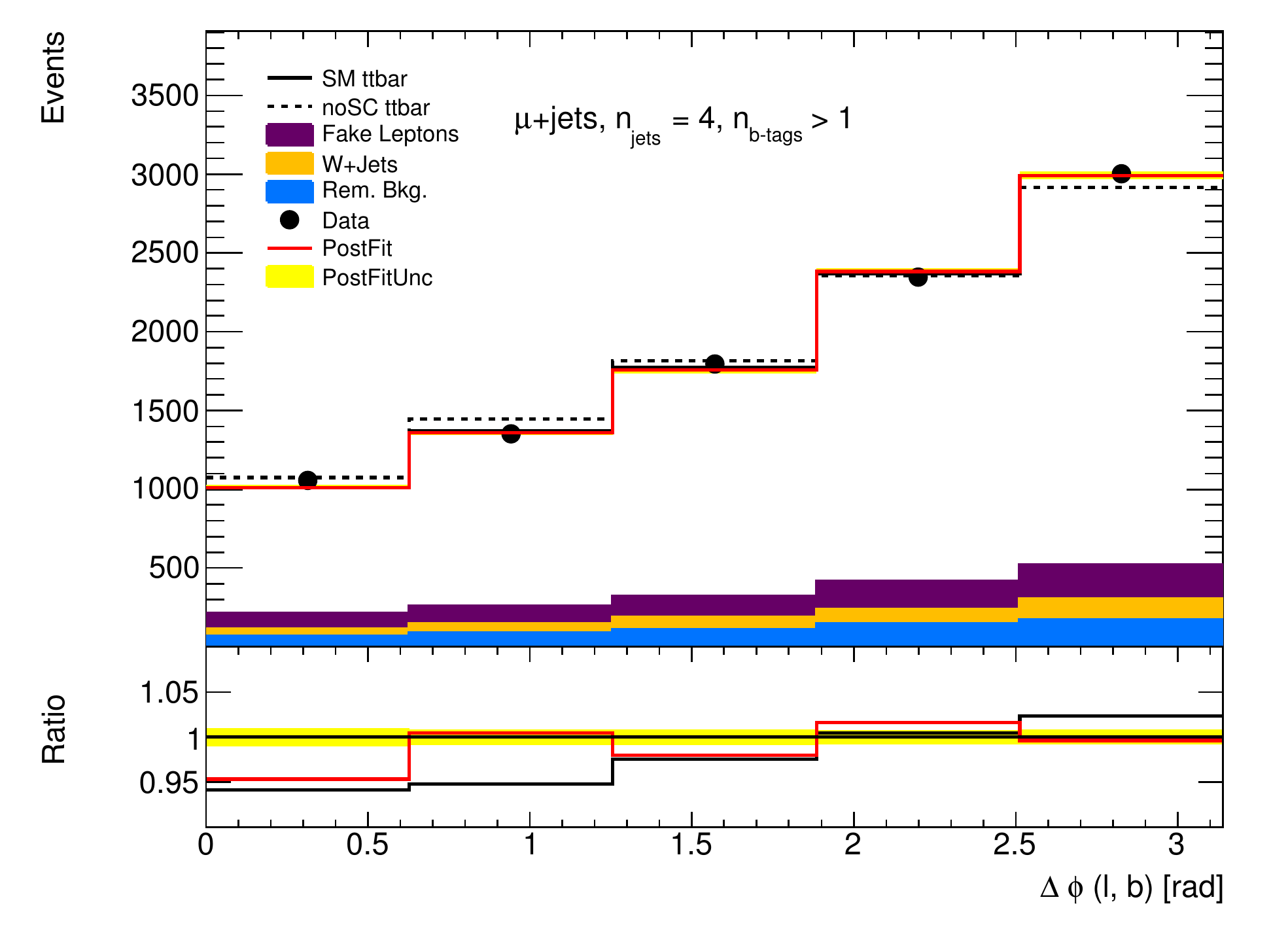} \\
\includegraphics[width=0.45\textwidth]{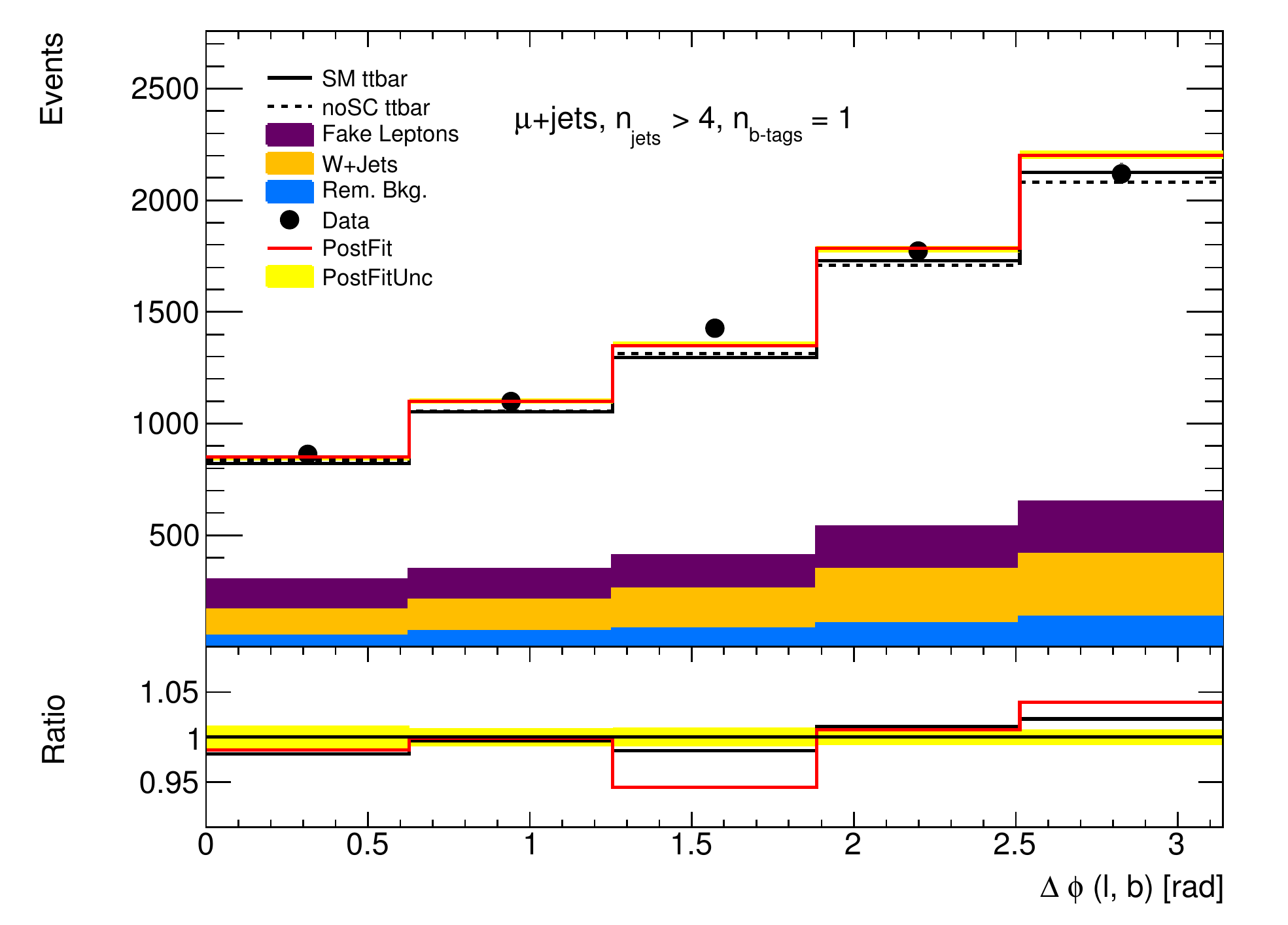} 
\includegraphics[width=0.45\textwidth]{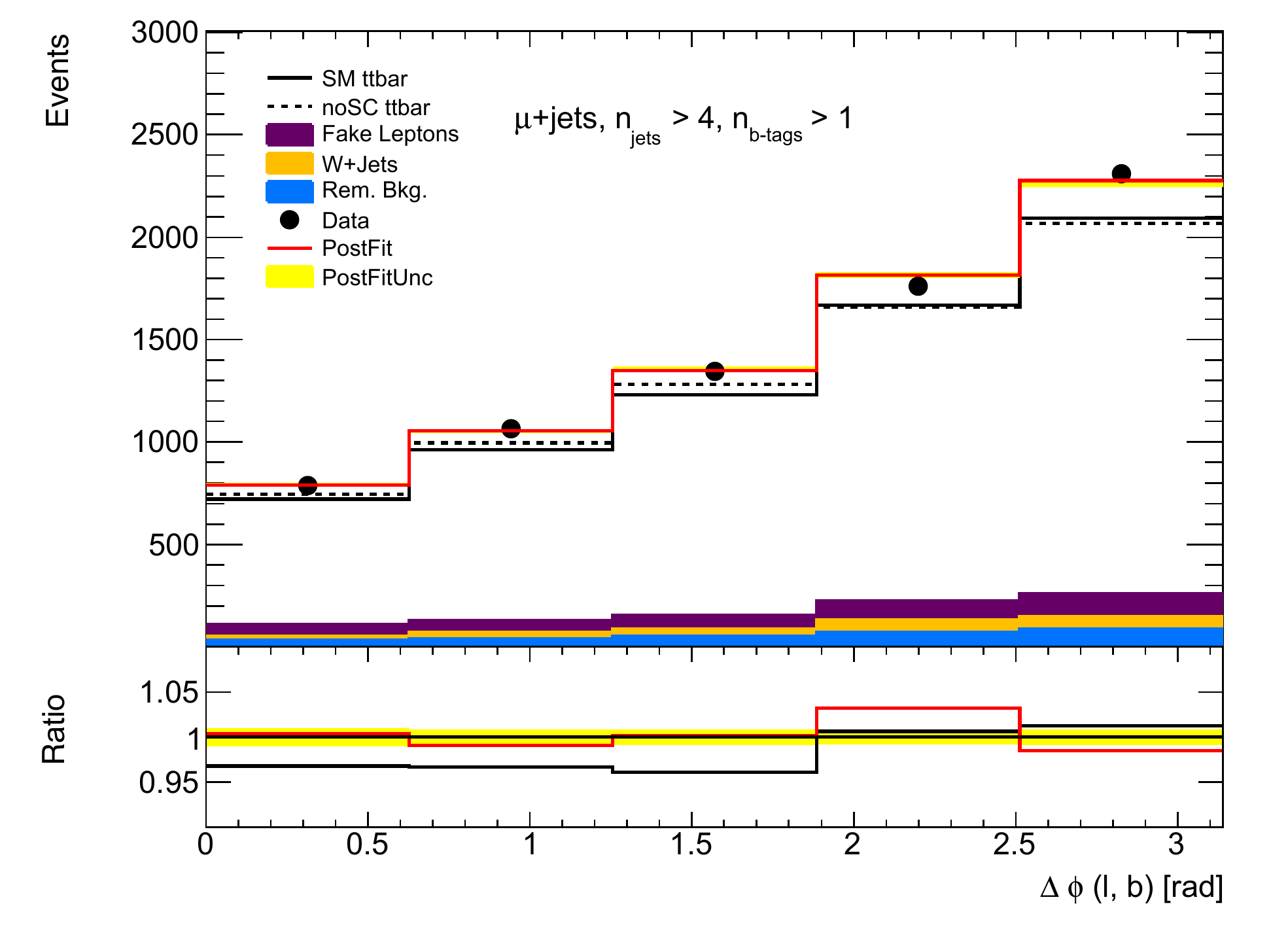} 
\end{center}
\caption{Prediction of the SM spin correlation and uncorrelated \ttbar\ pairs (black dashed and dotted) compared to data (black dots) and the best-fit result (red line) including uncertainties (yellow area). The ratios of SM and uncorrelated prediction (black line) as well as best-fit to data (red line) are shown. These plots show the four \mujets\ channels using the \bQ\ as analyser. 
}
\label{fig:sensitivity_mu_bQ}
\end{figure} 

It can be noticed that, in general, the fit is able to properly describe the data. These good fit results need a modification of the predicted yields for the signal and the background processes. This is achieved by using the degrees of freedom provided by the priors on the background estimation and the nuisance parameters. 

One trend that is useful, in particular for the discussion of the consistency, is a small slope visible for both the \dQ\ and the \bQ\ post-fit ratios between post-fit results and data. It comes along with the fact that the individual fit results for the \dQ\ and the \bQ\ tend to deviate from $\fsm = 1.0$ into opposite directions and to a larger extent than the combined fit allows. This discussion will go into detail in Section \ref{sec:unc_discussion}.

\subsection{Posterior PDFs}
Priors are set on the background yield estimations to constrain the fit. These are chosen to be Gaussian with a width corresponding to the evaluated normalization uncertainty. In this section the probability density functions (PDFs) of the priors are compared to those of the posteriors. The comparison allows checking if the fit either constrains the prediction (resulting in a narrower posterior) or if it prefers a normalization different than the predicted (leading to a shifted mean). Both effects are expected to be small. The posteriors for the full combination fit are shown in Figure \ref{fig:par_posteriors}. In addition to the priors and posteriors of the background yields the posterior of the jet multiplicity correction factor (see Section \ref{sec:jetmultcorr}) is shown.
\begin{figure}[htbp]
\begin{center}
\includegraphics[height=0.9\textheight]{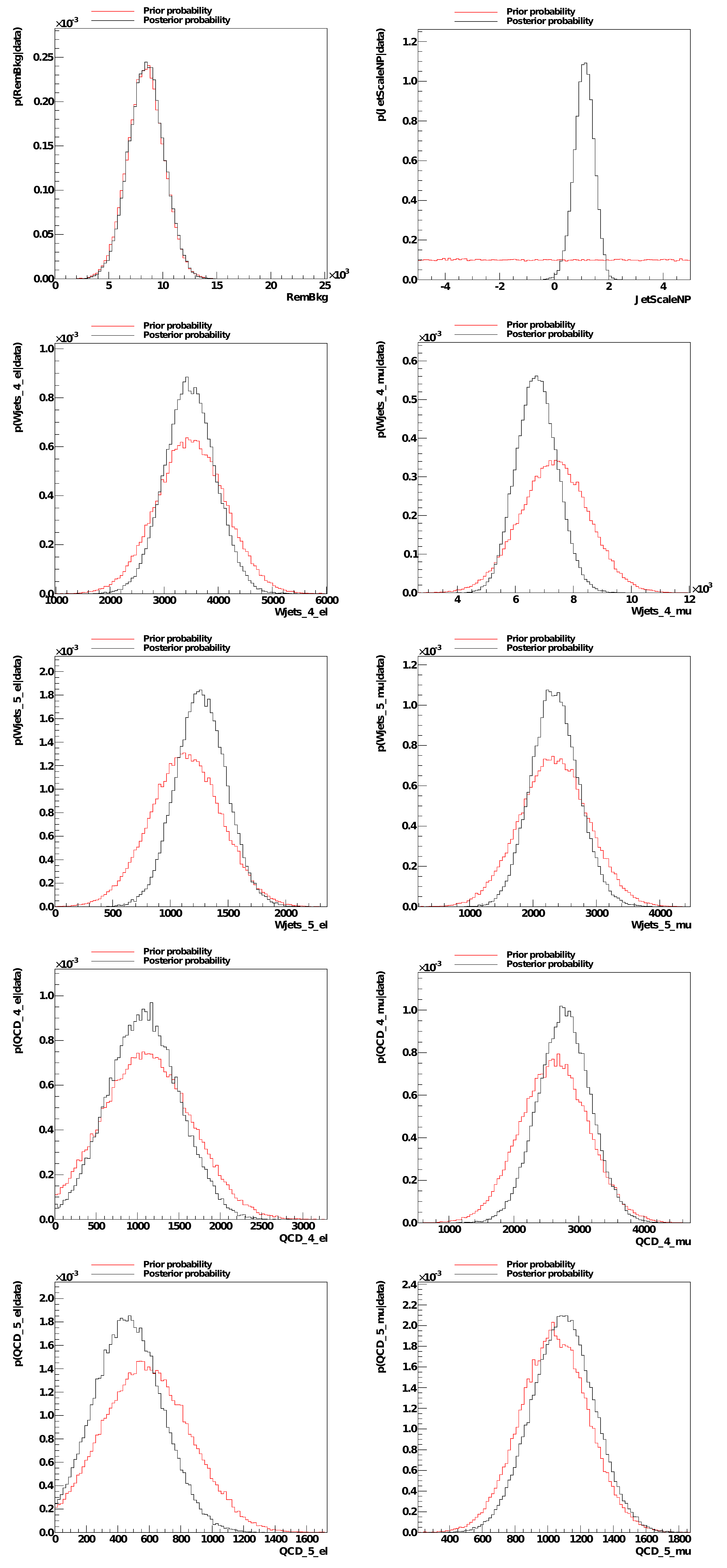}
\end{center}
\caption{Prior and posterior distributions for the fit parameters describing the background yields and the jet multiplicity correction for the full combination of the \dQ\ and the \bQ\ analysers.}
\label{fig:par_posteriors}
\end{figure} 
The corresponding posterior distributions for the \dQ\ and \bQ\ combinations can be found in the Appendix \ref{sec:app_posteriors}. No significant deviation of the mean and width values of the priors are observed. 

There is only a constant prior on the \ttbar\ cross section. It was tested that a Gaussian prior corresponding to the theory uncertainty on the cross section does not lead to an improved precision of the measurement. Hence, it can be extracted directly from the fit without a bias. The results of the \ttbar\ cross section scaling parameter $c$ (introduced in Section \ref{sec:fit}) are shown in Table \ref{tab:xsec_postfit}.

\begin{table}[htbp]
\begin{center}
\begin{tabular}{|c|c|c|c|}
\hline
 & \dQ\ & \bQ\ & Combination \\
 \hline
 \hline
$c$ & $0.97 \pm 0.04$ & $0.99 \pm 0.04$ & $0.97 \pm 0.04$\\

\hline
\end{tabular}
\end{center}
\caption{Fit results of the \ttbar\ cross section scaling parameter $c$. Uncertainties include statistical uncertainties and uncertainties due to nuisance parameters. }
\label{tab:xsec_postfit}
\end{table} 
The results are compatible with the SM expectation of $c=1.0$. 

\subsection{Nuisance Parameter Postfit Values}
\label{sec:NP_postfit}
Nuisance parameters are implemented via Gaussian priors. Their central values are set to zero as the current modelling is the best estimate by definition. The width of the NP priors is set to one, corresponding to one standard deviation. 
It is expected that the fit is able to constrain the uncertainties used as NPs. To estimate the possible constraint the fit was performed replacing the data with the simulated SM expectation (\newword{Asimov dataset} \cite{asimov}).

The expected constraint\footnote{A NPs is constrained if its posterior width in smaller than one.} of the systematic uncertainties is indicated by the grey bars in Figure \ref{fig:NP_postfit_comb} for the full combination fit. Results for the individual \dQ\ and the \bQ\ combinations can be found in the Appendix \ref{sec:app_NPs}.

\begin{figure}[htbp]
\begin{center}
\includegraphics[width=\textwidth]{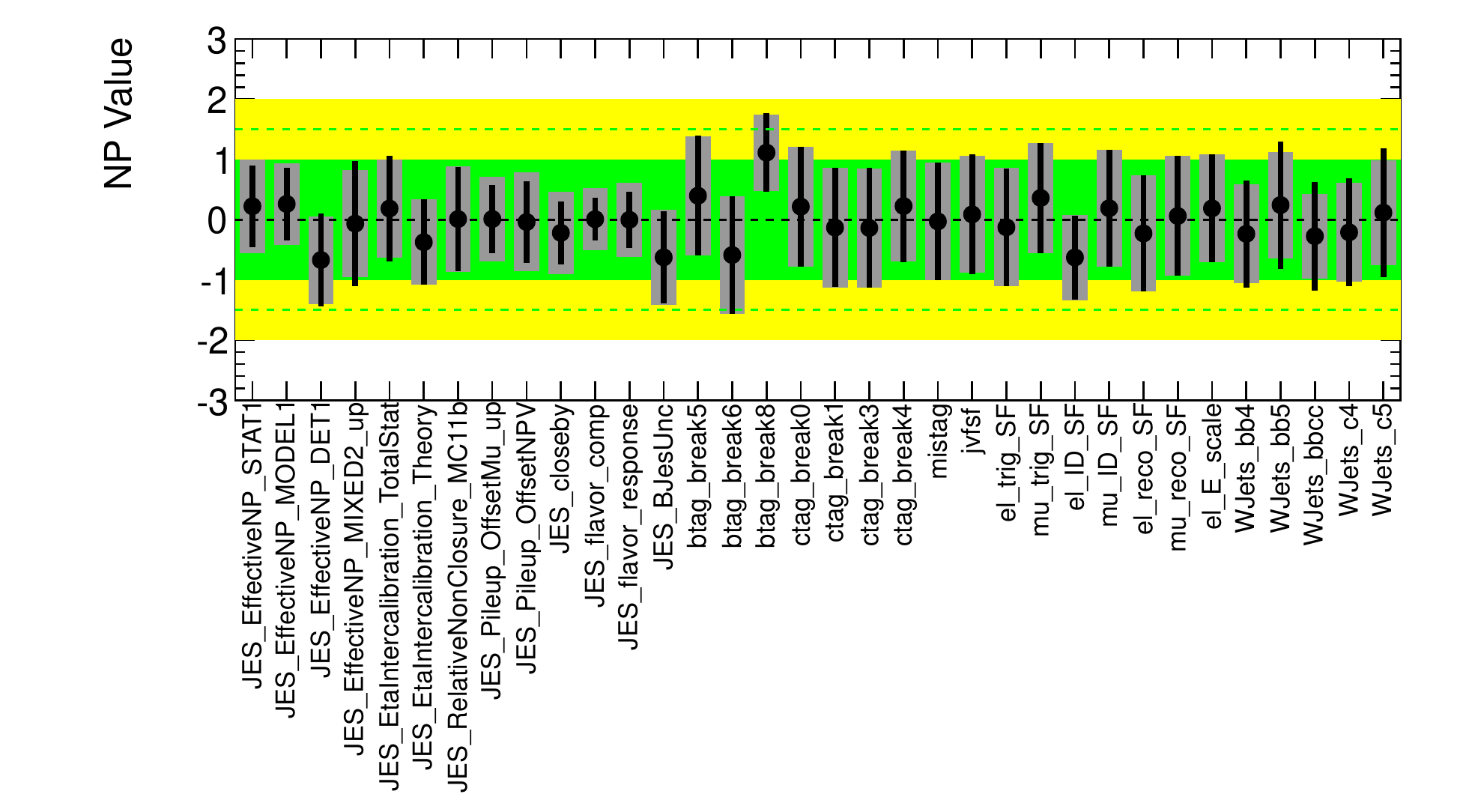}
\end{center}
\caption{Postfit values of the nuisance parameters (black lines) for the full combination. The grey vertical bands behind the lines show the expected uncertainties on the nuisance parameters. }
\label{fig:NP_postfit_comb}
\end{figure} 
In Figure \ref{fig:NP_postfit_comb}, all expected constraints are shifted. Their mean values are set to the ones of the NP post-fit results. This allows comparing the expected (grey band) to the measured (black line) constraints of the NPs. Before shifting the expected constraints, it was checked that the fit of the SM expectation leads to central values of zero for all NPs. 

The observed constraints are compatible with the expected ones. They are slightly higher in case of JES close-by, JES flavour composition and JES flavour response and slightly lower for the $W$+jets uncertainties. These differences arise from the \dQ\ combination (JES components) and the \bQ\ combination ($W$+jets components), shown in figures \ref{fig:NP_postfit_dQ} and \ref{fig:NP_postfit_bQ}.

Some of the NPs are highly correlated as seen in the correlation matrix in Figure \ref{fig:par_corr_matrix_comb} for the full combination fit. The values for the correlation coefficients vary between $- 0.58$ and $+0.45$.
The matrix includes, next to the nuisance parameters, all other fit parameters. They are listed in Table \ref{tab:all_pars_comb}.

\begin{table}[htbp]
\footnotesize
\begin{center}
\begin{tabular}{|c|c|}
\hline
Parameter & Name \\
\hline
\hline
$p_0$ &  $\frac{1}{2}(N_{SM \ttbar} + N_{unc. \ttbar})$ \\
$p_1$ &  $\frac{1}{2}(N_{SM \ttbar} - N_{unc. \ttbar})$ \\
$p_2$ &  $N_{\text{rem. backg.}, e+\text{jets}} + N_{\text{rem. backg.}, \mu+\text{jets}}$ \\
$p_3$ &  $N_{W+\text{jets}, n_{\text{jets}} = 4, e+\text{jets}}$ \\
$p_4$ &  $N_{W+\text{jets}, n_{\text{jets}} \geq 5, e+\text{jets}}$ \\
$p_5$ &  $N_{QCD, n_{\text{jets}} = 4, e+\text{jets}}$ \\
$p_6$ &  $N_{QCD, n_{\text{jets}} \geq 5, e+\text{jets}}$ \\
$p_7$ &  $N_{W+\text{jets}, n_{\text{jets}} = 4, \mu+\text{jets}}$ \\
$p_8$ &  $N_{W+\text{jets}, n_{\text{jets}} \geq 5, \mu+\text{jets}}$ \\
$p_9$ &  $N_{QCD, n_{\text{jets}} = 4, \mu+\text{jets}}$ \\
$p_{10}$ &  $N_{QCD, n_{\text{jets}} \geq 5, \mu+\text{jets}}$ \\
$p_{11}$ & Jet Multiplicity Correction \\
$p_{12}$ & JES/EffectiveNP\_Stat1 \\
$p_{13}$ & JES/EffectiveNP\_Model1 \\
$p_{14}$ & JES/EffectiveNP\_Det1 \\
$p_{15}$ & JES/EffectiveNP\_Mixed2\\
$p_{16}$ & JES/Intercal\_TotalStat \\
$p_{17}$ & JES/Intercal\_Theory \\
$p_{18}$ & JES/RelativeNonClosureMC11b \\
$p_{19}$ & JES/PileUpOffsetMu \\
$p_{20}$ & JES/PileUpOffsetNPV \\
$p_{21}$ & JES/Closeby \\
$p_{22}$ & JES/FlavorComp \\
$p_{23}$ & JES/FlavorResponse \\
$p_{24}$ & JES/BJES \\
$p_{25}$ & btag/break5 \\
$p_{26}$ & btag/break6 \\
$p_{27}$ & btag/break8 \\
$p_{28}$ & ctag/break0 \\
$p_{29}$ & ctag/break1 \\
$p_{30}$ & ctag/break3 \\
$p_{31}$ & ctag/break4 \\
$p_{32}$ & mistag \\
$p_{33}$ & JVF \\
$p_{34}$ & el/Trigger \\
$p_{35}$ & mu/Trigger\\
$p_{36}$ & el/ID \\
$p_{37}$ & mu/ID \\
$p_{38}$ & el//Reco \\
$p_{39}$ & el/E\_scale \\
$p_{40}$ & mu//Reco \\
$p_{41}$ & WJets/bb4 \\
$p_{42}$ & WJets/bb5 \\
$p_{43}$ & WJets/bbcc \\
$p_{44}$ & WJets/c4\\
$p_{45}$ & WJets/c5\\
\hline
\end{tabular}
\end{center}
\caption{List of all fit parameters. }
\label{tab:all_pars_comb}
\end{table}

\begin{figure}[htbp]
\begin{center}
\includegraphics[width=0.6\textwidth]{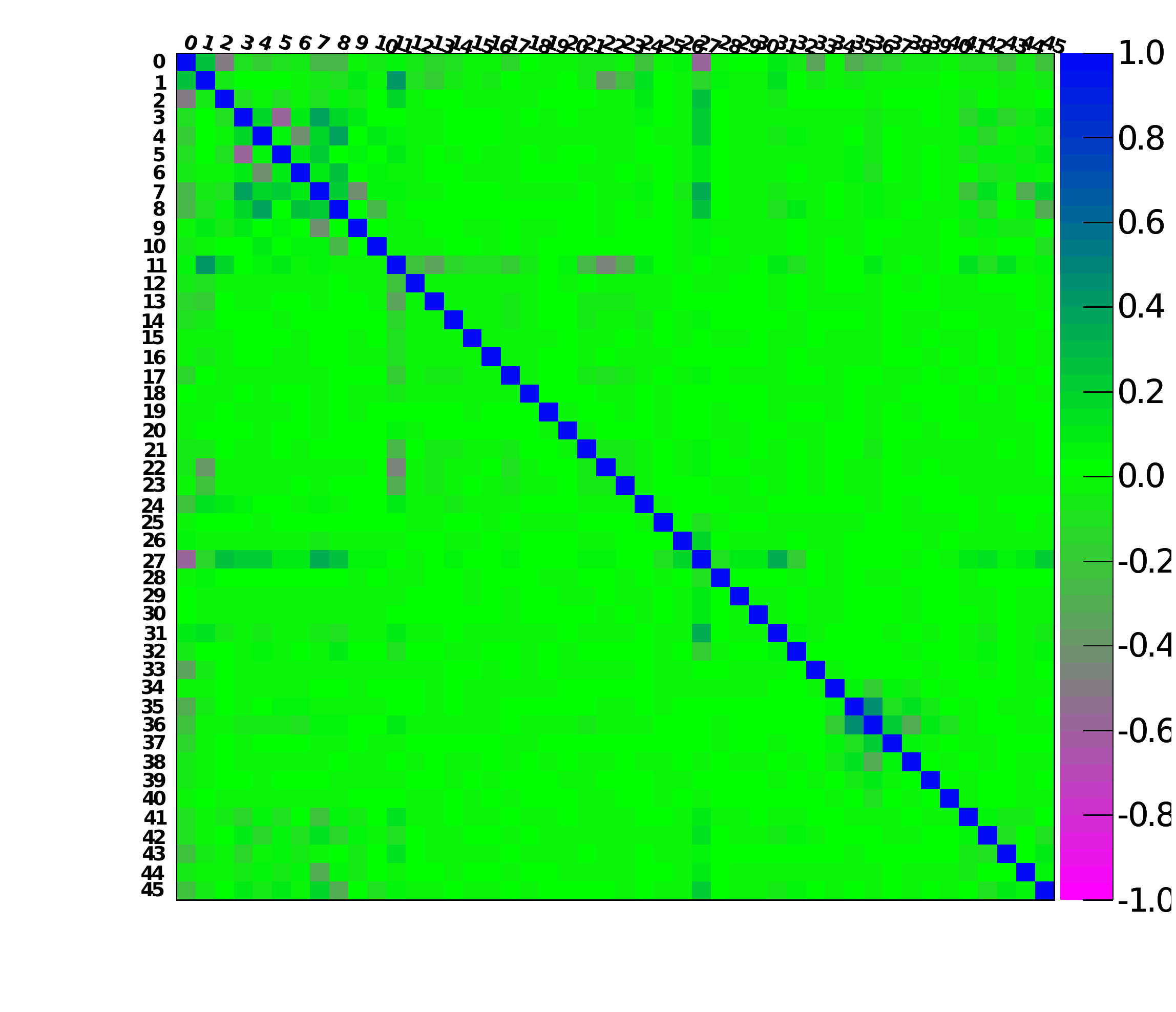}
\end{center}
\caption{Correlations between the fit parameters listed in Table \ref{tab:all_pars_comb} for the full combination of both analysers.}
\label{fig:par_corr_matrix_comb}
\end{figure} 
Good examples are the anticorrelation between the dominating \btag\ SF NP (27) and the yield as well as the jet multiplicity correction NP (11) and the JES components (12-22).

\section{Discussion of Uncertainties}
\label{sec:unc_discussion}
Uncertainties play a crucial role in this measurement. They limit the precision of the result and and can give a clear hint to further improvements. This section concludes the chapter of results by discussing the dominating uncertainties and explaining their effects.

\subsection{Dominating Uncertainties}
\label{sec:dom_unc}
\subsubsection{Uncertainties From Ensemble Testing}
A summary of all uncertainties evaluated via ensemble tests is listed in Table \ref{tab:sys_ensembles}. The dominating uncertainties are the renormalization/factorization scale, the top quark \pt\ (all affecting both \dQ\ and \bQ\ analysers), the PDFs, as well as the parton showering and the initial and final state radiation (affecting the \bQ\ a lot more than the \dQ). 

All these uncertainties affect the kinematic configuration of the \ttbar\ pair and the spin analysers. Hence, the impact of the measured spin correlation is expected to be large. This is confirmed in both the measurements of CMS \cite{CMS_spin_prelim, CMS_spin_paper} and ATLAS \cite{ueberpaper}. The PDF uncertainty can be highlighted as it affects not only the kinematics but also the initial state composition and the production mechanism. The relation of gluon fusion to quark/antiquark annihilation directly changes the spin configuration. In Figure \ref{fig:gluon_PDF} the effect of varied PDFs is illustrated. Two default PDF sets (CT10 and HERAPDF) are compared as well as their spread due to evaluation of the error sets. Both sets are plotted at the scale of the top quark mass ($Q^2 = m_t^2$).

\begin{figure}[htbp]
\begin{center}
			\subfigure[]{
\includegraphics[width=0.45\textwidth]{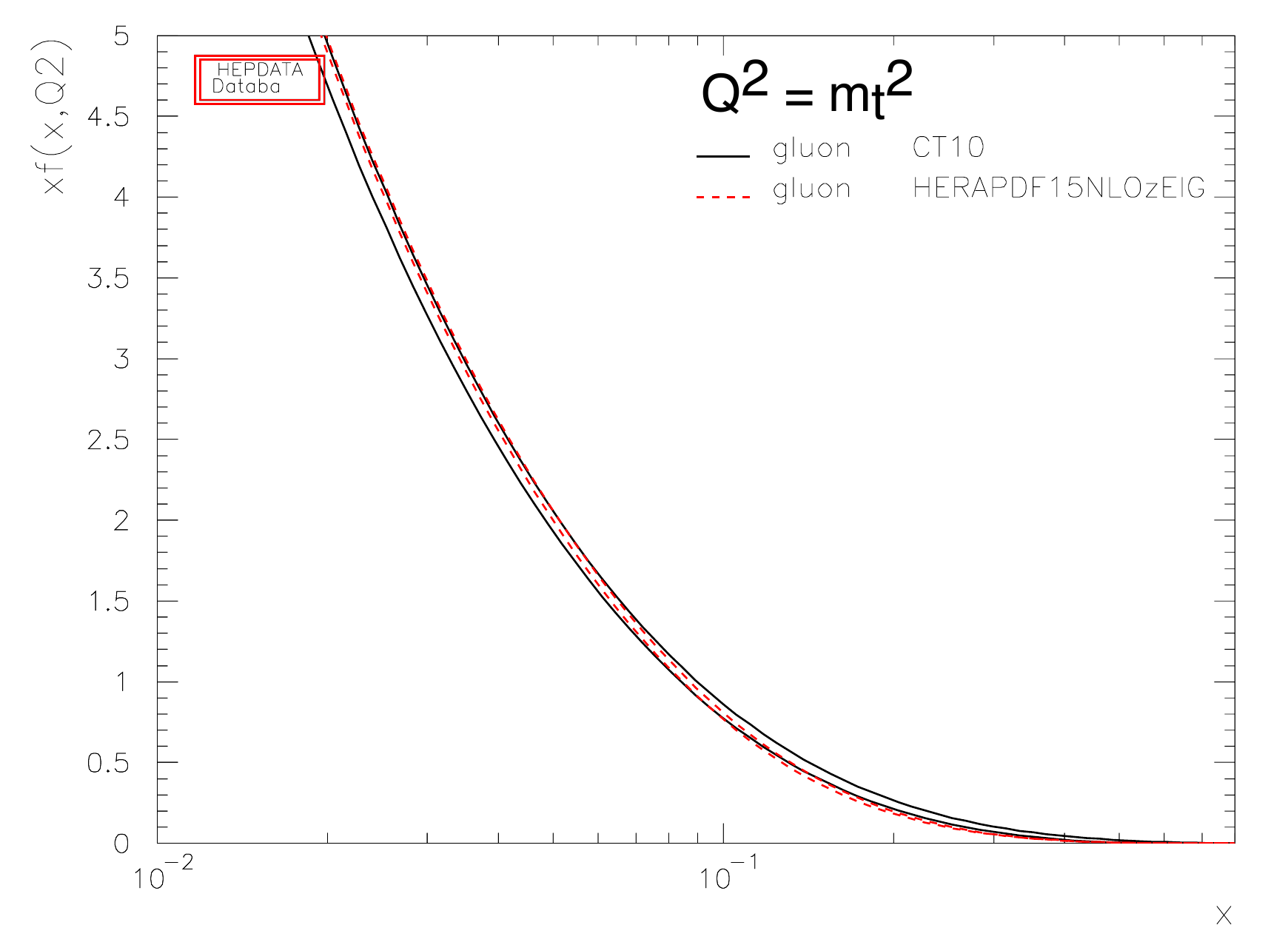}
\label{fig:pdf_gluon}
}
			\subfigure[]{
\includegraphics[width=0.45\textwidth]{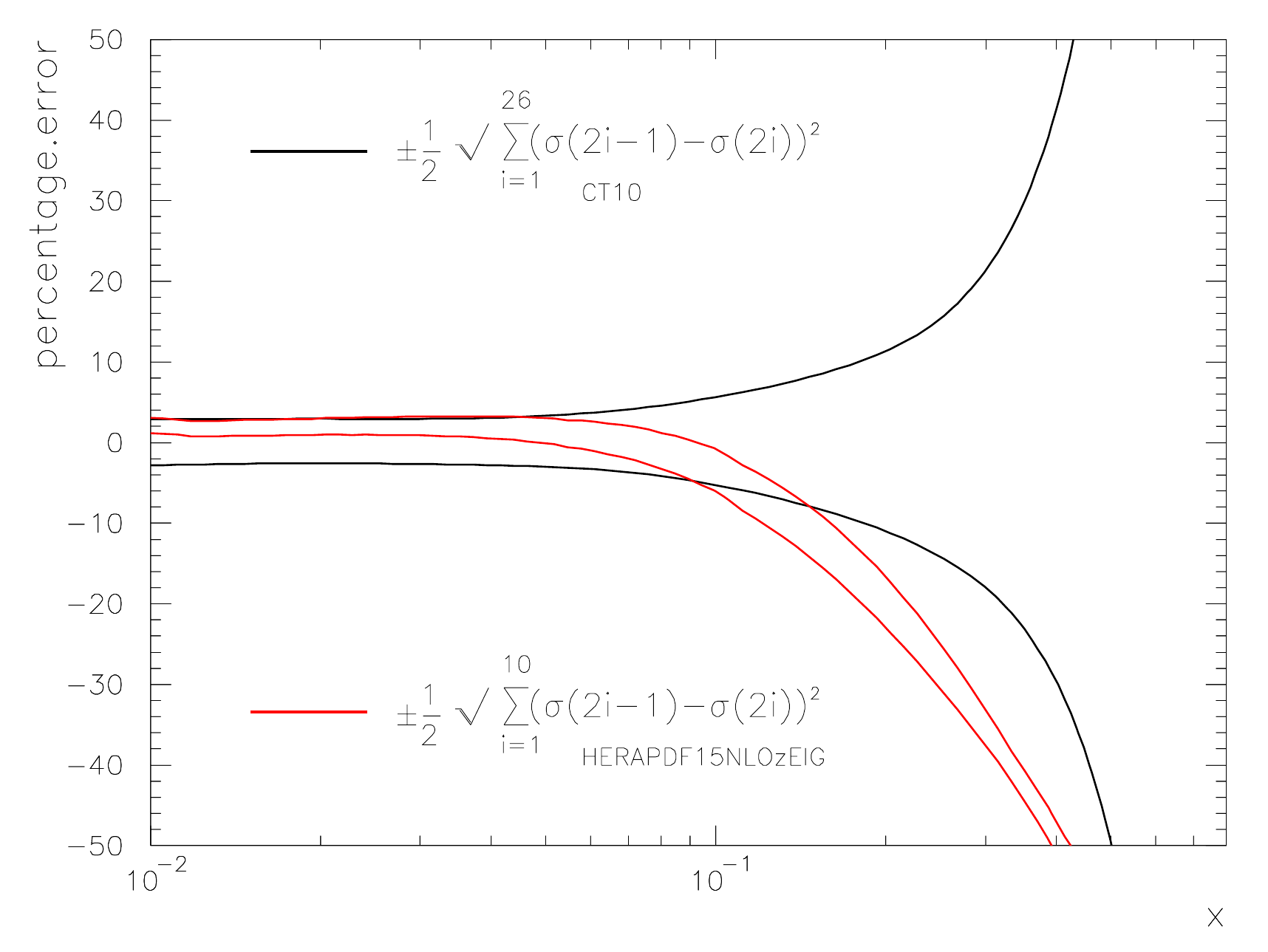}
\label{fig:pdf_gluon_unc}
}
\end{center}
\caption{\subref{fig:pdf_gluon} PDF distribution as a function of the momentum fraction $x$ for gluons. For both the CT10 and the HERPDF set the variations within the error sets are indicated by the two lines. \subref{fig:pdf_gluon_unc} Relative deviations to the central value of the CT10, caused by the variations of the CT10 and the HERAPDF error sets.}
\label{fig:gluon_PDF}
\end{figure} 

Two of the uncertainties should be emphasized as they have large effects which do not cancel in the combination. The first one is the initial / final state radiation. As seen in Table \ref{tab:sys_ensembles}, the \bQ\ is affected to much larger extent. In Figure \ref{fig:ISRFSR_comp} the effect on the ISR/FSR variation on the \dphi\ distributions is shown.

\begin{figure}[htbp]
\begin{center}
			\subfigure[]{
\includegraphics[width=0.45\textwidth]{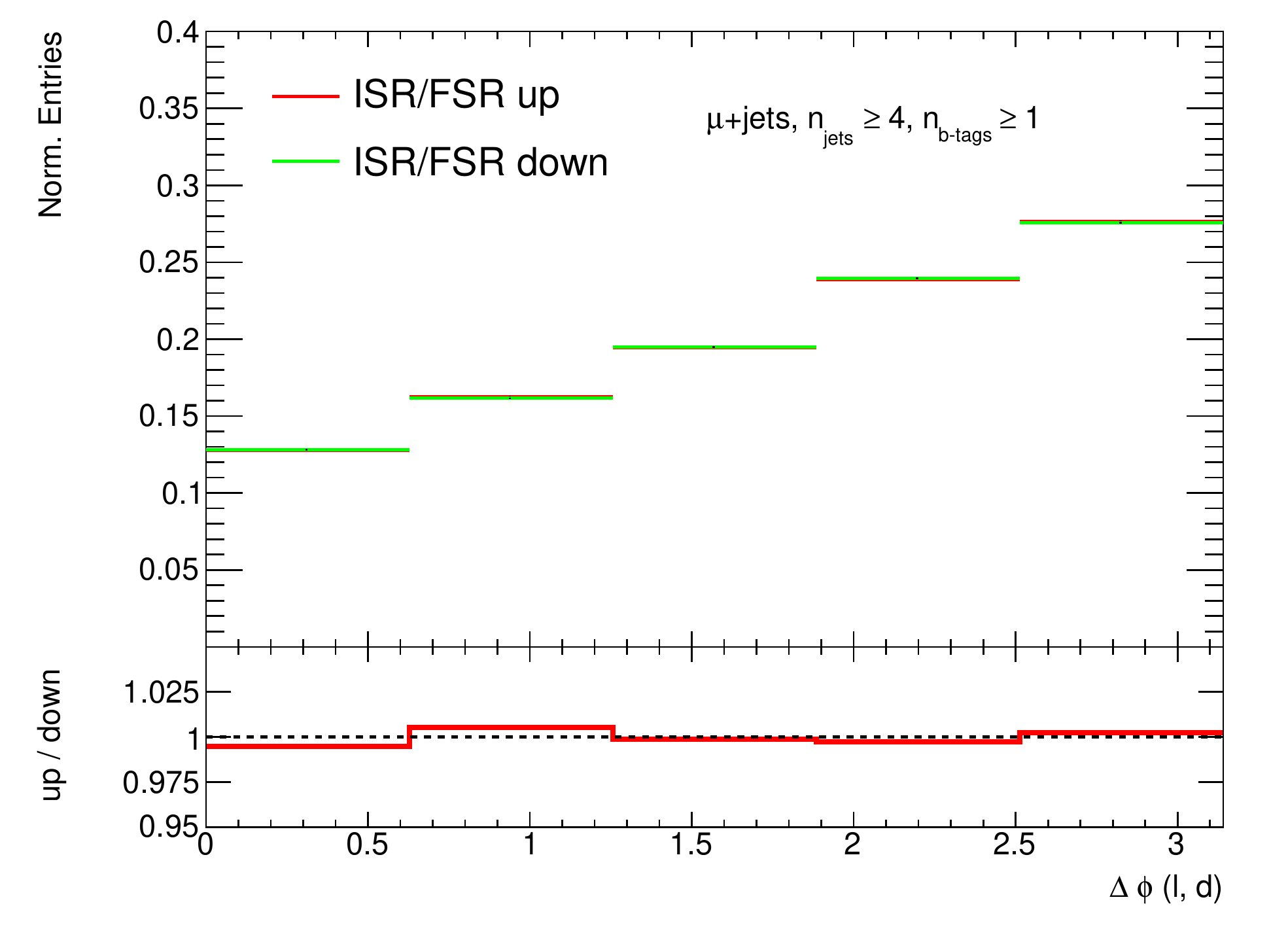}
\label{fig:ISRFSR_dQ}
}
			\subfigure[]{
\includegraphics[width=0.45\textwidth]{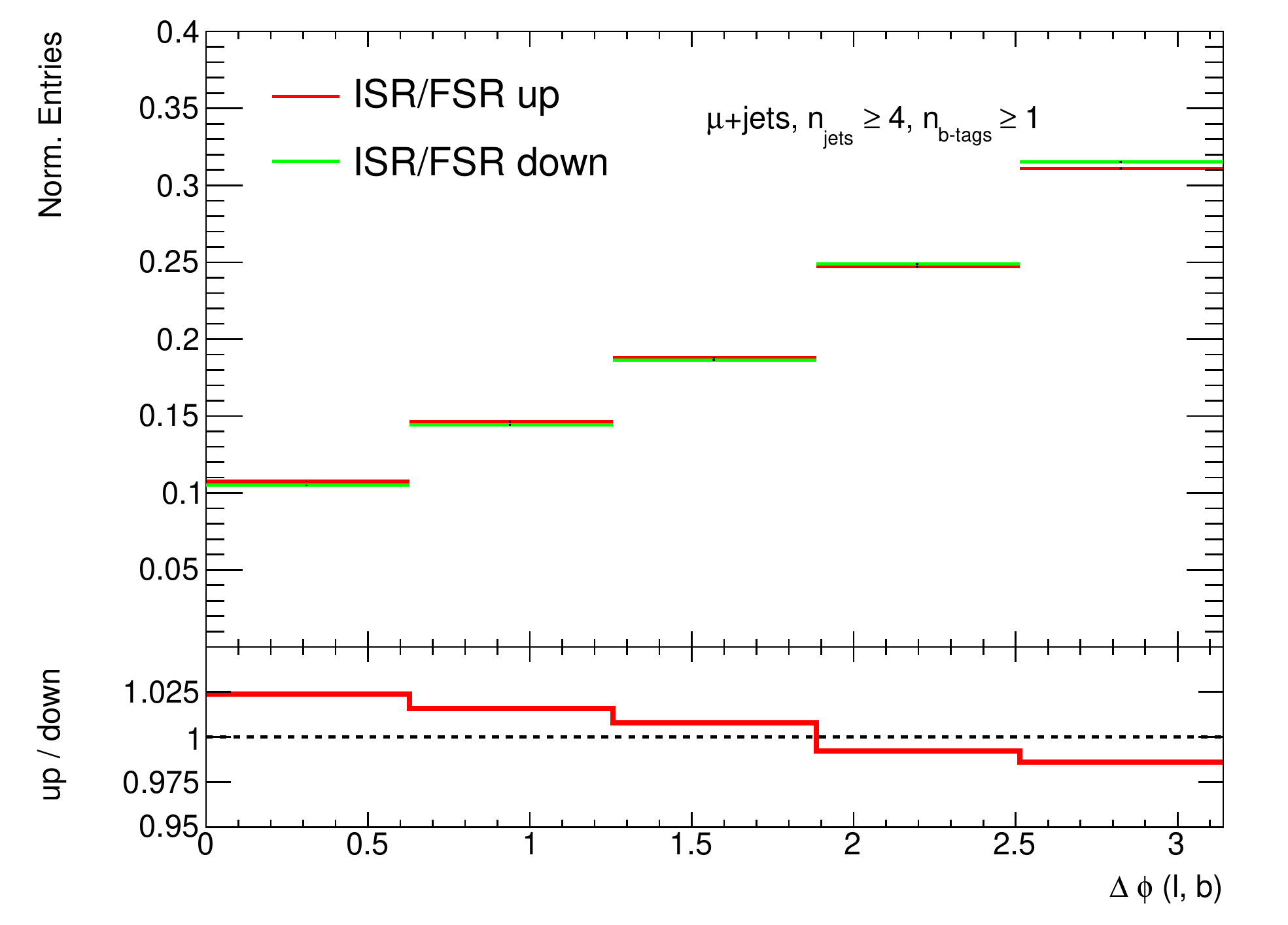}
\label{fig:ISRFSR_bQ}
}
\end{center}
\caption{\subref{fig:ISRFSR_dQ} \dphidQ\ distribution in the \mujets\ channel for the samples with increased and decreased initial and final state radiation. \subref{fig:ISRFSR_bQ} $\Delta \phi (l, b)$ distribution in the \mujets\ channel for the samples with increased and decreased initial and final state radiation.}
\label{fig:ISRFSR_comp}
\end{figure} 

 The \dQ\ is not affected while the \bQ\ shows a slope in the ISR/FSR up/down ratio. This slope is interpreted by the fit as a deviation in the spin correlation and leads to a large uncertainty. It is expected that the \bQ\ is affected by the FSR to a much larger extent. The reason is the larger phase space available for FSR radiation due to the \bQ 's larger \pt\ (see Figure \ref{fig:klfitter_control2}). 

Next to ISR/FSR, the modelling of the parton shower has a large impact on the measured \fsm\ using the \bQ. The compared showering generators, \herwig\ and \pythia, base on different showering models (cluster fragmentation vs. string model). Not only kinematics are affected, but also the flavour composition of the $b$-jets. Figure \ref{fig:btag_shower} shows the number of \btag ged jets for \powheg+\herwig\ and \powheg+\pythia. 
\begin{figure}[htbp]
\begin{center}

\includegraphics[width=0.6\textwidth]{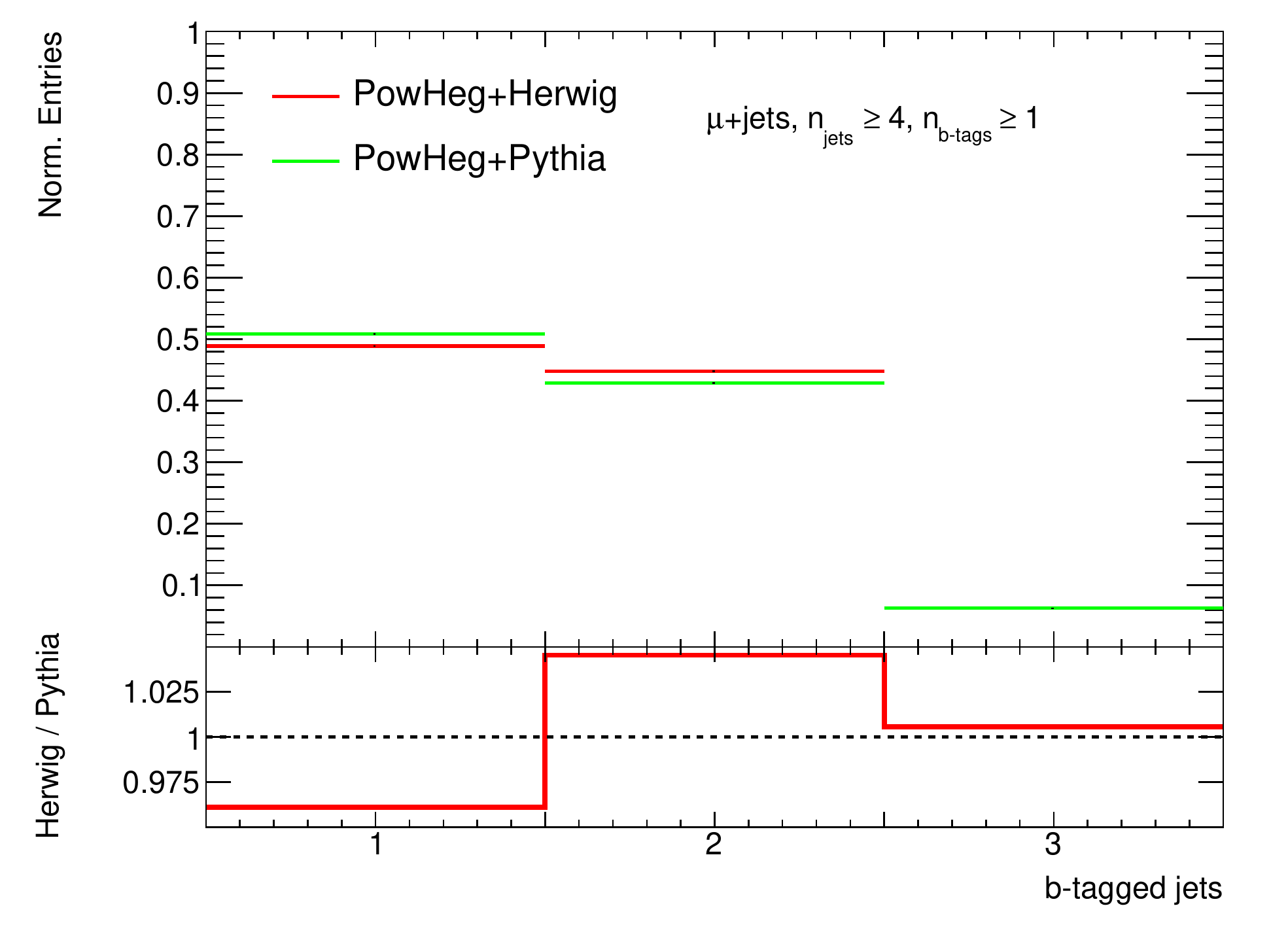}
\end{center}
\caption{Number of \btag ged jets for \powheg+\pythia\ and \powheg+\herwig\ in the \mujets\ channel. Events containing $\tau$ leptons were vetoed and both samples were reweighted to the same top quark \pt\ spectrum.}
\label{fig:btag_shower}
\end{figure} 
A clear difference is visible. For this plot, the distributions of the different generators are reweighted to the same top quark \pt\ spectrum. Also, $\tau$ leptons are vetoed as their polarization was not properly handled by the generators.

\subsubsection{Uncertainties From Nuisance Parameters}
There are different ways of classifying the most significant nuisance parameters. The five most significant NPs are presented. In the first type of ranking, shown in Table \ref{tab:top5_NPs_fsm_comb}, the NPs with the largest effect on the measured value of \fsm\ are listed for the full combination fit. The ranking is created by performing the fit with all NPs included. After that, each NP under test is taken out of the fit. The different \fsm\ values are compared. The sign of the value represents the relative change when taking the nuisance parameter out of the fit.
\\
\begin{table}[htbp]
\begin{center}
\begin{tabular}{|c|c|}
\hline
NP &  relative change of \fsm\ \\
\hline
\hline
JES/BJES  & $+2.4$\,\%\\
JES/EffectiveNP\_Stat1  & $+1.7$\,\%\\
JES/EffectiveNP\_Model1  & $+1.6$\,\%\\
btag/break8  & +1.2\,\%\\
JES/EffectiveNP\_Det1  & $- 1.0$\,\%\\
\hline
\end{tabular}
\end{center}
\caption{Most significant nuisance parameters (in terms of change of \fsm) for the full combination fit. }
\label{tab:top5_NPs_fsm_comb}
\end{table}

Another ranking can be created by evaluating the effect on the total uncertainty, not the central value. Table  \ref{tab:top5_NPs_width_comb} shows the NPs with the largest effect on the total uncertainty (which might become either larger or smaller) for the full combination fit. 
\begin{table}[htbp]
\begin{center}
\begin{tabular}{|c|c|}
\hline
NP &  relative change of $\Delta \fsm$ \\
\hline
\hline
JES/FlavorComp & $- 7.6$\,\%\\
btag/break8 & $- 4.6$\,\%\\
JES/FlavorResponse & $- 3.4$\,\%\\
JES/EffectiveNP\_Det1 & $+ 1.5$\,\%\\
JES/EffectiveNP\_Model1 &$- 1.4$\,\%\\
\hline
\end{tabular}
\end{center}
\caption{Most significant nuisance parameters (in terms of change of the fit uncertainty) for the full combination fit.}
\label{tab:top5_NPs_width_comb}
\end{table}
The most significant uncertainties for the individual combinations of the \dQ\ and \bQ\ analysers can be found in the Appendix  \ref{sec:app_sigunc}.

As the measurement depends a lot on the modelling of jets, the large contribution of the JES components is expected. The large impact of the \btag ging uncertainty is a consequence of the utilization of the $b$-jets as analysers as well as of the dependence of the \dQ\ reconstruction on the \btag\ weight. 

\subsection{Expected Deviations}
The title of this section might be misleading. In case a deviation is really expected, it can be calibrated. To allow for reweightings and calibrations, the preceding measurements must have sufficiently high precision. Until changes in the top quark modelling are established, indications can be noticed. Such indications are listed in the following, concluding this chapter. The question is: Where did independent measurements indicate a preference of the data to a different modelling than the one implemented in this analysis? And if such deviations are observed: What would be the effect on the current measurement? Would it cause further tension between the \dQ\ and the \bQ\ analysers? Or would it bring the results closer together?
 
\subsubsection{Top Quark \pt}
As shown in Figure \ref{fig:toppt_xsec_vs_pt}, the top quark \pt\ seems to be modelled imperfectly in \mcatnlo. The data prefers a softer \pt\ spectrum, shifted to lower values. As shown in Figure \ref{fig:dphi_topptslices}, this would lead to a flatter \dphi\ distribution. The fit templates based on a harder top \pt\ spectrum will interpret this as a higher spin correlation for the \dQ\ and a lower spin correlation for the \bQ, respectively (see Figure \ref{fig:dphi_truth}). 
This is exactly what was observed in data by measuring \fsm: For the \dQ, the measured \fsm\ is higher than the SM prediction and for the \bQ\ it is lower. 

\subsubsection{PDF}
During the discussion of the PDF uncertainties their large impact and opposite effect on \dQ\ and \bQ\ analysers was stressed (Section \ref{sec:unc_PDF}). The question is: Does the data have a preferred PDF? In the ATLAS measurement of the differential top quark cross section \cite{top_xsec_diff_ATLAS_conf} the impact of the PDF on the top quark \pt\ is checked. As shown in Figure \ref{fig:toppt_PDF}, HERAPDF is preferred by the data. In particular, this is the case for large values of top quark \pt. Furthermore, the worst modelling seems to be given by the CT10 PDF. This motivated to check how HERAPDF would affect the measured values of \fsm.
To answer this question, the LHAPDF reweighting was repeated using HERAPDF. The results are shown in Figure  \ref{fig:herapdfcheck}. 
 \begin{figure}[htbp]
 	\centering
			\subfigure[]{
		\includegraphics[width=0.45\textwidth]{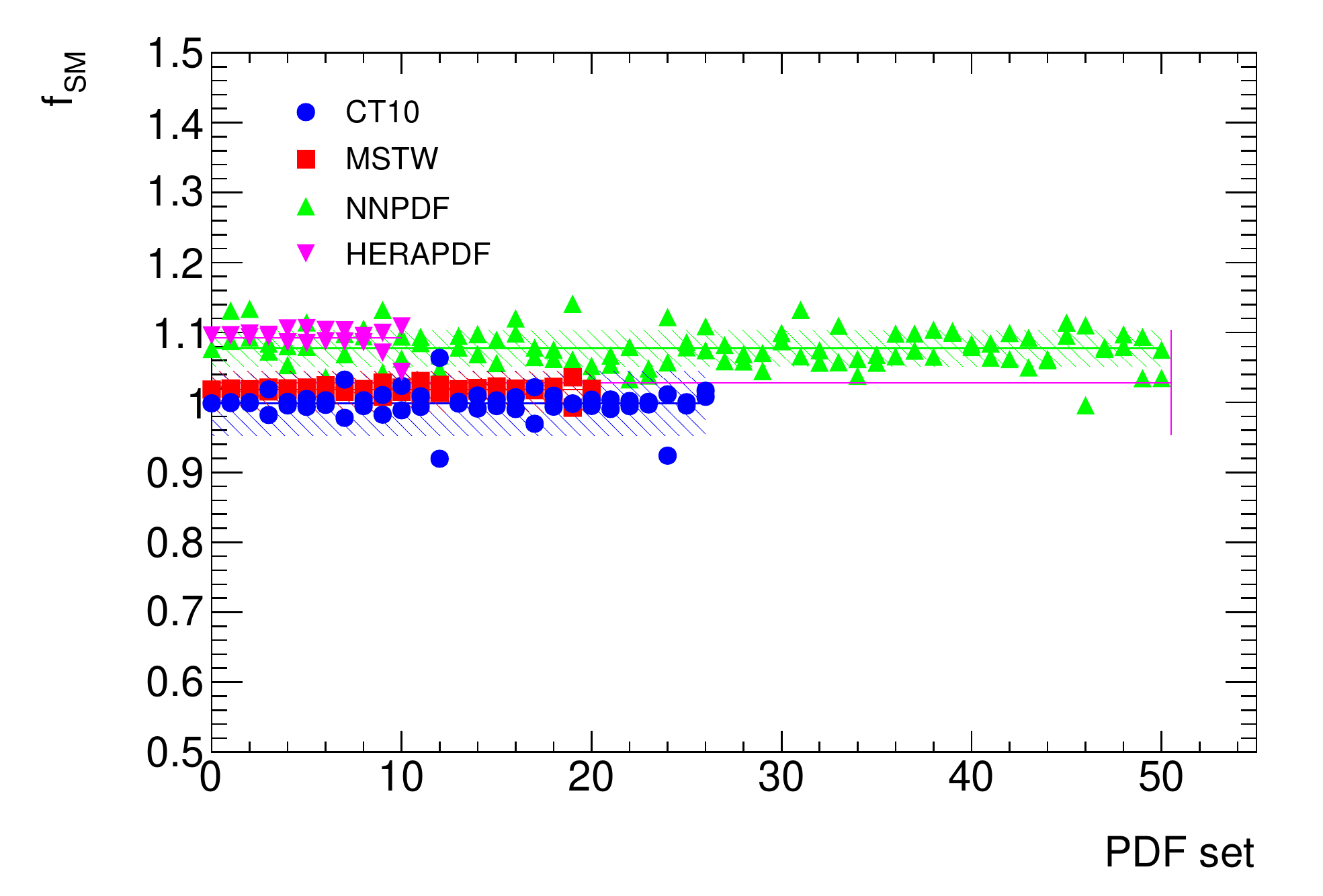}
			\label{fig:herapdfcheck_dQ}
		}
					\subfigure[]{
		\includegraphics[width=0.45\textwidth]{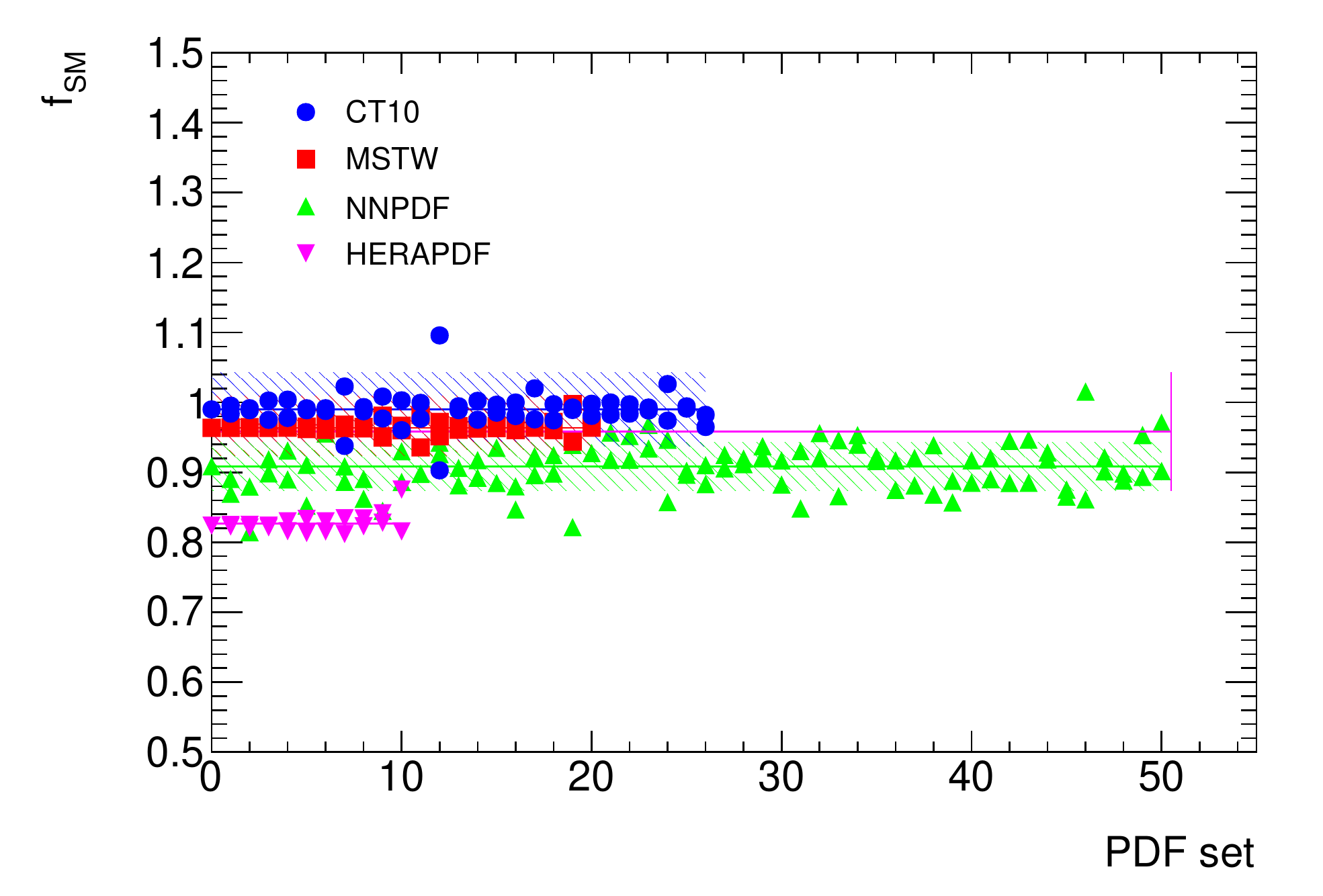}
			\label{fig:herapdfcheck_bQ}
		}		
\caption{Fit results for \fsm\ using pseudo data reweighted to different PDF sets and error sets. \subref{fig:herapdfcheck_dQ} Combined fit using the \dQ. \subref{fig:herapdfcheck_bQ} Combined fit using the \bQ.}
\label{fig:herapdfcheck}
\end{figure}
The shown fit values correspond to pseudo data created with distributions that are reweighted to a modified PDF.

In case the data was modelled with HERAPDF, a larger value of \fsm\ was fitted for the \dQ\ and a smaller value for the \bQ. This means that if the data prefers HERAPDF -- and the indications for that were shown in Figure \ref{fig:toppt_PDF} -- \fsm\ is expected to be fitted with $\fsm > 1.0$ for the \dQ\ and with $\fsm < 1.0$ for the \bQ. Indeed, this is what is measured.

\subsubsection{Generator Variation}

Concerning the uncertainties assigned to the \ttbar\ generator, effects coming from the parton showering, the renormalization/factorization scale, the underlying event, the ISR/FSR variation and the colour reconnection are evaluated. There has been no direct comparison of \mcatnlo\ to other \ttbar\ generators. The reasons are the following:
\begin{itemize}
\item The spin correlation is different for LO and NLO generators. Thus, only NLO generators should be used for comparison. 
\item Samples using \powheg+\herwig\ suffer from a bug in the $\tau$ lepton polarization.
\item All available \powheg\ samples suffer from an additional bug concerning spin correlation in the antiquark-gluon and gluon-antiquark production channel.
\item Samples with uncorrelated \ttbar\ spins are not available for any generator other than \mcatnlo.
\end{itemize}

However, it should be mentioned that fitting pseudo data created using \powheg+\herwig\ leads to values of \fsm\ deviating from the expectation of \fsm = 1.0. The results are shown in Table \ref{tab:comp_powheg}. 

One of the main differences between \powheg+\herwig\ and \mcatnlo\ is the top quark \pt\ spectrum.  Hence, \powheg+\herwig\ is reweighted to match the top \pt\ spectrum of \mcatnlo.

\begin{table}[htbp]
\begin{center}
\begin{tabular}{|c|c|c|c|}
\hline
Sample &  \multicolumn{3}{c|}{ \fsm } \\
{} &  \dQ\ & \bQ\ & Combination \\
\hline
\hline
\mcatnlo\ & 1.00 & 0.99 & 1.00\\
\powheg+\herwig\ (nominal) & 1.26 & 0.64 & 1.02\\
\powheg+\herwig\ (top \pt\ reweighted) & 1.15 & 0.73 & 0.99\\
\hline
\end{tabular}
\end{center}
\caption{Results from ensemble tests for pseudo data created with \mcatnlo, \powheg+\herwig\ (nominal) and \powheg+\herwig\ reweighted to the top \pt\ spectrum of \mcatnlo.}
\label{tab:comp_powheg}
\end{table}
Reweighting the top \pt\ spectrum reduces the deviations to the expected fit values of $\fsm=1.0$, but does not fully remove them. One aspect is the non-perfect reweighting. Reweighting techniques are not expected to replace event generations with a modified modelling. Furthermore, the \powheg+\herwig\ sample is known to suffer from bugs affecting the $\tau$ polarization and a small part of the spin correlation. To eliminate the effects of these bugs and the effects coming from the reconstruction, an additional comparison on parton level is done. 

Figure \ref{fig:powheg_parton} shows the differences in the shape of the \dphi\ distributions for both generators. Additional comparisons of the \dphi\ distributions on parton level before and after reweighting in top quark \pt\ are shown in Appendix \ref{sec:app_dphi_generators}.

\begin{figure}[htbp]
\begin{center}
\subfigure[]{
\includegraphics[width=0.45\textwidth]{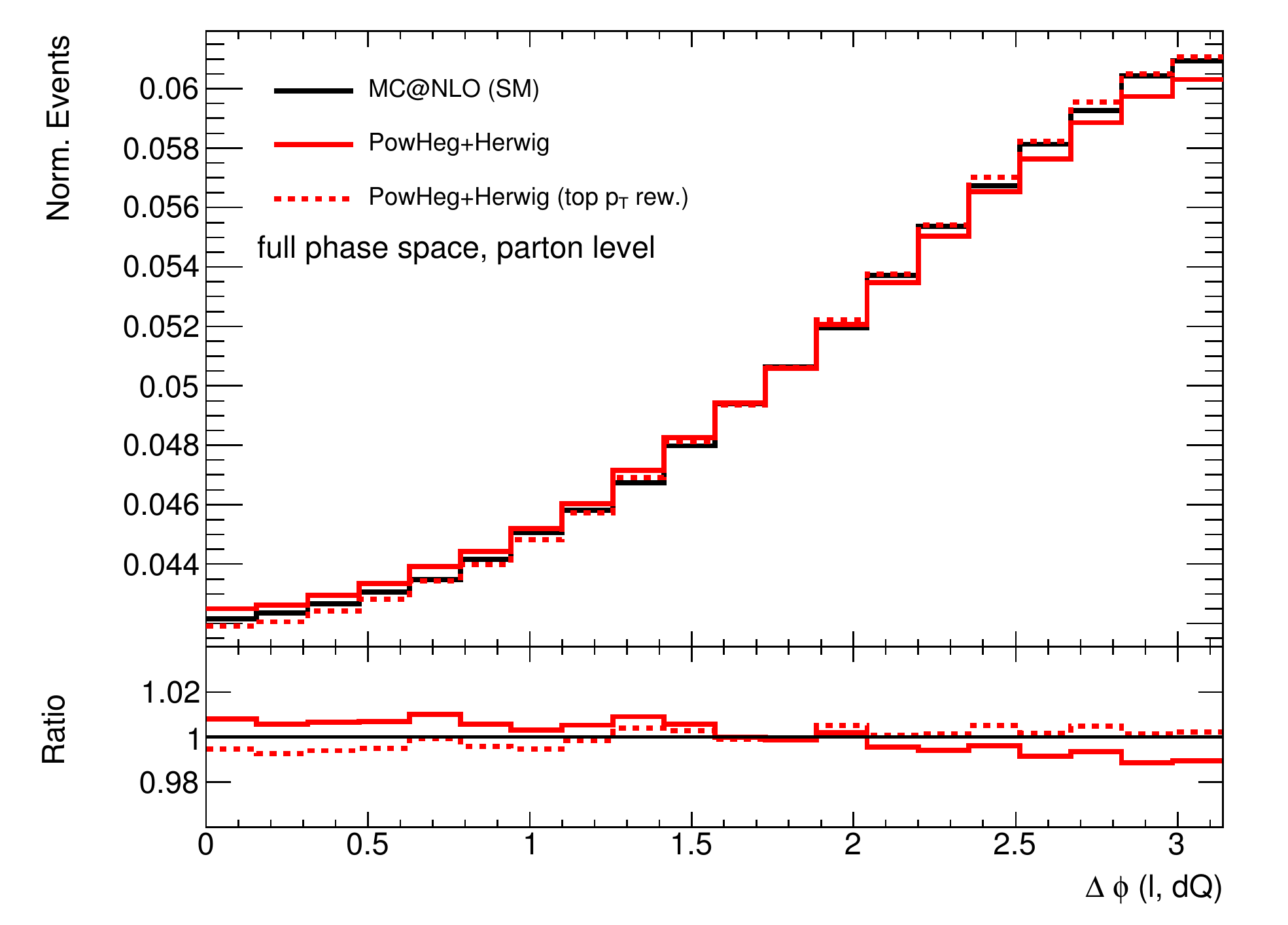}
\label{fig:powheg_parton_dQ}
}
\subfigure[]{
\includegraphics[width=0.45\textwidth]{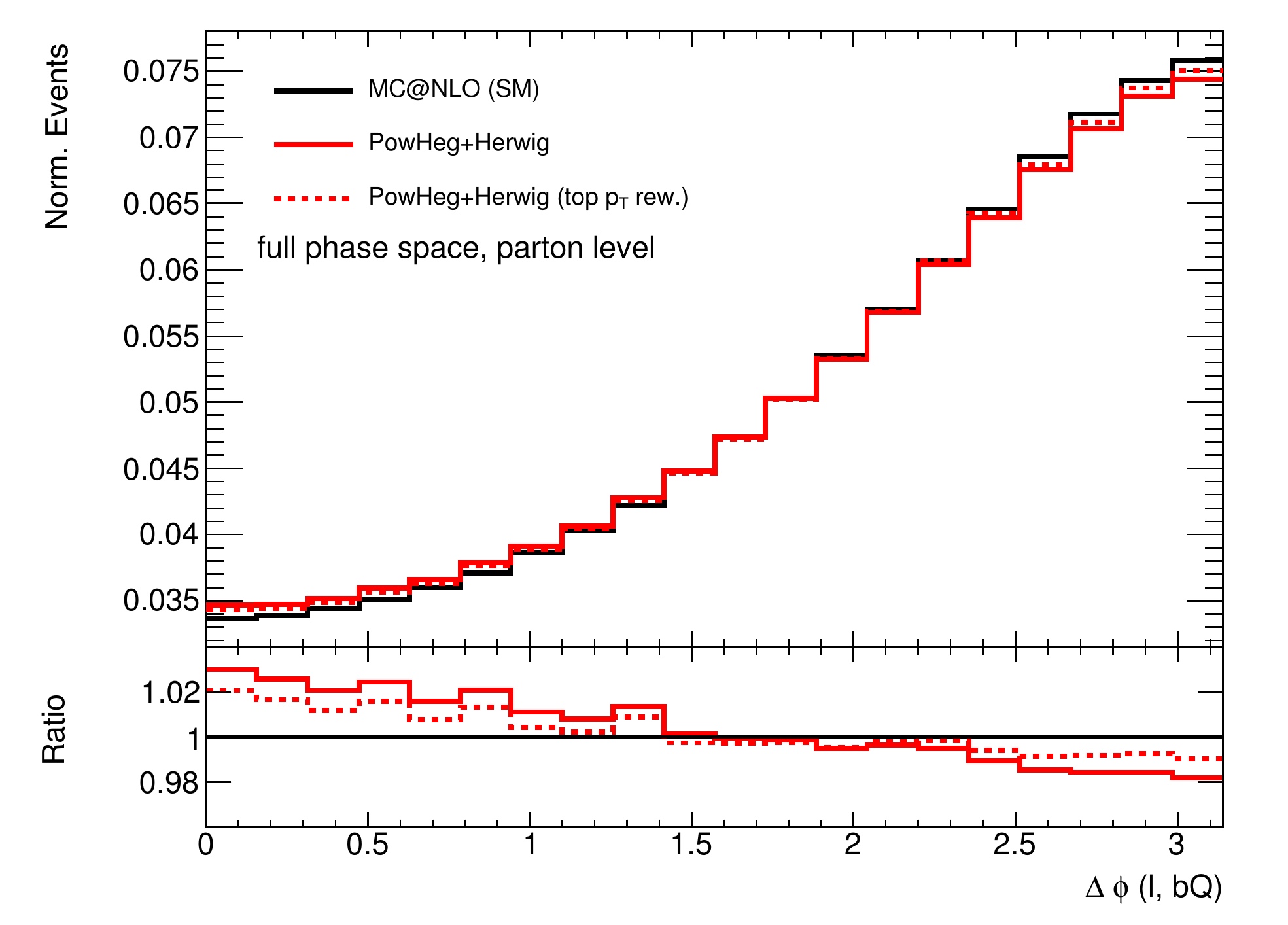}
\label{fig:powheg_parton_bQ}
}
\end{center}
\caption{Comparison of the \dphi\ distribution between \mcatnlo\ (black) and \powheg\ (red) using \subref{fig:powheg_parton_dQ} the \dQ\ and \subref{fig:powheg_parton_bQ} the \bQ\ as analyser. The dashed distribution shows the \powheg\ spectrum reweighted to match the top \pt\ distribution of \mcatnlo.}
\label{fig:powheg_parton}
\end{figure} 

The \dphi\ distribution is flatter for \powheg+\herwig. As shown in Figure \ref{fig:dphi_truth}, this results in different interpretations of a fitted \fsm: higher values of \fsm\ for the \dQ\ and lower ones for the \bQ\ analyser. Reweighting \powheg+\herwig\ to the top \pt\ spectrum of \mcatnlo\ leads to good agreement for the \dQ\ distribution, but not for the \bQ\ distribution. However, an improvement is observed. The reason for the remaining difference lies in the \bQ\ energy spectrum. This difference is still present after reweighting, even though the $W$ kinematics of the two generators do agree.

To conclude, \powheg+\herwig\ and \mcatnlo\ lead to different top \pt\ and \bQ\ energy spectra, both affecting the fitted value of \fsm. In the case that \powheg+\herwig\ is able to describe the data better than \mcatnlo\ -- and Figure \ref{fig:toppt_ratios} as well as the jet multiplicity distribution give indications for this assumption -- it could replace \mcatnlo\ for the fit to data. In this case the measured \fsm\ is expected to be fitted lower using the \dQ\ and higher using the \bQ. This would lead to a better compatibility of the \dQ\ and \bQ\ combinations.

\section{Spin Analyzer Consistency Checks}
The difference between the results of the \dQ\ and the \bQ\ combination immediately triggers the question whether the results are consistent. Two types of consistency are checked in this section: the one between the \ejets\ and the \mujets\ channel combinations as well as the one between the \dQ\ and the \bQ\ combinations.

The fit results for the lepton flavour comparison are shown in Table \ref{tab:lepflavcons} for a fit without nuisance parameters. The quoted uncertainties include the statistical and background normalization uncertainties.

\begin{table}[htbp]
\begin{center}
\begin{tabular}{|c|c|c|c|}
\hline
Lepton Channel & \dQ\ & \bQ\ & Combination \\
 \hline
 \hline
\ejets\ & $1.57 \pm 0.21$ & $0.44 \pm 0.29$ & $1.19 \pm 0.17$\\
\mujets\ & $1.61 \pm 0.19$& $0.34 \pm 0.23$ & $1.11 \pm 0.15$\\
Combination& $1.58 \pm 0.14$& $0.39 \pm 0.18$ & $1.14 \pm 0.11$ \\
\hline
\end{tabular}
\end{center}
\caption{Fit results of \fsm\ including statistical uncertainties. The fit has been split into lepton flavours for this cross check. }
\label{tab:lepflavcons}
\end{table} 
The results are in good agreement across the different lepton flavours. The consistency of the \dQ\ and the \bQ\ results are evaluated as well. It is important to consider the correlation between the spin analysers. Studying the effect of the top quark \pt, as done in Section \ref{sec:toppt_unc}, demonstrates the anticorrelation of certain uncertainties. Changes affecting the shape, which are common for both the \dQ\ and the \bQ\ channels, are interpreted differently in terms of spin correlation. 

The consistency check for the \dQ\ and the \bQ\ channels is done the following way:
\begin{itemize}
\item For  each uncertainty $i$, a random number $s_i$ according to a Gaussian distribution, centred at 0 with a width 1, is drawn. 
\item Each bin of each template of the signal and background predictions is modified by the relative change $r_i$ expected by the corresponding uncertainty $i$ multiplied with the random number $s_i$. For $s_i > 0$, the systematic up variation is taken and for $s_i < 0$, the down fluctuation.
\item After each systematic variation a Poissonian fluctuation of the template bins is applied on top to take the statistical uncertainty into account. 
\item Ensembles using the same variations of systematic effects are produced for all \dQ\ and \bQ\ templates. This ensures the correct propagation of the uncertainties' correlation to the spin analysers. 
\item The ensemble tests are performed and for each ensemble, \fsm\ (\dQ) is plotted against \fsm\ (\bQ).
\item The result from the fit on data is added to the two-dimensional distribution and compared to the spread of results from the ensemble tests.
\end{itemize}
The 100,000 ensembles were fitted without nuisance parameters. Hence, the result of \fsm\ from the fit to data, for which no nuisance parameters were used in the fit, is shown as a comparison.\\

By using only the statistical uncertainty, the nuisance parameter uncertainties, the renormalization/factorization scale uncertainty and the top \pt\ uncertainty, the results of the two analysers are already consistent within the 99.5\,\% confidence level interval as shown in Figure \ref{fig:CompCheck_all_NP_plusTopPt_plusScale}. The test was repeated without the top \pt\ uncertainty but with the uncertainty coming from ISR/FSR and PS/fragmentation instead. This result is shown in Figure \ref{fig:CompCheck_all_NP_plusScale_plusPS_plusISRFSR}. The compatibility of the two results for the single analysers is also confirmed using this set of uncertainties.
\begin{figure}[htbp]
\begin{center}
			\subfigure[]{
\includegraphics[width=0.45\textwidth]{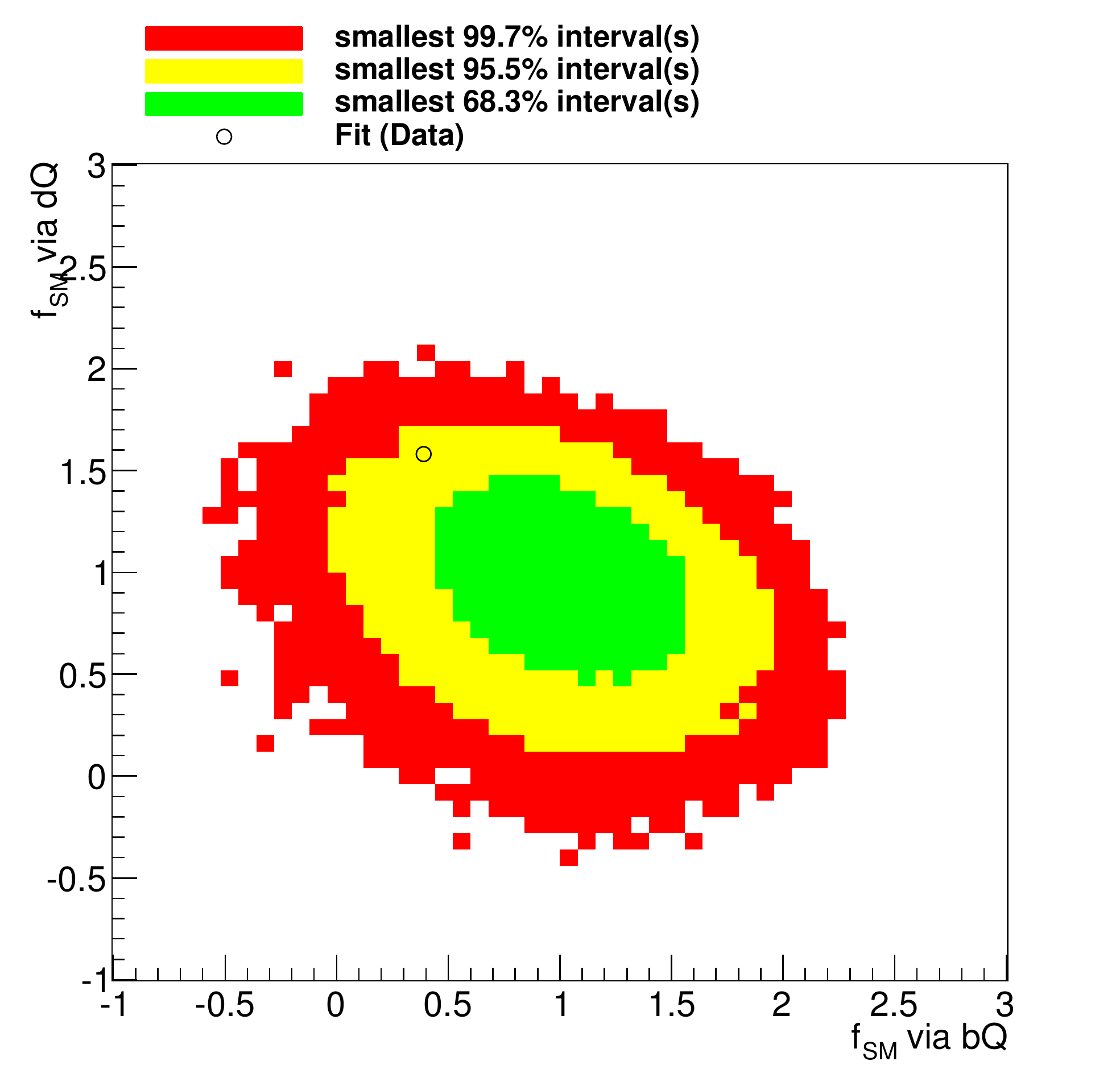}
\label{fig:CompCheck_all_NP_plusTopPt_plusScale}
}
			\subfigure[]{
\includegraphics[width=0.45\textwidth]{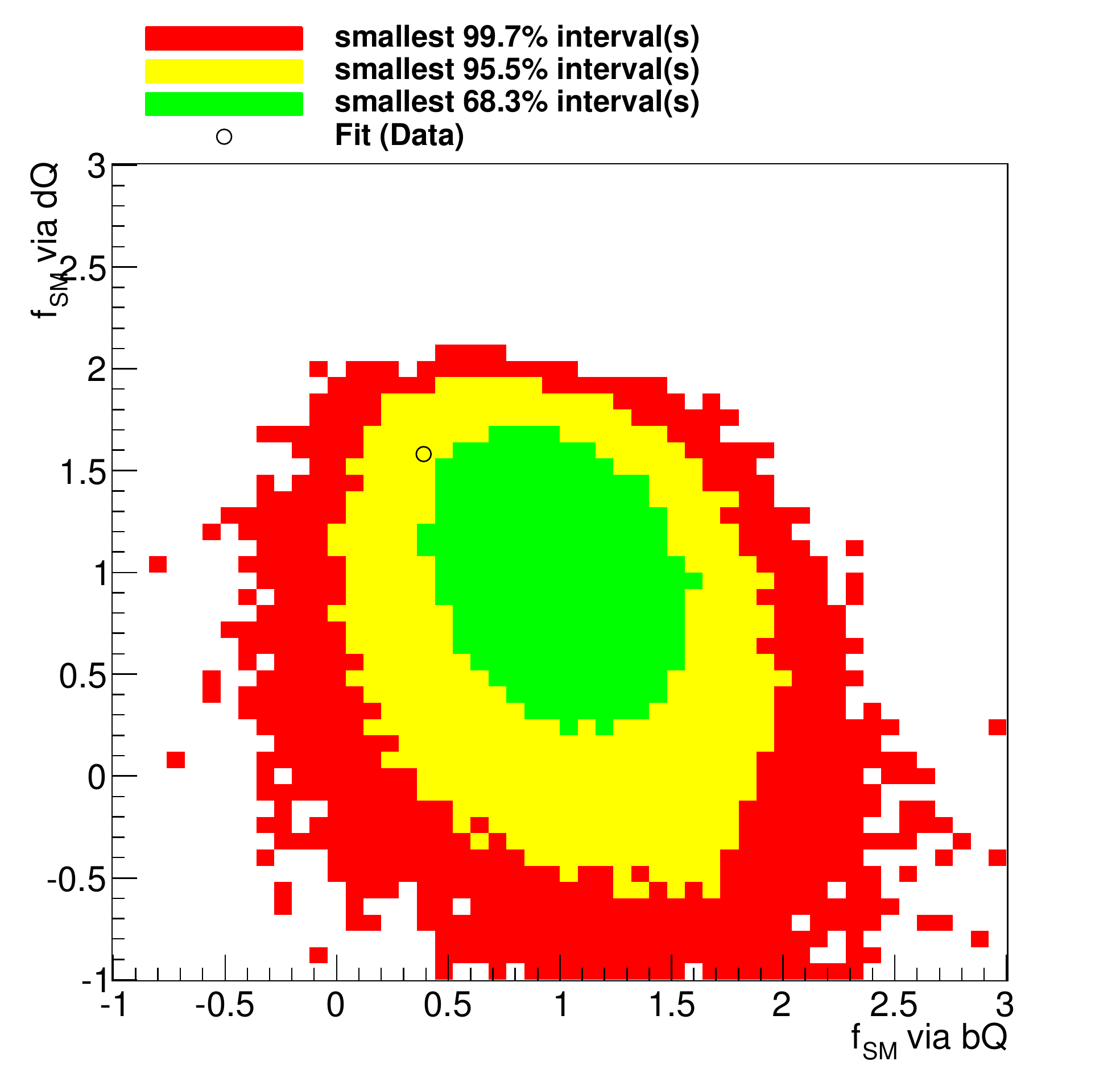}
\label{fig:CompCheck_all_NP_plusScale_plusPS_plusISRFSR}
}
\end{center}
\caption{\subref{fig:CompCheck_all_NP_plusTopPt_plusScale} Compatibility check for the results of the \dQ\ and the \bQ\ combination. Only statistical uncertainties, the nuisance parameter uncertainties, the renormalization/factorization scale and the top \pt\ were used as uncertainties (left). \subref{fig:CompCheck_all_NP_plusScale_plusPS_plusISRFSR} As a cross check the test was repeated without the top \pt\ uncertainty, but with ISR/FSR and PS/hadronisation uncertainty added.}
\label{fig:cons_check}
\end{figure}

\chapter[Summary, Conclusion and Outlook]{Summary, Conclusion and Outlook}
The aim to measure the \ttbar\ spin correlation at $\sqrt{s} = 7\,\TeV$ in the \ljets\ channel was ambitious. Hadronic spin analysers are hard to identify, in particular in events with high jet multiplicities, such as at the LHC. 

Motivated as a precision test of the Standard Model and a search for hints suggesting new physics, uncertainties needed to be kept low. 

This chapter concludes the thesis by presenting the results, comparing them to other measurements, and by drawing conclusions. Finally, a discussion about future measurements of \ttbar\ spin correlation provides ideas about what to do next. The presented results are interesting by themselves and give a glance on the spin properties of the top quark: Does it interact as a particle with a spin of $\frac{1}{2}$, produced by gluon fusion and quark/antiquark annihilation, decaying via the weak interaction before bound states can be formed? The answer is: yes.

The detailed studies of systematic effects, and in particular the comparison of the results that were measured to results that are expected by motivated changes in the top quark modelling, give a straight-forward recipe for a next-generation \ttbar\ spin correlation measurement.

\section{Summary of Results}
The \ttbar\ spin correlation was measured in the \ljets\ decay mode. By performing a template fit of the distributions of the azimuthal angle \dphi\ between the charged lepton and hadronic analysers, the degree of \ttbar\ spin correlation, as predicted by the SM, \fsm, was measured. 
Two different hadronic analysers were used: The \dQ\ and the \bQ. A kinematic fit was utilized to assign jets to the model partons, which induced them. To separate the up- and down-type quark jets from the hadronically decaying $W$ boson, \btag ging weight distributions and transverse momenta were utilized.
The data was split into events from the \ejets\ and from the \mujets\ channel, into jet multiplicity bins of $n_{\text{jets}} = 4$ and $n_{\text{jets}} \geq 5$ as well as into subsets for different numbers of jets tagged as b-jets ($n_{\text{b-tags}} = 1$ and $n_{\text{b-tags}} \geq 2$). 

All eight channels were fitted for both the \dQ\ and the \bQ. A combination of the analysers was also performed, constraining the systematic uncertainties to a large extent. The results obtained are
\begin{align*}
&\fsm ( \text{\dQ} ) &= &1.53 \pm 0.14\,\text{(stat.)} \pm 0.32\,\text{(syst.)}  \\ 
&\fsm ( \text{\bQ} ) &= &0.53 \pm 0.18\,\text{(stat.)} \pm 0.49\,\text{(syst.)}  \\ 
&\fsm (\text{comb.}) &= &1.12 \pm 0.11\,\text{(stat.)} \pm 0.22\,\text{(syst.)}  
\end{align*}

The results for both analysers are found to be consistent with the SM and with each other. This is possible due to the large systematic uncertainties and the asymmetric effects of the uncertainties on both analysers which are highly anti-correlated: while the \dQ\ will fit values with $\fsm > 1.0$, the \bQ\ will fit $\fsm < 1.0$ and vice versa. The combination of both analysers leads to a significant reduction of the uncertainties and to a better agreement with the SM. The result of the combined fit of the \dphi\ distributions is shown in Figure \ref{fig:postfit_stack}. 

\begin{figure}[htbp]
\begin{center}
\subfigure[]{
\includegraphics[width=0.45\textwidth]{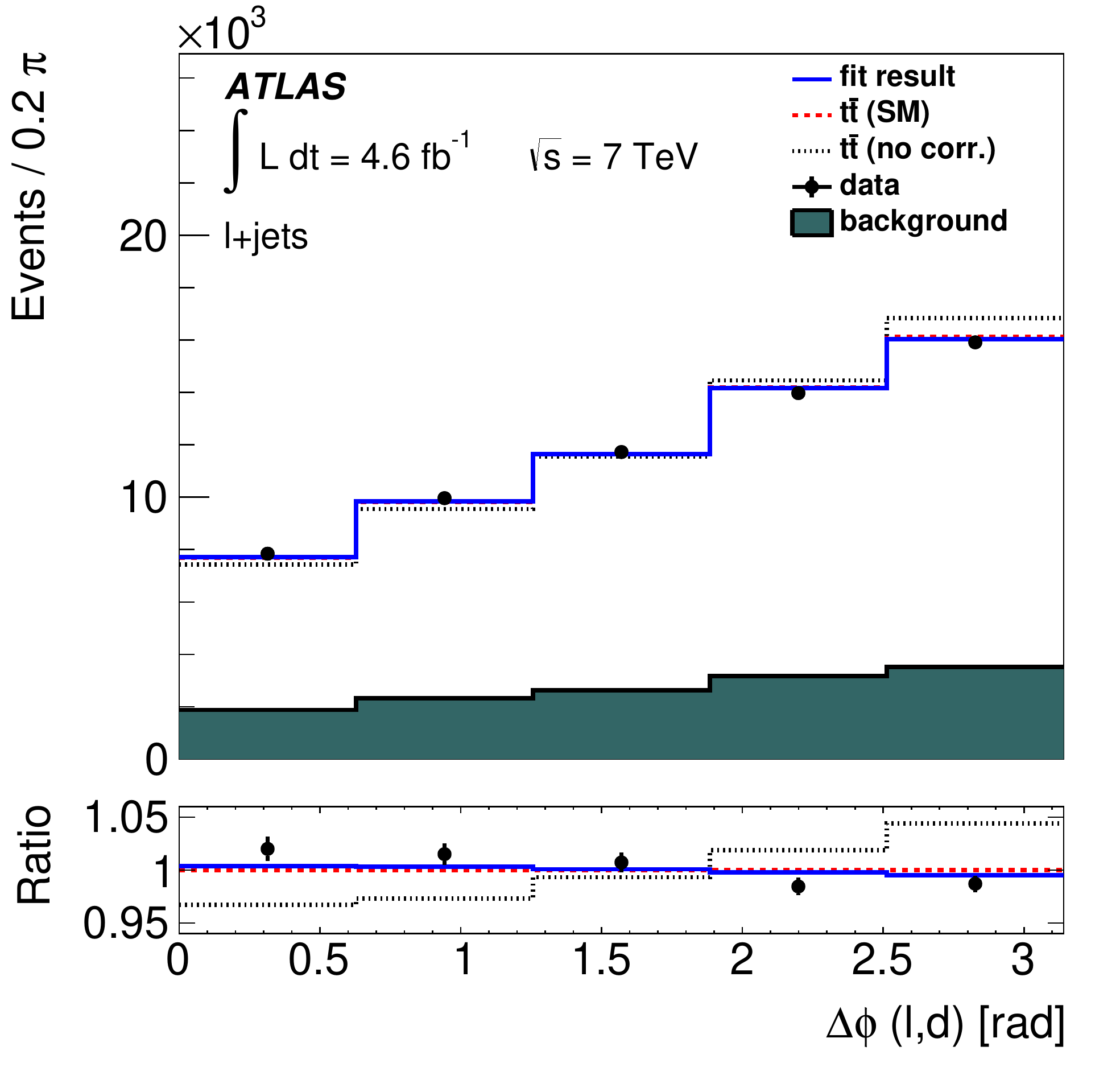}
\label{fig:postfit_stack_d}
}
\subfigure[]{
\includegraphics[width=0.45\textwidth]{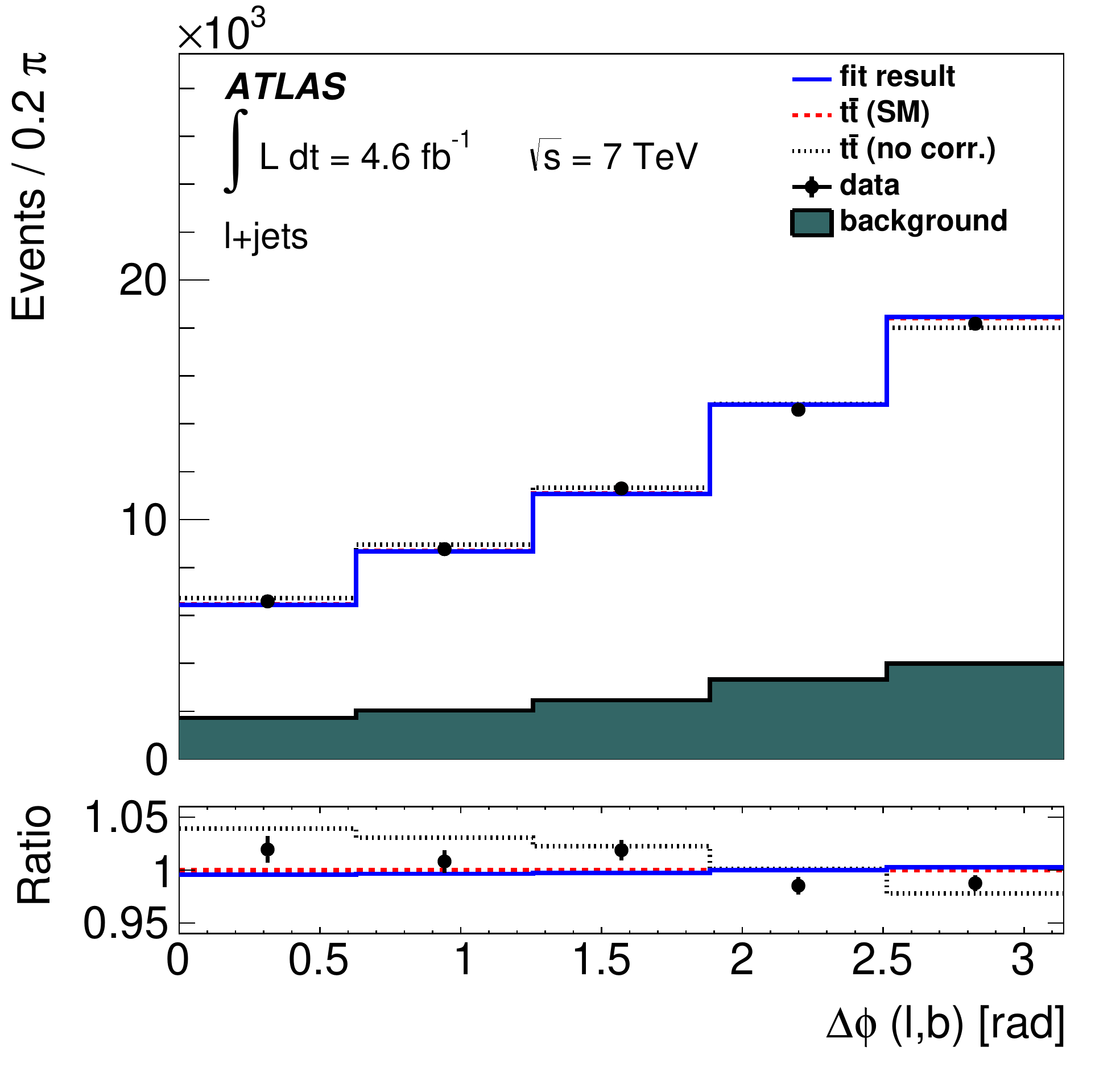}
\label{fig:postfit_stack_b}
}
\end{center}
\caption{Distributions of the stacked \subref{fig:postfit_stack_d} \dphidQ\ and \ref{fig:postfit_stack_b} \dphibQ\ distributions for the combined fit \cite{ueberpaper}. The result of the fit to data (blue) is compared to the templates for background plus \ttbar\ signal with SM spin correlation (red dashed) and without spin correlation (black dotted). The ratios of the data (black points), of the best fit (blue solid) and of the uncorrelated \ttbar\ prediction to the SM prediction are also shown.}
\label{fig:postfit_stack}
\end{figure}

To compare the results from this thesis to other measurements of \ttbar\ spin correlation, other measurement's results were transformed into \fsm\ by dividing the measured spin correlation by the SM expectation. An overall summary is given in Figure \ref{fig:megasummary}. It includes results from \cite{CDF_spin_ljets, CDF_spin_dilep, D0_spin_ljets, D0_spin_dilep_angles, D0_spin_dilep_MEM, ueberpaper, CMS_spin_paper}, using different \ttbar\ decay modes, observables of \com\ energies.

\begin{figure}[htbp]
\begin{center}

\includegraphics[width=\textwidth]{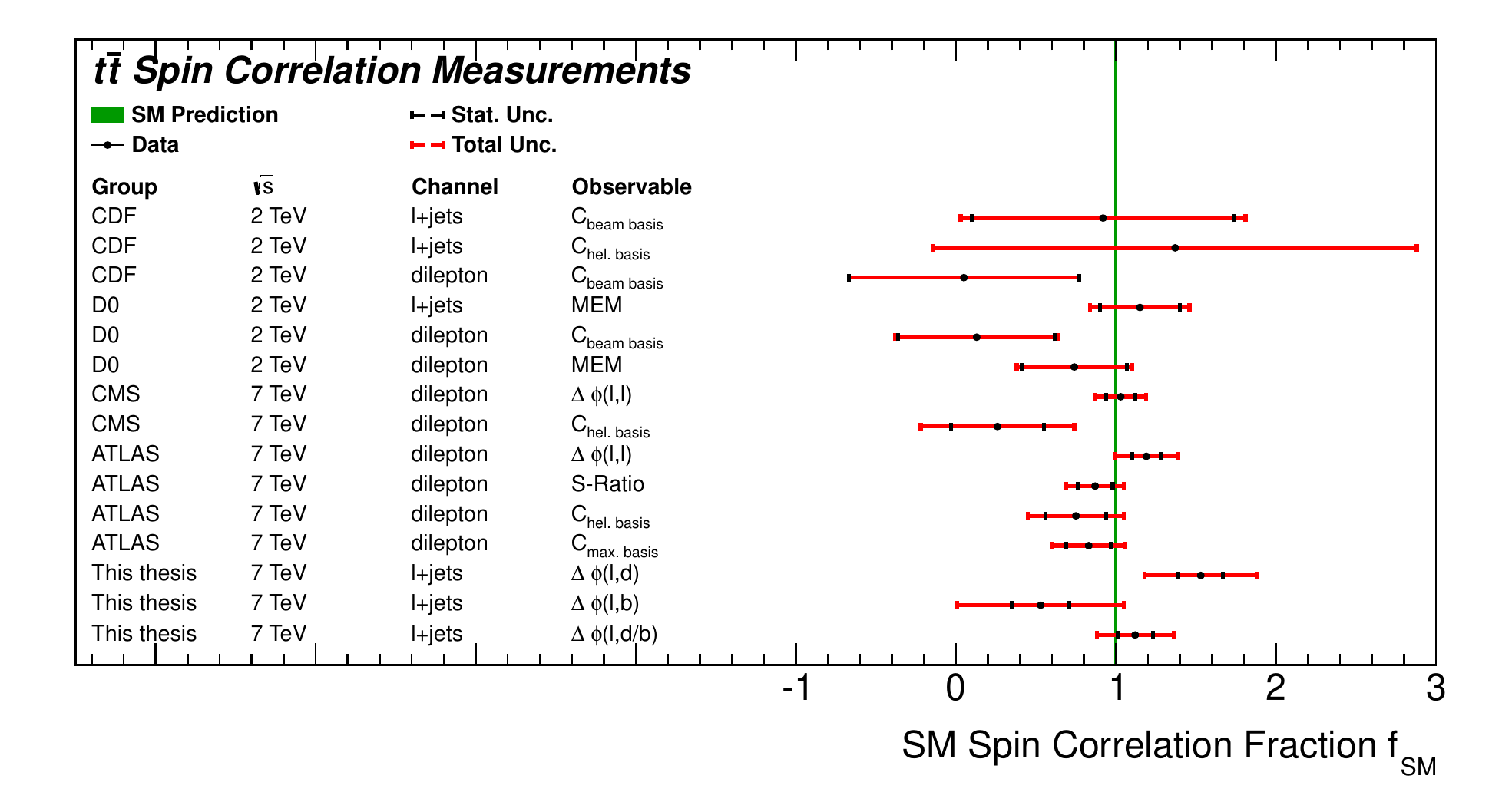}

\end{center}
\caption{Comparison of \ttbar\ spin correlation measurements. The results of \cite{CDF_spin_ljets, CDF_spin_dilep, D0_spin_ljets, D0_spin_dilep_angles, D0_spin_dilep_MEM, ueberpaper, CMS_spin_paper} using different observables have been divided by their SM expectations to compare a common \fsm.}
\label{fig:megasummary}
\end{figure} 
It can be noticed that all measurements of $C$ in the dilepton channel consistently observed less spin correlation than predicted. A second notice concerns the \dphi\ measurements. At ATLAS, \dphi\ lead to $\fsm > 1$ for both the dilepton result and the \ljets\ result using the \dQ\ as analysers. The deviation of the \dQ\ combination is likely to be caused by a mismodelling of the \ttbar\ kinematics as it was discussed in detail. Such a mismodelling, in particular concerning the top quark \pt, would lead to a deviation of the dilepton result in the same direction as for the \dQ. This was indeed observed as ATLAS measured $\fsm = 1.19\,\pm\,0.09\,\text{(stat.)}\,\pm\,0.18\,\text{(syst.)}$ \cite{ueberpaper}. 
The following section is dedicated to the question: To which conclusions do the \fsm\ values, measured in this thesis, lead?

\section{Conclusion}
Only advanced methods of \dQ\ reconstruction allowed a measurement of the \ttbar\ spin correlation in the \ljets\ channel. It is the first published measurement of \ttbar\ spin correlation in the \ljets\ channel at the LHC \cite{ueberpaper}. 

The obtained results are consistent with the SM prediction. Both utilized spin analysers, the \dQ\ and the \bQ, suffer differently from the effects of systematic uncertainties. The measurement helped to understand these effects and to build the basis for future measurements. A combination of the results leads to a significant reduction of the systematic uncertainties. It allows disentangling effects due to an imperfect modelling from effects caused by a modified spin configuration. 

The results of the \dQ\ and \bQ\ analyser combinations show deviations from the SM expectation in different directions. Given the uncertainties and their asymmetric effects on both spin analysers, the results were found to be compatible with the SM and with each other. 

It was found by independent measurements that the data prefers a \ttbar\ modelling which is different than the one implemented in this thesis. This concerns the used PDF, the generator and the top quark \pt\ spectrum. The effects of these alternative models on the spin correlation measurement were checked. All suggested modifications, preferred by data in independent measurements, lead to a lower result of \fsm\ for the \dQ\ and a higher for the \bQ. This is a clear indication that motivated modifications in the \ttbar\ modelling lead to a better agreement of the measured values of \fsm\ -- for both the \dQ\ and the \bQ\ combination -- with the SM. Furthermore, the individual \dQ\ and \bQ\ results would be even more consistent in case of the changed modelling. Details on these tests are shown in the Appendix \ref{sec:app_alt_models}.

The results presented in this thesis are in good agreement with the SM prediction and other \ttbar\ correlation results. Before strong implications on BSM physics can be deduced, the systematic uncertainties need to be reduced further. A trend of a higher spin correlation for the \dQ\ and a lower for the \bQ\ was observed. 

It can be checked if the measured results give a first indication for new physics phenomena. Concluding the spin physics requires reducing both the mismodelling effects and the uncertainties. BSM modifications in the \ttbar\ production with a SM \ttbar\ decay would lead to a coherent modification of the \fsm\ results for both the \dQ\ and the \bQ. This was not observed. Instead, the \dQ\ result was higher and the \bQ\ result lower than the expectation.
 
BSM physics in the decay would affect the two analysers differently. A popular model for a modified top quark decay is $t \rightarrow H^{+}b$, so the replacement of the $W$ vector boson by a scalar charged Higgs boson \cite{Hplus}. In the following it is assumed that such a decay mode occurs in only one of the two top quarks.\footnote{It is also possible that both top quarks decay via a charged Higgs boson, but unlikely as the effect is not at leading order.}

Such a decay modifies the spin analysing power $\alpha$ of the associated \bQ\ from $-0.39$ to $+1.0$ (see also \ref{fig:BSM_alpha} and \cite{Eriksson2007}). Hence, $\Delta \phi (l,b)$ would look more like the distribution of uncorrelated \ttbar\ pairs. At first sight, this matches the measured \fsm\ result using the \bQ. However, the \bQ\ under study was the one from the hadronically decaying top quark. This would require the $H^{\pm}$ to belong to the model of a top quark decaying into three jets. 

Such a decay of the $H^{\pm}$ into two jets would not be the preferred one. Instead, couplings to $\tau$ leptons would be preferred due to the large mass of the $\tau$ \cite{Eriksson2007}. This would not match the decay signature used for the reconstruction in this thesis. Hence, implications of a $t \rightarrow H^{+}b$ process would not be visible in $\Delta \phi (l,b)$, but in a measurement of $\Delta \phi (b,\bar{b})$.

Instead of affecting the \bQ\ and the \dQ\ in the \dphi\ distributions, the dominant effect of a charged Higgs boson would be on the side of the charged lepton due to the large $\tau$ coupling. Neither were $\tau$ leptons reconstructed explicitly, nor are the effects on the secondary leptons from decayed $\tau$ leptons known. Hence, no conclusions on a possible charged Higgs boson in \ttbar\ decays can be drawn. This is true in particular in the context of the large systematic uncertainties on the individual \dQ\ and \bQ\ spin analysers results. 

\section{Outlook}
This thesis concludes with proposals concerning future \ttbar\ spin correlation measurements at hadron colliders. 

\subsection{Reduction of Systematic Uncertainties}
The LHC provided a large number of \ttbar\ pairs, already for the 2011 dataset. Systematic uncertainties limited the presented analysis. Reducing them should have highest priority before repeating the measurement at $\sqrt{s}=8\,\TeV$. As for many analysis, the jet energy scale uncertainty had a strong impact, too. Improving the calibration would be a great benefit. 

A clear dependence on the modelling of the kinematics of the \ttbar\ pairs was shown. Recent measurements of the differential \ttbar\ cross section showed a preference of the top quark \pt\ spectrum and a parton distribution function which are different from to the currently implemented ones. Hence, a change of the default generator is suggested. This also concerns the jet multiplicity mismodelling of \mcatnlo\ which made it necessary to add another degree of freedom to the fit in order to deal with this problem. Next to changing the default MC generator setup, the uncertainties on the \ttbar\ modelling should be further investigated and tried to be reduced, too.
 
A larger dataset allows the application of further cuts to improve the purity of the sample. It was also shown that the reconstruction is stable in terms of pile-up which will increase with higher luminosities. 

\subsection{Dileptonic \ttbar\ Event Suppression}
A common principle in \ttbar\ analyses is to optimize the event selection in a way that the background is reduced and the signal contribution maximized. For \ttbar\ analyses in the \ljets\ channel, where the properties of the \ttbar\ topology assuming a \ljets\ topology are analysed, \ttbar\ contribution from the dilepton channel is a non-negligible background component. As shown in Table \ref{tab:cutflow}, the dileptonic \ttbar\ events represent more than 10\,\% of the \ttbar\ signal. 

A reduction of the dilepton contribution should be considered. Some quantities have different distributions for \ttbar\ events in the dilepton and in the \ljets\ channel and allow to separate the signatures. As an example, Figure \ref{fig:dilep_MET} shows the missing transverse momentum, \etmiss. The transverse $W$ boson mass (Figure \ref{fig:dilep_mtw}) and the likelihood from \KLFitter\ (Figure \ref{fig:dilep_LH}) show a good discrimination between the two \ttbar\ decay channels as well.

\begin{figure}[htbp]
\begin{center}
\subfigure[]{
\includegraphics[width=0.4\textwidth]{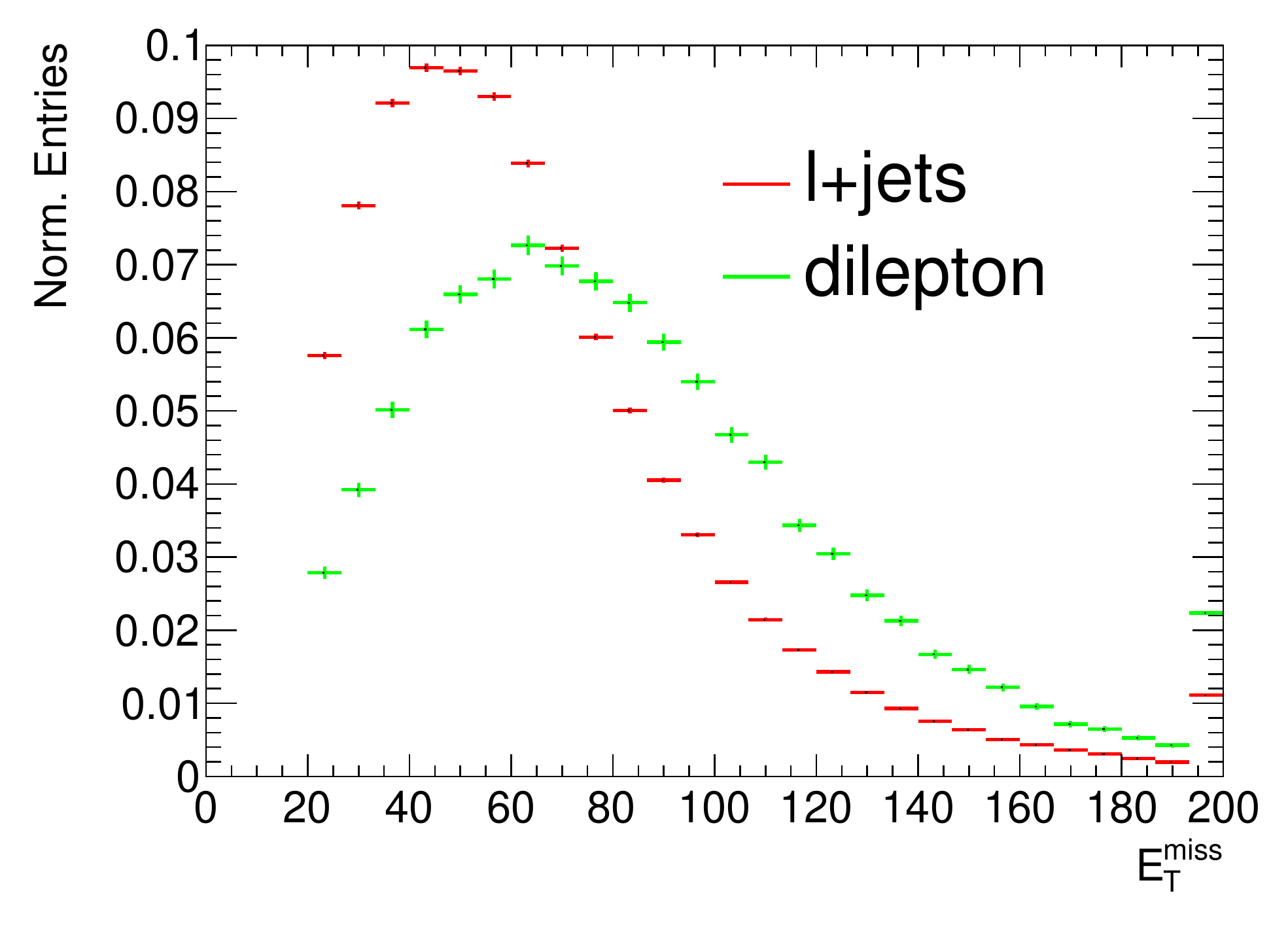}
\label{fig:dilep_MET}
}\\
\subfigure[]{
\includegraphics[width=0.4\textwidth]{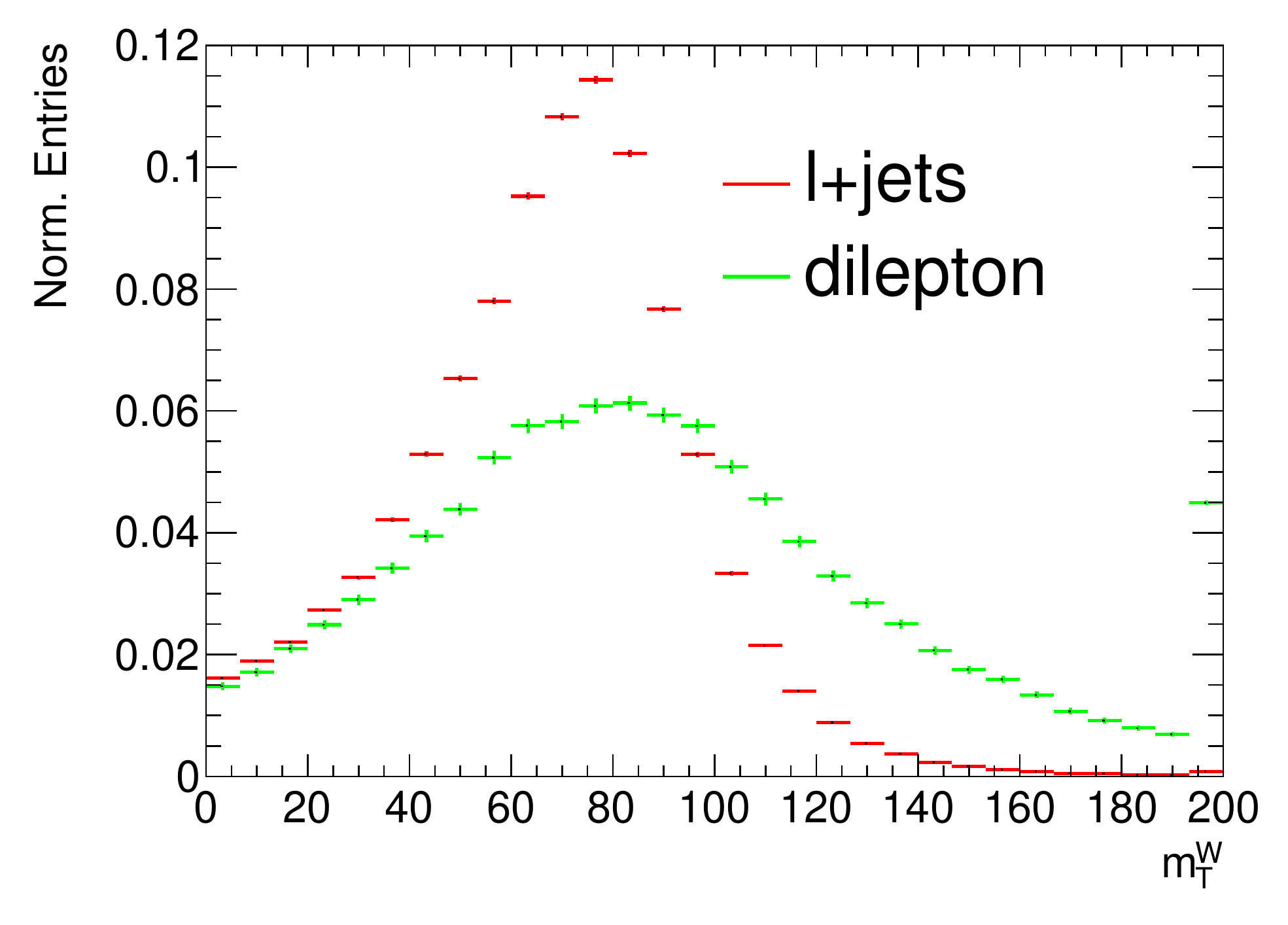}
\label{fig:dilep_mtw}
}
\subfigure[]{
\includegraphics[width=0.4\textwidth]{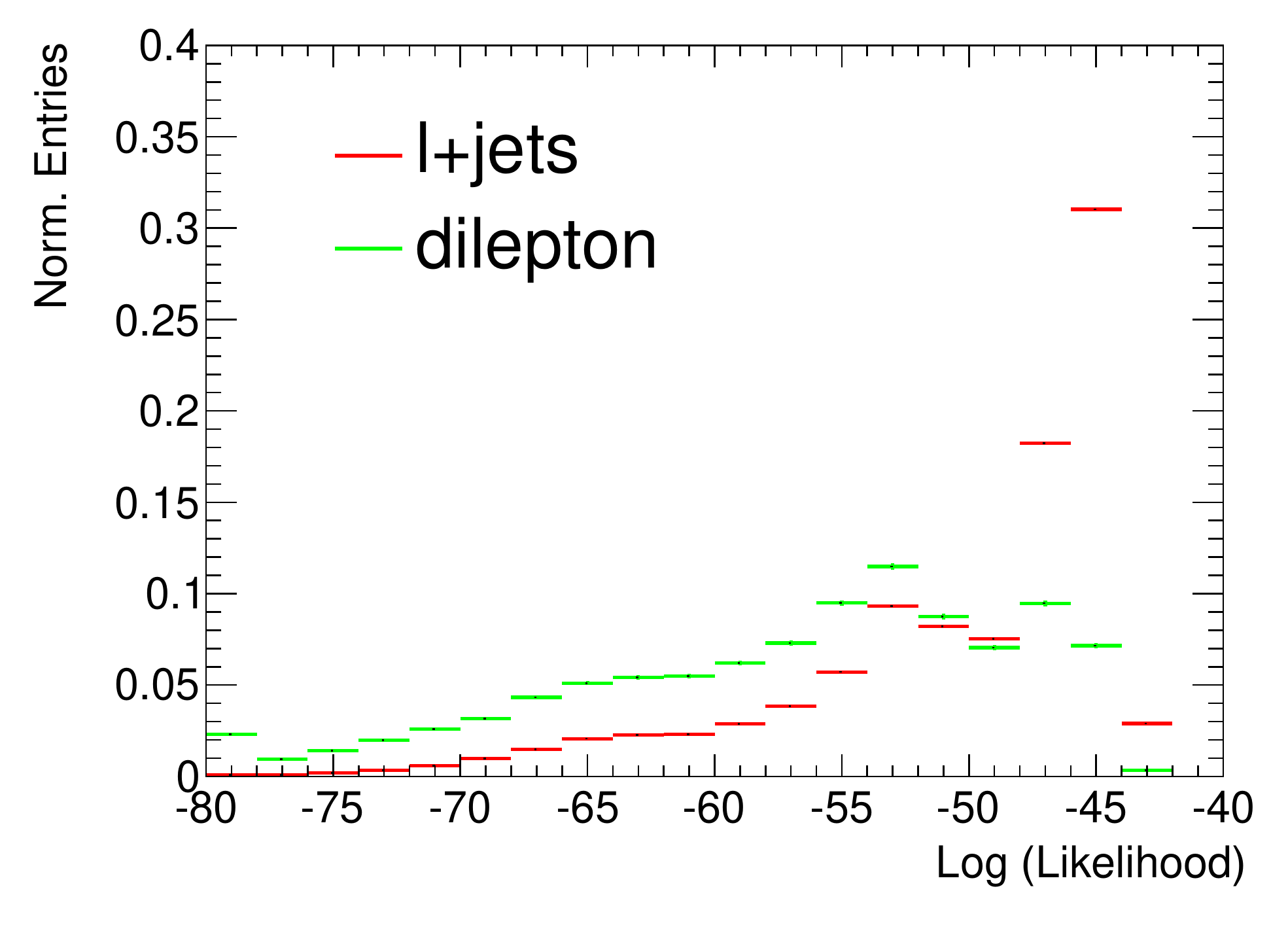}
\label{fig:dilep_LH}
}
\end{center}
\caption{Normalized distributions of \subref{fig:dilep_MET} the \etmiss, \subref{fig:dilep_mtw} transverse $W$ boson mass and the \subref{fig:dilep_LH} logarithm of the likelihood from {\tt KLFitter}. The distributions are shown for reconstructed simulated quantities of \ttbar\ pairs decaying into the \ljets\ channel and the dilepton channel.}
\label{fig:dilep_sep}
\end{figure} 

It should also be considered to veto a second lepton with a looser cut than the ones used in this analysis (see Section \ref{sec:electrons} and \ref{sec:muons}). 

\subsection{Usage of Jet Charge for \ttbar\ Reconstruction}
The jet \pt\ distributions and flavour composition was used to separate jets originating from light up- and down-type quarks. This allowed to reconstruct the \dQ\ as spin analyser. During the optimization studies for the \dQ\ reconstruction other methods were checked, too. 

A promising utility for jet discrimination is the \textit{jet charge}\index{Jet charge}. It makes use of the fact that the charge of the quark is propagated to the hadrons to which a jet fragments. By determining the hadron charges one can infer back on the original quark charge \cite{jetcharge}. Such a method was successfully used in \cite{top_charge_ATLAS, top_charge_CMS, top_charge_D0, top_charge_CDF} to measure the charge of the top quark. 

As both the up- and down-type quark from the $W$ boson have a charge of the same sign, the jet charge technique has not been used in this thesis. 
However, future measurements could benefit from an improved \ttbar\ reconstruction due to the usage of jet charge. In particular, the correct assignment of the two \bQ s to their parent top quarks can be improved. While studies in the \ljets\ channel can make use of other supportive reconstruction techniques, the dilepton channel could benefit a lot. 

In the following the feasibility of the jet charge is briefly demonstrated. Two methods of jet charge are used. For the ``MaxPtTrackCharge'' the jet charge corresponds to the charge of the track within the jet that has the highest \pt. For another approach a weighted sum of charges of the tracks within a jet is created. In Figure \ref{fig:jetcharge_uQ} the jet charges of jets matched to up and anti-up quarks is shown. A clear separation is visible. 

\begin{figure}[htbp]
\begin{center}
\subfigure[]{
\includegraphics[width=0.45\textwidth]{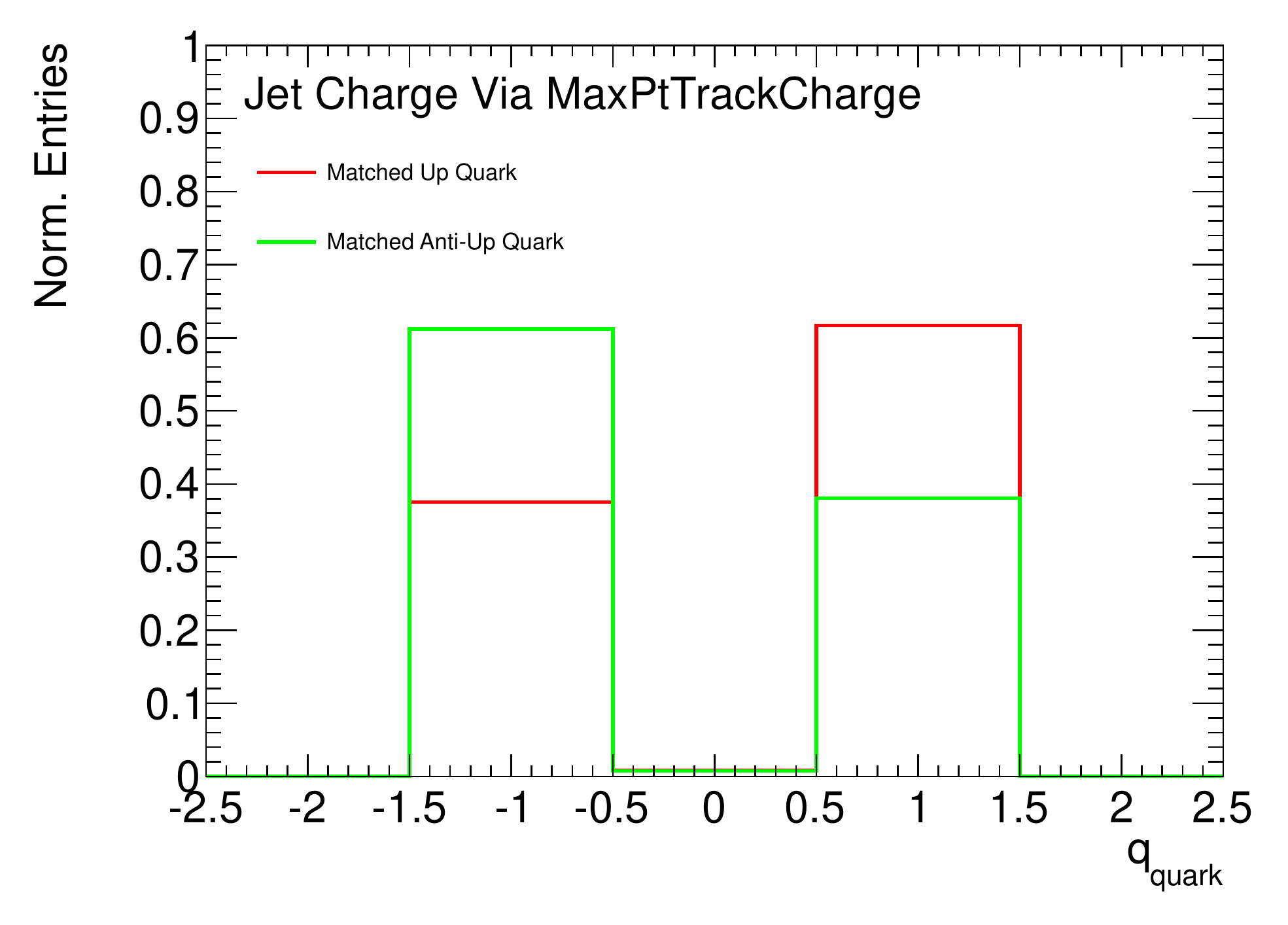}
\label{fig:jetcharge_uQ_max}
}
\subfigure[]{
\includegraphics[width=0.45\textwidth]{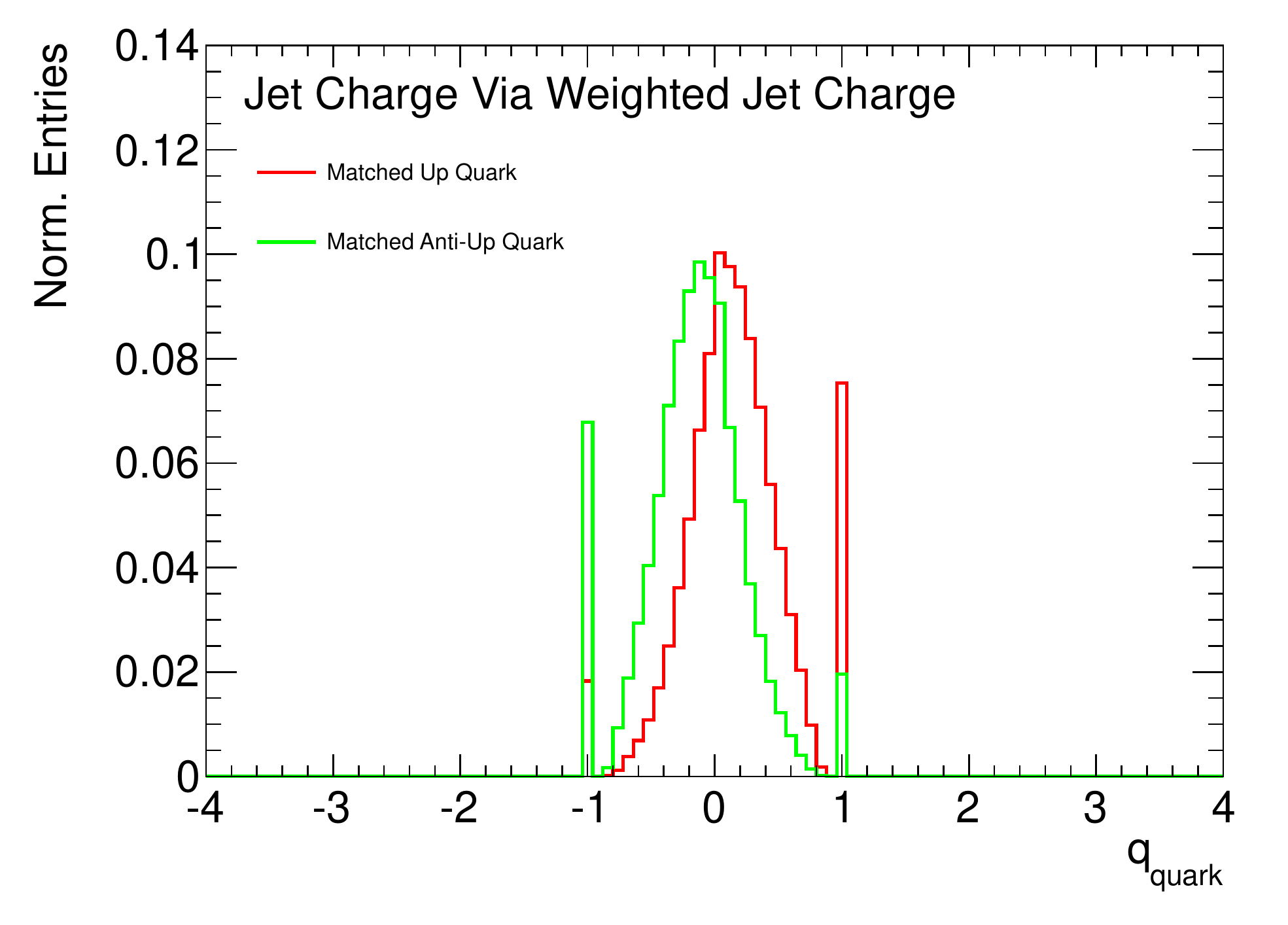}
\label{fig:jetcharge_uQ_weighted}
}
\end{center}
\caption{Charge of a jet matched to an up and anti-up quark using the \subref{fig:jetcharge_uQ_max} charge of the jet track with the highest \pt\ and \subref{fig:jetcharge_uQ_weighted} the weighted charge using all tracks. }
\label{fig:jetcharge_uQ}
\end{figure} 

The description of the weighted jet charge as well as further reconstruction optimization tests can be found in the Appendix \ref{sec:app_jetcharge}.
Studies of the jet charge technique in the context of \ttbar\ pairs produced in association with a Higgs boson can be found in \cite{veit_MA}.

\subsection{Future Measurements of \ttbar\ Spin Correlation}
A strategy for future measurements of \ttbar\ spin correlation in the \ljets\ channel is suggested. 

The Monte Carlo generator used to produce the \ttbar\ signal should be chosen such that no known mismodelling is included. In cases with a clear preference of the data, it should be followed. This also concerns the parton distribution functions and the parton shower modelling. 

The \ttbar\ reconstruction can be further improved by adding information from the jet charges. As several quantities are available to properly map jets from the \ttbar\ decay to the model partons, a multi-variate reconstruction algorithm is a promising way. 

Both the production and the decay of \ttbar\ pairs need further studies to carefully probe the Standard Model and to look for new physics effects beyond it. Studies in the \ljets\ channel will keep playing an important role. A larger dataset will allow choosing subsets with a high signal purity. Also, promising differential analyses will be possible. Furthermore, moving to higher \com\ energies allows to probe new production mechanism compositions due to the increasing dominance of the gluon fusion.

In this thesis the recipe for a powerful reconstruction in the \ljets\ channel was given and will help to establish the next-generation \ttbar\ spin correlation measurement. 
\chapter*{Danksagung}
\addcontentsline{toc}{chapter}{Danksagung}
\begin{otherlanguage}{german}

Alles selbst geschrieben? Aber klar doch. Im Ernst. Man hat ja schon so seine Anspr\"uche an sich selbst. 
Nur hei{\ss}t \glqq selbst geschrieben\grqq\ nicht gleich \glqq alles alleine hinbekommen\grqq. Denn w\"ahrend so einer Promotion muss der Mensch auch mal Maschine sein. Und daf\"ur braucht er  Unterst\"utzung von Freunden und Familie. Die hatte ich stets, und daf\"ur bin ich sehr dankbar. Ob vor Ort oder aus der Ferne, sie waren immer f\"ur mich da. Auch, wenn wir uns in der letzten Zeit nicht allzu oft sehen konnten. Ganz besonderer Dank gilt meiner Familie. Ihr habt mir immer Kraft und Unterst\"utzung gegeben. Mit dem Lemmer-Clan im R\"ucken kann einem nichts passieren! 

Meine wunderbaren Mitbewohner Joana, Johannes, Steffi, Jarka, Alex, Pascal, Konrad, Andrea und Jan machten meine WGs zu einem richtigen zu Hause. W\"art ihr nicht gewesen, h\"atte ich in der ein oder anderen schweren Stunde vielleicht schon die Koffer gepackt. Ich hatte das Gl\"uck, in G\"ottingen nicht nur richtig schnell Anschlu{\ss}, sondern auch richtig gute Freunde zu finden. Danke Folkert, Jan, Joana, Johanna, Lena, Lena, Maike, Marie, Sebastian und Steffi f\"ur die wundersch\"one Zeit!

Ohne Doktorvater kein Doktorsohn. Ich danke Arnulf Quadt, dass er mich als Quereinsteiger in die Teilchenphysik aufgenommen hat. Er hat mich viel gelehrt und lie{\ss} mir die Freiheit, sowohl am CERN unter optimalen Bedingungen zu forschen als auch beim Science Slam einen etwas unkonventionellen Weg der Wissensvermittlung zu gehen. 

Kevin Kr\"oninger und Lisa Shabalina sorgten f\"ur eine ausgezeichnete t\"agliche Betreuung, waren jederzeit hilfsbereit zur Stelle und hielten die Stimmung im B\"uro immer ganz weit oben. Mein Institut war w\"ahrend all der Zeit eine richtig starke Truppe und gro{\ss}e Unterst\"utzung. Heidi Afshar, Heike Ahrens, Lucie Hamdi, Gabriela Herbold, Bernadette Tyson und Christa Wohlfahrt sorgten daf\"ur, dass hinter den Kulissen alles reibungslos ablief. J\"org Meyer hielt die IT am Leben und versorgte mich mit wertvollem Wissen zur Physik und zu Computing. 

Meinen Freunden und Kollegen aus dem II. Physikalischen Institut danke ich f\"ur all den klugen Rat (ob zur Physik oder dar\"uber hinaus) und die Unterhaltung bei der Arbeit und vor allem auch drumherum. Danke insbesondere an Andrea, Anna, Chris, Cora, Fabian, Johannes, Katha, Martina, Matze und Philipp f\"ur Bier, Wein und fr\"ohlich sein!
Danke Andrea, du gute Seele des Instituts, dass du mir dabei geholfen hast, meine Analyse die ersten Schritte gehen zu lassen.

Mein herzlicher Dank gilt auch der gesamten ATLAS Kollaboration und dem LHC Beschleuniger-Team. Nur durch eine gewaltige Teamleistung unter Mitwirkung vieler flei{\ss}iger Menschen konnte ein so gro{\ss}artiges Experiment entstehen. 

Nicht zu vergessen sind meine alten Lehrmeister aus Schul- und Uni-Zeiten. Ganz besonders danken m\"ochte ich Volker Kreuter, Joachim Steinm\"uller und Volker Metag, vor allem auch f\"ur die vielen M\"oglichkeiten, die sie mir geboten haben.

Danke Andrea, Joahnnes, Katha und Kevin, dass ihr euch am Ende nochmal Zeit genommen habt, \"uber meine Arbeit zu schauen. 

Ein herzliches Dankesch\"on auch an die Unfallchirurgie der Uniklinik G\"ottingen, die mich in der Nacht vor Abgabe dieser Arbeit noch zusammengen\"aht hat. Gut gemacht, sieht fast wieder so aus wie vorher.

Zu guter Letzt auch ein gro{\ss}er Dank an dich, Dana. Du bereicherst t\"aglich mein Leben. Ich bin froh, dass ich dich an meiner Seite habe.

\end{otherlanguage}

\bibliographystyle{atlasnote}
\bibliography{boris}

\providecommand{\href}[2]{#2}\begingroup\raggedright\begin{thebibliography}{100}

\bibitem{top_disc_D0}
{D0} Collaboration, {\em {Search for high mass top quark production in
  $p\bar{p}$ collisions at $\sqrt{s} = 1.8$ TeV}\/},
\href{http://dx.doi.org/10.1103/PhysRevLett.74.2422}{Phys.Rev.Lett. {\bf 74}
  (1995)  2422}.

\bibitem{top_disc_CDF}
{CDF} Collaboration, {\em {Observation of top quark production in $\bar{p}p$
  collisions}\/},
\href{http://dx.doi.org/10.1103/PhysRevLett.74.2626}{Phys.Rev.Lett. {\bf 74}
  (1995)  2626}.

\bibitem{topmass_evolution}
C.~Quigg, {\em {Unanswered Questions in the Electroweak Theory}\/},
\href{http://dx.doi.org/10.1146/annurev.nucl.010909.083126}{Ann.Rev.Nucl.Part.Sci.
  {\bf 59} (2009)  505}.

\bibitem{gfitter}
M.~Baak et al., {\em {The Electroweak Fit of the Standard Model after the
  Discovery of a New Boson at the LHC}\/},
\href{http://dx.doi.org/10.1140/epjc/s10052-012-2205-9}{Eur.Phys.J. {\bf C72}
  (2012)  2205}.

\bibitem{periodictable}
D.~Mendelejeff, {\em Die periodische Gesetzm{\"a}{\ss}igkeit der chemischen
  Elemente\/},  Ann.Chem.Pharm. {\bf VIII Supp.} (1871)  133.

\bibitem{elements}
R.~D. Loss and J.~Corish, {\em Names and symbols of the elements with atomic
  numbers 114 and 116 (IUPAC Recommendations 2012)\/},  Pure Appl.Chem. {\bf
  84} (2012) no.~7, 1669.

\bibitem{maxwell}
J.~C. Maxwell, {\em {A dynamical theory of the electromagnetic field}\/},
\href{http://dx.doi.org/10.1098/rstl.1865.0008}{Phil.Trans.Roy.Soc.Lond. {\bf
  155} (1865)  459}.

\bibitem{electron}
J.~Thomson, {\em {Cathode rays}\/},
\href{http://dx.doi.org/10.1080/14786449708621070}{Phil.Mag. {\bf 44} (1897)
  293}.

\bibitem{rutherford}
H.~Geiger and E.~Marsden, {\em The laws of deflexion of a particles through
  large angles\/},
  \href{http://dx.doi.org/10.1080/14786440408634197}{Phil.Mag. Series 6 {\bf
  25} (1913) no.~148, 604}.

\bibitem{prout1}
W.~Prout, {\em On the relation between the specific gravities of bodies in
  their gaseous state and the weights of their atoms\/},  Ann.Phil. {\bf 6}
  (1815)  321.

\bibitem{prout2}
W.~Prout, {\em Correction of a mistake in the essay on the relation between the
  specific gravities of bodies in their gaseous state and the weights of their
  atoms\/},  Ann.Phil. {\bf 7} (1816)  111.

\bibitem{neutron}
J.~Chadwick, {\em The Existence of a Neutron\/},
  \href{http://dx.doi.org/10.1098/rspa.1932.0112}{Proc.Roy.Soc.Lond. Series A
  {\bf 136} (1932) no.~830, 692}.

\bibitem{positron}
C.~Anderson, {\em {The Positive Electron}\/},
\href{http://dx.doi.org/10.1103/PhysRev.43.491}{Phys.Rev. {\bf 43} (1933)
  491}.

\bibitem{lattes1}
C.~Lattes, H.~Muirhead, G.~Occhialini, and C.~Powell, {\em {PROCESSES INVOLVING
  CHARGED MESONS}\/},
\href{http://dx.doi.org/10.1038/159694a0}{Nature {\bf 159} (1947)  694}.

\bibitem{lattes2}
C.~Lattes, G.~Occhialini, and C.~Powell, {\em {Observations on the Tracks of
  Slow Mesons in Photographic Emulsions. 1}\/},
\href{http://dx.doi.org/10.1038/160453a0}{Nature {\bf 160} (1947)  453}.

\bibitem{lattes3}
C.~Lattes, G.~Occhialini, and C.~Powell, {\em {Observations on the Tracks of
  Slow Mesons in Photographic Emulsions. 2}\/},
\href{http://dx.doi.org/10.1038/160486a0}{Nature {\bf 160} (1947)  486}.

\bibitem{muon}
J.~Street and E.~Stevenson, {\em {New Evidence for the Existence of a Particle
  of Mass Intermediate Between the Proton and Electron}\/},
\href{http://dx.doi.org/10.1103/PhysRev.52.1003}{Phys.Rev. {\bf 52} (1937)
  1003}.

\bibitem{eightfold1}
Y.~Ne'eman, {\em {Derivation of strong interactions from a gauge
  invariance}\/},
\href{http://dx.doi.org/10.1016/0029-5582(61)90134-1}{Nucl.Phys. {\bf 26}
  (1961)  222}.

\bibitem{eightfold2}
M.~Gell-Mann, {\em {Symmetries of baryons and mesons}\/},
\href{http://dx.doi.org/10.1103/PhysRev.125.1067}{Phys.Rev. {\bf 125} (1962)
  1067}.

\bibitem{quarkmodel}
M.~Gell-Mann, {\em {A Schematic Model of Baryons and Mesons}\/},
\href{http://dx.doi.org/10.1016/S0031-9163(64)92001-3}{Phys.Lett. {\bf 8}
  (1964)  214}.

\bibitem{dis1}
E.~D. Bloom et al., {\em {High-Energy Inelastic e p Scattering at 6-Degrees and
  10-Degrees}\/},
\href{http://dx.doi.org/10.1103/PhysRevLett.23.930}{Phys.Rev.Lett. {\bf 23}
  (1969)  930}.

\bibitem{dis2}
M.~Breidenbach et al., {\em {Observed Behavior of Highly Inelastic
  electron-Proton Scattering}\/},
\href{http://dx.doi.org/10.1103/PhysRevLett.23.935}{Phys.Rev.Lett. {\bf 23}
  (1969)  935}.

\bibitem{dis3}
G.~Miller et al., {\em {Inelastic electron-Proton Scattering at Large Momentum
  Transfers}\/},
\href{http://dx.doi.org/10.1103/PhysRevD.5.528}{Phys.Rev. {\bf D5} (1972)
  528}.

\bibitem{ATLAS_higgs_discovery}
{ATLAS} Collaboration, {\em {Observation of a new particle in the search for
  the Standard Model Higgs boson with the ATLAS detector at the LHC}\/},
\href{http://dx.doi.org/10.1016/j.physletb.2012.08.020}{Phys.Lett. {\bf B716}
  (2012)  1}.

\bibitem{CMS_higgs_discovery}
{CMS} Collaboration, {\em {Observation of a new boson at a mass of 125 GeV with
  the CMS experiment at the LHC}\/},
\href{http://dx.doi.org/10.1016/j.physletb.2012.08.021}{Phys.Lett. {\bf B716}
  (2012)  30}.

\bibitem{Higgs1}
F.~Englert and R.~Brout, {\em {Broken Symmetry and the Mass of Gauge Vector
  Mesons}\/},
\href{http://dx.doi.org/10.1103/PhysRevLett.13.321}{Phys.Rev.Lett. {\bf 13}
  (1964)  321}.

\bibitem{Higgs2}
P.~W. Higgs, {\em {Broken symmetries, massless particles and gauge fields}\/},
\href{http://dx.doi.org/10.1016/0031-9163(64)91136-9}{Phys.Lett. {\bf 12}
  (1964)  132}.

\bibitem{Higgs3}
P.~W. Higgs, {\em {Broken Symmetries and the Masses of Gauge Bosons}\/},
\href{http://dx.doi.org/10.1103/PhysRevLett.13.508}{Phys.Rev.Lett. {\bf 13}
  (1964)  508}.

\bibitem{Higgs4}
G.~Guralnik, C.~Hagen, and T.~Kibble, {\em {Global Conservation Laws and
  Massless Particles}\/},
\href{http://dx.doi.org/10.1103/PhysRevLett.13.585}{Phys.Rev.Lett. {\bf 13}
  (1964)  585}.

\bibitem{Higgs5}
P.~W. Higgs, {\em {Spontaneous Symmetry Breakdown without Massless Bosons}\/},
\href{http://dx.doi.org/10.1103/PhysRev.145.1156}{Phys.Rev. {\bf 145} (1966)
  1156}.

\bibitem{Higgs6}
T.~Kibble, {\em {Symmetry breaking in nonAbelian gauge theories}\/},
\href{http://dx.doi.org/10.1103/PhysRev.155.1554}{Phys.Rev. {\bf 155} (1967)
  1554}.

\bibitem{tevatron}
R.~R. Wilson, {\em {The Tevatron}\/},
\href{http://dx.doi.org/10.1063/1.3037746}{Phys.Today {\bf 30N10} (1977)  23}.

\bibitem{D0}
{D0} Collaboration, {\em {An Experiment at D0 to Study anti-Proton - Proton
  Collisions at 2-TeV: Design Report}\/},
FERMILAB-PUB-83-111-E.

\bibitem{CDF}
{CDF} Collaboration, R.~Blair et al., {\em {The CDF-II detector: Technical
  design report}\/},
FERMILAB-PUB-96-390-E.

\bibitem{GWS1}
S.~Glashow, {\em {Partial Symmetries of Weak Interactions}\/},
\href{http://dx.doi.org/10.1016/0029-5582(61)90469-2}{Nucl.Phys. {\bf 22}
  (1961)  579}.

\bibitem{GWS2}
S.~Weinberg, {\em {A Model of Leptons}\/},
\href{http://dx.doi.org/10.1103/PhysRevLett.19.1264}{Phys.Rev.Lett. {\bf 19}
  (1967)  1264}.

\bibitem{GWS3}
S.~Glashow, J.~Iliopoulos, and L.~Maiani, {\em {Weak Interactions with
  Lepton-Hadron Symmetry}\/},
\href{http://dx.doi.org/10.1103/PhysRevD.2.1285}{Phys.Rev. {\bf D2} (1970)
  1285}.

\bibitem{georgi72}
H.~Georgi and S.~L. Glashow, {\em Unified Weak and Electromagnetic Interactions
  without Neutral Currents\/},
  \href{http://dx.doi.org/10.1103/PhysRevLett.28.1494}{Phys. Rev. Lett. {\bf
  28} (1972)  1494}.

\bibitem{politzer73}
H.~D. Politzer, {\em Reliable Perturbative Results for Strong Interactions?\/},
   \href{http://dx.doi.org/10.1103/PhysRevLett.30.1346}{Phys. Rev. Lett. {\bf
  30} (1973)  1346}.

\bibitem{asymp_freedom2}
H.~D. Politzer, {\em Asymptotic freedom: An approach to strong interactions\/},
   \href{http://dx.doi.org/http://dx.doi.org/10.1016/0370-1573(74)90014-3}{Physics
  Reports {\bf 14} (1974) no.~4, 129}.

\bibitem{asymp_freedom}
D.~J. Gross and F.~Wilczek, {\em {Ultraviolet Behavior of Nonabelian Gauge
  Theories}\/},
\href{http://dx.doi.org/10.1103/PhysRevLett.30.1343}{Phys.Rev.Lett. {\bf 30}
  (1973)  1343}.

\bibitem{weinberg2004}
S.~Weinberg, {\em {The Making of the standard model}\/},
\href{http://dx.doi.org/10.1140/epjc/s2004-01761-1}{Eur.Phys.J. {\bf C34}
  (2004)  5}.

\bibitem{hooft71}
G.~'t~Hooft, {\em {Renormalizable Lagrangians for Massive Yang-Mills
  Fields}\/},
\href{http://dx.doi.org/10.1016/0550-3213(71)90139-8}{Nucl.Phys. {\bf B35}
  (1971)  167}.

\bibitem{hooft72a}
G.~'t~Hooft and M.~Veltman, {\em {Regularization and Renormalization of Gauge
  Fields}\/},
\href{http://dx.doi.org/10.1016/0550-3213(72)90279-9}{Nucl.Phys. {\bf B44}
  (1972)  189}.

\bibitem{hooft72b}
G.~'t~Hooft and M.~Veltman, {\em {Combinatorics of gauge fields}\/},
\href{http://dx.doi.org/10.1016/S0550-3213(72)80021-X}{Nucl.Phys. {\bf B50}
  (1972)  318}.

\bibitem{PDG}
{Particle Data Group} Collaboration, {\em {Review of Particle Physics
  (RPP)}\/},
\href{http://dx.doi.org/10.1103/PhysRevD.86.010001}{Phys.Rev. {\bf D86} (2012)
  010001}.

\bibitem{m_top_world}
{ATLAS Collaboration, CDF Collaboration, CMS Collaboration, D0} Collaboration,
  {\em {First combination of Tevatron and LHC measurements of the top-quark
  mass}\/},  ATLAS-CONF-2014-008, CDF-NOTE-11071, CMS-PAS-TOP-13-014,
  D0-NOTE-6416,
\href{http://arxiv.org/abs/1403.4427}{{\tt arXiv:1403.4427 [hep-ex]}}.

\bibitem{halzenmartin}
F.~Halzen and A.~D. Martin, {\em {Quarks And Leptons: An Introductory Course In
  Modern Particle Physics}}.
\newblock Wiley,
1984.
\newblock

\bibitem{alpha_s_h1a}
{H1} Collaboration, {\em {Jet Production in ep Collisions at High $Q^2$ and
  Determination of alpha(s)}\/},
\href{http://dx.doi.org/10.1140/epjc/s10052-009-1208-7}{Eur.Phys.J. {\bf C65}
  (2010)  363}.

\bibitem{alpha_s_h1b}
{H1} Collaboration, {\em {Jet Production in ep Collisions at Low $Q^2$ and
  Determination of $\alpha_S$}\/},
\href{http://dx.doi.org/10.1140/epjc/s10052-010-1282-x}{Eur.Phys.J. {\bf C67}
  (2010)  1}.

\bibitem{alpha_s_zeus}
{ZEUS} Collaboration, {\em {Inclusive-jet photoproduction at HERA and
  determination of alphas}\/},
\href{http://dx.doi.org/10.1016/j.nuclphysb.2012.06.006}{Nucl.Phys. {\bf B864}
  (2012)  1}.

\bibitem{alpha_s_d0a}
{D0} Collaboration, {\em {Measurement of angular correlations of jets at
  $\sqrt{s}=1.96$ TeV and determination of the strong coupling at high momentum
  transfers}\/},
\href{http://dx.doi.org/10.1016/j.physletb.2012.10.003}{Phys.Lett. {\bf B718}
  (2012)  56}.

\bibitem{alpha_s_d0b}
{D0} Collaboration, {\em {Determination of the strong coupling constant from
  the inclusive jet cross section in $p\bar{p}$ collisions at $\sqrt{s}$=1.96
  TeV}\/},
\href{http://dx.doi.org/10.1103/PhysRevD.80.111107}{Phys.Rev. {\bf D80} (2009)
  111107}.

\bibitem{alpha_s}
{CMS} Collaboration, {\em {Measurement of the ratio of the inclusive 3-jet
  cross section to the inclusive 2-jet cross section in pp collisions at
  $\sqrt{s}$ = 7 TeV and first determination of the strong coupling constant in
  the TeV range}\/},
\href{http://dx.doi.org/10.1140/epjc/s10052-013-2604-6}{Eur.Phys.J. {\bf C73}
  (2013)  2604}.

\bibitem{CKM}
M.~Kobayashi and T.~Maskawa, {\em {CP Violation in the Renormalizable Theory of
  Weak Interaction}\/},
\href{http://dx.doi.org/10.1143/PTP.49.652}{Prog.Theor.Phys. {\bf 49} (1973)
  652}.

\bibitem{cabbibo}
N.~Cabibbo, {\em {Unitary Symmetry and Leptonic Decays}\/},
\href{http://dx.doi.org/10.1103/PhysRevLett.10.531}{Phys.Rev.Lett. {\bf 10}
  (1963)  531}.

\bibitem{wu}
C.~Wu et al., {\em {Experimental Test of Parity Conservation in Beta Decay}\/},
\href{http://dx.doi.org/10.1103/PhysRev.105.1413}{Phys.Rev. {\bf 105} (1957)
  1413}.

\bibitem{W_disc_UA1}
{UA1} Collaboration, {\em {Experimental Observation of Isolated Large
  Transverse Energy Electrons with Associated Missing Energy at $\sqrt{s}$ =
  540 GeV}\/},
\href{http://dx.doi.org/10.1016/0370-2693(83)91177-2}{Phys.Lett. {\bf B122}
  (1983)  103}.

\bibitem{W_disc_UA2}
{UA2} Collaboration, {\em {Observation of Single Isolated Electrons of High
  Transverse Momentum in Events with Missing Transverse Energy at the CERN
  $\bar{p}p$ Collider}\/},
\href{http://dx.doi.org/10.1016/0370-2693(83)91605-2}{Phys.Lett. {\bf B122}
  (1983)  476}.

\bibitem{Z_disc_UA1}
{UA1} Collaboration, {\em {Experimental Observation of Lepton Pairs of
  Invariant Mass Around 95GeV/$c^2$ at the CERN SPS Collider}\/},
\href{http://dx.doi.org/10.1016/0370-2693(83)90188-0}{Phys.Lett. {\bf B126}
  (1983)  398}.

\bibitem{Z_disc_UA2}
{UA2} Collaboration, {\em {Evidence for $Z^0$ $\rightarrow$ $e^{+} e^{-}$ at
  the CERN $\bar{p}p$ Collider}\/},
\href{http://dx.doi.org/10.1016/0370-2693(83)90744-X}{Phys.Lett. {\bf B129}
  (1983)  130}.

\bibitem{gmn1}
M.~Gell-Mann, {\em The interpretation of the new particles as displaced charge
  multiplets\/},  \href{http://dx.doi.org/10.1007/BF02748000}{Il Nuovo Cimento
  {\bf 4} (1956) no.~2, 848}.

\bibitem{gmn2}
T.~Nakano and K.~Nishijima, {\em {Charge Independence for V-particles}\/},
\href{http://dx.doi.org/10.1143/PTP.10.581}{Prog.Theor.Phys. {\bf 10} (1953)
  581}.

\bibitem{peskin}
M.~E. Peskin and D.~V. Schroeder, {\em {An Introduction to quantum field
  theory}}.
\newblock Addison-Wesley,
1995.
\newblock

\bibitem{higgs_ATLAS_spin_evidence}
{ATLAS} Collaboration, {\em {Evidence for the spin-0 nature of the Higgs boson
  using ATLAS data}\/},
\href{http://dx.doi.org/10.1016/j.physletb.2013.08.026}{Phys.Lett. {\bf B726}
  (2013)  120}.

\bibitem{darkmatter}
D.~Clowe et al., {\em A Direct Empirical Proof of the Existence of Dark
  Matter\/},  Astrophys.J.Lett. {\bf 648} (2006) no.~2, L109.

\bibitem{neutrino_oscillations}
M.~Gonzalez-Garcia and M.~Maltoni, {\em {Phenomenology with Massive
  Neutrinos}\/},
\href{http://dx.doi.org/10.1016/j.physrep.2007.12.004}{Phys.Rept. {\bf 460}
  (2008)  1}.

\bibitem{susy1}
H.~Miyazawa, {\em {Baryon Number Changing Currents}\/},
\href{http://dx.doi.org/10.1143/PTP.36.1266}{Prog.Theor.Phys. {\bf 36} (1966)
  no.~6, 1266}.

\bibitem{susy2}
P.~Ramond, {\em {Dual Theory for Free Fermions}\/},
\href{http://dx.doi.org/10.1103/PhysRevD.3.2415}{Phys.Rev. {\bf D3} (1971)
  2415}.

\bibitem{susy3}
Y.~Golfand and E.~Likhtman, {\em {Extension of the Algebra of Poincare Group
  Generators and Violation of p Invariance}\/},
JETP Lett. {\bf 13} (1971)  323.

\bibitem{susy4}
A.~Neveu and J.~Schwarz, {\em {Factorizable dual model of pions}\/},
\href{http://dx.doi.org/10.1016/0550-3213(71)90448-2}{Nucl.Phys. {\bf B31}
  (1971)  86}.

\bibitem{susy5}
A.~Neveu and J.~Schwarz, {\em {Quark Model of Dual Pions}\/},
\href{http://dx.doi.org/10.1103/PhysRevD.4.1109}{Phys.Rev. {\bf D4} (1971)
  1109}.

\bibitem{susy6}
J.-L. Gervais and B.~Sakita, {\em {Field Theory Interpretation of Supergauges
  in Dual Models}\/},
\href{http://dx.doi.org/10.1016/0550-3213(71)90351-8}{Nucl.Phys. {\bf B34}
  (1971)  632}.

\bibitem{susy7}
D.~Volkov and V.~Akulov, {\em {Is the Neutrino a Goldstone Particle?}\/},
\href{http://dx.doi.org/10.1016/0370-2693(73)90490-5}{Phys.Lett. {\bf B46}
  (1973)  109}.

\bibitem{susy8}
J.~Wess and B.~Zumino, {\em {A Lagrangian Model Invariant Under Supergauge
  Transformations}\/},
\href{http://dx.doi.org/10.1016/0370-2693(74)90578-4}{Phys.Lett. {\bf B49}
  (1974)  52}.

\bibitem{susy9}
J.~Wess and B.~Zumino, {\em {Supergauge Transformations in Four-Dimensions}\/},
\href{http://dx.doi.org/10.1016/0550-3213(74)90355-1}{Nucl.Phys. {\bf B70}
  (1974)  39}.

\bibitem{dglap1}
G.~Altarelli and G.~Parisi, {\em {Asymptotic Freedom in Parton Language}\/},
\href{http://dx.doi.org/10.1016/0550-3213(77)90384-4}{Nucl.Phys. {\bf B126}
  (1977)  298}.

\bibitem{dglap2}
Y.~L. Dokshitzer, {\em {Calculation of the Structure Functions for Deep
  Inelastic Scattering and e+ e- Annihilation by Perturbation Theory in Quantum
  Chromodynamics.}\/},
Sov.Phys.JETP {\bf 46} (1977)  641.

\bibitem{dglap3}
V.~Gribov and L.~Lipatov, {\em {Deep inelastic e p scattering in perturbation
  theory}\/},
Sov.J.Nucl.Phys. {\bf 15} (1972)  438.

\bibitem{herapdf}
{H1 and ZEUS} Collaboration, {\em {Combined Measurement and QCD Analysis of the
  Inclusive e$^{\pm}$ p Scattering Cross Sections at HERA}\/},
\href{http://dx.doi.org/10.1007/JHEP01(2010)109}{JHEP {\bf 1001} (2010)  109}.

\bibitem{CT10}
H.-L. Lai et al., {\em {New parton distributions for collider physics}\/},
\href{http://dx.doi.org/10.1103/PhysRevD.82.074024}{Phys.Rev. {\bf D82} (2010)
  074024}.

\bibitem{NNPDF}
R.~D. Ball et al., {\em {A first unbiased global NLO determination of parton
  distributions and their uncertainties}\/},
\href{http://dx.doi.org/10.1016/j.nuclphysb.2010.05.008}{Nucl.Phys. {\bf B838}
  (2010)  136}.

\bibitem{MSTW}
A.~Martin et al., {\em {Parton distributions for the LHC}\/},
\href{http://dx.doi.org/10.1140/epjc/s10052-009-1072-5}{Eur.Phys.J. {\bf C63}
  (2009)  189}.

\bibitem{croninfitch}
J.~Christenson, J.~Cronin, V.~Fitch, and R.~Turlay, {\em {Evidence for the 2
  $\pi$ Decay of the $K^0_2$ Meson}\/},
\href{http://dx.doi.org/10.1103/PhysRevLett.13.138}{Phys.Rev.Lett. {\bf 13}
  (1964)  138}.

\bibitem{tau_disc}
M.~L. Perl et al., {\em {Evidence for Anomalous Lepton Production in e+ - e-
  Annihilation}\/},
\href{http://dx.doi.org/10.1103/PhysRevLett.35.1489}{Phys.Rev.Lett. {\bf 35}
  (1975)  1489}.

\bibitem{b_disc}
S.~Herb et al., {\em {Observation of a Dimuon Resonance at 9.5 GeV in 400-GeV
  Proton-Nucleus Collisions}\/},
\href{http://dx.doi.org/10.1103/PhysRevLett.39.252}{Phys.Rev.Lett. {\bf 39}
  (1977)  252}.

\bibitem{gamma_top2}
C.~S. Li, R.~J. Oakes, and T.~C. Yuan, {\em {QCD corrections to $t \to W^{+}
  b$}\/},
\href{http://dx.doi.org/10.1103/PhysRevD.43.3759}{Phys.Rev. {\bf D43} (1991)
  3759}.

\bibitem{CODATA}
P.~J. Mohr, B.~N. Taylor, and D.~B. Newell, {\em {CODATA Recommended Values of
  the Fundamental Physical Constants: 2010}\/},
\href{http://dx.doi.org/10.1103/RevModPhys.84.1527}{Rev.Mod.Phys. {\bf 84}
  (2012)  1527}.

\bibitem{hadr_time_scale}
I.~I. Bigi et al., {\em {Production and Decay Properties of Ultraheavy
  Quarks}\/},
\href{http://dx.doi.org/10.1016/0370-2693(86)91275-X}{Phys.Lett. {\bf B181}
  (1986)  157}.

\bibitem{top_depol_timescale}
Y.~Grossman and I.~Nachshon, {\em {Hadronization, spin, and lifetimes}\/},
\href{http://dx.doi.org/10.1088/1126-6708/2008/07/016}{JHEP {\bf 0807} (2008)
  016}.

\bibitem{Mahlon2010}
G.~Mahlon and S.~J. Parke, {\em {Spin Correlation Effects in Top Quark Pair
  Production at the LHC}\/},
\href{http://dx.doi.org/10.1103/PhysRevD.81.074024}{Phys.Rev. {\bf D81} (2010)
  074024}.

\bibitem{fact_theorem}
J.~C. Collins and D.~E. Soper, {\em {The Theorems of Perturbative QCD}\/},
\href{http://dx.doi.org/10.1146/annurev.ns.37.120187.002123}{Ann.Rev.Nucl.Part.Sci.
  {\bf 37} (1987)  383}.

\bibitem{fact_theorem2}
J.~C. Collins, D.~E. Soper, and G.~F. Sterman, {\em {Factorization of Hard
  Processes in QCD}\/},
Adv.Ser.Direct.High Energy Phys. {\bf 5} (1988)  1.

\bibitem{ttbar_xsec_theo}
S.~Moch and P.~Uwer, {\em {Theoretical status and prospects for top-quark pair
  production at hadron colliders}\/},
\href{http://dx.doi.org/10.1103/PhysRevD.78.034003}{Phys.Rev. {\bf D78} (2008)
  034003}.

\bibitem{ttbar_xsec_theo2}
M.~Czakon, P.~Fiedler, and A.~Mitov, {\em {Total Top-Quark Pair-Production
  Cross Section at Hadron Colliders Through $\mathcal{O}(\alpha^{4}_{S})$}\/},
\href{http://dx.doi.org/10.1103/PhysRevLett.110.252004}{Phys.Rev.Lett. {\bf
  110} (2013) no.~25, 252004}.

\bibitem{Czakon:2011xx}
M.~Czakon and A.~Mitov, {\em {Top++: A Program for the Calculation of the
  Top-Pair Cross-Section at Hadron Colliders}\/},
\href{http://arxiv.org/abs/1112.5675}{{\tt arXiv:1112.5675 [hep-ph]}}.

\bibitem{top_xsec1}
{CDF and D0} Collaboration, {\em {Combination of measurements of the top-quark
  pair production cross section from the Tevatron Collider}\/},
\href{http://dx.doi.org/10.1103/PhysRevD.89.072001}{Phys.Rev. {\bf D89} (2014)
  072001}.

\bibitem{top_xsec2}
{ATLAS} Collaboration, {\em {Measurement of the cross section for top-quark
  pair production in $pp$ collisions at $\sqrt{s}=7$ TeV with the ATLAS
  detector using final states with two high-pt leptons}\/},
\href{http://dx.doi.org/10.1007/JHEP05(2012)059}{JHEP {\bf 1205} (2012)  059}.

\bibitem{top_xsec3}
{CMS} Collaboration, {\em {Measurement of the $t\bar{t}$ production cross
  section in the dilepton channel in $pp$ collisions at $\sqrt{s}=7$ TeV}\/},
\href{http://dx.doi.org/10.1007/JHEP11(2012)067}{JHEP {\bf 1211} (2012)  067}.

\bibitem{top_xsec4}
{ATLAS} Collaboration, {\em {Measurement of the ttbar production cross-section
  in pp collisions at $\sqrt{s}$ = 7 TeV using kinematic information of
  lepton+jets events}\/},
ATLAS-CONF-2011-121.

\bibitem{top_xsec5}
{CMS} Collaboration, {\em {Measurement of the $t\bar{t}$ production cross
  section in $pp$ collisions at $\sqrt{s}=7$ TeV with lepton + jets final
  states}\/},
\href{http://dx.doi.org/10.1016/j.physletb.2013.02.021}{Phys.Lett. {\bf B720}
  (2013)  83}.

\bibitem{top_xsec6}
{ATLAS} Collaboration, {\em {Combination of ATLAS and CMS top-quark pair cross
  section measurements using up to 1.1 fb${}^{-1}$ of data at 7 TeV}\/},
ATLAS-CONF-2012-134.

\bibitem{top_xsec7}
{ATLAS} Collaboration, {\em {Measurement of the $t\bar{t}$ production
  cross-section in $pp$ collisions at $\sqrt{s}=8$ TeV using $e\mu$ events with
  $b$-tagged jets}\/},
ATLAS-CONF-2013-097.

\bibitem{top_xsec8}
{ATLAS} Collaboration, {\em {Measurement of the top quark pair production cross
  section in the single-lepton channel with ATLAS in proton-proton collisions
  at 8 TeV using kinematic fits with b-tagging}\/},
ATLAS-CONF-2012-149.

\bibitem{top_xsec9}
{CMS} Collaboration, {\em {Top pair cross section in e/mu+jets at 8 TeV}\/},
CMS-PAS-TOP-12-006.

\bibitem{top_xsec10}
{CMS} Collaboration, {\em {Measurement of the $t \bar{t}$ production cross
  section in the dilepton channel in pp collisions at $\sqrt{s}$ = 8 TeV}\/},
\href{http://dx.doi.org/10.1007/JHEP02(2014)024}{JHEP {\bf 1402} (2014)  024}.

\bibitem{top_summary_plots}
{ATLAS} Collaboration.
  \url{https://atlas.web.cern.ch/Atlas/GROUPS/PHYSICS/CombinedSummaryPlots/TOP/}.
\newblock As of 26.03.2014.

\bibitem{st_xsec_theo_s}
N.~Kidonakis, {\em {NNLL resummation for s-channel single top quark
  production}\/},
\href{http://dx.doi.org/10.1103/PhysRevD.81.054028}{Phys.Rev. {\bf D81} (2010)
  054028}.

\bibitem{st_xsec_theo_t}
N.~Kidonakis, {\em {Next-to-next-to-leading-order collinear and soft gluon
  corrections for t-channel single top quark production}\/},
\href{http://dx.doi.org/10.1103/PhysRevD.83.091503}{Phys.Rev. {\bf D83} (2011)
  091503}.

\bibitem{st_xsec_theo_Wt}
N.~Kidonakis, {\em {Two-loop soft anomalous dimensions for single top quark
  associated production with a $W^{-}$ or $H^{-}$}\/},
\href{http://dx.doi.org/10.1103/PhysRevD.82.054018}{Phys.Rev. {\bf D82} (2010)
  054018}.

\bibitem{st_xsec1}
{CDF and D0} Collaboration, {\em {Observation of s-channel production of single
  top quarks at the Tevatron}\/},
\href{http://dx.doi.org/10.1103/PhysRevLett.112.231803}{Phys.Rev.Lett. {\bf
  112} (2014)  231803}.

\bibitem{st_xsec2}
{CDF} Collaboration, {\em Measurement of Single Top Quark Production in 7.5
  fb${}^{-1}$ of CDF Data Using Neural Networks\/},  CDF Conference Note 10794.

\bibitem{st_xsec3}
{D0} Collaboration, {\em {Evidence for s-channel single top quark production in
  $p\bar{p}$ collisions at $\sqrt{s}$ = 1.96 TeV}\/},
\href{http://dx.doi.org/10.1016/j.physletb.2013.09.048}{Phys.Lett. {\bf B726}
  (2013)  656}.

\bibitem{st_xsec6}
{ATLAS} Collaboration, {\em {Search for s-Channel Single Top-Quark Production
  in $pp$ Collisions at $\sqrt{s}$ = 7 TeV}\/},
ATLAS-CONF-2011-118.

\bibitem{st_xsec5}
{ATLAS} Collaboration, {\em {Measurement of the $t$-channel single top-quark
  production cross section in $pp$ collisions at $\sqrt{s}=7$ TeV with the
  ATLAS detector}\/},
\href{http://dx.doi.org/10.1016/j.physletb.2012.09.031}{Phys.Lett. {\bf B717}
  (2012)  330}.

\bibitem{st_xsec7}
{ATLAS} Collaboration, {\em {Evidence for the associated production of a $W$
  boson and a top quark in ATLAS at $\sqrt{s}=7$ TeV}\/},
\href{http://dx.doi.org/10.1016/j.physletb.2012.08.011}{Phys.Lett. {\bf B716}
  (2012)  142}.

\bibitem{st_xsec8}
{CMS} Collaboration, {\em {Measurement of the single-top-quark $t$-channel
  cross section in $pp$ collisions at $\sqrt{s}=7$ TeV}\/},
\href{http://dx.doi.org/10.1007/JHEP12(2012)035}{JHEP {\bf 1212} (2012)  035}.

\bibitem{st_xsec9}
{CMS} Collaboration, {\em {Evidence for associated production of a single top
  quark and W boson in pp collisions at $\sqrt{s}$ = 7 TeV}\/},
\href{http://dx.doi.org/10.1103/PhysRevLett.110.022003}{Phys.Rev.Lett. {\bf
  110} (2013)  022003}.

\bibitem{st_tchan_ATLAS_new}
{ATLAS} Collaboration, {\em {Measurement of the Inclusive and Fiducial
  Cross-Section of Single Top-Quark $t$-Channel Events in $pp$ Collisions at
  $\sqrt{s}$ = 8 TeV}\/},
ATLAS-CONF-2014-007.

\bibitem{st_xsec11}
{ATLAS} Collaboration, {\em {Measurement of the cross-section for associated
  production of a top quark and a W boson at $\sqrt{s}=8$ TeV with the ATLAS
  detector}\/},
ATLAS-CONF-2013-100.

\bibitem{st_xsec10}
{CMS} Collaboration, {\em {Measurement of the single top s-channel cross
  section at 8 TeV}\/},
CMS-PAS-TOP-13-009.

\bibitem{st_tchan_CMS_new}
{CMS} Collaboration, {\em {Measurement of the t-channel single-top-quark
  production cross section and of the $\left|V_{tb}\right|$ CKM matrix element
  in pp collisions at $\sqrt{s}$ = 8 TeV}\/},
  \href{http://arxiv.org/abs/1403.7366}{{\tt arXiv:1403.7366 [hep-ex]}}.
Submitted to JHEP.

\bibitem{st_xsec12}
{CMS} Collaboration, {\em {Observation of the associated production of a single
  top quark and a W boson in pp collisions at $\sqrt{s}$ = 8 TeV}\/},
  \href{http://arxiv.org/abs/1401.2942}{{\tt arXiv:1401.2942 [hep-ex]}}.
Submitted to Phys. Rev. Lett.

\bibitem{st_xsec4}
{ATLAS Collaboration and CMS} Collaboration, {\em {Combination of single
  top-quark cross-sections measurements in the t-channel at $\sqrt{s}$=8 TeV
  with the ATLAS and CMS experiments}\/},
CMS-PAS-TOP-12-002, ATLAS-CONF-2013-098.

\bibitem{neutrinoweighting}
{D0} Collaboration, {\em {Measurement of the top quark mass using dilepton
  events}\/},
\href{http://dx.doi.org/10.1103/PhysRevLett.80.2063}{Phys.Rev.Lett. {\bf 80}
  (1998)  2063}.

\bibitem{m_top_tevatron}
{Tevatron Electroweak Working Group, CDF Collaboration, D0} Collaboration, {\em
  {Combination of CDF and D0 results on the mass of the top quark using up to
  8.7 fb$^{-1}$ at the Tevatron}\/},  FERMILAB-CONF-13-164-PPD-TD,
\href{http://arxiv.org/abs/1305.3929}{{\tt arXiv:1305.3929 [hep-ex]}}.

\bibitem{m_top_LHC}
{ATLAS} Collaboration, {\em {Combination of ATLAS and CMS results on the mass
  of the top-quark using up to 4.9 fb$^{-1}$ of $\sqrt{s}=7$ TeV LHC data}\/},
ATLAS-CONF-2013-102.

\bibitem{m_top_world_input1}
{CDF} Collaboration, {\em {Precision Top-Quark Mass Measurements at CDF}\/},
\href{http://dx.doi.org/10.1103/PhysRevLett.109.152003}{Phys.Rev.Lett. {\bf
  109} (2012)  152003}.

\bibitem{m_top_world_input2}
{CDF} Collaboration, {\em {Top quark mass measurement using the template method
  at CDF}\/},
\href{http://dx.doi.org/10.1103/PhysRevD.83.111101}{Phys.Rev. {\bf D83} (2011)
  111101}.

\bibitem{m_top_world_input3}
{CDF} Collaboration, {\em {Measurement of the Top Quark Mass in the
  All-Hadronic Mode at CDF}\/},
\href{http://dx.doi.org/10.1016/j.physletb.2012.06.007}{Phys.Lett. {\bf B714}
  (2012)  24}.

\bibitem{m_top_world_input4}
{CDF} Collaboration, {\em {Top-quark mass measurement in events with jets and
  missing transverse energy using the full CDF data set}\/},
\href{http://dx.doi.org/10.1103/PhysRevD.88.011101}{Phys.Rev. {\bf D88} (2013)
  no.~1, 011101}.

\bibitem{m_top_world_input5}
{D0} Collaboration, {\em {Precise measurement of the top-quark mass from
  lepton+jets events at D0}\/},
\href{http://dx.doi.org/10.1103/PhysRevD.84.032004}{Phys.Rev. {\bf D84} (2011)
  032004}.

\bibitem{m_top_world_input6}
{D0} Collaboration, {\em {Measurement of the top quark mass in $p \bar{p}$
  collisions using events with two leptons}\/},
\href{http://dx.doi.org/10.1103/PhysRevD.86.051103}{Phys.Rev. {\bf D86} (2012)
  051103}.

\bibitem{m_top_world_input7}
{ATLAS} Collaboration, {\em {Measurement of the Top Quark Mass from
  $\sqrt{s}=7$ TeV ATLAS Data using a 3-dimensional Template Fit}\/},
ATLAS-CONF-2013-046.

\bibitem{m_top_world_input8}
{ATLAS} Collaboration, {\em {Measurement of the Top Quark Mass in Dileptonic
  Top Quark Pair Decays with $\sqrt{s}=7$ TeV ATLAS Data}\/},
ATLAS-CONF-2013-077.

\bibitem{m_top_world_input9}
{CMS} Collaboration, {\em {Measurement of the top-quark mass in $t\bar{t}$
  events with lepton+jets final states in $pp$ collisions at $\sqrt{s}=7$
  TeV}\/},
\href{http://dx.doi.org/10.1007/JHEP12(2012)105}{JHEP {\bf 1212} (2012)  105}.

\bibitem{m_top_world_input10}
{CMS} Collaboration, {\em {Measurement of the top-quark mass in $t\bar{t}$
  events with dilepton final states in $pp$ collisions at $\sqrt{s}=7$ TeV}\/},
\href{http://dx.doi.org/10.1140/epjc/s10052-012-2202-z}{Eur.Phys.J. {\bf C72}
  (2012)  2202}.

\bibitem{m_top_world_input11}
{CMS} Collaboration, {\em {Measurement of the top-quark mass in all-jets
  $t\bar{t}$ events in pp collisions at $\sqrt{s}$=7 TeV}\/},
\href{http://dx.doi.org/10.1140/epjc/s10052-014-2758-x}{Eur.Phys.J. {\bf C74}
  (2014)  2758}.

\bibitem{top_charge_CDF}
{CDF} Collaboration, {\em {Exclusion of exotic top-like quarks with -4/3
  electric charge using jet-charge tagging in single-lepton $t\bar{t}$ events
  at CDF}\/},
\href{http://dx.doi.org/10.1103/PhysRevD.88.032003}{Phys.Rev. {\bf D88} (2013)
  no.~3, 032003}.

\bibitem{top_charge_D0}
{D0} Collaboration, {\em {Experimental discrimination between charge 2e/3 top
  quark and charge 4e/3 exotic quark production scenarios}\/},
\href{http://dx.doi.org/10.1103/PhysRevLett.98.041801}{Phys.Rev.Lett. {\bf 98}
  (2007)  041801}.

\bibitem{top_charge_CMS}
{CMS} Collaboration, {\em {Constraints on the Top-Quark Charge from Top-Pair
  Events}\/},
CMS-PAS-TOP-11-031.

\bibitem{top_charge_ATLAS}
{ATLAS} Collaboration, {\em {Measurement of the top quark charge in $pp$
  collisions at $\sqrt{s} =$ 7 TeV with the ATLAS detector}\/},
\href{http://dx.doi.org/10.1007/JHEP11(2013)031}{JHEP {\bf 1311} (2013)  031}.

\bibitem{top_asymm_theo}
J.~H. K{\"u}hn and G.~Rodrigo, {\em {Charge asymmetry of heavy quarks at hadron
  colliders}\/},
\href{http://dx.doi.org/10.1103/PhysRevD.59.054017}{Phys.Rev. {\bf D59} (1999)
  054017}.

\bibitem{top_asymm_theo2}
J.~H. K{\"u}hn and G.~Rodrigo, {\em {Charge asymmetry of heavy quarks at hadron
  colliders}\/},
\href{http://dx.doi.org/10.1103/PhysRevD.59.054017}{Phys.Rev. {\bf D59} (1999)
  054017}.

\bibitem{top_asymm_CDF}
{CDF} Collaboration, {\em {Measurement of the top quark forward-backward
  production asymmetry and its dependence on event kinematic properties}\/},
\href{http://dx.doi.org/10.1103/PhysRevD.87.092002}{Phys.Rev. {\bf D87} (2013)
  092002}.

\bibitem{top_asymm_lep_CDF}
{CDF} Collaboration, {\em Combination of Leptonic $A_{FB}$ of $t\bar{t}$ at
  CDF\/},  CDF Conference Note 11035.

\bibitem{top_asymm_D0}
{CDF} Collaboration, {\em {Measurement of the top quark forward-backward
  production asymmetry and its dependence on event kinematic properties}\/},
\href{http://dx.doi.org/10.1103/PhysRevD.87.092002}{Phys.Rev. {\bf D87} (2013)
  092002}.

\bibitem{top_asymm_lep_D0}
{D0} Collaboration, {\em {Measurement of the forward-backward asymmetry in the
  distribution of leptons in $t\bar{t}$ events in the lepton+jets channel}\/},
  FERMILAB-PUB-14-041-E,
\href{http://arxiv.org/abs/1403.1294}{{\tt arXiv:1403.1294 [hep-ex]}}.

\bibitem{top_asymm_LHC}
{ATLAS Collaboration and CMS} Collaboration, {\em {Combination of ATLAS and CMS
  $t\bar{t}$ charge asymmetry measurements using LHC proton-proton collisions
  at $\sqrt{s}=7$ TeV}\/},
ATLAS-CONF-2014-012, CMS-PAS-TOP-14-006.

\bibitem{top_asymm_BSM1}
J.~Aguilar-Saavedra and M.~Perez-Victoria, {\em {Simple models for the top
  asymmetry: Constraints and predictions}\/},
\href{http://dx.doi.org/10.1007/JHEP09(2011)097}{JHEP {\bf 1109} (2011)  097}.

\bibitem{top_asymm_BSM2}
J.~Aguilar-Saavedra and M.~Perez-Victoria, {\em {Asymmetries in t $\bar{t}$
  production: LHC versus Tevatron}\/},
\href{http://dx.doi.org/10.1103/PhysRevD.84.115013}{Phys.Rev. {\bf D84} (2011)
  115013}.

\bibitem{top_asymm_D0_old}
{D0} Collaboration, {\em {Forward-backward asymmetry in top quark-antiquark
  production}\/},
\href{http://dx.doi.org/10.1103/PhysRevD.84.112005}{Phys.Rev. {\bf D84} (2011)
  112005}.

\bibitem{top_asymm_ATLAS}
{ATLAS} Collaboration, {\em {Measurement of the top quark pair production
  charge asymmetry in proton-proton collisions at $\sqrt{s}$ = 7 TeV using the
  ATLAS detector}\/},
\href{http://dx.doi.org/10.1007/JHEP02(2014)107}{JHEP {\bf 1402} (2014)  107}.

\bibitem{top_asymm_CMS}
{CMS} Collaboration, {\em {Inclusive and differential measurements of the $t
  \bar{t}$ charge asymmetry in proton-proton collisions at 7 TeV}\/},
\href{http://dx.doi.org/10.1016/j.physletb.2012.09.028}{Phys.Lett. {\bf B717}
  (2012)  129}.

\bibitem{ttgamma_CDF}
{CDF} Collaboration, {\em {Evidence for $t\bar{t}\gamma$ Production and
  Measurement of $\sigma_{t\bar{t}\gamma} / \sigma_{t\bar{t}}$}\/},
\href{http://dx.doi.org/10.1103/PhysRevD.84.031104}{Phys.Rev. {\bf D84} (2011)
  031104}.

\bibitem{ttgamma_ATLAS}
{ATLAS} Collaboration, {\em {Measurement of the inclusive $t\bar{t}\gamma$
  cross section with the ATLAS detector}\/},
ATLAS-CONF-2011-153.

\bibitem{ttgamma_CMS}
{CMS} Collaboration, {\em {Measurement of the inclusive top-quark pair + photon
  production cross section in the muon + jets channel in pp collisions at 8
  TeV}\/},
CMS-PAS-TOP-13-011.

\bibitem{ttZ_ATLAS}
{ATLAS} Collaboration, {\em {Search for $t\bar{t}Z$ production in the three
  lepton final state with $4.7$ ${\rm fb}^{-1}$ of $\sqrt{s}=7$ TeV $pp$
  collision data collected by the ATLAS detector}\/},
ATLAS-CONF-2012-126.

\bibitem{ttV_CMS}
{CMS} Collaboration, {\em {Measurement of associated production of vector
  bosons and top quark-antiquark pairs at $\sqrt{s}$ = 7 TeV}\/},
\href{http://dx.doi.org/10.1103/PhysRevLett.110.172002}{Phys.Rev.Lett. {\bf
  110} (2013)  172002}.

\bibitem{WHel_theo2}
G.~L. Kane, G.~Ladinsky, and C.~Yuan, {\em {Using the Top Quark for Testing
  Standard Model Polarization and CP Predictions}\/},
\href{http://dx.doi.org/10.1103/PhysRevD.45.124}{Phys.Rev. {\bf D45} (1992)
  124}.

\bibitem{WHel_theo}
A.~Czarnecki, J.~G. Korner, and J.~H. Piclum, {\em {Helicity fractions of W
  bosons from top quark decays at NNLO in QCD}\/},
\href{http://dx.doi.org/10.1103/PhysRevD.81.111503}{Phys.Rev. {\bf D81} (2010)
  111503}.

\bibitem{WHel_tevatron}
{CDF Collaboration, D0} Collaboration, {\em {Combination of CDF and D0
  measurements of the $W$ boson helicity in top quark decays}\/},
\href{http://dx.doi.org/10.1103/PhysRevD.85.071106}{Phys.Rev. {\bf D85} (2012)
  071106}.

\bibitem{WHel_LHC}
{ATLAS} Collaboration, {\em {Combination of the ATLAS and CMS measurements of
  the W-boson polarization in top-quark decays}\/},
ATLAS-CONF-2013-033.

\bibitem{gammatop_D0}
{D0} Collaboration, {\em {An Improved determination of the width of the top
  quark}\/},
\href{http://dx.doi.org/10.1103/PhysRevD.85.091104}{Phys.Rev. {\bf D85} (2012)
  091104}.

\bibitem{gammatop_CMS}
{CMS} Collaboration, {\em {Measurement of the ratio B($t \rightarrow Wb$)/B($t
  \rightarrow Wq$) in $pp$ collisions at $\sqrt{s}$ = 8 TeV}\/},
  \href{http://arxiv.org/abs/1404.2292}{{\tt arXiv:1404.2292 [hep-ex]}}.
Submitted to Phys. Lett. B.

\bibitem{Bernreuther2001}
W.~Bernreuther et al., {\em {Top quark spin correlations at hadron colliders:
  Predictions at next-to-leading order QCD}\/},
\href{http://dx.doi.org/10.1103/PhysRevLett.87.242002}{Phys.Rev.Lett. {\bf 87}
  (2001)  242002}.

\bibitem{leadingpole}
A.~Aeppli, G.~J. van Oldenborgh, and D.~Wyler, {\em {Unstable particles in one
  loop calculations}\/},
\href{http://dx.doi.org/10.1016/0550-3213(94)90195-3}{Nucl.Phys. {\bf B428}
  (1994)  126}.

\bibitem{Baumgart2013}
M.~Baumgart and B.~Tweedie, {\em {A New Twist on Top Quark Spin
  Correlations}\/},
\href{http://dx.doi.org/10.1007/JHEP03(2013)117}{JHEP {\bf 1303} (2013)  117}.

\bibitem{Bernreuther2010}
W.~Bernreuther and Z.-G. Si, {\em {Top quark spin correlations and polarization
  at the LHC: standard model predictions and effects of anomalous top chromo
  moments}\/},
\href{http://dx.doi.org/10.1016/j.physletb.2013.06.051}{Phys.Lett. {\bf B725}
  (2013) no.~1-3, 115}.

\bibitem{single_top_pol}
G.~Mahlon and S.~J. Parke, {\em {Single top quark production at the LHC:
  Understanding spin}\/},
\href{http://dx.doi.org/10.1016/S0370-2693(00)00149-0}{Phys.Lett. {\bf B476}
  (2000)  323}.

\bibitem{Bernreuther1996}
W.~Bernreuther, A.~Brandenburg, and P.~Uwer, {\em {Transverse polarization of
  top quark pairs at the Tevatron and the large hadron collider}\/},
\href{http://dx.doi.org/10.1016/0370-2693(95)01475-6}{Phys.Lett. {\bf B368}
  (1996)  153}.

\bibitem{Mahlon1996}
G.~Mahlon and S.~J. Parke, {\em {Angular correlations in top quark pair
  production and decay at hadron colliders}\/},
\href{http://dx.doi.org/10.1103/PhysRevD.53.4886}{Phys.Rev. {\bf D53} (1996)
  4886}.

\bibitem{Hara1991}
Y.~Hara, {\em {Angular Correlation of Charged Leptons From T Anti-t Produced in
  the Gluon Fusion}\/},
\href{http://dx.doi.org/10.1143/PTP.86.779}{Prog.Theor.Phys. {\bf 86} (1991)
  779}.

\bibitem{Arens1993}
T.~Arens and L.~Sehgal, {\em {Azimuthal correlation of charged leptons produced
  in p anti-p $\rightarrow$ t anti-t + ...}\/},
\href{http://dx.doi.org/10.1016/0370-2693(93)90433-I}{Phys.Lett. {\bf B302}
  (1993)  501}.

\bibitem{Bernreuther2004}
W.~Bernreuther, A.~Brandenburg, Z.~Si, and P.~Uwer, {\em {Top quark pair
  production and decay at hadron colliders}\/},
\href{http://dx.doi.org/10.1016/j.nuclphysb.2004.04.019}{Nucl.Phys. {\bf B690}
  (2004)  81}.

\bibitem{Stelzer1996}
T.~Stelzer and S.~Willenbrock, {\em {Spin correlation in top quark production
  at hadron colliders}\/},
\href{http://dx.doi.org/10.1016/0370-2693(96)00178-5}{Phys.Lett. {\bf B374}
  (1996)  169}.

\bibitem{Mahlon1997}
G.~Mahlon and S.~J. Parke, {\em {Maximizing spin correlations in top quark pair
  production at the Tevatron}\/},
\href{http://dx.doi.org/10.1016/S0370-2693(97)00987-8}{Phys.Lett. {\bf B411}
  (1997)  173}.

\bibitem{Parke1996}
S.~J. Parke and Y.~Shadmi, {\em {Spin correlations in top quark pair production
  at $e^{+} e^{-}$ colliders}\/},
\href{http://dx.doi.org/10.1016/0370-2693(96)00998-7}{Phys.Lett. {\bf B387}
  (1996)  199}.

\bibitem{Uwer2005}
P.~Uwer, {\em {Maximizing the spin correlation of top quark pairs produced at
  the Large Hadron Collider}\/},
\href{http://dx.doi.org/10.1016/j.physletb.2005.01.005}{Phys.Lett. {\bf B609}
  (2005)  271}.

\bibitem{ueberpaper}
{ATLAS} Collaboration, {\em {Measurements of spin correlation in top-antitop
  quark events from proton-proton collisions at $\sqrt{s}=7$ TeV using the
  ATLAS detector}\/},  \href{http://arxiv.org/abs/1407.4314}{{\tt
  arXiv:1407.4314 [hep-ex]}}.
Submitted to Phys. Rev. D.

\bibitem{Jezabek1994}
M.~Jezabek and J.~H. K{\"u}hn, {\em {V-A tests through leptons from polarized
  top quarks}\/},
\href{http://dx.doi.org/10.1016/0370-2693(94)90779-X}{Phys.Lett. {\bf B329}
  (1994)  317}.

\bibitem{Jezabek1989}
M.~Jezabek and J.~H. K{\"u}hn, {\em {Lepton Spectra from Heavy Quark Decay}\/},
\href{http://dx.doi.org/10.1016/0550-3213(89)90209-5}{Nucl.Phys. {\bf B320}
  (1989)  20}.

\bibitem{Brandenburg2002}
A.~Brandenburg, Z.~Si, and P.~Uwer, {\em {QCD corrected spin analyzing power of
  jets in decays of polarized top quarks}\/},
\href{http://dx.doi.org/10.1016/S0370-2693(02)02098-1}{Phys.Lett. {\bf B539}
  (2002)  235}.

\bibitem{Czarnecki1990}
A.~Czarnecki, M.~Jezabek, and J.~H. K{\"u}hn, {\em {Lepton Spectra From Decays
  of Polarized Top Quarks}\/},
\href{http://dx.doi.org/10.1016/0550-3213(91)90082-9}{Nucl.Phys. {\bf B351}
  (1991)  70}.

\bibitem{Hubaut2005}
F.~Hubaut et al., {\em {ATLAS sensitivity to top quark and $W$ boson
  polarization in $t \bar{t}$ events}\/},
\href{http://dx.doi.org/10.1140/epjcd/s2005-02-009-9}{Eur.Phys.J. {\bf C44S2}
  (2005)  13}.

\bibitem{Barger1989}
V.~D. Barger, J.~Ohnemus, and R.~Phillips, {\em {Spin Correlation Effects in
  the Hadroproduction and Decay of Very Heavy Top Quark Pairs}\/},
\href{http://dx.doi.org/10.1142/S0217751X89000297}{Int.J.Mod.Phys. {\bf A4}
  (1989)  617}.

\bibitem{ATLAS_spin_paper}
{ATLAS} Collaboration, {\em {Observation of spin correlation in $t \bar{t}$
  events from pp collisions at $\sqrt{s}$ = 7 TeV using the ATLAS detector}\/},
\href{http://dx.doi.org/10.1103/PhysRevLett.108.212001}{Phys.Rev.Lett. {\bf
  108} (2012)  212001}.

\bibitem{CMS_spin_paper}
{CMS} Collaboration, {\em {Measurements of $t\bar{t}$ spin correlations and
  top-quark polarization using dilepton final states in pp collisions at
  $\sqrt{s}$ = 7 TeV}\/},
\href{http://dx.doi.org/10.1103/PhysRevLett.112.182001}{Phys.Rev.Lett. {\bf
  112} (2014)  182001}.

\bibitem{CMS_spin_prelim}
{CMS} Collaboration, {\em {Measurement of Spin Correlations in ttbar
  production}\/},
CMS-PAS-TOP-12-004.

\bibitem{Baumgart2011}
M.~Baumgart and B.~Tweedie, {\em {Discriminating Top-Antitop Resonances using
  Azimuthal Decay Correlations}\/},
\href{http://dx.doi.org/10.1007/JHEP09(2011)049}{JHEP {\bf 1109} (2011)  049}.

\bibitem{Tavares2011}
G.~Marques~Tavares and M.~Schmaltz, {\em {Explaining the $t$-$\bar{t}$
  asymmetry with a light axigluon}\/},
\href{http://dx.doi.org/10.1103/PhysRevD.84.054008}{Phys.Rev. {\bf D84} (2011)
  054008}.

\bibitem{Frampton1987}
P.~H. Frampton and S.~L. Glashow, {\em {Chiral Color: An Alternative to the
  Standard Model}\/},
\href{http://dx.doi.org/10.1016/0370-2693(87)90859-8}{Phys.Lett. {\bf B190}
  (1987)  157}.

\bibitem{Arai2009}
M.~Arai et al., {\em {Influence of $Z^\prime$ boson on top quark spin
  correlations at the LHC}\/},
Acta Phys.Polon. {\bf B40} (2009)  93.

\bibitem{Fitzpatrick2007}
A.~L. Fitzpatrick et al., {\em {Searching for the Kaluza-Klein Graviton in Bulk
  RS Models}\/},
\href{http://dx.doi.org/10.1088/1126-6708/2007/09/013}{JHEP {\bf 0709} (2007)
  013}.

\bibitem{Randall1999}
L.~Randall and R.~Sundrum, {\em {A Large mass hierarchy from a small extra
  dimension}\/},
\href{http://dx.doi.org/10.1103/PhysRevLett.83.3370}{Phys.Rev.Lett. {\bf 83}
  (1999)  3370}.

\bibitem{Gao2010}
J.~Gao et al., {\em {Next-to-leading order QCD corrections to the heavy
  resonance production and decay into top quark pair at the LHC}\/},
\href{http://dx.doi.org/10.1103/PhysRevD.82.014020}{Phys.Rev. {\bf D82} (2010)
  014020}.

\bibitem{Bernreuther2013}
W.~Bernreuther and Z.-G. Si, {\em {Top quark spin correlations and polarization
  at the LHC: standard model predictions and effects of anomalous top chromo
  moments}\/},
\href{http://dx.doi.org/10.1016/j.physletb.2013.06.051}{Phys.Lett. {\bf B725}
  (2013) no.~1-3, 115}.

\bibitem{Fajfer2012}
S.~Fajfer, J.~F. Kamenik, and B.~Melic, {\em {Discerning New Physics in
  Top-Antitop Production using Top Spin Observables at Hadron Colliders}\/},
\href{http://dx.doi.org/10.1007/JHEP08(2012)114}{JHEP {\bf 1208} (2012)  114}.

\bibitem{stop}
Z.~Han et al., {\em {(Light) Stop Signs}\/},
\href{http://dx.doi.org/10.1007/JHEP08(2012)083}{JHEP {\bf 1208} (2012)  083}.

\bibitem{Eriksson2007}
D.~Eriksson et al., {\em {New angles on top quark decay to a charged Higgs}\/},
\href{http://dx.doi.org/10.1088/1126-6708/2008/01/024}{JHEP {\bf 0801} (2008)
  024}.

\bibitem{top_pol_D0}
{D0} Collaboration, {\em {Measurement of Leptonic Asymmetries and Top Quark
  Polarization in $t\bar{t}$ Production}\/},
\href{http://dx.doi.org/10.1103/PhysRevD.87.011103}{Phys.Rev. {\bf D87} (2013)
  011103}.

\bibitem{ATLAS_top_pol}
{ATLAS} Collaboration, {\em {Measurement of top quark polarization in
  top-antitop events from proton-proton collisions at $\sqrt{s}$ = 7 TeV using
  the ATLAS detector}\/},
\href{http://dx.doi.org/10.1103/PhysRevLett.111.232002}{Phys.Rev.Lett. {\bf
  111} (2013)  232002}.

\bibitem{D0_spin_dilep_angles}
{D0} Collaboration, {\em {Measurement of spin correlation in $t\bar{t}$
  production using dilepton final states}\/},
\href{http://dx.doi.org/10.1016/j.physletb.2011.05.077}{Phys.Lett. {\bf B702}
  (2011)  16}.

\bibitem{D0_spin_dilep_MEM}
{D0} Collaboration, {\em {Measurement of spin correlation in $t\bar{t}$
  production using a matrix element approach}\/},
\href{http://dx.doi.org/10.1103/PhysRevLett.107.032001}{Phys.Rev.Lett. {\bf
  107} (2011)  032001}.

\bibitem{D0_spin_ljets}
{D0} Collaboration, {\em {Evidence for spin correlation in $t\bar{t}$
  production}\/},
\href{http://dx.doi.org/10.1103/PhysRevLett.108.032004}{Phys.Rev.Lett. {\bf
  108} (2012)  032004}.

\bibitem{CDF_spin_dilep}
{CDF} Collaboration, {\em Measurement of $t\bar{t}$ Spin Correlations
  Coefficient in 5.1 fb$^{-1}$ Dilepton Candidates\/},  CDF Conference Note
  10719.

\bibitem{CDF_spin_ljets}
{CDF} Collaboration, {\em Measurement of $t\bar{t}$ Helicity Fractions and Spin
  Correlation Using Reconstructed Lepton+Jets Events\/},  CDF Conference Note
  10211.

\bibitem{LHC}
L.~Evans and P.~Bryant, {\em {LHC Machine}\/},
\href{http://dx.doi.org/10.1088/1748-0221/3/08/S08001}{JINST {\bf 3} (2008)
  S08001}.

\bibitem{LS1}
M.~Lamont, {\em {Status of the LHC}\/},
\href{http://dx.doi.org/10.1088/1742-6596/455/1/012001}{J.Phys.Conf.Ser. {\bf
  455} (2013)  012001}.

\bibitem{top_xsec_lhccomb_atlas}
{ATLAS} Collaboration, {\em {Combination of ATLAS and CMS top-quark pair cross
  section measurements using up to 1.1 fb${}^{-1}$ of data at 7 TeV}\/},
ATLAS-CONF-2012-134.

\bibitem{top_xsec_lhccomb_cms}
{CMS} Collaboration, {\em {Combination of ATLAS and CMS top-quark pair cross
  section measurements using proton-proton collisions at $\sqrt{s}$ = 7
  TeV}\/},
CMS-PAS-TOP-12-003.

\bibitem{LHC2}
S.~Myers, {\em The engineering needed for particle physics\/},
  \href{http://dx.doi.org/10.1098/rsta.2011.0053}{Phil.Trans.Roy.Soc.Lond. {\bf
  A370} (2012) no.~1973, 3887}.

\bibitem{ATLAS}
{ATLAS} Collaboration, {\em {The ATLAS Experiment at the CERN Large Hadron
  Collider}\/},
\href{http://dx.doi.org/10.1088/1748-0221/3/08/S08003}{JINST {\bf 3} (2008)
  S08003}.

\bibitem{CMS}
{CMS} Collaboration, {\em {The CMS experiment at the CERN LHC}\/},
\href{http://dx.doi.org/10.1088/1748-0221/3/08/S08004}{JINST {\bf 3} (2008)
  S08004}.

\bibitem{ALICE}
{ALICE} Collaboration, {\em {The ALICE experiment at the CERN LHC}\/},
\href{http://dx.doi.org/10.1088/1748-0221/3/08/S08002}{JINST {\bf 3} (2008)
  S08002}.

\bibitem{LHCb}
{LHCb} Collaboration, {\em {The LHCb Detector at the LHC}\/},
\href{http://dx.doi.org/10.1088/1748-0221/3/08/S08005}{JINST {\bf 3} (2008)
  S08005}.

\bibitem{MOEDAL}
{MoEDAL} Collaboration, {\em {Technical Design Report of the MoEDAL
  Experiment}\/},  CERN-LHCC-2009-006. MoEDAL-TDR-001.

\bibitem{LHCf}
{LHCf} Collaboration, {\em {The LHCf detector at the CERN Large Hadron
  Collider}\/},
\href{http://dx.doi.org/10.1088/1748-0221/3/08/S08006}{JINST {\bf 3} (2008)
  S08006}.

\bibitem{TOTEM}
{TOTEM} Collaboration, {\em {TOTEM: Technical design report. Total cross
  section, elastic scattering and diffraction dissociation at the Large Hadron
  Collider at CERN}\/},
CERN-LHCC-2004-002.

\bibitem{pixelperf}
{ATLAS} Collaboration, {\em {Performance of the ATLAS Inner Detector Track and
  Vertex Reconstruction in the High Pile-Up LHC Environment}\/},
ATLAS-CONF-2012-042.

\bibitem{transrad}
B.~Dolgoshein, {\em {Transition radiation detectors}\/},
\href{http://dx.doi.org/10.1016/0168-9002(93)90846-A}{Nucl.Instrum.Meth. {\bf
  A326} (1993)  434}.

\bibitem{IBL}
M.~Capeans et al., {\em {ATLAS Insertable B-Layer Technical Design Report}\/},
  CERN-LHCC-2010-013. ATLAS-TDR-19.

\bibitem{solenoid}
{ATLAS} Collaboration, {\em {ATLAS central solenoid: Technical design
  report}\/},
CERN-LHCC-97-21.

\bibitem{barrel_toroid}
{ATLAS} Collaboration, {\em {ATLAS barrel toroid: Technical design report}\/},
CERN-LHCC-97-19.

\bibitem{lumi_7tev}
{ATLAS} Collaboration, {\em {Luminosity Determination in $pp$ Collisions at
  $\sqrt{s}=7$ TeV Using the ATLAS Detector at the LHC}\/},
\href{http://dx.doi.org/10.1140/epjc/s10052-011-1630-5}{Eur.Phys.J. {\bf C71}
  (2011)  1630}.

\bibitem{LUCID}
P.~Jenni and M.~Nessi, {\em {ATLAS Forward Detectors for Luminosity Measurement
  and Monitoring}\/},  CERN-LHCC-2004-010. LHCC-I-014.

\bibitem{ZDC}
{ATLAS} Collaboration, {\em {Zero degree calorimeters for ATLAS}\/},
CERN-LHCC-2007-01.

\bibitem{BCM}
V.~Cindro et al., {\em {The ATLAS beam conditions monitor}\/},
\href{http://dx.doi.org/10.1088/1748-0221/3/02/P02004}{JINST {\bf 3} (2008)
  P02004}.

\bibitem{ATLAS_lumi}
{ATLAS} Collaboration, {\em {Luminosity Determination Using the ATLAS
  Detector}\/},
ATLAS-CONF-2010-060.

\bibitem{improved_lumi}
{ATLAS} Collaboration, {\em {Improved Luminosity Determination in $\mathbf{pp}$
  Collisions at $\mathbf{\sqrt s = 7}$ TeV using the ATLAS Detector at the
  LHC}\/},
ATLAS-CONF-2012-080.

\bibitem{ALFA}
{ATLAS ALFA} Collaboration, {\em {ALFA: Absolute Luminosity For ATLAS:
  Development of a scintillating fibre tracker to determine the absolute LHC
  luminosity at ATLAS}\/},
\href{http://dx.doi.org/10.1016/j.nuclphysbps.2009.10.110}{Nucl.Phys.Proc.Suppl.
  {\bf 197} (2009)  387}.

\bibitem{vdmscan}
S.~van~der Meer, {\em {Calibration of the Effective Beam Height in the ISR}\/},
CERN-ISR-PO-68-31.

\bibitem{beamshape}
H.~Burkhardt and P.~Grafstrom, {\em {Absolute luminosity from machine
  parameters}\/},
CERN-LHC-PROJECT-REPORT-1019.

\bibitem{lemmerbuch}
B.~Lemmer, \href{http://dx.doi.org/10.1007/978-3-642-37714-3}{{\em Bis(s) ins
  Innere des Protons}}.
\newblock Springer Berlin Heidelberg, 2014.

\bibitem{atlas_public_eventdisplays}
\url{https://twiki.cern.ch/twiki/bin/view/AtlasPublic/EventDisplayPublicResults}.
\newblock As of 28.11.2013.

\bibitem{reco_electrons}
{ATLAS} Collaboration, {\em {Electron performance measurements with the ATLAS
  detector using the 2010 LHC proton-proton collision data}\/},
\href{http://dx.doi.org/10.1140/epjc/s10052-012-1909-1}{Eur.Phys.J. {\bf C72}
  (2012)  1909}.

\bibitem{electron_reco}
{ATLAS} Collaboration, {\em {Electron reconstruction and identification
  efficiency measurements with the ATLAS detector using the 2011 LHC
  proton-proton collision data}\/},
\href{http://dx.doi.org/10.1140/epjc/s10052-014-2941-0}{Eur.Phys.J. {\bf C74}
  (2014)  2941}.

\bibitem{reco_el_eff}
{ATLAS} Collaboration, {\em {Electron identification efficiency dependence on
  pileup}\/},  ATL-COM-PHYS-2011-1636. Approved public plot.

\bibitem{electron_calib}
{ATLAS} Collaboration, {\em {Electron and photon energy calibration with the
  ATLAS detector using LHC Run 1 data}\/},
  \href{http://arxiv.org/abs/1407.5063}{{\tt arXiv:1407.5063 [hep-ex]}}.
Submitted to EPJ C.

\bibitem{reco_muons}
{ATLAS} Collaboration, {\em {Muon reconstruction efficiency in reprocessed 2010
  LHC proton-proton collision data recorded with the ATLAS detector}\/},
ATLAS-CONF-2011-063.

\bibitem{muon_performance}
{ATLAS} Collaboration, {\em {Measurement of the muon reconstruction performance
  of the ATLAS detector using 2011 and 2012 LHC proton-proton collision
  data}\/},  \href{http://arxiv.org/abs/1407.3935}{{\tt arXiv:1407.3935
  [hep-ex]}}.
Submitted to EPJ C.

\bibitem{topo_clusters}
W.~Lampl et al., {\em {Calorimeter clustering algorithms: Description and
  performance}\/},
ATL-LARG-PUB-2008-002.

\bibitem{antikt}
M.~Cacciari, G.~P. Salam, and G.~Soyez, {\em {The Anti-k(t) jet clustering
  algorithm}\/},
\href{http://dx.doi.org/10.1088/1126-6708/2008/04/063}{JHEP {\bf 0804} (2008)
  063}.

\bibitem{fastjet}
M.~Cacciari and G.~P. Salam, {\em {Dispelling the $N^{3}$ myth for the $k_t$
  jet-finder}\/},
\href{http://dx.doi.org/10.1016/j.physletb.2006.08.037}{Phys.Lett. {\bf B641}
  (2006)  57}.

\bibitem{JES}
{ATLAS} Collaboration, {\em {Jet energy measurement and its systematic
  uncertainty in proton-proton collisions at $\sqrt{s}=7$ TeV with the ATLAS
  detector}\/},  \href{http://arxiv.org/abs/1406.0076}{{\tt arXiv:1406.0076
  [hep-ex]}}.
Submitted to EPJ C.

\bibitem{reco_jets}
{ATLAS} Collaboration, {\em {Jet energy measurement with the ATLAS detector in
  proton-proton collisions at $\sqrt{s}=7$ TeV}\/},
\href{http://dx.doi.org/10.1140/epjc/s10052-013-2304-2}{Eur.Phys.J. {\bf C73}
  (2013)  2304}.

\bibitem{jvf_d0}
{D0} Collaboration, {\em {Measurement of the $p \bar{p} \to t \bar{t}$
  production cross section at $\sqrt{s}$ = 1.96-TeV in the fully hadronic decay
  channel.}\/},
\href{http://dx.doi.org/10.1103/PhysRevD.76.072007}{Phys.Rev. {\bf D76} (2007)
  072007}.

\bibitem{jvf_website}
\url{https://twiki.cern.ch/twiki/bin/view/AtlasPublic/JetEtmissApproved2011PileupOffsetAndJVF}.
\newblock As of 02.03.2014.

\bibitem{jvf_2012}
{ATLAS} Collaboration, {\em {Pile-up subtraction and suppression for jets in
  ATLAS}\/},
ATLAS-CONF-2013-083.

\bibitem{b-tagging}
{ATLAS} Collaboration, {\em {Commissioning of the ATLAS high-performance
  b-tagging algorithms in the 7 TeV collision data}\/},
ATLAS-CONF-2011-102.

\bibitem{b-tagging_calib}
{ATLAS} Collaboration, {\em {Calibrating the b-Tag Efficiency and Mistag Rate
  in 35 pb$^{-1}$ of Data with the ATLAS Detector}\/},
ATLAS-CONF-2011-089.

\bibitem{b-tagging_dilepcal}
{ATLAS} Collaboration, {\em {Measuring the b-tag efficiency in a top-pair
  sample with 4.7 fb$^{-1}$ of data from the ATLAS detector}\/},
ATLAS-CONF-2012-097.

\bibitem{reco_MET}
{ATLAS} Collaboration, {\em {Performance of Missing Transverse Momentum
  Reconstruction in Proton-Proton Collisions at 7 TeV with ATLAS}\/},
\href{http://dx.doi.org/10.1140/epjc/s10052-011-1844-6}{Eur.Phys.J. {\bf C72}
  (2012)  1844}.

\bibitem{Geant4}
{GEANT4} Collaboration, S.~Agostinelli et al., {\em {GEANT4: A Simulation
  toolkit}\/},
\href{http://dx.doi.org/10.1016/S0168-9002(03)01368-8}{Nucl.Instrum.Meth. {\bf
  A506} (2003)  250}.

\bibitem{atlas_public_lumi}
\url{https://twiki.cern.ch/twiki/bin/view/AtlasPublic/LuminosityPublicResults}.
\newblock As of 27.11.2013.

\bibitem{IPAC2013}
Z.~Dai et al. {Proceedings, 4th International Particle Accelerator Conference
  (IPAC 2013)}, 2013.

\bibitem{MCatNLO}
S.~Frixione and B.~R. Webber, {\em {The MC@NLO event generator}\/},
\href{http://arxiv.org/abs/hep-ph/0207182}{{\tt arXiv:hep-ph/0207182
  [hep-ph]}}.

\bibitem{MCatNLO2}
S.~Frixione, P.~Nason, and B.~R. Webber, {\em {Matching NLO QCD and parton
  showers in heavy flavor production}\/},
\href{http://dx.doi.org/10.1088/1126-6708/2003/08/007}{JHEP {\bf 0308} (2003)
  007}.

\bibitem{MCatNLO3}
S.~Frixione et al., {\em {Angular correlations of lepton pairs from vector
  boson and top quark decays in Monte Carlo simulations}\/},
\href{http://dx.doi.org/10.1088/1126-6708/2007/04/081}{JHEP {\bf 0704} (2007)
  081}.

\bibitem{MCatNLO4}
S.~Frixione et al., {\em {The MC@NLO 4.0 Event Generator}\/},
\href{http://arxiv.org/abs/1010.0819}{{\tt arXiv:1010.0819 [hep-ph]}}.

\bibitem{herwig}
G.~Corcella et al., {\em {HERWIG 6: An Event generator for hadron emission
  reactions with interfering gluons (including supersymmetric processes)}\/},
\href{http://dx.doi.org/10.1088/1126-6708/2001/01/010}{JHEP {\bf 0101} (2001)
  010}.

\bibitem{jimmy}
J.~Butterworth, J.~R. Forshaw, and M.~Seymour, {\em {Multiparton interactions
  in photoproduction at HERA}\/},
\href{http://dx.doi.org/10.1007/s002880050286}{Z.Phys. {\bf C72} (1996)  637}.

\bibitem{ATLAS_gentune}
{ATLAS} Collaboration, {\em {New ATLAS event generator tunes to 2010 data}\/},
ATL-PHYS-PUB-2011-008.

\bibitem{Cacciari:2011hy}
M.~Cacciari et al., {\em {Top-pair production at hadron colliders with
  next-to-next-to-leading logarithmic soft-gluon resummation}\/},
\href{http://dx.doi.org/10.1016/j.physletb.2012.03.013}{Phys.Lett. {\bf B710}
  (2012)  612}.

\bibitem{Baernreuther:2012ws}
P.~B{\"a}rnreuther, M.~Czakon, and A.~Mitov, {\em {Percent Level Precision
  Physics at the Tevatron: First Genuine NNLO QCD Corrections to $q \bar{q} \to
  t \bar{t} + X$}\/},
\href{http://dx.doi.org/10.1103/PhysRevLett.109.132001}{Phys.Rev.Lett. {\bf
  109} (2012)  132001}.

\bibitem{Czakon:2012zr}
M.~Czakon and A.~Mitov, {\em {NNLO corrections to top-pair production at hadron
  colliders: the all-fermionic scattering channels}\/},
\href{http://dx.doi.org/10.1007/JHEP12(2012)054}{JHEP {\bf 1212} (2012)  054}.

\bibitem{Czakon:2012pz}
M.~Czakon and A.~Mitov, {\em {NNLO corrections to top pair production at hadron
  colliders: the quark-gluon reaction}\/},
\href{http://dx.doi.org/10.1007/JHEP01(2013)080}{JHEP {\bf 1301} (2013)  080}.

\bibitem{Czakon:2013goa}
M.~Czakon, P.~Fiedler, and A.~Mitov, {\em {Total Top-Quark Pair-Production
  Cross Section at Hadron Colliders Through $\mathcal{O}(\alpha^4_S)$}\/},
\href{http://dx.doi.org/10.1103/PhysRevLett.110.252004}{Phys.Rev.Lett. {\bf
  110} (2013) no.~25, 252004}.

\bibitem{Botje:2011sn}
M.~Botje, J.~Butterworth, A.~Cooper-Sarkar, A.~de~Roeck, J.~Feltesse, et al.,
  {\em {The PDF4LHC Working Group Interim Recommendations}\/},
\href{http://arxiv.org/abs/1101.0538}{{\tt arXiv:1101.0538 [hep-ph]}}.

\bibitem{Martin:2009bu}
A.~Martin et al., {\em {Uncertainties on $\alpha_{S}$ in global PDF analyses
  and implications for predicted hadronic cross sections}\/},
\href{http://dx.doi.org/10.1140/epjc/s10052-009-1164-2}{Eur.Phys.J. {\bf C64}
  (2009)  653}.

\bibitem{Gao:2013xoa}
J.~Gao et al., {\em {The CT10 NNLO Global Analysis of QCD}\/},
\href{http://dx.doi.org/10.1103/PhysRevD.89.033009}{Phys.Rev. {\bf D89} (2014)
  033009}.

\bibitem{Aliev:2010zk}
M.~Aliev et al., {\em {HATHOR: HAdronic Top and Heavy quarks crOss section
  calculatoR}\/},
\href{http://dx.doi.org/10.1016/j.cpc.2010.12.040}{Comput.Phys.Commun. {\bf
  182} (2011)  1034}.

\bibitem{NLO_generators}
P.~Nason and B.~Webber, {\em {Next-to-Leading-Order Event Generators}\/},
\href{http://dx.doi.org/10.1146/annurev-nucl-102711-094928}{Ann.Rev.Nucl.Part.Sci.
  {\bf 62} (2012)  187}.

\bibitem{diagramremoval}
S.~Frixione et al., {\em {Single-top hadroproduction in association with a W
  boson}\/},
\href{http://dx.doi.org/10.1088/1126-6708/2008/07/029}{JHEP {\bf 0807} (2008)
  029}.

\bibitem{acermc}
B.~P. Kersevan and E.~Richter-Was, {\em {The Monte Carlo event generator AcerMC
  versions 2.0 to 3.8 with interfaces to PYTHIA 6.4, HERWIG 6.5 and ARIADNE
  4.1}\/},
\href{http://dx.doi.org/10.1016/j.cpc.2012.10.032}{Comput.Phys.Commun. {\bf
  184} (2013)  919}.

\bibitem{pythia}
T.~Sjostrand, S.~Mrenna, and P.~Z. Skands, {\em {PYTHIA 6.4 Physics and
  Manual}\/},
\href{http://dx.doi.org/10.1088/1126-6708/2006/05/026}{JHEP {\bf 0605} (2006)
  026}.

\bibitem{mrstcal}
A.~Sherstnev and R.~Thorne, {\em {Parton Distributions for LO Generators}\/},
\href{http://dx.doi.org/10.1140/epjc/s10052-008-0610-x}{Eur.Phys.J. {\bf C55}
  (2008)  553}.

\bibitem{mrstcal2}
A.~Martin, W.~Stirling, R.~Thorne, and G.~Watt, {\em {Update of parton
  distributions at NNLO}\/},
\href{http://dx.doi.org/10.1016/j.physletb.2007.07.040}{Phys.Lett. {\bf B652}
  (2007)  292}.

\bibitem{alpgen}
M.~L. Mangano et al., {\em {ALPGEN, a generator for hard multiparton processes
  in hadronic collisions}\/},
\href{http://dx.doi.org/10.1088/1126-6708/2003/07/001}{JHEP {\bf 0307} (2003)
  001}.

\bibitem{cteq6l}
J.~Pumplin et al., {\em {New generation of parton distributions with
  uncertainties from global QCD analysis}\/},
  \href{http://dx.doi.org/10.1088/1126-6708/2002/07/012}{JHEP {\bf 0207} (2002)
   012},
\href{http://arxiv.org/abs/hep-ph/0201195}{{\tt arXiv:hep-ph/0201195
  [hep-ph]}}.

\bibitem{Mangano:2006rw}
M.~L. Mangano et al., {\em {Matching matrix elements and shower evolution for
  top-quark production in hadronic collisions}\/},
\href{http://dx.doi.org/10.1088/1126-6708/2007/01/013}{JHEP {\bf 0701} (2007)
  013}.

\bibitem{matrixmethod}
{D0} Collaboration, {\em {Measurement of the $t \bar{t}$ production cross
  section in $p \bar{p}$ collisions at $\sqrt{s}$ = 1.96-TeV using kinematic
  characteristics of lepton + jets events}\/},
\href{http://dx.doi.org/10.1103/PhysRevD.76.092007}{Phys.Rev. {\bf D76} (2007)
  092007}.

\bibitem{muon_reco_eff}
{ATLAS} Collaboration, {\em {Preliminary results on the muon reconstruction
  efficiency, momentum resolution, and momentum scale in ATLAS 2012 pp
  collision data}\/},
ATLAS-CONF-2013-088.

\bibitem{jetmult}
{ATLAS} Collaboration, {\em {Measurement of the jet multiplicity in top
  anti-top final states produced in 7 TeV proton-proton collisions with the
  ATLAS detector}\/},
ATLAS-CONF-2012-155.

\bibitem{klfitter}
J.~Erdmann et al., {\em {A likelihood-based reconstruction algorithm for
  top-quark pairs and the KLFitter framework}\/},
\href{http://dx.doi.org/10.1016/j.nima.2014.02.029}{Nucl.Instrum.Meth. {\bf
  A748} (2014)  18}.

\bibitem{BAT}
A.~Caldwell, D.~Kollar, and K.~Kr{\"o}ninger, {\em {BAT: The Bayesian Analysis
  Toolkit}\/},
\href{http://dx.doi.org/10.1016/j.cpc.2009.06.026}{Comput.Phys.Commun. {\bf
  180} (2009)  2197}.

\bibitem{WHel_ATLAS}
{ATLAS} Collaboration, {\em {Measurement of the W boson polarization in top
  quark decays with the ATLAS detector}\/},
\href{http://dx.doi.org/10.1007/JHEP06(2012)088}{JHEP {\bf 1206} (2012)  088}.

\bibitem{mtop_ATLAS_template}
{ATLAS} Collaboration, {\em {Measurement of the top quark mass with the
  template method in the $t \bar{t}$ $\rightarrow$ lepton + jets channel using
  ATLAS data}\/},
\href{http://dx.doi.org/10.1140/epjc/s10052-012-2046-6}{Eur.Phys.J. {\bf C72}
  (2012)  2046}.

\bibitem{top_xsec_diff_ATLAS}
{ATLAS} Collaboration, {\em {Measurements of top quark pair relative
  differential cross-sections with ATLAS in $pp$ collisions at $\sqrt{s}=7$
  TeV}\/},
\href{http://dx.doi.org/10.1140/epjc/s10052-012-2261-1}{Eur.Phys.J. {\bf C73}
  (2013)  2261}.

\bibitem{ttH_ATLAS}
{ATLAS} Collaboration, {\em {Search for the Standard Model Higgs boson produced
  in association with top quarks in proton-proton collisions at $\sqrt{s}$ = 7
  TeV using the ATLAS detector}\/},
ATLAS-CONF-2012-135.

\bibitem{mistag}
{ATLAS} Collaboration, {\em {Measurement of the Mistag Rate with 5 fb${}^{-1}$
  of Data Collected by the ATLAS Detector}\/},
ATLAS-CONF-2012-040.

\bibitem{etmiss}
{ATLAS} Collaboration, {\em {Performance of Missing Transverse Momentum
  Reconstruction in Proton-Proton Collisions at 7 TeV with ATLAS}\/},
\href{http://dx.doi.org/10.1140/epjc/s10052-011-1844-6}{Eur.Phys.J. {\bf C72}
  (2012)  1844}.

\bibitem{fritz_BA}
F.~Pasternok, {\em Studies of Systematic Uncertainties of Transfer Functions
  used in the KLFitter for Top Quark Reconstruction\/},
  II.Physik-UniG{\"o}-BSc-2011/02. Bachelor Thesis, G{\"o}ttingen University.

\bibitem{Jezabek1994b}
M.~Jezabek, {\em {Top quark physics}\/},
\href{http://dx.doi.org/10.1016/0920-5632(94)90677-7}{Nucl.Phys.Proc.Suppl.
  {\bf 37B} (1994)  197}.

\bibitem{bisector}
{UA2} Collaboration, {\em {Measurement of Jet Production Properties at the CERN
  anti-p p Collider}\/},
\href{http://dx.doi.org/10.1016/0370-2693(84)91822-7}{Phys.Lett. {\bf B144}
  (1984)  283}.

\bibitem{JER}
{ATLAS} Collaboration, {\em {Jet energy resolution in proton-proton collisions
  at $\sqrt{s}=7$ TeV recorded in 2010 with the ATLAS detector}\/},
\href{http://dx.doi.org/10.1140/epjc/s10052-013-2306-0}{Eur.Phys.J. {\bf C73}
  (2013)  2306}.

\bibitem{jetreco}
{ATLAS} Collaboration, {\em {Jet energy resolution and selection efficiency
  relative to track jets from in-situ techniques with the ATLAS Detector Using
  Proton-Proton Collisions at a Center of Mass Energy $\sqrt{s}$ = 7 TeV}\/},
ATLAS-CONF-2010-054.

\bibitem{lhcprimer}
J.~M. Campbell, J.~Huston, and W.~Stirling, {\em {Hard Interactions of Quarks
  and Gluons: A Primer for LHC Physics}\/},
\href{http://dx.doi.org/10.1088/0034-4885/70/1/R02}{Rept.Prog.Phys. {\bf 70}
  (2007)  89}.

\bibitem{Berends1991}
F.~A. Berends et al., {\em {On the production of a W and jets at hadron
  colliders}\/},
\href{http://dx.doi.org/10.1016/0550-3213(91)90458-A}{Nucl.Phys. {\bf B357}
  (1991)  32}.

\bibitem{lhapdf}
M.~Whalley, D.~Bourilkov, and R.~Group, {\em {The Les Houches accord PDFs
  (LHAPDF) and LHAGLUE}\/},
\href{http://arxiv.org/abs/hep-ph/0508110}{{\tt arXiv:hep-ph/0508110
  [hep-ph]}}.

\bibitem{top_xsec_diff_ATLAS_conf}
{ATLAS} Collaboration, {\em {Measurement of top-quark pair differential
  cross-sections in the $l$+jets channel in $pp$ collisions at $\sqrt{s}=7$ TeV
  using the ATLAS detector}\/},
ATLAS-CONF-2013-099.

\bibitem{color_reconnection}
J.~Gallicchio and M.~D. Schwartz, {\em {Seeing in Color: Jet
  Superstructure}\/},
\href{http://dx.doi.org/10.1103/PhysRevLett.105.022001}{Phys.Rev.Lett. {\bf
  105} (2010)  022001}.

\bibitem{perugia_tune}
P.~Z. Skands, {\em {Tuning Monte Carlo Generators: The Perugia Tunes}\/},
\href{http://dx.doi.org/10.1103/PhysRevD.82.074018}{Phys.Rev. {\bf D82} (2010)
  074018}.

\bibitem{tauola}
S.~Jadach, J.~H. K{\"u}hn, and Z.~Was, {\em {TAUOLA: A Library of Monte Carlo
  programs to simulate decays of polarized tau leptons}\/},
\href{http://dx.doi.org/10.1016/0010-4655(91)90038-M}{Comput.Phys.Commun. {\bf
  64} (1990)  275}.

\bibitem{asimov}
G.~Cowan et al., {\em {Asymptotic formulae for likelihood-based tests of new
  physics}\/},
\href{http://dx.doi.org/10.1140/epjc/s10052-011-1554-0}{Eur.Phys.J. {\bf C71}
  (2011)  1554}.

\bibitem{Hplus}
V.~D. Barger and R.~Phillips, {\em {Hidden Top Quark With Charged Higgs
  Decay}\/},
\href{http://dx.doi.org/10.1103/PhysRevD.41.884}{Phys.Rev. {\bf D41} (1990)
  884}.

\bibitem{jetcharge}
R.~Field and R.~Feynman, {\em {A Parametrization of the Properties of Quark
  Jets}\/},
\href{http://dx.doi.org/10.1016/0550-3213(78)90015-9}{Nucl.Phys. {\bf B136}
  (1978)  1}.

\bibitem{veit_MA}
V.~P. Dahlke, {\em Reconstruction studies for $t\bar{t}H\left( H \rightarrow
  b\bar{b}\right)$ events with the ATLAS experiment at the LHC\/},
  II.Physik-UniG{\"o}-MSc-2013/06. Master Thesis, G{\"o}ttingen University.

\end{thebibliography}\endgroup

\cleardoublepage
\printindex
\cleardoublepage

\addappheadtotoc
\appendix
\appendixpage
\noappendicestocpagenum


\chapter{Spin Correlation Matrices}
\label{sec:app_spincorrmat}

The spin correlation matrix $\widehat{C}_{i\bar{i}}$ was introduced in Section \ref{sec:SC}. In \cite{Baumgart2013} these matrices were calculated at leading-order QCD. They are expressed in terms of the top production velocity $\beta$ and production angle $\Theta$ in the $t\bar t$ \com\ frame. The special property of the off-diagonal basis is the $\widehat{C}_{33}$ element, which is equal to unity, independent of the production kinematics. 
\\[1cm]
{$q\bar q \to t\bar t$}

\begin{eqnarray*}
{\rm off}\mbox{-}{\rm diagonal \; basis:} \;\;\;
\widehat{C} & \,=\, & \left( \begin{matrix}   
\frac{\beta^2\cos^2 \Theta}{2-\beta^2\sin^2 \Theta} & 0 & 0 \\
0 & \frac{-\beta^2\sin^2 \Theta}{2-\beta^2\sin^2 \Theta} & 0 \\
0 & 0 & 1
\end{matrix} \right) \nonumber \\
{\rm helicity \; basis:} \;\;\;
\widehat{C} & \,=\, & \left( \begin{matrix}   
\frac{(2-\beta^2)\sin^2 \Theta}{2-\beta^2\cos^2 \Theta} & 0 & \frac{-\sin^2\Theta}{\gamma(2-\beta^2\cos^2 \Theta)} \\
0 & \frac{-\beta^2\cos^2 \Theta}{2-\beta^2\cos^2 \Theta} & 0 \\
\frac{-\sin2\Theta}{\gamma(2-\beta^2\cos^2 \Theta)} & 0 & \frac{2\cos^2 \Theta + \beta^2\cos^2 \Theta}{2-\beta^2\cos^2 \Theta}
\end{matrix} \right)
\end{eqnarray*}
{}
\\[1cm]
{$gg \to t\bar t$}\\[0.3cm]
helicity basis:
\begin{eqnarray*}
\widehat{C} & \,=\, & \left( \begin{matrix}   
\frac{-1 + \beta^2(2-\beta^2)(1+\sin^4\Theta)}{1 + 2\beta^2\cos^2 \Theta - \beta^4(1+\sin^4\Theta)} & 0 & \frac{-\beta^2\sin2\Theta \cos^2 \Theta}{\gamma(1 + 2\beta^2\cos^2 \Theta - \beta^4(1+\sin^4\Theta))} \\
0 & \frac{-1 + 2\beta^2 - \beta^4(1+\sin^4\Theta)}{1 + 2\beta^2\cos^2 \Theta - \beta^4(1+\sin^4\Theta)} & 0 \\
\frac{-\beta^2\sin2\Theta \cos^2 \Theta}{1 + 2\beta^2\cos^2 \Theta - \beta^4(1+\sin^4\Theta)} & 0 & \frac{-1 + \beta^2\left(\beta^2(1+\sin^4\Theta) + (\sin^2 2\Theta)/2\right)}{1 + 2\beta^2\cos^2 \Theta - \beta^4(1+\sin^4\Theta)}
\end{matrix} \right)
\end{eqnarray*}

 
\chapter{Used Datasets}
\label{app:datasets}
\begin{sidewaystable}[htbp]
\begin{center}
{\tiny
\begin{tabular}{|l|l|}
\hline
Dataset & Function\\
\hline
data11\_7TeV.$*$.physics\_Egamma.merge.NTUP\_TOPEL.$*$\_p694\_p722\_p937\/ & Data for \ejets\ \\
\hline
data11\_7TeV.$*$.physics\_Muons.merge.NTUP\_TOPMU.$*$\_p694\_p722\_p937\/ & Data for \mujets\ \\
\hline
mc11\_7TeV.105200.T1\_McAtNlo\_Jimmy.merge.NTUP\_TOP.e835\_s1272\_s1274\_r3043\_r2993\_p937/ & \ttbar\ signal (SM spin corr.)\\
\hline
mc11\_7TeV.117200.T1\_McAtNlo\_Jimmy.merge.NTUP\_TOP.e944\_s1310\_s1300\_r3043\_r2993\_p937/ & \ttbar\ signal (no spin corr.)\\
\hline
mc11\_7TeV.107680.AlpgenJimmyWenuNp0\_pt20.merge.NTUP\_TOP.e825\_s1299\_s1300\_r3043\_r2993\_p937/ & $W \rightarrow e \nu + 0p$ \\
\hline
mc11\_7TeV.107681.AlpgenJimmyWenuNp1\_pt20.merge.NTUP\_TOP.e825\_s1299\_s1300\_r3043\_r2993\_p937/ & $W \rightarrow e \nu + 1p$ \\
\hline
mc11\_7TeV.107682.AlpgenJimmyWenuNp2\_pt20.merge.NTUP\_TOP.e825\_s1299\_s1300\_r3043\_r2993\_p937/ & $W \rightarrow e \nu + 2p$ \\
\hline
mc11\_7TeV.107683.AlpgenJimmyWenuNp3\_pt20.merge.NTUP\_TOP.e825\_s1299\_s1300\_r3043\_r2993\_p937/ & $W \rightarrow e \nu + 3p$ \\
\hline
mc11\_7TeV.107684.AlpgenJimmyWenuNp4\_pt20.merge.NTUP\_TOP.e825\_s1299\_s1300\_r3043\_r2993\_p937/ & $W \rightarrow e \nu + 4p$ \\
\hline
mc11\_7TeV.107685.AlpgenJimmyWenuNp5\_pt20.merge.NTUP\_TOP.e825\_s1299\_s1300\_r3043\_r2993\_p937/ & $W \rightarrow e \nu + 5p$ \\
\hline
mc11\_7TeV.107690.AlpgenJimmyWmunuNp0\_pt20.merge.NTUP\_TOP.e825\_s1299\_s1300\_r3043\_r2993\_p937/ & $W \rightarrow \mu \nu + 0p$ \\
\hline
mc11\_7TeV.107691.AlpgenJimmyWmunuNp1\_pt20.merge.NTUP\_TOP.e825\_s1299\_s1300\_r3043\_r2993\_p937/ & $W \rightarrow \mu \nu + 1p$ \\
\hline
mc11\_7TeV.107692.AlpgenJimmyWmunuNp2\_pt20.merge.NTUP\_TOP.e825\_s1299\_s1300\_r3043\_r2993\_p937/ & $W \rightarrow \mu \nu + 2p$ \\
\hline
mc11\_7TeV.107693.AlpgenJimmyWmunuNp3\_pt20.merge.NTUP\_TOP.e825\_s1299\_s1300\_r3043\_r2993\_p937/ & $W \rightarrow \mu \nu + 3p$ \\
\hline
mc11\_7TeV.107694.AlpgenJimmyWmunuNp4\_pt20.merge.NTUP\_TOP.e825\_s1299\_s1300\_r3043\_r2993\_p937/ & $W \rightarrow \mu \nu + 4p$ \\
\hline
mc11\_7TeV.107695.AlpgenJimmyWmunuNp5\_pt20.merge.NTUP\_TOP.e825\_s1299\_s1300\_r3043\_r2993\_p937/ & $W \rightarrow \mu \nu + 5p$ \\
\hline
mc11\_7TeV.107700.AlpgenJimmyWtaunuNp0\_pt20.merge.NTUP\_TOP.e835\_s1299\_s1300\_r3043\_r2993\_p937/ & $W \rightarrow \mu \nu + 0p$ \\
\hline
mc11\_7TeV.107701.AlpgenJimmyWtaunuNp1\_pt20.merge.NTUP\_TOP.e835\_s1299\_s1300\_r3043\_r2993\_p937/ & $W \rightarrow \mu \nu + 1p$ \\
\hline
mc11\_7TeV.107702.AlpgenJimmyWtaunuNp2\_pt20.merge.NTUP\_TOP.e835\_s1299\_s1300\_r3043\_r2993\_p937/ & $W \rightarrow \mu \nu + 2p$ \\
\hline
mc11\_7TeV.107703.AlpgenJimmyWtaunuNp3\_pt20.merge.NTUP\_TOP.e835\_s1299\_s1300\_r3043\_r2993\_p937/ & $W \rightarrow \mu \nu + 3p$ \\
\hline
mc11\_7TeV.107704.AlpgenJimmyWtaunuNp4\_pt20.merge.NTUP\_TOP.e835\_s1299\_s1300\_r3043\_r2993\_p937/ & $W \rightarrow \mu \nu + 4p$ \\
\hline
mc11\_7TeV.107705.AlpgenJimmyWtaunuNp5\_pt20.merge.NTUP\_TOP.e835\_s1299\_s1300\_r3043\_r2993\_p937/ & $W \rightarrow \mu \nu + 5p$ \\
\hline
mc11\_7TeV.107280.AlpgenJimmyWbbFullNp0\_pt20.merge.NTUP\_TOP.e887\_s1310\_s1300\_r3043\_r2993\_p937/ & $W \rightarrow e/\mu\tau \nu + b\bar{b}+0p$ \\
\hline
mc11\_7TeV.107281.AlpgenJimmyWbbFullNp1\_pt20.merge.NTUP\_TOP.e887\_s1310\_s1300\_r3043\_r2993\_p937/ & $W \rightarrow e/\mu\tau \nu + b\bar{b}+1p$ \\
\hline
mc11\_7TeV.107282.AlpgenJimmyWbbFullNp2\_pt20.merge.NTUP\_TOP.e887\_s1310\_s1300\_r3043\_r2993\_p937/ & $W \rightarrow e/\mu\tau \nu + b\bar{b}+2p$ \\
\hline
mc11\_7TeV.107283.AlpgenJimmyWbbFullNp3\_pt20.merge.NTUP\_TOP.e887\_s1310\_s1300\_r3043\_r2993\_p937/ & $W \rightarrow e/\mu\tau \nu + b\bar{b}+3p$ \\
\hline
mc11\_7TeV.117284.AlpgenWccFullNp0\_pt20.merge.NTUP\_TOP.e887\_s1310\_s1300\_r3043\_r2993\_p937/ & $W \rightarrow e/\mu\tau \nu + c\bar{c}+0p$ \\
\hline
mc11\_7TeV.117285.AlpgenWccFullNp1\_pt20.merge.NTUP\_TOP.e887\_s1310\_s1300\_r3043\_r2993\_p937/ & $W \rightarrow e/\mu\tau \nu + c\bar{c}+1p$ \\
\hline
mc11\_7TeV.117286.AlpgenWccFullNp2\_pt20.merge.NTUP\_TOP.e887\_s1310\_s1300\_r3043\_r2993\_p937/ & $W \rightarrow e/\mu\tau \nu + c\bar{c}+2p$ \\
\hline
mc11\_7TeV.117287.AlpgenWccFullNp3\_pt20.merge.NTUP\_TOP.e887\_s1310\_s1300\_r3043\_r2993\_p937/ & $W \rightarrow e/\mu\tau \nu + c\bar{c}+3p$ \\
\hline
mc11\_7TeV.117293.AlpgenWcNp0\_pt20.merge.NTUP\_TOP.e887\_s1310\_s1300\_r3043\_r2993\_p937/ & $W \rightarrow e/\mu\tau \nu + c+0p$ \\
\hline
mc11\_7TeV.117294.AlpgenWcNp1\_pt20.merge.NTUP\_TOP.e887\_s1310\_s1300\_r3043\_r2993\_p937/ & $W \rightarrow e/\mu\tau \nu + c+1p$ \\
\hline
mc11\_7TeV.117295.AlpgenWcNp2\_pt20.merge.NTUP\_TOP.e887\_s1310\_s1300\_r3043\_r2993\_p937/ & $W \rightarrow e/\mu\tau \nu + c+2p$ \\
\hline
mc11\_7TeV.117296.AlpgenWcNp3\_pt20.merge.NTUP\_TOP.e887\_s1310\_s1300\_r3043\_r2993\_p937/ & $W \rightarrow e/\mu\tau \nu + c+3p$ \\
\hline
mc11\_7TeV.117297.AlpgenWcNp4\_pt20.merge.NTUP\_TOP.e887\_s1310\_s1300\_r3043\_r2993\_p937/ & $W \rightarrow e/\mu\tau \nu + c+4p$ \\
\hline
\end{tabular}
}
\end{center}
\caption{Datasets used for the analysis (data, \ttbar\ and $W+\text{jets}$).}
\label{tab:samples1}
\end{sidewaystable}

\begin{sidewaystable}[htbp]
\begin{center}
{\tiny
\begin{tabular}{|l|l|}
\hline
Dataset & Function\\
\hline
\hline
mc11\_7TeV.109300.AlpgenJimmyZeebbNp0\_nofilter.merge.NTUP\_TOP.e835\_s1310\_s1300\_r3043\_r2993\_p937/ & $Z\rightarrow ee + b\bar{b} + 0p$ \\
mc11\_7TeV.109301.AlpgenJimmyZeebbNp1\_nofilter.merge.NTUP\_TOP.e835\_s1310\_s1300\_r3043\_r2993\_p937/ &  $Z\rightarrow ee + b\bar{b} + 1p$ \\
mc11\_7TeV.109302.AlpgenJimmyZeebbNp2\_nofilter.merge.NTUP\_TOP.e835\_s1310\_s1300\_r3043\_r2993\_p937/ & $Z\rightarrow ee + b\bar{b} + 2p$ \\
mc11\_7TeV.109303.AlpgenJimmyZeebbNp3\_nofilter.merge.NTUP\_TOP.e835\_s1310\_s1300\_r3043\_r2993\_p937/ & $Z\rightarrow ee + b\bar{b} + 3p$ \\
mc11\_7TeV.109305.AlpgenJimmyZmumubbNp0\_nofilter.merge.NTUP\_TOP.e835\_s1310\_s1300\_r3043\_r2993\_p937/ &  $Z\rightarrow \mu\mu + b\bar{b} + 0p$ \\
mc11\_7TeV.109306.AlpgenJimmyZmumubbNp1\_nofilter.merge.NTUP\_TOP.e835\_s1310\_s1300\_r3043\_r2993\_p937/ & $Z\rightarrow \mu\mu + b\bar{b} + 1p$ \\
mc11\_7TeV.109307.AlpgenJimmyZmumubbNp2\_nofilter.merge.NTUP\_TOP.e835\_s1310\_s1300\_r3043\_r2993\_p937/ & $Z\rightarrow \mu\mu + b\bar{b} + 2p$ \\
mc11\_7TeV.109308.AlpgenJimmyZmumubbNp3\_nofilter.merge.NTUP\_TOP.e835\_s1310\_s1300\_r3043\_r2993\_p937/ & $Z\rightarrow \mu\mu + b\bar{b} + 3p$ \\
mc11\_7TeV.109310.AlpgenJimmyZtautaubbNp0\_nofilter.merge.NTUP\_TOP.e835\_s1310\_s1300\_r3043\_r2993\_p937/ & $Z\rightarrow \tau\tau+ b\bar{b} + 0p$ \\
mc11\_7TeV.109311.AlpgenJimmyZtautaubbNp1\_nofilter.merge.NTUP\_TOP.e835\_s1310\_s1300\_r3043\_r2993\_p937/ & $Z\rightarrow \tau\tau+ b\bar{b} + 1p$ \\
mc11\_7TeV.109312.AlpgenJimmyZtautaubbNp2\_nofilter.merge.NTUP\_TOP.e835\_s1310\_s1300\_r3043\_r2993\_p937/ & $Z\rightarrow \tau\tau+ b\bar{b} + 2p$ \\
mc11\_7TeV.109313.AlpgenJimmyZtautaubbNp3\_nofilter.merge.NTUP\_TOP.e835\_s1310\_s1300\_r3043\_r2993\_p937/ &$Z\rightarrow \tau\tau+ b\bar{b} + 3p$ \\
mc11\_7TeV.107650.AlpgenJimmyZeeNp0\_pt20.merge.NTUP\_TOP.e835\_s1299\_s1300\_r3043\_r2993\_p937/ & $Z\rightarrow ee + 0p$ \\
mc11\_7TeV.107651.AlpgenJimmyZeeNp1\_pt20.merge.NTUP\_TOP.e835\_s1299\_s1300\_r3043\_r2993\_p937/ & $Z\rightarrow ee + 1p$ \\
mc11\_7TeV.107652.AlpgenJimmyZeeNp2\_pt20.merge.NTUP\_TOP.e835\_s1299\_s1300\_r3043\_r2993\_p937/ & $Z\rightarrow ee + 2p$ \\
mc11\_7TeV.107653.AlpgenJimmyZeeNp3\_pt20.merge.NTUP\_TOP.e835\_s1299\_s1300\_r3043\_r2993\_p937/ & $Z\rightarrow ee + 3p$ \\
mc11\_7TeV.107654.AlpgenJimmyZeeNp4\_pt20.merge.NTUP\_TOP.e835\_s1299\_s1300\_r3043\_r2993\_p937/ & $Z\rightarrow ee + 4p$ \\
mc11\_7TeV.107655.AlpgenJimmyZeeNp5\_pt20.merge.NTUP\_TOP.e835\_s1299\_s1300\_r3043\_r2993\_p937/ & $Z\rightarrow ee + 5p$ \\
mc11\_7TeV.107660.AlpgenJimmyZmumuNp0\_pt20.merge.NTUP\_TOP.e835\_s1299\_s1300\_r3043\_r2993\_p937/ & $Z\rightarrow \mu\mu + 0p$ \\
mc11\_7TeV.107661.AlpgenJimmyZmumuNp1\_pt20.merge.NTUP\_TOP.e835\_s1299\_s1300\_r3043\_r2993\_p937/ & $Z\rightarrow \mu\mu + 1p$ \\
mc11\_7TeV.107662.AlpgenJimmyZmumuNp2\_pt20.merge.NTUP\_TOP.e835\_s1299\_s1300\_r3043\_r2993\_p937/ & $Z\rightarrow \mu\mu + 2p$ \\
mc11\_7TeV.107663.AlpgenJimmyZmumuNp3\_pt20.merge.NTUP\_TOP.e835\_s1299\_s1300\_r3043\_r2993\_p937/ & $Z\rightarrow \mu\mu + 3p$ \\
mc11\_7TeV.107664.AlpgenJimmyZmumuNp4\_pt20.merge.NTUP\_TOP.e835\_s1299\_s1300\_r3043\_r2993\_p937/ & $Z\rightarrow \mu\mu + 4p$ \\
mc11\_7TeV.107665.AlpgenJimmyZmumuNp5\_pt20.merge.NTUP\_TOP.e835\_s1299\_s1300\_r3043\_r2993\_p937/ & $Z\rightarrow \mu\mu + 5p$ \\
mc11\_7TeV.107670.AlpgenJimmyZtautauNp0\_pt20.merge.NTUP\_TOP.e835\_s1299\_s1300\_r3043\_r2993\_p937/ & $Z\rightarrow \tau\tau+ 0p$ \\
mc11\_7TeV.107671.AlpgenJimmyZtautauNp1\_pt20.merge.NTUP\_TOP.e835\_s1299\_s1300\_r3043\_r2993\_p937/ & $Z\rightarrow \tau\tau+ 1p$ \\
mc11\_7TeV.107672.AlpgenJimmyZtautauNp2\_pt20.merge.NTUP\_TOP.e835\_s1299\_s1300\_r3043\_r2993\_p937/ & $Z\rightarrow \tau\tau+ 2p$ \\
mc11\_7TeV.107673.AlpgenJimmyZtautauNp3\_pt20.merge.NTUP\_TOP.e835\_s1299\_s1300\_r3043\_r2993\_p937/ & $Z\rightarrow \tau\tau+ 3p$ \\
mc11\_7TeV.107674.AlpgenJimmyZtautauNp4\_pt20.merge.NTUP\_TOP.e835\_s1299\_s1300\_r3043\_r2993\_p937/ & $Z\rightarrow \tau\tau+ 4p$ \\
mc11\_7TeV.107675.AlpgenJimmyZtautauNp5\_pt20.merge.NTUP\_TOP.e835\_s1299\_s1300\_r3043\_r2993\_p937/ & $Z\rightarrow \tau\tau+ 5p$ \\
\hline
\end{tabular}
}
\end{center}
\caption{Datasets used for the analysis ($Z$+jets).}
\label{tab:samples2}
\end{sidewaystable}

\begin{sidewaystable}[htbp]
\begin{center}
{\tiny
\begin{tabular}{|l|l|}
\hline
Dataset & Function\\
\hline
\hline

mc11\_7TeV.116250.AlpgenJimmyZeeNp0\_Mll10to40\_pt20.merge.NTUP\_TOP.e959\_s1310\_s1300\_r3043\_r2993\_p937/ & $Z\rightarrow ee + 0p$ $10~\text{GeV} < m_{ll}<40~\text{GeV}$\\
mc11\_7TeV.116251.AlpgenJimmyZeeNp1\_Mll10to40\_pt20.merge.NTUP\_TOP.e959\_s1310\_s1300\_r3043\_r2993\_p937/ & $Z\rightarrow ee + 1p$ $10~\text{GeV} < m_{ll}<40~\text{GeV}$\\
mc11\_7TeV.116252.AlpgenJimmyZeeNp2\_Mll10to40\_pt20.merge.NTUP\_TOP.e944\_s1310\_s1300\_r3043\_r2993\_p937/ & $Z\rightarrow ee + 2p$ $10~\text{GeV} < m_{ll}<40~\text{GeV}$\\
mc11\_7TeV.116253.AlpgenJimmyZeeNp3\_Mll10to40\_pt20.merge.NTUP\_TOP.e944\_s1310\_s1300\_r3043\_r2993\_p937/ & $Z\rightarrow ee + 3p$ $10~\text{GeV} < m_{ll}<40~\text{GeV}$\\
mc11\_7TeV.116254.AlpgenJimmyZeeNp4\_Mll10to40\_pt20.merge.NTUP\_TOP.e944\_s1310\_s1300\_r3043\_r2993\_p937/ & $Z\rightarrow ee + 4p$ $10~\text{GeV} < m_{ll}<40~\text{GeV}$\\
mc11\_7TeV.116255.AlpgenJimmyZeeNp5\_Mll10to40\_pt20.merge.NTUP\_TOP.e944\_s1310\_s1300\_r3043\_r2993\_p937/ & $Z\rightarrow ee + 5p$ $10~\text{GeV} < m_{ll}<40~\text{GeV}$\\
mc11\_7TeV.116260.AlpgenJimmyZmumuNp0\_Mll10to40\_pt20.merge.NTUP\_TOP.e959\_s1310\_s1300\_r3043\_r2993\_p937/ & $Z\rightarrow \mu\mu + 0p$ $10~\text{GeV} < m_{ll}<40~\text{GeV}$\\
mc11\_7TeV.116261.AlpgenJimmyZmumuNp1\_Mll10to40\_pt20.merge.NTUP\_TOP.e959\_s1310\_s1300\_r3043\_r2993\_p937/ & $Z\rightarrow \mu\mu + 1p$ $10~\text{GeV} < m_{ll}<40~\text{GeV}$\\
mc11\_7TeV.116262.AlpgenJimmyZmumuNp2\_Mll10to40\_pt20.merge.NTUP\_TOP.e944\_s1310\_s1300\_r3043\_r2993\_p937/ & $Z\rightarrow \mu\mu + 2p$ $10~\text{GeV} < m_{ll}<40~\text{GeV}$\\
mc11\_7TeV.116263.AlpgenJimmyZmumuNp3\_Mll10to40\_pt20.merge.NTUP\_TOP.e944\_s1310\_s1300\_r3043\_r2993\_p937/ & $Z\rightarrow \mu\mu + 3p$ $10~\text{GeV} < m_{ll}<40~\text{GeV}$\\
mc11\_7TeV.116264.AlpgenJimmyZmumuNp4\_Mll10to40\_pt20.merge.NTUP\_TOP.e944\_s1310\_s1300\_r3043\_r2993\_p937/ & $Z\rightarrow \mu\mu + 4p$ $10~\text{GeV} < m_{ll}<40~\text{GeV}$\\
mc11\_7TeV.116265.AlpgenJimmyZmumuNp5\_Mll10to40\_pt20.merge.NTUP\_TOP.e944\_s1310\_s1300\_r3043\_r2993\_p937/ & $Z\rightarrow \mu\mu + 5p$ $10~\text{GeV} < m_{ll}<40~\text{GeV}$\\
mc11\_7TeV.116270.AlpgenJimmyZtautauNp0\_Mll10to40\_pt20.merge.NTUP\_TOP.e959\_s1310\_s1300\_r3043\_r2993\_p937/ & $Z\rightarrow \tau\tau + 0p$ $10~\text{GeV} < m_{ll}<40~\text{GeV}$\\
mc11\_7TeV.116271.AlpgenJimmyZtautauNp1\_Mll10to40\_pt20.merge.NTUP\_TOP.e959\_s1310\_s1300\_r3043\_r2993\_p937/ & $Z\rightarrow \tau\tau + 1p$ $10~\text{GeV} < m_{ll}<40~\text{GeV}$\\
mc11\_7TeV.116272.AlpgenJimmyZtautauNp2\_Mll10to40\_pt20.merge.NTUP\_TOP.e959\_s1310\_s1300\_r3043\_r2993\_p937/ & $Z\rightarrow \tau\tau + 2p$ $10~\text{GeV} < m_{ll}<40~\text{GeV}$\\
mc11\_7TeV.116273.AlpgenJimmyZtautauNp3\_Mll10to40\_pt20.merge.NTUP\_TOP.e959\_s1310\_s1300\_r3043\_r2993\_p937/ & $Z\rightarrow \tau\tau + 3p$ $10~\text{GeV} < m_{ll}<40~\text{GeV}$\\
mc11\_7TeV.116274.AlpgenJimmyZtautauNp4\_Mll10to40\_pt20.merge.NTUP\_TOP.e959\_s1310\_s1300\_r3043\_r2993\_p937/ & $Z\rightarrow \tau\tau + 4p$ $10~\text{GeV} < m_{ll}<40~\text{GeV}$\\
mc11\_7TeV.116275.AlpgenJimmyZtautauNp5\_Mll10to40\_pt20.merge.NTUP\_TOP.e959\_s1310\_s1300\_r3043\_r2993\_p937/ & $Z\rightarrow \tau\tau + 5p$ $10~\text{GeV} < m_{ll}<40~\text{GeV}$\\
mc11\_7TeV.105985.WW\_Herwig.merge.NTUP\_TOP.e825\_s1310\_s1300\_r3043\_r2993\_p937/ & $WW$ \\
mc11\_7TeV.105986.ZZ\_Herwig.merge.NTUP\_TOP.e825\_s1310\_s1300\_r3043\_r2993\_p937/ & $ZZ$ \\
mc11\_7TeV.105987.WZ\_Herwig.merge.NTUP\_TOP.e825\_s1310\_s1300\_r3043\_r2993\_p937/ & $WZ$ \\
mc11\_7TeV.117360.st\_tchan\_enu\_AcerMC.merge.NTUP\_TOP.e835\_s1310\_s1300\_r3043\_r2993\_p937/ & single top, t-chan, $e\nu$ \\
mc11\_7TeV.117361.st\_tchan\_munu\_AcerMC.merge.NTUP\_TOP.e835\_s1310\_s1300\_r3043\_r2993\_p937/ & single top, t-chan, $\mu\nu$ \\
mc11\_7TeV.117362.st\_tchan\_taunu\_AcerMC.merge.NTUP\_TOP.e825\_s1310\_s1300\_r3043\_r2993\_p937/ & single top, t-chan, $\tau\nu$ \\
mc11\_7TeV.108343.st\_schan\_enu\_McAtNlo\_Jimmy.merge.NTUP\_TOP.e825\_s1310\_s1300\_r3043\_r2993\_p937/ & single top, s-chan, $e\nu$ \\
mc11\_7TeV.108344.st\_schan\_munu\_McAtNlo\_Jimmy.merge.NTUP\_TOP.e825\_s1310\_s1300\_r3043\_r2993\_p937/ & single top, s-chan, $\mu\nu$ \\
mc11\_7TeV.108345.st\_schan\_taunu\_McAtNlo\_Jimmy.merge.NTUP\_TOP.e835\_s1310\_s1300\_r3043\_r2993\_p937/ & single top, s-chan, $\tau\nu$ \\
mc11\_7TeV.108346.st\_Wt\_McAtNlo\_Jimmy.merge.NTUP\_TOP.e835\_s1310\_s1300\_r3043\_r2993\_p937/ & single top, Wt-chan, $e\nu$  \\
\hline
\end{tabular}
}
\end{center}
\caption{Datasets used for the analysis ($Z$+jets, single top and diboson).}
\label{tab:samples3}
\end{sidewaystable}

\begin{sidewaystable}[htbp]
\begin{center}
{\tiny
\begin{tabular}{|l|l|}
\hline
Dataset & Function\\
\hline
\hline

mc11\_7TeV.105861.TTbar\_PowHeg\_Pythia.merge.NTUP\_TOP.e873\_a131\_s1353\_a139\_r2900\_p937/ & PowHeg+Pythia\\
mc11\_7TeV.105860.TTbar\_PowHeg\_Jimmy.merge.NTUP\_TOP.e1198\_a131\_s1353\_a139\_r2900\_p937/ & PowHeg+Herwig\\
mc11\_7TeV.117520.AlpGenPythia\_P2011radHi\_KTFac05CTEQ5L\_ttbarlnqqNp0.merge.NTUP\_TOP.e1608\_a131\_s1353\_a145\_r2993\_p937/ & ISR/FSR up\\
mc11\_7TeV.117521.AlpGenPythia\_P2011radHi\_KTFac05CTEQ5L\_ttbarlnqqNp1.merge.NTUP\_TOP.e1608\_a131\_s1353\_a145\_r2993\_p937/ & ISR/FSR up\\
mc11\_7TeV.117522.AlpGenPythia\_P2011radHi\_KTFac05CTEQ5L\_ttbarlnqqNp2.merge.NTUP\_TOP.e1608\_a131\_s1353\_a145\_r2993\_p937/ & ISR/FSR up\\
mc11\_7TeV.117523.AlpGenPythia\_P2011radHi\_KTFac05CTEQ5L\_ttbarlnqqNp3.merge.NTUP\_TOP.e1608\_a131\_s1353\_a145\_r2993\_p937/ & ISR/FSR up\\
mc11\_7TeV.117524.AlpGenPythia\_P2011radHi\_KTFac05CTEQ5L\_ttbarlnqqNp4INC.merge.NTUP\_TOP.e1608\_a131\_s1353\_a145\_r2993\_p937/ & ISR/FSR up\\
mc11\_7TeV.117525.AlpGenPythia\_P2011radHi\_KTFac05CTEQ5L\_ttbarlnlnNp0.merge.NTUP\_TOP.e1608\_a131\_s1353\_a145\_r2993\_p937/ & ISR/FSR up\\
mc11\_7TeV.117526.AlpGenPythia\_P2011radHi\_KTFac05CTEQ5L\_ttbarlnlnNp1.merge.NTUP\_TOP.e1608\_a131\_s1353\_a145\_r2993\_p937/ & ISR/FSR up\\
mc11\_7TeV.117527.AlpGenPythia\_P2011radHi\_KTFac05CTEQ5L\_ttbarlnlnNp2.merge.NTUP\_TOP.e1608\_a131\_s1353\_a145\_r2993\_p937/ & ISR/FSR up\\
mc11\_7TeV.117528.AlpGenPythia\_P2011radHi\_KTFac05CTEQ5L\_ttbarlnlnNp3.merge.NTUP\_TOP.e1608\_a131\_s1353\_a145\_r2993\_p937/ & ISR/FSR up\\
mc11\_7TeV.117529.AlpGenPythia\_P2011radHi\_KTFac05CTEQ5L\_ttbarlnlnNp4INC.merge.NTUP\_TOP.e1608\_a131\_s1353\_a145\_r2993\_p937/ & ISR/FSR up\\
mc11\_7TeV.117530.AlpGenPythia\_P2011radLo\_KTFac2CTEQ5L\_ttbarlnqqNp0.merge.NTUP\_TOP.e1608\_a131\_s1353\_a145\_r2993\_p937/ & ISR/FSR down\\
mc11\_7TeV.117531.AlpGenPythia\_P2011radLo\_KTFac2CTEQ5L\_ttbarlnqqNp1.merge.NTUP\_TOP.e1608\_a131\_s1353\_a145\_r2993\_p937/ & ISR/FSR down\\
mc11\_7TeV.117532.AlpGenPythia\_P2011radLo\_KTFac2CTEQ5L\_ttbarlnqqNp2.merge.NTUP\_TOP.e1608\_a131\_s1353\_a145\_r2993\_p937/ & ISR/FSR down\\
mc11\_7TeV.117533.AlpGenPythia\_P2011radLo\_KTFac2CTEQ5L\_ttbarlnqqNp3.merge.NTUP\_TOP.e1608\_a131\_s1353\_a145\_r2993\_p937/ & ISR/FSR down\\
mc11\_7TeV.117534.AlpGenPythia\_P2011radLo\_KTFac2CTEQ5L\_ttbarlnqqNp4INC.merge.NTUP\_TOP.e1608\_a131\_s1353\_a145\_r2993\_p937/ & ISR/FSR down\\
mc11\_7TeV.117535.AlpGenPythia\_P2011radLo\_KTFac2CTEQ5L\_ttbarlnlnNp0.merge.NTUP\_TOP.e1608\_a131\_s1353\_a145\_r2993\_p937/ & ISR/FSR down\\
mc11\_7TeV.117536.AlpGenPythia\_P2011radLo\_KTFac2CTEQ5L\_ttbarlnlnNp1.merge.NTUP\_TOP.e1608\_a131\_s1353\_a145\_r2993\_p937/ & ISR/FSR down\\
mc11\_7TeV.117537.AlpGenPythia\_P2011radLo\_KTFac2CTEQ5L\_ttbarlnlnNp2.merge.NTUP\_TOP.e1608\_a131\_s1353\_a145\_r2993\_p937/ & ISR/FSR down\\
mc11\_7TeV.117538.AlpGenPythia\_P2011radLo\_KTFac2CTEQ5L\_ttbarlnlnNp3.merge.NTUP\_TOP.e1608\_a131\_s1353\_a145\_r2993\_p937/ & ISR/FSR down\\
mc11\_7TeV.117539.AlpGenPythia\_P2011radLo\_KTFac2CTEQ5L\_ttbarlnlnNp4INC.merge.NTUP\_TOP.e1608\_a131\_s1353\_a145\_r2993\_p937/ & ISR/FSR down\\

mc11\_7TeV.117428.TTbar\_PowHeg\_Pythia\_P2011.merge.NTUP\_TOP.e1683\_a131\_s1353\_a145\_r2993\_p937/ & PowHeg+Pythia (P2011 tune)\\
mc11\_7TeV.117430.TTbar\_PowHeg\_Pythia\_P2011noCR.merge.NTUP\_TOP.e1683\_a131\_s1353\_a145\_r2993\_p937/ & PowHeg+Pythia (P2011, no CR)\\
mc11\_7TeV.117429.TTbar\_PowHeg\_Pythia\_P2011mpiHi.merge.NTUP\_TOP.e1683\_a131\_s1353\_a145\_r2993\_p937/ & PowHeg+Pythia (P2011, more UE)\\

mc11\_7TeV.110006.McAtNloJimmy\_CT10\_ttbar\_mudown\_LeptonFilter.merge.NTUP\_TOP.e1468\_a131\_s1353\_a145\_r2993\_p937/ & Ren./Fact. Scale Up\\
mc11\_7TeV.110007.McAtNloJimmy\_CT10\_ttbar\_muup\_LeptonFilter.merge.NTUP\_TOP.e1468\_a131\_s1353\_a145\_r2993\_p937/ & Ren./Fact. Scale Down\\
\hline

\end{tabular}
}
\end{center}
\caption{Datasets used for the analysis (modelling uncertainties).}
\label{tab:samples4}
\end{sidewaystable}
 
\chapter{Pretag Yields}
\label{sec:app_pretag_yields}

\begin{table}[htbp]
\begin{center}
\begin{tabular}{
 |   l  |
    S[table-format=5.1]@{\,\( \pm \)\,}
    S[table-format=5.1] |
    S[table-format=-5.1]@{\,\( \pm \)\,}
    S[table-format=5.1]|
    } 
    \hline
$n_{\text{jets}} \geq 4$, pretag & \multicolumn{2}{c|}{\ejets} & \multicolumn{2}{c|}{\mujets} \\
\hline
\hline
$W+$jets (DD/MC) & 12930  &  1550 & 27880  &  3070\\
$Z+$jets (MC) & 2860  &  1370 & 2870  &  1380\\
Fake leptons (DD) & 2310  &  1160 & 5560  &  1110\\
Single top (MC) & 1460  &  70 & 2420  &  100\\
Diboson (MC) & 230  &  10 & 370  &  20\\
\hline
Total (non-\ttbar) & 19780  &  2370 & 39110  &  3540\\
\hline
\ttbar\ (MC, l+jets) & 17280  &  1030 & 28800  &  1710\\

\ttbar\ (MC, dilepton) & 2360  &  140 & 3530  &  210\\
\hline
Total expected & 39420  &  2590 & 71420  &  3940\\
\hline
Observed & \multicolumn{2}{c|}{40550} & \multicolumn{2}{c|}{70740}\\
\hline
\end{tabular}
\end{center}
\caption{Number of selected data events and background composition for $n_{\text{jets}} \geq 4$ without the \btag ging requirement. For data driven backgrounds statistical uncertainties are quoted. The uncertainties on the cross sections determine the uncertainties for the Monte Carlo driven backgrounds.}
\label{tab:cutflow_pretag}
\end{table}
\include{appendices/CPs}
 

\chapter{KLFitter Likelihood Components}
\label{sec:app_lhcomp}
The value of the (logarithm of the) likelihood (LH) of \KLFitter\ is a useful quantity to judge the quality of an event reconstruction. However, the likelihood is a complex quantity and needs to be understood properly prior to any interpretation. 

Instead of checking the likelihood for a global event quality, also its individual components (Breit-Wigner functions for the masses and transfer functions for the energy and momentum resolutions) can be checked. This allows judging the quality of certain objects. 

Figure \ref{fig:TF_dQ} shows the values of the \dQ\ transfer function component. In case the \dQ\ does not match, the fit needs to vary the energy of the \dQ\ candidate up the the tails to reach a proper \ttbar\ event topology. This effect gets even larger in case the whole $W$ boson (including the up-type quark) does not match. 

A similar effect can be observed for the Breit-Wigner function of the hadronic $W$ boson mass (Figure \ref{fig:BW_Whad}). The gap between the peak at $\log(LH) \approx -10$ and the tail starting at $\log(LH) \approx -15$ illustrates the interplay of the TFs and the Breit-Wigner functions. Due to the narrow width of the Breit-Wigner functions, the fit prefers the top quark and the $W$ boson masses to be on the resonance. As a consequence, the transfer functions get values off their peak. This is shown in Figure \ref{fig:TF_light_vs_BW_Whad}, where the sum of the TF values of the two light quarks from the $W$ boson decay plotted for different values of the Breit-Wigner component. Far away from the Breit-Wigner peak, the slope of the Gaussian TFs dominate the Breit-Wigner peak: while the Breit-Wigner values are far away from their peak, the TFs stay in theirs. 

The interplay between the hadronic top quark and $W$ boson mass is shown in Figure \ref{fig:BW_Thad_vs_BW_Whad}. Different areas in the plot indicate the misreconstruction of the hadronically decaying $W$ boson, the $b$-jet of the hadronically decaying top quark or of both. 

How the shape of the likelihood distributions is affected by the (mis)match of certain model partons is shown in figure \ref{fig:LH_dist}.

\begin{figure}[htbp]
\begin{center}
\subfigure[]{
\includegraphics[width=0.45\textwidth]{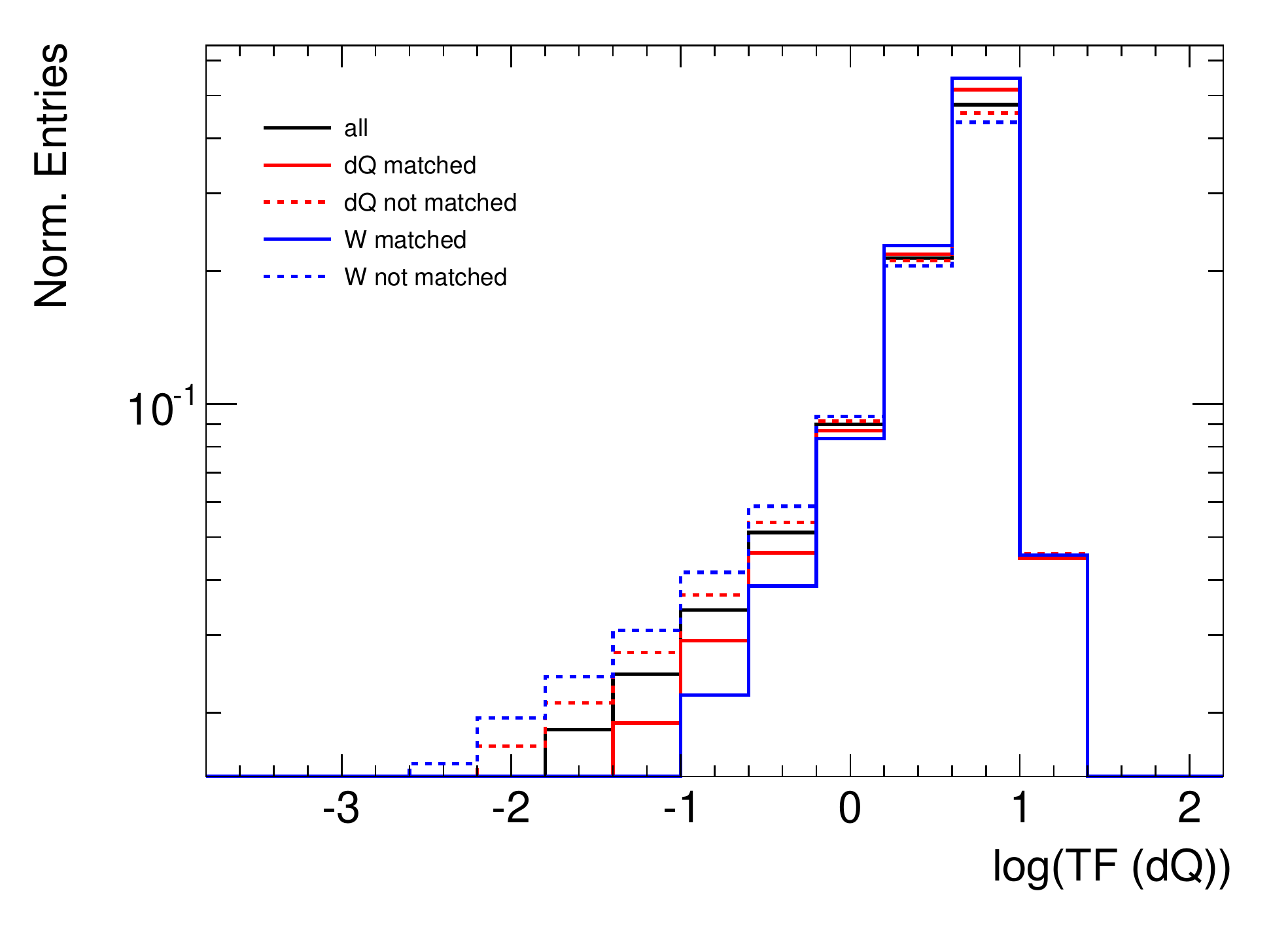}
\label{fig:TF_dQ}
}
			\subfigure[]{
\includegraphics[width=0.45\textwidth]{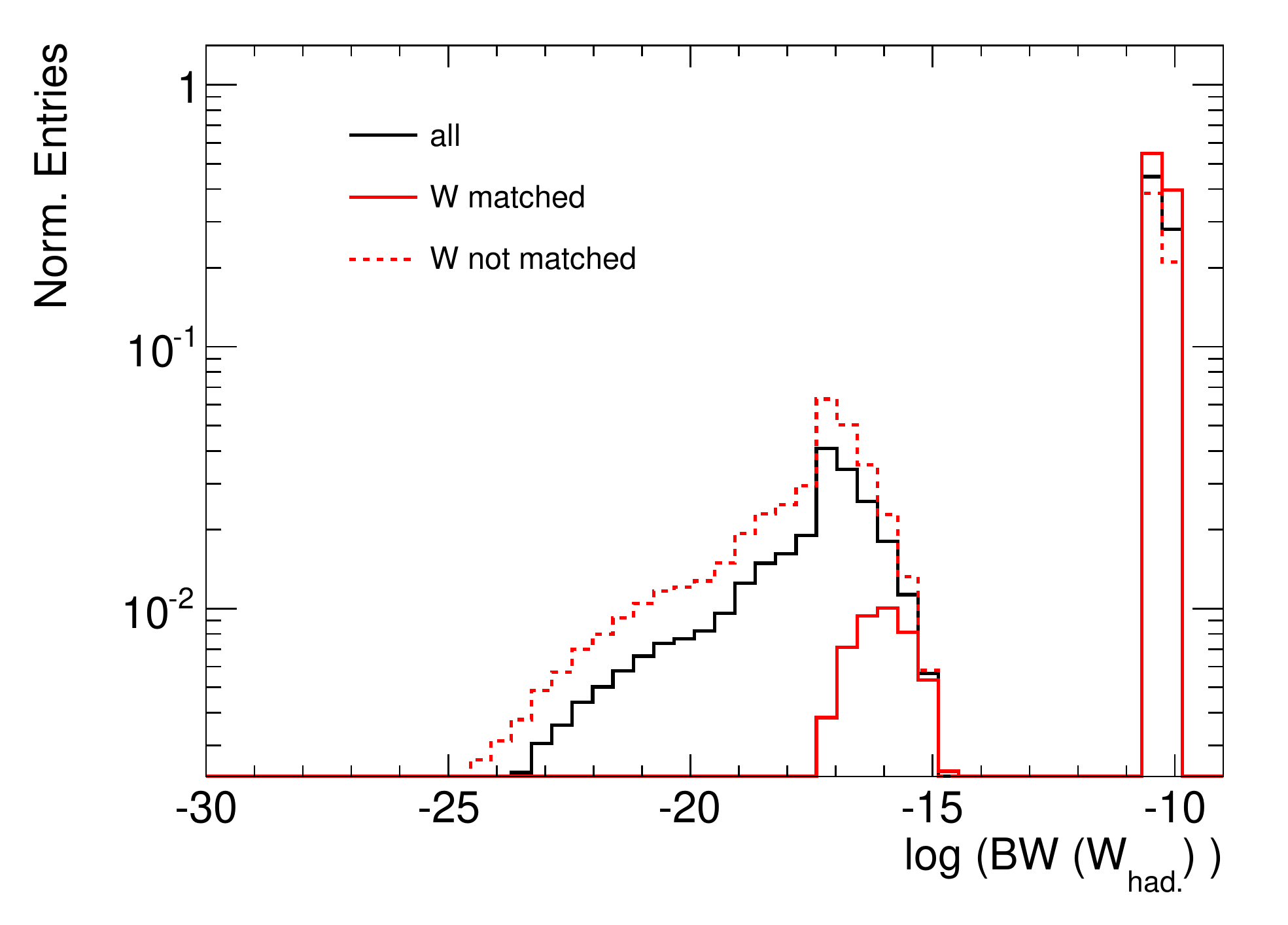}
\label{fig:BW_Whad}
}
			\subfigure[]{
\includegraphics[width=0.40\textwidth]{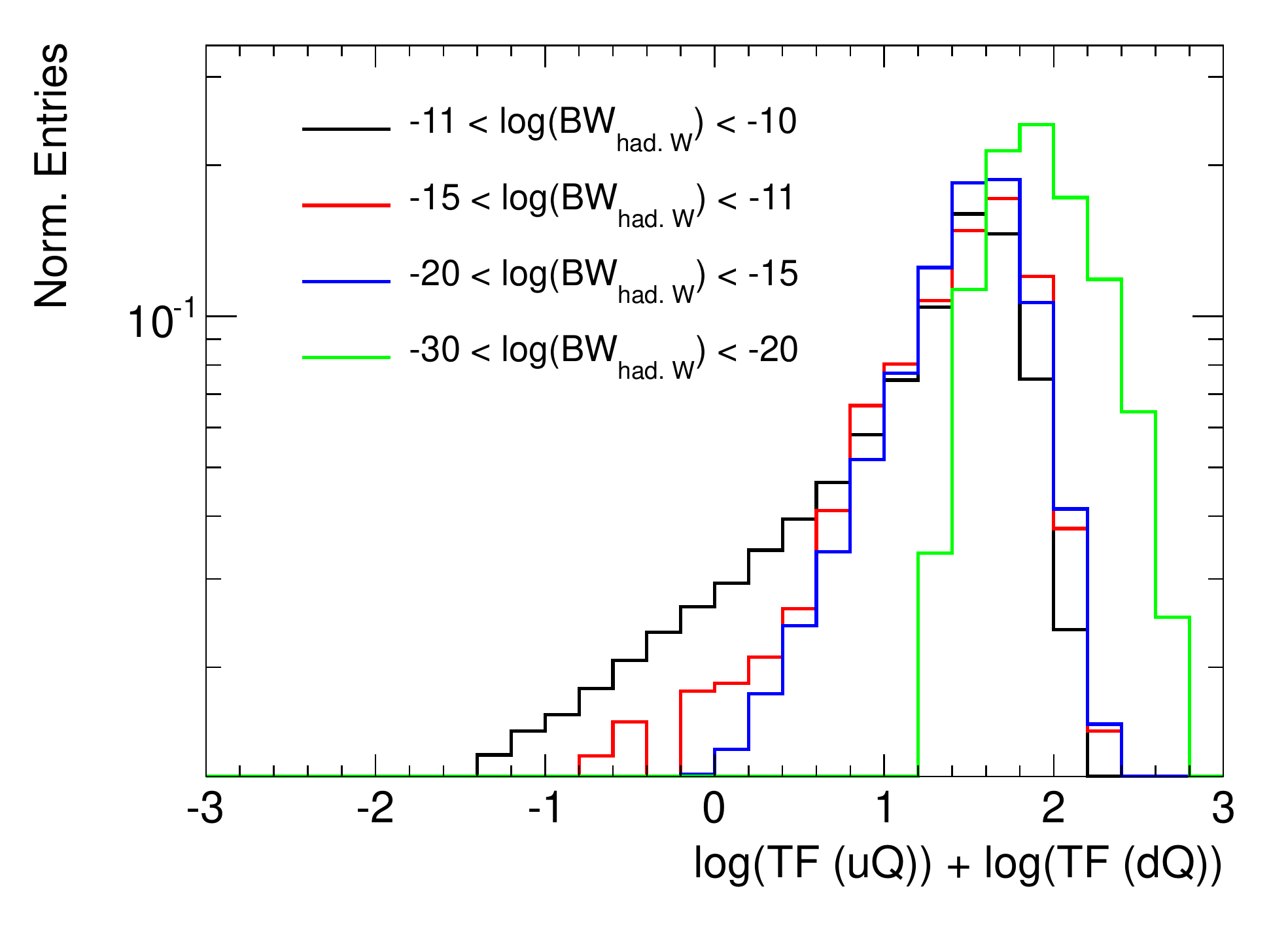}
\label{fig:TF_light_vs_BW_Whad}
}
			\subfigure[]{
\includegraphics[width=0.45\textwidth]{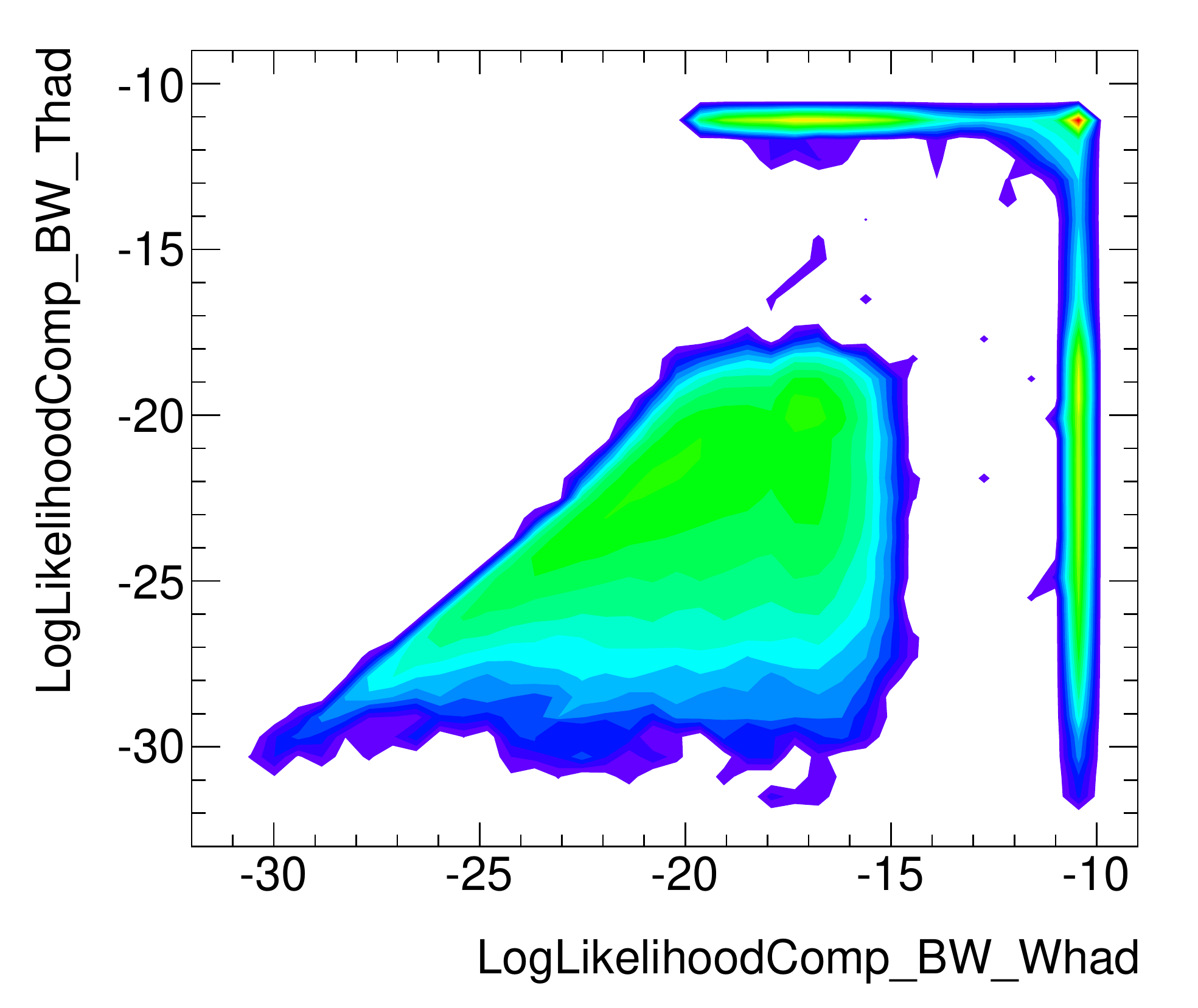}
\label{fig:BW_Thad_vs_BW_Whad}
}

\end{center}
\caption{Components of the \KLFitter\ likelihood. \subref{fig:TF_dQ} \dQ\ TF for (un)matched \dQ\ jets and $W$ bosons. \subref{fig:BW_Whad} Distribution of the Breit-Wigner function of the hadronic $W$ boson. \subref{fig:TF_light_vs_BW_Whad} Sum of the logarithm of the light quark jets' TF values for different values of the Breit-Wigner function of the hadronic $W$ boson. \subref{fig:BW_Thad_vs_BW_Whad} Breit-Wigner functions for the hadronic top quark and $W$ boson mass.}
\label{fig:LH_components}
\end{figure}

\begin{figure}[htbp]
\begin{center}
\includegraphics[width=0.65\textwidth]{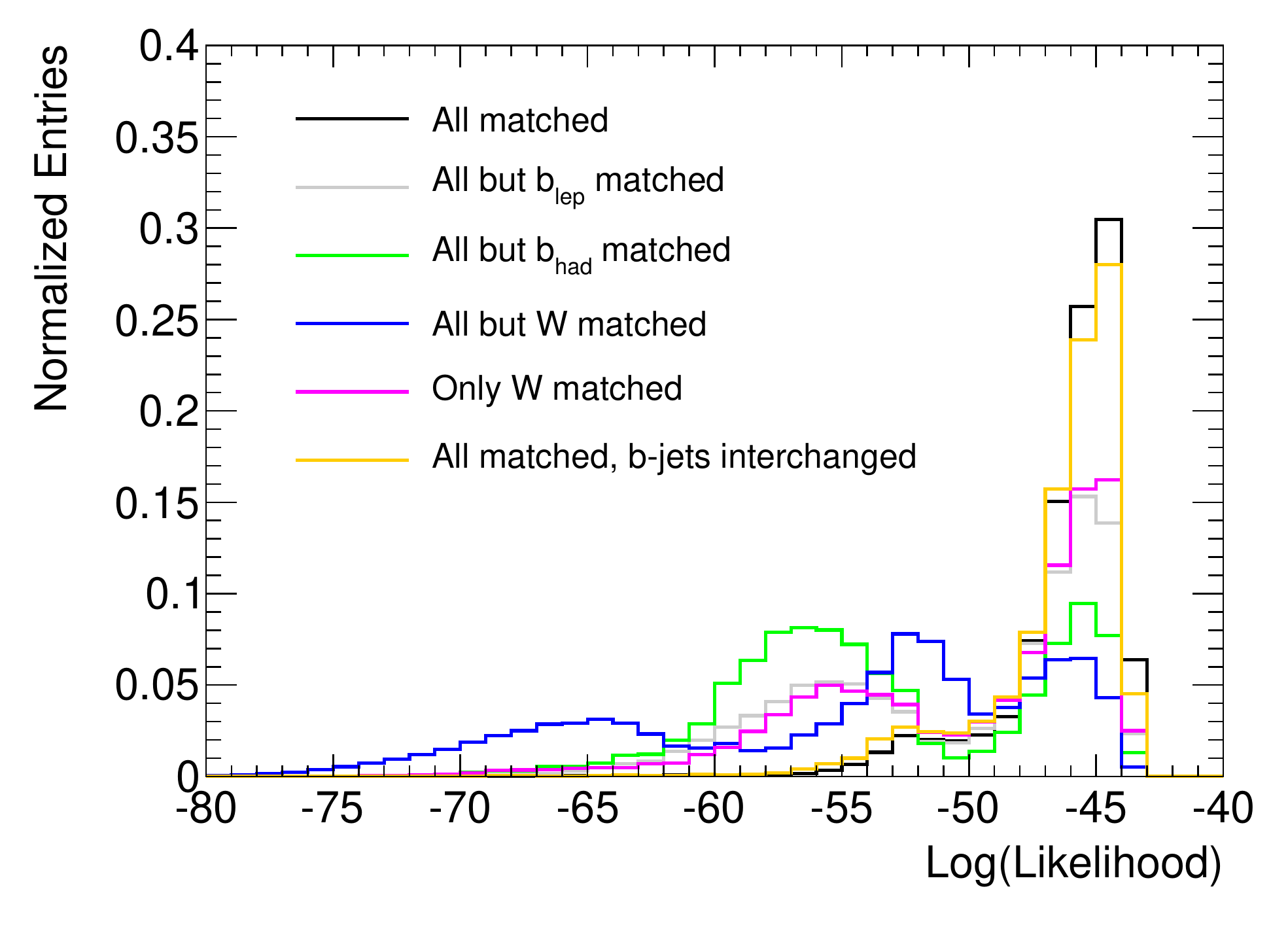}
\label{fig:LH_dist}
\end{center}
\caption{Distribution of the logarithm of the \KLFitter\ likelihood for the permutation with the highest event probability. Different (mis)matching scenarios are shown.}
\end{figure}

 

\chapter{Down-Type Quark \pt\ Spectrum in \powheg+\pythia}
\label{sec:ptpow}
In contrast to \mcatnlo, \powheg+\pythia\ is able to properly model the jet \pt\ spectra. Figure \ref{fig:app_jetpt} shows the \pt\ spectra of the \dQ\ jet. 
\vfill

\begin{figure}[ht]
\begin{center}
\includegraphics[width=0.3\textwidth]{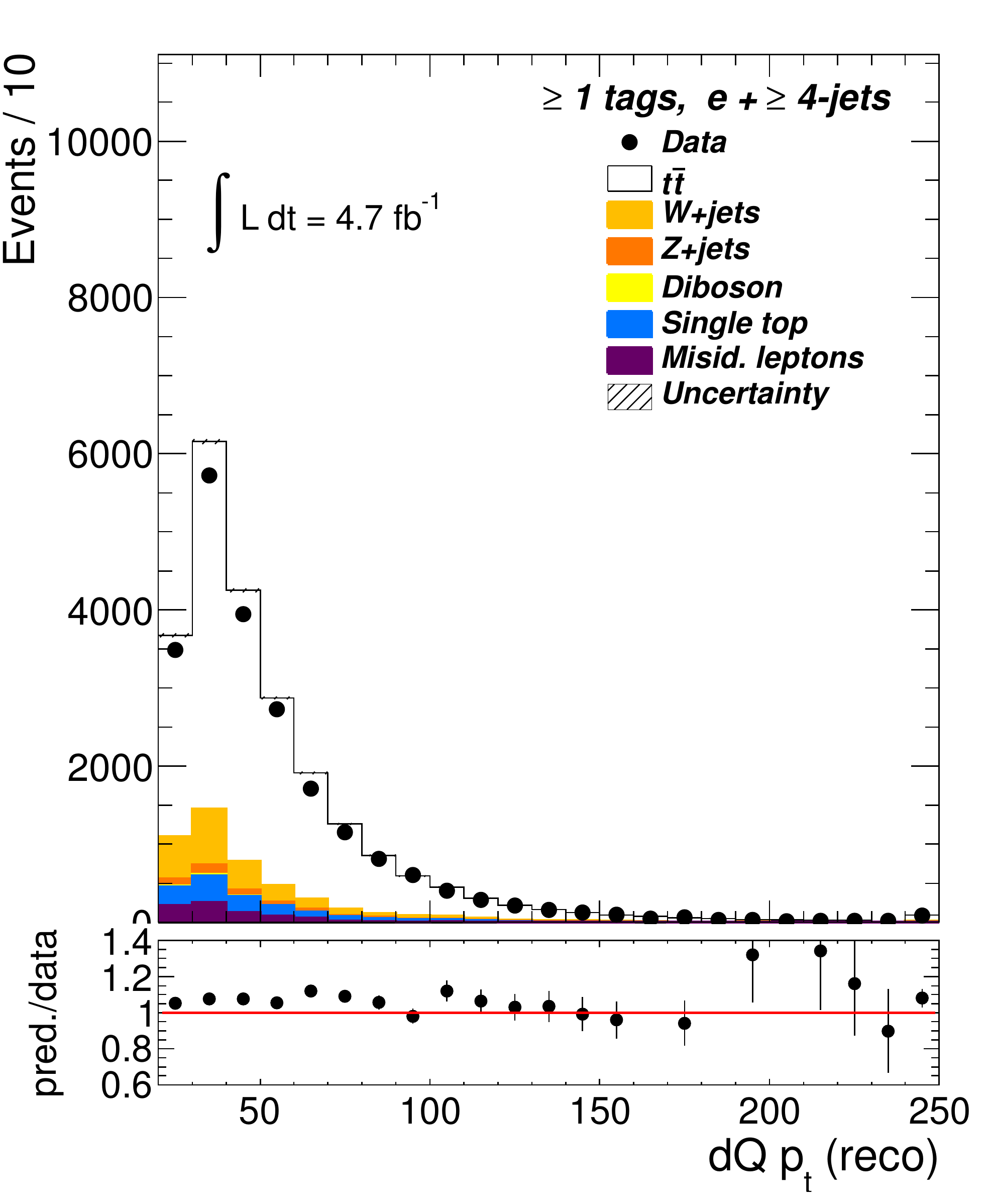}
\includegraphics[width=0.3\textwidth]{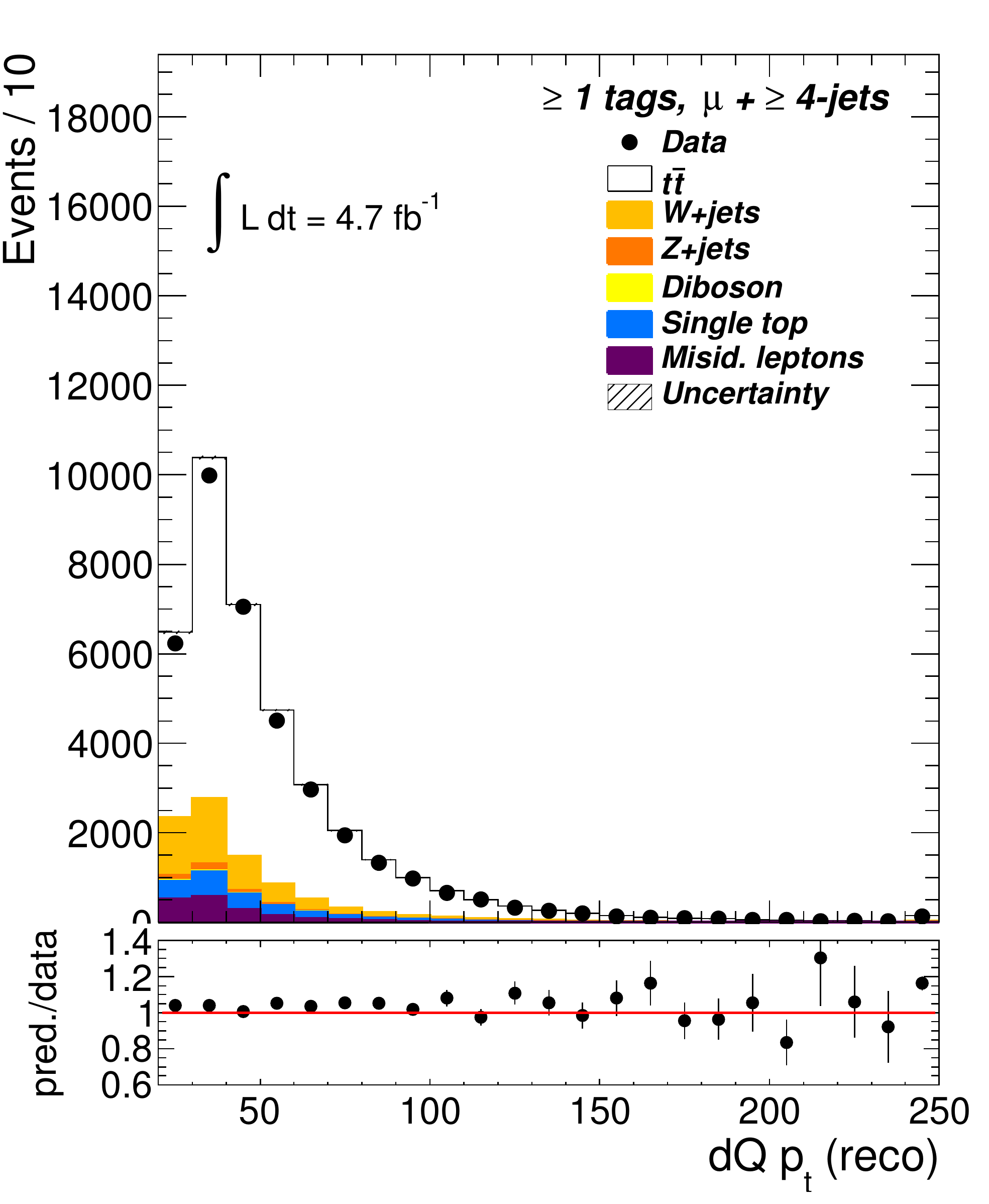}\\
\includegraphics[width=0.3\textwidth]{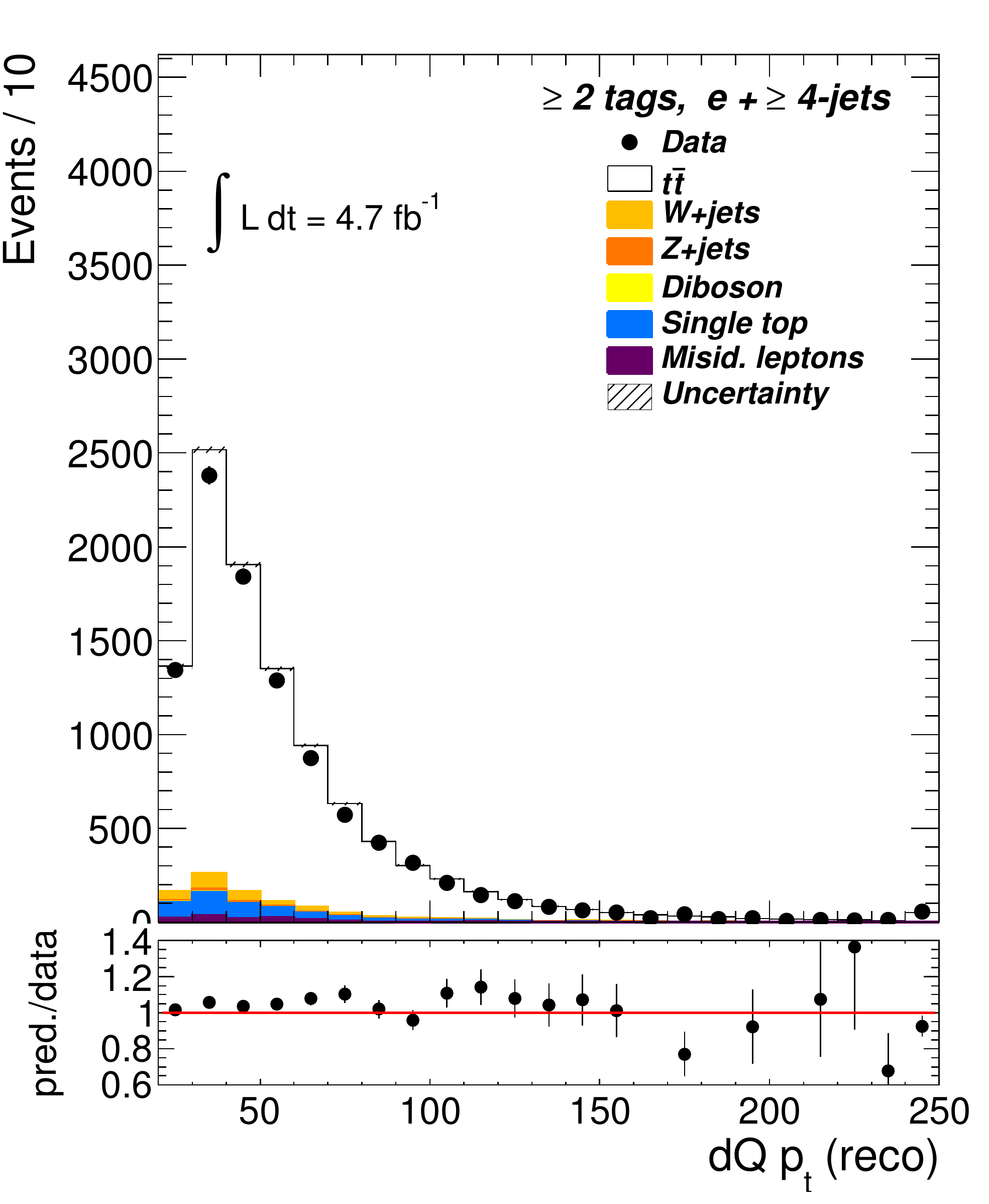}
\includegraphics[width=0.3\textwidth]{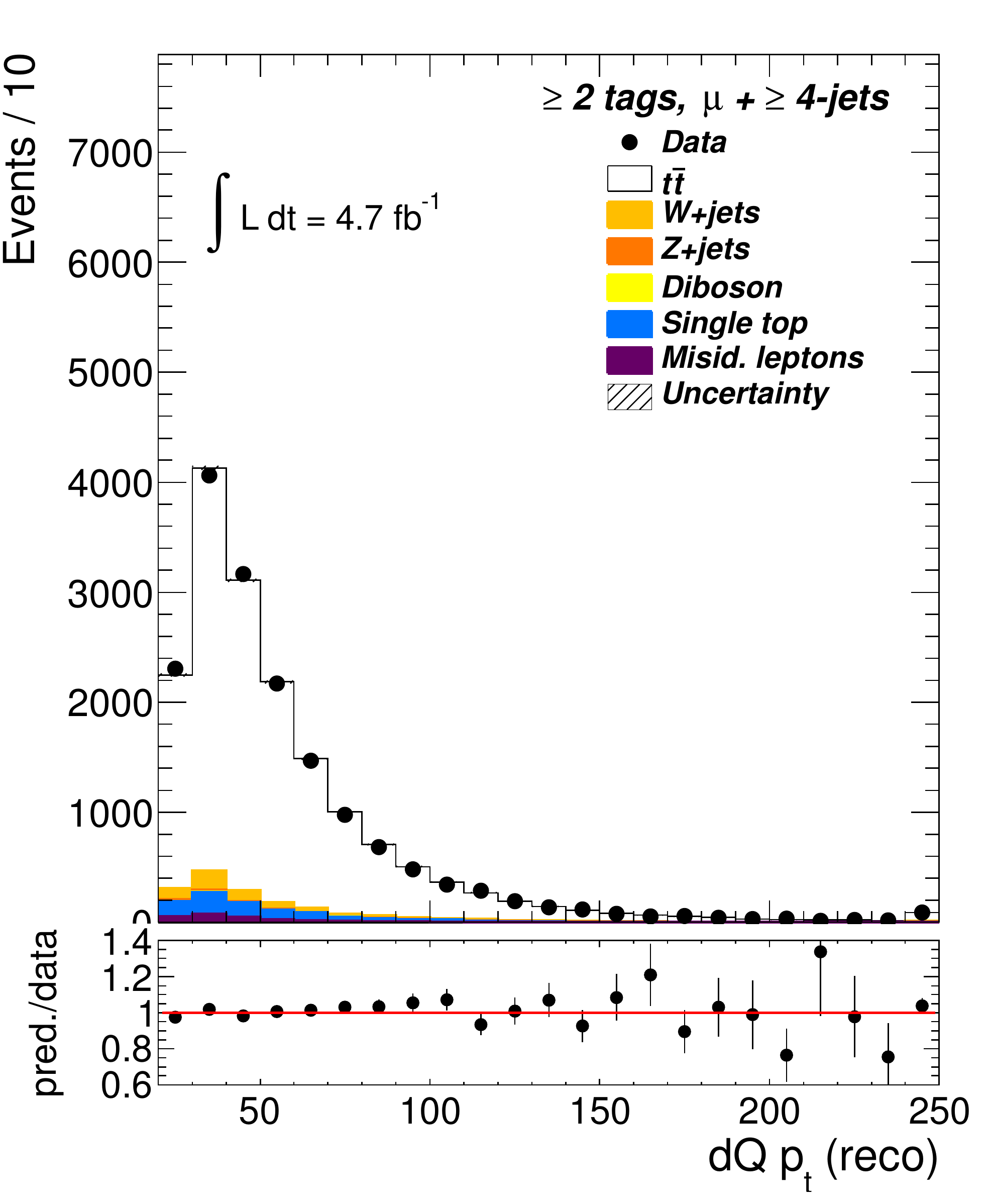}
\end{center}
\caption{\pt\ spectrum of the reconstructed \dQ\ jet with the selection requirement of $n_{\text{jets}} \geq 4$ and $n_{\text{b-tags}} = 1$ (upper row) and $n_{\text{b-tags}} \geq 2$ (lower row) in the \ejets\ (left) and the \mujets\ channel (right). \powheg+\pythia\ was used as generator. }
\label{fig:app_jetpt}
\end{figure}

 

\chapter{Posterior Distributions of Fit Parameters}
\label{sec:app_posteriors}
\begin{figure}[htbp]
\begin{center}
\includegraphics[height=0.9\textheight]{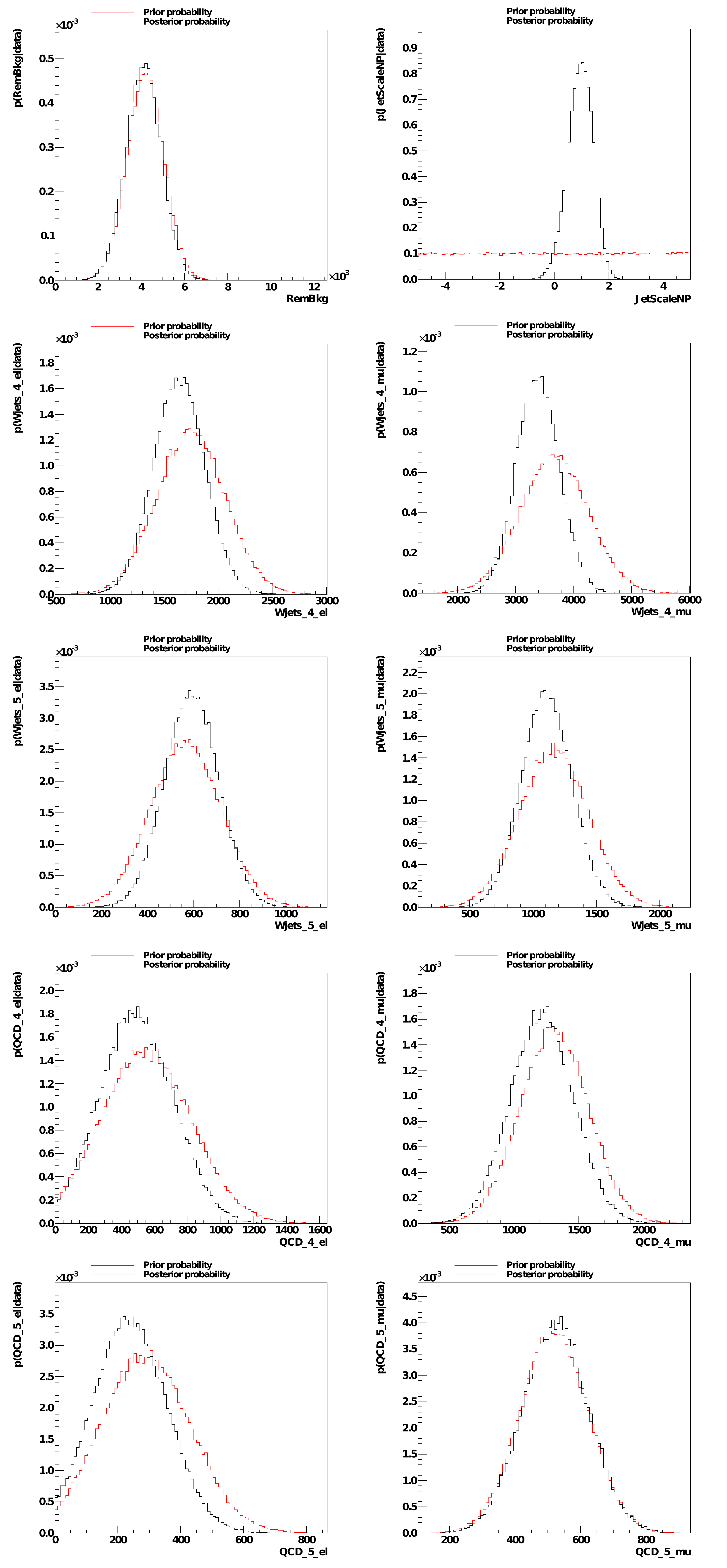}
\end{center}
\caption{Prior and posterior distributions for the fit parameters describing the background yields and the jet multiplicity correction for the combination of the \dQ\ analysers.}
\label{fig:par_posteriors_dQ}
\end{figure} 
\begin{figure}[htbp]
\begin{center}
\includegraphics[height=0.9\textheight]{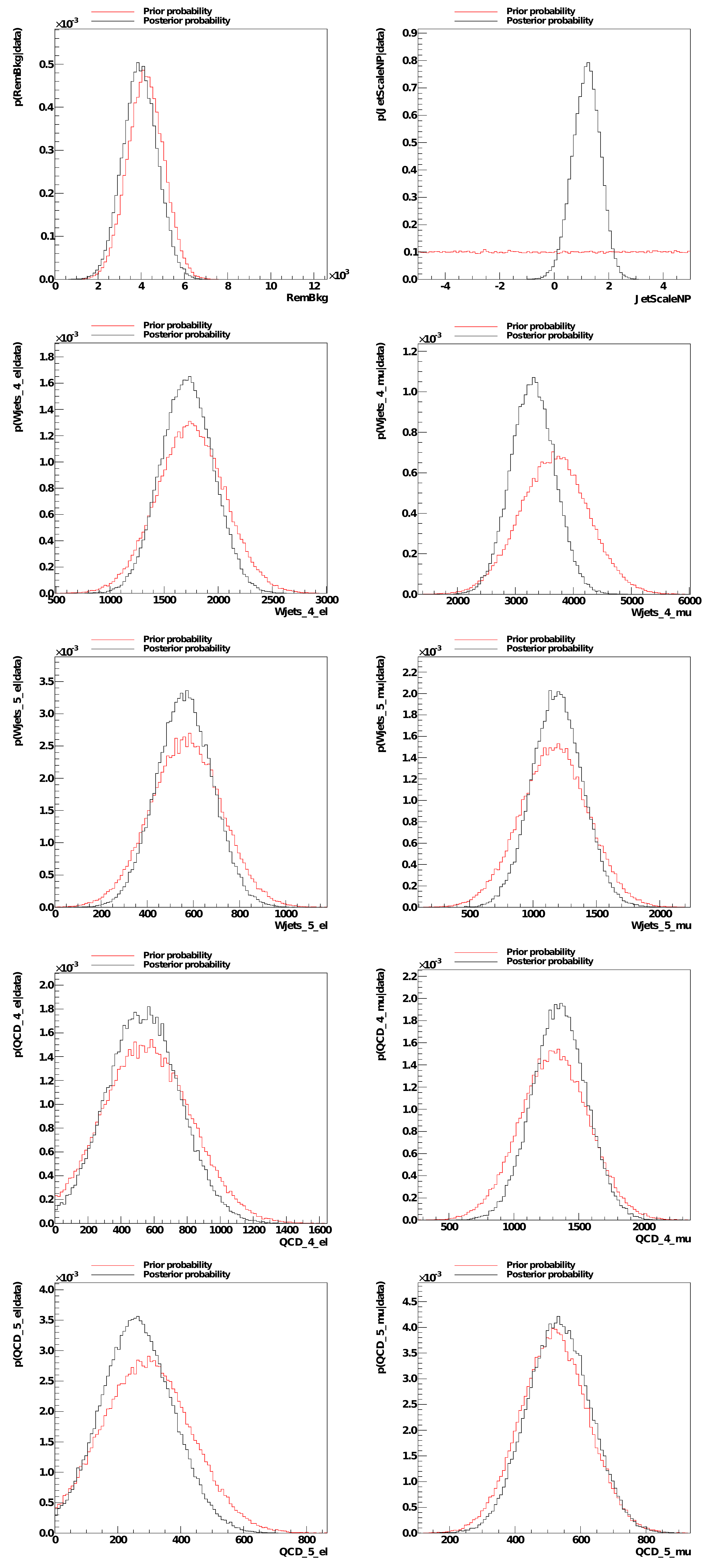}
\end{center}
\caption{Prior and posterior distributions for the fit parameters describing the background yields and the jet multiplicity correction for the combination of the \bQ\ analysers.}
\label{fig:par_posteriors_bQ}
\end{figure} 
 

\chapter{Postfit Values of Nuisance Parameters}
\label{sec:app_NPs}
\begin{figure}[H]
\begin{center}
\includegraphics[width=\textwidth]{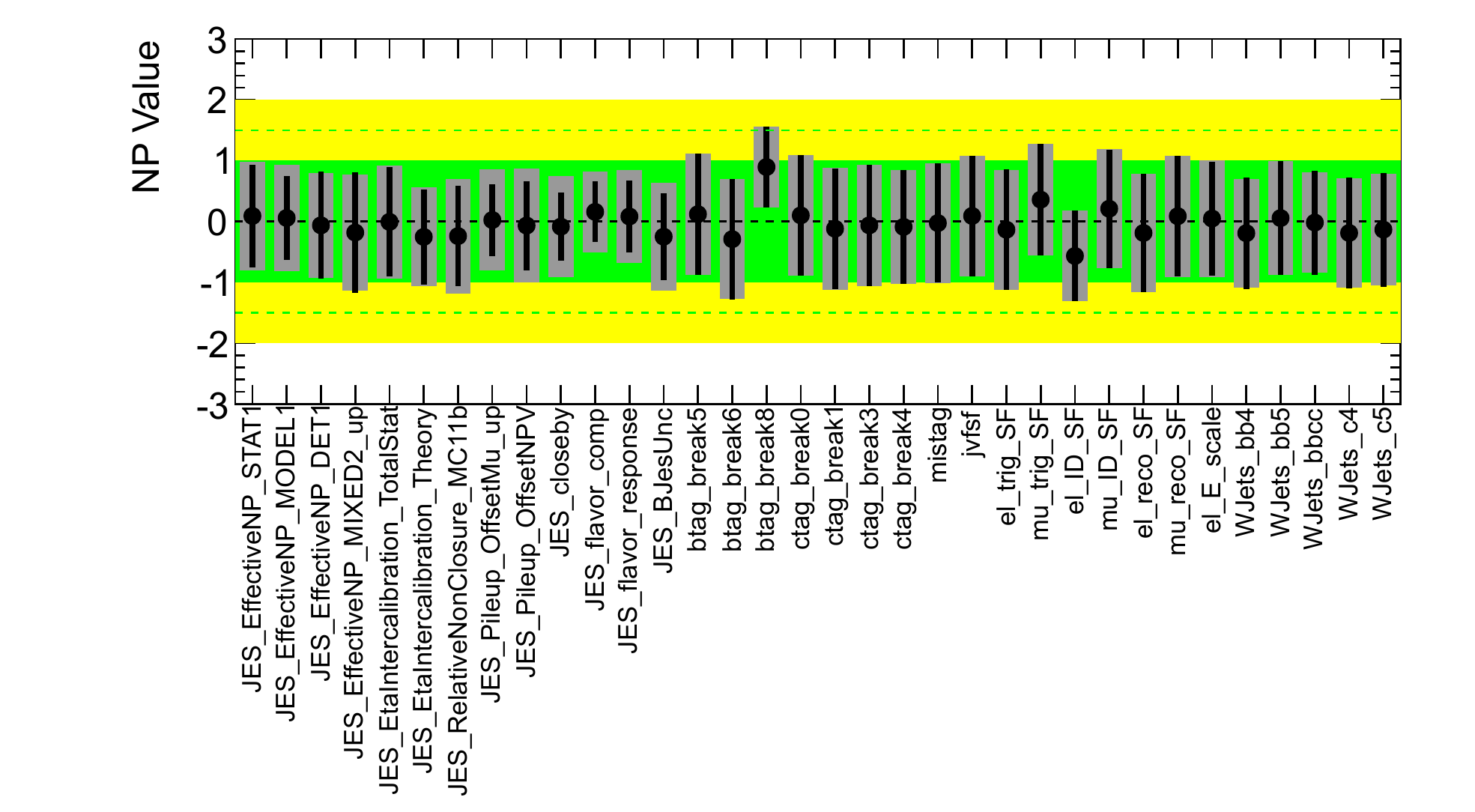}
\end{center}
\caption{Postfit values of the nuisance parameters (black lines) for the \dQ\ combination. The grey bars areas behind the lines show the expected uncertainties on the nuisance parameters.}
\label{fig:NP_postfit_dQ}
\end{figure} 

\begin{figure}[H]
\begin{center}
\includegraphics[width=\textwidth]{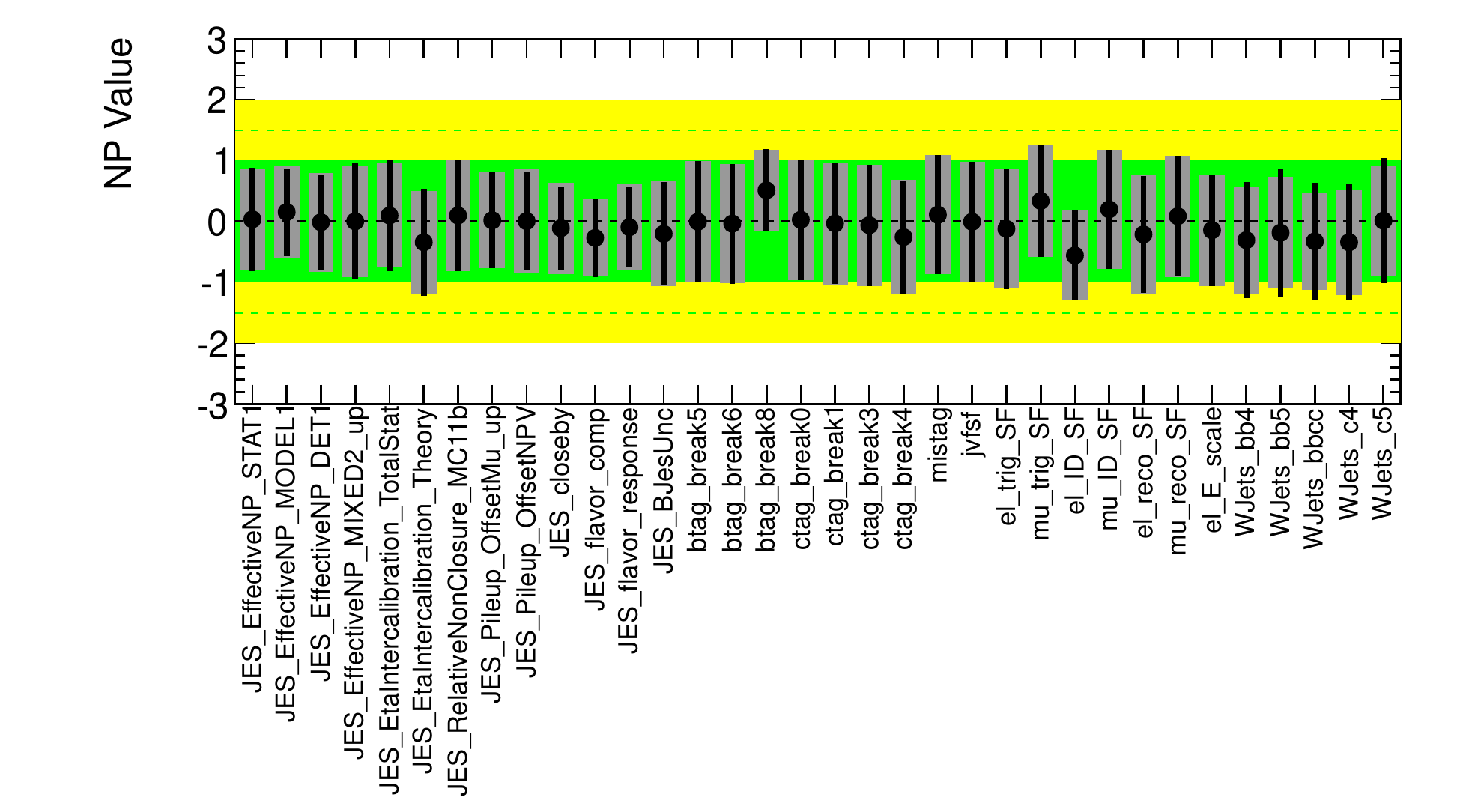}
\end{center}
\caption{Postfit values of the nuisance parameters (black lines) for the \bQ\ combination. The grey bars areas behind the lines show the expected uncertainties on the nuisance parameters.}
\label{fig:NP_postfit_bQ}
\end{figure} 
\include{appendices/lepflavons}
 
\chapter{Most Significant Uncertainties}
\label{sec:app_sigunc}
This section contains lists of the most important uncertainties used as nuisance parameters for the combinations of the individual spin analysers.
\begin{table}[htbp]
\begin{center}
\begin{tabular}{|c|c|}
\hline
NP &  relative change of \fsm\ \\
\hline
\hline
btag/break8  & + 1.8\,\%\\
JES/BJES  & + 1.3\,\%\\
JES/Intercal\_TotalStat & + 1.1\,\%\\
JES/EffectiveNP\_Model1  & + 1.0\,\%\\
JES/EffectiveNP\_Stat1  & + 0.9\,\%\\
\hline
\end{tabular}
\end{center}
\caption{Most significant nuisance parameters (in terms of change of \fsm) for the combined fit of \dQ\ analysers. }
\label{tab:top5_NPs_fsm_dQ}
\end{table}

\begin{table}[htbp]
\begin{center}
\begin{tabular}{|c|c|}
\hline
NP &  relative change of \fsm\ \\
\hline
\hline
JES/RelativeNonClosureMC11b  & $-$ 3.1\,\%\\
JES/EffectiveNP\_Det1  & $-$ 2.6\,\%\\
JES/BJES  & $-$ 2.4\,\%\\
JES/EffectiveNP\_Stat1  & + 2.4\,\%\\
ctag/break4 & + 2.4\,\%\\
\hline
\end{tabular}
\end{center}
\caption{Most significant nuisance parameters (in terms of change of \fsm) for the combined fit of \bQ\ analysers. }
\label{tab:top5_NPs_fsm_bQ}
\end{table}

\begin{table}[htbp]
\begin{center}
\begin{tabular}{|c|c|}
\hline
NP &  relative change of fit uncertainty \\
\hline
\hline
JES/FlavorComp & $-$ 10.4\,\%\\
btag/break8 & $-$ 4.2\,\%\\
JES/FlavorResponse & $-$ 3.5\,\%\\
JES/EffectiveNP\_Model1 & $-$ 3.0\,\%\\
btag/break5  & $-$ 1.7\,\%\\
\hline
\end{tabular}
\end{center}
\caption{Most significant nuisance parameters (in terms of change of the fit uncertainty) for the combined fit of \dQ\ analysers.}
\label{tab:top5_NPs_width_dQ}
\end{table}

\begin{table}[htbp]
\begin{center}
\begin{tabular}{|c|c|}
\hline
NP &  relative change of fit uncertainty \\
\hline
\hline
JES/FlavorComp & $-$ 5.8\,\%\\
JES/Intercal\_TotalStat & + 2.3\,\%\\
el/ID & $-$ 2.1\,\%\\
ctag/break4 & $-$ 1.9\,\%\\
el/trigger\_SF & $-$ 1.6\,\%\\
\hline
\end{tabular}
\end{center}
\caption{Most significant nuisance parameters (in terms of change of the fit uncertainty) for the combined fit of \bQ\ analysers.}
\label{tab:top5_NPs_width_bQ}
\end{table}
 

\chapter{\texorpdfstring{$\Delta \phi$}{Delta Phi} for Different MC Generators}
\label{sec:app_dphi_generators}
The SM prediction of the \dphi\ distributions varies for different generators. To separate the effects of reconstruction and modeling of the hard scattering process, parton level results are shown in Figure \ref{fig:dphi_gen_nom} without any selection cuts. The same plots without the sample of uncorrelated \ttbar\ pairs and a zoomed ratio is shown in Figure \ref{fig:dphi_gen_nom_onlySM}. It is known that the top quark \pt\ distributions, which varies for the generators (see Figure \ref{fig:truth_top_pt}). Hence, the samples were reweighted to the top quark \pt\ spectrum of \mcatnlo. The result is shown in Figure \ref{fig:dphi_gen_rew_onlySM}. The good agreement between the top quark \pt\ spectrum of \powheg+\pythia\ and \mcatnlo+\herwig\ is visible as well as a residual effect, independent of the top quark \pt. 

\begin{figure}[htbp]
\begin{center}
\subfigure[]{
\includegraphics[width=0.45\textwidth]{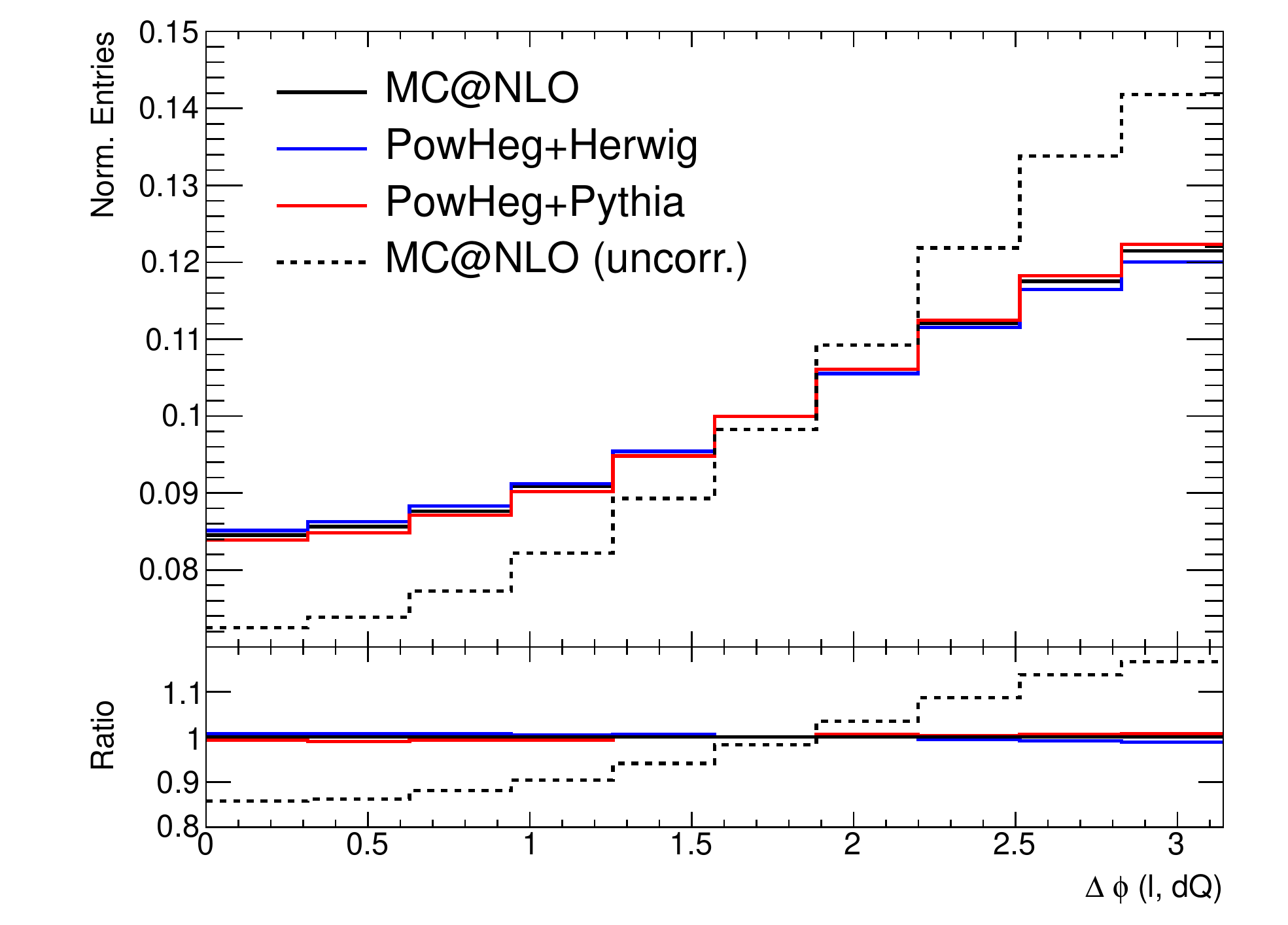}
\label{fig:dphi_gen_dQ_nom}
}
\subfigure[]{
\includegraphics[width=0.45\textwidth]{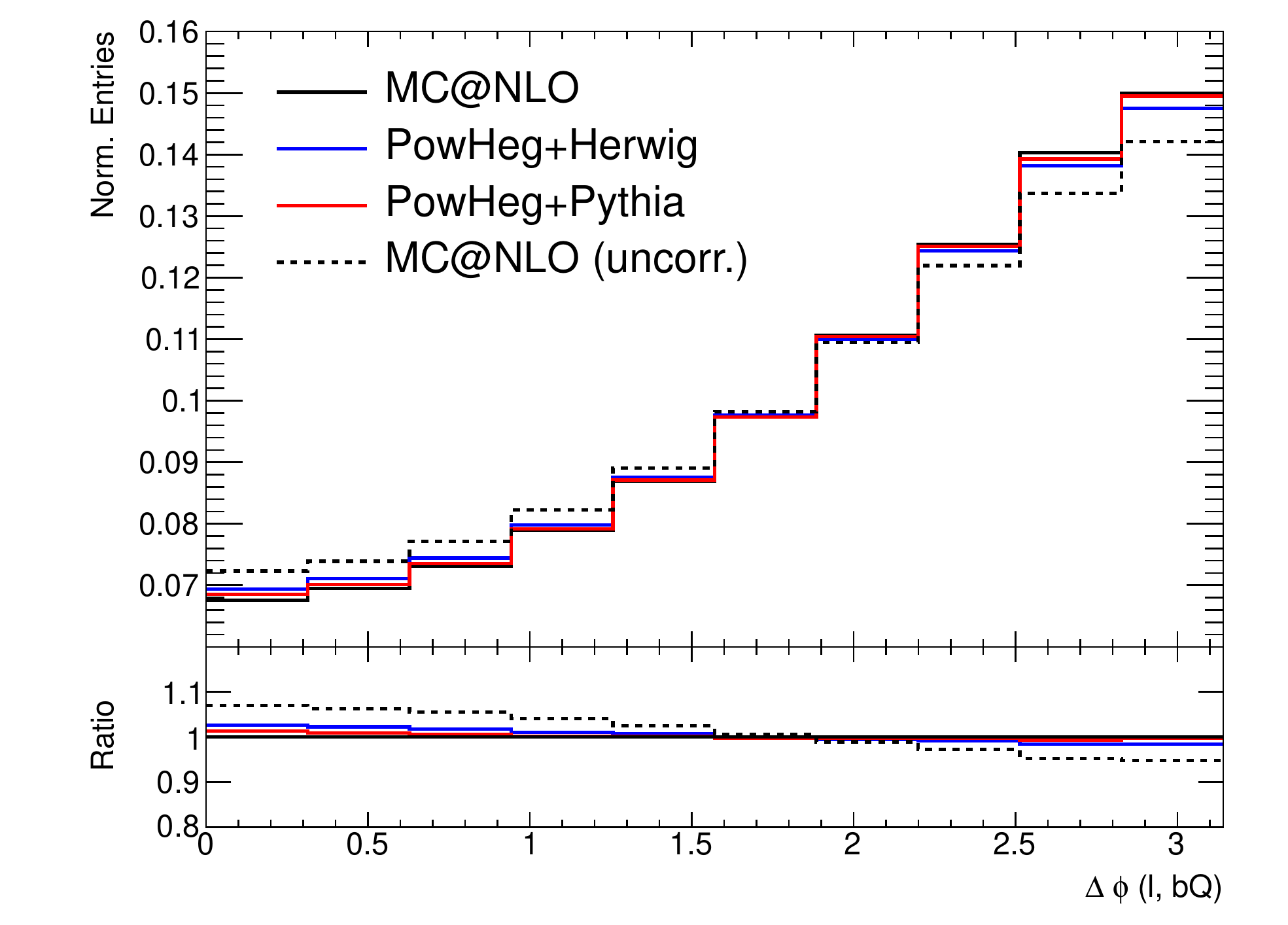}
\label{fig:dphi_gen_bQ_nom}
}
\end{center}
\caption{SM predictions of the \dphi\ distributions on parton level using different MC generators. The charged lepton was used as spin analyzer together with the \subref{fig:dphi_gen_dQ_nom} \dQ\ and \subref{fig:dphi_gen_bQ_nom} \bQ. The variations are compared to the \mcatnlo\ sample using uncorrelated \ttbar\ pairs.}
\label{fig:dphi_gen_nom}
\end{figure}

\begin{figure}[htbp]
\begin{center}
\subfigure[]{
\includegraphics[width=0.45\textwidth]{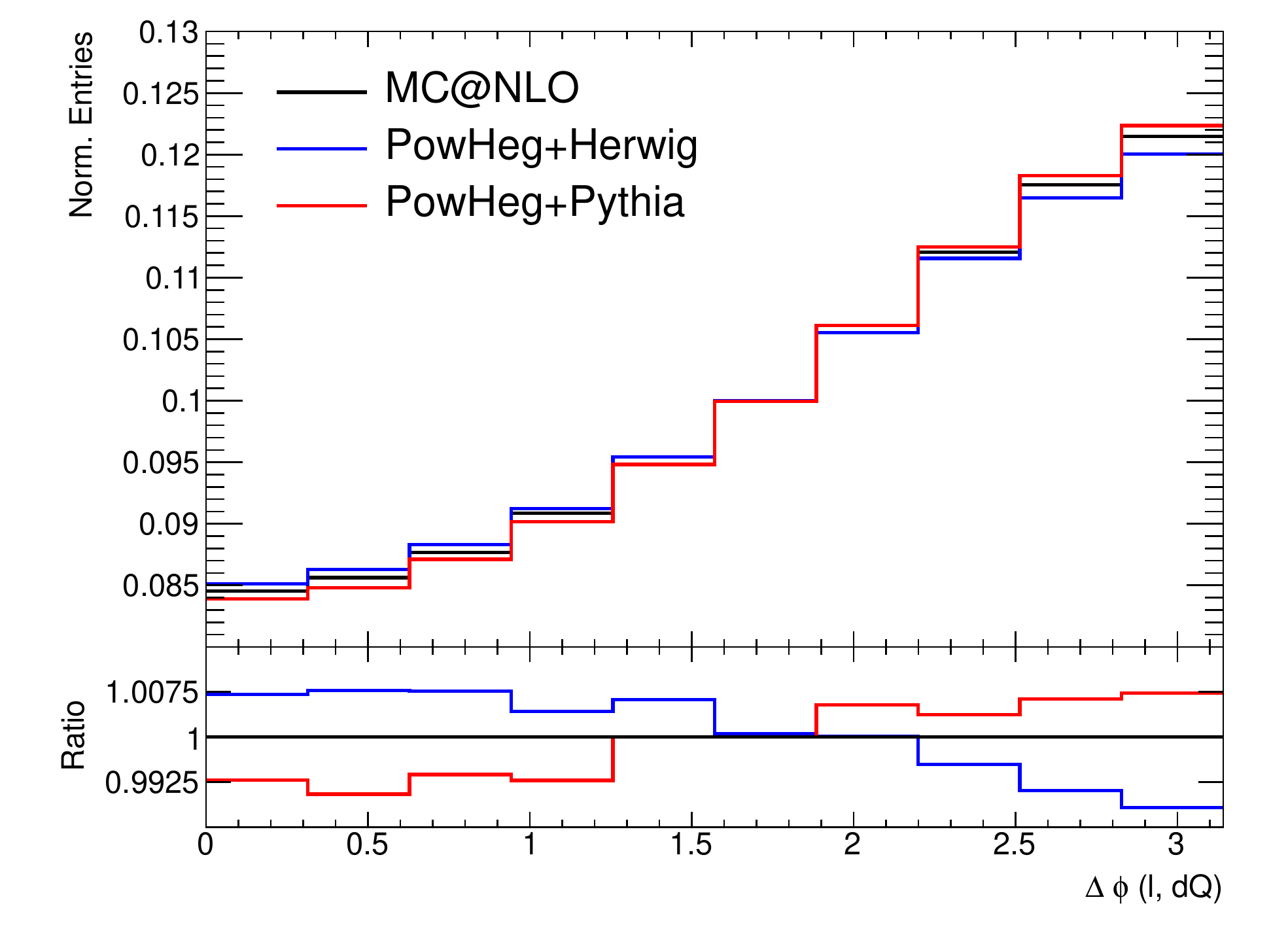}
\label{fig:dphi_gen_dQ_nom_onlySM}
}
\subfigure[]{
\includegraphics[width=0.45\textwidth]{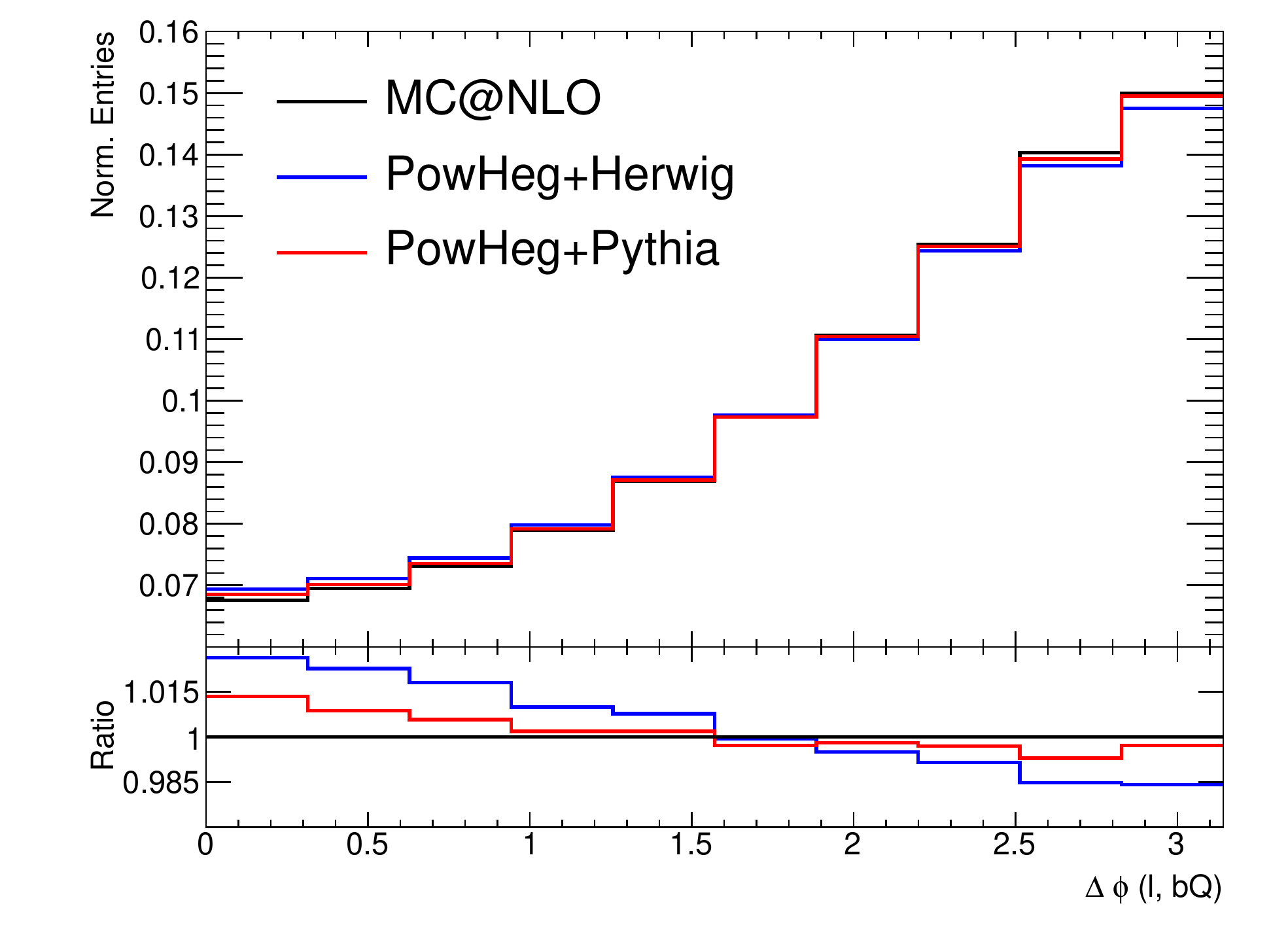}
\label{fig:dphi_gen_bQ_nom_onlySM}
}
\end{center}
\caption{SM predictions of the \dphi\ distributions on parton level using different MC generators. The charged lepton was used as spin analyzer together with the \subref{fig:dphi_gen_dQ_nom_onlySM} \dQ\ and \subref{fig:dphi_gen_bQ_nom_onlySM} \bQ.}
\label{fig:dphi_gen_nom_onlySM}
\end{figure} 

\begin{figure}[htbp]
\begin{center}
\subfigure[]{
\includegraphics[width=0.45\textwidth]{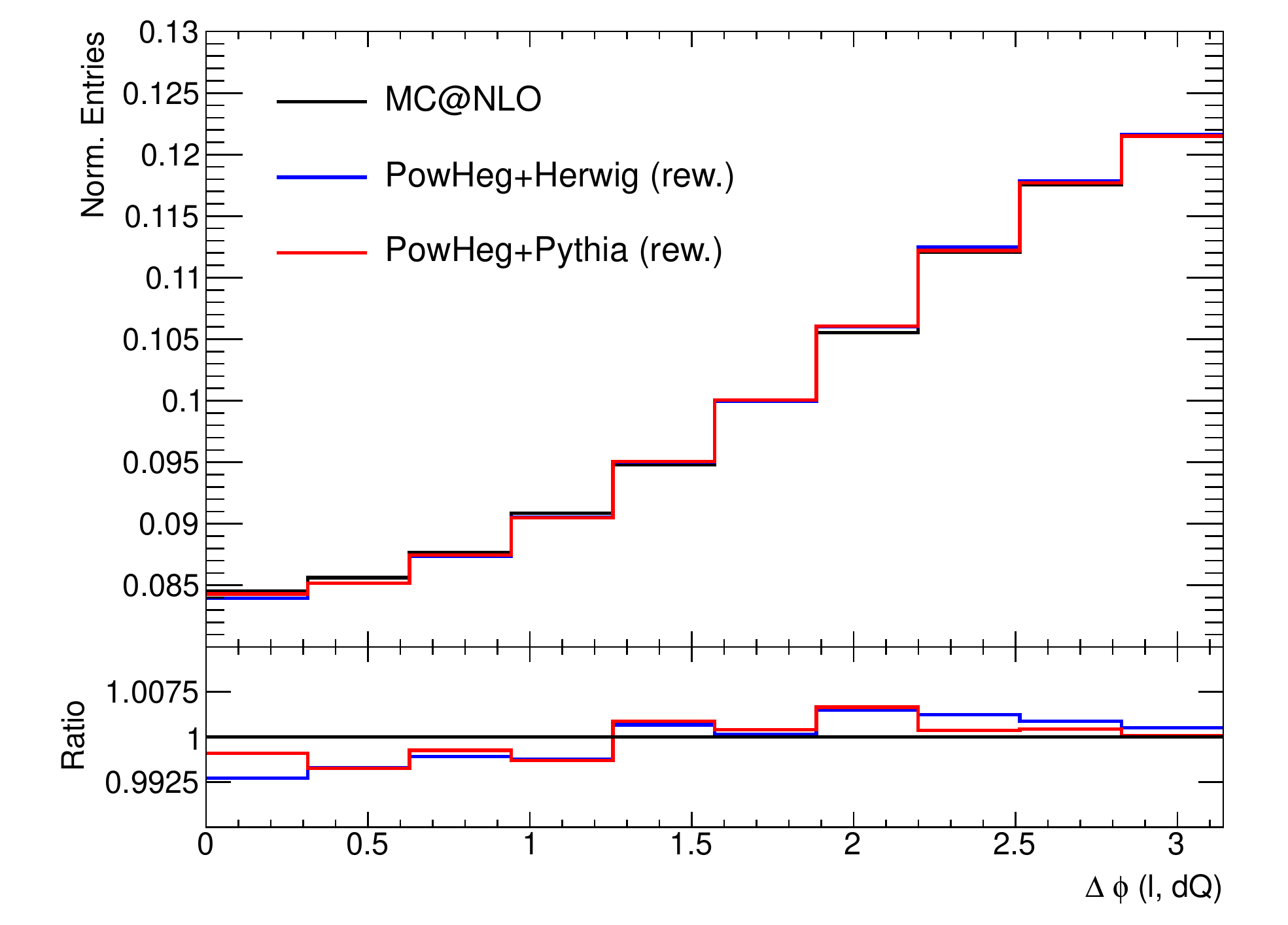}
\label{fig:dphi_gen_dQ_rew_onlySM}
}
\subfigure[]{
\includegraphics[width=0.45\textwidth]{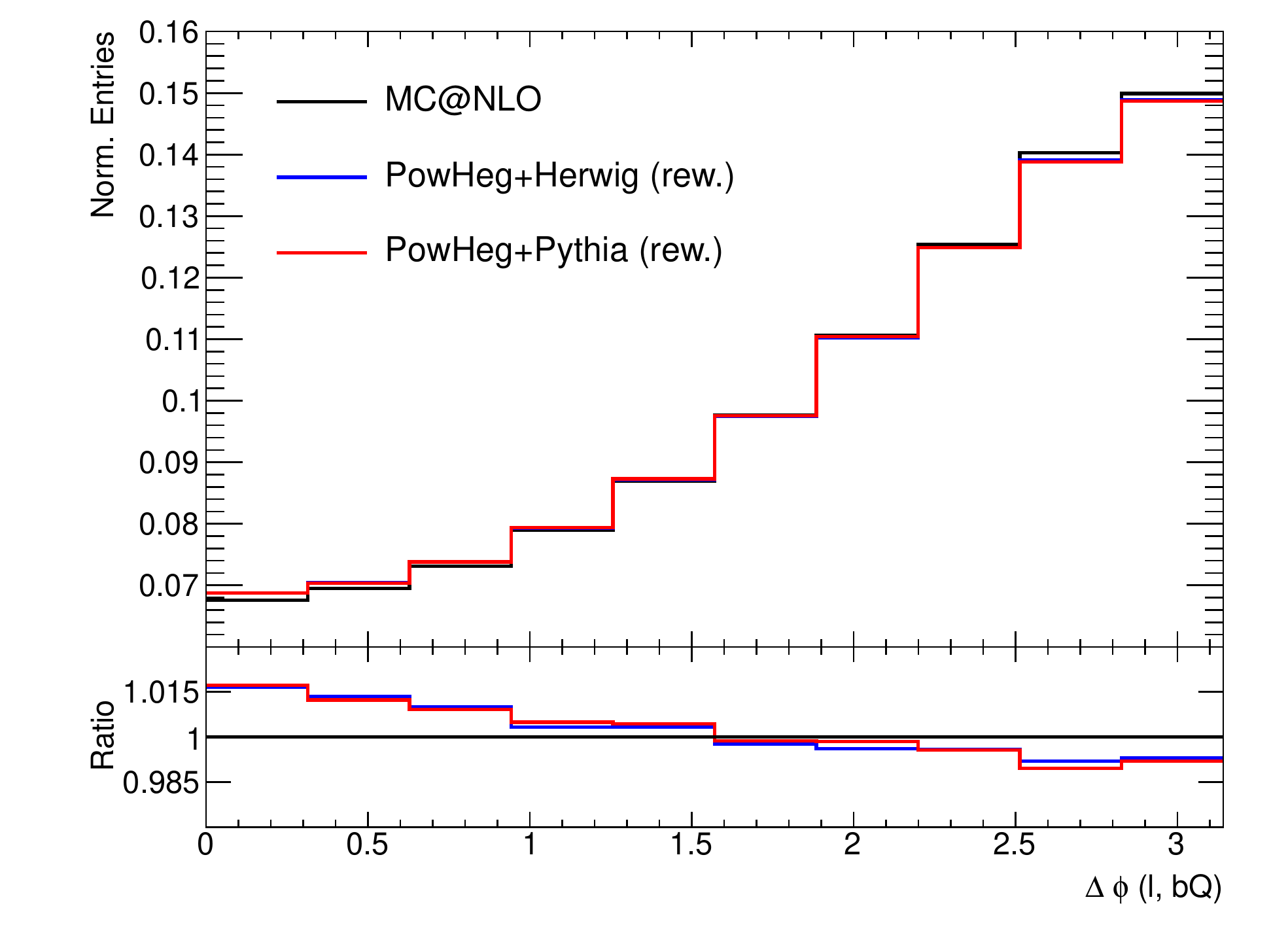}
\label{fig:dphi_gen_bQ_rew_onlySM}
}
\end{center}
\caption{SM predictions of the \dphi\ distributions on parton level using different MC generators. The charged lepton was used as spin analyzer together with the \subref{fig:dphi_gen_dQ_rew_onlySM} \dQ\ and \subref{fig:dphi_gen_bQ_rew_onlySM} \bQ. The samples were reweighted to match the top quark \pt\ spectrum of \mcatnlo.}
\label{fig:dphi_gen_rew_onlySM}
\end{figure} 
 

\chapter{Alternative \texorpdfstring{$t \bar{t}$}{Top/Anti-Top Quark} Modeling}
\label{sec:app_alt_models}

Table \ref{tab:exp_changes} shows several ``what if?'' scenarios. The modifications are applied to the pseudo data. The effects on the fitted \fsm\ results are shown. 

\begin{table}[htbp]
\begin{center}
\begin{tabular}{|c|c|c|}
\hline
Change in Pseudo Data &  \multicolumn{2}{c|}{ Fitted \fsm } \\
{} &  \dQ\ & \bQ\ \\
\hline
\hline
--- & 1.00 & 1.00 \\
\hline
Replacing \mcatnlo\ with \powheg+\herwig & 1.26 & 0.64 \\
\hline
Reweighting top \pt\ to measured spectrum & 1.17 & 0.76 \\
\hline
Reweighting PDF from CT10 to HERAPDF & 1.09 & 0.83 \\
\hline
\end{tabular}
\end{center}
\caption{Effects of changes on the default signal MC. The quoted results of \fsm\ were obtained by fitting to pseudo data created with the default MC and applied changes.}
\label{tab:exp_changes}
\end{table}

The changes must be read as 'If the data would be more like the suggested change, the following results for \fsm\ are expected'.
It is remarkable that all tested changes would explain a larger \fsm\ for the \dQ\ and a lower for the \bQ. This means on the other hand: replacing the fitting templates (and not the pseudo data) with a modified version would lead to a lower \fsm\ for the \dQ\ and a higher for the \bQ\ when fitting the data. 
 
\chapter{Jet Charge}
\label{sec:app_jetcharge}
Jets consist of tracks leaving signatures in the ID. There are several options to assign a charge to a jet. For example, the charge of the track with the highest \pt\ can be chosen. A track is part of a jet if $\Delta R\left( \text{track, jet}\right) < 0.25$.  It is also possible to create a weighted charge using all tracks and their momentum contribution to the total jet momentum. The weighted jet charge is determined via
\begin{align}
q_{\text{jet}} = \frac{\sum_i q_i \left| \vec{j} \cdot \vec{p}_i \right|^k}{\sum_i \left| \vec{j} \cdot \vec{p}_i \right|^k}
\end{align}
using the jet momentum vector $\vec{j}$, the momentum vectors $\vec{p}_i$ of all tracks of the jet, the track charges $q_i$ and a weighting factor $k$. The weighting factor was set to 0.5 as it was done in \cite{top_charge_ATLAS}. 

Tracks taken into account for the charge determination need to pass certain quality criteria. These are the following:
\begin{itemize}
\item Transverse momentum of the track must be at least 1 GeV.
\item The absolute value of the impact parameter in the transverse plane, $d_0$, must not be larger than $2\,\text{mm}$.
\item The absolute value of the distance to the primary vertex in z-direction, $z_0$, multiplied with the sine of the track angle to the z-axis must not be larger than 10 mm ($\left| z_0 \cdot sin\left( \theta\right)\right| \leq 2\,\text{mm}$).
\item The track fit quality must be sufficient ($\chi^2 / \text{nDOF} \leq 2.5$).
\item The track must have at least one hit in the pixel detector. 
\item The track must have at least six hits in the silicon tracker.
\end{itemize}
In the case where a jet contains no track, a jet charge of zero is assigned.

The separation of up and anti-up quarks was demonstrated in Figure \ref{fig:jetcharge_uQ}.
Another test is supposed to check the correct assignment of a $b$-jet to the leptonically decaying top quark. The charges of the $b$-jet and the charged lepton should be of opposite sign. The distributions of the squared sum of the lepton and the presumed $b$-jet from the leptonically decaying top is shown in Figure \ref{fig:jetcharge_blep_opt}. A preference of opposite sign charges in case of a matched jet is visible. 
\begin{figure}[htbp]
\begin{center}
\subfigure[]{
\includegraphics[width=0.45\textwidth]{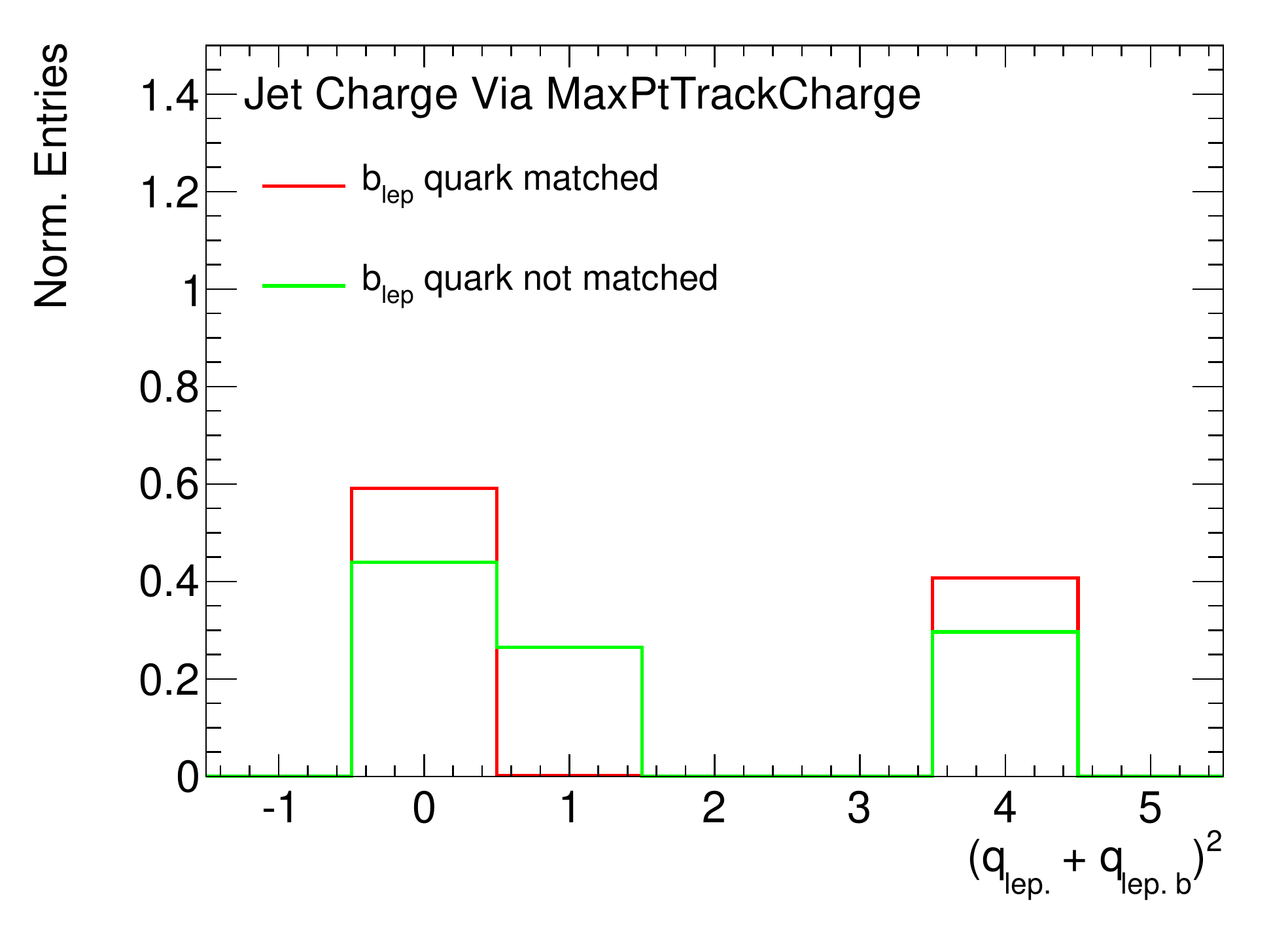}
\label{fig:jetcharge_blep_opt_max}
}
\subfigure[]{
\includegraphics[width=0.45\textwidth]{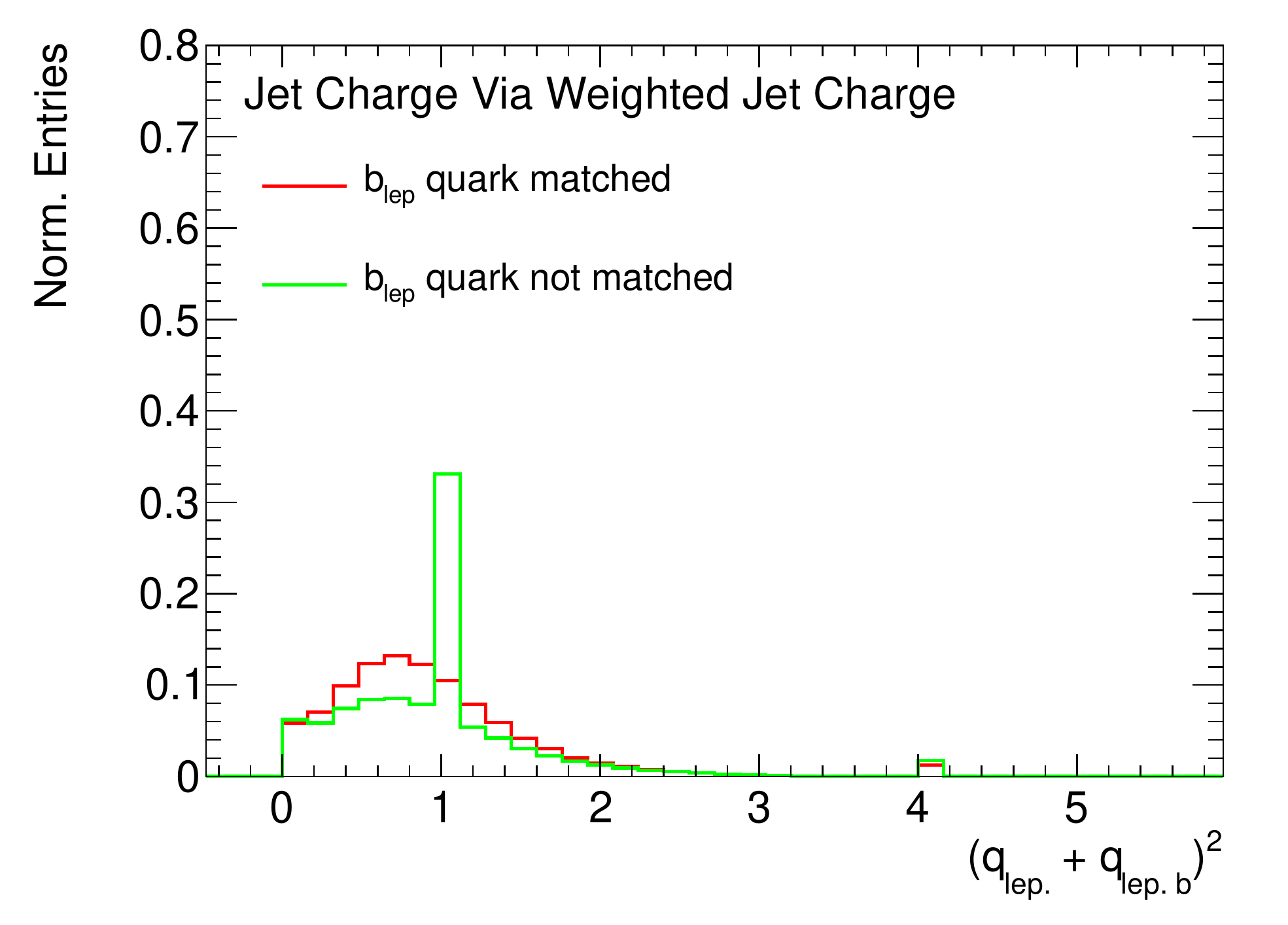}
\label{fig:jetcharge_blep_opt_weighted}
}
\end{center}
\label{fig:jetcharge_blep_opt}
\caption{Squared sum of the charges of the charged lepton and the $b$-quark from the leptonically decaying top quark using \subref{fig:jetcharge_blep_opt_max} the charge of the jet track with the highest \pt\ and \subref{fig:jetcharge_blep_opt_weighted} the weighted charge using all tracks. }
\end{figure} 
Finally, a test concerning the correct reconstruction of the hadronically decaying top quark was made. The quantity $\chi \equiv \left(q_{dQ} + q_{uQ} \right) \cdot \left(q_{dQ} + q_{uQ} - q_{had. b} \right)$ is plotted in Figure \ref{fig:jetcharge_thad_opt}. It uses the charges of the light up- and down-type jets as well as the charge of the $b$-jet of the hadronically decaying top quark. A correct assignment of all jets and a correct jet charge determination should lead to a maximized $\chi$. This is due to the same sign of the light jet charges and the opposite sign of the light jet and $b$-jet charges. 
For fully matched hadronically decaying top quarks a trend to high values of $\chi$ is clearly visible. 

\begin{figure}[htbp]
\begin{center}
\subfigure[]{
\includegraphics[width=0.45\textwidth]{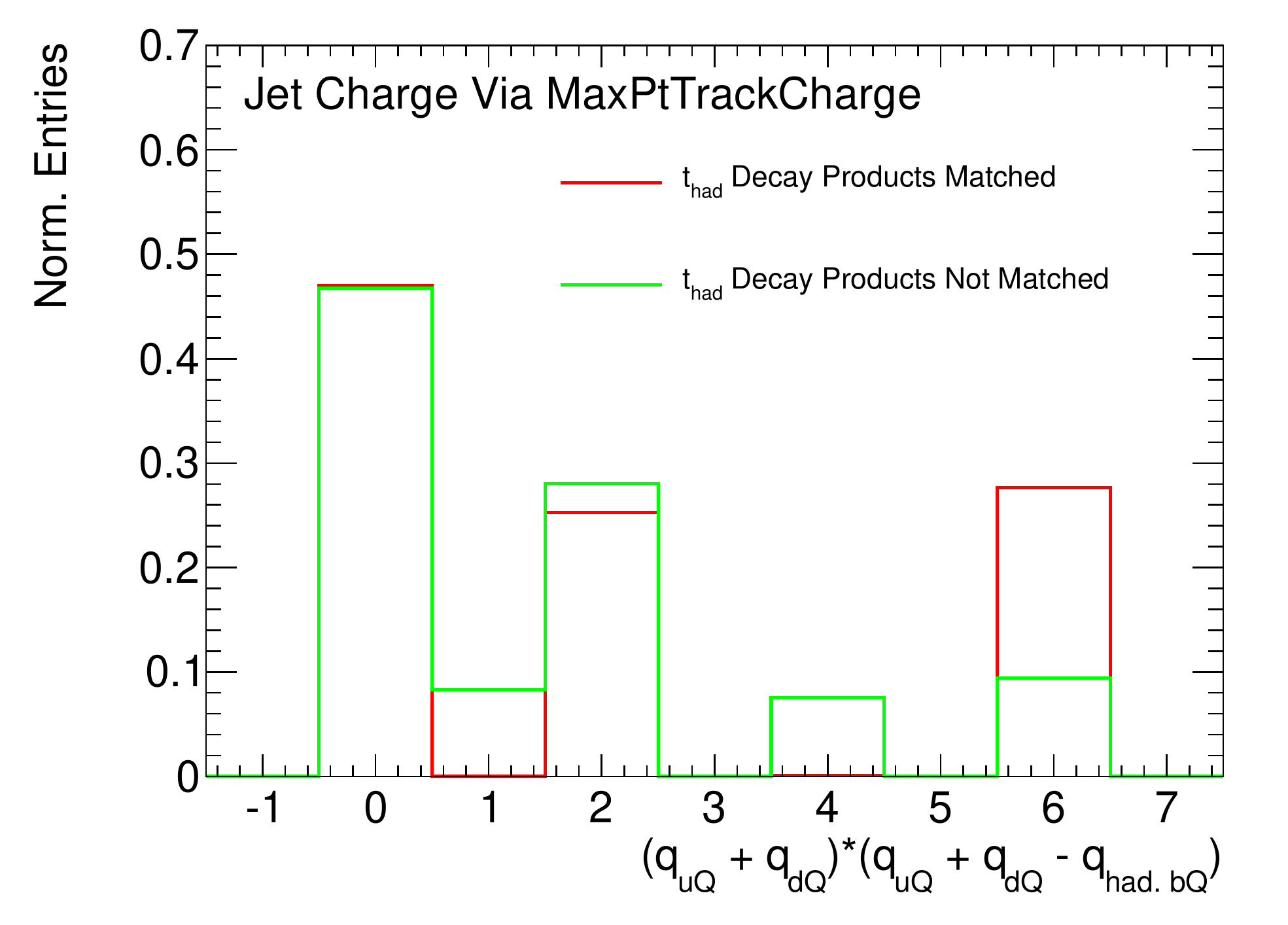}
\label{fig:jetcharge_thad_opt_max}
}
\subfigure[]{
\includegraphics[width=0.45\textwidth]{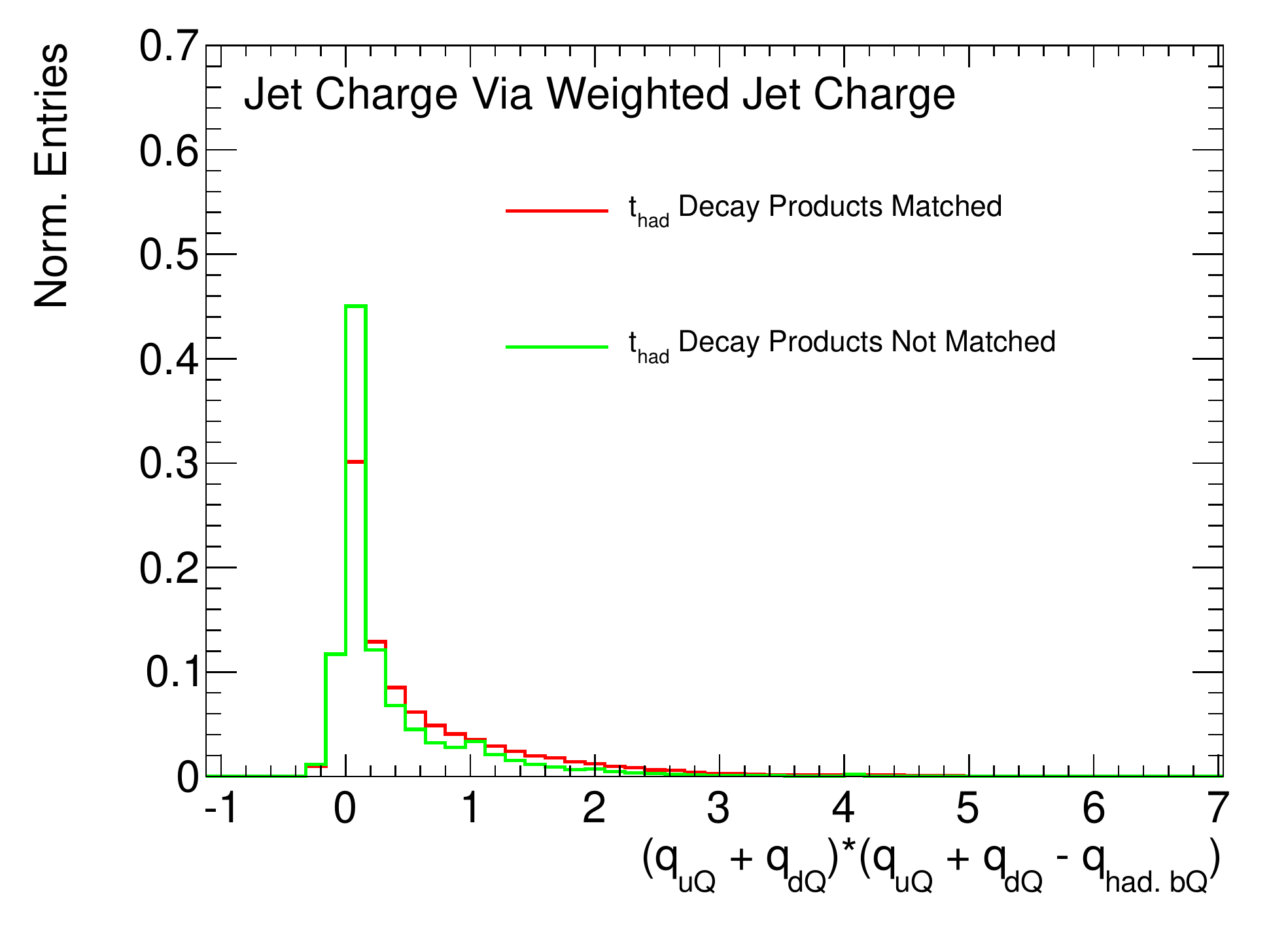}
\label{fig:jetcharge_thad_opt_weighted}
}
\end{center}
\caption{Distribution of the quantity $\left(q_{dQ} + q_{uQ} \right) \cdot \left(q_{dQ} + q_{uQ} - q_{had. b} \right)$ involving the charges of the light up- and down-type quark jet and the $b$-jet from of the hadronically decaying top quark. The used jet charges were \subref{fig:jetcharge_thad_opt_max} the charge of the jet track with the highest \pt\ and \subref{fig:jetcharge_thad_opt_weighted} the weighted charge using all tracks. }
\label{fig:jetcharge_thad_opt}
\end{figure} 

\cleardoublepage
\addcontentsline{toc}{part}{CV}


\section*{Supplementary material}

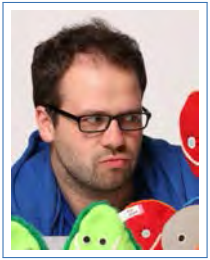

# Boris Lemmer

*Curriculum Vitae*

| | |
|---|---|
| Date of Birth | 16.06.1984 |
| Place of Birth | Gießen |
| Nationality | German |

## Education

| | |
|---|---|
| 2012 | **CERN**, *Meyrin*, Research Stay. |
| 2010-2014 | **Georg-August-Universität**, *Göttingen*, PhD Studies, Supervisor: Prof. A. Quadt.<br>Qualification: Dr. rer. nat. |
| Winter 2007/08 | **Umeå Universitet**, *Umeå*, Semester Abroad. |
| 2006-2008 | **Justus-Liebig-Universität**, *Gießen*, Mathematics Diploma Program, Minor subjects: Numerical Analysis and Experimental Physics |
| 2004-2009 | **Justus-Liebig-Universität**, *Gießen*, Physics Diploma Program, Minor subjects: Mathematics and Nuclear Physics.<br>Qualification: Dipl.-Phys. |
| 1994-2003 | **Landgraf-Ludwigs-Gymnasium**, *Gießen*, High School.<br>Qualification: Allgemeine Hochschulreife |

## Diploma Thesis

| | |
|---|---|
| Title | *Measurement of the Excitation Function of $\omega$ Photoproduction on Carbon and Niobium* |
| Referees | Prof. V. Metag, Prof. U. Mosel |
| Description | Indications for possible mass and width modifications of the $\omega$ meson embedded in a hadronic medium were investigated by measuring the photoproduction cross section as a function of the incoming beam energy. |

## Awards and Scholarships

| | |
|---|---|
| 2011 | Foundation Council Award of the Georg-August-Universität Göttingen - Category: Science and Publicity |
| 2011-2013 | PhD Scholarship from the German National Academic Foundation |
| 2010 | Participant of the Lindau Nobel Laureate Meeting |
| 2008-2009 | Scholarship from the German National Academic Foundation |

## Teaching

| | |
|---|---|
| 2012 | **Lecturer / Dozent**, *Teilchenphysik*, B. Lemmer, CERN. As part of the German Teachers Programme |
| WS 2010/11 | **TA / Übungsgruppenleiter**, *Einführung in die Kern- und Teilchenphysik*, A. Frey, Göttingen University. |
| SS 2010 | **TA / Übungsgruppenleiter**, *Atomphysik*, A. Quadt, Göttingen University. |
| WS 2009/10 | **Lecturer / Dozent**, *Physik für Medizinisch-Techn. Assistenten*, M. Thiel / B. Lemmer, Gießen University Medical Center. |
| WS 2009/10 | **Lab Assistant / Praktikumsbetreuer**, *Physikpraktikum für Mediziner*, R. Novotny, Gießen University. |
| SS 2009 | **Lab Assistant / Praktikumsbetreuer**, *Physikpraktikum für Mediziner*, R. Novotny, Gießen University. |
| SS 2009 | **TA / Korrekteur**, *Analysis II für Lehramtsstudenten*, M. Väth, Gießen University. |
| WS 2008/09 | **Lab Assistant / Praktikumsbetreuer**, *Physikpraktikum für Mediziner*, R. Novotny, Gießen University. |
| WS 2008/09 | **TA / Korrekteur**, *Mathematik für Physiker I*, B. Lani-Wayda, Gießen University. |
| SS 2008 | **TA / Übungsgruppenleiter**, *Theorie der höheren Mechanik*, U. Mosel, Gießen University. |
| SS 2008 | **TA / Übungsgruppenleiter**, *Analysis II*, B. Lani-Wayda, Gießen University. |
| WS 2007/08 | **TA / Korrekteur**, *Analysis I*, B. Lani-Wayda, Gießen University. |
| SS 2007 | **TA / Tutor**, *Experimentalphysik II*, V. Metag, Gießen University. |
| SS 2007 | **TA / Korrekteur**, *Analysis IIIB*, B. Lani-Wayda, Gießen University. |
| WS 2006/07 | **TA / Korrekteur**, *Analysis IIIA*, B. Lani-Wayda, Gießen University. |

## Schools, Conferences and Workshops

| | |
|---|---|
| 2014 | Talk at the Spring Meeting of the German Physical Society, Mainz |
| 2014 | Poster at LHCC Meeting at CERN, Meyrin |
| 2013 | Talk at the Phenomenology 2013 Symposium, Pittsburgh |
| 2013 | Talk at the Spring Meeting of the German Physical Society, Dresden |
| 2012 | 5th International Workshop on Top Quark Physics, Winchester |
| 2012 | Talk at the Spring Meeting of the German Physical Society, Göttingen |
| 2011 | Spring Meeting of the German Physical Society, Karlsruhe |

| 2011 | CTEQ Summer School, Madison |
|------|------------------------------|
| 2010 | Talk at the Spring Meeting of the German Physical Society, Bonn |
| 2009 | Quarkonium Working Group Meeting, Nara |
| 2009 | Talk at the Lecture Week of the Graduate School Hadrons and Nuclei, Copenhagen |
| 2009 | Talk at the Crystal Ball Collaboration Meeting, Edinburgh |
| 2009 | Summer School of the German National Academic Foundation, Rot an der Rot |
| 2009 | Lecture Week of the Graduate School Hadrons and Nuclei, Turin |
| 2009 | DESY Summer School, Hamburg |

## Publications

- The ATLAS Collaboration, Measurements of spin correlation in top-antitop quark events from proton-proton collisions at $\sqrt{s} = 7$ TeV using the ATLAS detector, Submitted to Phys. Rev. D
- J. Erdmann et al., A likelihood-based reconstruction algorithm for top-quark pairs and the KLFitter framework, Nuclear Instruments and Methods in Physics Research Section A 748 (2014) 18
- B. Lemmer, Bis(s) ins Innere des Protons, Springer Spektrum (2013)
- M. Thiel et al., In-medium modifications of the $\omega$ meson near the production threshold, Eur. Phys. J. A (2013) 49: 132
- The ATLAS Collaboration, Observation of Spin Correlation in top/antitop Events from $pp$ Collisions at $\sqrt{s} = 7$ TeV Using the ATLAS Detector, Phys. Rev. Lett. 108 (2012) 212001
- V. Metag et al., Experimental approaches for determining in-medium properties of hadrons from photo-nuclear reactions, Progress in Particle and Nuclear Physics, Volume 67, Issue 2 (2012) 530

## Miscellaneous

| 2011 | German Science Slam Champion |
|------|------------------------------|
| 2006-2010 | Town Municipal (Stadtverordneter) in Staufenberg (Hesse) |
| 2010-2014 | Several Outreach Activities (Rent-a-Scientist, Physikshow "Zauberhafte Physik", CERN Tour Guide) |

## Languages

| German | **Native speaker** |
|--------|--------------------|
| English | **Fluent** |
| Swedish | **Basic communication skills** |



\end{document}
\end